 \def\includebiggraphics#1 {}  % big graphics OFF

\documentclass{JHEP3}
\usepackage[all]{xy}
\usepackage{latexsym,graphics,epsfig}
\usepackage{rotate}
\usepackage{pb-diagram}
\usepackage{pst-grad}
\usepackage{pstricks,}
\usepackage{pst-plot}
\usepackage[tiling]{pst-fill}
\usepackage{amsfonts, amssymb, amsmath}
\title{From Loop Space Mechanics to Nonabelian Strings}

\author{Urs Schreiber 
\\ Fachbereich Physik
\\ Universit{\"a}t Duisburg-Essen 
\\ Essen, 45117, Germany
\\ E-mail: \email{Urs.Schreiber@uni-essen.de}}
\newcommand{\R}{{\mathbb R}}
\newcommand{\C}{{\mathbb C}}
\newcommand{\N}{{\mathbb N}}
\newcommand{\Z}{{\mathbb Z}}

\def\bigdisunion {\bigsqcup}

\newcommand{\maps}{\colon}
\def\ker {{\rm ker}}
\def\image {{\rm im}}

\def\stackto#1 { \, {\stackrel{#1}{\longrightarrow}}\, }
\def\stackTo#1 { {\stackrel{#1}{\Longrightarrow}} }
\newcommand{\To}{\Longrightarrow}

\newcommand{\id}{{\rm id}}
\newcommand{\tr}{{\rm tr}}
\newcommand{\ad}{{\rm ad}}
\newcommand{\inv}{{\rm inv}}

\newcommand{\U}{{\rm U}}
\renewcommand{\u}{\mathfrak{u}}
\newcommand{\SO}{{\rm SO}}

\newcommand{\SU}{{\rm SU}}
\newcommand{\Aut}{{\rm Aut}}
\newcommand{\AUT}{\mathcal{AUT}}

\newcommand{\Spin}{{\rm Spin}}
\newcommand{\String}{{\rm String}}

\newcommand{\LG}{{\cal L}_kG}

\newcommand{\PG}{{\cal P}_kG}
\newcommand{\Pg}{{\cal P}_k\g}
\newcommand{\OG}{\Omega G}
\newcommand{\Og}{\Omega \mathfrak{g}}
\newcommand{\wOG}{\widehat{\Omega G}}
\newcommand{\wOg}{\widehat{\Omega \mathfrak{g}}}
\newcommand{\wOkG}{\widehat{\Omega_k G}}
\newcommand{\wOkg}{\widehat{\Omega_k \g}}
\newcommand{\cG}{\mathcal{G}}

\newcommand{\g}{\mathfrak{g}}
\newcommand{\h}{\mathfrak{h}}
\renewcommand{\u}{{\mathfrak u}}
\newcommand{\lietwoalg}{{\mathfrak c}}
\newcommand{\lietwogrp}{C}

\def\semidir {\rtimes}

\newcommand{\Set}{\mathrm{Set}}
\newcommand{\Top}{\mathrm{Top}}
\newcommand{\Cat}{\mathrm{Cat}}
\newcommand{\Grp}{\mathrm{Grp}}
\newcommand{\Lie}{\mathrm{Lie}}
\newcommand{\Alg}{\mathrm{Alg}}
\newcommand{\Diff}{\mathrm{Diff}}
\newcommand{\twoDiff}{2C^\infty}
\newcommand{\Hilb}{\mathrm{Hilb}}
\newcommand{\Vect}{\mathrm{Vect}}
\def\LMod #1{{#1}\!-\!\mathrm{Mod}}
\def\RMod #1{\mathrm{Mod}\!-\!{#1}}
\def\BiMod #1#2{{#1}\!-\!\mathrm{Mod}\!-\!{#2}}
\newcommand{\Ch}{\mathrm{\mathbf{Ch}}}
\def\Derived #1{\mathbf{D}\of{{#1}}}

\def\GTor {G\!\!-\!\!\mathrm{Tor}}
\def\GTwoTor {G_2\!-\!2\mathrm{Tor}}

\newcommand{\Ob}{\mathrm{Ob}}
\newcommand{\Mor}{\mathrm{Mor}}

\def\paths {P}

\renewcommand{\P}{\mathcal{P}}

\def\con {{\rm con}}

\def\hol {{\rm hol}}

\def\trans {{\rm Trans}}

\def\A {\mathbf{a}}
\def\C {\mathbf{c}}

\def\p {\mathfrak{p}}
\def\g {\mathfrak{g}}
\def\f {\mathfrak{f}}

\def\covering {{\mathcal{U}}}

\renewcommand{\a}{\alpha}
\renewcommand{\b}{\beta}
\renewcommand{\c}{\gamma}

\newcommand{\B}{{\cal B}}
\newcommand{\T}{{\cal E}}
\newcommand{\Der}{{\rm Der}}

\newcommand{\TT}{{\U(1)}}

\def\Cech {{{\v C}ech }}

\def\annotation #1 {\marginpar{\footnotesize {#1}}}

   \def\twogroup {\mathcal{G}}

   \def\lietwoalgebra {\mathcal{L}}

   \def\crossedmodule {\mathcal{C}}

   \def\manifold {{M}}

   \def\twoB {{B}}

   \def\twoF {{F}}

   \def\twoU {{U}}

   \def\twoP {{P}}

\newtheorem{theorem}{Theorem}[section] 
\newtheorem{definition}{Definition}[section]{}
{}
\newtheorem{proposition}{Proposition}[section]
\newtheorem{lemma}[theorem]{Lemma}

\newtheorem{example}[theorem]{Example}

\newtheorem{corollary}[theorem]{Corollary}

\newcommand{\et}{\hspace{-0.08in}{\bf .}\hspace{0.1in}}

\makeatletter
\newif\ifmydraft
 \mydraftfalse
\def\joburl {http://www-stud.uni-essen.de/~sb0264/\jobname .pdf}
\def\clock{%
  \@tempcnta=\time\relax\@tempcntb=\@tempcnta\divide\@tempcnta by60
  \edef\hour{\ifnum\@tempcnta<10\relax0\fi\number\@tempcnta}
  \multiply\@tempcnta by60\advance\@tempcntb by-\@tempcnta
  \edef\minute{\ifnum\@tempcntb<10\relax0\fi\number\@tempcntb}
}
\def\@oddhead{%
  \name{pag\thepage}
  \ifmydraft
    \hbox to0pt{\tiny draft last modified \number\month/\number\day/\number\year\ --\ \hour:\minute,
      current version at {\tt \href{\joburl}{\joburl}}}
    \hbox to0pt{\hskip-\oddsidemargin\vbox to0pt{\vskip-\topmargin\parbox[t]{210mm}{\epsfxsize210mm%
           \vss\vskip-1.5ex\epsffile{draft.eps}}}\hss}\hfil\if@draft\copy\drft@box\fi 
  \fi
}
\makeatother
\clock
\renewcommand{\Im}{\mathrm{Im}}

\def\comment #1{}
\def\cf {{\it cf. }}

\def\refer #1{{(\ref{#1})}}
\def\fullref #1{\ref{#1} (p.\pageref{#1})}
\def\refdef #1{{(def. \ref{#1})}}

\def\bra #1{\left\langle{#1}\right|}
\def\ket #1{\left|{#1}\right\rangle}
\def\bracket #1#2{\left\langle{#1}|{#2}\right\rangle}

\def\of #1{\!\left({#1}\right)}

\def\grad {\nabla}
\def\gradOp {\hat\grad}

\def\set #1{\left\lbrace{#1}\right\rbrace}
\def\id {\mathrm{id}}

\def\brackets #1{\left[{#1}\right]}
\def\braces #1{\left\lbrace{#1}\right\rbrace}

\def\order #1{\mathcal{O}\of{#1}}

\def\commutator #1#2{\brackets{{#1},{#2}}}
\def\antiCommutator #1#2{\braces{{#1},{#2}}}
\def\superCommutator #1#2{\brackets{{#1},{#2}}_\involution}

\def\adjoint #1{{{{#1}^{\dagger}}}}

\def\defas {\equiv}
\def\shallbe {\stackrel{!}{=}}
\def\equalby #1{\stackrel{\refer{#1}}{=}}

\def\iso{\cong}

\def\isomorphic {\simeq}

\def\onbComp {\mathrm{e}}
\def\onbAnnihilator {{\mbox{$\hat \onbComp$}}}
\def\onbCreator {\mbox{$\adjoint{\onbAnnihilator}$}}
\def\coordAnnihilator {{\mbox{$\hat {\mathrm{c}}$}}}
\def\coordCreator {{\adjoint{\coordAnnihilator}}}

\def\extd {\mathbf{{d}}}

\def\coextd {\adjoint{\extd}}

\def\involution {\mathbf \iota}

\def\Dirac {\mathbf{D}}

\def\endofproof {\hfill$\Box$\newline\newline}

\def\eigenspace #1#2 {\mathrm{eig}\of{#1,#2}}

\def\fatDelta {\mbox{\boldmath$\Delta$}}

\newlength{\skiplength}
\def\skiph #1{\settowidth{\skiplength}{#1}\hspace{\skiplength}}

\def\inner {\!\cdot\!}

\abstract{
  Lifting supersymmetric quantum mechanics
  to loop space
  yields the superstring. 
  A particle charged under 
  a fiber bundle thereby turns into a string
  charged under a 2-bundle, or gerbe. 
  This stringification is nothing but
  categorification.  \\
  We 
  look at supersymmetric quantum mechanics on loop space and
  demonstrate
  how deformations here give rise to
  superstring background fields and boundary states, and,
  when generalized, to local nonabelian connections on loop
  space.
  In order to get a global description of these connections we
  introduce and study categorified global holonomy in the
  form of 2-bundles with 2-holonomy.
  We show how these relate
  to nonabelian gerbes and go beyond by obtaining
  global nonabelian surface holonomy, thus providing
  a class of action functionals for nonabelian strings.
  The examination of the differential formulation,
  which is adapted to the study of nonabelian
  $p$-form gauge theories, gives rise to
  generalized nonabelian Deligne hypercohomology.
  The (possible) relation of this to
  strings in Kalb-Ramond backgrounds,
  to M2/M5-brane systems,
  to spinning strings and to the derived category description of
  D-branes is discussed. In particular, there is a 2-group related
  to the
  $\String$-group which should be the right structure 2-group for
  the global description of spinning strings.
  \\
  $\,$\\
  \hspace{4cm }(July 2005)
}

\begin{document}

\addtocontents{toc}{\setcounter{page}{2}}
\addtocontents{toc}{\vspace{-2.7em}}

\newpage

\parbox{6cm}{
  \flushleft
  ``Stringification is the conversion of an object to a string [\dots].''
  \\
{\it F. Ribeiro} (programmer) \cite{Ribeiro:2005}
}

\part{Overview}
\label{Introduction}

In modern formal theoretical physics a certain idea has been found to be 
very fruitful: {\bf stringification}. 
Whenever one faces a theory describing particles, one may ask
if this theory arises as a limit of a theory where the particles are
really one-dimensional strings stretching between their endpoints.

In modern mathematics a certain idea has been found to be very fruitful:
{\bf categorification}. 
Whenever one faces a theory of some algebraic structure
describing certain objects, one may ask if this 
lifts to a structure where objects are replaced by morphisms going between
their source and target.

The domains of applicability of these two procedures have a nontrivial 
intersection where the physics of particles is described by algebra.

This happens in particular when (supersymmetric) quantum mechanics is 
formulated  in terms of spectral triples in Connes' 
noncommutative spectral geometry (NCG) \cite{Connes:1994}.

Here the configuration space of the particle
is encoded in the algebra $A$ of (complex valued) 
continuous functions over it. 
This is represented as an operator algebra on a graded Hilbert space 
$\mathcal{H}$,
whose elements describe states of the particle. On this space
is defined an odd-graded nilpotent operator $D$
(the ``Dirac operator'' or
``supercharge'') which encodes the dynamics of the particle.

This picture of (supersymmetric) quantum mechanics as well its
suggestive relation to the RNS superstring,
which was pointed out in the second halfs of \cite{Witten:1982,Witten:1985},
has been particularly emphasized in
\cite{FroehlichGrandjeanRecknagel:1996,FroehlichGrandjeanRecknagel:1997}.
It was noted that superstring dualities find a natural formulation in
terms of spectral geometry
\cite{LizziSzabo:1997,LizziSzabo:1998,Lizzi:1999}
and further hints for a deeper conceptual rooting of perturbative superstrings
in spectral noncommutative geometry were discussed for instance in
\cite{Chamseddine:1997,Chamseddine:1997b}. 

Attention to these arguably more conceptual ideas was soon 
dwarfed by the popularity gained by the noncommutative aspect of
NCG that was eventually realized to be ubiquitous in
string theory: Open strings in Kalb-Ramond backgrounds
were found to give rise to noncommutative field theories
\cite{DouglasHull:1998,SeibergWitten:1999} and matrix theory formulations of
nonperturbative string dynamics
\cite{BanksFischlerShenkerSusskind:1996,ConnesDouglasSchwarz:1998,Banks:1999}
resolved smooth spaces by noncommutative matrix algebras.
Last not least, string field theory with its noncommutative
star product had long been regarded as a manifestation of noncommutative
geometry in string theory \cite{Witten:1986}.

\vskip 1em

On the other hand there is more to noncommutative spectral geometry
than just noncommutativity (and in fact a better terminology would
be `\emph{not-necessarily commutative} spectral geometry').

For instance the fact
that the spectrum of the supercharge in supersymmetric quantum mechanics
contains important information about geometric properties of the systems's
configuration space (e.g. by way of Morse theory
\cite{Witten:1982} or index theorems) suggests that similarly for instance
the spectrum of the string's supercharge should contain interesting
information about the  configuration space of the string, which is a
{\bf loop space} (for closed strings) over target space.
And indeed \cite{Witten:1988} relates this index to elliptic cohomology.
This generalized form of cohomology seems alternatively to be obtainable from
ordinary `point geometry' by the method of categorification
\cite{BaasDundasRognes:1993,StolzTeichner:2004}.

It is at this point that one may reasonably suspect that there could
be a deeper principle behind lifting spectral geometry and
supersymmetric quantum mechanics from points to strings.

\addtocontents{toc}{\vspace{-0.7em}}

\clearpage

\section{Preliminaries}
\label{preliminaries of part I}

The term {\bf nonabelian strings} is supposed to make one think of
a generalization of the following situation
\cite{Baez:2002}:

An ordinary particle is a point
\[
\xymatrix{
   \bullet 
}
\]
which traces out a wordline as time goes by
\[
\xymatrix{
   \bullet \ar@/^1pc/[rr]
&& \bullet
}
\,.
\]
When the particle is charged, there is a connection in some bundle,
which, locally, associates a group element $g\in G$ to any such path
\[
\xymatrix{
   \bullet \ar@/^1pc/[rr]^{g}
&& \bullet
}
 \,.
\]
This happens in such a way that when the particle is transported
a little further
\[
\xymatrix{
   \bullet \ar@/^1pc/[rr]
&& \bullet \ar@/^1pc/[rr]
&& \bullet
}
\]
the composition of paths corresponds to multiplication of
group elements
\[
\xymatrix{
   \bullet \ar@/^1pc/[rr]^{g}
&& \bullet \ar@/^1pc/[rr]^{g'}
&& \bullet
}
\,.
\]
It is the associativity of the product in the group that
makes this procedure well defined over longer paths
\[
\xymatrix{
   \bullet \ar@/^1pc/[rr]^{g}
&& \bullet \ar@/^1pc/[rr]^{g'}
&& \bullet \ar@/^1pc/[rr]^{g''}
&& \bullet
}
\]
and the existence of inverses which corresponds to the reversal of paths
\[
\xymatrix{
   \bullet
&& \bullet \ar@/_1pc/[ll]_{g^{-1}}
}
\,.
\]
The theory of {\bf fiber bundles} with connection tells us how these local 
consideration fit into a global picture.
When the elements $g$ come from a nonabelian group this would be
the situation of a {\bf nonabelian particle} which we wish to generalize.

So suppose now that the particle is replaced by a string, which at one moment
in time itself already looks like this:
\[
\xymatrix{
   \bullet \ar@/^1pc/[rr]
&& \bullet
}\,,
\]
where the arrow is to remind us of some sort of orientation that we
might want to keep track of.
Now, as time goes by, this
piece of string traces out a worldsheet:
\[
\xymatrix{
   \bullet \ar@/^1pc/[rr]_{}="0"
           \ar@/_1pc/[rr]_{}="1"
           \ar@{=>}"0";"1"
&& \bullet
}
\,.
\]
Is it possible to associate some sort of algebraic object $f$ to this
worldsheet
\[
\xymatrix{
   \bullet \ar@/^1pc/[rr]_{}="0"
           \ar@/_1pc/[rr]_{}="1"
           \ar@{=>}"0";"1"^{f}
&& \bullet
}
\]
such that we can make sense of composing pieces of such worldsheet
horizontally
\[
\xymatrix{
   \bullet \ar@/^1pc/[rr]_{}="0"
           \ar@/_1pc/[rr]_{}="1"
           \ar@{=>}"0";"1"
&& \bullet \ar@/^1pc/[rr]_{}="2"
           \ar@/_1pc/[rr]_{}="3"
           \ar@{=>}"2";"3"
&& \bullet
}
\]
and vertically
\[
\xymatrix{
   \bullet \ar@/^2pc/[rr]_{}="0"
           \ar[rr]_{}="1"
           \ar@{=>}"0";"1"
           \ar@/_2pc/[rr]_{}="2"
           \ar@{=>}"1";"2"
&& \bullet
}
\]
and such that multiple compositions like
\[
\xymatrix{
   \bullet \ar@/^2pc/[rr]_{}="0"
           \ar[rr]_{}="1"
           \ar@{=>}"0";"1"
           \ar@/_2pc/[rr]_{}="2"
           \ar@{=>}"1";"2"
&& \bullet \ar@/^2pc/[rr]_{}="3"
           \ar[rr]_{}="4"
           \ar@{=>}"3";"4"
           \ar@/_2pc/[rr]_{}="5"
           \ar@{=>}"4";"5"
&& \bullet
}
\]
are well defined? This is what we would call a theory of
{\bf nonabelian strings}. Can we find a global description of this
situation that would generalize that of bundles with connection?

We will discuss here that, indeed, one can. This leads to 
the notion of what we call 
{\bf 2-bundles with 2-connection and 2-holonomy}, which generalize
ordinary fiber bundles with connection from the case of points to 
the case of strings.

\subsection{Motivations}
\label{motivations}

There are several motivations for being interested in these kinds
of questions:

\subsubsection{The Kalb-Ramond Field}

First of all there is the well-known abelian situation which we would like
to reobtain as a special case of the above idea: 

In string theory
there is a field, called the Kalb-Ramond field, which locally looks
like a 2-form $B$ taking values in the real numbers. 
Given a piece
of worldsheet,$\Sigma$,  one can (locally) associate the group element
\[
  \Sigma \mapsto \hol\of{\Sigma}
  \defas \exp\of{i \int_\Sigma B} \in U\of{1}
\]
to it. The action functional of the string in this background has the form
\begin{eqnarray}
  \label{action functional of string in KR background}
  \exp\of{i S\of{\Sigma}} = \exp\of{i S_\mathrm{kinetic}\of{\Sigma}}
  \hol\of{\Sigma}
  \,.
\end{eqnarray}
Since $U\of{1}$ is abelian, the order in which the
different $\hol\of{\Sigma}$ are being multiplied does not
matter.

From considerations of ``\emph{worldsheet anomalies}''
it is well known that the 2-form $B$ globally has to be described by
a structure called an \emph{abelian gerbe} 
\cite{FreedWitten:2000}
and how this nontrivially affects the
computation of the \emph{global} definition of $\hol$. A general formalism of
nonabelian strings should reproduce all this in appropriate
special cases. The formalism which is going to be presented in the following
does so. It is however not only more general than that, but also provides
a more natural (namely ``diagrammatic'') language for computing 
these $B$-field surface holonomies.

Before proceeding, consider an open string ending on a
stack of D-branes (a couple of D-branes on top of each
other)
in the presence of the Kalb-Ramond field.
\begin{figure}[h]
\begin{center}
\begin{picture}(200,200)
  \includegraphics{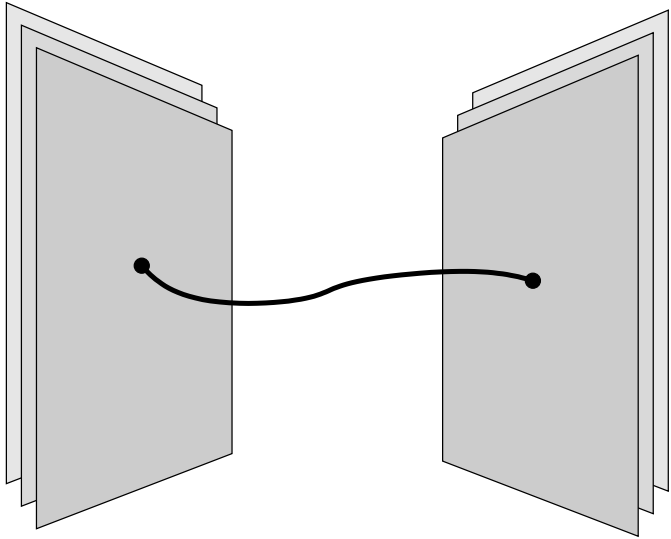}
\end{picture}
\end{center}
\caption{ {\bf An open string stretching between stacks of D-branes}.
The bulk of the string couples to an abelian 2-form. The boundary of the
string, its endpoints, couple to a nonabelian 1-form. }
\end{figure}
There is a general argument saying that  
\begin{itemize}
\item
an object with $p$-dimensional
worldvolume coupled to some \emph{abelian} $p$-form 
\item
can have coupled to its 
$(p-1)$-dimensional boundary a \emph{nonabelian} $(p-1)$-form.
\end{itemize}
This argument is sufficient to deduce from the presence of the 
abelian Kalb-Ramond field alone that the boundary of the open string
may couple to a possibly nonabelian 1-form. 
Hence an open abelian string has nonabelian endpoints.
This is precisely the well-known
statement that there are possibly nonabelian bundles living on
stacks of D-branes.

A nice account of these facts and the following consequence can be found
in \cite{AschieriJurco:2004}.

\subsubsection{Open Membranes on 5-Branes}

The above scenario can be ``lifted to M-theory''. Assuming for
simplicity that the stack of D-branes that we started with were
D4-branes, this lifts the dimension of everything by one unit:
the former open string now becomes an open membrane (the M2-brane) 
while  the 4-branes become 5-branes (the M5-branes). 
The former nonabelian endpoint
of the string on the stack of branes now becomes an
end\emph{string}.
\begin{figure}[h]
\begin{center}
\begin{picture}(200,200)
  \includegraphics{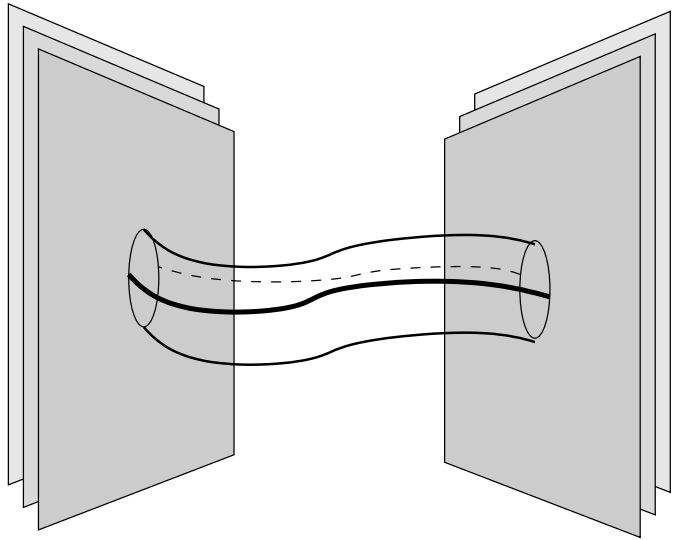}
\end{picture}
\end{center}
\caption{ {\bf An open membrane stretching between stacks of M-branes}.
The bulk of the membrane couples to an abelian 3-form. The boundary of the
membrane, its end\emph{strings}, are expected to couple to a
nonabelian 2-form. }
\end{figure}
This heuristic picture alone already suggests that this 
endstring is a candidate for a nonabelian string in the above sense. 
Since the bulk of the membrane couples to the abelian supergravity 3-form,
the above argument also leads to the conclusion that the
boundary of the membrane should couple to a nonabelian 2-form.

Further arguments for the existence of nonabelian strings on
stacks of 5-branes have been given
(see \S\fullref{more motivation: Open Membranes on 5-Branes}), 
but still these systems remain
notoriously mysterious. A good conceptual understanding of the
formal properties of a theory of nonabelian strings should 
certainly help to shed light on these questions. 

We will show how to compute global nonabelian 2-holonomy\footnote{
 We are using the term ``holonomy'' where some people might rather
say ``parallel transport''. These people would use ``holonomy''
for the parallel transport around a closed loop only, while we
use the term for parallel transport along any path. When
we want to emphasize that we are talking about the holonomy of
a closed curve we will speak of \emph{monodromy}.
}
$\hol\of{\Sigma}$ for a given surface $\Sigma$
under certain conditions.
This immediately allows to write down candidate action principles for 
nonabelian strings of the form
\[
  \exp\of{i S\of{\Sigma}}
  =
  \exp\of{iS_{\mathrm{kinetic}}\of{\Sigma}}
  \mathrm{Tr}\of{\hol\of{\Sigma}}
  \,.
\] 
This is precisely of the same general form as in the abelian case 
\refer{action functional of string in KR background}. The only difference
to be taken care of in the nonabelian case is that a suitable
operation
$\mathrm{Tr}$ analogous to the
ordinary trace in some representation of the gauge group 
used in ordinary gauge theory.
Whether or not these action principles could pertain to strings on
5-branes is not understood yet, though.

One important consistency check is related to what is called the
$N^3$-scaling behaviour on theories of 5-branes. 
It is known that the entropy of ordinary gauge theory asymptotically
scales with the square of the rank of the
Lie algebra of the gauge group. In the stringy picture this can be 
thought of as being related to the $\sim N^2$ ways in which the
two endpoints
of an open string can be attached to $N$ D-branes.

Now, even though the effective field theories on 5-branes are not well
understood, there are indirect arguments which indicate that the entropy
of these theories should asymptotically scale as $N^3$, i.e. with the
cube of the number of 5-branes involved. 

There is a simple heuristic picture making this plausible: The membrane
has a certain particularly stable state, called a BPS state, in which it
has three disconnected boundary components and hence looks like a 
pair of pants. Hence in this state there are $\sim N^3$ different 
possibilities to attach the boundaries of the membrane to one of 
$N$ 5-branes. 

Any formalism of nonabelian strings applicable to M2/M5-brane systems
will have to account for this property, somehow. In the
formalism developed here there seem to be mechanisms related to that.
But this requires further investigation. 
For more discussion see 
\S\fullref{does 2-Holonomy capture n-cube scaling behaviour}.

\subsubsection{Spinning Strings}

Even though configurations of M2- and M5-branes are thought to be
the fundamental objects in M-theory, these scenarios may look 
rather exotic. There is however also a much more general way in which
nonabelian strings should play a crucial role in string theory.

Whether or not an ordinary particle is charged, it may carry spin.
There has to be a spinor bundle with connection which describes
how the spin of the particle transforms as it is transported 
along its worldline.

Superstrings are much like continuous lines of spinning particles. 
Hence a good global description of spinning strings has to take
into account how their spinor degrees of freedom transform
as they are parallel transported along their worldsheets.
\begin{figure}[h]
\begin{center}
\begin{picture}(320,80)
  \includegraphics{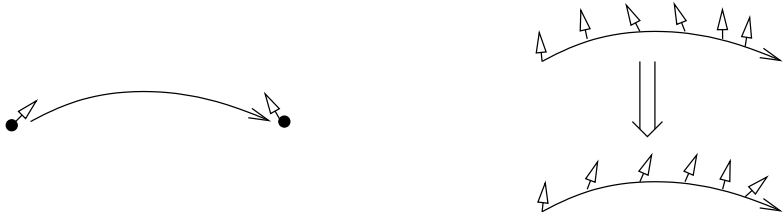}
\end{picture}
\end{center}
\caption{{\bf Parallel transport of spinning strings}, depicted in
  the cartoon on the right, is much like the parallel transport of a 
  line of spinning point particles, indicated on the left.}
\end{figure}
The supercharges of the various flavors of string are generalized
Dirac operators on loop space. Given a spinor bundle
\[
  \begin{array}{l}
    E \\ \Bigg\downarrow  \Spin\of{n} \\ M
  \end{array}
\]
over spacetime $M$, one can take loops everywhere and get an 
$L\Spin\of{n}$-bundle
\[
  \begin{array}{l}
    LE \\ \;\Bigg\downarrow  L\Spin\of{n} \\ LM
  \end{array}
  \,.
\]
Due to the Virasoro anomaly, this is however not sufficient for
the description of superstrings. 
What is needed is instead a lift of
the structure group to a Kac-Moody central extension $\hat L\Spin$
of this loop group
\[
  \begin{array}{l}
    \hat LE \\ \;\;\Bigg\downarrow  \hat L\Spin\of{n} \\ LM
  \end{array}
  \,.
\]
This is possible only if the first Pontryagin class of the original
spin bundle $E$ over $M$ vanishes. In this situation the above can
be reformulated by saying that it is possible to lift the structure
group of $E$ from $\Spin\of{n}$ to a group called $\String\of{n}$.

The topological group $\String\of{n}$ is defined to be a group which has the
same homotopy type as $\Spin\of{n}$ except that $\pi_3\of{\String\of{n}}$
vanishes. 

These considerations play a role for instance in the computation of
the index of the Dirac operator on loop spaces. It is natural to ask
if there is a way to capture this somewhat intricate situation with
a good concept of nonabelian strings. This indeed turns out to be the
case and we will explain how. 

More background on spinning strings is recalled in 
\S\fullref{more details on Spinning Strings}

\subsubsection{Mathematical Motivations}
\label{Mathematical Motivations}

There are several aspects of ``higher gauge theory''
that are interesting by themselves, for purely mathematical reasons.
For quite a while people have already studied aspects of 
nonabelian surface holonomy for simplicial surfaces in the 
(comparably simple) case that we call a ``trivial 2-bundle'' here. For instance
topological invariants of knotted surfaces are obtained
from counting the number of ``flat 2-connections'' that one
can put on triangulations of these surfaces. There has also
been an application of surface holonomy to the four-color theorem
\cite{Attal:2005}.

\paragraph{Categorification.}
\label{motivation: categorification}

More generally, the developments presented here fit into a 
grand framework called {\bf categorification}, which lifts 
mathematical concepts from sets to `stringified' sets, called
categories. From this point of view the nonabelian strings to be
discussed here are but a tiny aspect of an immense
structure that mathematicians and physicists 
(maybe unwittingly) are beginning to explore.

Like a set, a category consists of a collection of {\bf objects},
but unlike a set there are in addition {\bf morphisms}
going between pairs of objects in a category. 
While a map between sets is just a function, a map between
categories is called a {\bf functor}. Such a functor takes
morphisms to morphisms, respecting their composition.
While the image of two functions can only be equal or not, 
the image of two functors, being line-like, can be
``congruent'' (can be translated into each other) without 
being equal. 
In this case one says
there is a {\bf natural transformation} between these
functors.

Given any algebraic structure, we can hence {\bf categorify}
it by using the following dictionary \cite{BaezDolan:1998}:
\begin{center}
\begin{tabular}{ccc}
  {\bf sets} &$\longrightarrow$& {\bf categories}
  \\
  \hline
  objects &$\longrightarrow$& morphisms
  \\
  functions &$\longrightarrow$& functors
  \\
  equations &$\longrightarrow$& natural transformations
  \,.
\end{tabular}
\end{center}

This is just the first step in an infinite series of 
categorification steps. Morphisms themselves can be regarded
as objects again. The morphisms between these are then 
{\bf 2-morphisms}. We have already encountered this situations
in the diagrams at the beginning of
\S\fullref{preliminaries of part I}.
For instance one can think of a surface $\Sigma$
(for instance a piece of worldsheet) as a 2-morphism
\[
\xymatrix{
   x \ar@/^1pc/[rr]^{\gamma_1}_{}="0"
           \ar@/_1pc/[rr]_{\gamma_2}_{}="1"
           \ar@{=>}"0";"1"^{\Sigma}
&& y
}
\]
between two 1-morphsism $\gamma_1$ and $\gamma_2$, which
themselves are
paths stretching between the objects $x$ and $y$, which are nothing
but points.

More details on concepts from category theory
are summarized in  \S\fullref{more details on category theory}.

\subsubsection{Category Theoretic Description of Strings}
\label{Derived Category Description of Open Strings on D-Branes}

Despite its simplicity, the idea of thinking of a string 
as a morphism in some category 
(\cf \S\fullref{motivation: categorification}), i.e. thinking of a 
{\bf string as a categorified point particle}, contains in it
the seed for essentially all the developments to be discussed here. 
While this point of view is, among string theorists, rather exotic,
there are directions of string research where its ramifications 
already play a major role. 

This is
\begin{enumerate}

\item
  the category theoretic description of conformal and topological
  2-dimensional field theories following G. Segal's 
  conception of these issues \cite{Segal:2004},

\item
  the description of states of open strings on D-branes in terms of
  what are called {\bf derived categories}.

\end{enumerate}

There are some obvious vague relations of the approach presented in
part \ref{Nonabelian Strings} to the first point.
The relation to the second point
appears to be more subtle but might also be much deeper. We will
not try here (nor would there be the space to do so) to
elaborate on that in adequate detail. But a brief
discussion is given in
\S\fullref{2-NCG and Derived Category Description of D-Branes}.

\newpage
\subsection{Outline}

The material presented here is mainly a collection of
the content of the papers
\cite{
   Schreiber:2004,
   Schreiber:2004b,
   Schreiber:2004c},
which make up part \ref{SQM on Loop Space},
and
 \cite{
   Schreiber:2004e,
   BaezSchreiber:2004,
   BaezCransSchreiberStevenson:2005}
as well as two papers in preparation
\cite{Schreiber:2005d,Schreiber:2005}, constituting
part \ref{Nonabelian Strings},
equipped with further results
and with background information such as to provide a
coherent picture of the unifying idea underlying these.
Of these, \cite{BaezSchreiber:2004} is a collaboration with
John Baez,
\cite{BaezCransSchreiberStevenson:2005} is a collaboration with
John Baez, Alissa Crans and Danny Stevenson.

An overview over the material of part \ref{SQM on Loop Space}
is given in \S\fullref{introduction: SQM on loop space} 
and over that of part 
\ref{Nonabelian Strings} is given in 
\S\fullref{introduction: nonabelian strings}.

The presentation is supposed to be largely self-contained.
Throughout part III we make freely use of concepts of
($n$-)category theory, which the reader can find reviewed in
\S\fullref{more details on category theory}.

An electronic version of this document is available at\\
\href{http://golem.ph.utexas.edu/string/archives/000578.html}
{http://golem.ph.utexas.edu/string/archives/000578.html}.

\vskip 2em

There are several roads that lead to the considerations presented here.
The one we are going to follow starts with supersymmetric quantum mechanics.

\begin{itemize}

\item \underline{\S\fullref{Morse-theory-like Deformations on Loop Space}
 \cite{Schreiber:2004}
}

A system of supersymmetric quantum mechanics (SQM)
is specified by
giving a $C^*$-algebra $A$ of observables, called the ``position operators'',
which is represented on a graded Hilbert space $\mathcal{H}$, together
with hermitean operators $\set{D^i}_{i = 1,2,\dots,N}$
of odd grade that satisfy the superalgebra
\[
  \antiCommutator{D^i}{D^j} = 2\delta^{ij} H
\]
and hence give rise to the {\bf Hamiltonian} $H$. 
(We shall be cavalier with technical fine print here, which can be
dealt with by the usual standard methods.)
For the special case
$N=2$ it is convenient to go over to the nilpotent polar combinations
\begin{eqnarray*}
  \extd &\defas& D^1 + i D^2
  \\
  \coextd &\defas& D^1 - i D^2
  \,.
\end{eqnarray*}
The notation here is to be suggestive of the archetypical case of
SQM, where $\mathcal{H}$ is the Hilbert space of suitable sections of the exterior
bundle over some configuration space
$\mathcal{M}$ equipped with the
Hodge inner product
\[
  \bracket{\alpha}{\beta} = \int_M \alpha \wedge \star \beta
  \,,
\]
and where $A$ is the algebra of functions on $M$ and $\extd$ the
de Rham operator.
This can
be regared as the point-particle limit of the RR-sector of the
RNS superstring.
\begin{figure}[h]
\begin{center}
\begin{picture}(440,200)
\includegraphics{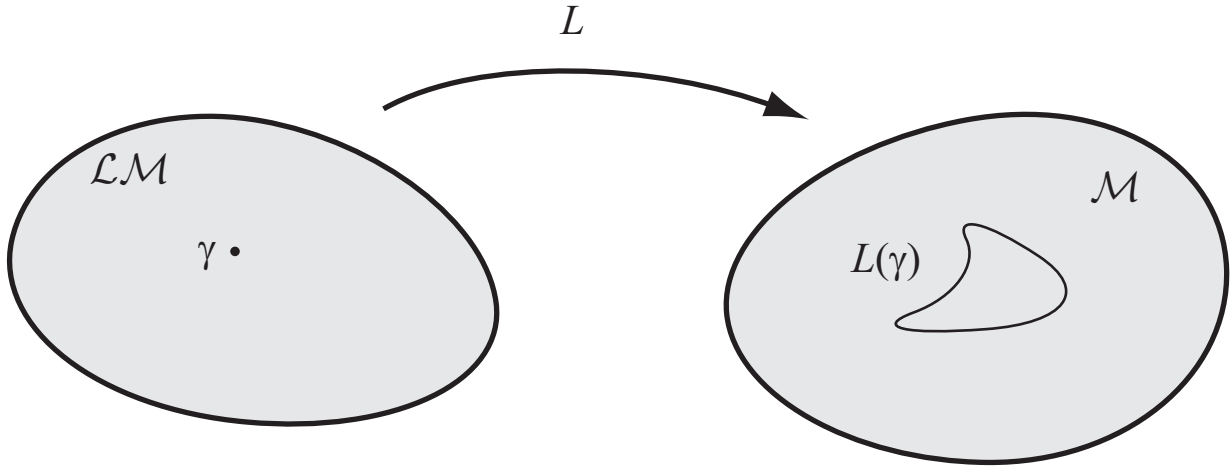}
\end{picture}
\end{center}
\caption{{\bf A point $\gamma$ in loop space $\mathcal{LM}$} maps to a loop 
  $L\of{\gamma}$ in target space $\mathcal{M}$. Loop space is the
  bosonic part of the configuration space of the closed string.
  The full configuration space of the type II superstring is the
  exterior bundle over loop space.
\label{point in loop space}}
\end{figure}

We can ``turn on background fields'' by deforming the supercharges
by invertible operators $e^W$ as
\begin{eqnarray*}
  \extd &\to& e^{-W} \extd e^W\\
  \coextd &\to& e^{W^\dagger} \coextd e^{-W^\dagger}
  \,.
\end{eqnarray*}
For instance, when $W \in A$ is just a function this turns on a potential
$|\nabla W|^2$. This is because the anticommutator of the deformed
supercharges, the deformed Hamiltonian, is the original Hamiltonian, plus
a potential term of the form $|\nabla W|^2$, plus fermionic terms 
(i.e. terms containing
differential form creators and annihilators). 
As another example, when
$W = B \in \bigwedge^2 T^*\mathcal{M}$ is the operator of
exterior multiplication with a 2-form, the turns on a torsion term
$T = \extd B$.

Now let $\mathcal{LM}$ be the free loop space over
$\mathcal{M}$.
\begin{figure}[h]
\begin{center}
\begin{picture}(400,200)
\includegraphics{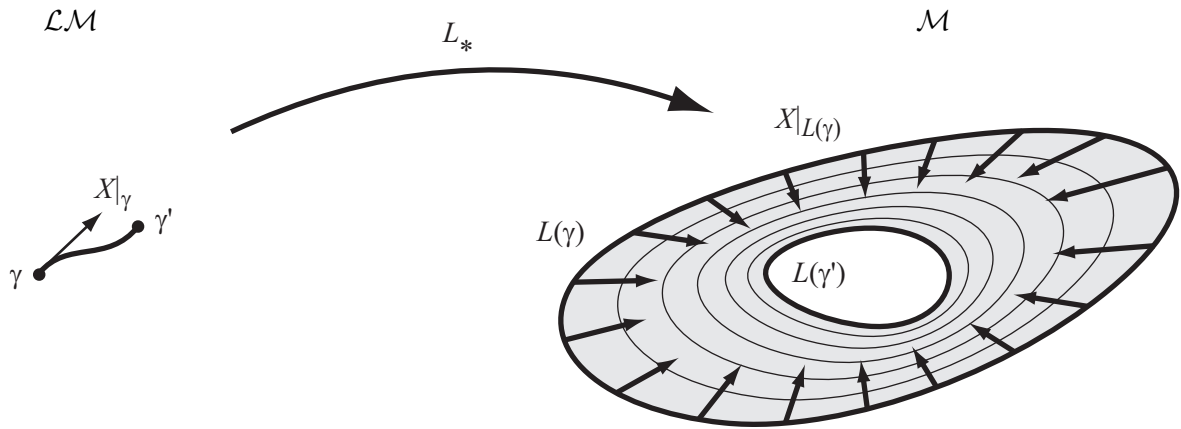}
\put(-430,150){$\mathcal{LM}$}
\put(-100,150){$\mathcal{M}$}
\end{picture}
\end{center}
\caption{{\bf A trajectory in loop space} $\mathcal{LM}$ maps to a surface
  in target space $\mathcal{M}$.
  String dynamics can be regarded as point dynamics in loop space. 
  Using deformations of string supercharges
  %, discussed in %part \ref{SQM on Loop Space},  
  one obtains local 
  connection 1-forms on loop space.
  Their line holonomy gives rise to a 
  notion of local surface holonomy $\hol_i$ in patches
  $U_i \subset \mathcal{M}$ of target space.
\label{path in loop space figure}
}
\end{figure}
We have an exterior derivative
$\extd$ over loop space. 
In local coordinates this looks like
\[
  \extd = \int d\sigma\; \extd \gamma^\mu\of{\sigma}\wedge\;
  \frac{\delta }{\delta \gamma^\mu\of{\sigma}}
  \,,
\]
where $\extd \gamma^\mu\of{\sigma}\wedge$ is the operator of exterior
multiplication by the loop space 1-form $\extd \gamma^\mu\of{\sigma}$,
while $\frac{\delta }{\delta \gamma^\mu\of{\sigma}}$ is the functional
derivative.

Hence we can try to lift the above SQM 
framework from configuration spaces of points to those of strings. 
When appropriately dealing with subtleties induced by the
infinite dimensionality of $\mathcal{LM}$ one finds that $\extd$ is related to the
fermionic super-Virasoro genrerators $G$ and $\bar G$ describing the
superstring as
\[
  \extd \propto G + i \bar G
  \,.
\]
Similar deformations of $\extd$ as for the point particle case can 
be shown to account for all the massless NS background fields of the
string. 

In particular, switching on the $B$-field leads to a deformation
\[
  \extd \to \extd + \int \mathrm{ev}^*\of{B}\wedge + \cdots
\]
where the second term denotes a 1-form on loop space obtained by
taking the 2-form $B$ on target space, pulling it back to
$\mathcal{LM}$ with the
{\bf evaluation map}
\begin{eqnarray*}
  \mathrm{ev} : \mathcal{LM} \times S^1 &\to& \mathcal{M}
  \\
  (\gamma,\sigma) &\mapsto& \gamma\of{\sigma}
\end{eqnarray*}
and integrating over $S^1$. This has the interpretation of 
an abelian local connection 1-form on loop space. 
Taking its holonomy over a curve in loop space reproduces the integral
of $B$ over the corresponding surface in target space.

It is well known that globally this surface holonomy can be 
obtained from what is called an abelian gerbe with
connection and curving.

There are more interesting deformations that one may consider.
Some correspond to gauge transformations of traget space fields,
other to superstring dualities.

\item \underline{\S\fullref{Worldsheet Invariants and Boundary States}
  \cite{Schreiber:2004b,Schreiber:2004d}
}

A large class of deformations, those induced by
so-called {\bf worldsheet invariants}, has no effect at all on the
supersymmetry generators. Still, they act nontrivially on states and
indeed can be shown to be related to the {\bf boundary states}
describing D-branes with gauge fields.

\item 
\underline{\S\fullref{Connections on Loop Space from Worldsheet Deformations}
 \cite{Schreiber:2004e}
 }

There is a straightforward common generalization of 
those deformations which
induce an abelian 1-form connection on loop space 
and those that correspond
to the boundary state 
describing a D-brane with a nonabelian connection 1-form
turned on. 
\begin{figure}[h]
  \begin{center}
    \begin{picture}(325,260)
      \includegraphics{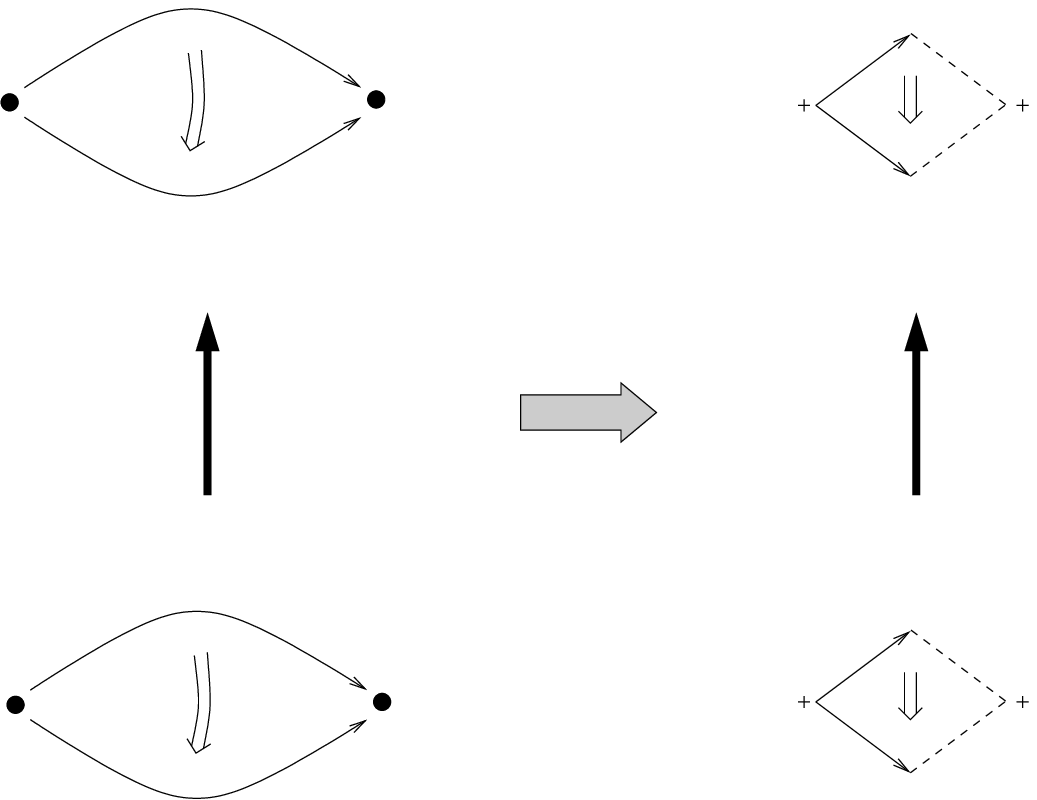}
      \put(-350,-16){$\underbrace{\hspace{6.7cm}}_{\mbox{\footnotesize integral picture}}$}
      \put(-88,-16){$\underbrace{\hspace{5.5cm}}_{\mbox{\footnotesize differential picture}}$}
      \put(-230,110){$\hol_i$}
      \put(-137,109){\footnotesize diff.}
      \put(-25,110){$\con_i$}
      \put(-340,25){$\P_2\of{U_i}$}
      \put(-330,198){$G_2$}
      \put(18,25){$\p_2\of{U_i}$}
      \put(21,198){$\g_2$}
      \put(-298,34){$x$}
      \put(-188,34){$y$}
      \put(-245,60){$\gamma_1$}
      \put(-245,-7){$\gamma_2$}
      \put(-257,27){$[\Sigma]$}
      \put(-10,174){\begin{picture}(0,0)
        \put(-245,60){$\hol_i\of{\gamma_1}$}
        \put(-245,-9){$\hol_i\of{\gamma_2}$}
        \put(-267,27){$\hol_i\of{\Sigma}$}
      \end{picture}}
    \end{picture}
  \end{center}
\vskip 1em
\caption{{\bf Local 2-holonomy and local 2-connection}
  is the higher dimensional generalization (``categorification'')
  of local holonomy and local connection.
  Local 2-holonomy is a 2-functor $\hol_i$ that maps surface elements
  in a 2-path 2-groupoid $\P_2\of{U_i}$ to
  elements of a categorified Lie group (Lie 2-group) $G_2$.
  %This is discussed in {\S\fullref{Local 2-Holonomy and Transitions}}.
  Differentially, this comes from a 2-connection $\con_i$ which can be
  realized as a 2-functor from the 2-path 2-algebroid $\p_2\of{U_i}$
  to the Lie 2-algebra $\g_2$. Such a local 2-connection is specified
  by a 1-form $A$ and a 2-form $B$, taking values in $\g_2$.
  %This differential version is explained in \S\fullref{infinitesimal strict 2-bundles}.
  \label{figure: local 2-holonomy and local 2-connection}
  }
\end{figure}
It leads to a deformation
\[
  \extd \to \extd + \int W_A\of{\mathrm{ev}^*\of{B}}\wedge + \cdots\,,
\]
where now $B \in \Omega^2\of{M,\h}$ takes values in a possibly
nonabelian Lie algebra $\h$ and where $W_A$ denotes the parallel
transport of $B$ to the origin of the loop by means of a 
connection 1-form $A \in \Omega^1\of{M,\g}$ taking values in a
Lie algebra $\g$ which acts on $\h$. In order for this to be 
meaningful, $\g$ and $\h$ have to form what is called a
{\bf differential crossed module} $(\g,\h,d\alpha,dt)$ where
$\alpha$ and $t$ are Lie algebra homomorphisms
\begin{eqnarray*}
  &&d\alpha : \g \to \mathrm{Der}\of{\h}
  \\
  &&dt : \h \to g
\end{eqnarray*}
satisfying a certain compatibility condition.

The physics described by this deformation can no longer be that
of D-branes. The connection 1-form on loop space is now nonabelian and hence
integrating it over a curve in loop space yields a nonabelian
group element associated to the corresponding worldsheet
in target space. This suggests that it describes nonabelian
strings. For this to make good sense, a global description of these
nonabelian connections is necessary. It turns out that the
above connection 1-forms are precisely those that appear in
a \emph{categorified} 
version (\cf \S\fullref{motivation: categorification})
of ordinary fiber bundles, called 2-bundles,
which should hence provide precisely this global description.

\vskip 1em

The relation of the above loop space formalism to categorification
was suspected when it was found that for the above nonabelian
connection 1-form to yield a reparameterization invariant
surface holonomy in target space and to behave sensibly
under gauge transformations, the following relation between the
2-form $B$ and the field strength $F_A$ of the 1-form needs to hold:
\[
  dt\of{B} + F_A = 0
  \,.
\]
It turned out that this relation was encountered before, in the
study of categorified lattice gauge theory by Girelli and Pfeiffer.

We will discuss this condition in detail in the main text.
The reader is  reminded that what are, conventionally, called $B$
and $A$ now can no longer be the fields of the same name in the
context of abelian strings on stacks of D-branes.

\item \underline{\S\fullref{2-Groups, Loop Groups  and String}
\cite{BaezCransSchreiberStevenson:2005}
}

In categorifying gauge theory, one of the crucial steps is to find a
categorification of the concept of gauge group. The result of
applying the dictionary in
\S\fullref{motivation: categorification}
to the definition of an ordinary group is called
a {\bf 2-group} or ``gr-category''.

In ordinary gauge theory one associates group elements to pieces
of worldlines. In categorified gauge theory one instead associates
morphisms of a 2-group to pieces of worldsheet.

All perturbative superstrings, regardless of the backgrounds
that they propagate in, carry spin degrees of
freedom. Hence there should be a 2-group related to the
$\Spin$-group which describes the parallel transport of
spinning strings.
\begin{figure}[h]
\begin{center}
\begin{picture}(400,400)
\includegraphics{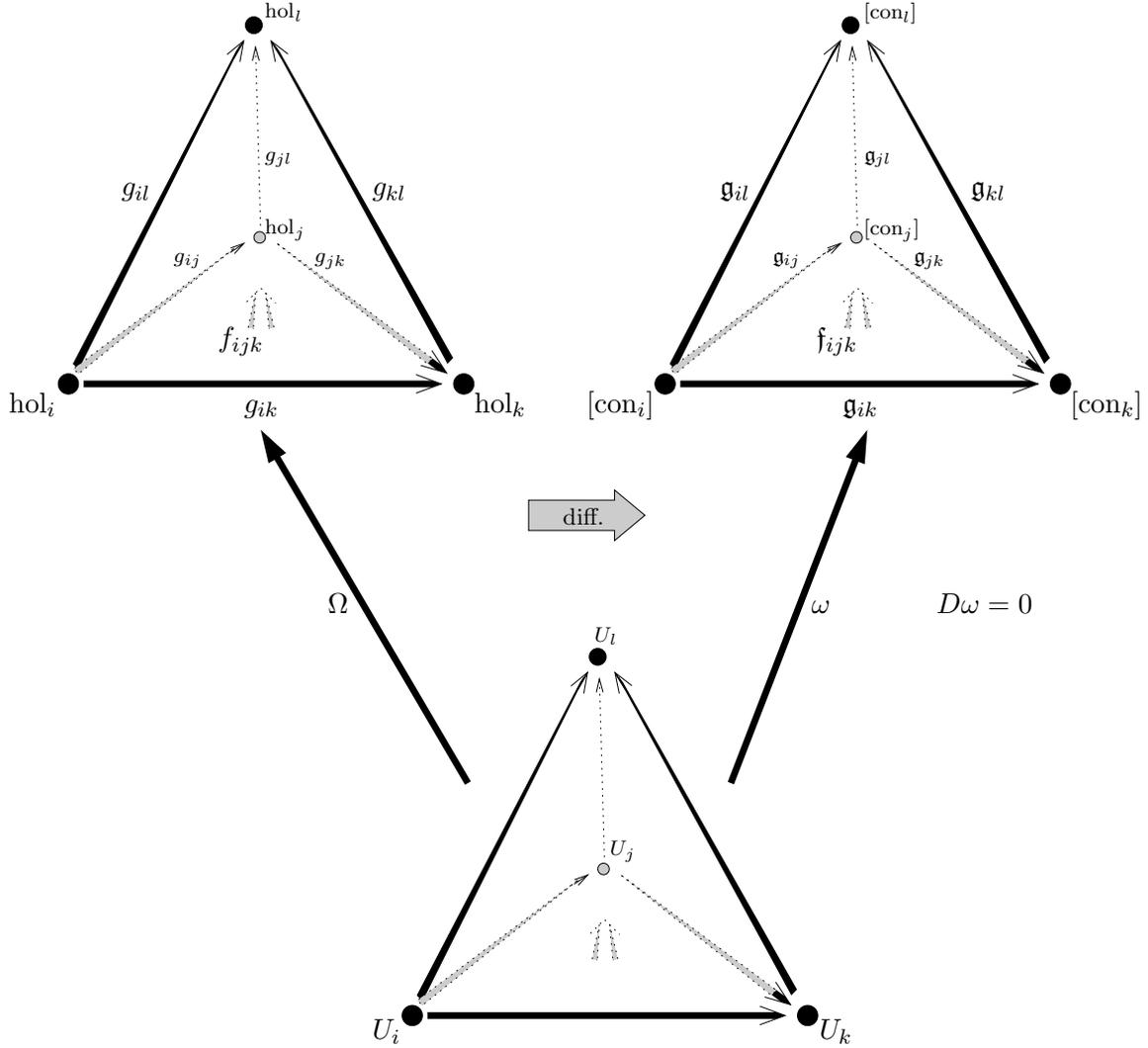}
\put(-95,0){\begin{picture}(0,0)
  \put(-173,-3){$U_i$}
  \put(-2,-3){$U_k$}
  \put(-82,69){${}_{U_j}$}
  \put(-88,151){${}_{U_l}$}
\end{picture}}
\put(-227,240){\begin{picture}(0,0)
  \put(-117,56){${}_{g_{ij}}$}
  \put(-63,56){${}_{g_{jk}}$}
  \put(-90,-3){$g_{ik}$}
  \put(-100,22){$f_{ijk}$}
  \put(-137,80){$g_{il}$}
  \put(-41,80){$g_{kl}$}
  \put(-82,95){${}_{g_{jl}}$}
  \put(-180,-3){$\hol_i$}
  \put(-2,-3){$\hol_k$}
  \put(-82,67){${}_{\hol_j}$}
  \put(-82,150){${}_{\hol_l}$}
\end{picture}}
\put(2,240){\begin{picture}(0,0)
  \put(-117,56){${}_{\g_{ij}}$}
  \put(-63,56){${}_{\g_{jk}}$}
  \put(-90,-3){$\g_{ik}$}
  \put(-100,22){$\f_{ijk}$}
  \put(-137,80){$\g_{il}$}
  \put(-41,80){$\g_{kl}$}
  \put(-82,95){${}_{\g_{jl}}$}
  \put(-188,-3){$[\con_i]$}
  \put(-3,-3){$[\con_k]$}
  \put(-82,67){${}_{[\con_j]}$}
  \put(-82,150){${}_{[\con_l]}$}
\end{picture}}
\put(-285,160){$\Omega$}
\put(-100,160){$\omega$\hspace{1.3cm} $D\omega = 0$}
\put(-195,194){\footnotesize diff.}
\end{picture}
\end{center}
\caption{\label{figure: 2-bundle as simpl map} A {\bf 2-Bundle with 2-Holonomy}
   over an ordinary base space
   $B$ is, when locally trivialized
   with respect to a good covering
   $\covering = \bigsqcup\limits_{i \in I} U_i$ of $B$,
     an assignment
   $\Omega$
   of a) local 2-holonomy 2-functors $\hol_i$ to patches $U_i$,
   b) of pseudo-natural transformations
   $\hol_i \stackto{g_{ij}} \hol_j$
   to double overlaps $U_{ij}$,
   and c) of modifications
   $g_{ik} \stackto{f_{ijk}} g_{ij} \circ g_{jk}$ of such transformations
   to triple overlaps $U_{ijk}$, such that the tetrahedron on the left
   2-commutes. (There is a 2-morphisms in every face of this tetrahedron,
   but for convenience only one of them is displayed.)
   %This is the content of \S\fullref{principal 2-Bundles}.
   Differentially, this is an assignment $\omega$ of 2-connections
   $\con_i$
   to patches $U_i$ and of 1-morphisms $\g_{ij}$ and 2-morphisms
   $\f_{ijk}$ between these to double and triple overlaps, respectively,
   such that the tetrahedron on the right 2-commutes.
   This is equivalent to saying that $\omega$ is a cocycle with respect to a generalized
   (nonabelian ) Deligne coboundary operator $D$.
   Gauge transformations correspond to homotopies of the
   map $\Omega$, which in the differential picture comes from
   shifts by $D$-exact elements: $\omega \to \omega + D\lambda$.
  % This differential picture is described in \S\fullref{The Differential Picture: Morphisms between p-Algebroids}.
  }
\end{figure}
It turns out that a known Lie 2-algebra,
which is called $\mathfrak{spin}_1$ and which in a subtle way is
what is called ``non-strict'', is categorically equivalent to the
Lie 2-algebra of a Lie 2-group called
$\mathcal{P}_1 \Spin\of{n}$, which has the right
properties to do just that.

\item \underline{\S\fullref{principal 2-Bundles}
\cite{BaezSchreiber:2004}
}

Ordinary gauge theory requires the notion of a
principal fiber bundle. This is
a total space $E$ together with a projection $E \to M$ of this
space onto spacetime $M$, such that over contractible patches
$U_i \subset M$ of spacetime the total space looks like
$E|_{U_i} \simeq U_i \times G$, i.e. like spacetime with a
copy of the
``gauge group'' $G$ attached to each point.

When the categorification dictionary displayed in
\S\fullref{motivation: categorification}
is applied to this
structure, one ends up with a category $E$, a category $M$ and a
functor $E \to M$, such that $E$ locally looks like $U_i \times G$,
where $G$ is now a 2-group. This is called a {\bf 2-bundle}
\cite{Bartels:2004}.

In order to find how a 2-bundle describes nonabelian strings, one
needs to furthermore categorify the notion of \emph{connection of
a bundle} such that it admits a categorification of the notion of
\emph{holonomy of a connection}.

One nice way to describe the concept of an ordinary connection
on an ordinary principal bundle uses the idea of a functor. One can regard the
set of \emph{paths} (worldlines) in spacetime as a category whose
objects are all the points of spacetime and whose morphisms are
all paths between pairs of these points. One can also regard
an ordinary (gauge) group as a category with a single object and
one morphism for every group element. An ordinary connection is then
nothing but a functor $\hol_i$ from the category of paths in contractible
patches $U_i$ of spacetime
to the gauge group
(\cf \S\fullref{elements of cat theory: functors}).
This is just the formal version of the
familiar statement that a connection allows to do
``parallel transport'' along any given path.

On double overlaps
$U_{ij} = U_i \cap U_j$ of two contractible patches $U_i$
and $U_j$ the parallel tranports induced by $\hol_i$ and $\hol_j$
are related by a \emph{gauge transformation} $g_{ij}$.
In terms of functors
this is nothing but a natural transformation
(\cf \S\fullref{elements of cat theory: natural transformations})
between $\hol_i$ and
$\hol_j$.
On triple overlaps these transformations have to satisfy the
familar consistency identity $g_{ij}\circ g_{jk} = g_{ik}$.

Now, a 2-connection with 2-holonomy on a 2-bundle is the
categorification of this situation.
So it is locally on $U_i$ a 2-functor
(\cf \S\fullref{elements of cat theory: 2-functors and pseudo-nat trafos})
$\hol_i$\,, which assigns
elements of a 2-group $G_2$ to surface elements in $U_i$.
On double overlaps\,, $\hol_i$ and $\hol_j$ are related by
a pseudo-natural transformation $g_{ij}$ of 2-functors. On a triple
intersection the natural transformations $g_{ik}$ and
$g_{ij}\circ g_{jk}$ are themselves related by a morphism
$f_{ijk}$ between natural transformations, called a
modification. These $f_{ijk}$
finally satisfy a certain consistency condition on
quadruple overlaps.

It turns out that in the case that the 
gauge 2-group $G_2$ has a property called ``strictness'',
a 2-connection with 2-holonomy  locally comes from
a 2-functor $\hol_i$ which itself is determined precisely
by the connection 1-form
\[
  \mathcal{A}\of{\gamma} = \int\limits_\gamma W_A\of{\mathrm{ev}^*\of{B}}
\]
on the space of paths $\gamma$ that we encountered before in the
context of deformations of SQM on loop space.
It turns out that the condition
\[
  F_A + dt\of{B} = 0
\]
arises as consequence of \emph{functoriality}, i.e. of
the fact that functors respect composition of morphisms.

\item \underline{\S\fullref{the integral picture}
\cite{Schreiber:2005}
}
With the concept of 2-connection with 2-holonomy
in a principal 2-bundle thus
available, it is now possible to compute the surface holonomy
of any given surface with respect to this 2-connection.

It turns out that it is possible to glue the local
2-holonomy 2-functors $\hol_i$ on every patch $U_i$ into
a global 2-holonomy 2-functor by using 2-group elements
that enter the definition of the transition morphisms
$g_{ij}$ and $f_{ijk}$. This is indicated in
figure \ref{figure: global 2-holonomy}.

One can give a more intrinsic description of this situation,
one that does not make recourse to a choice of good covering,
in terms of a single global 2-functor
\[
  \hol \maps \P_2\of{M} \to \GTwoTor
\]
that maps 2-paths in all of base space not to the structure
2-group $G_2$, but to the category of $G_2$-2-\emph{torsors}.
For an ordinary group $G$, a (left) $G$-torsor is a space which has a
free and transitive (left) $G$-action, i.e. which is a
left $G$-space, and which furthermore is isomorphic to $G$
as a $G$-space. But not necessarily canonically isomorphic. 
The
fiber of an ordinary principal $G$-bundle is a $G$-torsor. 
It is well known how an ordinary principal bundle with connection
is given by a global 1-holonomy 1-functor
\[
  \hol \maps \P_1\of{M} \to \GTor
\]
from paths in base space to the category 
$\GTor$
of $G$-torsors. This category has $G$-torsors as objects and
$G$-torsor morphisms as morphism. This are maps between torsors
that are compatible with the left $G$-action. 

More precisely, given a $G$-bundle $E$, we have the smooth category
$\trans_1\of{E}$ whose objects are the fibers $E_x$ of $E$, regarded as
$G$-torsors, and whose morphisms are the $G$-torsor morphisms between
these fibers. When we forget about the smooth structure of $\trans_1\of{E}$
we can regard it as a subcategory of $\GTor$ and our global 1-holonomy
1-functor looks like
\[
  \hol \maps \P_1\of{M} \to \trans_1\of{E}
  \,.
\]

A 2-torsor is the obvious categorification of the concept of a
torsor. There is a 2-category $\GTwoTor$ of $G_2$-2-torsors.
Similarly, when $E$ is a principal $G_2$ 2-bundle with connection and
holonomy it is specified by a global 2-holonomy 2-functor
\[
  \hol \maps \P_2\of{M} \to \trans_2\of{E}
  \,,
\]
where now $\trans_2\of{E}$ is the 2-category whose objects are the
fibers $E_x$ of the $G_2$-2-bundle $E$, regarded as $G_2$-2-torsors.

This is the most elegant description of principal 2-bundles with
2-connection and 2-holonomy that we are discussing here.

\clearpage
\begin{figure}[h]
\begin{center}
\begin{picture}(350,530)
\includegraphics{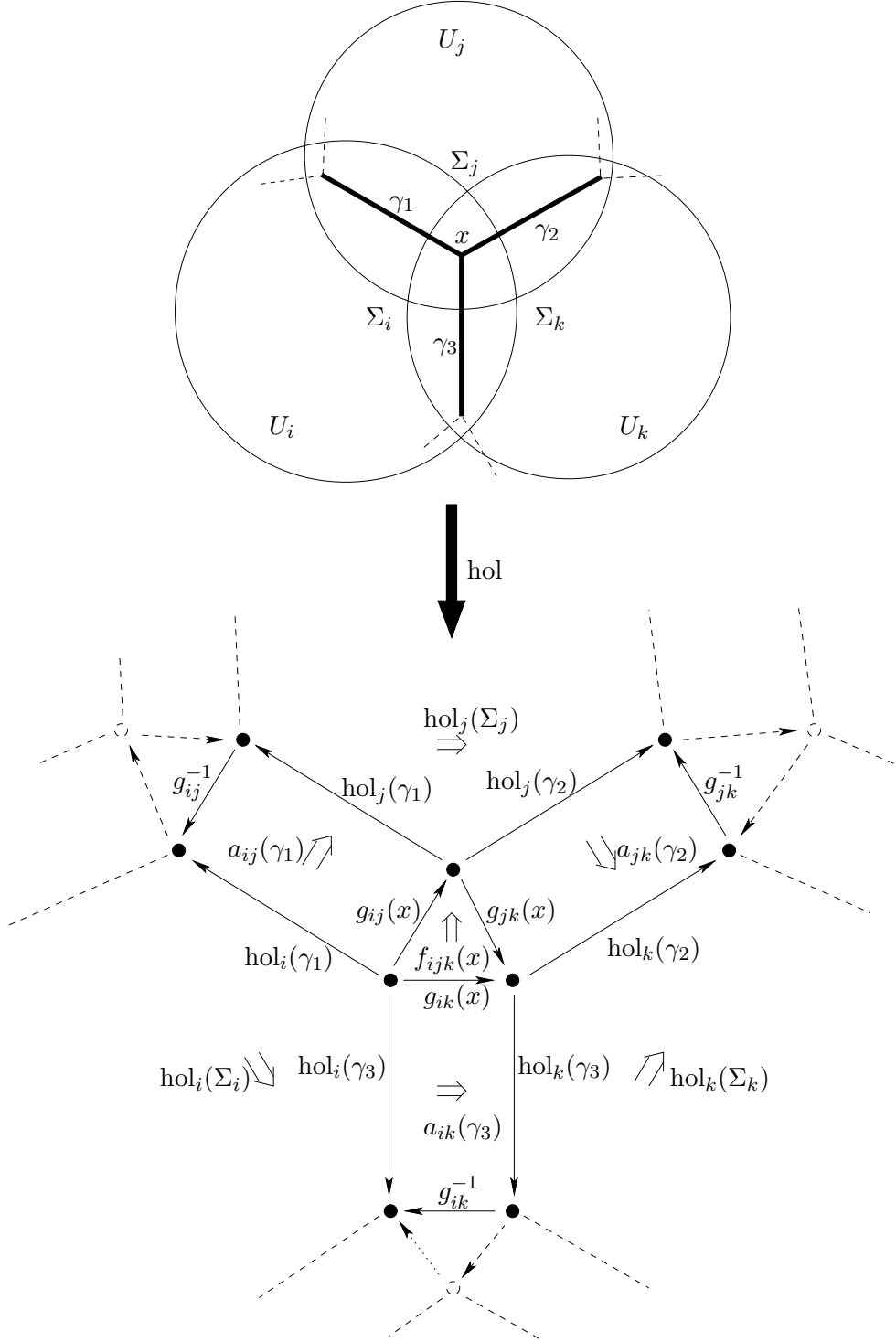}
\put(-14,16){
\begin{picture}(0,0)
\put(-250,360){$U_i$}
\put(-105,360){$U_k$}
\put(-180,520){$U_j$}
\put(-210,405){$\Sigma_i$}
\put(-140,405){$\Sigma_k$}
\put(-175,470){$\Sigma_j$}
\put(-173,439){$x$}
\put(-182,395){$\gamma_3$}
\put(-140,442){$\gamma_2$}
\put(-200,454){$\gamma_1$}
\put(-168,300){$\hol$}
\put(-186,124){$g_{ik}\of{x}$}
\put(-180,43){$g_{ik}^{-1}$}
\put(-214,160){$g_{ij}\of{x}$}
\put(-160,160){$g_{jk}\of{x}$}
\put(-190,140){$f_{ijk}\of{x}$}
\put(-186,70){$a_{ik}\of{\gamma_3}$}
\put(-238,95){$\hol_i\of{\gamma_3}$}
\put(-147,95){$\hol_k\of{\gamma_3}$}
\put(-260,140){$\hol_i\of{\gamma_1}$}
\put(-110,144){$\hol_k\of{\gamma_2}$}
\put(-220,210){$\hol_j\of{\gamma_1}$}
\put(-160,213){$\hol_j\of{\gamma_2}$}
\put(-267,185){$a_{ij}\of{\gamma_1}$}
\put(-106,185){$a_{jk}\of{\gamma_2}$}
\put(-290,213){$g_{ij}^{-1}$}
\put(-70,213){$g_{jk}^{-1}$}
\put(-295,90){$\hol_i\of{\Sigma_i}$}
\put(-84,90){$\hol_k\of{\Sigma_k}$}
\put(-186,239){$\hol_j\of{\Sigma_j}$}
\end{picture}
}
\end{picture}
\end{center}
\caption{
  {\bf Global surface holonomy} of a surface $\Sigma$
  is obtained from the local 2-holonomy
  2-functors $\hol_i$ by suitably gluing them together.
  First triangulate $\Sigma$ such that each face
  $\Sigma_i$
  sits in a single patch $U_i$. Then assign the local 2-holonomy
  $\hol_i\of{\Sigma_i}$ to these faces. Certain
  2-group elements $a_{ij}\of{\gamma}$ (coming from the transition
  on double overlaps)
  are assigned to edges $\gamma$ and
  2-group elements $f_{ijk}\of{x}$
  (coming from the transition on triple overlaps)
  are assigned to  vertices $x$
  of the triangulation.
  The global 2-holonomy is then the well-defined composition of all these
  2-group elements. In a special simple case this reproduces the
  well-knonw formula for surface holonomy in abelian gerbes
  with connection and curving.
  %Global 2-holonomy is the content of  \S\fullref{the integral picture}.
  \label{figure: global 2-holonomy}
}
\end{figure}
\clearpage
Recalling that the exponentiated action functional for a
nonabelian particle is the kinetic term times the holonomy along the
worldline, we can thus write down exponentiated action
functionals for \emph{nonabelian strings} by multiplying the
usual kinetic term with the above notion of surface holonomy
over the worldsheet of the string.
\[
  \exp\of{iS\of{\Sigma}}
  =
  \exp\of{iS_{\rm kinetic}\of{\Sigma}}
  \, 
   \mathrm{Tr}\of{
   \hol\of{\Sigma}
   }
  \,,
\]
where $\mathrm{Tr}$ is a suitable operation that maps morphisms
of a 2-group to complex numbers in a gauge invariant way.

\item \underline{\S\fullref{The Differential Picture: Morphisms between p-Algebroids}
  \cite{Schreiber:2005d}
}

Instead of working with $p$-holonomy functors $\hol_i$ 
that associate $p$-group elements to $p$-dimensional volumes, one
can go to the differential description of these. This leads
to functors that associate Lie $p$-algebra morphisms to $p$-forms
and provides a complementary perspective on the above issues, which
for instance provides a powerful formalism for writing down
action principles for higher $p$-forms such as $B$
\cite{Strobl:2004}. It also provides
a nonabelian generalization of Deligne hypercohomology, which 
allows to conveniently handle $p$-bundles with $p$-connection
and $p$-holonomy using cohomological methods.

\end{itemize}

\newpage

\subsection{Acknowledgments}

  This research first and foremost
  owes its existence and nature to the 
  beneficial working environment provided by 
  Prof. R. Graham, who offered the liberty to do
  autonomous research together with 
  his valuable guidance and advice.
  The idea of applying SQM deformation methods
  to supergravity systems goes back to him. This 
  idea I had ample chance to investigate
  in my master thesis, and building on that 
  I could apply these deformations 
  to the string's worldsheet supergravity
  (\S\ref{SQM on Loop Space}), which
  is the starting point for
  most of the considerations reported here.

  Sections
  \S\fullref{nonabelian strings: preliminaries},
  \S\fullref{2-Groups, Loop Groups  and String} and
  \S\fullref{principal 2-Bundles} are
  due to a very fruitful and most inspiring
  collaboration with John Baez. The integration
  of the loop space formalism from the first half of
  this work (\S\ref{SQM on Loop Space})
  into a theory of 2-connections on
  2-bundles is ongoing joint work with him.
  It is the
  timely appearance of Toby Bartels' definition of
  2-bundles in \cite{Bartels:2004}
  which was crucial for making our collaboration
  possible in the first place.

  Even though I am solely responsible for a couple of
  further developments on 2-bundles which are presented here,
  like the discussion of global 2-holonomy
  and of 2-gauge transformations in
  principal 2-bundles 
  (\S\ref{the integral picture}),
   of strict 3-groups and 3-bundles 
  (\S\ref{Strict 3-Groups} and \S\fullref{Global 3-Holonomy in 3-Bundles}) 
  and  of vector 2-bundles 
  (\S\ref{Vector 2-Bundles}), all of these have benefited from
  discussion with John Baez and would hardly have been 
  conceived without his influence on my thinking.

  The research which lead to the results concerning the 2-group $\PG$
  in \S\ref{2-Groups, Loop Groups  and String}
  was to some extent motivated by
  a comment by Edward Witten regarding a possible
  relation of elliptic cohomology to our 2-connections, 
  as well as by the announcement 
  of a talk by Andr{\'e} Henriques on a relation between
  the Lie 2-algebra $\g_k$ (\S\ref{ghbar.section}) and
  the group $\Spin(n)$. The results presented in
  this section are due to joint work with
  John Baez, Alissa Crans and Danny Stevenson and 
  taken from our paper \cite{BaezCransSchreiberStevenson:2005}.
  This work started while I had the chance to visit
  John Baez's group at UC Riverside in February 2005. I am
  most grateful for this kind invitation and for the intensive
  discussions we had there, many aspects of which have
  found their way into the  presentation given here.
  I also learned a lot from many exchanges of ideas with
  Danny Stevenson since then, who helped me open the door to
  the world of gerbes and bundle gerbes.

  I am much obliged to several other people who have
  shown interest in results of my work by
  inviting me, giving me the opportunity to talk about
  my ideas and providing valuable feedback. 
  I had the opportunity to visit
  (in this order)
  Ioannis Giannakis
  at Rockefeller University in New York, talking about
  deformations of conformal field theories; 
  Hermann Nicolai at the Albert-Einstein-Institute in 
  Golm, and Rainald Flume at the University of Bonn,
  who were interested in my papers on string quantization
  related to Pohlmeyer and DDF invariants;
  Paolo Aschieri and Branislav Jur{\v c}o, who I met
  at University of Torino where we talked about
  nonabelian gerbes and 2-bundles; Christoph Schweigert
  at University of Hamburg, who listened to what I had to say 
  about 2-bundles and 2-connections, Branislav Jur{\v c}o
  once again who also invited me to University of Munich
  for further discussion; and Thomas Strobl
  at University of Jena.

  Apart from those people that I had the chance to meet in person,
  there are several with whom I had 
  helpful discussion by electronic means
  on various topics related to my work. In this context I want
  to thank Jacques Distler for setting up the weblog 
  \emph{The String Coffee Table} \cite{StringCoffeeTable:2004} 
  for this purpose, and for equipping it with the high 
  standard of technology for math on the web that it has.
  I am indebted to all participants of discussions on this weblog.

  For the first part of my research 
  this includes most notably Eric Forgy.
  Our intensive and very enjoyable collaboration on discrete
  differential geometry by means of deformed spectral triples
  has lead to the preprint \cite{ForgySchreiber:2004}, several
  aspects of which reappear, in one guise or another, in the
  discussion of CFT deformations in
  \S\ref{Morse-theory-like Deformations on Loop Space}.
  I am deeply indebted to Eric for taking genuine interest in my
  ideas and for all our very constructive discussions.
  Last not least, I thank him for
  providing  figures
  \ref{point in loop space},
  \ref{path in loop space figure},
  \ref{figure 8 figure}
  and
  \ref{loop rep Killing figure}.
  I could hardly ever have created these myself.

  When I thought about the issues that are now discussed in
  \S\ref{2-NCG and Derived Category Description of D-Branes},
  it was Aaron Bergman who provided a lot
  of help with pointers to the literature
  on various aspects of derived categories in string theory and
  discussion of technical
  details, as well as on the underlying principle that one
  might or might not suspect here.

  I also benefited from comments by   
  Robert Helling, Andrew Neitzke and Lubo{\v s} Motl 
  on $N^3$-scaling behaviour 
  in 5-brane theories. 

  Other participants of discussions about categorified gauge theory
  that I am grateful for are
  Orlando Alvarez,
  Jens Fjelstad, 
  Amitabha Lahiri, 
  and 
  Thomas Larsson. With Jens
  Fjelstad I had some interesting personal discussion about 
  the nature of 2-curvature in 2-bundles, which has
  become part of the exposition in \S\fullref{2-Transition of Curvature}.

  \vskip 1em

  I have received valuable comments while this document was being
  proofread from John Baez, Robert Helling and Branislav Jur{\v c}o. 
  Of course
  all remaining imperfections are mine. Many thanks to Axel Pelster
  for his help with formatting issues.

 \vskip 1em

  Finally, I heartily thank Philip Kuhn for his concern about my
  water balance and for the most kind continuous supply with
  herbal tea that he provided. Without him this
  research might have decayed to dust before it was even finished.
 \vskip 2em

  This work was supported by SFB/TR 12.

\newpage
\section{SQM on Loop Space}
\label{introduction: SQM on loop space}

Start by considering ordinary 
{\bf supersymmetric quantum mechanics},
consisting of 
a graded Hilbert space $\mathcal{H}$ on which an algebra 
$A$ of `position operators'
and $N = 1,2,\dots$ odd-graded, self-adjoint
`{\bf Dirac operators}' or 
`{\bf supercharges}' $D^i$ are represented, which determine the 
Hamiltonian $H$ by the relation
\[
  \antiCommutator{D^i}{D^j} = 2\delta_{ij} H\,,
\]
called the $D=1$, $N=1,2,\dots$ (Poincar\'e) {\bf supersymmetry algebra}.

The triple $\set{A,\mathcal{H},D^i}$ is alternatively known
as a {\bf spectral triple} and can be seen as an algebraic
description of the geometry of configuration space.

For $N=2$, in particular, the nilpotent linear
combinations $\extd \propto D^1 + i D^2$ and
$\coextd \propto D^1 - i D^2$ are of interest. Given any
1-parameter family $\exp\of{W\of{t}}$ of invertible operators
on $\mathcal{H}$, the deformation
\begin{eqnarray*}
  \extd &\to& e^{-W}\circ \extd \circ e^W
  \\
  \coextd &\to&  e^{W^\dagger} \circ \coextd \circ e^{-W^\dagger}
\end{eqnarray*}
preserves the superalgebra and hence defines a new system of
supersymmetric quantum mechanics.

Note that this is a global similarity transformation which
leaves the physics unaffected \emph{only if}
the deformation operator
$W$ is anti-hermitean, in which case the above describes
gauge transformations.

The standard example of supersymmetric quantum mechanics is the case 
where $\extd$ is the
exterior derivative on some manifold $M$, $A$ is the algebra of
continuous (real- or complex-valued) functions on $M$ and 
$\mathcal{H}$ is the Hilbert space of suitably well-behaved 
sections of the exterior bundle over $M$, equipped with the Hodge 
scalar product $\bracket{\alpha}{\beta} = \int \alpha \wedge \star \beta$.
Choosing the deformation $W$ to be in $A$ introduces a scalar
potential $|\nabla W|^2$ (a `background field'!) 
into the Hamiltonian $H$, which is famously related
to the Morse theory of $M$.
This setup (for $W=0$) 
can be thought of as giving the point-particle limit
of the R-R sector of the RNS superstring.

This already suggests that there is nothing more natural than
replacing $M$ with $LM$, the {\bf free loop space} over $M$,
$\extd$ with the exterior derivative on $LM$, and so on.
In other words this amounts to switching from the spectral
triple for the configuration space $M$ of a particle to
that of the configuration space $LM$ of a closed string.

When we think of loop space as 
locally coordinatized by the set 
$\set{\gamma^\mu(\sigma)} \defas \set{\gamma^{(\mu,\sigma)}}$
of coordinates, where $\gamma : [0,2\pi] \to \manifold$ is a parameterized
loop, then for instance the exterior derivative locally
reads
\[
  \extd
   =
   \int d\sigma\; \extd\gamma^\mu\of{\sigma}\wedge \;
   \frac{\delta}{\delta \gamma^\mu\of{\sigma}}
  \,,
\]
where
$\frac{\delta}{\delta \gamma^\mu\of{\sigma}}$ is the
functional derivative.

\begin{figure}[h]
\begin{center}
\begin{picture}(420,200)
\includebiggraphics{Figure8.eps}
\end{picture}
\end{center}
\caption{{\it A vector on loop space} does not necessarily induce a vector field on a loop in target space
\label{figure 8 figure}}
\end{figure}

Taking care of issues related to the infinite-dimensionality
of $LM$ one finds that the {\bf super-Virasoro} generators 
represented on
the Hilbert space of the closed superstring 
(i.e. the left- and right-moving parts $T$ and $\bar T$
of the worldsheet energy-momentum tensor as well as the corresponding
supercurrent with components $G$ and $\bar G$)
indeed provide a 
supersymmetric quantum mechanics on loop space in the above sense.
For instance for a purely gravitational background the polar combination
\[
  G_0 + i \bar G_0 \propto \extd_K
\]
is proportional to the exterior derivate $\extd$ on loop space
summed with the operator $K$ of inner multiplication with the 
generator of reparameterizations of loops.

\begin{figure}[h]
\begin{center}
\begin{picture}(380,200)
\includebiggraphics{Killing.eps}
\end{picture}
\end{center}
\caption{{\it The reparameterization Killing vector} on parameterized loop space always exists
\label{loop rep Killing figure}}
\end{figure}

\subsection{Deformations and Background Fields}

One may hence ask what deformation operators $W$ do to this
system, i.e. what dynamics the deformed operator
\begin{eqnarray*}
  \extd_K 
  \to
  &&
  e^{-W} \, \extd_K \, e^W
  \\
  &\propto&
  e^{-W} \, (G_0 + i \bar G_0) \, e^W
\end{eqnarray*}
describe.

It turns out that all massless NS-NS {\bf background fields}
of the superstring can be encoded in a suitable deformation
$W$ of the loop space spectral triple.

For instance when choosing 
\[
  W_B\of{\gamma}
    \propto 
  \int_\gamma d\sigma \;
  B_{\mu\nu}\of{\gamma\of{\sigma}} 
  \extd\gamma^\mu\of{\sigma} 
  \wedge \extd\gamma^\nu\of{\sigma} \wedge
\] 
for a (possibly only locally defined) 2-form $B$ on target space
and with $\gamma$ a point in loop space,
the deformed super-Virasoro operators are those
that otherwise follow from a canonical analysis of the 
supersymmetric $\sigma$-model for the Kalb-Ramond background
described by $B$, i.e. from the
supersymmetric $\sigma$-model with action
\[
  S
  =
  \frac{T}{2}
  \int
    d^2 \xi
    d^2 \theta
    \;
    \left(
      G_{\mu\nu} + B_{\mu\nu}
    \right)
    D_+ {\bf X}^\mu D_- {\bf X}^\nu
    \,,
\]
where $\theta$ is a Grassmann variable and $\mathbf{X}$
the worldsheet superfield.

In particular, the deformed $\extd_K$ reads
\[
  e^{-W_B}\extd_K e^{W_B}
  =
  \extd_K
  - iT \int_\gamma d\sigma\; B_{\mu\nu}\gamma'^\nu \extd \gamma^\mu
  +
  \frac{1}{6}\int_\gamma d\sigma \; (\extd B)_{\alpha\beta\rho}
    d\gamma^\mu\wedge d\gamma^\nu \wedge d\gamma^\rho
  \,.
\]
The first new term on the right
is the $B$-field pulled back to and integrated
over the given loop. The resulting loop space 1-form has the
interpretation of a local {\bf connection 1-form on loop space}.
The other is the field strength of $B$, which
is interpretable as a torsion term.
The proper global framework for these quantities is well known
to be that of {\bf abelian gerbes} with connection and
curving, which we will reproduce in
part III as a special case of 2-bundles with 2-connection.

By expanding the above deformations 
to first order in the background fields
it is found that they produce the well-known 
so-called {\bf canonical deformations} of 2D conformal field
theories. 

Moreover, deformation operators $W$ which are anti-hermitean
should give rise to gauge transformations of our system, since for
them (and only for them) the above deformation degenerates to 
a global similarity transformation. 
Indeed,
such operators can be shown to describe {\bf gauge transformations}
of background fields as well as {\bf T-duality} operations on the
background.

\subsection{Worldsheet Invariants and Boundary States}

Another special role is played by deformations $e^W$ which commute with
$\extd_K$ and all of its modes. 

Among them are the {\bf worldsheet invariants}, namely
those observables which commute with \emph{all} the super-Virasoro
generators. Traditionally these are known in their incarnation as  
{\bf DDF invariants}. These can be shown to be essentially equivalent
to the set of what are called (supersymmetric) {\bf Pohlmeyer invariants}.

Deforming by such operators evidently does not lead to any 
effective deformations at all
when conjugating $\extd_K$. However, they are still of interest
as deformation operators (apart from their main interest as 
invariant observables of the string): 

It can be seen that the 
constant 0-form $\mathbf{1}$ on loop space is nothing but the
{\bf boundary state} describing the bare, space-filling
$D9$-brane. It turns out that $\extd_K$-closed deformations $W$ give
rise to {\bf boundary state deformations}
\[
  \mathbf{1} \to e^{W}\mathbf{1}
  \,.
\]

One finds that the deformation 
\[
  \mathrm{Tr}\, \mathrm{P}
  \exp\of{
    \int\limits_0^{2\pi}
     d\sigma\,
    \left(
      i 
      A_\mu \gamma^{\prime \mu} 
      + 
      \frac{1}{2T}(F_A)_{\mu\nu}
      \extd \gamma^\mu
      \wedge
      \extd\gamma^\nu
      \wedge
    \right)
  }
  \mathbf{1}
\] 
which assigns to each element of loop space its 
supersymmetric Wilson line with respect to some gauge field $A$,
corresponds to the boundary state obtained by 
turning on that gauge field on a stack of D-branes.
The (supersymmetric) Pohlmeyer invariants, which themselves 
have the rough form of Wilson lines, give rise to such
boundary states when applied to the 0-form $\mathbf{1}$.

All this holds classically in general, while at the quantum
level one encounters the usual divergences which should vanish
(as has been checked to low order) when
the background fields satisfy their equations of motion.

This way there is a nice correspondence between algebraic
deformations of spectral triples on loop space and various
aspects of known string physics.

\subsection{Local Connections on Loop Space from Worldsheet Deformations}

From the loop space perspective there is a natural generalization
of the above inhomogenous differential form on loop space, namely
\[
  (e^{W})_{A,B}
  \defas
  \mathrm{Tr}\, \mathrm{P}
  \exp\of{
    \int\limits_0^{2\pi}
     d\sigma\,
    \left(
      i 
      A_\mu \gamma^{\prime \mu} 
      + 
      \frac{1}{2T}(F_A + B)_{\mu\nu}
      \extd \gamma^\mu
      \wedge
      \extd\gamma^\nu
      \wedge
    \right)
  }
  \,,
\] where $B$ is a Lie-algebra-valued 2-form. 
For nonvanishing $B$ this no longer commutes with $\extd_K$.
Instead one finds that
\[
  (e^W)^{-1}_{A,B}(\extd_K (e^W)_{A,B})
  =
  iT \oint_A (B)
  +
  (\mbox{terms of grade $> 1$})
\]
where the term on the right denotes the loop space 1-form obtained
by pulling $B$ back to the given loop and intergrating it over that
loop while continuously parallel transporting it to the basepoint using
the algebra-valued 1-form $A$.
This can be interpreted as a 
{\bf nonabelian connection 1-form} on loop space.

Hence this cannot describe the boundary state for a
fundamental string on a D-brane anymore. There is also
no nonabelian 2-form field living on D-branes.

There are, however, nonabelian 2-forms expected to arise on
stacks of M5-branes, where they should couple to the endstrings
of open membranes.

A closer examination of the above loop-space connections
reveals certain features that are known from the theory
of {\bf 2-groups}, which are a categorified
(stringified) version of an ordinary group.
This indicates that these constructions have to be
thought of as arising in a theory of {\bf categorified
gauge theory}. And indeed, this turns out to be the case.
The deeper investigation of the above structure requires
however
to step back and look at the larger picture that is emerging here.
This is the content of \S\ref{introduction: nonabelian strings}.

\newpage

\section{Nonabelian Strings}
\label{introduction: nonabelian strings}

We begin our overview of nonabelian strings by
give a pedagogical introduction to the concept
of 2-group (which is well known to algebraists
but hardly known among physicists) in
\S\fullref{introduction: 2-Groups, Loop Groups and the String-Group}.
Then we give an overview of the new results concerning the 2-group
which is related to
spinning strings, summarizing
\S\fullref{2-Groups, Loop Groups  and String}.
In \S\fullref{introduction: principal 2-bundles} the definition of
a principal 2-bundle, following \cite{Bartels:2004}, 
and the derivation of the basic cocycle condition is discussed.
The main definitions and results of the theory of 2-bundles
with 2-holonomy are then given in
\S\fullref{introduction: global 2-holonomy}, summarizing the
discussion in \S\fullref{principal 2-Bundles}
and \S\fullref{the integral picture}. Finally an overview
of the
differential approach to these issues is given in
\S\fullref{introduction: the differential picture}, summarizing
\S\fullref{The Differential Picture: Morphisms between p-Algebroids}.

\subsection{2-Groups, Loop Groups and the $\String$-Group}
\label{introduction: 2-Groups, Loop Groups and the String-Group}

The concept of a 2-group is a basic ingredient for all of the dicussion
to follow. It is in principle well-known and well-understood, and,
at least for the case of strict 2-groups, which we will mostly make
use of, easy to deal with. Before discussing results about
``nonabelian strings'', i.e. about nonabelian surface holonomy, it
should be worthwhile to give the non-expert reader an accessible
introduction to the essence of the concept. This is the aim of the
following subsection.

\subsubsection{Heuristic Motivation of 2-Groups}
\label{Heuristic Motivation of 2-Groups}

For illustration purposes, first consider the case of ordinary lattice gauge theory, where one is 
looking at a graph whose edges are
labeled by group elements of some possibly non-abelian group $G$. 
These group elements specify a holonomy of some $G$-connection along the given edge.
In order to compute the holonomy associated with a concatenation of elementary edges
one simply multiplies the associated group elements in the given order. 
Due to the associativity of the group product, the total holonomies
obtained this way are well-defined in that they 
do not depend on which edges were concatenated first and which later.

This may seem quite trivial, as certainly it is, but it contains in it the seed of a 
non-trivial generalization to higher order holonomies.

Suppose we have not just a graph but a 2-complex and not just edges are labeled with group elements, 
but faces, are, too. (Assume for the moment, for simplicity of
exposition, that both,
edges and faces, are labeled by elements of the same group $G$.) The group label of any
elementary face can naturally be addressed as the \emph{surface holonomy} of that face.

Is there a way, in analogy to the above line holonomies, that we can associate a total surface holonomy
to a connected collection of elementary faces? 

It is immediately clear that the associativity of the group product, which is inherently linear in nature,
alone is no longer sufficient to capture the ``2-associativity'' implicit in the different ways 
in which elementary faces can be composed. Given a square of four faces, for instance, we can first glue them
horizontally along their vertical boundaries and then vertically, along their horizontal boundaries
-- or the other way around. The resulting total surface is of course the same in both cases, but when the
group $G$ is not abelian there is obviously no equally unique way to associate with it a product of the
respective four surface labels.

On the other hand, if we just had, say, vertical composition of faces
in a linear fashion, there would be no
problem. In that case we could just multiply the associated group elements in the respective order.
\begin{center}
\begin{picture}(270,100)
\includegraphics{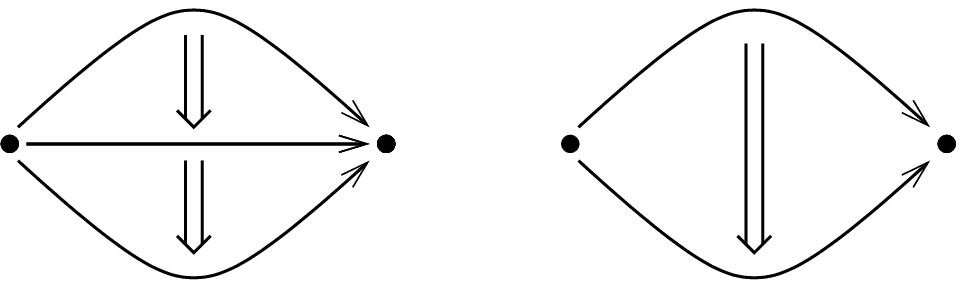}
\put(-223,85){$g_1$}
\put(-223,-7){$g_3$}
\put(-250,45){$g_2$}
\put(-214,55){$f_1$}
\put(-214,20){$f_2$}
\put(-143,37){$=$}
\put(-54,37){$(f_1 \circ f_2)$}
\put(-60,85){$g_1$}
\put(-60,-7){$g_3$}
\end{picture}
\end{center}
We write this \emph{vertical product} of surface elements
as
\begin{eqnarray}
  \label{definition of vertical product}
  \left(\begin{array}{c} f_1 \\ \circ \\ f_2\end{array}\right)
  &\defas&
  f_1 \circ f_2
  \;\defas\;
  f_2 f_1
  \,.
\end{eqnarray}
(Note that here and elsewhere we follow the convention popular in 
category theoretic literature of writing the composition of
arrows $f_1 \circ f_2$ in \emph{literal order} instead of the other
way around.)
On the right we here have the ordinary product in the group $G$. The
order of the factors is purely conventional and could have been choosen
the other way around.

With a vertical product in hand, the task of finding a consistent definition for general
surface composition can be solved by defining a consistent way by which horizontally added faces are
\emph{inserted} into the vertical string of faces. In other words, a procedure is needed which
allows to consistently ``squash'' a collection of elementary faces until it becomes a linear
``vertical'' string of faces whose surface holonomies can be multiplied unambiguously.

The ``squashing'' involves moving surface group labels along the edges of the 1-complex, and this
is naturally described by ``parallel transporting'' them with respect to the edge holonomies.
So if we move a surface label $f$ along a directed edge labeled by $g$, it should become $g^{-1}f g$.

This means that given two horizontally adjacent surface elements with group labels 
$f_1$ and $f_2^\prime$
and an edge $g_1$ along the upper boundary of $f_1$ to $f_2^\prime$,
as well as an edge $g_2$
along the lower boundary of $f_1$ to $f_2^\prime$,
\begin{center}
\begin{picture}(270,160)
\includegraphics{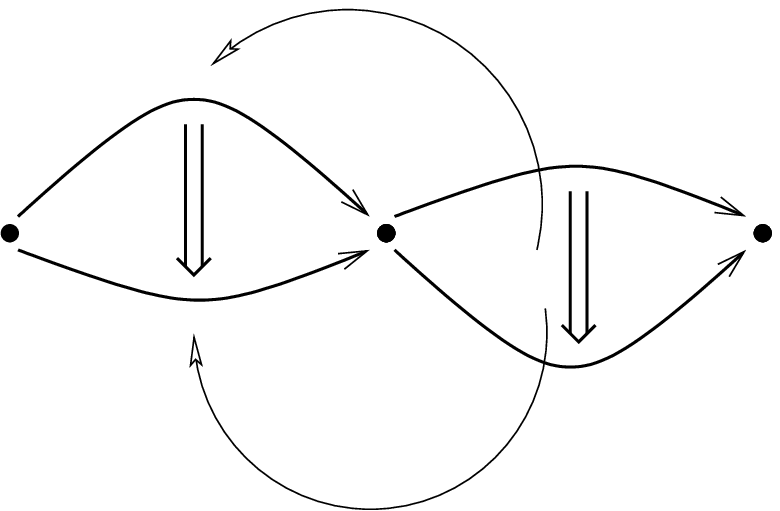}
\put(-203,110){$g_1$}
\put(-203,60){$g_2$}
\put(-162,87){$f_1$}
\put(-50,68){$f_2^\prime$}
\put(-140,150){$f_1\cdot f_2^\prime = (g_1 f_2^\prime g_1^{-1})\circ f_1$}
\put(-140,-10){$f_1\cdot f_2^\prime = f_1 \circ (g_2 f_2^\prime g_2^{-1})$}
\end{picture}
\end{center}
\vskip 1em
we can form the \emph{horizontal} product 
$f_1 \cdot f_2^\prime$ of $f_1$ and 
$f_2^\prime$ by 
\begin{itemize}
\item \emph{either} first moving $f_2^\prime$ along $g_1^{-1}$ \emph{upper} 
boundary of $f_1$ (such that the target edge of $f_2'$ coincides with the
source edge of $f_1$)
where 
it becomes $g_1 f_2^\prime g_1^{-1}$ and
where it can be
vertically multiplied with $f_1$ to produce $(g_1 f_2^\prime g_1^{-1})\circ f_1 = f_1\; g_1 f_2^\prime g_1^{-1}$.
\item
\emph{or} first moving $f_2^\prime$ along $g_2^{-1}$ to the \emph{lower} boundary of $f_1$, where it becomes 
$g_2 f_2^\prime g_2^{-1}$ and where it is vertically multiplied with $f_1$ in the order
$f_1 \circ (g_2 f_2^\prime g_2^{-1}) =  g_2 f_2^\prime g_2^{-1}\; f_1$.
\end{itemize}

In order that the total resulting surface holonomy be well defined,
both these results have to agree,
which gives a crucial \emph{consistency condition} on the group labels of edges and surfaces:
\begin{eqnarray}
  \label{2-associativity conditions 1}
  f_1\; g_1 f_2^\prime g_1^{-1}
  &=&
  g_2 f_2^\prime g_2^{-1}\; f_1
  \,.
\end{eqnarray}
This is fulfilled when the source edge $g_1$ and the target edge $g_2$
of $f_1$ are related by
\[
  g_2 = f_1 g_1
  \,.
\]
(There can be more general solutions. But only this one leads to the full
structure of a 2-group, as explained in the next section.)
When this condition is satisfied
the computation of total surface holonomy of a collection
of elementary faces is independent of the order in which
vertical \refer{definition of vertical product} composition and horizontal composition
\begin{eqnarray}
  \label{definition of horizontal product}
  f_1 \cdot f_2^\prime
  &\defas&
  f_1 \; g_1 f_2^\prime g_1^{-1}
\end{eqnarray}
is applied, and hence in this case we can associate
a well-defined surface holonomy to a collection of elementary faces. 

It is helpful to think of this conditions as expressing a higher order form of ordinary associativity
(which ensures well defined line holonomies), that we could call
\emph{2-associativity}.

Note that both horizontal and vertical products are associative by themselves. 
For the vertical product this is just the
associativity of the group product, while for the horizontal product it is not quite as trivial but
can be easily checked.
But in both cases this is a \emph{linear} (1-dimensional) notion of associativity. 

In order to see how \refer{2-associativity conditions 1} encodes a 2-dimensional notion
of associativity,
consider computing the total surface holonomy
of four faces $f_1$, $f_1^\prime$, $f_2$ and $f_2^\prime$, composed vertically
\emph{and} horizontally
\begin{center}
\begin{picture}(200,120)
\includegraphics{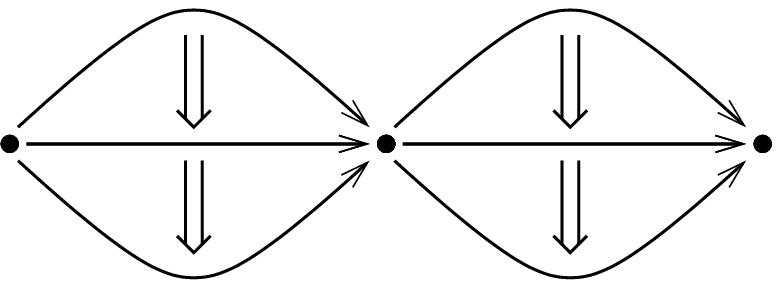}
\put(-170,85){$g_1$}
\put(-61,85){$g_1^\prime$}
\put(-170,-7){$g_3$}
\put(-61,-10){$g_3^\prime$}
\put(-198,45){$g_2$}
\put(-88,45){$g_2^\prime$}
\put(-162,55){$f_1$}
\put(-52,55){$f_1^\prime$}
\put(-162,20){$f_2$}
\put(-52,20){$f_2^\prime$}
\end{picture}
\end{center}
The fact that
the order of composing these faces is irrelevant is expressed by the equation
\begin{eqnarray}
  \label{2-associativity conditions 2}
  \left(
    f_1 \circ f_2
  \right)
  \cdot
  \left(
    f_1^\prime \circ f_2^\prime
  \right)
  &=&
  \left(
    f_1 \cdot f_1^\prime
  \right)
  \circ
  \left(
    f_2 \cdot f_2^\prime
  \right)
  \,,
\end{eqnarray}
which is the form in which the 2-associativity condition usually appears in the 2-group literature
(where it is called the 'exchange law'). It is instructive to emphasize the 2-dimensional
character of this equation by actually writing the vertical product along the vertical
as in \refer{definition of vertical product},
so that \refer{2-associativity conditions 2} becomes
\begin{eqnarray}
  \left(\begin{array}{c}f_1 \\ \circ \\ f_2\end{array}\right)
  \cdot
  \left(\begin{array}{c}f_1^\prime \\ \circ \\ f_2^\prime\end{array}\right)
  &=&
  \begin{array}{c}\left(f_1 \cdot f_1^\prime\right) \\ \circ \\ \left(f_2\cdot f_2^\prime\right)\end{array}
  \begin{array}{c}\\ \\ \;\;\,.\end{array}  
\end{eqnarray}

It is easily checked 
by using \refer{definition of vertical product} and \refer{definition of horizontal product}
that 
this is equivalent to the relation \refer{2-associativity conditions 1} which we used before.

More generally, edges and surfaces need not be labeled by
elements of the same group $G$. We can assume that, while edges are
labeled with elements of $G$, surfaces are labeled with elements of a 
group $H$. In order to generalize the definition of the horizontal
product to this case we need an action of $G$ on $H$ which mimics the
adjoint action of $G$ on itself. Furthermore, in order to generalize
the relation between the source and the target edge, one needs a way
to send an element of $H$ to an element of $G$. 

The structure needed is known as a {\bf crossed module}
$(G,H,\alpha,t)$ of two groups $G$ and $H$. Here
\[
  \alpha \maps G \to \Aut\of{H}
\]
is a group homomorphism from $G$ to the automorphisms of $H$ and
\[
  t \maps H \to G
\]
is a homomorphism from $H$ to $G$. The horizontal product in this
more general case then reads
\[
  f_1 \cdot f_2 = f_1 \alpha\of{g_1}\of{f_2}
\]
and the relation between the source and the target edge becomes
\[
  g_2 = t\of{f_1}g_1
  \,.
\]
In order for all this to be consistent there are the following two
compatibility conditions between $\alpha$ and $t$:
\begin{eqnarray*}
  &&\alpha\of{t\of{h}}\of{h'} = hh'h^{-1}
  \\
  &&
  t\of{\alpha\of{g}\of{h}} = gt\of{h}g^{-1}
  \,,
\end{eqnarray*}
which express the idea that $\alpha\of{g}$ is a generalization of
conjugation by $g$.

\subsubsection{2-Groups as Categorified Groups}
\label{introduction: 2-Groups as Categorified Groups}

The above discussion, emphasizing the idea that the horizontal product
involves parallel transport of surface labels along edges, gives
a rough heuristic approach to 2-groups and their role in 2-holonomy
theory. But more formally 2-groups arise as the {\bf categorification}
of the concept of an ordinary group. Since the inner workings
of 2-groups are
important for much of the discussion to follow, and
since their derivation nicely illustrates the concept of
categorification, we here want to spell this out in detail.

The following makes use of categories and functors between categories.
The reader unfamiliar with these concepts is urged to skip to
\S\fullref{more details on category theory} where a brief introduction 
to basic elements of category theory is provided.

\vskip 1em

An ordinary group is defined to be a
\emph{set} $G$ together with \emph{functions}
\[
  G \stackto{s} G
\]
(inversion)
and
\[
  G \times G \stackto{m} G
  \,,
\]
(multiplication)
which satisfy the \emph{equations}
\[
  m\of{g,s\of{g}} = 1 = m\of{s\of{g},g}
\]
and
\[
  m\of{s_1,m\of{s_2,s_3}}
  =
  m\of{m\of{s_1,s_2},s_3}
  \,.
\]

Using the dictionary discussed in
\S\fullref{motivation: categorification}
this is categorified by saying that there is a \emph{category}
$\mathcal{G}$ together with a \emph{functor}
\[
  \mathcal{G} \times \mathcal{G}
  \stackto{m}
  \mathcal{G}
\]
such that the above equations become natural isomorphisms.

The special case where all these natural isomorphisms are actually
identities is called the \emph{strict} case. A \emph{strict 2-group}
is hence a category with a product functor as above which satisfies the
usual axioms of a group ``on the nose''.

By going through the above axioms of a strict 2-group
$\twogroup$
one can work out how it is described
\emph{in terms of two ordinary groups}:

First of all consider all the identity morphisms going from an object
$g \in \twogroup$ to  itself: $g \stackto{\mathrm{Id}} g$. Restricted to
these the axioms for the product functor $m : \twogroup \to \twogroup$
reduce to the axioms of an ordinary group product. Call this group $G$.
Hence for every
element in $G$ there is an object in $\twogroup$ and the product between
the corresponding identity morphisms is
\[
  \begin{array}{ccc}
    g &  & g'\\
    1\Big\downarrow \skiph{$1$}& \cdot & 1\Big\downarrow\skiph{$1$} \\
    g &  &  g'
  \end{array}
  =
  \begin{array}{c}
    gg' \\
    1\Big\downarrow \skiph{1}\\
    gg'
  \end{array}
  \,,
\]
where we indicate the product functor $m$ by a dot `$\cdot$'.

Next consider the nontrivial morphisms which start at the identity element
$1 \in G$, i.e. which are of the form $ 1 \stackto{f} g$. Obviously,
these form a group under the product $m$ themselves, since the product of any two of
them is a again a morphism starting at the identity. Call this group $H$
and write
\[
  \begin{array}{ccc}
    1 &  & 1\\
    h\Big\downarrow\skiph{$h$} & \cdot & h'\Big\downarrow\skiph{h'} \\
    g &  &  g'
  \end{array}
  =
  \begin{array}{c}
    1 \\
    hh'\Big\downarrow\skiph{hh'} \\
    gg'
  \end{array}
  \,,
\]
where $h,h'\in H$

Given any morphism $g \stackto{f} g'$, let $t$ denote the operation of
sending it to its \emph{target} object, i.e.
\[
  t\of{g \stackto{f} g'} \defas g'
  \,.
\]
Applying this to the above equation shows that $t$ restricts on those
morphisms that start at the identity object to a group homomorphism
\[
  t \maps H \to G
  \,.
\]
We can conjugate every morphism in $H$ with an arbitrary identity morphisms
and stay in $H$:
\[
  \begin{array}{ccccc}
    g &  & 1 & & g^{-1}\\
    1\Big\downarrow\skiph{$1$} & \cdot & h\Big\downarrow\skiph{h}  & \cdot & 1\Big\downarrow\skiph{$1$}\\
    g &  &  t\of{h} & & g^{-1}
  \end{array}
  \defas
  \begin{array}{c}
    1 \\
    \alpha\of{g}\of{h}\Big\downarrow\skiph{$\alpha\of{g}\of{h}$} \\
    gt\of{h}g^{-1}
  \end{array}
  \,.
\]
Since this is just conjugation in our 2-group it obviously gives an
automorphism of $H$ and hence the $\alpha$ appearing in the above formula
is a group homomorphism from $G$ to $\Aut\of{H}$:
\[
  \alpha \maps G \to \Aut\of{H}
  \,.
\]
The homomorphisms $t$ and $\alpha$ have to satisfy
certain compatibility conditions. The first of these is
\[
  t\of{\alpha\of{g}\of{h}} = g t\of{h}g^{-1}\\
  \,,
\]
which follows immediately from the above considerations.
The other one is
\[
  \alpha\of{t\of{h}}\of{h'} = hh'h^{-1}
  \,.
\]
This is a consequence of the fact that the multiplication $m$
in the 2-group is a functor. For consider the left-hand side,
which is given by
\[
  \begin{array}{c}
    1 \\
    \alpha\of{t\of{h}}\of{h'}\Big\downarrow\skiph{$\alpha\of{t\of{h}}\of{h'}$} \\
    t\of{hh'h^{-1}}
  \end{array}
  =
  \begin{array}{ccccc}
    t\of{h} &  & 1 & & t\of{h}^{-1}\\
    1\Big\downarrow\skiph{$1$} & \cdot & h'\Big\downarrow\skiph{$h'$}  & \cdot & 1\Big\downarrow\skiph{$1$}\\
    t\of{h} &  &  t\of{h'} & & t\of{h}^{-1}
  \end{array}
  \,.
\]
Since the product functor has to respect the composition of morphisms,
we can extend the diagram on the right by an identity morphism as
follows:
\[
  \begin{array}{ccccc}
    t\of{h} &  & 1 & & t\of{h}^{-1}\\
    1\Big\downarrow\skiph{$1$} & \cdot & h'\Big\downarrow\skiph{$h'$}  & \cdot & 1\Big\downarrow\skiph{1}\\
    t\of{h} &  &  t\of{h'} & & t\of{h}^{-1}
  \end{array}
  =
  \begin{array}{ccccc}
    1 & & 1 & & 1\\
    h\Big\downarrow\skiph{$h$} & \cdot & 1\Big\downarrow\skiph{$1$}  & \cdot & h^{-1}\Big\downarrow\skiph{$h^{-1}$}\\
    t\of{h} &  & 1 & & t\of{h}^{-1}\\
    1\Big\downarrow\skiph{1} & \cdot & h'\Big\downarrow\skiph{h'}  & \cdot & 1\Big\downarrow\skiph{$1$}\\
    t\of{h} &  &  t\of{h'} & & t\of{h}^{-1}
  \end{array}
  \,.
\]
Composing these morphisms before multiplying them then yields
\[
  \cdots
  =
  \begin{array}{ccccc}
    1 &  & 1 & & 1\\
    h\Big\downarrow\skiph{$h$} & \cdot & h'\Big\downarrow\skiph{$h'$}  & \cdot & h^{-1}\Big\downarrow\skiph{$h^{-1}$}\\
    t\of{h} &  &  t\of{h'} & & t\of{h}^{-1}
  \end{array}
  =
  \begin{array}{c}
    1 \\
    hh' h^{-1}\Big\downarrow\skiph{$hh' h^{-1}$} \\
    t\of{hh'h^{-1}}
  \end{array}
  \,.
\]
This is equivalent to the above consistency condition.

Now we can generalize to arbitrary morphisms. Due to the group structure on
our category $\twogroup$, every morphism can be written as a morphism
$1 \stackto{h} t\of{h}$
starting at the identity element and multiplied
(from the right, say)
with an identity morphism on
an object $g$. We will denote these morphisms by
pairs $(g,h)$:
\[
  (g,h)
  \defas
  \begin{array}{c}
    g \\
    h\Big\downarrow\skiph{$h$} \\
    t\of{h}g
  \end{array}
  \defas
	  \begin{array}{ccc}
    1 &  & g\\
    h\Big\downarrow\skiph{$h$} & \cdot & 1\Big\downarrow\skiph{$1$} \\
    t\of{h} &  &  g
  \end{array}
  \,.
\]
Given this definition and what we already know about conjugation in our
2-group, it is easy to work out the product of general morphisms as follows:
\begin{eqnarray*}
   \begin{array}{ccc}
    g &  & g'\\
    h\Big\downarrow\skiph{$h$} & \cdot & h'\Big\downarrow\skiph{$h'$} \\
    t\of{h}g &  & t\of{h'}g'
  \end{array}
  &=&
   \begin{array}{ccccccc}
    1 &  & g &  & 1 & & g'\\
    h\Big\downarrow\skiph{$h$}
       & \cdot &
       1\Big\downarrow\skiph{$1$}
       & \cdot &
       h'\Big\downarrow\skiph{$h'$}
       & \cdot &
       1\Big\downarrow\skiph{$1$}
        \\
    t\of{h} &  & g & & t\of{h'} & & g'
  \end{array}
  \\
  &=&
   \begin{array}{ccccccccccc}
    1 &  & g &  & 1 & & g^{-1} & & g & &g'\\
    h\Big\downarrow\skiph{$h$}
       & \cdot &
       1\Big\downarrow\skiph{$1$}
       & \cdot &
       h'\Big\downarrow\skiph{$h'$}
       & \cdot &
       1\Big\downarrow\skiph{$1$}
       & \cdot &
       1\Big\downarrow\skiph{$1$}
       & \cdot &
       1\Big\downarrow\skiph{$1$}
        \\
    t\of{h} &  & g & & t\of{h'} & & g^{-1} & & g & & g'
  \end{array}
  \\
  &=&
   \begin{array}{ccccc}
    1 &  & 1 &  & gg'\\
    h\Big\downarrow\skiph{$h$}
       & \cdot &
       \alpha\of{g}\of{h'}\Big\downarrow \skiph{$\alpha\of{g}\of{h'}$}
       & \cdot &
       1\Big\downarrow\skiph{$1$}
        \\
    t\of{h} &  & gt\of{h'}g^{-1} & & gg'
  \end{array}
  \\
  &=&
   \begin{array}{c}
    gg'\\
       h\alpha\of{g}\of{h'}\Big\downarrow \skiph{$h\alpha\of{g}\of{h'}$}
        \\
    t\of{h}gt\of{h'}g'
  \end{array}
  \,.
\end{eqnarray*}
Hence we find the rule for horizontal multiplication
\[
  (g,h) \cdot (g',h')
  =
  (gg', h \alpha\of{g}\of{h'})
  \,.
\]
This is the mulötiplication operation in the
\emph{semidirect product} of groups $G \ltimes H$,
which we have interpreted in terms of parallel transport in
the previous subsection
\S\fullref{Heuristic Motivation of 2-Groups}.

Finally we need to work out what the result of \emph{composing} two
morphisms is. For this we again need to make use of the fact that the product is
a functor and that it respects the composition of morphisms.

Starting with the composition
\[
  \begin{array}{c}
    g \\
    h\Big\downarrow\skiph{$h$} \\
    t\of{h}g\\
    h'\Big\downarrow\skiph{$h'$}
    \\
    t\of{h'h}g
  \end{array}
\]
we can horizontally split this to obtain
\[
  \cdots =
  \begin{array}{ccc}
    1 & & g \\
    1\Big\downarrow\skiph{$1$} & \cdot & h\Big\downarrow\skiph{$h$} \\
    1 & & t\of{h}g\\
    h'\Big\downarrow\skiph{$h'$} & \cdot &  1\Big\downarrow\skiph{$1$} \\
    t\of{h'} & & t\of{h}g
  \end{array}
\]
and then use vertical composition to get
\[
  \cdots =
  \begin{array}{ccc}
    1 & & g \\
    h'\Big\downarrow\skiph{$h'$} & \cdot & h\Big\downarrow\skiph{$h$} \\
    t\of{h'} & & t\of{h}g\\
  \end{array}
  \,.
\]
Performing the product operation now yields
\[
  \cdots =
  \begin{array}{c}
    g \\
    h'h\Big\downarrow\skiph{hh'} \\
    t\of{h'h}g
  \end{array}
  \,.
\]
Hence the vertical composition of the morphism $(g,h)$ with
the morphism $(t\of{h}g,h')$ is
simply the morphism $(g,h'h)$.

\vskip 1em

Given the concept of a \emph{2-category}, which is briefly
discussed in \S\fullref{2-categories}, it is clear that we 
can think of a 2-group as a 2-category with a single object
$\bullet$. This is essentially just a consequence of how we
can think of an ordinary group as a category with a single
object, as explained in \S\fullref{introduction: categories: examples}.

So, where we had an object $g$ in the above disucssion, we can
think of it as a morphism 
\[
\xymatrix{
   \bullet \ar@/^1pc/[rr]^{g}="0"
&& \bullet 
}
\]
starting and ending at the single object $\bullet$. When doing
so the morphisms
\[
  \cdots =
  \begin{array}{c}
    g \\
    h\Big\downarrow\skiph{h} \\
    g'
  \end{array}
\]
become 2-morphisms
\[
\xymatrix{
   \bullet \ar@/^1pc/[rr]^{g}="0"
           \ar@/_1pc/[rr]_{g'}="1"
           \ar@{=>}_h"0";"1"
&& \bullet 
}
\]
and the product functor 
\[
  \begin{array}{ccc}
    g_1 & & g_2 \\
    h_1\Big\downarrow\skiph{$h_1$} & \cdot & h_2\Big\downarrow\skiph{$h_2$} \\
    g'_1 & & g'_2
  \end{array}
  \,.
\]
becomes nothing but the 
\emph{horizontal composition} of such 2-morphisms
\[
\xymatrix{
   \bullet \ar@/^1pc/[rr]^{g_1}="0"
           \ar@/_1pc/[rr]_{g'_1}="1"
           \ar@{=>}_h"0";"1"
&& \bullet \ar@/^1pc/[rr]^{g_2}="2"
           \ar@/_1pc/[rr]_{g'_2}="3"
           \ar@{=>}_{h_2}"2";"3"
&& \bullet
}
\,.
\]
The functoriality of the product, i.e. the fact that it respects
(vertical) composition is then nothing but the \emph{exchange law}
of 2-categories, which says that the order of horizontal and
vertical compositon in diagrams like
\[
\xymatrix{
   a \ar@/^2pc/[rr]^{g_1}="0"
           \ar[rr]^<<<<<<{g_2}_{}="1"
           \ar@{=>}^{f_1}"0";"1"
           \ar@/_2pc/[rr]_{g_3}="2"
           \ar@{=>}^{f_2}"1";"2"
&& b \ar@/^2pc/[rr]^{g'_1}="3"
           \ar[rr]^<<<<<<{g'_2}_{}="4"
           \ar@{=>}^{f'_1}"3";"4"
           \ar@/_2pc/[rr]_{g'_3}="5"
           \ar@{=>}^{f'_2}"4";"5"
&& c
}
\]
is irrelevant.

It is this 2-categorical language that we will mostly use when
talking about 2-groups.

\subsubsection{Lie 2-Algebras}
\label{introduction: Lie 2-Algebras}

Just like an ordinary Lie group has a Lie algebra, a Lie 2-group has
a {\bf Lie 2-algebra}, the categorification of an
ordinary Lie algebra. We need some basic understanding of Lie 2-algebras
for the discussion of the 2-group realization of the $\String$-group 
below in \S\fullref{introduction: the 2-group PG and its relation to string}
as well as for the differential formalism to be introduced in
\S\fullref{introduction: the differential picture}.

The theory of Lie 2-algebras has been worked out in
\cite{BaezCrans:2003}.
By performing the process of categorification by
internalization (\S\fullref{section: Internalization}),
which is the precise formulation of the categorification dictionary
in \S\fullref{motivation: categorification},
one finds that a (semistrict) Lie 2-algebra $\mathcal{L}$ is a category whose
objects $x$ are elements of a vector space $V_0$ and whose
morphisms $x\stackto{\vec f} y$ are labeled by elements
$\vec f$ of another vector space $V_1$.

In order to emphasize the relation between Lie 2-algebras and Lie
2-groups we could draw such a morphism as follows:
\[
\xymatrix{
   \bullet \ar@/^1pc/[rr]^{x}_{}="0"
           \ar@/_1pc/[rr]_{y}_{}="1"
           \ar@{=>}"0";"1"^{\vec f}
&& \bullet
}
\]
The source $x$ and target $y$ of this morphism are related by 
a map
\[
  d \maps V_1 \to V_0
\]
as follows:
\[
  y = x + d\vec f
  \,.
\]
Therefore we can specify any Lie 2-algebra morphism by a couple
$(x,\vec f)$, where $x$ is its source and $\vec f$ is called its
{\bf arrow part}.

Composition of such morphisms turns out to be given 
simply by the addition of their
arrow parts:
\[
  (x,\vec f) \circ (x+d\vec f,\vec g)
  =
  (x, \vec f + \vec g)
  \,.
\]

So far this data defines a {\bf 2-vector space}. A Lie 2-algebra is
a 2-vector space with extra structure, the {\bf Lie bracket functor}.
\[
  [\cdot,\cdot] \maps \mathcal{L}\times \mathcal{L} \to \mathcal{L}
  \,.
\]
This turns out to be expressible in terms of a linear map
\begin{eqnarray*}
  l_2 &\maps& V_0 \times V_0 \to V_0\\
  l_2 &\maps& V_0 \times V_1 \to V_1
\end{eqnarray*}
which is antisymmetric on $V_0 \times V_0$
and is given by
\[
  \commutator{(x_1,\vec f_1)}{(x_2,\vec f_2)}
  =
  \left(
    l_2\of{x_1,x_2},\;
    l_2\of{x_1,\vec f_2} - l_2\of{x_2,\vec f_1} + l_2\of{d\vec f_1,\vec f_2}
  \right)
  \,.
\]
Note how this is like a bracket operation on $V_0$ together with a
bracket operation on $V_1$ which is `twisted' by elements of $V_0$.

In the special case that $l_2$ is a Lie bracket on $V_0$,
that $l_2\of{d\cdot,\cdot}$ is a Lie bracket on
$V_1$ and that $V_0$ acts on $V_1$ by derivations,
all this data defines a
{\bf differential crossed module}\,, and 
$\mathcal{L}$ is the Lie 2-algebra of a \emph{strict} Lie 2-group
and is itself called a strict Lie 2-algebra.

More general Lie 2-algebras of the above form are weaker than that and are called
\emph{semistrict}. 
For them the map $l_2$ fails to satisfy the Jacobi identity, which would read
\begin{eqnarray*}
  &&
  l_2\of{x,l_2\of{y,z}}
  +
  l_2\of{z,l_2\of{x,y}}
  +
  l_2\of{y,l_2\of{z,x}}
  =
  0
  \,.
\end{eqnarray*}
This failure is measured by a trilinear antisymmetric map
\[
  l_3 \maps V_0^3 \to V_1
\]
as follows:
\begin{eqnarray*}
  &&
  l_2\of{x,l_2\of{y,z}}
  +
  l_2\of{z,l_2\of{x,y}}
  +
  l_2\of{y,l_2\of{z,x}}
  =
  dl_3\of{x,y,z}
  \,.
\end{eqnarray*}
In this sense the strictness property of a Lie algebra is weakened. But this
weakening requires a consistency condition, a coherence law. This
somewhat intricate law
is discussed below in \S\fullref{Lie 2-algebras} and in more detail
in \cite{BaezCrans:2003} and here we will not bother to write it down.

\vskip 1em

Given the above definitions
it can be easily shown that every strict Lie 2-group has its
strict Lie 2-algebra.

Just like a strict 2-group is defined by a crossed module $(G,H,\alpha,t)$
of groups (\cf end of \S\fullref{Heuristic Motivation of 2-Groups}), 
a strict Lie 2-algebra is defined by a differential
crossed module $(\g,\h,d\alpha,dt)$.

But it turns out to be quite nontrivial to find explicit realizations of
semistrict Lie 2-algebras that are not strict and, moreover, to find
which, if any, weak 2-group these come from.

\subsubsection{The 2-Group $\PG$ and its Relation to $\String\of{n}$}
\label{introduction: the 2-group PG and its relation to string}

Baez and Crans have introduced in \cite{BaezCrans:2003} a familiy
of non-strict Lie 2-algebras $\g_k$ for every ordinary Lie algebra $\g$
which are only very slightly non-strict.
Interestingly, even though the weakening in this case superficially
looks trivial, it turns out that these Lie 2-algebras $\g_k$ give
rise to the highly non-trivial 2-groups $\PG$.

They are defined as follows:

Let $k$ be any real number.
The vector space of objects of the Lie 2-algebra $\g_k$ is simply 
that of the Lie algebra $\g$ itself 
\[
  V_0 = \g
  \,.
\]
The vector space of morphism is just the 1-dimensional one over the
real numbers
\[
  V_1 = \R
  \,.
\]
The map $V_1 \stackto{d} V_0$ is defined to be trivial
\[
  d\of{\R} = 0
\]
as is the action of $V_0$ on $V_1$:
\[
  l_2\of{V_0,V_1} = 0
  \,.
\]
The interesting aspect of this is that,
due to the triviality of $d$, the above
conditions say that $l_2$ must be the ordinary Lie bracket on $V_0 = \g$
\[
  l_2\of{x,y} = \commutator{x}{y}\,,\hspace{1cm} \forall\, x,y \in \g
  \,,
\] 
even though $l_3$ may be nontrivial. By solving the coherence law for $l_3$ in the
present case one finds that one can indeed choose it to be non-trivial by setting
\[
  l_3\of{x,y,z} = k \langle x, \commutator{y}{z}\rangle
  \,.
\]
Here $\langle\cdot,\cdot\rangle$ is the Killing
form on $\g$ with respect to some normalization.
The real number $k$ appearing here is the one parameterizing the family
$\g_k$ of Lie 2-algebras.

Despite its simplicity, it turns out that $\g_k$ does not
exponentiate to a Lie 2-group \cite{BaezLauda:2003}.

But since $\g_k$ is a category, not just a set, one has to take care
of the following issue: Any two sets 
with additional structure are `essentially equal' if they are
isomorphic in a way that respects this structure.
This is because sets live in the 1-category $\Set$. But categories themselves
live in the 2-category $\Cat$. For them to be `essentially equal' they don't
have to be isomorphic but have to be just what is called
{\bf equivalent as categories}. This is explained in more detail in
\S\fullref{2-categories}.

A mere isomorphism between two categories would amount to having two 
functors between these categories which are mutually inverse. An equivalence
of these two categories means that these two functors are not necessarily 
inverse, but that their composition is naturally isomorphic to the
identity functor. This is one level weaker than the concept of direct
isomorphism. 

For this reason it makes good sense to search for Lie 2-algebras which
are \emph{equivalent} in the category-theoretic sense to $\g_k$.

By some general reasoning one can find that a promising candidate for
such an equivalent Lie 2-algebra is the \emph{strict} Lie 2-algebra
which comes from the crossed module 
\[
  \Pg = (P_0 \g, \widehat{\Omega_k\g},d\alpha,dt)
  \,.
\] 
Here $P_0 \g$ is the Lie algebra of smooth based paths in $\g$ with
pointwise Lie bracket, while $\widehat{\Omega_k \g}$ is the Kac-Moody
centrally extended Lie algebra of smooth based loops in $\g$ with
pointwise Lie bracket and $d\alpha$ and $dt$ are the natural operations
on these. 

One can then indeed check that the infinite-dimensional Lie 2-algebra
$\Pg$ is equivalent to $\g_k$. This result involves some rather nontrivial
numerical `coincidences' and is presented in detail in 
\S\fullref{equivalence.section}. Interestingly, the nontrivial
$l_3$ that is present in $\g_k$  is related to boundary terms 
of the Kac-Moody cocycle coming from the paths in $P_0 \g$.

In other words, this result says that we can understand the non-strict 
Lie 2-algebra $\g_k$
as being that Lie 2-algebra which is obtained by starting with 
the strict but `large' Lie 2-algebra $\P_k\g$ and then taking isomorphism
classes of objects.

The relevance of this result is that the Lie 2-algebra $\P_k\g$, being strict,
does have a corresponding strict Lie 2-group -- but only if $k$ is integer.
This latter condition is a consequence of the infinite-dimensionality
of $\P_k\g$. But it is in fact nothing but the well-known
condition that the
\emph{level} $k$ of a Kac-Moody centrally extended loop group has to be
an integer.

The strict Lie 2-group corresponding to
$\P_k \g$ is called $\PG$ and is given by the crossed module
\[
  \PG
  =
  (P_0 G, \widehat{\Omega_k G}, \alpha,t)
  \,,
\]
where $P_0 G$ is the group of smooth based paths in $G$ and
$\widehat{\Omega_k G}$ is the Kac-Moody centrally extended group
of smooth based loops in $G$. The details of how this group can
be constructed are reproduced in \S\fullref{PG.construction.section}.

Following the above given motivation, it is now interesting to see how,
for the case $G = \Spin\of{n}$, the 2-group $\PG$ `is' (related to) the
group $\String\of{n}$.

This requires some explanation, but in a simple form it can already be
made plausible as follows:

First note that there is a simple way in which we can construct a
Lie 2-algebra $\mathrm{b}\u\of{1}$ from $\u\of{1}$
by `lifting' $\u\of{1}$ to the space of morphisms, i.e. by
letting $V_0 = 0$ and $V_1 = \u\of{1} \simeq i \R$.
Similarly, we can regard any ordinary Lie algebra $\g$ as a 
Lie 2-algebra by setting $V_0 = \g$ and $V_1 = 0$.

There is a morphism of Lie 2-algebras $\g_k \to \g$ which just forgets
about the label of the morphisms in $\g_k$.
Similarly, there is a morphism of Lie 2-algebras $\mathrm{b}\u\of{1} \to \g_k$
which is just the injection of morphisms based at $x=0$. 
Obviously we have an exact sequence
\[
  0 \to \mathrm{b}\u\of{1} \to \g_k \to \g \to 0
  \,.
\]
This simple sequence can be thought of, in a sense to be made precise shortly, 
as the infinitesimal version of the sequence of groups mentioned before:
\[
  1 \to K\of{\Z,2} \to \String\of{n} \to \Spin\of{n} \to 1
\]
(for the case that $\g = \mathrm{Lie}\of{\Spin\of{n}}$). This means
that $\g_k$ is in a certain sense the infinitesimal version of the sought-after
group $\String\of{n}$.

Essentially the same statement can be made as follows: 
If we think of the ordinary group $G$ as a 2-group with 
just identity morphisms,
then there is an obvious morphism of 2-groups $\PG \stackto{\pi} G$ which sends
every path in $P_0 G$ to its endpoint and just forgets about the morphisms.
This map has as strict kernel the 2-group called $\LG$ which is the sub-2-group
of $\PG$ where all objects are closed paths.
Hence there is a strictly exact sequence 
of 2-groups
\[
  1 \to \LG \stackto{\iota} \PG \stackto{\pi} G \to 1
  \,. 
\]
This sequence again can be identified in a certain sense with the sequence
defining the group $\String\of{n}$ 
for the case that $\g = \mathrm{Lie}\of{\Spin\of{n}}$, which implies that
in this sence $\PG$ is like $\String\of{n}$.

The sense in which this is true is the following:

Every topological category $C$ (one whose set of objects and morphisms are topological
spaces such that all operations in the category are continuous functions)
gives rise to a simplicial space, called the \emph{nerve} of this category.
By using the topology on $C$ this simplicial space can be turned into a 
topological space called the `geometric realization' of the simplicial space.
This topological space is called $|C|$.

Here is how to think of the space $|C|$:
The points in $|C|$ are all the objects of our category $C$.
The edges in $|C|$ are the morphism in $C$. The triangles in $|C|$ are given by
pairs of composable morphsism in $C$. And so on:
$(p+1)$-simplices in $|C|$ are given
by $p$-tuples of composable morphism in $C$. 

When the category $C$ is a topological 2-group, the group multiplication
defined on it makes $|C|$ into an ordinary topological group.
In fact, the operation $|\cdot|$ is a functor from the category of 
topological 2-groups to that of topological groups. 

Hence when this functor is applied to the above sequence of 2-groups
we obtain the sequence
\[
  1 \to |\mathcal{L}_1G| \stackto{|\iota|} |\P_1G| \stackto{\pi} G \to 1
\]
of topological groups, where one uses the fact that $|G| = G$.

Finally, from an old result by Segal it follows that 
$|\mathcal{L}_1 G|$ is
in fact the Eilenberg-MacLane space $K\of{\Z,2}$ space.
As is explained in \S\fullref{more details on Spinning Strings},
this implies that for $G =\Spin\of{n}$ we have
$|\P_1 G| \simeq \String\of{n}$.

\subsection{Principal 2-Bundles}
\label{introduction: principal 2-bundles}

With the concept of 2-group in hand, we can now define and
study the categorification of principal bundles so as to obtain
2-bundles with {\bf structure 2-group}.

The process of \emph{categorification} essentially amounts 
to taking any algebraic structure, expressing it in terms of 
diagrams which depict morphisms in some 1-category, and then
re-interpreting the very same diagrams as depicting morphisms
in a suitable 2-category. 

The study of categorified bundles, which was initiated by 
Toby Bartels \cite{Bartels:2004}, is, as far as bare bundles
without any extra structure and properties are concerned,
actually the simplest example of this procedure.

This is because an ordinary bundle is nothing but some space $E$,
called the {\bf total space}, some space $B$, called the
{\bf base space}, together with a map
\[
  E \stackto{p} B
\]
called the projection map.
Better yet, $E$ and $B$ are objects in some category $C$ and
$p$ is a morphism in that category. This is the simplest
``diagram'' that one can think of.

For instance, if $C$ is the category of topological spaces, then
$E$ and $B$ are any two topological spaces and $p$ is any continuous map
from $E$ to $B$. Or, if $C$ is the category of smooth spaces then
$E$ and $B$ are any two smooth spaces and $p$ is a smooth map
between them. 

If we forget about all extra structure for the moment we can think
of $C$ as the category $\Set$ whose objects are (small) sets and whose
morphisms are nothing but functions between sets. Then $E$ and $B$ are any
two sets and $p$ is any function of sets between $E$ and $B$. 

Now, the category $\Set$ of all (small) sets has a natural categorification,
namely the 2-category $\Cat$ of all categories. The objects of $\Cat$
are categories, the morphisms of $\Cat$ are functors between categories
and the 2-morphisms of $\Cat$ are natural transformations between these
functors (\cf \S\fullref{2-categories}). 

This means that if we interpret the diagram
\[
  E \stackto{p} B
\]
as a morphism in $\Cat$ it describes
\begin{itemize}
  \item
   a category $E$
  \item
   a category $B$
  \item
   a functor $p \maps E \to B$
  \,.
\end{itemize}
This is what is called a {\bf 2-bundle}.

Like an ordinary (1-)bundle is just a map of elements of one set $E$ 
to elements of another set $B$, a 2-bundle is a map of \emph{morphisms}
(together with their source and target objects, of course) 
of the category $E$ to morphisms of the category $B$, such that
composition of morphisms is respected. 

This is easy enough. Now let us add extra properties to the
notion of ``bundle'' and categorify them, too. The bundles that play a
role in gauge theory are {\bf locally trivializable fiber bundles}
within the category of smooth spaces.
This means that they have the property that 
the pre-image $E|_{U_i} = p^{-1}\of{U_i}$ of a contractible patch
$U_i$ of the smooth base space $B$ is isomorphic, via an isomorphism
$t_i$, to the cartesian
product of $U_i$ with another smooth space, $F$, called the
{\bf typical fiber}, such that this diagram commutes:
\[ 
\xymatrix @!0
{ E|_{\twoU_i}
 \ar [dddrr]_{p}
  \ar[rrrr]^{t_i} 
  & &  & &
  U_i \times F
  \ar[dddll]^{}
  \\ \\ &  \\ & &
   U_i 
 }
 \,.
\]
(Here the arrow on the right is just the projection from $U_i \times F$
onto the $U_i$ factor.)

Let us again think of this as a diagram in $\Set$.
The fact that this diagram commutes really means that there is
a 2-morphism from the 1-morphism 
$E|_{U_i} \stackto{t_i} U_i \times F \to U_i$ to the 1-morphism
$E|_{U_i} \stackto{p} U_i$. But since $\Set$ is a 1-category, all its
2-morphisms are identity 2-morphisms, which implies the
intended equality
\begin{eqnarray*}
&&E|_{U_i} \stackto{t_i} U_i \times F \to U_i
\\&=&
E|_{U_i} \stackto{p} U_i
 \,.
\end{eqnarray*}
Categorifying this diagram means taking the same diagram, but interpreting
it as a diagram in $\Cat$. Then the diagram says that there is a 
\emph{functor}
$E|_{U_i} \stackto{t_i} U_i \times F \to U_i$ and a \emph{functor}
$E|_{U_i} \stackto{p} U_i$, since 1-morphisms in $\Cat$ are functors
between categories. 

But $\Cat$, being a 2-category, also has 2-morphisms between its 
1-morphisms, given by natural transformations between functors.
So in $\Cat$ the implcit 2-morphism expressing the commutativity
of the above diagram becomes a (invertible) natural transformation
$\lambda$
\begin{eqnarray*}
&&E|_{U_i} \stackto{t_i} U_i \times F \to U_i
\\&\overset{\lambda}{\Rightarrow}&
E|_{U_i} \stackto{p} U_i
 \,.
\end{eqnarray*}

This is how categorification works. Of course, one can now refine 
this construction by adding additional structure. 
Usually we want our ordinary bundles 
to be not just sets with functions between sets, but to be smooth
spaces with smooth maps between them. This amounts to interpreting the
above diagrams not in $\Set$, but in $\Diff$, the category of smooth
spaces, whose objects are smooth spaces and whose morphisms are
smooth maps.

If we want to have something similar in the categorified case, we need
a 2-category
$\twoDiff$,
whose objects are ``smooth categories'', whose morphisms
are ``smooth functors'' and whose 2-morphisms are ``smooth natural
transformations'' between smooth functors. 

A smooth category can be defined as a category whose diagrammatic
definition is interpreted not in $\Set$, but in $\Diff$. This means that
it is a category which has not just a \emph{set} of objects and a 
\emph{set} of morphisms, but where the objects and morphisms both form a
smooth space and where all operations in the category, such as composition
of morphisms, is given by smooth maps between smooth spaces. Since such
a smooth category can be regarded as the categorification of the concept
of a smooth space, it is called a {\bf 2-space} \cite{Bartels:2004}.

The archetypical example of this is the category obtained
from the space of paths in some smooth space $U$. Objects are all the
points of $U$ and morphisms are all the paths of $U$. Physicists should
think of such a category as a configuration space for a (open) string,
being the categorification of an ordinary space $U$, which can be 
regarded as the configuration space of a particle.

Hence the locally trivializable 2-bundles that we shall be concerned 
with are given by the diagrams
\[
  E \stackto{p} B
\]
and
\begin{eqnarray*}
&&E|_{U_i} \stackto{t_i} U_i \times F \to U_i
\\&\overset{\lambda}{\Rightarrow}&
E|_{U_i} \stackto{p} U_i
 \,,
\end{eqnarray*}
interpreted as diagrams in the 2-category $\twoDiff$ of smooth spaces.

This game can now be played further. Next we will want to consider the
categorification of the concept {\bf principal fiber bundle}.
A principal fiber bundle is a locally trivializable bundle of the
above kind such that the typical fiber $F$ is a Lie group,
$F = G$, and such that the transition 
$g_{ij} \defas t_i^{-1} \circ t_j|_{U_{ij}}$ from
the trivialization over $U_i$ to the trivialization over $U_j$ is
given by multiplication in the group $G$. 

We already know from 
\S\fullref{introduction: 2-Groups as Categorified Groups}
how to categorify this part. Hence a {\bf principal 2-bundle} is 
a locally trivializable 2-bundle of the above sort such that the
typical fiber category is a Lie 2-goup $\twogroup$.

One can now essentially copy all the considerations concerning 
ordinary principal bundles from the textbook, while contiuously
expressing everything in terms of diagrams in $\Diff$ and 
re-interpreting all these diagrams in $\twoDiff$, thus obtaining the
theory of principal 2-bundles. 
Up to some details this amounts essentially to applying the
dictionary from \S\fullref{motivation: categorification} to all
the familiar phenomena.

For instance in ordinary principal bundles we have the
transition \emph{functions}
\[
 g_{ij} \maps U_{ij} \to G
\]
from double overlaps of patches of the base space to the
structure group $G$, and these functions satisfy an
\emph{equation}
\[
  g_{ik} = g_{ij} \cdot g_{jk}
\]
on triple overlaps $U_{ijk}$. According to the dictionary in 
\S\fullref{motivation: categorification} a principal 2-bundle
will have transition \emph{functors}
\[
 g_{ij} \maps U_{ij} \to \twogroup
\]
from a double intersection of patches of the base \emph{2-space}
to the structure 2-group, such that on triple intersections there
is a \emph{natural isomorphism} $f_{ijk}$
\[
  g_{ik} \overset{f_{ijk}}{\Rightarrow} g_{ij} \cdot g_{jk}
  \,.
\]
Functions are replaced by functors, and equations between functions
are replaced by natural isomorphisms between functors. The
existence of the natural isomorphism $f_{ijk}$ is characterized
by the following naturality diagram:
\begin{center}
  \begin{picture}(370,150)
    \includegraphics{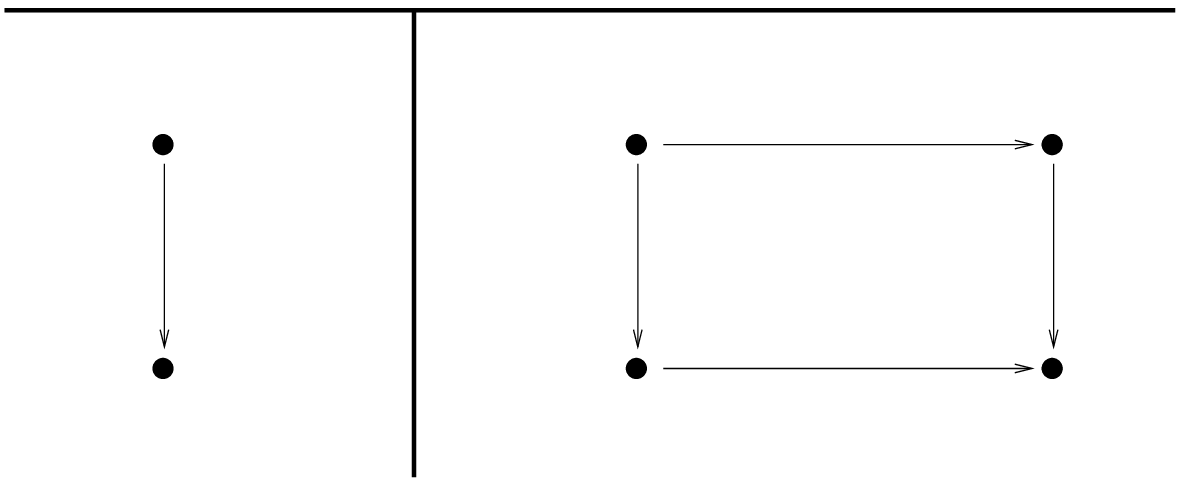}
    \put(-296,142){$U_{ijk}$}
    \put(-100,142){$\twogroup$}
    \put(-296,107){$x$}
    \put(-296,20){$y$}
    \put(-303,65){$\gamma$}
    \put(-180,107){$g_{ik}\of{x}$}
    \put(-180,20){$g_{ik}\of{y}$}
    \put(-186,65){$g_{ik}\of{\gamma}$}
    \put(-50,107){$g_{ij}\cdot g_{jk}\of{x}$}
    \put(-50,20){$g_{ij}\cdot g_{jk}\of{y}$}
    \put(-33,65){$g_{ij}\cdot g_{jk}\of{\gamma}$}
    \put(-120,104){$f_{ijk}\of{x}$}
    \put(-120,20){$f_{ijk}\of{y}$}
  \end{picture}
\end{center}

Here $x \stackto{\gamma} y$ is a morphism in the 2-space $U_{ijk}$
which is mapped by $g_{ik}$ to the morphism
$g_{ik}\of{x} \stackto{g_{ik}\of{\gamma}} g_{ik}\of{y}$
in the 2-group $\twogroup$. Similarly for
$g_{ij}\cdot g_{jk}$, where $\cdot$ denotes the product functor in
the 2-group.

We shall mostly be interested in this diagram for the special case that
the base 2-space $B$ of our 2-bundle is really just an ordinary space,
i.e. a smooth category all of whose morphisms are identity morphisms.

In this case the above naturality diagram reduces to
\begin{eqnarray}
  \label{nat diagram for transition on catdisc base space}
  \begin{picture}(370,150)
    \includegraphics{nattraf.eps}
    \put(-296,142){$U_{ijk}$}
    \put(-100,142){$\twogroup$}
    \put(-296,107){$x$}
    \put(-296,20){$x$}
    \put(-305,65){$\mathrm{Id}$}
    \put(-180,107){$g_{ik}\of{x}$}
    \put(-180,20){$g_{ik}\of{x}$}
    \put(-170,65){$\mathrm{Id}$}
    \put(-50,107){$g_{ij}\cdot g_{jk}\of{x}$}
    \put(-50,20){$g_{ij}\cdot g_{jk}\of{x}$}
    \put(-33,65){$\mathrm{Id}$}
    \put(-120,104){$f_{ijk}\of{x}$}
    \put(-120,20){$f_{ijk}\of{x}$}
  \end{picture}
\end{eqnarray}
Obviously, this diagram commutes in any case. Hence the existence of
the natural isomorphism $f_{ijk}$ becomes tantamount to the
mere existence of
a morphism $g_{ik}\of{x} \stackto{f_{ijk}\of{x}} g_{ij}\cdot g_{jk}\of{x}$
in $\twogroup$. If we think of $\twogroup$ as a 2-category with a
single object $\bullet$, then the diagram expressing this is the
following:
\begin{center}
 \begin{picture}(140,110)
  \includegraphics{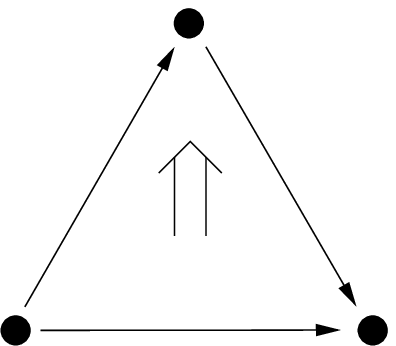}
  \put(-100,50){$g_{ij}$}
  \put(-25,50){$g_{jk}$}
  \put(-70,-5){$g_{ik}$}
  \put(-70,24){$f_{ijk}$}
  \end{picture}
  \vskip 1em
\end{center}
From the properties of (strict) 2-groups that were discussed in
\S\fullref{introduction: 2-Groups as Categorified Groups}, we know that
$f_{ijk}$ is represented by a pair
\[
  (g_{ik} \in G, f_{ijk} \in H)\,,
\]
where $G$ and $H$ are the two groups of the crossed module that
describes the 2-group $\twogroup$, and where we 
conveniently denote the element in $H$
by the same letter as the morphism itself. For this morphism to have 
$g_{ij}\cdot g_{jk}$ as its target, the relation
\[
  g_{ij}g_{jk} = t\of{f_{ijk}}g_{ik}
\]
has to hold. In the context of 
nonabelian gerbes this is known as one of the cocycle relations
describing these structures.  It generalizes the
ordinary relation $g_{ij}g_{jk} = g_{ik}$ of an ordinary bundle.

But now there must be another equation governing the
transition of the $f_{ijk} \in H$ themselves. This is given
by a {\bf coherence law} in the 2-bundle.

Recall that categorification implied the weakening of equations
to mere natural isomorphisms. We had indicated how this can be 
understood as replacing identity 2-morphisms in a 1-category like
$\Set$ with nontrivial 2-morphisms in a 2-category like $\Cat$.
But there are also identity 3-morphisms in $\Set$, which however
remain ``invisible'' since they only go between identity 2-morphisms.
But in $\Cat$ we can have identity 3-morphisms between non-identity
2-morphisms. These give rise to coherence laws. These laws ensure
that the weakening that takes place in categorification is consistent.

In the present case this means the following. Since $f_{ijk}$ is invertible,
the above triangular diagram can be read as a prescription for reducing
$g_{ij}g_{jk}$ to $g_{ik}$ by using $\bar f_{ijk}$, the inverse of
$f_{ijk}$. This is similar to a product operation and we want this operation
to be \emph{associative}. In other words, the transformation of
$g_{ij}g_{jk}g_{kl}$ to $g_{il}$ on quadruple overlaps $U_{ijkl}$
must be well defined. This is expressed by the following equation:
\begin{eqnarray*}
\begin{picture}(300,130)
\includegraphics{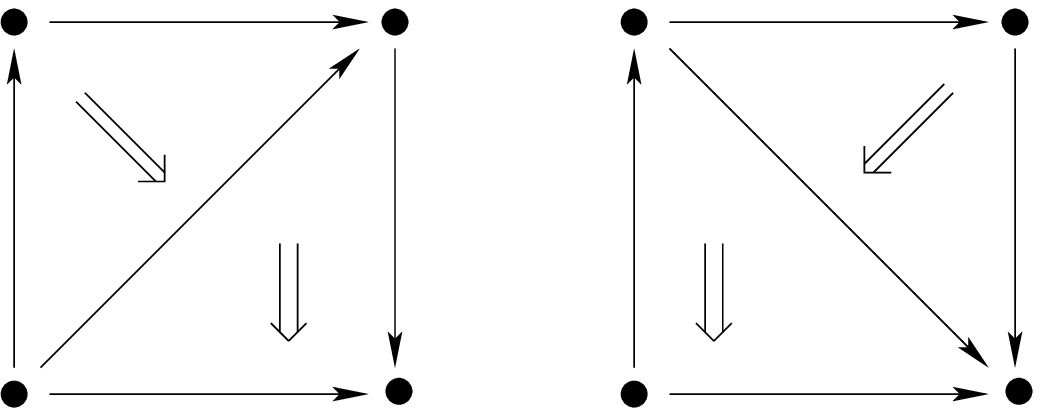}
\put(-7,2){
\begin{picture}(300,150)
\put(-245,115){${g}_{jk}$}
\put(-65,115){${g}_{jk}$}
\put(-245,-5){${g}_{il}$}
\put(-65,-5){${g}_{il}$}
\put(-307,55){${g}_{ij}$}
\put(-127,55){${g}_{ij}$}
\put(-180,55){${g}_{kl}$}
\put(-1,55){${g}_{kl}$}
\put(-152,55){$=$}
\put(-280,30){${g}_{ik}$}
\put(-40,21){${g}_{jl}$}
\put(-58,80){$\bar f_{jkl}$}
\put(-82,30){$\bar f_{ijl}$}
\put(-234,30){$\bar f_{ikl}$}
\put(-254,77){$\bar f_{ijk}$}
\end{picture}
}
\end{picture}
\end{eqnarray*}
When working out what this equation says in terms of ordinary
group elements one finds the relation 
\[
  f_{ijk}f_{jkl} = \alpha\of{g_{ij}}\of{f_{jkl}}f_{ijl}
  \,.
\]
This is known in the theory of nonabelian gerbes as the basic cocycle
relation for $f_{ijk}$.

In order to better see what this has to do with identity 3-morphisms,
note that the two sides of the above equation can be glued along their
common boundary to produce a tetrahedron like this:
\begin{center}
\begin{picture}(180,170)
  \includegraphics{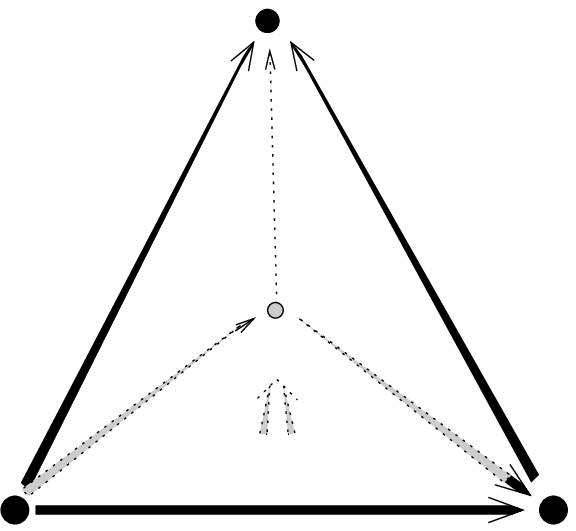}
  \put(-117,56){${}_{g_{ij}}$}
  \put(-63,56){${}_{g_{jk}}$}
  \put(-90,-3){$g_{ik}$}
  \put(-100,22){$f_{ijk}$}
  \put(-137,80){$g_{il}$}
  \put(-41,80){$g_{kl}$}
  \put(-82,95){${}_{g_{jl}}$}
\end{picture}
\end{center}
\vskip 2em
The above equation is equivalent to saying that there is a 3-morphism inside
this tetrahedron, going between the two 2-morphisms that constitute the
boundary of the tetrahedron. Since we are working with 2-bundles that
live in the 2-category $\twoDiff$, this 3-morphism has to be an
identity 3-morphism and hence the 2-morphisms on the boundary have to
satisfy an equation.

\vskip 2em

This are the basic ideas concerning principal 2-bundles.
Next we discuss how to define 2-connections with 2-holonomy on these
2-bundles. That will allow us to describe nonabelian strings.

\subsection{Global 2-Holonomy}
\label{introduction: global 2-holonomy}

In 1985, Alvarez had stated
\cite{Alvarez:1985}, 
motivated by topological field theory,
a procedure for computing global surface holonomy 
for what would now be called abelian gerbes with connection and
curving, or, as we now know, 
equivalently
2-bundles over ordinary base spaces with
structure group given by the crossed module 
$(G = 1, H = U\of{1},\alpha = \mathrm{trivial}, t = \mathrm{trivial})$.
This formula, which also appeared in \cite{Gawedzki:1988},
uses a covering of base space, works in local patches and
glues things appropriately.

This procedure was later found to be precisely the right one
to describe the coupling of the fundamental string to the Kalb-Ramond
field. In the context of open strings attached to D-branes
it is discussed for instance in 
\cite{Kapustin:1999,GawedzkiReis:2002}.
In \cite{CareyJohnsonMurray:2004} it is shown how this procedure
is equivalent to a more intrinsic definition
of holonomy of an abelian gerbe as defined for instance in
\cite{Chatterjee:1998,MackaayPicken:2000}.

In order to describe ``nonabelian strings'' we are interested in
a generalization of this formula to more general cases and in an
understanding of its mechanism in terms of categorified holonomy
in 2-bundles. 

\vskip 2em

There are several equivalent definitions of the notion of an 
ordinary connection on an ordinary bundle. The categorifications of
all these definitions need not be equivalent, however. What we are
interested in is a notion of categorified connection (2-connection)
that allows to define a notion of categorified parallel 
transport and categorified holonomy (2-holonomy) and hence a notion of 
nonabelian strings.
Therefore the definition of connection that we want to categorify is that
which characterizes a connection as something that allows to do
parallel transport. Furthermore, we want to work with local
trivializations in order to obtain explicit algorithms for how to
compute global surface holonomies from a set of local differential 
forms on base space. 

So the definition of an ordinary connection that is best suited for our needs
is the following:

\emph{
  Given a base manifold $M$ with good covering $\covering \to M$ and
  a Lie group $G$, a {\bf connection on a principal $G$-bundle} 
  over $M$ is
  \begin{itemize}
    \item
     on each patch $U_i$ a functor
     \[
       \hol_i \maps \P_1\of{U_i} \to G
     \]
     from the groupoid 
     $\P_1\of{U_i}$ of paths in $U_i$ 
     (\cf \S\fullref{introduction: categories: examples})
     to the group $G$ (regarded as
     a category with a single object),
    \item
     on each double overlap $U_{ij}$ an invertible 
     \emph{morphism of functors}
     \[
       \hol_i \stackto{g_{ij}} \hol_j
       \,,
     \]
     i.e. a natural isomorphism between the restrictions of $\hol_i$
     and $\hol_j$ to $U_{ij}$,
    \item
    on each triple overlap an equation
    \[
      g_{ij} \circ g_{jk} = g_{ik}
    \]
    between these natural isomorphisms.
  \end{itemize}
}
In the following \S\fullref{introduction: 1-connections in 1-bundles}
we briefly point out how this does indeed encode the familiar 
properties of a $G$-connection.

The above definition has a straightforward elegant categorification.
In fact, it is just as easy to categorify it once as to categorify
it once again. Therefore we would like to state the general 
notion of {\bf $p$-holonomy} in the above sense, for arbitrary integer $p$.

An $n$-category has objects, (1-)morphisms between objects, 
2-morphism between 1-morphisms, 3-morphisms between 2-morphisms,
and so on, up to $n$-morphisms between $(n-1)$-morphisms
(\cf \S\fullref{elements of cat theory: 2-functors and pseudo-nat trafos}). 

Up to
technical details it is obvious what a 
{\bf $p$-category of $p$-paths} in a patch $U_i$ should look like:
Its objects should be points in $U_i$, its morphisms should be paths
in $U_i$, its 2-morphisms should be bounded surfaces in $U_i$, and so on,
up to $p$-dimensional hypervolumes in $U_i$.
When done correctly, all the $n$-morphisms in such a category are
invertible up to equivalence, and hence we really have a
{\bf $p$-groupoid of $p$-paths} in $U_i$. 
This we shall call $\P_p\of{U_i}$. 

Similarly, like we have 1-groups and 2-groups, we can consider
$p$-groups for general integer $p$. Let us pick some such $p$-group
and call it $G_p$, the {\bf structure $p$-group}. 
(In fact, the definition of $p$-holonomy that
we are about to give works just as well if $G_p$ is just a 
$p$-groupoid.)

A $p$-connection has to be something that, locally, 
labels $n$-dimensional
hypervolumes in a patch $U_i$ by some $n$-morphisms of the structure
$p$-group in a way that is compatible with the composition of
such hypervolumes. This is nothing but a $p$-functor
\[
  \hol_i \maps \P_p\of{U_i} \to G_p
  \,.
\]

Two (1-)functors can be related by a morphism of 1-functors
(= a
natural transformation), if their images are ``homotopic'', i.e. if their
images can be translated into each other inside the target category.
Similarly, there are morphisms of this sort between $p$-functors, sometimes
called pseudo-natural transformations. But, for $p > 1$, there can
in addition be 2-morphisms between these pseudo-natural transformations, and 
3-morphisms between these, and so on, up to $p$-morphisms between
$(p-1)$-morphisms of $p$-functors. 

This gives rise to the following conception of $p$-connection:

\emph{
  Given a base manifold $M$ with good covering $\covering \to M$ and
  a Lie $p$-group(oid) $G_p$, a
  {\bf $p$-connection with $p$-holonomy
   on a locally trivialized principal $G_p$-$p$-bundle} 
  over $M$ is
  \begin{itemize}
    \item
     on each patch $U_i$ a $p$-functor
     \[
       \hol_i \maps \P_p\of{U_i} \to G_p
     \]
     from the $p$-groupoid 
     $\P_p\of{U_i}$ of $p$-paths in $U_i$ to the $p$-group(oid) 
     $G_p$,
    \item
     on each double overlap $U_{ij}$ a
     \emph{morphism of $p$-functors}
     \[
       \hol_i \stackto{g_{ij}} \hol_j
       \,,
     \]
    \item
    on each triple overlap a 2-morphisms of $p$-functors
    \[
      g_{ik} \stackto{f_{ijk}} g_{ij} \circ g_{jk}\,,
    \]
    \item
      in general, on each $n$-fold overlap an $n$-morphisms of 
      $p$-functors between the $(n-1)$-morphisms of $p$-functors
      on the respective $(n-1)$-fold intersections, where the $n$-morphism
      consitutes the interior and the $(n-1)$-morphisms the
      faces of an $n$-simplex.
  \end{itemize}
}

Another way to summarize this is to say that, roughly, a locally
trivialized principal $p$-bundle with $p$-connection is
a simplicial map from an abstract $p$-simplex to the
$p$-category of local holonomy $p$-functors.
See figure \ref{figure: 2-bundle as simpl map}
(p. \pageref{figure: 2-bundle as simpl map}).
This makes it obvious
that a {\bf gauge transformation} from one local trivialization
to another is nothing but a natural transformation of this
simplicial map.

This is the general idea, though we shall mostly be
concerned with the cases $p=1$ and $p=2$ and just a little bit
with $p=3$. We shall show for $p=2$ how this definition encodes
all the cocycle relations of a $p$-bundle with $p$-connection and
$p$-holonomy, and how the local $p$-functors $\hol_i$ glue together
to give a globally defined $p$-holonomy.

The most elegant way to see this is by realizing how the above
construction is really the local trivialization of a global
$p$-holonomy $p$-functor
\[
  \hol \maps \P_p\of{M} \to G_p-p\mathrm{Tor}
\]
from $p$-paths in the base manifold $M$ to the $p$-category of 
$G_p$-p-torsors. More precisely, for a given principal $G_p$-bundle
$E\to M$ over a categorically trivial base space $M$ (i.e. for $M$ an 
ordinary manifold), let $\trans_p\of{E}$ be the 
smooth category whose objects
are the fibers $E_x$, $x\in M$, regarded as $G_p$-p-torsors and whose
$n$-morphisms are the $p$-torsor $n$-morphisms between these. Then a
smooth $G_p$-$p$-bundle $E\to M$ 
with $p$-connection and $p$-holonomy should be
a smooth functor
\[
  \hol \maps \P_p\of{M} \to \trans_p\of{E}
  \,.
\]
We shall show that this is the case for $p=1$ and $p=2$ in 
\S\fullref{p-Functors from p-Paths to p-Torsors}.

It turns out that the above notion of $p$-connection has a
\emph{differential} reformulation, which is quite useful and
much easier to handle in the general case. This is
discussed in \S\fullref{introduction: the differential picture}.

\subsubsection{1-Connections with 1-Holonomy in 1-Bundles}
\label{introduction: 1-connections in 1-bundles}

To start with, let us check how the above definition works in the
familiar case of 1-bundles.

A connection in an ordinary bundle $E\to M$ locally gives rise to a 
functor from the groupoid of paths in the base space to the
structure group, regarded as a category.

Given a patch $U_i \subset M$, 
we denote by $\P_1\of{U_i}$ the groupoid of
paths in $U_i$. The objects of this groupoid are points in $U_i$,
while the morphisms are 
{\bf \emph{thin} homotopy} equivalence classes of smooth
paths between these points. 
Thin homotopy is homotopy induced by \emph{degenerate surfaces}. 
Hence dividing
out by thin homotopy divides out by reparameterizations of paths
and removes \emph{zig-zag moves}, i.e. of pieces of path that 
retrace themselves.
If $G$ is the structure group,
regarded as a category with a single object and all morphisms
invertible, then any smooth functor
\[
  \hol_i \maps \P_1\of{U_i} \to G
\]
defines a connection on the trivial bundle $E|_{U_i} \to U_i$. 
Let us call this functor the local {\bf holonomy (1-)functor}
on $U_i$.

In a well known way the specification of any such functor is
equivalent to choosing a 1-form
\[
  A_i \in \Omega^1\of{U_i,\g = \mathrm{Lie}\of{G}}
  \,.
\]

A gauge transformation is nothing but a natural isomorphism 
\[
  \hol_i \stackto{h} \widetilde \hol_i  
\]
between two such functors. Any such gauge transformation is given by
a group-valued function
\[
  h_i \in C^\infty\of{U_i,G}
\]
and its action on the 1-form $A_i$ coming with $\hol_i$ is
\[
  A_i \to h_i A_i h_i^{-1}  + h_i \extd h_i^{-1}
  \,.
\]

In particular, the holonomy functors on overlapping patches
are related by a gauge transformation induced by the 
transition function of the local trivialization. Hence if
\[
  \hol_i \maps \P_1\of{U_i} \to G
\]
and
\[
  \hol_j \maps \P_1\of{U_j} \to G
\]
are holonomy functors on $U_i$ and $U_j$, respectively, then
their restrictions to the double overlap 
$U_{ij} = U_i \cap U_j$ are naturally isomorphic
\[
  \hol_i|_{U_{ij}} \stackto{g_{ij}} \hol_j|_{U_{ij}}
  \,,
\] 
where the natural isomorphism is induced by the transition function
\[
  g_{ij} \maps U_{ij} \to G
  \,.
\]
In terms of diagrams this means that we have a commuting naturality 
diagram of the following form:
\begin{center}
\begin{picture}(60,200)
  \includegraphics{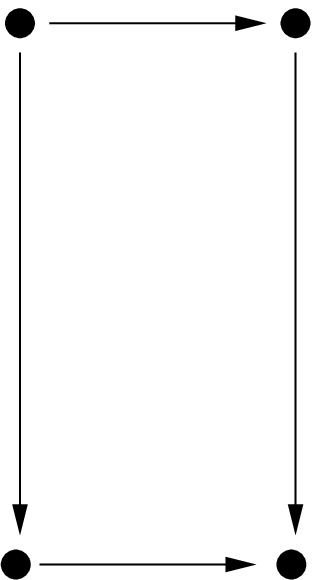}
  \put(-60,167){$g_{ij}\of{x}$}
  \put(-60,-5){$g_{ij}\of{y}$}
  \put(-125,80){$\hol_i\of{[\gamma]}$}
  \put(-1,80){$\hol_j\of{[\gamma]}$}
\end{picture}
\end{center}
Here $[\gamma] \maps x \to y$ is a morphism in $\P_1\of{U_{ij}}$, namely
(the class of) a parameterized path $\gamma$ in $U_{ij}$.

By the above, this implies for the connection 1-forms $A_i$ and
$A_j$ the relation
\[
  A_i = g_{ij} A_j g_{ij}^{-1} + g_{ij}\extd g_{ij}^{-1}
  \,.
\]
This is called the {\bf transition law} or {\bf cocycle condition} for the
connection 1-form.

As is well known, the local holonomy functors $\hol_i$ can be glued
to give a global holonomy functor. Given any path in the base manifold,
this amounts to cutting the path into segments that sit in single
patches, computing the local holonomy of the paths in these patches
and then gluing them by insertions of the transition function $g_{ij}$,
as indicated in the following figure:

\begin{center}
\begin{picture}(200,217)
\includegraphics{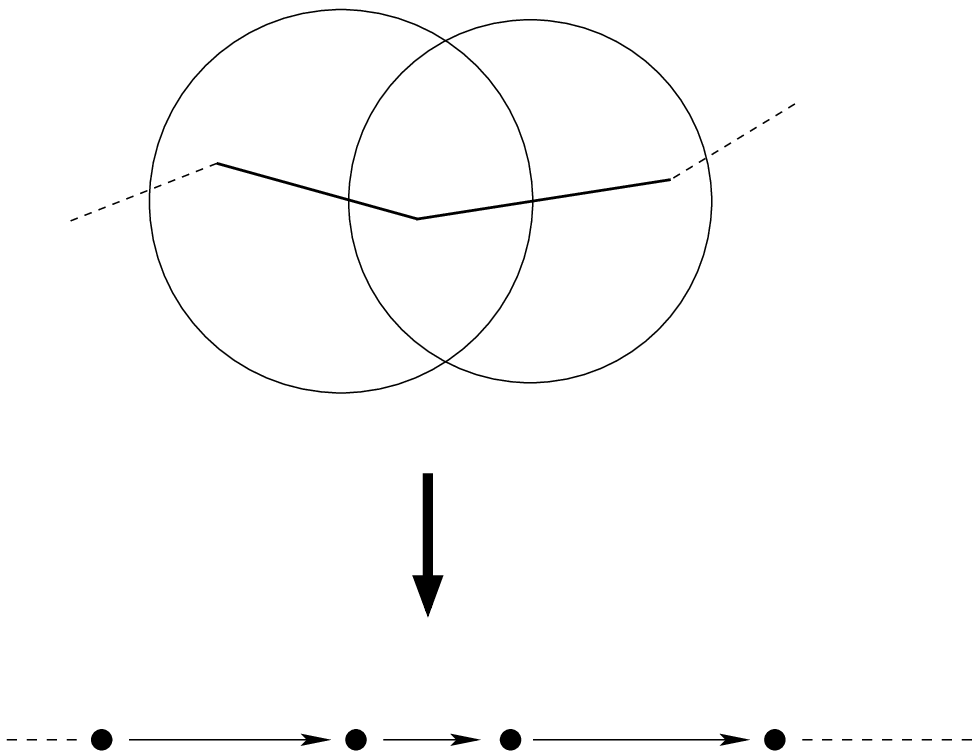}
\put(-200,112){$U_i$}
\put(-126,114){$U_j$}
\put(-203,170){$\gamma_1$}
\put(-121,168){$\gamma_2$}
\put(-235,10){$\hol_i\of{\gamma_1}$}
\put(-117,10){$\hol_j\of{\gamma_2}$}
\put(-173,10){$g_{ij}\of{x}$}
\put(-167,147){$x$}
\put(-154,60){$\hol$}
\end{picture}
%\caption{\label{globallinehol eps}
%{\bf Computing global line holonomy}
%}
\end{center}

In formulas this says that the total holonomy is given by the product
\[
  \cdots \hol_i\of{\gamma_1} g_{ij}\of{x}\hol_j\of{\gamma_2}\cdots
  \,.
\]
This is gauge invariant because under a gauge transformation 
we have
\begin{eqnarray*}
  \hol_i\of{x \stackto{\gamma} y} 
   &\mapsto& 
    h_i\of{x}\hol_i\of{\gamma}h^{-1}\of{y}
  \\
  g_{ij}\of{x} 
  &\mapsto&
  h_i\of{x}g_{ij}\of{x}h_j^{-1}\of{x}
  \,.
\end{eqnarray*}
There is a deeper principle behind this property, which we will 
describe in \S\fullref{the integral picture}.

\subsubsection{2-Connections with 2-Holonomy in 2-Bundles}
\label{introduction: 2-connections in 2-bundles}

We lift this scenario to 2-bundles by categorifying everything
in sight.

This requires first of all to define a 2-path 2-groupoid
$\P_2{U_i}$ on every patch $U_i$. The objects of 
$\P_2\of{U_i}$ are the points $x\in U_i$ and the
morphisms are parameterized paths $x \stackto{\gamma} y$ between them. 
There are now also 2-morphisms 
\[
  [\Sigma] \maps \gamma_1 \to \gamma_2
\]
between paths with
coinciding endpoints. These 2-morphisms are given by
thin homotopy equivalence classes of homotopies between 
the two given paths, i.e. by equivalence classes of surfaces 
which interpolate between two given paths.

We already know that, upon categorifying, 
the structure group $G$ becomes a 2-group
$\twogroup$, which we can regard as a 2-category with a 
single object and with all 1-morphisms and 2-morphisms invertible.

This implies that the categorified local holonomy functors 
$\hol_i$ should be smooth
2-functors from the 2-categories $\P_2\of{U_i}$ to the
structure 2-group:
\[
  \hol_i \maps \P_2\of{U_i} \to \twogroup
  \,.
\]
\begin{eqnarray*}
\hol_i\of{
\xymatrix{
   x \ar@/^1pc/[rr]^{\gamma_1}_{}="0"
           \ar@/_1pc/[rr]_{}_{\gamma_2}="1"
           \ar@{=>}"0";"1"^{[\Sigma]}
&& y
}
}
&\defas&
\xymatrix{
   \bullet \ar@/^1pc/[rr]^{\hol_i\of{\gamma_1}_{}}="0"
           \ar@/_1pc/[rr]_{}_{\hol_i\of{\gamma_2}}="1"
           \ar@{=>}"0";"1"^{\hol_i\of{\Sigma}}
&& \bullet
}
\end{eqnarray*}

One expects that these functors are again specified by differential
forms on $U_i$. We will show that, indeed, specifying such $\hol_i$
is equivalent to specifying a 1-form
\[
  A_i \in \Omega^1\of{U_i,\g}
\]
and a 2-form
\[
  B_i \in \Omega^2\of{U_i,\h}
\]
where $\g = \mathrm{Lie}\of{G}$ and $\h = \mathrm{Lie}\of{H}$
are the Lie algebras belonging to the groups $G$ and $H$ that
constitute the crossed module $(G,H,\alpha,t)$ coming from the
strict 2-group $\mathcal{G}$. 

More precisely, we will show that $\hol_i$ comes from a family
of connection 1-forms
\[
  \mathcal{A}_i \in \Omega^1\of{P_s^t\of{U_i},\h}
\]
on the spaces $P_s^t\of{U_i}$ of paths
in $U_i$ with endpoints $s,t$.
These are given by the formula
\[
  \mathcal{A}_i\of{\gamma}
  =
  \int_\gamma
  \alpha\of{W_{A_i}}\of{\mathrm{ev}^*\of{B_i}}
  \,.
\]
Here
\begin{eqnarray*}
  \mathrm{ev} \maps P_s^t\of{U_i}\times [0,1] &\to& U_i
  \\
      (\gamma,\sigma) & \mapsto & \gamma\of{\sigma}
\end{eqnarray*}
is the \emph{evaluation map} which sends a path $\gamma \in P_s^t\of{U_i}$
and a parameter value $\sigma$ to the position 
$\gamma\of{\sigma} \in U_i$ of the
path at that parameter value, and $W_{A_i}$ denotes the holonomy of $A_i$
along $\gamma$, from the integration parameter to the endpoint.

So this formula tells us to pull back the 2-form $B_i$ 
from $U_i$ to $P_s^t\of{U_i} \times [0,1]$ using the evaluation map,
and then to integrate the result over the path $\gamma$.
The term $\alpha\of{W_{A_i}}$ in this formula indicates that,
while doing this integration, we are to use the ordinary line holonomy
\[
  W_{A_i}\of{\gamma} = \hol_i\of{\gamma}
\]
to continuously parallel transport $\mathrm{ev}^*(B_i)$ to
the endpoint of the path. 

But it turns out that not all combinations $(A_i,B_i)$ correspond to
holonomy 2-functors $\hol_i \maps \P_2\of{U_i} \to \twogroup$.
Instead, the 1-forms $A_i$ and 2-forms $B_i$ that correspond to 
holonomy 2-functors satisfy the relation
\[
  F_{A_i} + dt\of{B_i}  = 0
  \,.
\]
Following 
\cite{BreenMessing:2001} 
we say that the {\bf fake curvature} has to vanish.
This relation can be shown to encode the functoriality of $\hol_i$, i.e.
the fact that $\hol_i$ respects the combined horizontal and vertical 
composition of surfaces in $\P_2\of{U_i}$.

While a gauge transformation for an ordinary (1-)holonomy is the
same as a natural isomorphism between (1-)functors, a gauge transformation
for the above holonomy 2-functors is a pseudo-natural 
isomorphism 
(\cf \S\fullref{elements of cat theory: 2-functors and pseudo-nat trafos}).

This means that for every surface 
$\gamma \stackto{\Sigma} \tilde \gamma \in 
  \Mor_2\of{\P_2\of{U_{ij}}}$
in a double overlap, there are 2-group elements
\begin{eqnarray*}
  a_{ij}\of{\gamma}, a_{ij}\of{\tilde \gamma}
\end{eqnarray*}
such that the images of $\Sigma$ under $\hol_i$ and $\hol_j$
are related by the following 2-commuting diagram:
\begin{center}
\begin{picture}(240,160)
 \includegraphics{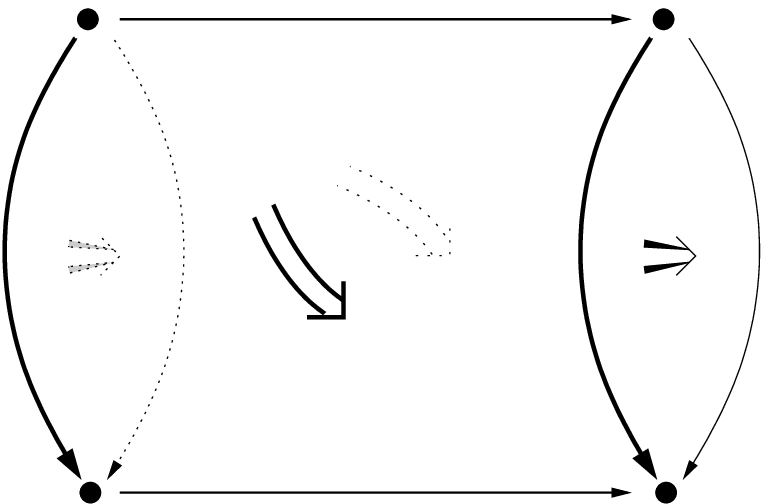}
 \put(-3,-17){
 \begin{picture}(0,0)
 \put(122,0){
  \begin{picture}(0,0)
   \put(-378,70){$\hol_i\of{\gamma}$}
   \put(-293,70){${}_{\hol_i\of{\tilde \gamma}}$}
   \put(-333,105){$\hol_i\of{\Sigma}$}
   \put(-290,106){$a_{ij}\of{\gamma}$}
   \put(-250,118){${}_{a_{ij}\of{\tilde\gamma}}$}
   \put(-250,163){$g_{ij}\of{x}$}
   \put(-250,10){$g_{ij}\of{y}$}
  \end{picture}
 }
 \put(288,0){
  \begin{picture}(0,0)
   \put(-378,70){$\hol_j\of{\gamma}$}
   \put(-293,70){${}_{\hol_j\of{\tilde \gamma}}$}
   \put(-333,105){$\hol_j\of{\Sigma}$}
  \end{picture}
 }
 \end{picture}
}
\end{picture}
\end{center}

We will show that in terms of the differential forms $(A_i,B_i)$
that determine the 2-functors $\hol_i$, the 2-commutativity of this
diagram implies that there exist 1-forms
\[
  a_{ij} \in \Omega^1\of{U_{ij},\h}
\]
such that the following equations hold
\begin{eqnarray*}
  A_i &=& g_{ij}A_j g_{ij}^{-1} + g_{ij}\extd g_{ij}^{-1}
          - dt\of{a_{ij}}
  \\
  B_i &=& \alpha\of{g_{ij}}\of{B_j} + \extd a_{ij} + a_{ij} \wedge a_{ij}
  \,.
\end{eqnarray*}

These are the transition laws (cocycle relations) for $A_i$ and $B_i$.

For this to be consistent, there is a condition on the $a_{ij}$.
This is expressed by the
requirement that on triple overlaps $U_{ijk}$
the following diagram 2-commutes, expressing the
existence of a 2-morphism of 2-functors
(a modification):

\begin{center}
\begin{picture}(200,280)
\includegraphics{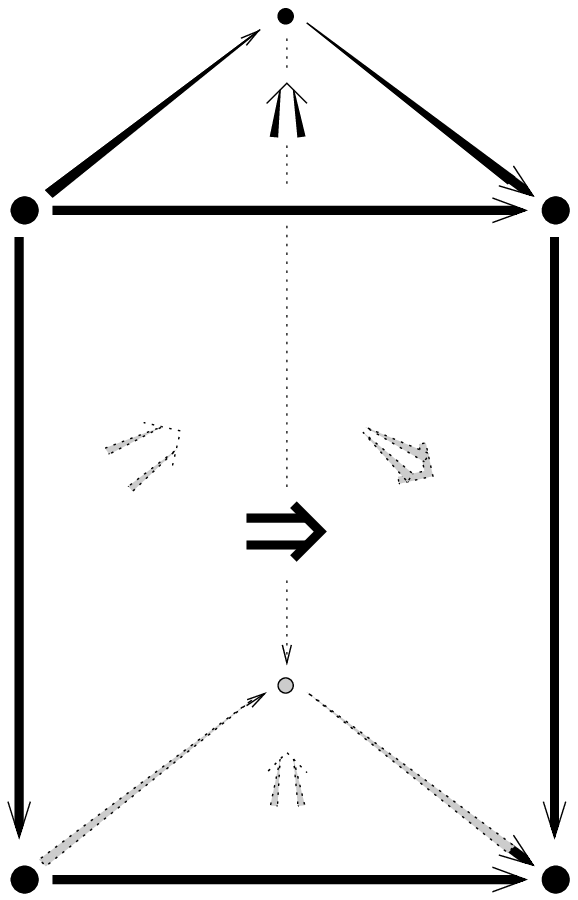}
  \put(-3,100){$\hol_k\of{\gamma}$}
  \put(-194,100){$\hol_i\of{\gamma}$}
  \put(-82,160){${}_{\hol_j\of{\gamma}}$}
  \put(-100,-7){$g_{ik}\of{y}$}
  \put(-100,187){$g_{ik}\of{x}$}
  \put(-138,43){${}_{g_{ij}\of{y}}$}
  \put(-138,237){${}_{g_{ij}\of{x}}$}
  \put(-47,43){${}_{g_{jk}\of{y}}$}
  \put(-47,237){${}_{g_{jk}\of{x}}$}
  \put(-95,23){${}_{f_{ijk}\of{y}}$}
  \put(-95,215){${}_{f_{ijk}\of{x}}$}
  \put(-100,90){$a_{ik}\of{\gamma}$}
  \put(-150,124){${}_{a_{ij}\of{\gamma}}$}
  \put(-50,117){${}_{a_{jk}\of{\gamma}}$}
\end{picture}
\end{center}

It can be shown that this is the case when in terms of local
data the following equation holds:
\begin{eqnarray*}
    f_{ijk}d\alpha\of{A_i}\of{f_{ijk}^{-1}}
    +
    f_{ijk}\extd f_{ijk}^{-1}
    -
    a_{ij}
    -
    g^1_{ij}\of{a_{jk}}
    +
    f_{ijk} a_{ik} f_{ijk}^{-1}
   &=& 
   0
  \,.
\end{eqnarray*}

This is the cocycle condition on the $a_{ij}$.

These cocycle relations have previously been found
by other methods in the
context of nonabelian gerbes in
\cite{BreenMessing:2001, AschieriCantiniJurco:2004}.
Here we obtain them for the
special case that the fake curvature vanishes. The notion
of 2-connection for nonabelian gerbes should come from
categorifying a different definition of ordinary connections
than we used here, one that does not refer to holonomy.
The vanishing of the fake curvature, as we have mentioned,
is a result of our insistence on having a notion of 2-holonomy.

\vskip 1em

It can now be shown that the local 2-holonomy functors $\hol_i$
can be glued together to a global functor that computes global
2-holonomy.

In generalization of the situation for ordinary bundles, this
requires covering the surface with patches $U_i$ and
picking a trivalent graph on the surface such that each face
comes to sit in a single patch, each edge in a double overlap and
each vertex in a triple overlap.
This procedure is familiar from the theory of surface holonomy
for abelian gerbes \cite{Alvarez:1985,MackaayPicken:2000,GawedzkiReis:2002}.
Then to each face of the
graph we can associate the surface holonomy computed by the
local holonomy 2-functor, and the resulting 2-group elements
are glued along their common boundaries by means of
$f_{ijk}$ and $a_{ij}$, which generalize the
transition function $g_{ij}$ known from ordinary bundles.
The resulting action of the global 2-holonomy 2-functor is
illustrated by figure \ref{figure: global 2-holonomy}
(p. \pageref{figure: global 2-holonomy}).

While this may look complicated at first sight, we would like to
emphasize the direct analogy of this procedure to the one for
ordinary 1-connections in 1-bundles, described in
\S\fullref{introduction: 1-connections in 1-bundles}. In fact, staring at
figure \ref{figure: global 2-holonomy}
for a moment reveals that it simply says that the local
surface holonomies $\hol_i$ have to be glued in the only possible
way using the 2-group elements $f_{ijk}$ and $a_{ij}$, just like
global 1-holonomy was obtained by gluing local 1-holonomies in the
only possible way using the group element $g_{ij}$.

In the context of abelian gerbes with connection and
curving a formula for how to compute a global surface holonomy is
well known \cite{Alvarez:1985,MackaayPicken:2000,GawedzkiReis:2002}. 
It is easily seen that this formula arises from the
above diagrammatic prescription in the special case where the
structure 2-group $G_p$ comes from a crossed module of the form
$G_2 = (1,H,\alpha = \mathrm{trivial}, t = \mathrm{trivial})$, with
$H$ an abelian Lie group. The above gives a diagrammatic understanding
of this formula and generalizes it to more general
nonabelian strict 2-groups.

In fact, all the diagrams that we display essentially apply directly to the
much more general case where $G_p$ is any weak coherent 2-group or even
just a coherent 2-groupoid. The minor refinement necessary to describe
the weak case is discussed in an example in
\S\fullref{Weak Principal 2-Bundles}. What is however
much harder for non-strict structure 2-groups is to find the
expression of the respective 2-holonomies and their transition relations
in terms of local differential forms. These are much more conveniently
found using the differential formulation of $p$-bundles with
$p$-connections that is discussed in
\S\ref{introduction: the differential picture} below.

\subsection{The Differential Picture: Nonabelian Deligne Hypercohomology}
\label{introduction: the differential picture}

The above discussion focused on and was motivated by the desire to
write down global surface holonomies, i.e. to associate $p$-group elements
to $p$-paths. As section \S\fullref{section: Path Space}
will demonstrate, there
is quite some gymnastics in path-space
differential geometry required when these ``integral''
notions are to be translated into local differential forms, like $A_i$
and $B_i$.

However, there are situations in which one will be more interested 
in these
local differential forms than in the 2-holonomy that they give rise
to. Most notably, once we are interested not so much in the dynamics
of strings in the background of the fields described by these
local forms, but in the dynamics of these background fields themselves,
the indirect definition of these fields in terms of $p$-holonomy
$p$-functors becomes unwieldy.

For instance, the holonomy 1-functor which was
discussed in \S\fullref{introduction: 1-connections in 1-bundles}
does allow to write down the Yang-Mills action, at least on the lattice,
using Wilson's prescription, but in the continuum limit we will want
to use the action functional in the form 
$\int_{U_i} \mathrm{Tr}\of{F_{A_i} \wedge \star F_{A_i}}$, making use of
$A_i$, which is only somewhat indirectly encoded by
$\hol_i$.

Similarly constructing interesting and sensible action functionals 
for the differential forms
that appear in 2-bundles with 2-connection 
is already a rather more delicate issue, and it turns out that the
``integral formalism'' described above is a clumsy tool 
for attacking such issues.

Recently it had been noticed, for instance in \cite{Strobl:2004}, 
that, apparently, using differential graded algebras (dg-algebras) 
a natural and much more powerful language for dealing with 
higher nonabelian $p$-form gauge theories can be obtained. 
While superficially this approach may look rather unrelated to the
considerations presented here, they are in fact closely related, 
as illustrated in figures
\ref{figure: local 2-holonomy and local 2-connection}
(p. \pageref{figure: local 2-holonomy and local 2-connection}) and
\ref{figure: 2-bundle as simpl map}
(p. \pageref{figure: 2-bundle as simpl map}).

The point is that, like a Lie group can be ``differentiated''
to a Lie algebra, there should be a differential analog of the
local $p$-holonomy $p$-functor  $\hol_i$, called a 
{\bf local $p$-connection $p$-morphism}, which associates 
$n$-morphisms of a Lie $n$-algebra to differential $n$-forms.
We write this as
\[
  \con_i \maps \p_p\of{U_i} \to \g_p
  \,,
\]
where $\p_p\of{U_i}$ is an algebroid called the $p$-path
$p$-algebroid and $\g_p$ is the Lie $p$-algebra associated to the
structure $p$-group $G_p$.

The local $p$-connection $p$-functors $\con_i$ glue together on
multiple overlaps just as before. Hence on double overlaps
there is a 1-morphisms of local $p$-connection $p$-morphisms,
on triple overlaps there is a 2-morphism between such 
1-morphisms, and so on.
\begin{center}
\begin{picture}(180,170)
  \includegraphics{3dtetrahedron.eps}
  \put(-117,56){${}_{\g_{ij}}$}
  \put(-63,56){${}_{\g_{jk}}$}
  \put(-90,-3){$\g_{ik}$}
  \put(-100,22){$\f_{ijk}$}
  \put(-137,80){$\g_{il}$}
  \put(-41,80){$\g_{kl}$}
  \put(-82,95){${}_{\g_{jl}}$}
  \put(-193,-3){$[\con_i]$}
  \put(3,-3){$[\con_k]$}
  \put(-82,67){${}_{[\con_j]}$}
  \put(-82,150){${}_{[\con_l]}$}
\end{picture}
\end{center}
\vskip 2em
It is known that these Lie $p$-algebras, at least as long as they
are what is called ``semistrict'', have a dual description in terms of
dg-algebras. In this language the local $p$-connection becomes a
morphism of dg-algebras, which is known as a \emph{chain map}.
A 2-morphism between such 1-morphisms is what is known as a 
\emph{chain homotopy} between chain maps, and so on.

This has the following interesting consequence: 

There is a natural
nilpotent operator $Q$ acting on the space of dg-algebra $n$-morphisms,
which is essentially the commutator with the differentials of the
target and the source dg-algebra. Two $(n-1)$-morphisms 
of dg-algebras are related by an $n$-morphisms of dg-algebras 
precisely if
they differ by a $Q$-exact term.

Moreover, there is another nilpotent operator, $\delta$,
which sends an $n$-morphism to the linear combination of its source and
target $(n-1)$-morphisms, which label (recall the discussion in
\S\fullref{introduction: global 2-holonomy}) the faces of an $n$-simplex.
More precisely, this $\delta$ is nothing but the 
\emph{{\v C}ech coboundary operator} 
on the complex of sheaves of dg-algebra $n$-morphisms. 

The point is that the differential characterization of a $p$-bundle 
with $p$-connection, which,
according to \S\fullref{introduction: global 2-holonomy}, is
the assignment, $\omega$, of dg-algebra $n$-morphisms 
to $n$-faces of a $p$-simplex,
is concisely encoded in the equation
\[
  (\delta + Q)\omega = 0
  \,.
\]
When the details are unraveled, this equation says nothing but that
the $(n-1)$-morphisms that $\omega$ associates to the $(n-1)$-faces
of an $n$-simplex are the source and target of the $n$-morphism
associated to the simplex itself.

In other words, every $\omega$ in the kernel of the nilpotent operator
\[
  D = \delta + Q
\]
specifies the differential version of the local trivialization of 
a $p$-bundle with $p$-connection.

Moreover, a gauge transformation in that local trivialization 
corresponds to shifting $\omega$ by a $D$-exact term
\[
  \omega \to \omega + D\lambda
  \,.
\]
This way gauge equivalence classes of the differential version of
$p$-bundles with $p$-connection can be characterized by 
\emph{cohomology classes} of the operator $D$.

In the special case that the target Lie $p$-algebra 
$\g_p$ is abelian and strict, one finds that the above operator $D$
reduces to what is known as the {\bf Deligne coboundary operator}.
The cohomology of this operator is well known to describe 
abelian gerbes with connection and curving and hence
abelian 2-bundles with 2-connection. Our generalized operator $D$
should hence be called a {\bf nonabelian Deligne coboundary operator}.
A better term might be {\bf generalized Deligne coboundary operator},
since general $\g_p$ may differ from
strict abelian Lie $p$-algebras not only in being non-abelian, but
also in being semistrict or being $p$-algebroids instead of 
$p$-algebras.

However, the nonabelian Deligne coboundary operator captures
terms in addition to those seen by the ordinary abelian Deligne
operator only to linear order. It lives in a fiber to
the tangent bundle to the
space of all $p$-connections instead of on all of that space. Hence,
in general, a cohomology class of the generalized nonabelian
Deligne coboundary operator does not classify an ``integral''
$p$-bundle with $p$-connection.

We will demonstrate in 
\S\fullref{Infinitesimal 1-Bundles with 1-Connection} 
and \S\fullref{infinitesimal strict 2-bundles}, 
for the cases where $\g_p$
is the differential version of a strict 1-group or of a strict 2-group
$G_p$,
how the equation $D\omega = 0$ does indeed encode the differential
(linearized) version
of all the cocycle relations that were discussed in
\S\fullref{introduction: 1-connections in 1-bundles} and
\S\fullref{introduction: 2-connections in 2-bundles}, and how
a shift $\omega \to \omega + D\lambda$ does indeed describe the
differential version of the gauge transformation laws for these
cases.

Therefore the differential picture and the integral picture of 
$p$-bundles with $p$-connection are somewhat complementary.
While in the differential picture all equations are obtainable only
``infinitesimally'', it can easily deal with situations that are
hard or even impossible to treat using the integral formulation.

For instance, there are non-strict Lie $p$-algebras that are known
not to be integrable to any Lie $p$-group. One example for these
was the family of Lie 2-algebras called $\g_k$ in 
\S\fullref{introduction: the 2-group PG and its relation to string}.
As was discussed there, while $\g_k$ itself is not integrable, it
is equivalent, in the category-theoretic sense, to an infinite-dimensional
strict Lie 2-algebra which is. The differential formalism discussed above
now allows to use $\g_k$ itself as the ``structure 2-algebra'' of a
differential 2-bundle, and to analyze the classification of these
differential 2-bundles by studying the
respective generalized Deligne cohomology
classes - even though there is no integral 2-bundle directly related to
this. 
The discussion of this 
in \S\fullref{Semistrict Infinitesimal gk-2-Bundles with 2-Connection}
concludes the investigations to be presented here.

\clearpage
\section{More Background and More on Motivations}

Much more can be said concerning the motivations, background
and related literature of the
ideas presented here than was done in \S\fullref{motivations}.
The following is an attempt to give a somewhat more detailed
account
\begin{itemize}
  \item
    of the literature on membranes attached to 5-branes in
   \S\fullref{Literature on M2M5},
  \item
    of $n^3$-scaling in these theories and how this could be
    described by 2-bundles with 2-connections in
\S\fullref{does 2-Holonomy capture n-cube scaling behaviour},
  \item
    of the nature of spinning strings in
    \S\fullref{more details on Spinning Strings},
  \item
    of the concepts of ($n$-)category theory
    in \S\fullref{more details on category theory},
  \item
    of the possible relations between 2-bundles
    and the derived category description of open strings
    on D-branes in
    \S\fullref{2-NCG and Derived Category Description of D-Branes}.
\end{itemize}

\subsection{Open Membranes on 5-Branes}
\label{more motivation: Open Membranes on 5-Branes}

\subsubsection{Literature}
\label{Literature on M2M5}

The target space theories which give rise to non-abelian 2-forms are not at all well
understood
\cite{Ganor:1996}.
One expects \cite{BekaertHenneauxSevrin:2000,Hofman:2002}
that they involve stacks of 5-branes on which open membranes may end
\cite{Lechner:2004,BraxMourad:1997,Strominger:1995}. This has recently been
made more precise \cite{AschieriJurco:2004}
using anomaly cancellation on M5-branes and the language of nonabelian gerbes
developed in \cite{AschieriCantiniJurco:2004}. 
The boundary of these membranes appear as strings, 
\cite{Witten:1997,Townsend:1996}, 
(self-dual strings \cite{CallanMaldacena:1997,Gibbons:1997,HoweLambertWest:1997},
``little strings'' \cite{Aharony:1999},
fundamental strings or D-strings 
\cite{LosevMooreSamsonShtashvili:1997}) 
in the world-volume theory of the 5-branes \cite{Strominger:1995}, generalizing \cite{Ganor:1996}
the way how open string endpoints appear as ``quarks'' in the world-volume theory of D-branes.
Just like a nonabelian 1-form couples to these ``quarks'', i.e. to the boundary of an open string,
a (possibly non-abelian) 2-form should couple
\cite{Kalkkinen:1999}
 to the boundary of an open membrane 
\cite{Townsend:1996,BraxMourad:1997b,KawamotoSasakura:2000,BergshoeffBermanVanDerSchaarSundell:2001}, 
i.e. a to string on the (stack of) 5 branes. 
One proposal for how such a non-abelian $B$ field might be induced by a stack of branes
has been made in \cite{Kalkkinen:1999}. A more formal derivation of the non-abelian
2-forms arising on stacks of M5 branes is given in \cite{AschieriJurco:2004}.
General investigations into the possible nature of such non-abelian 2-forms
have been done for instance in \cite{Lahiri:2001,Lahiri:2003}.

(From the point of view of the effective 6-dimensional
supersymmetric worldvolume theory of the 5-branes these 2-form field(s) come either
from a tensor multiplet or from a gravitational multiplet of the 
worldvolume supersymmetry representation \cite{SeibergWitten:1996}.)

This analogy suggests that there is a \emph{single} Chan-Paton-like factor associated to
each string living on the stack of 5 branes, indicating which of the $N$ branes in the stack it 
is associated with. This Chan-Paton factor should be the degree of freedom that the non-abelian
$B$-field acts on.

Hence the higher-dimensional generalization of ordinary gauge theory should, in terms of strings,
involve the steps upwards the dimensional ladder indicated in table \ref{dimensional ladder}.

\TABULAR[h]{|lcl|}{
  \hline
  (1-)gauge theory && 2-gauge theory\\
  \hline\hline
  string ending on D-brane &$\to$& membrane ending on NS brane\\ \hline 
  ``quark'' on D-brane &$\to$& string on NS brane\\ \hline
  nonabelian 1-form gauge field &$\to$&  nonabelian 2-form gauge field $B$\\ \hline
  \parbox{6cm}{
  coupling to the boundary of a 1-brane (string)} 
   &$\to$& 
  \parbox{6cm}{coupling to the boundary of a 2-brane (membrane)}\\ \hline
  \parbox{6cm}{Chan-Paton factor indicating which D-brane in the stack the ``quark'' sits on}
  &$\to$&
  \parbox{6cm}{
  Chan-Paton-like factor indicating wich NS brane in the stack the membrane boundary string sits on.}
  \\ \hline
}{\label{dimensional ladder} Expected relation between 1-form and 2-form gauge theory in stringy terms}

These considerations receive substantiation by the fact that,
indeed, the contexts
in which nonablian 2-forms have been argued to arise naturally are the
worldsheet theories on these NS 5-branes
\cite{Hofman:2002,BekaertHenneauxSevrin:2000,Witten:1997,LosevMooreSamsonShtashvili:1997,Ganor:1996,Witten:1995b}.

The study of little strings, tensionless strings and  $N=(2,0)$ QFTs in six dimensions is
involved, and no good understanding of any non-abelian 2-form from this target space
perspective has emerged so far.
However, a compelling connection is the relation of these 6-dimensional theories,
upon compactification, to
Yang-Mills theory in 4-dimensions, where the 1-form gauge field of the Yang-Mills theory
arises as one component of the 2-form in the 6-dimensional theory \cite{Witten:1995b}.

For this to work the dimension $d=5+1$ of the world-volume theory of the 5-branes plays a crucial role,
because here the 2-form $B$ can have and does have \emph{self-dual} field strength $H = \star H$
\cite{Townsend:1996,Witten:1995b}
(related to the existence of the self-dual strings in 6 dimensions
first discussed in \cite{DuffLu:1993}).

But this means that there cannot be any ordinary non-topological action of the form $dH\wedge \star dH$
for the $B$-field, and that
furthermore the dynamical content of the $B$ field would essentially be that of a 1-form $\alpha$
\cite{Witten:1995b}: Namely when the 1+5 dimensional field theory is compactified on a circle
and $B$ is rewritten as 
\begin{eqnarray}
  B = B_{ij}\,dx^i \wedge dx^j + \alpha_i dx^i \wedge dx^6 \hspace{1cm}\mbox{for $i,j \in \set{1,2,3,4,5}$}
  \,.
  \nonumber
\end{eqnarray}
with $\partial_6 B = 0$, then $dB = \star dB$ implies that in five dimensions $B$ is just dual
to $\alpha$
\begin{eqnarray}
  d^{(5)}B = \star^{(5)} d\alpha
  \,.
\end{eqnarray}
In particular, since the compactified theory should give possibly non-abelian Yang-Mills with $\alpha$
the gauge field \cite{Witten:1995b} it is natural to expect 
\cite{Hofman:2002} that in the uncompactified theory there must be a non-abelian $B$ field.
Since there is no Lagrangian description of the brane's worldvolume theory 
\cite{Witten:1996,BekaertHenneauxSevrin:2000} 
it is hard to make this explicit. This is one reason why it seems helpful to consider the
worldsheet theory of strings propagating in the 6-dimensional brane volume.
The non-abelian
Yang-Mills theory in the context of NS 5-branes considered in \cite{Witten:1997} uses $n$ D4-branes
suspended between two NS5-branes. The former can however be regarded as a single M5-brane wrapped $n$
times around the $S^1$ (\cf p. 34 of \cite{Witten:1997}).

In \cite{BermanHarvey:2004} it is argued that, while the worldvolume theory
on a stack of 5-branes with non-abelian 2-form fields is not known, it cannot
be a local field theory. This harmonizes with the attempts in \cite{Ganor:1996}
to define it in terms of ``nonabelian surface equations'' which are supposed to generalize the
well-known Wilson loop equations of ordinary Yang-Mills theory to Wilson \emph{surfaces}.
These Wilson surfaces become ordinary Wilson loops in loop space, and should
be closely related to the notion of 2-holonomy presented here.

We should emphasize that 
it is an open question whether gerbes and 2-bundles 
are the right language to describe
stacks of 5-branes, since their physics is not understood well enough at 
this point.

\subsubsection{$n^3$-Scaling}
\label{does 2-Holonomy capture n-cube scaling behaviour}

Albeit the effective field theories on stacks of 5-branes
are not very well understood, it it known that they have
a property called {\bf $n^3$-scaling behaviour}
\cite{KlebanovTseytlin:1996,
GubserKlebanov:1997,
GubserKlebanovArkady:1997,
HenningsonSkenderis:1998,
BastianelliFrolovTseytlin:2000,Berman:2003}.

An ordinary gauge theory gives rise to an entropy which asymptotically
scales like the \emph{square} of the 
rank of
the gauge group, i.e. its rank. In stringy language this means that the entropy 
asymptotically scales
with the square of the number of coincident D-branes. 
This can be directly understood as being due to strings which can stretch
between $\sim n^2$ possible pairs of these branes, which again is
reflected in the $\sim n^2$ entries of the matrix representing the
gauge connection. 

It turns out, however, that the entropy of theories describing
stacks of M5-branes scales with the \emph{cube} of the number of
branes in the stack. 
Even though the implication of this phenomenon for the 
conceptual nature of the effective field theory on these
branes has remained rather mysterious, 
there are several ways to understand from the string theory
picture how this comes about: 

Like open strings stretch between D-branes, there are open 
membranes stretching between M5-branes. These membranes happen
to have a BPS state in which their spatial configuration is 
that of a `pair of pants'. Therefore, like open strings stretch
between pairs of branes, open BPS membranes can stretch between
\emph{triples} of branes 
\cite{AharonyHananyKol:1997,KolRahmfeld:1998}.

Indeed, in \cite{BerensteinLeigh:1998} 
the $n^3$-scaling of theories on stacks of 5-branes has simply
been interpreted as being due to the $\sim n^3$ possible triples
of 5-branes between which the `pair-of-pants' BPS state of the
M2-brane can stretch. 

In light of the above interpretation of
$n^2$-scaling in ordinary gauge theories this strongly suggests that
there should be a generalized form of gauge connection in
the effective theories of stacks of 5-branes which, in one way
or another, generalize the matrix representing the gauge connection
to a cubic array of numbers, i.e. to some tensor of total rank three.

An interesting question is therefore if the formalism  of 2-holonomy
presented here can capture such a phenomenon. While this is
an open question not further studied here, we would like to mention
two possibly interesting speculations concerning this point.

\begin{enumerate}

\item {\bf Correlators in 2D TFTs.}

It is well known that
topological field theories (TFTs) in two dimensions
on triangulated manifolds are in 1-1 correspondence with
semisimple associative algebras 
$\mathcal{A}$ \cite{FukumaHosonoKawai:1992}.
Let $C_{ab}{}^c$ be the structure constants of such an algebra
in a given basis.
The partition
function of the corresponding TFT on a given surface $\Sigma$
is computed by choosing a
triangulation of $\Sigma$
with oriented edges, assigning a copy of $C$ to each triangle
with each index associated to one of the edges, raising and lowering
these indices with the algebra's Killing metric $\kappa_{ab}$
according to whether the respective edge is ingoing or outgoing,
and then contracting all pairs of indices belonging to the
same edge.

The associativity of the algebra as well as the non-degeneracy
of $\kappa$ can be seen to ensure that the number obtained this
way is independent of the triangulation chosen. It is a topological
invariant of $\Sigma$.

One can regard any rank 3 tensor $V \in \otimes^3 \mathcal{A}$ as a
``vertex operator'' for such a theory. The $N$-point function
\[
  \langle V_1 V_2 \dots V_N \rangle_\Sigma
\]
can be defined by picking a triangulation of $\Sigma$ with
$n$ non-adjacent triangles removed, assigning copies of $C$ to
this triangulation as before and assigning the $V_i$ to the
triangles that have been removed, contracting all indices as before.

In particular, when we have a 2-form $B$ on $\Sigma$ which takes
values in $\otimes^3 \mathcal{A}$, we can form the correlator
\begin{eqnarray*}
  h_\Sigma\of{B}
  &\defas&
  \left\langle
    \exp\of{\int_\Sigma B}
  \right\rangle_\Sigma
  \,,
\end{eqnarray*}
which is naturally interpreted as a form of surface holonomy of $B$
over $\Sigma$. Essentially this construction has been proposed in
\cite{Akhmedov:2005}.

Note how this can be regarded as a direct generalization of a
similar formulation of ordinary line holonomy. We could define
a trivial 1-dimensional TFT on the lattice by picking some vector space
$\mathcal{A}$,
assigning the identity operator on $\mathcal{A}$ to intervals of
a segmentation of a 1-dimensional manifold,
letting vertices be rank-2 tensors in $\mathcal{A}$
and letting the gauge connection $A$ be
a vertex-valued 1-form. Then ordinary line holonomy could be written as
\[
  \left\langle
    \mathrm{P}\exp\of{\int A}
  \right\rangle
\]
with the correlator taking care of the index contraction.

There should be an abelian 2-group describing the above
concept of surface holonomy, and
the formalism described in part III should allow to get a globally
defined notion of surface holonomy of the above kind for the
general situation where the 2-form $B$ is only locally defined.

Even though this setup is ``abelian'', it does have the interesting
property that the degrees of freedom encoded in the
$\bigotimes^3 \mathcal{A}$-valued 2-form $B$ scale with the
\emph{cube} of the dimension
of the vector space associated with the algebra $\mathcal{A}$,
corresponding to the fact that a simplex in two dimensions has
three 1-faces (edges).

\item {\bf Algebroid YM Theories.}

As we have remarked before, the definition of $p$-bundles with
$p$-holonomy as described in part III
(and sketched in figure
\ref{figure: 2-bundle as simpl map},
p. \pageref{figure: 2-bundle as simpl map}) is directly
applicable to cases where the ``structure $p$-group'' is really a
$p$-groupoid
(see \S\fullref{intro on categories: categories} for more on groupoids).
In the differential formulation
(\cf \S\fullref{The Differential Picture: Morphisms between p-Algebroids})
this corresponds to replacing the structure $p$-algebra by a
$p$-algebroid. Where an ordinary algebra has structure \emph{constants}
an algebroid has position-dependent structure \emph{functions},
in a sense. These are again potential candidates for causing $n^3$-scaling
behaviour.

Such structure functions
were apparently first considered in the context of higher
gauge theory in \cite{Hofman:2002}, which is reviewed in
\S\fullref{Gerstenhaber brackets and Hochschild Cohomology}.
The somewhat more systematic treatment using algebroids was
more recently discussed in \cite{Strobl:2004}, where some interesting
consequences are reported that might be of relevance for
nonabelian strings.

\end{enumerate}

In the present context
these hints are all that we are going to say about the
issue of $n^3$-scaling in higher gauge theory.

\subsection{Spinning Strings}
\label{more details on Spinning Strings}

In order to approach the issue of spinning strings, first
recall the situation for $\mathrm{Spin}(n)$.
 A (Riemannian) manifold $M$
\emph{is spin} or
\emph{admits a spin structure} if spinning
\emph{particles} can consistently propagate on it.

This is the case iff an $\mathrm{SO}(n)$-bundle 
\[
\begin{array}{c}
  E 
  \\
  \Big \downarrow
  \\
  M
\end{array}
\]
 over the manifold $M$ can be lifted to a
$\Spin(n)$-bundle, where 
$\Spin(n)$ is the central extension of 
$\mathrm{SO}(n)$ by $\mathbb{Z}/2$:
\[
  1 \to \Z_2 \to \Spin(n) \to \mathrm{SO}(n)
  \to 1
  \,.
\]
This is the case iff $M$ is orientable and the second 
Stiefel-Whitney class $w_2(E) \in H^1(M;\mathbb{Z}/2)$ vanishes.

The situation for $\String(n)$ is similar, but with everything 
lifted by one dimension.
A manifold 
\emph{is string} or 
\emph{admits a string structure} 
if spinning \emph{strings} can consistently propagate on it.

This is the case iff a principal loop-group $L\mathrm{SO}(n)$-bundle
\[
  \begin{array}{c}
    LE \\
    \Big\downarrow \\
    LM 
  \end{array}
\] 
over the free loop space $LM$ can be lifted to a 
$\widehat{L\mathrm{SO}(n)}$-bundle, 
where $\widehat{L\mathrm{SO}(n)}$ is a (Kac-Moody-)central extension of 
$L\mathrm{SO}(n)$ by $U(1)$:
\[
  1 
  \to 
  U(1)
  \to
  \widehat{L\mathrm{SO}(n)}
  \to
  L\mathrm{SO}(n)
  \to
  1
  \,.
\]
And this is the case iff the so-called 
\emph{string class} of $LM$ in $H^3(LM;\,\Z)$ vanishes.

These two conditions on the topology of $LM$ can equivalently be 
formulated in terms of $M$ itself: 

\begin{enumerate}

\item The vanishing of the string class in 
$H^3(LM;\Z)$ is equivalent to the vanishing of the 
first Pontryagin class $\frac{1}{2}p_1(E)$ of a vector bundle 
associated to a principal $\Spin(n)$-bundle $E \to M$. 

The string class in $H^3(LM;\mathbb{Z})$ is obtained from the
Pontryagin class $p_1/2$ by \emph{transgression}. 
This means that it is represented by the 3-form
\[
  \int_\gamma \mathrm{ev}^*(\xi)
  \,,
\]
where $\xi$ is a representative of $p_1/2$, $\mathrm{ev}^*$ is 
the pull-back by the evaluation map
\[
  \begin{array}{rcl}
    \mathrm{ev} : LM \times S^1 & \to & M \\
                 (\gamma,\sigma) & \mapsto & \gamma(\sigma)
   \end{array}
\]
and $\int_\gamma$ denotes the integral over the $S^1$-factor in 
$LM \times S^1$.

\item

 This again is equivalent to the existence of a lift of the structure
group of $E$ from $\Spin(n)$ to the topological group called
$\String(n)$.

The group
$\String(n)$ (or rather a 'realization' thereof) is defined as a
topological group all of whose homotopy groups equal those
of $\Spin(n)$, except for the third one, which has to vanish for
$\String(n)$:
\[
  \pi_k(\String(n))
  =
  \left\{
  \begin{array}{cl}
    \pi_k(\mathrm{Spin}(n)) & \mbox{for $k \neq 3$} \\
    1 & \mbox{for $k = 3$}
  \end{array}
  \right.
\]

This should be seen in the following context:

As is well known, the first homotopy groups $\pi_k$ of the orthogonal group
$O\of{n}$ for $n>8$ are given by the following table:

\begin{center}
\begin{tabular}{|c||c|c|c|c|c|c|c|c|}
  \hline
  k& 0           & $1$ & $2$ & $3$ & $4$ & $5$ & $6$ & $7$ \\
  \hline
  $\pi_k\of{O\of{n}}$  & $\Z/2$ & $\Z/2$ & $0$ & $\Z$ & $0$ & $0$ & $0$ & $\Z$\\
  \hline
\end{tabular}
\end{center}

The 0-th homotopy group $\pi_0 = \Z/2$ indicates that $O\of{n}$
is not connected but has
two connected components. One can `kill' this homotopy group by
going over to the connected component of the identity element, i.e.
to the special orthogonal group $SO\of{n}$, which has
$\pi_0\of{SO\of{n}} \simeq 0$.

The first homotopy group $\pi_1 = \Z/2$ indicates that $O\of{n}$ is not
simply connected. One can `kill' this homotopy group by
going over to its universal double cover, the group $\Spin\of{n}$, which has
$\pi_1\of{\Spin\of{n}} \simeq 0$.

All of $O\of{n}$, $SO\of{n}$ and $\Spin\of{n}$ are of course
semisimple Lie groups.
Every semisimple Lie group has nonvanishing $\pi_3$. Hence, if one
wishes to
continue with `killing' homotopy groups of $O\of{n}$ this way, one
will end up with a group that is no longer smooth. Instead its
group space will just be a topological space with the group operation being
a continuous map on this space. Such groups are called topological
groups.

It can be shown and is well known that an equivalent way to
define (a realization of) the group $\String(n)$ is as the
topological group which makes this sequence of groups exact:
\[
  1  \to
    K(\mathbb{Z},2)
     \to
    \mathrm{String}(n)
      \to
    \mathrm{Spin}(n)
      \to
    1
  \,.
\]
Here $K(\mathbb{Z},2)$ denotes (a realization of) the Eilenberg-MacLane 
space $K(\mathbb{Z},2)$, which is by definition a topological space 
all whose homotopy groups vanish, except for the second one, which 
is isomorphic to $\mathbb{Z}$. In general
\[
  \pi_k(K(G,n))
  \simeq
  \left\lbrace
    \begin{array}{cl}
      G & \mbox{for $k = n$} \\
      1 & \mbox{otherwise}
    \end{array}
  \right.
  \,,
\]
by definition.

\end{enumerate}

The importance of string structures in string theory results from 
the fact that superstrings are nothing but ``spinning strings'', 
i.e. fermions on loop space, and that their quantum equations of 
motion are nothing but a generalized Dirac  equation on loop space. 
(The 0-mode of the worldsheet supercharge is a generalized 
Dirac(-Ramond) operator on loop space (for the closed string).)

It hence follows by the above discussion that superstrings can propagate 
consistently only on manifolds which \emph{are string}, 
just like an ordinary point-like fermion can propagate consistently only 
on a manifold that is \emph{spin}.

More technically, the wavefunction of a point-like fermion is really 
a section of a $\widehat{\mathrm{SO}(n)} \simeq \mathrm{Spin}(n)$-bundle 
and hence such a bundle needs to exist over spacetime in order for the 
fermion to exists.

Similarly, the wavefunction of a fermionic string 
(spinning string) is really a section of a 
$\widehat{L\mathrm{SO}(n)}$-bundle over loop space, 
and hence such a bundle needs to exist over the loop space over spacetime 
for fermionic strings to exist.

For instance the worldsheet supercharge of the heterotic string is 
a Dirac operator on loop space for fermions that are also 
``charged'' under an $\mathrm{SO}(32)$- or $E_8 \times E_8$-bundle
\[
  \begin{array}{c}
    V \\
    \Big \downarrow \\
    M  
  \end{array}
  \,.
\]

In $K$-theory one can form the difference bundle
\[
  E = V - T
  \,,
\]
where $T$ is the tangent bundle and the condition for this bundle 
to admit a string structure is that the Pontryagin class 
vanishes, i.e. that
\[
  p_2(V) - p_1(T) = 0
  \,.
\]
This is in  fact the relation which follows from the cancellation of 
the perturbative anomaly of the effective 
$\mathrm{SO}(32)$- or $E_8 \times E_8$-field theory obtained from
these strings. Hence this famous anomaly is related to the fact 
that heterotic strings are spinors on loop space.

One of the earliest discussions of these issues is given in
\cite{Killingback:1987}.
Killingback discusses the spinning point particle, 
the spinning string, the obstructions to lifting the 
$L\mathrm{SO}(n)$-bundle on loop space to the central extension and 
the relation to the perturbative anomalies of the effective field 
theory of the heterotic string.

As a supplement to this there is the nice and more detailed 
discussion of the relation between $H^3(LM;\mathbb{Z})$ and $p_1/2$ 
as given in \cite{MurrayStevenson:2001}.

A more detailed discussion of the nature of Dirac operators on 
loop space with a review of Killinback's results is given in
\cite{Witten:1988}
which has the companion paper
\cite{Witten:1987}

Concerning the group $\mathrm{String}(n)$ the best available
reference is probably
\cite{StolzTeichner:2004}.

A discussion of the group $\mathrm{String}(n)$
and of string structures is given on the top of p. 5 of 
\cite{StolzTeichner:2004}
and then in the beginning of section 5 
of \cite{StolzTeichner:2004} on pp. 65.
The ``killing'' of homotopy groups is discussed on
p. 65 of \cite{StolzTeichner:2004}, 
the definition of $\mathrm{String}(n)$ by means of an
exact sequence is discussed on p. 66, and the relation to the
Pontryagin class is discussed on p. 67.

\clearpage
\subsection{Category Theory}
\label{more details on category theory}

For the physicist's convenience the following gives a
quick introduction to some elementary concepts of category theory
that are used in the main text.

\vskip 1em

{\bf Literature.}
The standard introductory textbook
for category theory is \cite{MacLane}, of which mainly only the
first few pages are needed here. The history of $n$-categorical
physics is treated in \cite{BaezLauda:2005}, which also serves as
a nice introduction to the elementary concepts.
A pedagogical discussion of categorification is given in
\cite{BaezDolan:1998}. Details on 2-categorical technology in the
context of categorified gauge theory are given in
\cite{Pfeiffer:2003,GirelliPfeiffer:2004}.

\subsubsection{Categories}
\label{intro on categories: categories}

The concept {\bf category} is a generalization of the concept
{\bf set}. A category $C$ is a set
$\Ob\of{C} =\set{a,b,\dots}$, called the set of
{\bf objects}, together with
a set $\Mor\of{C} = \set{r,s,\dots}$ of arrows, called the
set of {\bf morphisms}. A morphism is is something that
goes between two given objects
\[
\xymatrix{
   a \ar@/^1pc/[rr]^{r}
&& b
}\,.
\]
Morphisms can be anything.
In particular, they need not be
functions or maps between sets. But they can.
All one requires is that given
two morphisms
\[
\xymatrix{
   a \ar@/^1pc/[rr]^{r}
&& b
}
\]
and
\[
\xymatrix{
   b \ar@/^1pc/[rr]^{s}
&& c
}
\]
which are {\bf composable}, i.e. where the target object $b$
of one  matches the source object of the other, there is a uniquely
defined morphism
\[
\xymatrix{
   a \ar@/^1pc/[rr]^{r \circ s}
&& c
}
\defas
\xymatrix{
   a \ar@/^1pc/[rr]^{r}
&& b \ar@/^1pc/[rr]^{s}
&& c
}
\]
obtained by {\bf composition}.
Furthermore, this composition $\circ$
is supposed to be associative
\[
\xymatrix{
   a \ar@/^1pc/[rr]^{(r \circ s) \circ t}
&& d
}
\defas
\xymatrix{
   a \ar@/^1pc/[rr]^{r \circ (s \circ t)}
&& d
}
\defas
\xymatrix{
   a \ar@/^1pc/[rr]^{r}
&& b \ar@/^1pc/[rr]^{s}
&& c \ar@/^1pc/[rr]^{t}
&& d
}
  \,,
\]
and for every object there must be an
{\bf identity morphism}
\[
\xymatrix{
   a \ar@/^1pc/[rr]^{\mathrm{Id}}
&& a
}
\]
such that its composition with any other morphism
equals that other morphism:
\begin{eqnarray*}
  &&
\xymatrix{
   a \ar@/^1pc/[rr]^{\mathrm{Id}}
&& a \ar@/^1pc/[rr]^{r}
&& b
}
\\
  &=&
\xymatrix{
   a \ar@/^1pc/[rr]^{r}
&& b \ar@/^1pc/[rr]^{\mathrm{Id}}
&& b
}
  \\
  &=&
\xymatrix{
   a \ar@/^1pc/[rr]^{r}
&& b
}
  \,.
\end{eqnarray*}

If a morphism
\[
\xymatrix{
   a \ar@/^1pc/[rr]^{r}
&& b
}
\]
is invertible, i.e. if there is another morphism
\[
\xymatrix{
   b \ar@/^1pc/[rr]^{r^{-1}}
&& a
}
\]
such that
\[
\xymatrix{
   a \ar@/^1pc/[rr]^{r}
&& b \ar@/^1pc/[rr]^{r^{-1}}
&& a
}
=
\xymatrix{
   a \ar@/^1pc/[rr]^{\mathrm{Id}}
&& a
}
\]
and
\[
\xymatrix{
   b \ar@/^1pc/[rr]^{r^{-1}}
&& a \ar@/^1pc/[rr]^{r}
&& b
}
=
\xymatrix{
   b \ar@/^1pc/[rr]^{\mathrm{Id}}
&& b
}\,,
\]
then $r$ is called an {\bf isomorphism}.

\paragraph{Examples.}
\label{introduction: categories: examples}

One large class of categories that everybody is familiar with
consists of
categories whose objects are sets with a given extra structure and
whose morphisms are maps between sets that preserve this structure.

So there is the category
\begin{itemize}
  \item
   $\Set$, whose objects are all (small) sets and whose morphisms are
     maps between these sets.
  \item
    $\Top$, whose objects are topological spaces
    (i.e. sets equipped with a topology) and whose morphisms are
    continuous maps between topological spaces.
  \item
    $\Diff$, whose objects are smooth spaces (i.e. sets equipped with
    a smooth stucture) and whose morphisms are smooth maps between
    such spaces.
  \item
   $\Vect$, whose objects are (finite dimensional) vector spaces
   and whose morphisms are linear maps between these.
\end{itemize}
In all these cases, the composition of morphisms is nothing but
the composition of the respective maps that they represent.

Other categories are rather different in character. For instance
consider any oriented graph $(V,E)$, with $V$ a set of vertices and
$E$ a set of oriented edges between pairs of vertices. The free
category over $(V,E)$, called $C_{(V,E)}$ is the category whose objects 
are the
vertices in $V$ and whose morphisms are all the paths of edges that
one obtains by concatenating edges in $V$. Composition of morphisms
here is concatenation of edge paths.

We may want to think of a graph $(V,E)$ as an approximation to a
path space. Let $M$ be any smooth space and let $P\of{M}$ be the
free path space over $M$, defined to be the space of all smooth maps
\[
  \gamma \maps [0,1] \to M
  \,.
\]
Here $M$ is like the continuum version of $E$, while $P\of{M}$ is like
a continuum version of $V$. In order to get a category out of this,
by interpreting a path $\gamma$ as a morphism between the source object
$\gamma\of{0}$ and the target object $\gamma\of{1}$,
we need to define what the
composition of two such paths $\gamma_1$ and $\gamma_2$ is supposed to be.
Of course we want this
to be the path obtained by first tracing out $\gamma_1$ and then
tracing out $\gamma_2$. This can be made into an
associative operation by going over to thin homotopy equivalence classes
of paths, which implies that we forget about the parameterization
of these paths. The resulting category is called the
{\bf path groupoid} $\P_1\of{M}$.

Other types of categories have morphisms which are ``arrows''
in a much more abstact sense. For instance,
given any set $\set{a,b,\dots}$ with a partial ordering
$\cdot \geq \cdot$, we obtain a category whose
objects are elements of this set and which has precisely one morphism
\[
\xymatrix{
   a \ar@/^1pc/[rr]^{}
&& b
}
\]
from
element $a$ to element $b$ precisely if $a \geq b$ with respect to the
partial ordering.

The example of relevance in our context is the category $\mathcal{O}\of{M}$
of open subsets of a topological space $M$ with the partial ordering that
given by inclusion $U_1 \supset U_2$ of subset $U_2$ in subset $U_1$.

A related example is the category called the
{\bf {\v C}ech-groupoid}.
Given any smooth space $M$ with good covering
$\covering = \bigsqcup\limits_{i \in I} U_i$
(i.e. a covering of $M$ by open sets such that every finite intersection
of these is contractible),
its {\v C}ech groupoid is the category whose objects are all pairs
\[
  (x,i)
\]
with $x \in U_i$, and which has precisely one morphism
\[
\xymatrix{
   (x,i) \ar@/^1pc/[rr]^{}
&& (x,j)
}
\]
whenever $x$ is a point in the double overlap $U_{ij} = U_i \cap U_j$.

This category is called a groupoid because all of its morphisms
are invertible. 

A {\bf groupoid} is a category all whose morphisms are invertible,
i.e. all whose morphisms are isomorphisms.

Since, by definition
of the {\v C}ech-groupoid, there is a unique morphism
between any two such pairs and because, by definition of a category,
there must be an identity morphism from each object to itself, it follows
that the composition of
\[
\xymatrix{
   (x,i) \ar@/^1pc/[rr]^{}
&& (x,j)
}
\]
with
\[
\xymatrix{
   (x,j) \ar@/^1pc/[rr]^{}
&& (x,i)
}
\]
equals the identity morphism
\[
\xymatrix{
   (x,i) \ar@/^1pc/[rr]^{\mathrm{Id}}
&& (x,i)
}
  \,.
\]

The reason for the term ``groupoid'' is that in the case
that such a category has just a single object, it is the same as an
ordinary group:

A {\bf group} is a category with just a single object and all morphisms
invertible.

So a group is a groupoid with a single object.

Namely with a single object, $\bullet$,
every morphism
\[
\xymatrix{
   \bullet \ar@/^1pc/[rr]^{g}
&& \bullet
}
\]
can be composed with any other
\[
\xymatrix{
   \bullet \ar@/^1pc/[rr]^{g'}
&& \bullet
}
\]
(since their source and target always match). Hence composition of
morphisms
\[
\xymatrix{
   \bullet \ar@/^1pc/[rr]^{g}
&& \bullet \ar@/^1pc/[rr]^{g'}
&& \bullet
}
\]
becomes an associative product map
\[
  g \circ g' \defas gg'
\]
from the set of morphisms to itself.
Since, by assumption, to every morphism there is an inverse morphism,
this product
operation is that of a group.

For illustration purposes, if one likes, one can think of the
morphisms here again as maps,
but that's not compulsory. So for instance given the
permutation group $S_p$, one might want to identitfy the single
object $\bullet$ with any given $p$-element set
\[
  \bullet = \set{1,2,3,\dots p}
\]
and identify any element $g \in S_p$ with the map
\[
\xymatrix{
   \set{1,2,\dots p}
   \ar@/^1pc/[rr]^{g}
&& \set{1,2,\dots p}
}
\]
that permutes these $p$ elements.

\paragraph{Functors.}
\label{elements of cat theory: functors}

Categories consist of objects and morphisms. A map from one category to
another should respect the composition of these morphisms.
In analogy to how a map between sets is called a function, such
a map between categories is called a {\bf functor}.

So given categories $C$ and $D$, a functor $F$
\[
\xymatrix{
   C \ar@/^1pc/[rr]^{F}
&& D
}
\]
is a map from the set of morphisms of $C$ to the set of
morphisms of $D$ such that
\[
  F\of{
    \xymatrix{
    a \ar@/^1pc/[rr]^{r}
    && b
    }
  }
  \circ
  F\of{
    \xymatrix{
    b \ar@/^1pc/[rr]^{s}
    && c
    }
  }
  =
  F\of{
    \xymatrix{
    a \ar@/^1pc/[rr]^{r \circ s}
    && c
    }
  }
  \,.
\]

For instance, a functor from a group $G$ to a group $H$
(regarded categories
with single objects and all morphisms invertible)
is  nothing but a group homomorphism $G \to H$.

Or consider a functor
\[
 F \maps C_{(V,E)} \to G
\]
from the free category
$C_{(V,E)}$ of a graph $(V,E)$ to any group $G$. Such a functor sends all
vertices in the graph to the unique object $\bullet$ in $G$ and labels
each edge path of the graph, say $x \stackto{\gamma_1} y \stackto{\gamma_2} z$,
with a given group element in $G$, such that
concatenation of edge paths corresponds to multiplication of their
group elements:
\[
\begin{array}{c}
\xymatrix{
   \bullet \ar@/^1pc/[rr]^{g}
&& \bullet  \ar@/^1pc/[rr]^{g'}
&& \bullet
}
\\
\\
\Big\Uparrow F
\\
\xymatrix{
   x \ar@/^1pc/[rr]^{\gamma_1}
&& y  \ar@/^1pc/[rr]^{\gamma_2}
&& z
}
\end{array}
\]

This is nothing but what happens in ordinary local holonomy in
lattice gauge theory. Paths are consistently labeled by group elements.

Another example relevant for gauge theory is that of a pre-sheaf.
A pre-sheaf over a topological space $M$ is the assignment of some set
$S\of{U}$ to each open subset $U \subset M$, together with a restriction map
$S\of{U_1} \to  S\of{U_2}$ whenever $U_1 \supset U_2$, such that restricting
first from $U_1$ to $U_2$ and then to $U_3$ is the same as
restricting directly to $U_3$.
This can be summarized by saying that a pre-sheaf is a functor
\[
  S \maps \mathcal{O}\of{M} \to \Set
\]
from
the category  of open subsets of $M$ to the category of all (small) sets.

For instance, if one lets $S\of{U}$ be the set of all continuous complex-valued
functions over $U$ and lets the restriction map be the restriction of the domain
of these functions, one obtains a pre-sheaf (and in fact a sheaf).

\paragraph{Natural Transformations.}
\label{elements of cat theory: natural transformations}

One should think of the functor $F \maps C \to D$ as something that produces
an image of any
diagram in $C$ in terms of a diagram in $D$. While the images of two functions
can be either equal or not, the images of two functors, being ``1-dimensional''
in a sense, can be ``congruent'' or ``homotopic'' without being equal.

Given two functors
\[
  F_1 \maps S \to T
\]
and
\[
  F_2 \maps S \to T
\]
one says that there is a {\bf natural transformation}
\[
  F_1 \stackto{\eta} F_2
\]
from $F_1$ to $F_2$, if for every object $a$ in the source category $S$ there
is morphism $\eta\of{a}$ in the target category $T$, such that all
diagrams of the following kind commute in $T$:
\begin{center}
  \begin{picture}(370,150)
    \includegraphics{nattraf.eps}
    \put(-296,142){$S$}
    \put(-100,142){$T$}
    \put(-296,107){$a$}
    \put(-296,20){$b$}
    \put(-303,65){$r$}
    \put(-180,107){$F_1\of{a}$}
    \put(-180,20){$F_1\of{b}$}
    \put(-186,65){$F_1\of{r}$}
    \put(-40,107){$F_2\of{a}$}
    \put(-40,20){$F_2\of{b}$}
    \put(-33,65){$F_2\of{r}$}
    \put(-110,104){$\eta\of{a}$}
    \put(-110,20){$\eta\of{b}$}
  \end{picture}
\end{center}
In other words, there is a natural transformation
$F_1 \stackto{\eta} F_2$ if the images of any morphism
\[
\xymatrix{
   a \ar@/^1pc/[rr]^{r}
&& b
}
\]
under $F_{1,2}$
\[
  F_{1,2}\of{
\xymatrix{
   a \ar@/^1pc/[rr]^{r}
&& b
}
}
\defas
\xymatrix{
   F_{1,2}\of{a} \ar@/^1pc/[rr]^{F_{1,2}\of{r}}
&& F_{1,2}\of{b}
}
\]
are related by
\[
\xymatrix{
   F_{1}\of{a} \ar@/^1pc/[rr]^{\eta\of{a}}
&& F_{2}\of{a} \ar@/^1pc/[rr]^{F_2\of{r}}
&& F_2\of{b}
}
=
\xymatrix{
   F_{1}\of{a} \ar@/^1pc/[rr]^{F_1\of{r}}
&& F_{1}\of{b} \ar@/^1pc/[rr]^{\eta\of{b}}
&& F_2\of{b}
}
\,.
\]
One should think of $\eta$ as being a ``translation'' of $F_1$ to
$F_2$ using the ``paths'' in $T$. We can also write the natural
transformation $\eta$ like this:
\[
\xymatrix{
   S \ar@/^1pc/[rr]^{F_1}="0"
           \ar@/_1pc/[rr]_{F_2}="1"
           \ar@{=>}_{\eta}"0";"1"
&& T
}
\]

If all morphisms $\eta\of{a}$ are isomorphisms, then $\eta$
itself is invertible as a natural transformation
and is called a {\bf natural isomorphism}.

For instance recall the functor $F \maps C_{(V,E)} \to G$ from the
free category of a graph $(V,E)$ to any group $G$, which we
noticed can be interepreted as assigning to each path of edges
(morphism in $C_{(V,E)}$) a holonomy (morphism in $G$).
A natural transformation of such
a functor is the assignment of group elements $\eta\of{x}$ to every
vertex $x \in V$, such that all diagrams of the following form commute:
\begin{center}
  \begin{picture}(370,150)
    \includegraphics{nattraf.eps}
    \put(-296,142){$C_{(V,E)}$}
    \put(-100,142){$G$}
    \put(-296,107){$x$}
    \put(-296,20){$y$}
    \put(-303,65){$\gamma$}
    \put(-186,65){$F_1\of{\gamma}$}
    \put(-33,65){$F_2\of{\gamma}$}
    \put(-110,104){$\eta\of{x}$}
    \put(-110,20){$\eta\of{y}$}
  \end{picture}
\end{center}
This implies that
\[
  \eta\of{x}F_2\of{\gamma}
  =
  F_1\of{\gamma}\eta\of{y}
  \,,
\]
where the product is that in the group $G$. Multiplying with
$(\eta\of{y})^{-1}$ from the right gives
\[
  F_2\of{\gamma} = \eta\of{x} F_2\of{\gamma}\eta^{-1}\of{y}
  \,.
\]
This is nothing but a {\bf gauge transformation} of the holonomy $F_1$.

Since a natural transformation goes between functors
and since composition of natural transformations is associative, 
these transformations can be
regarded as morphism of a category themselves. Given categories $C$ and $D$
we denote by $C^D$ the {\bf functor category} whose objects are
functors $D \to C$ and whose morphisms are natural transformations
between these functors.

\subsubsection{2-Categories}
\label{2-categories}

One remarkable thing about natural transformations is that they are
morphisms that go between functors -- which are morphisms themselves.

There is a category, called $\Cat$, whose objects are all (small)
categories and whose morphisms are functors between these categories.
But since there are now also natural transformations between these functors,
$\Cat$ is really a 2-category.

A {\bf 2-category} consists of a set of objects
\[
  a,b,\dots\,,
\]
together with a set of
morphisms
\[
\xymatrix{
   a \ar@/^1pc/[rr]^{r}="0"
&& b
}
\]
between these objects, together with a set of
{\bf 2-morphisms}
\[
\xymatrix{
   a \ar@/^1pc/[rr]^{r}="0"
           \ar@/_1pc/[rr]_{s}="1"
           \ar@{=>}_{\rho}"0";"1"
&& b
}
\]
between these morphisms. In what are called
{\bf strict 2-categories} the 1-morphisms as well as the 2-morphisms
have an associative composition operation. In addition to the obvious
``vertical'' composition of 2-morphisms
\[
  \xymatrix{
   a \ar@/^2pc/[rr]^{r}="0"
           \ar[rr]_{s}="1"
           \ar@{=>}^{\rho}"0";"1"
           \ar@/_2pc/[rr]_{t}="2"
           \ar@{=>}^{\sigma}"1";"2"
&& b
}
\]
there is now also a ``horizontal'' composition
\[
\xymatrix{
   a \ar@/^1pc/[rr]^{r}="0"
           \ar@/_1pc/[rr]_{s}="1"
           \ar@{=>}_\rho"0";"1"
&& b \ar@/^1pc/[rr]^{r'}="2"
           \ar@/_1pc/[rr]_{s'}="3"
           \ar@{=>}_{\sigma}"2";"3"
&& c
}
\,.
\]
One demands that horizontal and vertical compositon are
compatible in that the order in which they are applied in diagrams
like
\[
\xymatrix{
   a \ar@/^2pc/[rr]^{r}="0"
           \ar[rr]^{s}="1"
           \ar@{=>}_{\rho}"0";"1"
           \ar@/_2pc/[rr]^{t}="2"
           \ar@{=>}_{\tilde \rho}"1";"2"
&& b \ar@/^2pc/[rr]^{r'}="3"
           \ar[rr]^{s'}="4"
           \ar@{=>}_\sigma"3";"4"
           \ar@/_2pc/[rr]^{t'}="5"
           \ar@{=>}_{\tilde \sigma}"4";"5"
&& c
}
\]
is immaterial. This is called the {\bf exchange law}.

For example the category called the path groupoid $\P_1\of{M}$ has a
natural extension to a 2-category $\P_2\of{M}$, called the 2-path 2-groupoid
$\P_2\of{M}$, whose 2-morphisms
\[
\xymatrix{
   x \ar@/^1pc/[rr]^{\gamma_1}="0"
           \ar@/_1pc/[rr]_{\gamma_1}="1"
           \ar@{=>}_{\Sigma}"0";"1"
&& y
}
\]
are surfaces interpolating between paths $\gamma_1$ and $\gamma_2$.
Horizontal and vertical composition of these 2-morphisms comes from
the ordinary horizontal and vertical gluing of these surfaces.

Another example for a 2-category is a 2-group. In
\S\fullref{introduction: 2-Groups as Categorified Groups}
we have introduced a 2-group as a 1-category with a group product
functor on it. Like an ordinary group can be regarded as a
1-category with a single object, such a 2-group can be regarded as
a 2-category with a single object.

The 2-categorical nature of the category $\Cat$ of all
(small) categories is what gives rise to the notion of
{\bf equivalence of categories} that in the present
context plays an important role in the study of the
2-group $\P_k G$ in
\S\fullref{introduction: the 2-group PG and its relation to string} and
\S\fullref{2-Groups, Loop Groups  and String}.

Namely, in the presence of 2-morphisms it becomes
unnatural to talk about strict invertibility of 1-morphisms.
In a 1-category, a (1-)morphism $r$ is invertible
(is an isomorphism) if there exists another (1-)morphism
$r^{-1}$, such that the (1-)morphism $r \circ r^{-1}$
\emph{equals} the identity 1-morphism. In a 2-category
these 1-morphisms are related by 2-morphisms, and hence
we do not ask if two 1-morphisms are equal, but if they are
\emph{(2-)isomorphic}, i.e. if they are related by an invertible
2-morphism.

So in a 1-category, two objects $a$ and $b$ are
isomorphic if there exist 1-morphisms
\[
\xymatrix{
   a \ar@/^1pc/[rr]^{r}="0"
   && b
   }
\]
and
\[
\xymatrix{
   b \ar@/^1pc/[rr]^{r^{-1}}="0"
   && a
   }
\]
such that their composition equals the identity morphism,
which means that there is are identity 2-morphisms
\[
\xymatrix{
   a \ar@/^1pc/[rr]^{r\circ r^{-1}}="0"
           \ar@/_1pc/[rr]_{\mathrm{Id}}="1"
           \ar@{=>}_{=}"0";"1"
&& a
}
\]
and
\[
\xymatrix{
   b \ar@/^1pc/[rr]^{r^{-1}\circ r}="0"
           \ar@/_1pc/[rr]_{\mathrm{Id}}="1"
           \ar@{=>}_{=}"0";"1"
&& b
}
\,.
\]
Obviously, in a 2-category, which has nontrivial 2-morphisms,
this should be generalized to the statement that two objects
$a$ and $b$ are \emph{equivalent} if there are 1-morphisms
$r$ and $\bar r$ together with \emph{invertible} 2-morphisms
$\xi$ and $\chi$ of the form
\[
\xymatrix{
   a \ar@/^1pc/[rr]^{r\circ \bar r}="0"
           \ar@/_1pc/[rr]_{\mathrm{Id}}="1"
           \ar@{=>}_{\xi}"0";"1"
&& a
}
\]
and
\[
\xymatrix{
   b \ar@/^1pc/[rr]^{\bar r\circ r}="0"
           \ar@/_1pc/[rr]_{\mathrm{Id}}="1"
           \ar@{=>}_{\chi}"0";"1"
&& b
}
\,.
\]

Hence the correct notion of ``sameness'' of categories is
equivalence in this sense. Two categories are equivalent, if
they are equivalent as objects of the 2-category $\Cat$.
This means that they are equivalent if there are
functors (1-morphisms in $\Cat$) going between them
whose compositions are naturally isomorphic
(related by an invertible 2-morphism in $\Cat$) to the
identity functor.

\paragraph{2-Functors and Pseudo-Natural Transformations.}
\label{elements of cat theory: 2-functors and pseudo-nat trafos}

In an obvious generalization of the concept of an ordinary functor,
a 2-functor is a map from a 2-category to another 2-category which
respects horizontal and vertical composition of 2-morphisms
(and hence also ordinary composition of 1-morphisms).  The obvious
generalization of a natural transformation between two functors is
called a {\bf pseudo-natural transformation}, and it again formalizes the
idea that the images of two 2-functors are ``congruent''.

Hence, given 2-functors

\[
  F_{1,2} \maps S \to T
\]
between 2-categories $S$ and $T$, a pseudo-natural transformation
\[
  \eta \maps F_1 \to F_2
\]
is the assignment of 1-morphisms $\eta\of{a}$ in $T$ to objects $a$ in
$S$ together with the assignment of 2-morphism $\eta\of{r}$ in $T$ to
1-morphisms $r$ in $S$, such that diagrams of the following form
{\bf 2-commute}, meaning that they commute at the level of 2-morphisms:

\begin{center}
\begin{picture}(440,200)
 \includegraphics{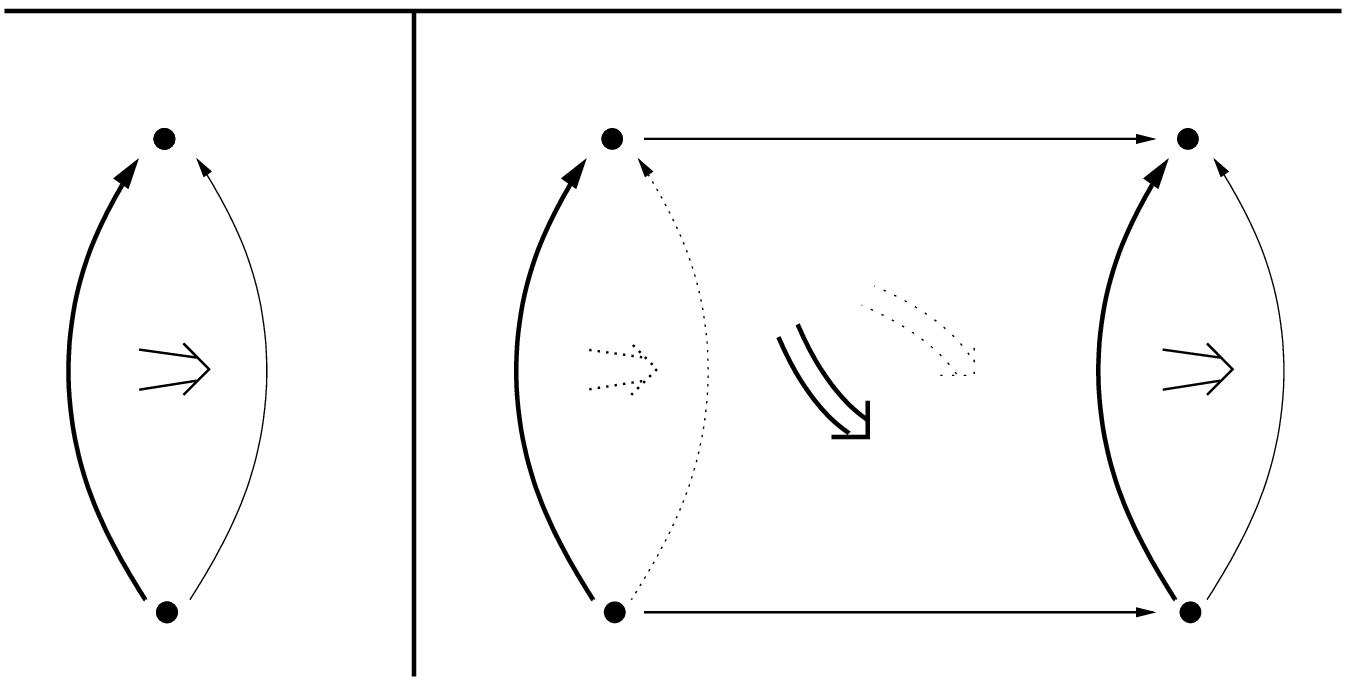}
 \put(-19,-2){
 \begin{picture}(0,0)
 \put(-10,0){
  \begin{picture}(0,0)
   \put(-355,70){$r$}
   \put(-290,70){${}_{s}$}
   \put(-320,105){$\rho$}
   \put(-322,168){$a$}
   \put(-322,7){$b$}
  \end{picture}
 }
 \put(122,0){
  \begin{picture}(0,0)
   \put(-373,70){$F_1\of{r}$}
   \put(-293,70){${}_{F_1\of{s}}$}
   \put(-333,105){$F_1\of{\rho}$}
   \put(-333,165){$F_1\of{a}$}
   \put(-333,10){$F_1\of{b}$}
   \put(-287,106){$\eta\of{r}$}
   \put(-250,118){${}_{\eta\of{s}}$}
   \put(-250,163){$\eta\of{a}$}
   \put(-250,10){$\eta\of{b}$}
  \end{picture}
 }
 \put(288,0){
  \begin{picture}(0,0)
   \put(-373,70){$F_2\of{r}$}
   \put(-293,70){${}_{F_2\of{s}}$}
   \put(-333,105){$F_2\of{\rho}$}
   \put(-333,165){$F_2\of{a}$}
   \put(-333,10){$F_2\of{b}$}
  \end{picture}
 }
 \end{picture}
}
  \put(-360,203){$S$}
  \put(-135,203){$T$}
\end{picture}
\end{center}

\vskip 2em

It should not come as a surprise that now we also have transformations
between pseudo-natural transformations (3-morphisms). These are called
{\bf modifications}. A modification
between pseudo-natural transformations $\eta_1$ and $\eta_2$
is the assignment of
a 2-morphism $\xi\of{a}$ in $T$ to every object $a$ in $S$, such that
these diagrams 2-commute:

\begin{eqnarray}
  \label{transition modification}
\begin{picture}(240,190)
 \includegraphics{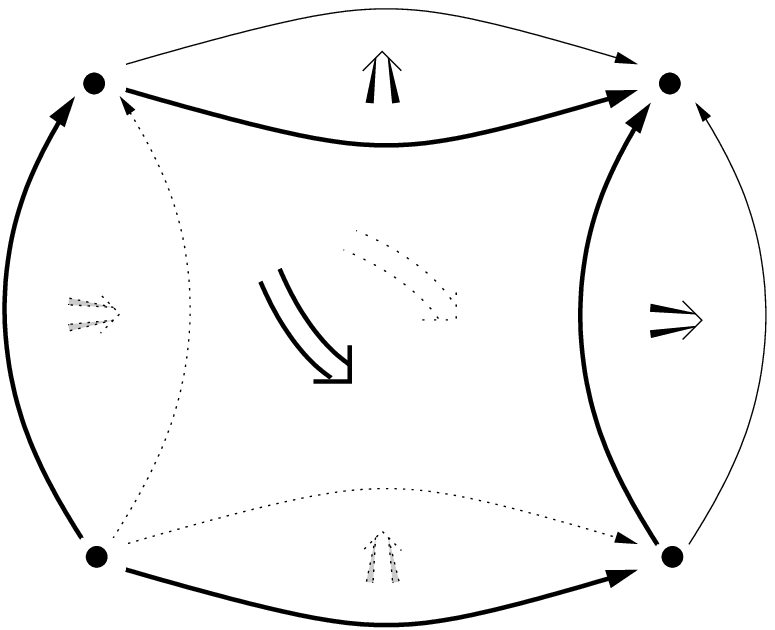}
 \put(0,20){\begin{picture}(0,0)
 \put(-3,-17){
 \begin{picture}(0,0)
 \put(122,0){
  \begin{picture}(0,0)
   \put(-374,70){$F_1\of{r}$}
   \put(-293,70){${}_{F_1\of{s}}$}
   \put(-333,100){$F_1\of{\rho}$}
   \put(-287,106){$\eta\of{r}$}
   \put(-250,118){${}_{\eta\of{s}}$}
   \put(-339,163){$F_1\of{a}$}
   \put(-339,5){$F_1\of{b}$}
   \put(-284,149){$\eta_1\of{a}$}
   \put(-256,183){${}_{\eta_2\of{a}}$}
   \put(-230,158){${}_{\xi\of{a}}$}
  \end{picture}
 }
 \put(125,-139){\begin{picture}(0,0)
   \put(-284,149){$\eta_1\of{b}$}
   \put(-256,183){${}_{\eta_2\of{b}}$}
   \put(-230,158){${}_{\xi\of{b}}$}
 \end{picture}}
 \put(288,0){
  \begin{picture}(0,0)
   \put(-374,70){$F_2\of{r}$}
   \put(-293,70){${}_{F_2\of{s}}$}
   \put(-333,100){$F_2\of{\rho}$}
   \put(-328,163){$F_2\of{a}$}
   \put(-328,5){$F_2\of{b}$}
  \end{picture}
 }
 \end{picture}
}
\end{picture}}
\end{picture}
\end{eqnarray}
Following Baez, we can write such 3-morphisms of 2-functors as
\begin{center}
\begin{picture}(200,135)
\includegraphics{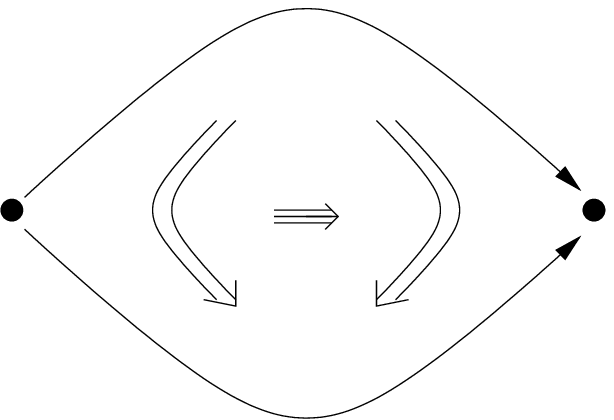}
\put(-186,58){$S$}
\put(5,58){$T$}
\put(-92,122){$F_1$}
\put(-92,-10){$F_2$}
\put(-144,58){$\eta_1$}
\put(-38,58){$\eta_2$}
\put(-92,68){$\xi$}
\end{picture}
\end{center}
\vskip 1em

It follows that all 2-functors between given 2-categories $C$
and $D$ form a 2-category $C^D$, called a {\bf 2-functor 2-category},
whose objects are 2-functors $D\to C$, whose morphisms are 
pseudo-natural transformations $\eta$ between such functors and whose
2-morphisms are modifications $\xi$ of pseudo-natural transformations.

\vskip 2em

It should be clear that one can keep on going to
$n$-categories and $n$-functors
for higher and higher $n$ this way. While the basic idea is obvious, each
step introduces more and more freedom to ``weaken'' relations, and the general
theory of $n$-categories is still under development, with the relation between
various alternative definitions and approaches remaining to be 
better understood.

\clearpage
\subsection{2-NCG and Derived Category Description of D-Branes}
\label{2-NCG and Derived Category Description of D-Branes}

A \emph{field} in physics is something that locally looks
like a function on spacetime, but which really is a
\emph{section of a bundle}. For instance a spinor field
looks like a spinor-valued function locally, but is really,
globally, a section of a spinor bundle.

The quanta of ordinary fields are ``point particles''.
Strings, on the other hand, are the quanta of what is called the
{\bf string field}
(e.g.
\cite{Witten:1986,Siegel:2001,Ohmori:2001,Berkovits:2001,TaylorZwiebach:2004}).

Often, these string fields are treated as nothing but
\emph{functions} on space-time that take values in a vector space spanned by
the excitation modes of the single first-quantized string.
One would expect that, more precisely, globally these string fields
should be generalized sections of some generalized notion of fiber bundle,
instead.

At least to some extent this is captured by using ordinary fiber bundles
on loop space, as we have reviewed in
\S\fullref{more details on Spinning Strings}. Given our discussion in
\S\fullref{introduction: the 2-group PG and its relation to string}
and
\S\fullref{2-Groups, Loop Groups  and String}
on how, in the case of (uncharged) spinors on loop space,
this is related to 2-bundles, it is tempting to guess that, more
generally, a string field should be a 2-section of some
2-vector 2-bundle.

This is to some degree motivated by our discussion of the relation of
the RNS string to supersymmetric quantum mechanics on loop space
in \S\fullref{introduction: SQM on loop space}.
Given the close relation of SQM to noncommutative spectral geometry (NCG),
we can consider
states of a supersymmetric particle to be sections of a vector bundle,
which arises as a finitely generated projective module of the algebra
$A$ of functions on configuration space. It seems to be a plausible
conjecture that there is a categorification of this scenario which
exhibits
string fields as sections of some sort of ``2-vector 2-bundle''
that arises as a module for a ``2-algebra of 2-functions''.
In fact, aspects of such a setup have been considered in the
literature, as discussed below.

In part III we will exclusively deal with principal 2-bundles,
since the
generalization to the categorification of associated and
vector bundles remains to be better understood.
But, as a motivation for the general philosophy relating
stringification with categorification that emerges from the
considerations presented here, and as an outlook for further
studies, we would like to sketch in the following some
existing approaches and some further observations concerning
vector 2-bundles, string fields, categorified supersymmetric
quantum mechanics and noncommutative geometry and a possible
relation to the description of D-brane states in terms
of derived categories.

\subsubsection{2-NCG}

Ordinary
Noncommtative Geometry (NCG)
starts with the {\bf Gelfand-Naimark theorem}, which
says that a topological space is equivalently encoded
in the $C^\ast$-algebra of continuous complex-valued
functions over it. In the present context we wish to think
of such a space as the configuration space of some particle.
Upon ``stringification'' this particle is expected to become a
linearly extended entity. Its
configurations, when suitably interpreted, include the position of
its endpoints together with a specification of how it stretches
from one endpoint to the other. The collection of this data,
a set of points (objects) and a set of strings (morphisms)
between them, may form a category.

Therefore a natural question is whether there is a generalization
of the Gelfand-Naimark theorem from sets to categories and if it
can serve as a basis for a categorification of all of NCG --
and how the result is related to string theory.

The answer to the first part of this question is positive, at least
in the case where the underlying spaces are discrete.
This, and the idea of categorified Hilbert spaces
(which would be the second ingredient in a categorified spectral
triple) was discussed in \cite{Baez:1997}.

And indeed, it seems that starting from such a categorified GN theorem
and following the logic of categorified NCG one does arrive at
descriptions of string physics, as discussed further in
\S\fullref{Vector 2-Bundles}.

The starting point for turning geometry into algebra is that
spaces may be characterized by algebras of functions over them.
For instance, topological spaces are characterized by
$C^\ast$-algebras
of continuous functions (the Gelfand-Naimark theorem)
and measure spaces by
von Neumann algebras of bounded measurable functions.

In each case points of the space $X$ are recovered in terms
homomorphisms from the algebra of functions
$K^X \defas \set{f : X \to K}$
to $K$ itself: For every $x \in X$ we get a homomorphism
$\tilde x \maps K^X \to K$ by setting
\begin{eqnarray*}
  \tilde x &:& K^X \to K
  \\ 
  && f \mapsto f\of{x}
  \,.
\end{eqnarray*}

When categorifying, spaces becomes 2-spaces
(categories whose point and morphism spaces are topological spaces, or
measure space, etc.) and functions become functors.

Let $Q$ be any 2-space and
let $\mathcal{K}$ be any monoidal category. The functor category
$\mathcal{K}^Q$
(\cf \S\fullref{elements of cat theory: natural transformations})
now indeed encodes not just the point space of
$Q$, but also the arrow space:

Every point $x \in \Ob\of{Q}$ gives rise to a functor $\tilde x$
defined by
\begin{eqnarray*}
  \tilde x &:& \mathcal{K}^Q \to \mathcal{K}
  \\
  &&(F \stackto{\eta} G) \mapsto 
  \left(
    F\of{x} \stackto{\eta\of{x}} G\of{x}
  \right)
\end{eqnarray*}
and every arrow $\gamma : x \to y$ in $\Mor\of{Q}$ gives rise to a natural transformation
$\tilde\gamma$ between these functors
\begin{eqnarray*}
  \tilde \gamma &:& \tilde x \to \tilde y
  \\
  && 
    \tilde x\of{F} \stackto{F\of{\gamma}} \tilde y\of{F}
  \,.
\end{eqnarray*}

This is best seen by looking at some naturality squares.
Here is a 2-space $Q$ together with a chain of three 
functors $F \to G \to H$ from $Q$ to $\mathcal{K}$:

%\begin{figure}[h]
\begin{center}
\begin{picture}(200,260)
\includegraphics{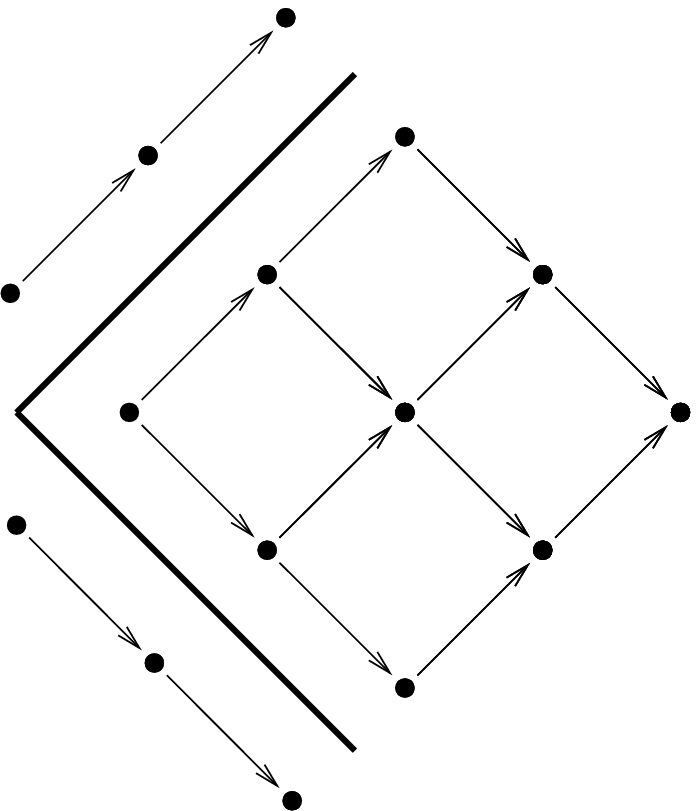}
\put(-207,153){$x$}
\put(-187,173){$\gamma_1$}
\put(-167,193){$y$}
\put(-197,223){$Q$}
\put(-147,213){$\gamma_2$}
\put(-127,233){$z$}
\put(-208,73){$F$}
\put(-185,59){$\eta$}
\put(-168,33){$G$}
\put(-220,5){$\mathcal{K}^Q$}
\put(-145,19){$\rho$}
\put(-128,-8){$H$}
\put(46,-38){
\begin{picture}(0,0)
\put(-207,153){${}_{F\of{x}}$}
\put(-192,170){${}_{F\of{\gamma_1}}$}
\put(-167,193){${}_{F\of{y}}$}
\put(-152,210){${}_{F\of{\gamma_2}}$}
\put(-127,233){${}_{F\of{z}}$}
\end{picture}
}
\put(85,-78){
\begin{picture}(0,0)
\put(-207,153){${}_{G\of{x}}$}
\put(-192,170){${}_{G\of{\gamma_1}}$}
\put(-167,193){${}_{G\of{y}}$}
\put(-152,210){${}_{G\of{\gamma_2}}$}
\put(-127,233){${}_{G\of{z}}$}
\end{picture}
}
\put(124,-118){
\begin{picture}(0,0)
\put(-207,153){${}_{H\of{x}}$}
\put(-192,170){${}_{H\of{\gamma_1}}$}
\put(-167,193){${}_{H\of{y}}$}
\put(-152,210){${}_{H\of{\gamma_2}}$}
\put(-127,233){${}_{H\of{z}}$}
\end{picture}
}
\put(60,-54){
\begin{picture}(0,0)
\put(-207,153){${}_{\eta\of{x}}$}
\put(-167,193){${}_{\eta\of{y}}$}
\put(-127,233){${}_{\eta\of{z}}$}
\end{picture}
}
\put(100,-94){
\begin{picture}(0,0)
\put(-207,153){${}_{\rho\of{x}}$}
\put(-167,193){${}_{\rho\of{y}}$}
\put(-127,233){${}_{\rho\of{z}}$}
\end{picture}
}
\end{picture}
%\caption{\label{funcduality eps}
%{\bf }
%}
\end{center}
%\end{figure}

By using the definition of $\tilde x$ and $\tilde \gamma$ from
above this can be relabelled equivalently to look like
a 2-space with three points $F$, $G$ and $H$ and a chain 
$\tilde x \to \tilde y \to \tilde z$ of three functors from
this to $\mathcal{K}$:

%\begin{figure}[h]
\begin{center}
\begin{picture}(200,260)
\includegraphics{funcduality.eps}
\put(-207,153){$\tilde x$}
\put(-187,173){$\tilde \gamma_1$}
\put(-167,193){$\tilde y$}
\put(-240,223){$\mathcal{K}^{\left(\mathcal{K}^Q\right)}$}
\put(-147,213){$\tilde \gamma_2$}
\put(-127,233){$\tilde z$}
\put(-208,73){$F$}
\put(-185,59){$\eta$}
\put(-168,33){$G$}
\put(-220,5){$\mathcal{K}^Q$}
\put(-145,19){$\rho$}
\put(-128,-8){$H$}
\put(46,-38){
\begin{picture}(0,0)
\put(-207,153){${}_{\tilde x\of{F}}$}
\put(-192,170){${}_{\tilde \gamma_1\of{F}}$}
\put(-167,193){${}_{\tilde y\of{F}}$}
\put(-152,210){${}_{\tilde \gamma_2\of{F}}$}
\put(-127,233){${}_{\tilde z\of{F}}$}
\end{picture}
}
\put(85,-78){
\begin{picture}(0,0)
\put(-207,153){${}_{\tilde x\of{G}}$}
\put(-192,170){${}_{\tilde \gamma_1\of{G}}$}
\put(-167,193){${}_{\tilde y\of{G}}$}
\put(-152,210){${}_{\tilde \gamma_2\of{G}}$}
\put(-127,233){${}_{\tilde z\of{G}}$}
\end{picture}
}
\put(124,-118){
\begin{picture}(0,0)
\put(-207,153){${}_{\tilde x\of{H}}$}
\put(-192,170){${}_{\tilde \gamma_1\of{H}}$}
\put(-167,193){${}_{\tilde y\of{H}}$}
\put(-152,210){${}_{\tilde \gamma_2\of{H}}$}
\put(-127,233){${}_{\tilde z\of{H}}$}
\end{picture}
}
\put(60,-54){
\begin{picture}(0,0)
\put(-207,153){${}_{\tilde x\of{\eta}}$}
\put(-167,193){${}_{\tilde y\of{\eta}}$}
\put(-127,233){${}_{\tilde z\of{\eta}}$}
\end{picture}
}
\put(100,-94){
\begin{picture}(0,0)
\put(-207,153){${}_{\tilde x\of{\rho}}$}
\put(-167,193){${}_{\tilde y\of{\rho}}$}
\put(-127,233){${}_{\tilde z\of{\rho}}$}
\end{picture}
}
\end{picture}
%\caption{\label{funcduality eps}
%{\bf }
%}
\end{center}
%\end{figure}

This duality is at the heart of the categorified Gelfand-Naimark
theorem \cite{Baez:1997}.

\subsubsection{Vector 2-Bundles}
\label{Vector 2-Bundles}

Physical fields are in general not just functions on
parameter space (spacetime/worldvolume) but
are {\bf sections of fiber bundles} over parameter space.
Similarly, the wave function itself is in general not a
function on configuration space, but a section of some
bundle over configuration space.
A central theme in categorified SQM is therefore necessarily that of
2-bundles.

It is well known (and reviewed in \cite{Aspinwall:2004,Lazaroiu:2003})
that states of open strings ending on
D-branes are described by certain {\bf derived categories}
(of coherent sheaves or of quiver representations).
In \S\fullref{Possible Relation of Vector 2-Bundles to String Theory}
it is discussed how this might naturally be interpretable in terms
of categorified wave functions taking values in a {\bf line 2-bundle}.

There are several equivalent descriptions of ordinary vector bundles. It
turns out that the categorification depends on which description one starts
with.

When using the definition which says that the typical fiber of a
vector bundle is a vector space one ends up with categorifying the
concept of a vector space itself. This was done in
\cite{BaezCrans:2003}. The {\bf 2-vector spaces} obtained this way are the
right concept for instance for discussing Lie 2-algebras
(\S\fullref{introduction: Lie 2-Algebras})
but they do not
seem to give rise to an interesting notion of vector 2-bundle.

The definition of vector bundles most natural in the NCG context
is that saying that a vector bundle
$E \to M$ is a finitely generated projective module of the
$C^\ast$­-algebra $\mathbb{C}^M$ of complex-valued continuous functions
on $M$. Categorifying this definition amounts to categorifying
$\mathbb{C}^M$ such that the result is what should be called a 2-ring.

\paragraph{Vector 2-Bundles as 2-Modules.}
\label{Vector 2-bundles as modules}

Aspects of this problem have notably been addressed in \cite{Baez:1997}.
There it was argued that a good categorification of
$\mathbb{C}^M$ is
a functor category $\Hilb^Q$, where $Q$ is some base 2-space replacing
$M$ and $\Hilb$ is the category of Hilbert spaces, replacing
$\mathbb{C}$.

A slightly different but very similar idea is used in
\cite{KapranovVoevodsky:1994}, where instead of the category
$\Hilb$ the category $\Vect$ is used. This amounts simply to
forgetting about the scalar product.

The crucial point is that the tensor product in $\Vect$ ($\Hilb$)
makes $\Vect^Q$ ($\Hilb^Q$) into a monoidal category
and indeed,
at least in the case studied in \cite{Baez:1997}, into something that
deserves to be called a {\bf 2-algebra}.

In the spirit of this concept of categorified function algebras
the authors of \cite{BaasDundasRognes:1993} defined a vector 2-bundle to be
something that locally is similar to
a bundle whose  typical fiber looks like $\Vect^n$, for some integer $n$.

Equivalence classes of ordinary vector bundles are
described by K-theory. Therefore one would expect that equivalence
classes of vector 2-bundles are described by some categorification
of K-theory, which perhaps should be related to the elliptic genus.

In \cite{BaasDundasRognes:1993} however it was found that equivalence
classes of the vector 2-bundles as defined there are not quite described
by elliptic cohomology, even though by something the authors call a
\emph{form} of elliptic cohomology.

On the other hand, this is maybe not too surprising. One should note
that $\Vect^Q$ is more like a categorification of functions taking
values in the natural numbers, than in the complex numbers
(compare the discussion in \cite{BaezDolan:1998}). In particular,
there are no additive or multiplicative inverses in $\Vect^Q$.
Due to that the ``transition functions'' in \cite{BaasDundasRognes:1993}
are in general not invertible, for instance. This should mean that
there must be a categorification of the notion
``vector bundle'' which
more faithfully captures the crucial properties of ordinary vector bundles.

\paragraph{How to categorify function algebras?}
\label{How to categorify function algebras}

The above disucssion suggests that some more thoughts on the
``right'' categorification of function algebras is in order
before vector 2-bundles can be addressed.
One possibility to improve on $\Vect^Q$ might be the following:

We had observed that $\Vect^Q$ is lacking additive and multiplicative
inverses.
Hence we could try to
enlarge $\Vect^Q$ by including such inverses,
in a way similar to how one gets from the natural numbers
to the integers and then the rational numbers.

In order to discuss this it turns out to be helpful to circumvent
a couple of problems for the moment by restricting attention to
base 2-spaces $Q$ whose sets of points and arrows are \emph{finite}.
In particular, restrict attention to categories $Q = C_{(V,E)}$ which
are free categories over finite directed graphs $(V,E)$
(\cf \S\fullref{introduction: categories: examples}). 
This serves as the categorification of the concept of a space
consisting of a finite number of points. 

In this case it is a simple fact that the functor 
category $\Vect^Q$ is the same as the category $\LMod{KQ}$ of (left, say)
modules of the {\bf path algebra} $KQ$ of $Q$,
\[
  \Vect^Q = \LMod{KQ}
  \,.
\]
Here the path algebra $KQ$ is the 
algebra freely generated by the set of morphisms in $Q$ with the
product between these generators defined to be their composition
when composable and zero otherwise.

This equivalent reformulation suggests to use the tensor product
over $KQ$ in order to form a monoidal category. By this reasoning
we are led to include multiplication and multiplicative inverses 
by going from
$\LMod{KQ}$ to $\BiMod{KQ}{KQ}$, the 
category of $KQ$ \emph{bimodules} (over $K$).
The multiplicative inverses in $\BiMod{KQ}{KQ}$
give rise to a group known as the \emph{Picard group} of $KQ$.

(In a more general context one might of course want to consider
different algebras $A$, $B$ and the category $\BiMod{A}{B}$.
By left monoidal multiplication 
the weakly invertible elements $T \in \BiMod{A}{B}$  
give rise (if they exist) to a 'tilting equivalence'
between $\LMod{A}$ and $\RMod{B}$, in which case
$A$ and $B$ are Morita equivalent.)

In a next step this category should be enlarged to allow a notion
of subtraction. This again implies that given any object $b$ there
should be a way to 'decompose' it into abjects $a$ and $c$. 
A diagramatic way to do this is by means of an
exact sequence $a \to b \to c$. A slightly more
general concept than this is that of a \emph{distinguished triangle}
in a triangulated category.

Using a triangulated category together with a {\bf stability condition}
on it \cite{Bridgeland:2003,GorodentscevKuleshovRudakov:2003} 
subtraction is implemented by taking
direct sums and then projecting the result onto the subset of
objects which are {\bf stable} with respect to this stability
condition. 

Now, a triangulated category is naturally obtained from $\LMod{KQ}$
by passing to its {\bf derived category}  
$\Derived{\LMod{KQ}}$.
The derived category $\Derived{C}$ of any additive category
$C$ is like the category $\Ch\of{C}$ of chain complexes in $C$
but modulo some identifications.

Hence we should choose a stability condition on
$\Derived{\LMod{KQ}}$ and
also pass from
$\BiMod{KQ}{KQ}$ to its derived category.
$\Derived{\BiMod{KQ}{KQ}}$.

(The weakly invertible objects in $\Derived{KQ-\mathrm{Mod}-KQ}$
are known as \emph{two-sided tilting complexes} and their
isomorphism classes form a group known as the
\emph{derived Picard group} of $KQ$ 
\cite{Yekutieli:1999,MiyachiYekutieli:2000,MiyachiYekutieli:2001,Yekutieli:2004}.)

\paragraph{Vector 2-Bundles as $\Derived{\BiMod{A}{A}}$-modules.}
\label{Vector 2-Bundles as D BiModAA-modules}

This way we have arrived at the proposal that a `good' categorification
of an ordinary  function algebra on the set $\Ob\of{Q}$ with values in 
$K$ would be 
to replace $\Ob\of{Q}$ by Q and the function algebra by the
monoidal category
$\Derived{\BiMod{KQ}{KQ}}$. Modules of this 
would locally look like 
$\left(\Derived{KQ-\mathrm{Mod}}\right)^n$ for
some integer $n$. These would be our proposed vector 2-bundles
over $Q$.

One might be worried that by going to derived categories
the simple idea that a categorified function on $\Ob\of{Q}$
is a functor on $Q$ is lost. However, this is not the case.
Namely, a chain complex of functors $Q \to \Vect$ is the same
as a functor $Q \to \mathrm{\mathbf{Ch}}\of{\Vect}$, 
\[
  \mathrm{\mathbf{Ch}}\of{\Vect^Q}
  =
  \left(
    \mathrm{\mathbf{Ch}}\of{\Vect}
  \right)^Q
\]
as 
the following diagram illustrates.

%\begin{figure}[h]
\begin{center}
\begin{picture}(290,180)
\includegraphics{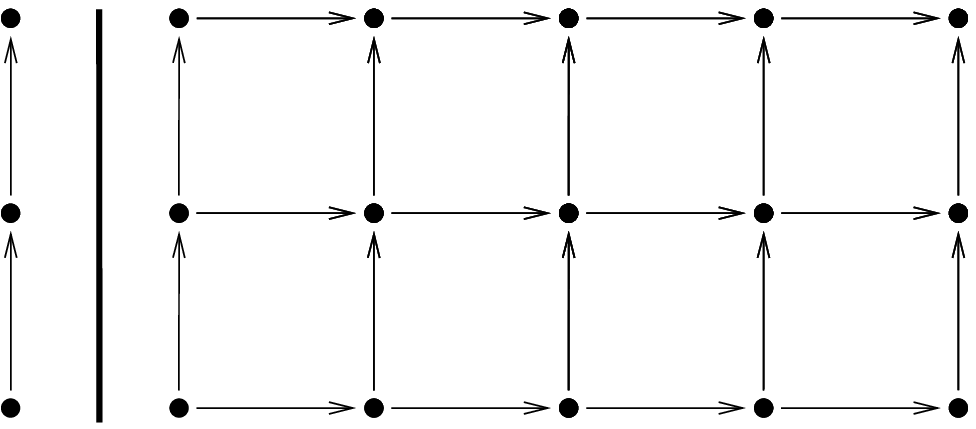}
\put(-289,3){$x$}
\put(-289,30){$\gamma_1$}
\put(-289,61){$y$}
\put(-289,87){$\gamma_2$}
\put(-289,117){$z$}
\put(92,11){
\begin{picture}(0,0)
\put(-289,3){${}_{F^0\of{x}}$}
\put(-292,30){${}_{F^0\of{\gamma_1}}$}
\put(-289,61){${}_{F^0\of{y}}$}
\put(-292,87){${}_{F^0\of{\gamma_2}}$}
\put(-289,117){${}_{F^0\of{z}}$}
\end{picture}
}
\put(148,11){
\begin{picture}(0,0)
\put(-289,3){${}_{F^1\of{x}}$}
\put(-292,30){${}_{F^1\of{\gamma_1}}$}
\put(-289,61){${}_{F^1\of{y}}$}
\put(-292,87){${}_{F^1\of{\gamma_2}}$}
\put(-289,117){${}_{F^1\of{z}}$}
\end{picture}
}
\put(204,11){
\begin{picture}(0,0)
\put(-289,3){${}_{F^2\of{x}}$}
\put(-292,30){${}_{F^2\of{\gamma_1}}$}
\put(-289,61){${}_{F^2\of{y}}$}
\put(-292,87){${}_{F^2\of{\gamma_2}}$}
\put(-289,117){${}_{F^2\of{z}}$}
\end{picture}
}
\put(276,11){
\begin{picture}(0,0)
\put(-289,3){${}_{0}$}
\put(-289,61){${}_{0}$}
\put(-289,117){${}_{0}$}
\end{picture}
}
\put(50,11){
\begin{picture}(0,0)
\put(-289,3){${}_{0}$}
\put(-289,61){${}_{0}$}
\put(-289,117){${}_{0}$}
\end{picture}
}
\put(122,8){
\begin{picture}(0,0)
\put(-289,3){${}_{d_1\of{x}}$}
\put(-289,61){${}_{d_1\of{y}}$}
\put(-289,117){${}_{d_1\of{z}}$}
\end{picture}
}
\put(180,8){
\begin{picture}(0,0)
\put(-289,3){${}_{d_2\of{x}}$}
\put(-289,61){${}_{d_2\of{y}}$}
\put(-289,117){${}_{d_2\of{z}}$}
\end{picture}
}
\end{picture}
%\caption{\label{funcduality eps}
%{\bf }
%}
\end{center}
%\end{figure}

Here $x \stackto{\gamma_1} y \stackto{\gamma_2} z$
is a morphism in $Q$. The $F$ on the right can be either regarded as
giving a functor $Q \to \Ch{\Vect}$ or a chain complex
\[
  0 \to F^1 \stackto{d_1} F^1 \stackto{d_2} F^2 \to 0
\]
of functors $F^i : Q \to \Vect$.
That this diagram commutes can be regarded as a consequence
of the definition of morphisms between functors (natural transformations)
or as a consequence of the definition of chain maps.
Moreover,
due to the rule for 'vertical' composition of natural
transformations (which is really going horizontal in this figure),
we have $d_1\of{x}\circ d_2\of{x} = 0$ and similarly for
$y$ and $z$ irrespective whether we regard this as a chain complex of
functors
or as a single functor into $\mathrm{\mathbf{Ch}}\of{\Vect}$.

This means that a
{\bf vector 2-bundle} $E$
according to the above proposal would be a 2-bundle
with typical
fiber $\left(\Ch{\Vect}\right)^n$, i.e. one with
local 2-trivializations
\[
\xymatrix @!0
{ p^{-1} U_i
 \ar [dddrr]_{p|_{p^{-1}U_i}}
  \ar[rrrr]^{t_i}
  & &  & &
  U_i \times \left(\Ch{\Vect}\right)^n
  \ar[dddll]^{}
  \\ \\ &  \\ & &
   U_i
 }
\]
(\cf \S\fullref{introduction: principal 2-bundles}).

A 2-section of this bundle resticted to $U_i \isomorphic Q_i$
would be a functor from $Q_i$ to
$\left(\Ch{\Vect}\right)^n$, i.e.
 an element in
\[
  \Ob\of{
    \left(\left(\Ch{\Vect}\right)^{Q_i}\right)^n
  }
  =
  \Ob\of{
      \left(
         \Derived{\LMod{KQ_i}}
      \right)^n
  }
  \,.
\]

On double overlaps $U_{ij} = Q_i \cap Q_j$ the 2-transition
\[
  \left.\bar t_i \circ t_j\right|_{U_{ij}} : 
  Q_{ij} \times (\Ch{\Vect})^n \to
  Q_{ij} \times (\Ch{\Vect})^n
\] 
must be an invertible $n\times n$ matrix of
two-sided tilting complexes in 
$\Derived{\BiMod{KQ}{KQ}}$.

More precisely, such a matrix should act componentwise 
by the usual formula
for matrix multiplication using the derived product operation
of $\Derived{\BiMod{KQ_{ij}}{KQ_{ij}}}$ on 
$\Derived{\LMod{KQ_{ij}}}$
and the direct sum operation in $\Derived{\LMod{KQ_{ij}}}$
followed by a stability projection using the stability condition
that we have chosen on $\Derived{\LMod{KQ_{ij}}}$ 
(in \S\fullref{How to categorify function algebras}).
These matrices would supersede the non-invertible transition matrices
in \cite{BaasDundasRognes:1993}.

The tractable special
case of a {\bf line 2-bundle} (i.e. a vector 2-bundle with $n=1$)
is already quite interesting:

\paragraph{Derived Category Description of D-Branes.}
\label{Possible Relation of Vector 2-Bundles to String Theory}

If the general relation between categorification and
stringification mentioned at the beginning of
\S\fullref{Introduction}, as well as the notion of vector 2-bundle
in \S\fullref{Vector 2-Bundles as D BiModAA-modules}
are any good, then a 2-section of a line 2-bundle
as described above should describe states of
string, somehow.

And this indeed turns out to be the case.

By the above reasoning a 2-section of a line 2-bundle is
locally an object of
$\Derived{\LMod{KQ}} = \Derived{\Vect^Q}$.
This is indeed known to describe states of string stretched
between D-branes, as reviewed in
\cite{Aspinwall:2004,Lazaroiu:2003}.

In this context the path category $Q$, or rather its underlying
graph, is really a quiver diagram encoding the precise nature of the moduli
space of the effective field theory on these D-branes, or equivalently
the nature of the transverse 'compact' dimensions. Hence it might seem that
the analogy argued for above breaks down in that $Q$ is not in any sense
a categorified spacetime. But it turns out that
viewed from a suitable perspective $Q$ does have to be identified with a
latticized spacetime after all, an effect known as
{\bf dimensional deconstruction}. In particular, when
$Q$ is like $\Z_\infty$ or $\Z_\infty \times \Z_\infty$ it
describes \cite{Arkani-HamedCohenKaplanKarchMotl:2003}
two compactified dimensions of theories on 5-branes
where 2-bundles are expected to play a role.

As discussed in \S\fullref{Vector 2-Bundles as D BiModAA-modules},
gauge transformations of such a 2-section of a line 2-bundle
have to be elements
of the weak Picard 2-group inside $\Derived{\BiMod{KQ}{KQ}}$.
And indeed, these elements are known to describe duality transformations
on these configurations, known as (fractional) Seiberg
duality.

The relation of Seiberg duality to tilting equivalences is
discussed in section 5.4 of \cite{BerensteinDouglas:2002}.
Its relation to monodromies in moduli space, which goes back
to \cite{FengHananyHeUranga:2002} and others, is briefly
reviewed in the introduction of \cite{FrancoHanany:2004}.

\newpage
\section{Conclusion}

The work we have presented consists of two parts. One is concerned with
lifting supersymmetric quantum mechanics and its deformation theory
to loop space and looking at string theory from this point of view. 
The other is concerned with constructing a global 
framework for this by categorifying the notion of
globally defined parallel transport and holonomy for (world)lines to 
obtain globally defined 2-holonomy and parallel transport for 
(world)surfaces.

In the last subsection, 
\S\fullref{2-NCG and Derived Category Description of D-Branes},
we sketched some observations that are suggestive of a deeper 
principle behind this. Based on the fact that supersymmetric quantum
mechanics is essentially nothing but spectral geometry of
configuration space in terms of spectral triples, the above considerations
suggest a grand scheme where stringy physics arises as a
\emph{categorified} supersymmetric quantum mechanics, or a categorified
spectral geometry of stringy configuration space.
While we did not attempt to seriously tackle this idea in its entirety,
it serves as a motivating principle for the
investigations and results that we presented.

We started by elaborating on known reformulations of the super Virasoro
constraints of the RNS supersting as Dirac-K{\"a}hler operators on loop
space, and showed how the known deformation principle for supersymmetric
quantum mechanics allows to describe, when lifted to loop space, 
background fields for the superstring in terms of algebraic deformations.
Without going into any mathematical rigour concerning this point we
pointed out how this can be understood as a deformation of spectral
triples for loop space geometry. 

In the process of studying how D-branes, 
expressed in terms of boundary states, fit into this picture, we 
studied aspects of the so-called Pohlmeyer invariants, which are
functionals on loop space that commute with all (super) Virasoro
constraints. After (re)discovering the relation of these
invariants to the more popular DDF invariants, we related them
to the boundary states that describe D-branes with nonabelian 
gauge fields turned on. 

This lead to the observation that a formally very natural 
generalization of these 
boundary states gives rise to a nonabelian connection 1-form on loop
space with certain curious propertie, that were previously encountered 
in the context of categorified gauge theory. 

Motivated by this we stepped back and began to investigate the global
picture that this construction would be part of, showing that
these connection 1-forms are precisely those that give rise to
what we call 2-connections with 2-holonomy in strict 2-bundles.
This was done in collaboration with John Baez.

In particular, we showed how the well-known cocycle relations for
nonabelian gerbes with connection and curving are reproduced 
using this theory, thereby providing a useful alternative
point of view on these structures, maybe 
in some sense a more transparent one.

This allowed us to construct what was previously known only in the
case of abelian gerbes or else in the nonabelian but simplicial and
globally trivial case, namely globally well-defined nonabelian 2-holonomy
in possibly nontrivial 2-bundles. 

We remarked that with such a 2-holonomy in hand there is an obvious way
to generalize the usual action functional for strings coupled to 
an abelian 2-form (described by the abelian 2-holonomy 
of an abelian gerbe) to the nonabelian case. 
To the best of our knowledge this gives for the first time a 
serious candidate formalism for capturing the dynamics of
strings coupled to nonabelian $2$-form fields. But here
we did not attempt to go into any details concerning 
possible relations of such nonabelian action functionals to known or
expected physics of such nonabelian 2-form fields.

On the other hand, in a collaboration with John Baez, Alissa Crans and
Danny Stevenson, we could show that there are 2-bundles whose
structure 2-group is related to the group
$\String\of{n}$, which have apparently all the necessary properties to
describe the global dynamics of spinning strings, i.e. which
capture the global issues of spinors on loop space. There are several
indications that our concept of 2-holonomy for these 2-bundles
is closely related to the work on spinning strings by Stolz and Teichner,
but here we only gave some hints concerning this point. 

A largely complementary way to describe $p$-bundles with $p$-connection
is their ``differential'' formulation. This gives rise to a nonabelian
and weakened generalization of the well-known Deligne hypercohology
description of abelian gerbes, which is a powerful formalism 
for deriving linearized cocycle conditions and 
gauge transformation laws in general semistrict $p$-bundles with
$p$-connection. Moreover, this formalism is well 
adapted to the study of the dynamics of nonabelian $p$-form field theories,
at least perturbatively. We end our present discussion with an
analysis of the
nonabelian Deligne hypercohomology description of the
semistrict differential version of the above 2-bundles with
structure 2-group related to $\String\of{n}$.

\addtocontents{toc}{\protect\newpage}

\clearpage

\parbox{6cm}{
``One cannot help but\\
feel that there are many\\
beautiful secrets hidden\\
in loop space.''\\
{\it A. Polyakov} \cite{Polyakov:1987}
}

\part{SQM on Loop Space}
\label{SQM on Loop Space}

In \S\ref{Morse-theory-like Deformations on Loop Space}
the RNS superstring is approached from the point of view
of supersymmetric quantum mechanics on loop space and, 
background fields are related to deformations. This is taken from
\cite{Schreiber:2004}.
For special cases of deformations this leads to the
discussion of worldsheet invariants and boundary states in
\S\fullref{Worldsheet Invariants and Boundary States}, which
is taken from \cite{Schreiber:2004b,Schreiber:2004d}.
A generalization of such boundary states is shown in
\S\fullref{Connections on Loop Space from Worldsheet Deformations}
to give rise to the nonabelian local connections on loop
space which then motivate the developments in part
\ref{Nonabelian Strings}.

\addtocontents{toc}{\vspace*{-.2em}}

\clearpage
\section{Deformations and Background Fields}
\label{Morse-theory-like Deformations on Loop Space}

\subsection{Introduction}

Supersymmetric field theories look like 
Dirac-K{\"a}hler systems when formulated in Schr{\"o}dinger representation.
This has been well studied in the special limits where only a finite number of
degrees of freedom are retained, such as the semi-classical quantization
of solitons in field theory (see e.g. \cite{HollowoodKingaby:2003}
for a brief introduction and further references). 
That this phenomenon is rooted in the general structure of supersymmetric field 
theory has been noted long ago in the second part of 
\cite{Witten:1982} (see also the second part of \cite{Witten:1985}). 
For 2 dimensional superconformal field theories describing
superstring worldsheets a way to exploit this fact for the construction of 
covariant target space Hamiltonians 
(applicable to the computation of curvature corrections of string spectra 
in nontrivial backgrounds) has been proposed in \cite{Schreiber:2003a}.
In the construction of these Hamiltonians a pivotal role is played 
by a new method for obtaining functional representations of
superconformal algebras (corresponding to non-trivial
target space backgrounds) by means of certain deformations of the 
superconformal algebra.

In \cite{Schreiber:2003a} the focus was on deformations which 
induce Kalb-Ramond backgrounds and
only the 0-mode of the superconformal algebra was considered explicitly (which
is sufficient for the construction of covariant target space Hamiltonians). Here
this deformation technique is developed in more detail for the full superconformal
algebra and for all massless bosonic string background fields. 
Other kinds of backgrounds can also be incorporated in principle and one goal 
of this discussion is to demonstrate the versatility of the new deformation technique
for finding explicit functional realizations of the two-dimensional superconformal algebra.

The setting for our formalism is the
representation of the superconformal algebra on the exterior bundle
over loop space (the space of maps from the circle into target space) 
by means of $K$-deformed
exterior (co)derivatives $\extd_K$, $\coextd_K$, where
$K$ is the Killing vector field on loop space which induces loop reparameterizations.

The key idea is that 
the form of the superconformal algebra is preserved under the 
deformation\footnote{
  Throughout this section we use the term ``deformation'' to mean the
  operation \refer{the fundamelental deformation idea} on the
  superconformal generators, the precise definition of which is given
  in \S\fullref{deformations}. These ``deformations'' are actually
  \emph{isomorphisms} of the superconformal algebra, but affect its
  representations in terms of operators on 
  the exterior bundle over loop space.
  In the literature one finds also other usages of the word ``deformation''
  in the context of superalgebras, 
  for instance for describing the map where the superbrackets 
  $\superCommutator{\cdot}{\cdot}$
  are transformed as
  \begin{eqnarray}
    \superCommutator{A}{B} \to \superCommutator{A}{B} + \sum\limits_{t=1}^\infty
    \omega_i\of{A,B}t^i
  \end{eqnarray}
  with $\omega_i\of{A,B}$ elements of the superalgebra and $t$ a real number 
  (see \cite{AzcarragaIzquierdoPiconVarela:2004}).
} 
\begin{eqnarray}
  \label{the fundamelental deformation idea}
  \extd_K &\to& e^{-\bf W}\,\extd_K \,e^{\bf W}
  \nonumber\\
  \coextd_K &\to& e^{{\bf W}^\dag}\,\coextd_K\, e^{-{\bf W}^\dag}
\end{eqnarray}
if $\bf W$ is an even graded operator that satisfies a certain consistency condition.

The canonical (functional) form of the superconformal generators for all 
massless NS and NS-NS backgrounds can neatly be expressed this way by deformation operators 
$\bf W$ that are bilinear in the fermions, as will be shown here. 
It turns out that there is one further bilinear in the fermions 
which induces a background that probably has to be interpreted as the RR 2-form as coupled to the
D-string.

It is straightforward to find further deformation operators and hence further backgrounds. 
While the normal ordering effects which affect the 
superconformal algebras and which would give rise to equations of motion for the
background fields are not investigated here,
there is still a consistency condition to be satisfied 
which constrains the admissible deformation operators. 

This approach for obtaining new superconformal algebras from existing ones by applying deformations 
is similar in spirit, but rather complementary, to the method of
'\emph{canonical deformations}' studied by Giannakis, Evans, Ovrut, Rama, Freericks, Halpern and  others
\cite{Giannakis:1999,OvrutRama:1992,EvansOvrut:1990,EvansOvrut:1989,FreericksHalpern:1988}. 
There, the superconformal generators $T$ and $G$ of one chirality are deformed to lowest order as
\begin{eqnarray}
  \label{canonical deformation in introduction}
  T\of{z} &\to& T\of{z} + \delta T\of{z}
  \nonumber\\
  G\of{z} &\to& G\of{z} + \delta G\of{z}
  \,.
\end{eqnarray}
Requiring the deformed generators to satisfy the desired algebra to first order shows 
that $\delta T$ and $\delta G$ must be bosonic and fermionic components 
of a weight 1 worldsheet superfield. 
(An adaption of this procedure to deformations of the BRST charge itself is discussed 
in \cite{Giannakis:2002}. Another related discussion of deformations of BRST operators
is given in \cite{Kato:1995}.)

The advantage of this method over the one discussed in the following is that it operates at the 
level of quantum SCFTs and has powerful CFT tools at its disposal, such as normal ordering and 
operator product expansion. The disadvantage is that it only applies perturbatively to 
first order in the background fields, and that these background fields always
appear with a certain gauge fixed.

On the other hand, the deformations discussed here which are induced by 
$\extd_K \to e^{-\bf W}\extd_K e^{\bf W} \sim e^{-\bf W}(iG + \bar G)e^{\bf W}$ 
preserve the superconformal algebra for arbitrarily large perturbations $\bf W$. 
The drawback is that normal ordering is non-trivially affected, too, and without 
further work the resulting superconformal algebra is only available  
on the level of (bosonic and fermionic) Poisson brackets. 

We show in \S\fullref{canonical deformations from d to dingens} that when
restricted to first order the deformations that we are considering reproduce the theory
of canonical deformations \refer{canonical deformation in introduction}.

Our deformation method is also technically different from but related to the 
\emph{marginal deformations} of conformal field theories 
(see \cite{FoersteRoggenkamp:2003} for a review and further references),
where one sends the correlation function $\langle A\rangle$ of some operator
$A$ to the deformed correlation function
\begin{eqnarray}
  \langle A\rangle^\lambda
  &\defas&
  \langle A \exp\of{\sum\limits_i \lambda_i \int {\cal O}_i\; {\rm dvol}}\rangle
  \,,
\end{eqnarray}
where ${\cal O}_i$ are fields of conformal weight $1$. This corresponds
to adding the integral over a field of unit weight to the action. How this relates to 
the algebraic deformations of the superconformal algebra considered here
is discussed in \S\fullref{review of first order canonical CFT deformations}.

The method discussed here generalizes the transformations 
studied in \cite{LizziSzabo:1998}, where strings are regarded from
the non-commutative geometry perspective. The main result of 
this approach (which goes back to \cite{FroehlichGawedzki:1993}
and \cite{EvansGiannakis:1996})
is that T-duality as well as mirror symmetry can nicely be encoded 
by means of automorphisms of the vertex operator algebra. In terms of
the above notation such automorphisms correspond to deformations
induced by \emph{anti-Hermitean} ${\bf W}^\dag = -{\bf W}$,
which induce pure gauge transformations on the algebra. 

The analysis given here generalizes the approach of \cite{LizziSzabo:1998}
in two ways: First, the use of Hermitean $\bf W$ in our formalism 
produces backgrounds which are not related by string dualities. Second, by 
calculating the functional form of the superconformal generators for
these backgrounds we can study the action of anti-Hermitean $\bf W$
on these more general generators and find the transformation of the
background fields under the associated target space duality.

In particular, we find a duality transformation which changes the sign of the
dilaton and interchanges $B$- and $C$-form fields. It would seem that
this must hence be related to S-duality. This question requires further analysis.

$\,$\\

The structure of this section is as follows:

In \S\ref{loop space} some technical preliminaries necessary for
the following discussion are given.
The functional loop space notation is introduced in 
\S\ref{loop space definitions},
some basic facts about loop space geometry are discussed (\S\ref{exterior geometry on loop space}),
the exterior derivative and coderivative on that space are introduced 
(\S\ref{exterior algebra over loop space}),
and some remarks on isometries of loop space are given in \S\ref{isometries on loop space}.

This is then applied in \S\ref{loop space super-Virasoro generators} 
to the general analysis
of deformations of the superconformal generators. First of all,
the purely
gravitational target space background is shown to be associated to the 
ordinary $K$-deformed loop space exterior derivative (\S\ref{purely gravitational}).
\S\ref{deformations} 
then discusses how general continuous classical deformations
of the superconformal algebra are obtained. As a first application,
\S\ref{gravitational background by algebra isomorphism}
shows how this can be used to get the previously
discussed superconformal generators for purely gravitational backgrounds from
those of flat space by a deformation.

Guided by the form of this deformation the following sections
systematically list and analyze the deformations which are
associated with the Kalb-Ramond, dilaton, and gauge field backgrounds
(\S\ref{loop space and B-field background}, \S\ref{Dilaton background},
\S\ref{A-field background}).
It turns out (\S\ref{C-field background}) 
that one further 2-form background can be obtained
in a very similar fashion, which apparently has to be interpreted 
as the S-dual coupling of the D-string to the $C_2$ 2-form background.

After having understood how the NS-NS backgrounds arise in our formalism
we turn in \S\fullref{canonical deformations and vertex operators} to a comparison of the method 
presented here 
with the well-known 'canonical deformations', which are briefly reviewed in
\S\fullref{review of first order canonical CFT deformations}. 
In \S\fullref{canonical deformations from d to dingens} it is shown how these canonical deformations
are reproduced by means of the methods discussed here and how our deformation
operator $\bf W$ relates to the vertex operators of the respective background fields.

Next the inner relations between the various deformations found are
further analyzed in \S\ref{relations between susy algebras}. 
First of all \S\ref{dK exact deformations} demonstrates how
$\extd_K$-exact deformation operators yield target space gauge
transformations. Then, in \S\ref{T-duality} 
the well known realization of
T-duality as an algebra isomorphism is adapted to the present context,
and in \S\ref{T-duality for various background fields} 
the action of a target space duality obtained
from a certain modified algebra isomorphism on the various
background fields is studied. It turns out that there are
certain similarities to the action of loop space Hodge duality,
which is discussed in \S\ref{Hodge duality on loop space}.

The appendix lists some results from the canonical analysis of the
D-string action, which are needed in the main text.

\newpage
\subsection{Loop Space}
\label{loop space}

In this section the technical setup is briefly established. The  
0-mode $\extd_K$ of the sum of the left- and the rightmoving 
supercurrents is represented as the $K$-deformed exterior derivative on loop space. 
Weak nilpotency of this $K$-deformed operator (namely nilpotency up to reparameterizations) 
is the essential property which implies that the modes of $\extd_K$ and its adjoint generate 
a superconformal algebra. In this sense the loop space perspective on superstrings highlights 
a special aspect of the super Virasoro constraint algebra which turns out to be pivotal 
for the construction of classical deformations of that algebra. 

The kinematical configuration space of the closed \emph{bosonic} string is loop space
${\cal L}\manifold$, the
space of parameterized loops in target space $\manifold$. 
As discussed in \S 2.1 of \cite{Schreiber:2003a}
the kinematical configuration space of the closed \emph{superstring} is therefore the superspace
over ${\cal L}\manifold$, which can be identified with the 1-form bundle
$\Omega^1\of{{\cal L}\manifold}$. Superstring states in Schr{\"o}dinger representation
are super-functionals on $\Omega^1\of{{\cal L}\manifold}$ and hence section of the form bundle
$\Omega\of{{\cal L}\manifold}$ over loop space. 

The main
technical consequence of the infinite dimensionality are the well known divergencies
of certain objects, such as the Ricci-Tensor and the Laplace-Beltrami operator, which
inhibit the naive implementation of quantum mechanics on ${\cal L}\manifold$. But of course
these are just the well known infinities that arise, when working in the Heisenberg (CFT)
instead of in the Schr{\"o}dinger picture, from operator ordering effects, and which
should be removed by imposing normal ordering. Since the choice of Schr{\"o}dinger or Heisenberg
picture is just one of language, the same normal ordering (now expressed in terms of 
functional operators instead of Fock space operators) takes care of infinities in
loop space. We will therefore not have much more to say about this issue here. The
main result of this section are various (deformed) 
representations of the super-Virasoro algebra on loop space
(corresponding to different spacetime backgrounds), and will be
derived in their classical
(Poisson-bracket) form without considering normal ordering effects.

A mathematical discussion of aspects of
loop space can for instance be found in \cite{Wurzbacher:1995,Wurzbacher:2001}.
A rigorous treatment of some of the objects discussed below is also given in
\cite{Arai:1993}.

\subsubsection{Definitions}
\label{loop space definitions}

Let $\left(\manifold,g\right)$ be a pseudo-Riemannian manifold, the target space,
 with metric $g$,  and let
${\cal L}\manifold$ be its loop space consisting of smooth maps of the \emph{parameterized}
circle with parameter $\sigma \sim \sigma + 2\pi$ into $\manifold$:
\begin{eqnarray}
  {\cal L}\manifold &\defas& C^\infty\of{S^1,\manifold} 
  \,.
\end{eqnarray}
The tangent space $T_X{\cal L}\manifold$ of ${\cal L}\manifold$ 
at a loop $X : S^1 \to \manifold$ is  the space of vector fields along that loop.
The metric on $\manifold$ induces a metric on $T_X{\cal L}\manifold$:
Let $g\of{p} = g_{\mu\nu}\of{p}dx^\mu \otimes dx^\nu$ 
be the metric tensor on $\manifold$.
Then we choose for the metric on ${\cal L}\manifold$ at a point $X$ the mapping
\begin{eqnarray}
  \label{metric on loop space}
  T_X{\cal L}\manifold \times T_X{\cal L}\manifold
 &\to&
  \R
  \nonumber\\
  (U,V) 
  &\mapsto&
  U\inner V
  \;=\;
  \int\limits_0^{2\pi} d\sigma\; g\of{X\of{\sigma}}\of{U\of{\sigma}, V\of{\sigma}}
  \nonumber\\
  &&
  \skiph{$U\inner V\;$}
  \;=\;
  \int\limits_0^{2\pi} d\sigma\; g_{\mu\nu}\of{X\of{\sigma}}U^\mu\of{\sigma}V^\nu\of{\sigma}
  \,.
\end{eqnarray}
For the intended applications $T{\cal L}\manifold$ is actually too small, since there will be need to deal with
distributional vector fields on loop space. Therefore one really considers 
$\bar T{\cal L}\manifold$, the completion of $T{\cal L}\manifold$ at each point
$X$ with respect to the norm induced by the inner product \refer{metric on loop space}.)
For brevity, whenever we refer to ``loop space'' in the following, we mean ${\cal L}\manifold$
equipped with the metric \refer{metric on loop space}. Also, the explicit integration region 
$\sigma \in (0,2\pi)$ will be implicit in the following.

To abbreviate the notation, let us introduce formal multi-indices $(\mu,\sigma)$ and
write equivalently 
\begin{eqnarray}
  \label{loop space multi-indices}
  U^\mu\of{\sigma}  
  &\defas&
  U^{(\mu,\sigma)}
\end{eqnarray}
for a vector $U\in T_X{\cal L}\manifold$, and similarly for higher-rank tensors on
loop space.

Extending the usual index notation to the infinite-dimensional setting in the 
obvious way, we also write:
\begin{eqnarray}
  \int U^\mu\of{\sigma} V_\mu\of{\sigma}
  &\defas&
  U^{(\mu,\sigma)}V_{(\mu,\sigma)}
  \,.
\end{eqnarray}
For this to make sense we need to know how to ``shift'' the continuous index
$\sigma$. Because of
\begin{eqnarray}
  \int d\sigma \; g_{\mu\nu}\of{X\of{\sigma}}U^\mu\of{\sigma}V^\nu\of{\sigma}
   &=&
  \int d\sigma\;d\sigma^\prime \;
    \delta\of{\sigma,\sigma^\prime}
   g_{\mu\nu}\of{X\of{\sigma}}U^\mu\of{\sigma}V^\nu\of{\sigma^\prime}
  \nonumber
\end{eqnarray}
it makes sense to write the metric tensor on loop space as
\begin{eqnarray}
  \label{concise metric on loop space}
  G_{(\mu,\sigma)(\nu,\sigma^\prime)}\of{X}
  &\defas&
  g_{\mu\nu}\of{X\of{\sigma}}\delta\of{\sigma,\sigma^\prime}
  \,.
\end{eqnarray}
Therefore 
\begin{eqnarray}
  \langle U, V\rangle
  &=&
  U^{(\mu,\sigma)}
  G_{(\mu,\sigma)(\nu,\sigma^\prime)}
  V^{(\nu,\sigma^\prime)}
\end{eqnarray}
and
\begin{eqnarray}
  V_{(\mu,\sigma)}
  &=&
  G_{(\mu,\sigma)(\nu,\sigma^\prime)}
  V^{(\nu,\sigma^\prime)}
  \nonumber\\
  &=&
  V_\mu\of{\sigma}
  \,.
\end{eqnarray}
Consequently, it is natural to write
\begin{eqnarray}
  \delta\of{\sigma,\sigma^\prime}
  &\defas&
  \delta^{\sigma^\prime}_\sigma 
  \;=\; \delta^\sigma_{\sigma^\prime} 
  \;=\;\delta_{\sigma,\sigma^\prime} \;=\;\delta^{\sigma,\sigma^\prime}
  \,.
\end{eqnarray}
A (holonomic) basis for $T_X{\cal L }\manifold$ may now be written as
\begin{eqnarray}
  \partial_{(\mu,\sigma)}
  &\defas&
  \frac{\delta}{\delta X^\mu\of{\sigma}}
  \,,
\end{eqnarray}
where the expression on the right denotes the functional derivative, so that
\begin{eqnarray}
  \partial_{(\mu,\sigma)} X^{(\nu,\sigma^\prime)}
  &=&
  \delta^{(\nu,\sigma^\prime)}_{(\mu,\sigma)}
  \nonumber\\
  &=&
  \delta^\nu_\mu\,\delta\of{\sigma,\sigma^\prime}
  \,.
\end{eqnarray}

By analogy, many concepts known from finite dimensional geometry carry over
to the infinite dimensional case of loop spaces. Problems arise when 
traces over the continuous ``index'' $\sigma$ are taken, like for contractions
of the Riemann tensor, which leads to undefined diverging expressions. 
It is expected that these are taken care of by the usual normal-ordering of quantum field theory.

\subsubsection{Differential Geometry on Loop Space}
\label{exterior geometry on loop space}

With the metric \refer{concise metric on loop space} on loop space in hand
\begin{eqnarray}
  \label{metric on loop space again}
  G_{(\mu,\sigma)(\nu,\sigma^\prime)}\of{X}
  &=&
  g_{\mu\nu}\of{X\of{\sigma}}\delta_{\sigma,\sigma^\prime}
\end{eqnarray}
the usual objects of differential geometry can be derived for loop space.
Simple calculations yield the Levi-Civita connection as well as the
Riemann curvature, which will be frequently needed later on. The exterior
algebra over loop space is introduced and the exterior derivative and its
adjoint,
which play the central role in the construction
of the super-Virasoro algebra in \S\fullref{purely gravitational},
are constructed in terms of operators on the exterior bundle.
Furthermore isometries on loop space are considered, both 
the one coming from reparameterization of loops as well as those induced from
target space. The former leads to the reparameterization constraint on strings,
while the latter is crucial for the Hamiltonian evolution on loop space
\cite{Schreiber:2003a}.

\paragraph{Basic geometric data.}
The inverse metric is obviously
\begin{eqnarray}
  G^{(\mu,\sigma)(\nu,\sigma^\prime)}\of{X}
  &=&
  g^{\mu\nu}\of{X\of{\sigma}} \delta\of{\sigma,\sigma^\prime}
  \,.
\end{eqnarray}
A vielbein field ${\bf e}^a = e^a{}_\mu \extd x^\mu$ on $\manifold$ gives rise
to a vielbein field ${\bf E}^{(a,\sigma)}$ on loop space:
\begin{eqnarray}
  \label{vielbein on loop space}
  E^{(a,\sigma)}{}_{(\mu,\sigma^\prime)}\of{X}
  &\defas&
  e^a{}_\mu\of{X\of{\sigma}}\delta^\sigma{}_{\sigma^\prime}
\end{eqnarray}
which satisfies
\begin{eqnarray}
  E^{(a,\sigma)}{}_{(\mu,\sigma^{\prime\prime})}
  E^{(b,\sigma)(\mu,\sigma^{\prime\prime})}
  &=&
  \eta^{ab}\delta^{\sigma,\sigma^\prime}
  \nonumber\\
  &\defas&
  \eta^{(a,\sigma)(b,\sigma^\prime)}
\end{eqnarray}
For the Levi-Civita connection one finds:
\begin{eqnarray}
  &&\Gamma_{(\mu\sigma)(\alpha \sigma^\prime)(\beta\sigma^{\prime\prime})}\of{X}
  \nonumber\\
  &=&
  \frac{1}{2}
  \left(
    \frac{\delta}{\delta X^\mu\of{\sigma}} G_{(\alpha,\sigma^\prime)(\beta,\sigma^{\prime\prime})}\of{X}
    +
    \frac{\delta}{\delta X^\beta\of{\sigma^{\prime\prime}}} G_{(\mu,\sigma)(\alpha,\sigma^\prime)}\of{X}
    -
    \frac{\delta}{\delta X^\alpha\of{\sigma^{\prime}}} G_{(\beta,\sigma^{\prime\prime})(\mu,\sigma)}\of{X}
  \right)
  \nonumber\\
  &=&
  \frac{1}{2}\left(
  (\partial_\mu G_{\alpha\beta})\of{X\of{\sigma^\prime}}
  \delta\of{\sigma,\sigma^\prime}\delta\of{\sigma^\prime,\sigma^{\prime\prime}}
  +
  (\partial_\beta G_{\mu\alpha})\of{X\of{\sigma}}
  \delta\of{\sigma^{\prime\prime},\sigma}\delta\of{\sigma,\sigma^{\prime}}  
  \right)
  \nonumber\\
  &&
  -
  \frac{1}{2}
  (\partial_\alpha G_{\beta\mu})\of{X\of{\sigma^{\prime\prime}}}
  \delta\of{\sigma^{\prime},\sigma^{\prime\prime}}\delta\of{\sigma^\prime,\sigma}  
  \nonumber\\
  &=&
  \Gamma_{\mu\alpha\beta}\of{X\of{\sigma}}\;
     \delta\of{\sigma,\sigma^\prime}\delta\of{\sigma^\prime,\sigma^{\prime\prime}}
  \,,
\end{eqnarray}
and hence
\begin{eqnarray}
  \label{loop space LC connection}
  \Gamma_{(\mu,\sigma)}{}^{(\alpha,\sigma^\prime)}{}_{(\beta,\sigma^{\prime\prime})}
  \of{X}
  &=&
  \Gamma_{\mu}{}^\alpha{}_\beta\of{X\of{\sigma}}\;
     \delta\of{\sigma,\sigma^\prime}\delta\of{\sigma^\prime,\sigma^{\prime\prime}}  
 \,.
\end{eqnarray}
The respective connection in an orthonormal basis is
\begin{eqnarray}
  \omega_{(\mu,\sigma)}{}^{(a\sigma^\prime)}{}_{(b,\sigma^{\prime\prime})}\of{X}
  &=&
  E^{(a,\sigma^\prime)}{}_{(\alpha,\rho)}\of{X}
  \left(
    \delta^{(\alpha,\rho)}_{(\beta,\rho^\prime)}
    \partial_{(\mu,\sigma)}
    +
    \Gamma_{(\mu,\sigma)}{}^{(\alpha,\rho)}{}_{(\beta,\rho^{\prime})}\of{X}
  \right)
  E^{(\beta,\rho^{\prime})}{}_{(b,\sigma^\prime)}\of{X}
  \nonumber\\
  &=&
  \omega_\mu{}^a{}_b\of{X\of{\sigma}}
  \delta\of{\sigma,\sigma^\prime}\delta\of{\sigma^\prime,\sigma^{\prime\prime}}
  \,.
\end{eqnarray}
From \refer{loop space LC connection} the Riemann tensor on loop space is obtained as
\begin{eqnarray}
  &&R_{(\mu,\sigma_1)(\nu,\sigma_2)}{}^{(\alpha,\sigma_3)}{}_{(\beta,\sigma_4)}\of{X}
  \nonumber\\
  &=&
  2
  \frac{\delta}{\delta X^{[(\mu,\sigma_1)}}
  \Gamma_{(\nu,\sigma_2)]}{}^{(\alpha,\sigma_3)}{}_{(\beta,\sigma_4)}
  +
  2
  \Gamma_{[(\mu,\sigma_1)}{}^{(\alpha,\sigma_3)}{}_{|(X,\sigma_5)|}
  \Gamma_{(\nu,\sigma_2)]}{}^{(X,\sigma_5)}{}_{(\beta,\sigma_4)}
  \nonumber\\
  &=&
  R_{\mu\nu}{}^\alpha{}_\beta\of{X\of{\sigma_1}}\;
  \delta\of{\sigma_1,\sigma_2}
  \delta\of{\sigma_2,\sigma_3}
  \delta\of{\sigma_3,\sigma_4}
  \,.
\end{eqnarray}
The Ricci tensor is formally
\begin{eqnarray}
  R_{(\mu,\sigma)(\nu,\sigma^\prime)}\of{X}
  &=&
  R_{(\kappa,\sigma^{\prime\prime})(\mu,\sigma)}{}^{(\kappa,\sigma^{\prime\prime})}{}_{(\nu,\sigma^\prime)}\of{X}
  \nonumber\\
  &=&
  R_{\mu\nu}\of{X\of{\sigma}}\delta\of{\sigma,\sigma^\prime} \;\delta\of{\sigma^{\prime\prime},\sigma^{\prime\prime}}
  \,,
\end{eqnarray}
which needs to be regularized. Similarly the curvature scalar is formally
\begin{eqnarray}
  R\of{X}
  &=&
  R_{(\mu,\sigma)}{}^{(\mu,\sigma)}\of{X}
  \nonumber\\
  &=&
  R\of{X\of{\sigma}} \delta^\sigma_\sigma \delta\of{\sigma^{\prime\prime},\sigma^{\prime\prime}}
  \,.
\end{eqnarray}

\paragraph{Exterior and Clifford algebra over loop space.}
\label{exterior algebra over loop space}

The anticommuting fields ${\cal E}^{\dag (\mu,\sigma)}$,
${\cal E}_{(\mu,\sigma)}$, satisfying the CAR
\begin{eqnarray}
 \antiCommutator
   {{\cal E}^{\dag(\mu,\sigma)}}
   {{\cal E}^{\dag (\nu,\sigma^\prime)}}
  &=& 0
 \nonumber\\
 \antiCommutator
   {{\cal E}_{(\mu,\sigma)}}
   {{\cal E}_{(\nu,\sigma^\prime)}}
  &=& 0
 \nonumber\\
 \antiCommutator
   {{\cal E}_{(\mu,\sigma)}}
   {{\cal E}^{\dag (\nu,\sigma^\prime)}}
  &=&
  \delta^{(\mu,\sigma)}_{(\nu,\sigma^\prime)}
  \,,
\end{eqnarray}
are assumed to exist over loop space, in analogy with
the creators and annihilators $\coordCreator^\mu$, $\coordAnnihilator_\mu$ 
on the exterior bundle in finite dimensions as described in  
appendix A of \cite{Schreiber:2003a}.
(For a mathematically rigorous treatment of the continuous CAR
compare \cite{Wurzbacher:2001} and references given there.)
From them the Clifford fields
\begin{eqnarray}
  \label{loop space cliffords}
  \Gamma_\pm^{(\mu,\sigma)}
  &\defas&
  {\cal E}^{\dag (\mu,\sigma)}
  \pm{\cal E}^{(\mu,\sigma)}
\end{eqnarray}
are obtained, which satisfy
\begin{eqnarray}
  \label{modes by commuting with number operator}
  \antiCommutator{\Gamma_\pm^{(\mu,\sigma)}}{\Gamma_\pm^{(\nu,\sigma^\prime)}}
  &=&
 \pm 2 G^{(\mu,\sigma)(\nu,\sigma^\prime)}
  \nonumber\\
  \antiCommutator{\Gamma_\pm^{(\mu,\sigma)}}{\Gamma_\mp^{(\nu,\sigma^\prime)}}
  &=&
  0
  \,.
\end{eqnarray}
Since the $\Gamma_\pm$ will be related to spinor fields on the string's worldsheet,
we alternatively use spinor indices $A, B,\ldots \in \set{1,2} \simeq \set{+,-}$ and write
\begin{eqnarray}
  \antiCommutator{
   \Gamma_A^{(\mu,\sigma)}}{\Gamma_B^{(\nu,\sigma^\prime)}}
 &=&
  2 s_A \delta_{AB} G^{(\mu,\sigma)(\nu,\sigma^\prime)}
  \,.
\end{eqnarray}
Here $s_A$ is defined by
\begin{eqnarray}
 s_+ \;=\; +1\,,\hspace{.2cm} s_-\;=\;-1
  \,.
\end{eqnarray}
The above operators will frequently be needed with respect to some orthonormal frame
$E^{(a,\sigma)}$:
\begin{eqnarray}
  \Gamma_A^{(a,\sigma)}
  &\defas&
  E^{(a,\sigma)}{}_{(\mu,\sigma^\prime)}\Gamma_A^{(\mu,\sigma^\prime)}
  \,.
\end{eqnarray}

Just like in the finite dimensional case, the following derivative operators
can now be defined:

The covariant derivative operator (\cf A.2 in \cite{Schreiber:2003a}) on 
the exterior bundle over loop space is
\begin{eqnarray}
  \gradOp_{(\mu,\sigma)}
  &=&
  \partial^c_{(\mu,\sigma)}
  -
  \Gamma_{(\mu,\sigma)}{}^{(\alpha,\sigma^\prime)}{}_{(\beta,\sigma^{\prime\prime})}
  {\cal E}^{\dag(\beta,\sigma^{\prime\prime})}
  {\cal E}_{(\alpha,\sigma^\prime)}
  \nonumber\\
  &=&
  \partial^c_{(\mu,\sigma)}
  -
  \int d\sigma^\prime\;d\sigma^{\prime\prime}
  \Gamma_{\mu}{}^\alpha{}_\beta\of{X\of{\sigma}}
  \delta\of{\sigma,\sigma^\prime}
  \delta\of{\sigma^\prime,\sigma^{\prime\prime}}
  {\cal E}^{\dag \beta}\of{\sigma^{\prime\prime}}
  {\cal E}_{\alpha}\of{\sigma^\prime}
  \nonumber\\
  &=&
  \partial^c_{(\mu,\sigma)}
  -
  \Gamma_{\mu}{}^\alpha{}_\beta\of{X\of{\sigma}}
  {\cal E}^{\dag \beta}\of{\sigma}
  {\cal E}_{\alpha}\of{\sigma}
\end{eqnarray}
or alternatively
\begin{eqnarray}
  \label{loop space grad op}
  \gradOp_{(\mu,\sigma)}
  &=&
  \partial_{(\mu,\sigma)}
  -
  \omega_{\mu}{}^a{}_b\of{X\of{\sigma}}
  {\cal E}^{\dag b}\of{\sigma}
  {\cal E}_{a}\of{\sigma}  
  \,.
\end{eqnarray}
One should note well the difference between the functional derivative 
$\partial^c_{(\mu,\sigma)}$ which commutes with the coordinate frame
forms ($[{\partial^c_{(\mu,\sigma)}}{{\cal E}^{\dag \nu}}] = 0$)
and the functional derivative $\partial_{(\mu,\sigma)}$ which instead
commutes with the ONB frame forms
($[{\partial_{(\mu,\sigma)}}{{\cal E}^{\dag a}}] = 0$). See
(A.29) of \cite{Schreiber:2003a} for more details.

In terms of these operators the exterior derivative and coderivative on loop space read, 
respectively (A.39)
\begin{eqnarray}
  \label{extds on loop space}
  \extd
  &=&
  {\cal E}^{\dag(\mu,\sigma)}\partial^c_{(\mu,\sigma)}
  \nonumber\\
  &=&
  {\cal E}^{\dag(\mu,\sigma)}\gradOp_{(\mu,\sigma)}
  \nonumber\\
  \coextd
  &=&
  -{\cal E}^{(\mu,\sigma)}\gradOp_{(\mu,\sigma)}
  \,.
\end{eqnarray}

We will furthermore need the form number operator
\begin{eqnarray}
  {\cal N}
  &=&
  {\cal E}^{\dag (\mu,\sigma)}{\cal E}_{(\mu,\sigma)}
\end{eqnarray}
as well as its \emph{modes}: Let $\xi : S^1 \to \mathbb{C}$ be a smooth function then
\begin{eqnarray}
  \label{mode of form number operator}
  {\cal N}_\xi &\defas&
  \int d\sigma\;
  \xi\of{\sigma} {\cal E}^{\dag \mu}\of{\sigma}{\cal E}_{\mu}\of{\sigma}
\end{eqnarray}
is the $\xi$-mode of the form number operator. Commuting it with the
exterior derivative yields the modes of that operator:
\begin{eqnarray}
  \extd_\xi
  &\defas&
  \commutator{{\cal N}_\xi}{\extd}
  \nonumber\\
  &=&
  \int d\sigma\;
  \xi\of{\sigma}
  \;
  {\cal E}^{\dag \mu}\of{\sigma}\gradOp_\mu\of{\sigma}
  \nonumber\\
  \coextd_\xi &\defas&
  -
  \commutator{{\cal N}_\xi}{\coextd}
  \nonumber\\
  &=&
  -
  \int d\sigma\;
  \xi\of{\sigma}
  \;
  {\cal E}^{\mu}\of{\sigma}\gradOp_\mu\of{\sigma}
  \,.
\end{eqnarray}
These modes will play a crucial role in \S\fullref{loop space super-Virasoro generators}.

\paragraph{Isometries.}
\label{isometries on loop space}

Regardless of the symmetries of the metric $g$ on $\manifold$, loop space
$\left({\cal L}\manifold,G\right)$ has an isometry generated by the
reparameterization flow vector field $K$, which is defined by:\footnote{
Here and in the following a prime indicates the derivative with respect to
the loop parameter $\sigma$: $X^\prime\of{\sigma} = \partial_\sigma X\of{\sigma}$.
}
\begin{eqnarray}
  \label{reparameterization Killing field on loop space}
  K^{(\mu,\sigma)}\of{X}
  &=&
  T
  \,
  X^{\prime\mu}\of{\sigma}
  \,.
\end{eqnarray}
(Here $T$ is just a constant which we include for later convenience.)
The flow generated by this
vector field obviously rotates the loops around. Since the metric \refer{metric on loop space again}
is ``diagonal'' in the parameter $\sigma$, this leaves the geometry of loop space invariant, and
the vector field $K$ satisfies Killing's equation
\begin{eqnarray}
  G_{(\nu,\sigma^\prime)(X,\sigma^{\prime\prime})}
  \nabla_{(\mu,\sigma)} K^{(X,\sigma^{\prime\prime})}
  +
  G_{(\mu,\sigma)(X,\sigma^{\prime\prime})}
  \nabla_{(\nu,\sigma^\prime)} K^{(X,\sigma^{\prime\prime})}
  &=&
  0\,,
\end{eqnarray}
as is readily checked.

The Lie-derivative along $K$ is (see section A.4 of \cite{Schreiber:2003a})
\begin{eqnarray}
  {\cal L}_K
  &=&
  \antiCommutator{{\cal E}^{\dag(\mu,\sigma)}\partial^c_{(\mu,\sigma)}}
  {{\cal E}_{(\nu,\sigma^\prime)}X^{\prime (\nu,\sigma^\prime)}}
  \nonumber\\
  &=&
  X^{\prime(\mu,\sigma)}
  \partial^c_{(\mu,\sigma)}
  +
  {\cal E}^{\dag(\mu,\sigma)}{\cal E}_{(\nu,\sigma^\prime)}
  \delta^\prime_{\sigma^\prime,\sigma}
  \nonumber\\
  &=&
  X^{\prime(\mu,\sigma)}
  \partial^c_{(\mu,\sigma)}
  +
  {\cal E}^{\dag\prime(\mu,\sigma)}{\cal E}_{(\mu,\sigma)}
  \,.
\end{eqnarray}
This operator will be seen to be an essential ingredient of the 
super-Virasoro algebra in \S\fullref{loop space super-Virasoro generators}.

Apart from the generic isometry \refer{reparameterization Killing field on loop space},
every symmetry of the target space manifold $\manifold$ gives rise to a family
of symmetries on ${\cal L}\manifold$:
Let $v$ be any Killing vector on target space,
\begin{eqnarray}
  \nabla_{(\mu}v_{\nu)} &=& 0
  \,,
\end{eqnarray}
then every vector $V$ on loop space of the form
\begin{eqnarray}
  \label{loop space Killing from target space Killing}
  V_\xi\of{X} 
  &=& V_\xi^{(\mu,\sigma)}\of{X}\partial_{(\mu,\sigma)}
  \;\defas\;
  v^\mu\of{X\of{\sigma}} \xi^\sigma \partial_{(\mu,\sigma)}
  \,,
\end{eqnarray}
where $\xi^\sigma = \xi\of{\sigma}$ is some differentiable function 
$S^1\to \mathbb{C}$,
is a Killing vector on loop space.
For the commutators one finds
\begin{eqnarray}
  \label{commutator of reparameterization with induced Killing vectors}
  \commutator{V_{\xi_1}}{V_{\xi_2}} &=& 0
  \nonumber\\
  \commutator{V_\xi}{K} &=& V_{\xi^\prime}
  \,.
\end{eqnarray}
The reparameterization Killing vector $K$ will be used to deform
the exterior derivative on loop space as discussed in \S 2.1.1 of 
\cite{Schreiber:2003a}, and a target space induced Killing vector $V_\xi$ will
serve as a generator of parameter evolution on loop space along the lines
of \S 2.2 of \cite{Schreiber:2003a}. There it was found in equation (88) that the
condition
\begin{eqnarray}
  \commutator{K}{V_\xi} &=& 0
\end{eqnarray}
needs to be satisfied for this to work. Due to 
\refer{commutator of reparameterization with induced Killing vectors}
this means that one needs to choose $\xi = {\rm const}$, i.e. use the
integral lines of $V_{\xi =1 }$ as the ``time''-parameter on loop space.
This is only natural: It means that every point on the loop is evolved
equally along the Killing vector field $v$ on target space.

\subsection{Superconformal Generators for Various Backgrounds}
\label{loop space super-Virasoro generators}

We now use the loop space technology to 
show that the loop space exterior derivative deformed by the
reparameterization Killing vector $K$ gives rise to the superconformal
algebra which describes string propagation in purely gravitational
backgrounds. General deformations of this algebra are introduced and
applying these we find representations
of the superconformal algebra that correspond to all the massless
NS and NS-NS background fields.

(Parts of this construction were already indicated in \cite{Schreiber:2003a},
but there only the 0-modes of the generators and 
only a subset of massless bosonic background fields was considered, without 
spelling out the full nature of the necessary constructions on loop space.)

\subsubsection{Purely Gravitational Background}
\label{purely gravitational}

In this subsection it is described how to obtain a representation of the classical super-Virasoro algebra
on loop space. For a trivial background the construction itself is relatively trivial and,
possibly in different notation, well known. The point that shall be emphasized here
is that the identification of super-Virasoro generators with modes of the 
deformed exterior(co-)derivative on loop space allows a convenient treatment of curved
backgrounds as well as more general non-trivial background fields.

As was discussed in \cite{Schreiber:2003a}, \S 2.1.1 
(which is based on \cite{Witten:1982,Witten:1985}), 
one may obtain from the exterior derivative and
its adjoint on a manifold the generators of a global $D=2$, $N=1$ superalgebra by
deforming with a Killing vector. The generic Killing vector field on loop space is
the reparameterization generator \refer{reparameterization Killing field on loop space}.
Using this to deform the exterior derivative and its adjoint as in 
equation (19) of \cite{Schreiber:2003a} yields the operators
\begin{eqnarray}
  \label{k-defomred extds on loop space}
  \extd_{K}
  &\defas&
  \extd + i{\cal E}_{(\mu,\sigma)}X^{\prime(\mu,\sigma)}
  \nonumber\\
  \coextd_{K}
  &\defas&
  \coextd - i{\cal E}^\dag_{(\mu,\sigma)}X^{\prime(\mu,\sigma)}  
  \,,
\end{eqnarray}
(where for convenience we set $T=1$ for the moment)
which generate a \emph{global} superalgebra. Before having a closer look at this algebra
let us first enlarge it to a local superalgebra by considering the \emph{modes}
defined by
\begin{eqnarray}
  \label{super-virasoro generators}
  \extd_{K,\xi}
  &\defas&
  \commutator{{\cal N}_\xi}{\extd^\ast_{K}}
  \nonumber\\
  \coextd_{K,\xi^\ast}
  &\defas&
  -\commutator{{\cal N}_\xi}{\coextd^\ast_{K}}  
  \,,
\end{eqnarray}
where $\cdot^\ast$ is the complex adjoint and ${\cal N}_\xi$ is the $\xi$-mode of the
form number operator discussed in \refer{mode of form number operator}.
They explicitly read
\begin{eqnarray}
  \label{modes of deformed extd coexted on loop space}
  \extd_{K,\xi}
  &=&
  \int d\sigma\;
  \xi\of{\sigma}
  \left(
    {\cal E}^{\dag\mu}\of{\sigma}
    \partial^c_\mu\of{\sigma}
    +
    i{\cal E}_\mu\of{\sigma}
    X^{\prime \mu}\of{\sigma}
  \right)
  \nonumber\\
  \coextd_{K,\xi}
  &=&
  -
  \int d\sigma\;
  \xi\of{\sigma}
  \left(
    {\cal E}^{\mu}\of{\sigma}
    \nabla_\mu\of{\sigma}
    +
    i{\cal E}^\dag_\mu\of{\sigma}
    X^{\prime \mu}\of{\sigma}
  \right)
  \,.  
\end{eqnarray}
Making use of the fact that $\extd_{K,\xi}$ is actually independent of the background metric,
it is easy to establish the algebra of these operators. We do this for the ``classical'' fields,
ignoring normal ordering effects and the anomaly:

The anticommutator  of the operators \refer{super-virasoro generators}
with themselves
defines the $\xi$-mode ${\cal L}_{K,\xi}$ of the Lie-derivative ${\cal L}_{K}$ along $K$:
\begin{eqnarray}
  \label{anticom of extds}
  \antiCommutator{\extd_{K,\xi_1}}{\extd_{K,\xi_2}}  
  &=&
  2i{\cal L}_{K,\xi_1\xi_2}
  \,,
\end{eqnarray}
where
\begin{eqnarray}
  \label{modes of the reparameterization Lie derivative}
  {\cal L}_{\xi}
  &=&
  \int d\sigma\;
  \left(
  \xi\of{\sigma}
    X^{\prime\mu}\of{\sigma}
   \partial_\mu^c\of{\sigma}
   +
   \sqrt{\xi}
   \left(
     \sqrt{\xi}
     {\cal E}^{\dag\mu}
   \right)^\prime
  \of{\sigma}
   {\cal E}_\mu\of{\sigma}
  \right)
  \,.
\end{eqnarray}
We say that a field $A\of{\sigma}$ has \emph{reparameterization weight}
$w$ if
\begin{eqnarray}
  \label{transformation under reparameterization}
  \superCommutator{{\cal L}_{\xi}}{A\of{\sigma}}
  &=&
  \left(
  \xi A^\prime + w \xi^\prime A
  \right)\of{\sigma}
  \nonumber\\
  \superCommutator{{\cal L}_{\xi_1}}{A_{\xi_2}}
  &=&
  A_{(w-1)\xi_1^\prime\xi_2 - \xi_1\xi_2^\prime}
  \,,
\end{eqnarray}
where $A_\xi \defas \int d\sigma\; \xi A$.
For the basic fields we find
\begin{eqnarray}
  \label{reparameterization weight of basic fields on loop space}
  w\of{X^\mu} &=& 0
  \nonumber\\
  w\of{X^{\prime\mu}} &=& 1
  \nonumber\\
  w\of{\partial^c_\mu} &=& 1
  \nonumber\\
  w\of{\Gamma_\pm^\mu} &=& 1/2
  \,.
\end{eqnarray}
Because of  $w\of{AB} = w\of{A} + w\of{B}$ it follows that
$\extd_{K,\xi}$ and $\coextd_{K,\xi}$ are modes of integrals over densities of
reparameterization weight $w = 3/2$. This implies in particular that
\begin{eqnarray}
  \label{comm of L extd}
  \commutator
   {{\cal L}_{\xi_1}}
   {\extd_{K,{\xi_2}}}
   &=&
  \extd_{K,(\frac{1}{2}\xi_1^\prime\xi_2 - \xi_1\xi_2^\prime)}
  \\
  \commutator{{\cal L}_{K,\xi_1}}{{\cal L}_{K,\xi_2}}
  &=&
  {\cal L}_{K,(\xi_1^\prime\xi_2 - \xi_1\xi_2^\prime)}
  \,.
\end{eqnarray}

By taking the adjoint of \refer{anticom of extds} and \refer{comm of L extd}
(or by doing the calculation explicitly),
analogous relations are found for $\coextd_{K,\xi}$:
\begin{eqnarray}
  \label{vir algebra for coextd}
  \antiCommutator{\coextd_{K,\xi_1}}{\coextd_{K,\xi_2}}
  &=&
  2i {\cal L}_{K,\xi_1\xi_2}
  \nonumber\\
  \label{comm L coextd}
  \commutator{
    {\cal L}_{K,\xi_1}
  }{\coextd_{K,\xi_2}}
  &=&
  \coextd_{K,(\frac{1}{2}\xi_1^\prime\xi_2-\xi_1\xi_2^\prime)}
  \,.
\end{eqnarray}
Equations \refer{anticom of extds}, 
\refer{comm of L extd}, and \refer{vir algebra for coextd} 
give us part of the sought-after algebra.
A very simple and apparently unproblematic but rather crucial step for finding the rest 
is to now define the \emph{modes of the deformed Laplace-Beltrami operator}
as the right hand side of
\begin{eqnarray}
  \label{extdxi coextdxi anticommutator}
  \antiCommutator{\extd_{K,\xi_1}}{\coextd_{K,\xi_2}}
  &=&
  \fatDelta_{K,\xi_1\xi_2}
  \,.
\end{eqnarray}
For this definition to make sense one needs to check that
\begin{eqnarray}
  \label{crucial condition}
  \antiCommutator{\extd_{K,\xi_1\xi_3}}{\coextd_{K,\xi_2}} &=& 
\antiCommutator{\extd_{K,\xi_1}}{\coextd_{K,\xi_2\xi_3}}\,.
\end{eqnarray} It is easy to
verify that this is indeed true for the operators as given in 
\refer{modes of deformed extd coexted on loop space}. 
However, in \S\fullref{deformations} it is found that this condition is a rather strong
constraint on the admissible perturbations of these operators, and the innocent
looking equation \refer{crucial condition} plays a pivotal role
in the algebraic construction of superconformal field theories in the
present context.

With $\fatDelta_{K,\xi}$ consistently defined as in 
\refer{extdxi coextdxi anticommutator}
all remaining brackets follow by using the Jacobi-identity:
\begin{eqnarray}
  \label{remaining brackets}
  \commutator{\frac{1}{2}\fatDelta_{K,\xi_1}}{\extd_{K,\xi_2}}
  &=&
  i\coextd_{K,(\frac{1}{2}\xi_1^\prime\xi_2 - \xi_1\xi_2^\prime)}
  \nonumber\\
  \commutator{\frac{1}{2}\fatDelta_{K,\xi_1}}{\coextd_{K,\xi_2}}
  &=&
  i\extd_{K,(\frac{1}{2}\xi_1^\prime\xi_2 - \xi_1\xi_2^\prime)}
  \nonumber\\
 \nonumber\\
  \commutator{\frac{1}{2}\fatDelta_{K,\xi_1}}{\frac{1}{2}\fatDelta_{K,\xi_2}}
  &=&
  -\, {\cal L}_{K,(\xi_1^\prime\xi_2 - \xi_1\xi_2^\prime)}
  \,.
\end{eqnarray}

In order to make the equivalence to the super-Virasoro algebra
of the algebra thus obtained
more manifest consider the modes of the $K$-deformed Dirac-K{\"a}hler operators 
on loop space:
\begin{eqnarray}
  \label{K-deformed loop space Dirac operator}
  \Dirac_{K,\pm} 
   &\defas& \extd_{K} \pm \coextd_{K}
  \nonumber\\
   &=&
  \Gamma_\mp^{(\mu,\sigma)}\left(\gradOp_{(\mu,\sigma)} \mp iTX^\prime_{(\mu,\sigma)}\right)
  \nonumber\\
  \Dirac_{K,\pm,\xi} &\defas& \extd_{K,\xi} \pm \coextd_{K,\xi}
  \,.
\end{eqnarray}
They are the supercharges which generate the super-Virasoro algebra in the usual chiral form
\begin{eqnarray}
  \label{suVir algebra in almost usual form}
  \antiCommutator{\Dirac_{K,\pm,\xi_1}}{\Dirac_{K,\pm,\xi_2}}
  &=&
  4
  \left(
    \pm
    \frac{1}{2}\fatDelta_{\xi_1\xi_2}
    + 
    i {\cal L}_{\xi_1\xi_2}
  \right)
  \nonumber\\
  \commutator{
    \pm
    \frac{1}{2}\fatDelta_{K,\xi_1}
    + 
    i {\cal L}_{\xi_1}
  }{\Dirac_{K,\pm,\xi_2}}
  &=&
  2 \Dirac_{K,\pm,\frac{1}{2}\xi_1^\prime \xi_2 - \xi_1 \xi_2^\prime}
  \nonumber\\
  \commutator{
    \pm
    \frac{1}{2}\fatDelta_{K,\xi_1}
    + 
    i {\cal L}_{\xi_1}
  }{
    \pm
    \frac{1}{2}\fatDelta_{K,\xi_2}
    + 
    i {\cal L}_{\xi_2}
}
  &=&
  2i
  \left(
    \pm
    \frac{1}{2}\fatDelta_{K,\xi_1^\prime\xi_2 - \xi_1 \xi_2^\prime}
    + 
    i {\cal L}_{\xi_1^\prime\xi_2 - \xi_1 \xi_2^\prime}
  \right)
  \,.
\end{eqnarray}
It is easily seen that this acquires the standard form when we set $\xi\of{\sigma} = e^{in\sigma}$
for $n\in \N$. In order to make the connection with the usual formulation more
transparent consider a flat target space.
If we define the operators
\begin{eqnarray}
  \label{functional definition of Pplusminus}
  {\cal P}_{\pm,(\mu,\sigma)}
  &\defas&
  \frac{1}{\sqrt{2T}}\left(-i\partial_{(\mu,\sigma)} \pm T X^\prime_{(\mu,\sigma)}\right)
\end{eqnarray}  
with commutator
\begin{eqnarray}
  \label{the functional commutator of Pplusminus}
  \commutator
   { {\cal P}_{A,(\mu,\sigma)} }
   { {\cal P}_{B,(\nu,\sigma^\prime)} }
  &=&
  i s_A \delta_{AB}\eta_{\mu\nu}\delta^\prime_{\sigma,\sigma^\prime}
  \,,
  \hspace{1cm}
  \mbox{for}\;g_{\mu\nu} = \eta_{\mu\nu}
\end{eqnarray}
we get, up to a constant factor, the usual modes
\begin{eqnarray}
  \Dirac_{K,\pm,\xi}
  &=&
  \sqrt{2T}i
  \int d\sigma\,
  \xi\of{\sigma}
  \Gamma_{\mp}^\mu\of{\sigma}
  {\cal P}_{\mu,\mp}\of{\sigma}
  \nonumber\\
  \Dirac^2_{K,\pm,\xi^2}
  &=&
  \pm 2T
  \int d\sigma\;
  \left(
  \xi^2\of{\sigma}
  {\cal P}_{\mp}\of{\sigma}\inner {\cal P}_{\mp}\of{\sigma}
  -
  \frac{i}{2}
  \xi\of{\sigma}\left(\xi\Gamma_{\mp}\right)^\prime\of{\sigma}\inner\Gamma_{\mp}\of{\sigma}
  \right)
  \,.
\end{eqnarray}

\subsubsection{Isomorphisms of the Superconformal Algebra}
\label{deformations}

The representation of the superconformal algebra as above is manifestly of the
form considered in \S 2.1.1 of \cite{Schreiber:2003a}. We can therefore now
study isomorphisms of the algebra along the lines of \S 2.1.2 of that
paper in order to obtain new SCFTs from known ones. 

From \S 2.1.2 of \cite{Schreiber:2003a} it follows 
that the general continuous isomorphism of the 0-mode sector ($\xi = 1$)
of the algebra \refer{suVir algebra in almost usual form} is induced by some operator
\begin{eqnarray}
  {\bf W} &=& \int d\sigma\; W\of{\sigma}
  \,,
\end{eqnarray}
where $W$ is a field on loop space of unit reparameterization weight 
\begin{eqnarray}
  \label{weight condition on deformation operator}
  w\of{W} = 1\,,
\end{eqnarray}
and looks like
\begin{eqnarray}
  \extd_{K,1} &\mapsto& \extd^{\bf W}_{K,1} \;\defas\; \exp\of{-{\bf W}}\extd_{K,1} \exp\of{{\bf W}}
  \nonumber\\
  \coextd_{K,1} &\mapsto& \coextd^{\bf W}_{K,1} \;\defas\; \exp\of{{\bf W}^\dag}\coextd_{K,1} \exp\of{-{\bf W}^\dag}
  \nonumber\\
  \fatDelta_{K,1} &\mapsto& \fatDelta^{\rm W}_{K,1} \;\defas\; \antiCommutator{\extd^{\bf W}_{K,1}}{\coextd^{\bf W}_{K,1}}
  \nonumber\\
  {\cal L}_1 &\mapsto& {\cal L}_1
  \,.
\end{eqnarray}
This construction immediately generalizes to the full algebra including all modes
\begin{eqnarray}
  \label{isomorphism of Virasoro algebra}
  \extd_{K,\xi} &\mapsto& \extd^{\bf W}_{K,\xi} \;\defas\; \exp\of{-{\bf W}}\extd_{K,\xi} \exp\of{{\bf W}}
  \nonumber\\
  \coextd_{K,\xi} &\mapsto& \coextd^{\bf W}_{K,\xi} \;\defas\; \exp\of{{\bf W}^\dag}\coextd_{K,\xi} \exp\of{-{\bf W}^\dag}
  \nonumber\\
  {\cal L}_\xi &\mapsto& {\cal L}_\xi
\end{eqnarray}
\emph{if}
the crucial relation
\begin{eqnarray}
  \label{crucial relation again}
  \fatDelta^{\bf W}_{K,\xi_1\xi_2} &=& \antiCommutator{\extd^{\bf W}_{K,\xi_1}}{\coextd^{\bf W}_{K,\xi_2}}
\end{eqnarray}
remains well defined, i.e. if \refer{crucial condition} remains true:
\begin{eqnarray}
  \label{crucial condition for deformed operators}
  \antiCommutator{\extd^{\bf W}_{K,\xi_1\xi_3}}{\coextd^{\bf W}_{K,\xi_2}}
  &=&
  \antiCommutator{\extd^{\bf W}_{K,\xi_1}}{\coextd^{\bf W}_{K,\xi_2\xi_3}}
  \,.
\end{eqnarray}

The form of these deformations follows from the fact that no matter which background fields
are turned on, the generator \refer{modes of the reparameterization Lie derivative} of
spatial reparameterizations (at fixed worldsheet time) remains the same, because the
string must be reparameterization invariant in any case. Preservation of the relation
$\mathbf{d}_K^2 = i {\cal L}_K$, which says that $\mathbf{d}_K$ is nilpotent up to
reparameterizations, then implies that $\mathbf{d}_K$ may transform under a similarity
transformation as in the first line of \refer{isomorphism of Virasoro algebra}.  The
rest of \refer{isomorphism of Virasoro algebra} then follows immediately. 

Since this is an important point, at the heart of the approach presented here, 
we should also reformulate it in a more conventional language. Let $L_m$,
$\bar L_m$, $G_m$, $\bar G_m$ be the holomorphic and
antiholomorphic modes of the super Virasoro algebra.
As discussed in \S\fullref{purely gravitational} we have
\begin{eqnarray}
  \label{loop supervirasor in terms of modes}
  \mathbf{\Delta}_{K,\xi} &\propto& L_m + \bar L_{-m}
  \nonumber\\
  \mathbf{\cal L}_{K,\xi} &\propto& L_m - \bar L_{-m}
  \nonumber\\
  \extd_{K,\xi} &\propto& i G_m + \bar G_{-m}
  \nonumber\\
  \coextd_{K,\xi} &\propto& -iG_m + \bar G_{-m} 
  \,,
\end{eqnarray}
with $\xi\of{\sigma} = e^{-im\sigma}$,
as well as
\begin{eqnarray}
  {\mathbf W} &\propto& \sum\limits_n W_n \bar W_n
  \,,
\end{eqnarray}
where $W_m$ and $\bar W_m$ are the modes of the holomorphic and antiholomorphic 
parts of $\mathbf{W}$, which have weight $h$ and $\bar h$, respectively. 
The goal is to find a deformation of \refer{loop supervirasor in terms of modes}
such that $L_m - \bar L_{-m}$ is preserved. Since this is the square of $\pm i G_m + \bar G_{-m}$
the latter may receive a similarity transformation which does not affect $L_m - \bar L_{-m}$
itself. Using $\commutator{L_m}{W_n} = ((h-1)m-n)W_{n+m}$ and similarly for the antiholomorphic part
we see that this is the case for
\begin{eqnarray}
  \label{deformations in mode algebra picture}
  i G_m + \bar G_{-m} &\to& \exp\of{-\sum\limits_n W_n \bar W_n}
  \left(i G_m + \bar G_{-m}\right)\exp\of{\sum\limits_n W_n \bar W_n}
 \nonumber\\
  -i G_m + \bar G_{-m} &\to& \exp\of{\sum\limits_m  \bar W_n^\dagger W_n^\dagger}
  \left(-i G_m + \bar G_{-m}\right)\exp\of{-\sum\limits_n \bar W_n^\dagger W_n^\dagger}
\end{eqnarray}
with 
\begin{eqnarray}
  \label{total unit weight condition on W}
  h + \bar h = 1
  \,,
\end{eqnarray}
because then
\begin{eqnarray}
  L_m - \bar L_{-m} &\to&
  \exp\of{-\sum\limits_n W_n \bar W_n}\left(L_m - \bar L_{-m}\right)
  \exp\of{\sum\limits_n W_n \bar W_n} = L_m - L_{-m}
  \,.
  \nonumber\\
\end{eqnarray}

The point of the loop-space formulation above is to clarify the nature of these deformations,
which in terms of the $L_m, \bar L_m, G_m,\bar G_m$ look somewhat peculiar. In the loop
space formulation it becomes manifest that we are dealing here with a generalization of
the deformations first considered in \cite{Witten:1982} for supersymmetric quantum mechanics, where
the supersymmetry generators are the exterior derivative and coderivative
and are sent by two different similarity transformations
to two new nilpotent supersymmetry generators. This and the relation to the
present approach to superstrings is discussed in detail in section 2.1 of
\cite{Schreiber:2003a}.

Every operator ${\bf W}$ which satisfies \refer{weight condition on deformation operator} 
and \refer{crucial relation again}
hence induces a classical algebra isomorphism of the superconformal algebra
\refer{suVir algebra in almost usual form}. (Quantum corrections to these algebras can be computed
and elimination of quantum anomalies will give background equations of motion, but
this shall not be our concern here.)
Finding such ${\bf W}$ is therefore
a task analogous to finding superconformal Lagrangians in 2 dimensions.

However, two different ${\bf W}$ need
not induce two different isomorphisms.
In particular, \emph{anti-Hermitean} ${\bf W}^\dag = -{\bf W}$ induce 
\emph{pure gauge} transformations in the sense that all algebra elements are
transformed by the \emph{same} unitary similarity transformation
\begin{eqnarray}
  \label{pure gauge deformations}
  {\bf X} &\mapsto& e^{-{\bf W}}{\bf X}e^{\bf W}
  \hspace{1cm}
  \mbox{for ${\bf X}\in \set{\extd_{K,\xi}, \coextd_{K,\xi}, \fatDelta_{K,\xi}, {\cal L}_{\xi}}$
  and ${\bf W}^\dag = - {\bf W}$}
  \,.
\end{eqnarray}
Examples for such unitary transformations are given in 
\S\fullref{A-field background} and \S\fullref{T-duality}.
They are related to background gauge transformations as well as
to string dualities.
For a detailed discussion of the role of such automorphism in the 
general framework
of string duality symmetries see \S 7 of \cite{FroehlichGawedzki:1993}.\\

In the next subsections deformations of the above form are studied in 
general terms and by way of specific examples.

\subsubsection{NS-NS Backgrounds}
\label{NS-NS backgrounds}

We start by deriving superconformal deformations corresponding to 
background fields in the NS-NS sector of the closed Type II string.
Since the conformal weight of an NS-NS vertex comes from a \emph{single} Wick
contraction with the superconformal generators, while that of a spin field, which
enters R-sector vertices, comes from a \emph{double} Wick contraction, the deformation
theory of NS-NS backgrounds is much more transparent than that of NS-R or R-NS
sectors,  as will be made clear in the following.

\paragraph{Gravitational background by algebra isomorphism.}
\label{gravitational background by algebra isomorphism}

First we reconsider the purely gravitational background from the point of view that
its superconformal algebra derives from
the superconformal algebra for \emph{flat} cartesian target space by a deformation
of the form \refer{isomorphism of Virasoro algebra}. For the point particle
limit this was discussed in equations (38)-(42) of \cite{Schreiber:2003a} and
the generalization to loop space is straightforward:
Denote by
\begin{eqnarray}
  \extd_{K,1}^{\eta}
  &\defas&
  {\cal E}^{\dag (\mu,\sigma)} \partial_{(\mu,\sigma)} + i {\cal E}_{(\mu,\sigma)} X^{\prime(\mu,\sigma)}
\end{eqnarray}
the $K$-deformed exterior derivative on \emph{flat} loop space and define the deformation
operator by
\begin{eqnarray}
  \label{deformation for gravitational background}
  {\bf W} &=& 
  {\cal E}^{\dag} \inner (\ln E) \inner {\cal E}
  \nonumber\\
  &=&
  \int d\sigma\; {\cal E}^{\dag}\of{\sigma} \inner (\ln e\of{X\of{\sigma}}) \inner {\cal E}
  \,,
\end{eqnarray}
where $\ln E$ is the logarithm of a vielbein \refer{vielbein on loop space} on loop space,
regarded as a matrix.
This ${\bf W}$ is constructed so as to satisfy
\begin{eqnarray}
  e^{{\bf W}}{\cal E}^{\dag a}\of{\sigma}e^{-{\bf W}}
  &=&
  \sum_\nu
  e^a{}_\nu
  {\cal E}^{\dag (b=\nu)}
  \,,
\end{eqnarray}
which yields
\begin{eqnarray}
  e^{{\bf W}}
  {\cal E}^{\dag \mu}\of{\sigma}
  e^{-{\bf W}}
  &=&
    e^{{\bf W}}
  e^\mu{}_a {\cal E}^{\dag a}\of{\sigma}
  e^{-{\bf W}}
  \nonumber\\
  &=&
  e^\mu{}_a e^a{}_\nu {\cal E}^{\dag (b=\nu)}
  \nonumber\\
  &=&
  {\cal E}^{\dag (b=\mu)}
  \,.
\end{eqnarray}
Since $e^{\bf W}$ interchanges between two different
vielbein fields which define two different metric tensors the index 
structure becomes a little awkward in the above equations. Since we won't need
these transformations for the further developments we don't bother to 
introduce special notation to deal with this issue more cleanly. The point
here is just to indicate that a $e^{\bf W}$ with the above properties does
exist. It replaces all $p$-forms with respect to $E$ by $p$-forms with respect
to the flat metric. One can easily convince oneself that hence 
the operator $\extd_K$ associated with the
metric $G = E^2$ is related to the operator $\extd_K^\eta$ for flat space by
\begin{eqnarray}
  \label{gravity deformation of extd}
  \extd_{K,\xi}
  &=&
  e^{-{\bf W}}\extd_{K,\xi}^{\eta} e^{{\bf W}}
  \,.
\end{eqnarray}
Therefore, indeed, ${\bf W}$ of \refer{deformation for gravitational background}
induces a gravitational field on the target space.

As was discussed on p. 10 of \cite{Schreiber:2003a} we need to require ${\rm det}\,e = 1$,
and hence 
\begin{eqnarray}
  \label{tracelessness of ln e}
  {\rm tr}\ln e &=& 0
\end{eqnarray}
in order that $\coextd_{K,\xi}^{\bf W} = (\extd_{K,\xi^\ast})^\dag$. This is just a condition
on the nature of the coordinate system with respect to which the metric is constructed in our
framework. As an abstract operator $\extd_{K,\xi}$ is of course
\emph{independent} of any metric, its representation in terms of the operators 
$X^{(\mu,\sigma)},\partial_{(\mu,\sigma)}, {\cal E^{\dag \mu}}, {\cal E}^\mu$ is not, which is what the
above is all about. 

Note furthermore, that
\begin{eqnarray}
  \label{symmetric and antisymmetric ln e}
  {\bf W}^\dag = \pm {\bf W}
  &\Leftrightarrow&
  (\ln e)^{\rm T} = \pm \ln e
  \,.
\end{eqnarray}
According to \refer{pure gauge deformations}
this implies that the antisymmetric part of $\ln e$ generates a pure gauge transformation
and \emph{only the (traceless) symmetric part} of $\ln e$ is responsible for a perturbation of the gravitational 
background. A little reflection shows that the gauge transformation induced by
antisymmetric $\ln e$ is a rotation of the vielbein frame. For further discussion of this
point see pp. 58 of \cite{Schreiber:2001}.

\paragraph{$B$-field background.}
\label{loop space and B-field background}

As in \S 2.1.3 of \cite{Schreiber:2003a} we now consider
the Kalb-Ramond $B$-field 2-form 
\begin{eqnarray}
  B &=& \frac{1}{2}B_{\mu\nu}dx^\mu \wedge dx^\nu
\end{eqnarray}
on target space with field strength $H = dB$. This induces on loop space the 2-form
\begin{eqnarray}
  B_{(\mu,\sigma)(\nu,\sigma^\prime)}\of{X}
  &=&
  B_{\mu\nu}\of{X\of{\sigma}}\delta_{\sigma,\sigma^\prime}
  \,.
\end{eqnarray}
We will study the deformation operator
\begin{eqnarray}
  \label{B-field deformation operator}
  {\bf W}^{(B)}\of{X}
  &\defas&
  \frac{1}{2}
  B_{(\mu,\sigma)(\nu,\sigma^\prime)}\of{X}
  {\cal E}^{\dag (\mu,\sigma)}{\cal E}^{\dag(\nu,\sigma^\prime)}
  \nonumber\\
  &\defas&
  \int d\sigma\;
  \frac{1}{2}B_{\mu\nu}\of{X\of{\sigma}}{\cal E}^{\dag \mu}\of{\sigma}
   {\cal E}^{\dag \nu}\of{\sigma}
\end{eqnarray}
on loop space
(which is manifestly of reparameterization weight 1)
and show that the superconformal algebra that it induces is indeed that found 
by a canonical treatment of the usual supersymmetric $\sigma$-model with 
gravitational and Kalb-Ramond background.  

When calculating the deformations \refer{isomorphism of Virasoro algebra} explicitly
for ${\bf W}$ as in \refer{B-field deformation operator} one finds
\begin{eqnarray}
  \label{b-field deformed loop space susy generators}
  \extd_{K,\xi}^{(B)}
  &\defas&
  \exp\of{-{\bf W}^{(B)}}
  \extd_{K,\xi}
  \exp\of{{\bf W}^{(B)}}
  \nonumber\\
  &=&
  \extd_{K,\xi} + 
  \commutator{\extd_{K,\xi}}{{\bf W}^{(b)}}
  \nonumber\\
  &=&
  \int d\sigma\;
  \xi
  \left(
  {\cal E}^{\dag\mu}\gradOp_\mu
  +
  i T{\cal E}_\mu X^{\prime\mu}
  +
  \frac{1}{6}
  H_{\alpha\beta\gamma}\of{X}
  {\cal E}^{\dag \alpha}{\cal E}^{\dag \beta}
   {\cal E}^{\dag \gamma}
  -
  iT{\cal E}^{\dag\mu}B_{\mu\nu}\of{X}X^{\prime\nu}
  \right)
  \nonumber\\
  \coextd_{K,\xi}^{(B)}
  &=&
  \exp\of{{\bf W}^{\dag(B)}}
  \coextd_K
  \exp\of{-{\bf W}^{\dag(B)}}
  \nonumber\\
  &=&
  -
  \int d\sigma\;
  \xi\of{\sigma}
  \left(
  {\cal E}^{\mu}\gradOp_\mu
  +
  i T{\cal E}^\dag_\mu X^{\prime\mu}
  +
  \frac{1}{6}
  H_{\alpha\beta\gamma}\of{X}
  {\cal E}^{\alpha}{\cal E}^{\beta}
   {\cal E}^{\gamma}
  -
  iT{\cal E}^{\mu}B_{\mu\nu}\of{X}X^{\prime\nu}
  \right)
  \,.
  \nonumber\\
  \,.
\end{eqnarray}
This is essentially equation (72) of \cite{Schreiber:2003a}, with the only difference
that here we have mode functions $\xi$ and an explicit realization of the 
deformation Killing vector.

In order to check that the above is a valid isomorphim condition \refer{crucial condition for deformed operators}
must be calculated. Concentrating on the potentially problematic terms one finds
\begin{eqnarray}
  \antiCommutator{\extd^{(B)}_{K,\xi_1}}{\coextd^{(B)}_{K,\xi_2}}
  &=&
  \int d\sigma\; \xi_1 \xi_2 (\cdots)
  \nonumber\\
  &&+
  \int d\sigma\, d\sigma^\prime
  \xi_1\of{\sigma}\xi_2\of{\sigma^\prime}
    i
    \left(
      {\cal E}^\dag_\mu\of{\sigma^\prime}
      -{\cal E}^\nu\of{\sigma^\prime} B_{\nu\mu}\of{X\of{\sigma^\prime}} 
    \right)
    {\cal E}^{\dag\mu}\of{\sigma}
    \delta^\prime\of{\sigma^\prime,\sigma}
\nonumber\\
  &&+
  \int d\sigma\, d\sigma^\prime
  \xi_1\of{\sigma}\xi_2\of{\sigma^\prime}
    i
    \left(
      {\cal E}_\mu\of{\sigma}
      -{\cal E}^{\dag\nu}\of{\sigma} B_{\nu\mu}\of{X\of{\sigma}} 
    \right)
    {\cal E}^{\mu}\of{\sigma^\prime}
    \delta^\prime\of{\sigma,\sigma^\prime}
  \nonumber\\
  &=&
  \int d\sigma\; \xi_1 \xi_2 (\cdots)
  -i
  \int d\sigma\; 
  \left(
    \xi^\prime_1 \xi_2 
    B_{\nu\mu}{\cal E}^\nu{\cal E}^{\dag \mu}
    +
    \xi_1 \xi_2^\prime
    B_{\nu\mu}{\cal E}^{\dag \nu}{\cal E}^\mu
  \right)
  \nonumber\\
  &=&
  \int d\sigma\; \xi_1 \xi_2 (\cdots)
  \,.
\end{eqnarray}
This expression therefore manifestly satisfies \refer{crucial condition for deformed operators}.

With hindsight this is no surprise, because 
\refer{b-field deformed loop space susy generators} are precisely the 
superconformal generators in functional form as found by canonical analysis of 
the non-linear supersymmetric $\sigma$-model 
\begin{eqnarray}
  \label{susy action of string in b-field background}
  S
  &=&
  \frac{T}{2}
  \int 
    d^2 \xi
    d^2 \theta
    \;
    \left(
      G_{\mu\nu} + B_{\mu\nu}
    \right)
    D_+ {\bf X}^\mu D_- {\bf X}^\nu
    \,,
\end{eqnarray}
where ${\bf X}^\mu$ are worldsheet superfields
\begin{eqnarray}
  {\bf X}^\mu\of{\xi,\theta_+,\theta_-}
  &\defas&
  X^\mu\of{\xi}
  + 
  i\theta_+ \psi_+^{\mu}\of{\xi}
  -
  i \theta_-\psi_-^{\mu}\of{\xi}
  +
  i
  \theta_+
  \theta_- F^\mu\of{\xi}
  \nonumber
\end{eqnarray}
and
$
  D_\pm
  \defas
  \partial_{\theta_{\pm}}
  -i
  \theta_\pm \partial_\pm
$
with
$
  \partial_\pm \defas
  \partial_0 \pm \partial_1
$
are the superderivaties. The calculation can be found in 
section 2 of \cite{Chamseddine:1997}.  
(In order to compare the final result, equations (32),(33)
of \cite{Chamseddine:1997}, with our \refer{b-field deformed loop space susy generators}
note that our fermions $\Gamma_\pm$ are related to the fermions $\psi_\pm$ of
\cite{Chamseddine:1997} by $\Gamma_\pm = (i^{(1\mp 1)/2}\sqrt{2T})\psi_\pm$.)

\paragraph{Dilaton background.}
\label{Dilaton background}
The deformation operator in \refer{deformation for gravitational background} which induces 
the gravitational background 
was of the form ${\bf W} = {\cal E}^\dag \inner M \inner {\cal E}$ with $M$ a traceless
symmetric matrix. If instead we consider a deformation of the same form but for pure trace we get
\begin{eqnarray}
  {\bf W}^{(D)}
  &=&
  -
  \frac{1}{2}
  \int d\sigma\;
  \Phi\of{X}
  {\cal E}^{\dag \mu}{\cal E}_\mu
  \,,
\end{eqnarray}
for some real scalar field $\Phi$ on target space.
This should therefore induce a dilaton background.
The associated superconformal generators are (we suppress the $\sigma$ dependence and the mode
functions $\xi$ from now on)
\begin{eqnarray}
  \label{dilaton deformation}
  \exp\of{-{\bf W}^{(D)}}\extd_{K}\exp\of{{\bf W}^{(D)}}
  &=&
    e^{\Phi/2}{\cal E}^{\dag \mu}
    \left(
      \gradOp_\mu
      -
      \frac{1}{2}(\partial_\mu \Phi){\cal E}^{\dag \nu}{\cal E}_\nu
    \right)
    +iT
    e^{-\Phi/2}
    X^{\prime \mu}
    {\cal E}_\mu
  \nonumber\\
  \exp\of{{\bf W}^{(D)}}\coextd_{K}\exp\of{-{\bf W}^{(D)}}
  &=&
  -
    e^{\Phi/2}{\cal E}^{ \mu}
    \left(
      \gradOp_\mu
      +
      \frac{1}{2}(\partial_\mu \Phi){\cal E}^{\dag \nu}{\cal E}_\nu
    \right)
    -iT
    e^{-\Phi/2}
    X^{\prime \mu}
    {\cal E}^\dag_\mu
  \,.
  \nonumber\\
\end{eqnarray}
It is readily seen that for this deformation equation \refer{crucial condition for deformed operators} 
is satisfied, so that these operators indeed generate a superconformal algebra.

The corresponding Dirac operators are
\begin{eqnarray}
  \label{Phi Dirac operators}
  \extd_K^{(\Phi)}
  \pm 
  \coextd_K^{(\Phi)}
  &=&
  \Gamma_\mp^\mu
  \left(
    e^{\Phi/2}
    \gradOp_\mu
    \mp
    iT
    e^{-\Phi/2}
    G_{\mu\nu}
    X^{\prime\mu}
  \right)
  \mp
  e^{\Phi/2}
  \Gamma_\pm^\mu
  (\partial_\mu \Phi){\cal E}^{\dag \nu}{\cal E}_\nu
  \,.
\end{eqnarray}
Comparison of the superpartners of $\Gamma_{\pm,\mu}$
\begin{eqnarray}
  \mp\frac{1}{2}
  \antiCommutator{  \extd_K^{(\Phi)}
  \pm 
  \coextd_K^{(\Phi)}
  }{
    \Gamma_{\mp,\mu}
  }
  &=&
    e^{\Phi/2}
    \partial_\mu
    \mp
    iT
    e^{-\Phi/2}
    G_{\mu\nu}
    X^{\prime\mu}
  + \mbox{fermionic terms}  
\end{eqnarray}
with equation \refer{Born-Infeld current for pure dilaton background} 
in appendix \S\fullref{Canonical analysis of bosonic D1 brane action} shows that
this has the form expected for the dilaton coupling of a 
D-string.

\paragraph{Gauge field background.}
\label{A-field background}

A gauge field background $A = A_\mu dx^\mu$ should express itself via
$B \to B - \frac{1}{T}F$, where $F = dA$ (e.g. \S 8.7 of 
\cite{  Polchinski:1998}), 
if we assume $A$ to be a $U\of{1}$ connection 
for the moment. Since the present discussion so far refers only
to closed strings and since closed strings have trivial coupling to $A$ it is to be
expected that an $A$-field background manifests itself as a pure gauge 
transformation in the present context. This motivates to investigate the
deformation induced by the anti-Hermitean
\begin{eqnarray}
  \label{A field deformation operator}
  {\bf W}
  &=&
  i A_{(\mu,\sigma)}\of{X}X^{\prime(\mu,\sigma)}
  \;=\;
  i \int d\sigma\;
  A_\mu\of{X\of{\sigma}}X^{\prime\mu}\of{\sigma}
  \,.
\end{eqnarray}
The associated superconformal generators are found to be
\begin{eqnarray}
  \extd_{K}^{(A)(B)}
  &=&
  \extd^{(B)}_{K} +
  i {\cal E}^{\dag \mu}F_{\mu\nu} X^{\prime\nu}
  \nonumber\\
  \coextd_{K}^{(A)(B)}­
  &=&
  \coextd^{(B)}_{K}
  - i {\cal E}^{\mu}F_{\mu\nu} X^{\prime\nu}
  \,.   
\end{eqnarray}
Comparison with \refer{b-field deformed loop space susy generators} shows that indeed
\begin{eqnarray}
  \label{combination of B and F}
  \extd_{K}^{(A)(B)}
  &=&
  \extd_K^{(B-\frac{1}{T}F)}
  \,,
\end{eqnarray}
so that we can identify the background induced by 
\refer{A field deformation operator} with that of the NS 
$U\of{1}$ gauge field.
 
Since
$\exp\of{\bf W}\of{X}$ is nothing but the Wilson loop of $A$ around $X$, it is
natural to conjecture that for a general (non-abelian) gauge field background
$A$ the corresponding deformation is the Wilson loop as well:
\begin{eqnarray}
  \extd_K^{(A)}
  &=&
  \left(
  {\rm Tr}{\cal P}e^{-i \int A_\mu X^{\prime\mu}}
  \right)
  \extd_K
  \left(
  {\rm Tr}{\cal P}e^{+i \int A_\mu X^{\prime\mu}}
  \right)
  \,,
\end{eqnarray}
where $\cal P$ indicates path ordering and $\rm Tr$ the trace in the
Lie algebra, as usual.

\paragraph{$C$-field background.}
\label{C-field background}

So far we have found deformation operators for all massless NS and NS-NS background fields. 
One notes a close similarity between the form of these deformation operators and the
form of the corresponding vertex operators (in fact, the deformation operators are
related to the vertex operators in the (-1,-1) picture. This is discussed in 
\S\fullref{canonical deformations from d to dingens}): 
The deformation operators 
for $G$, $B$ and $\Phi$ are bilinear in the form
creation/annihilation operators on loop space, with the bilinear form (matrix)
seperated into its traceless symmetric, antisymmetric and trace part.

Interestingly, though, there is one more 
deformation operator obtainable by such a bilinear in the form creation/annihilation 
operators. It is
\begin{eqnarray}
  {\bf W}^{(C)} &\defas&
  \frac{1}{2}\int d\sigma\; C_{\mu\nu}\of{X}{\cal E}^{\mu}{\cal E}^\nu
  \,,
\end{eqnarray}
i.e. the \emph{adjoint} of \refer{B-field deformation operator}. It induces the generators
\begin{eqnarray}
  \label{C2 background generators}
  {\extd}^{(C)}_{K,\xi}
  &=&
  \int d\sigma\;
  \xi
  \Big(
    {\cal E}^{\dag \mu}\gradOp_\mu
    +i{\cal E}_\mu X^{\prime \mu}
    -
    {\cal E}^\nu C_\nu{}^\mu \gradOp_\mu
    +
    \frac{1}{2}
    {\cal E}^{\dag \alpha} {\cal E}^\mu {\cal E}^\nu(\nabla_\alpha C_{\mu\nu})
    \nonumber\\
    &&
    -
    \frac{1}{2}
    C_\nu{}^\mu {\cal E}^\nu {\cal E}^\alpha {\cal E}^\beta
      (\nabla_\mu C_{\alpha\beta})
    + \frac{1}{2}C^\alpha{}_\beta{\cal E}^\beta{\cal E}^\mu{\cal E}^\nu(\nabla_\alpha C_{\mu\nu})
  \Big)
  \nonumber\\
  \nonumber\\
  {\coextd}^{(C)}_{K,\xi}
  &=&
  -
  \int d\sigma\;
  \xi
  \Big(
    {\cal E}^{\mu}\gradOp_\mu
    +
    i{\cal E}^\dag_\mu X^{\prime \mu}
    -
    {\cal E}^{\dag \nu} C_\nu{}^\mu \gradOp_\mu
    +
    \frac{1}{2}
    {\cal E}^{\alpha} {\cal E}^{\dag \mu} {\cal E}^{\dag \nu}(\nabla_\alpha C_{\mu\nu})
   \nonumber\\
   &&
   \;\;\;\;
   -\frac{1}{2}
   C_\nu{}^\mu {\cal E}^{\dag \nu}{\cal E}^{\dag\alpha}{\cal E}^{\dag \beta}
    (\nabla_\mu C_{\alpha\beta})
    + \frac{1}{2}C^{ \alpha}{}_\beta{\cal E}^{\dag \beta}{\cal E}^{\dag \mu}{\cal E}^{\dag \nu}
      (\nabla_\alpha C_{\mu\nu})
  \Big)
  \,.
\end{eqnarray}
Furthermore it turns out that this deformation, too, does respect
\refer{crucial condition for deformed operators}:  When we again concentrate only on the potentially
problematic terms we see that
\begin{eqnarray}
  \antiCommutator{{\extd}^{(C)}_{K,\xi_1}}{{\coextd}^{(C)}_{K,\xi_2}}
  &=&
 \int d\sigma \; \xi_1 \xi_2 (\cdots)
  \nonumber\\
  &&
  +
  \antiCommutator{-\int d\sigma\;\xi_1 {\cal E}^\nu C_\nu{}^\mu \gradOp_\mu}
  {-i \int d\sigma\; \xi_2 {\cal E}^{\dag}_\mu X^{\prime \mu}}
  \nonumber\\
  &&
  +
  \antiCommutator{\int d\sigma\;\xi_2 {\cal E}^{\dag \nu} C_\nu{}^\mu \gradOp_\mu}
  {i \int d\sigma\; \xi_1 {\cal E}_\mu X^{\prime \mu}}
  \nonumber\\
  &=&
  \int d\sigma\; \xi_1\xi_2 \left(\cdots\right)
  \nonumber\\
  &&+
  i
  \int d\sigma\;
  \left(
    \xi_1\xi_2^\prime
    \;
    {\cal E}^\dag_\mu C_\nu{}^\mu {\cal E}^\nu 
    +
    \xi_1^\prime \xi_2\;
    {\cal E}_\mu C_\nu{}^\mu {\cal E}^{\dag \nu}
  \right)
  \nonumber\\
  &=&
  \int d\sigma\; \xi_1\xi_2 \left(\cdots\right)  
  \,.
\end{eqnarray}
Therefore \refer{C2 background generators} do generate a superconformal algebra and hence define
an SCFT. 

What, though, is the physical interpretation of the field $C$ on spacetime?
It is apparently not the NS 2-form field, because the generators 
\refer{C2 background generators}
are different from \refer{b-field deformed loop space susy generators}
and don't seem to be unitarily equivalent.
A possible guess would therefore be that it is the \emph{RR 2-form} $C_2$, but
now coupled to a D-string instead of an F-string. 

The description of the F-string in an RR background would involve ghosts and spin fields,
which we do not discuss here. But the coupling of the D-string to the RR 2-form is very
similar to the coupling of the F-string to the Kalb-Ramond 2-form and does not
involve any spin fields. That's why the above deformation might allow an interpretation in 
terms of D-strings in RR 2-form backgrounds.

But this needs to be further examined. A hint in this direction is that
under a duality transformation which changes the sign of the dilaton, the
$C$-field is exchanged with the $B$-field. This is discussed in 
\S\fullref{T-duality for various background fields}.

\subsubsection{Canonical Deformations and Vertex Operators}
\label{canonical deformations and vertex operators}

With all NS-NS backgrounds under control (\S\fullref{NS-NS backgrounds}) 
we now turn to a more general analysis of the deformations of \S\fullref{deformations}
that puts the results of the previous subsections in perspective and shows
how general backgrounds are to be handled.

\paragraph{Review of first order canonical CFT deformations.}
\label{review of first order canonical CFT deformations}

Investigations of conformal deformations by way of adding terms to the conformal generators
go back as far as\footnote{We are grateful to M. Halpern for making us aware of this work.} 
\cite{FreericksHalpern:1988}, which builds on earlier insights 
\cite{BardakciHalpern:1971,ChodosThorn:1974} into continuous
families of conformal algebras.

It has been noted long ago \cite{OvrutRama:1992} that adding an integrated 
background vertex operator
$V$ (a worldsheet field of weight (1,1))
to the string's action to first order induces a perturbation
\begin{eqnarray}
  L_m &\to&
  L_m + \int d\sigma\; e^{-im\sigma}V\of{\sigma}
\end{eqnarray}
of the Virasoro generators
and a similar shift occurs for the supercurrent \cite{Giannakis:1999}.\\

While in \cite{OvrutRama:1992} this is discussed in CFT language it becomes
quite transparent in canonical language: From the string's worldsheet action
for gravitational $G_{\mu\nu}$, Kalb-Ramond $B_{\mu\nu}$ 
and dilaton $\Phi$ background one finds the classical stress-energy
tensor (\cf \S\fullref{Canonical analysis of bosonic D1 brane action})

\hspace{-2cm}\parbox{1cm}{
\begin{eqnarray}
  T\of{\sigma}
  &=&
  \frac{1}{2}
  G^{\mu\nu}
  \frac{1}{\sqrt{2T}}
  \left(
    e^{\Phi/2}P_\mu + T \left( e^{\Phi/2}B_{\mu\kappa} + e^{-\Phi/2}G_{\mu\kappa} \right)X^{\prime\kappa}
  \right)
  \frac{1}{\sqrt{2T}}
  \left(
    e^{\Phi/2}P_\nu + T \left( e^{\Phi/2}B_{\nu\kappa} + e^{-\Phi/2}G_{\nu\kappa} \right)X^{\prime\kappa}
  \right)\of{\sigma}
  \,,
  \nonumber\\
\end{eqnarray}
}

\hspace{-0.7cm}where $P_\mu$ is the canonical momentum to $X^\mu$.

When expanded in terms of small perturbations
\begin{eqnarray}
  G_{\mu\nu}\of{X}
  &=&
  \eta_{\mu\nu}
  +
  h_{\mu\nu}\of{X} + \cdots
  \nonumber\\
  B_{\mu\nu}\of{X}
  &=&
  0 +
  b_{\mu\nu}\of{X} + \cdots
  \nonumber\\
  \Phi\of{X} &=& 0 + \phi\of{X} + \cdots  
\end{eqnarray}
of the background fields this yields
\begin{eqnarray}
  \label{expanded stress enegery tensor}
  T &\approx&
  \frac{1}{2}
  \left(\eta^{\mu\nu} - h^{\mu\nu}\right)
  \left(
    {\cal P}_{+\mu}
    +
    \sqrt{\frac{T}{2}}
    b_{\mu\kappa}X^{\prime \kappa}
    +
    \sqrt{\frac{T}{2}}
    h_{\mu\kappa}
    X^{\prime \kappa}
    +
    \frac{1}{\sqrt{8T}}\phi P_\mu
    -
    \sqrt{\frac{T}{8}}\phi \eta_{\mu\kappa}X^{\prime\kappa}
  \right)
  \Bigg(
    \cdots
  \Bigg)_\nu
  \nonumber\\
  &=&
  \frac{1}{2}
  \eta_{\mu\nu}
  {\cal P}_{+}^\mu {\cal P}_{+}^\nu
  -
  \underbrace{
  \frac{1}{2}
  h_{\mu\nu}
  {\cal P}_{+}^\mu{\cal P}_{-}^\nu
  }_{\defas V_G}
  -
  \underbrace{
  \frac{1}{2}
  b_{\mu\nu}{\cal P}_+^\mu {\cal P}_-^\nu
  }_{\defas V_B}
  +
  \underbrace{
  \frac{1}{2}
  \phi
  \eta_{\mu\nu}{\cal P}_+^\mu {\cal P}_-^\nu
  }_{\defas V_\Phi}
  +
  \mbox{higher order terms}
  \,,
  \nonumber\\
\end{eqnarray}
where we have defined
\begin{eqnarray}
  {\cal P}^\mu_\pm\of{\sigma}
  &\defas&
  \frac{1}{\sqrt{2T}}
  \left(
    \eta^{\mu\nu}P_\nu \pm T X^{\prime \nu}
  \right)\of{\sigma}\,.
\end{eqnarray}
It must be noted that while the objects ${\cal P}_\pm$, which have Poisson-bracket 
\begin{eqnarray}
  \{{\cal P}^\mu_\pm\of{\sigma},{\cal P}^\nu_\pm\of{\sigma^\prime}\} 
  &=&
  \mp \eta^{\mu\nu}\delta^\prime\of{\sigma-\sigma^\prime}
  \,,
\end{eqnarray}
generate the current algebra of the free theory, they involve, via
$P_\mu = \delta S/\delta \dot X^\mu$, data of the perturbed background and are hence
not proportional to $\partial X$ and $\bar \partial X$.

Still, the first term in \refer{expanded stress enegery tensor} is the generator of the
Virasoro algebra which is associated with the $\mathrm{U}(1)$-currents ${\cal P}_\pm$,
while the following terms are the weight (1,1) vertices 
$V_G$, $V_B$, $V_\Phi$ (with respect to the first term) of
the graviton, 2-form and dilaton, respectively.

Hence in the sense that we regard the canonical coordinates and momenta as fundamental and
hence unaffected by the background perturbation, i.e.
\begin{eqnarray}
  \label{canonical data does not shift}
  X^\mu &\to& X^\mu
  \nonumber\\
  P_\mu &\to& P_\mu
  \,,
\end{eqnarray}
while only the `coupling constants' are shifted
\begin{eqnarray}
  \eta_{\mu\nu} &\to& \eta_{\mu\nu} + h_{\mu\nu}\,,\hspace{1cm}\mbox{etc.}
\end{eqnarray}
we can write
\begin{eqnarray}
  T &\to&
  T + V
  \,,
\end{eqnarray}
where $V$ denotes a collection of weight (1,1) vertices in the above sense.\footnote{
  As is discussed in \cite{OvrutRama:1992}, the 
  issue concerning \refer{canonical data does not shift} in CFT language translates into
the question whether one chooses to treat $\partial X$ and $\bar \partial X$ as free fields
in the perturbed theory and whether the $\partial X \,\partial X$-OPE is taken to receive a
perturbation or not.

For a further discussion of perturbations of SCFTs where this issue is addressed, see
\cite{Schreiber:2003a} and in particular section 2.2.4.
}

CFT deformations of this form are called \emph{canonical deformations} 
\cite{EvansOvrut:1990,EvansGiannakis:1991}.

The central idea of canonical first order deformations is that the (super-) Virasoro algebra
\begin{eqnarray}
  \commutator{T\of{\sigma}}{T\of{\sigma^\prime}}
  &=&
   2i T\of{\sigma^\prime}\delta^\prime\of{\sigma-\sigma^\prime}
   - iT^\prime\of{\sigma^\prime}\delta\of{\sigma-\sigma^\prime}
  +
  A\of{\sigma-\sigma^\prime}
  \nonumber\\
  \antiCommutator{T_F\of{\sigma}}{T_F\of{\sigma^\prime}}
  &=&
  -\frac{1}{2\sqrt{2}}T\of{\sigma^\prime}\delta\of{\sigma^\prime}
  +
  B\of{\sigma-\sigma^\prime}
  \nonumber\\
  \commutator{T\of{\sigma}}{T_F\of{\sigma^\prime}}
  &=&
  \frac{3i}{2}T_F\of{\sigma^\prime}\delta^\prime\of{\sigma-\sigma^\prime}
  -
  i T^\prime_F\of{\sigma^\prime}\delta\of{\sigma-\sigma^\prime}
\end{eqnarray}
(where $A$ and $B$ are the anomaly terms)
together with its chiral partner, generated by $\bar T$ and $\bar T_F$,
is preserved to first order under the perturbation
\begin{eqnarray}
  T\of{\sigma} &\to& T\of{\sigma} + \delta T\of{\sigma}
  \nonumber\\
  T_F\of{\sigma} &\to& T_F\of{\sigma} + \delta T_F\of{\sigma}  
\end{eqnarray}
if, in particular, 
\begin{eqnarray}
  \delta T\of{\sigma} &=& \Phi\of{\sigma}\bar\Phi\of{\sigma}
  \nonumber\\ 
  \delta F\of{\sigma} &=& \Phi_F\of{\sigma}\bar \Phi_F\of{\sigma}
\end{eqnarray}
with
\begin{eqnarray}
  \commutator{T\of{\sigma}}{\Phi\of{\sigma^\prime}}
  &=&
  i \Phi\of{\sigma^\prime}\delta^\prime\of{\sigma-\sigma^\prime} 
  - i 
  \Phi^\prime\of{\sigma^\prime}\delta\of{\sigma-\sigma^\prime}
  \nonumber\\
  \commutator{T\of{\sigma}}{\Phi_F\of{\sigma^\prime}}  
  &=&
  \frac{i}{2} \Phi\of{\sigma^\prime}\delta^\prime\of{\sigma-\sigma^\prime} 
  - i 
  \Phi^\prime\of{\sigma^\prime}\delta\of{\sigma-\sigma^\prime}
  \nonumber\\
  \commutator{T\of{\sigma}}{\bar \Phi\of{\sigma^\prime}}    
  &=& 
  0
  \nonumber\\
  \commutator{T\of{\sigma}}{\bar \Phi_F\of{\sigma^\prime}}    
  &=& 
  0
\end{eqnarray}
and analogous relations for $\delta \bar T$ and $\delta \bar T_F$. 

There are however
also more general fields $\delta T$, $\delta T_F$ of total weight $2$ and $3/2$, 
respectively, which preserve the above super-Virasoro algebra to first order 
\cite{BaggerGiannakis:2001}. But the weight $(1,1)$ part $\Phi\of{\sigma}\bar \Phi\of{\sigma}$
is special in that it corresponds directly to the vertex operator of the background
which is described by the deformed worldsheet theory. 
Further deformation fields of weight different from (1,1) are related to 
\emph{gauge} degrees of 
freedom of the background fields (\cf \cite{BaggerGiannakis:2001} and
the discussion below equation \refer{totally non-gauge delta T}).

\paragraph{Canonical deformations from $\extd_K \to e^{-\bf W}\extd_K e^{{\bf W}}$.}
\label{canonical deformations from d to dingens}

We would like to see how the deformation theory 
reviewed above
relates to the SCFT deformations that have been
studied in \S\fullref{loop space super-Virasoro generators}.

First recall from \S\fullref{purely gravitational} 
that the chiral bosonic fields in
our notation read
\begin{eqnarray}
  {\cal P}_\pm\of{\sigma}
  &\defas&
  \frac{1}{\sqrt{2T}}
  \left(
    -i
    \frac{\delta}{\delta X}
    \pm
    T X^\prime
  \right)
  \of{\sigma}
\end{eqnarray}
and that according to \S\fullref{exterior algebra over loop space} we
write the worldsheet fermions $\psi$, $\bar \psi$ as $\Gamma_\pm$, respectively, which are
normalized so that
$\antiCommutator{\Gamma^\mu_\pm\of{\sigma}}{\Gamma_\pm^\nu\of{\sigma^\prime}}
= \pm 2 g^{\mu\nu}\of{X\of{\sigma}}\delta\of{\sigma-\sigma^\prime}$, and we
frequently make use of the linear combinations
\begin{eqnarray}
  {\cal E}^{\dag \mu} &=& \frac{1}{2}\left(\Gamma_+^\mu + \Gamma_-^\mu\right)
  \nonumber\\
  {\cal E}^{\mu} &=& \frac{1}{2}\left(\Gamma_+^\mu - \Gamma_-^\mu\right)
  \,.
\end{eqnarray}

In this notation 	the supercurrents for the trivial background read
\begin{eqnarray}
  \label{T_F in terms of d_K}
  T_F\of{\sigma}
  &=&
  \frac{1}{\sqrt{2}}
  \Gamma_+\of{\sigma}
  {\cal P}_+\of{\sigma}
  \;=\;
  \frac{-i}{\sqrt{4T}}
  \left(
    \extd_K\of{\sigma}
    -
    \coextd_K\of{\sigma}
  \right)
  \nonumber\\
  \bar T_F\of{\sigma}
  &=&
  \frac{i}{\sqrt{2}}
  \Gamma_-\of{\sigma}
  {\cal P}_-\of{\sigma}
  \;=\;
  \frac{1}{\sqrt{4T}}
  \left(
    \extd_K\of{\sigma}
    +
    \coextd_K\of{\sigma}
  \right)  
  \,,
\end{eqnarray}
where the $K$-deformed exterior derivative and coderivative on loop space
are identified as
\begin{eqnarray}
  \extd_K &=& 
  \sqrt{T}
  \left(
    \bar T_F
    +
    i
    T_F
  \right)
  \nonumber\\
  \coextd_K &=& 
  \sqrt{T}
  \left(
    \bar T_F
    -
    i T_F
  \right)    
  \,.
\end{eqnarray}
According to \S\fullref{deformations} a consistent deformation of the 
superconformal algebra generated by $T_F$ and $\bar T_F$ is given by
sending
\begin{eqnarray}
  \label{canonical deformation of extd}
  \extd_K\of{\sigma}
  \;\to\;
  \extd^{(W)}_K\of{\sigma}
  &=&
  e^{-{\bf W}}\extd_K\of{\sigma}e^{\bf W}
  =
  \extd_K\of{\sigma}
  +
  \underbrace{
  \commutator{\extd_K\of{\sigma}}{{\bf W}}
  }_{\defas \delta \extd_K\of{\sigma}}
  +
  \cdots
  \nonumber\\
  \coextd_K\of{\sigma}
  \;\to\;
  \coextd^{(W)}_K\of{\sigma}
  &=&
  e^{{\bf W}^\dag}\coextd_K\of{\sigma}e^{-{\bf W}^\dagger}
  =
  \coextd_K\of{\sigma}
  +
  \underbrace{
  \commutator{{\bf W}^\dagger}{\coextd_K\of{\sigma}}
  }_{\defas \delta \coextd_k\of{\sigma}}
  +
  \cdots
\end{eqnarray}
for $\bf W$ some reparameterization invariant operator. From this one finds
$\delta T_F$ by using \refer{T_F in terms of d_K} 
\begin{eqnarray}
  \label{delta T_F and delta bar T_F}
  \delta T_F
  &=&
  -i
  \commutator{\bar T_F}{\frac{1}{2}\left({\bf W} + {\bf W}^\dagger\right)}
  +
  \commutator{T_F}{\frac{1}{2}\left({\bf W} - {\bf W}^\dagger\right)}
  \nonumber\\
  \delta \bar T_F
  &=&
  i
  \commutator{T_F}{\frac{1}{2}\left({\bf W} + {\bf W}^\dagger\right)}
  +
  \commutator{\bar T_F}{\frac{1}{2}\left({\bf W} - {\bf W}^\dagger\right)}  
\end{eqnarray}
which again gives
$\delta T$ by means of 
\begin{eqnarray}
  \label{first order delta T from anticommutators of T_F and delta T_F}
  \antiCommutator{T_F\of{\sigma}}{\delta T_F\of{\sigma^\prime}}
  +
  \antiCommutator{T_F\of{\sigma^\prime}}{\delta T_F\of{\sigma}}
  &=&
  -\frac{1}{2\sqrt{2}}
  \delta T\of{\sigma}
  \delta\of{\sigma-\sigma^\prime}
  \,.
\end{eqnarray}

Before looking at special cases one should note that this necessarily implies that
$\delta T_F$ is of total weight $3/2$ and that $\delta T$ is of total weight $2$. 
That is because $\bf W$, being reparameterization invariant, must be the integral 
(along the string at fixed worldsheet time) over a
field of unit total weight (\cf \refer{weight condition on deformation operator} and
\refer{total unit weight condition on W}) and
because supercommutation with $\extd_K$ or $\coextd_K$ increases the total weight
by $1/2$.

Furthermore, recall from \refer{pure gauge deformations} that the anti-hermitean part
$\frac{1}{2}\left({\bf W} - {\bf W}^\dagger\right)$ of the deformation operator ${\bf W}$
is responsible for \emph{pure gauge transformations} while the hermitean part
$\frac{1}{2}\left({\bf W} + {\bf W}^\dagger\right)$ induces true modifications of the 
background fields. Hence for a pure gauge transformation \refer{delta T_F and delta bar T_F}
yields
\begin{eqnarray}
  \delta T_F &=& \commutator{T_F}{{\bf W}}
  \nonumber\\
  \delta \bar T_F &=& \commutator{\bar T_F}{{\bf W}}
  \,,
  \hspace{1cm}
  \mbox{for ${\bf W}^\dagger = -{\bf W}^\dagger$}
  \,,
\end{eqnarray}
which of course comes from the global similarity transformation \refer{pure gauge deformations}
\begin{eqnarray}
  {\bf X} &\mapsto& e^{-{\bf W}}{\bf X}e^{{\bf W}}\,,
  \hspace{1cm}
  {\bf X} \in \set{T_F, \bar T_F, \cdots}
  \,.
\end{eqnarray}
On the other hand, for a strictly non-gauge transformation 
the transformation \refer{delta T_F and delta bar T_F} simplifies to
\begin{eqnarray}
  \delta T_F
  &=&
  -i
  \commutator{\bar T_F}{{\bf W}}
  \nonumber\\
  \delta \bar T_F
  &=&
  i
  \commutator{T_F}{{\bf W}}
  \,,
  \hspace{1cm}
  \mbox{for ${\bf W}^\dagger = + {\bf W}$}
  \,.
\end{eqnarray}
In the cases where $\bf W$ is antihermitean \emph{and} a $(1/2,1/2)$ field 
(as is in particular the case for the gravitational ${\bf W}^{(G)}$ of 
\S\fullref{gravitational background by algebra isomorphism},
the dilaton ${\bf W}^{(D)}$ of \S\fullref{Dilaton background} 
and the hermitean part of the
Kalb-Ramond ${\bf W}^{(B)} + {\bf W}^{(B)\dagger}$ of 
\S\fullref{loop space and B-field background} )
this, together with \refer{first order delta T from anticommutators of T_F and delta T_F} implies that 
\begin{eqnarray}
  \label{totally non-gauge delta T}
  \delta T
  &\propto&
  \antiCommutator{T_F}{\commutator{\bar T_F}{\bf W}}
\end{eqnarray}
is indeed of weight $(1,1)$, as discussed in the theory of canonical deformations 
\S\fullref{review of first order canonical CFT deformations}.
Furthermore, this shows explicitly that all contributions to $\delta T$ which are of
total weight 2 but \emph{not} of weight (1,1) must come from the antihermitean component 
$\frac{1}{2}\left({\bf W} - {\bf W}^\dagger\right)$ and hence must be associated with background
gauge transformations. (This proves in full generality the respective observation in 
\cite{BaggerGiannakis:2001} concerning 2-form field deformations.)\\

Finally, equation \refer{totally non-gauge delta T} clarifies exactly how the
deformation operators $\bf W$ are related to the \emph{vertex operators} of the
respective background fields, namely it shows that the hermitean part of $\bf W$ is proportional to the
vertex operator in the (-1,-1) picture (i.e. the pre-image under $\antiCommutator{T_F}{\commutator{\bar T_F}{\cdot}}$).\\

As an example, consider the deformation induced by a $B$-field background:

According to \S\fullref{loop space and B-field background} a Kalb-Ramond background is induced by
choosing
\begin{eqnarray}
  {\bf W}
  &=&
  \int d\sigma\;
  \frac{1}{2}
  B_{\mu\nu}\of{X\of{\sigma}}
  {\cal E}^{\dag \mu}
  {\cal E}^{\dag \nu}
\end{eqnarray}
which, using \refer{canonical deformation of extd} gives rise to
\begin{eqnarray}
  \delta \extd_K\of{\sigma}
  &=&
  \left(
  \frac{1}{6}
  H_{\alpha\beta\gamma}\of{X}
  {\cal E}^{\dag \alpha}{\cal E}^{\dag \beta}
   {\cal E}^{\dag \gamma}
  -
  iT{\cal E}^{\dag\mu}B_{\mu\nu}\of{X}X^{\prime\nu}
  \right)
  \of{\sigma}
  \nonumber\\
  \delta\coextd_{K}\of{\sigma}
  &=&
  \left(
    -\frac{1}{6}
    H_{\alpha\beta\gamma}\of{X}
    {\cal E}^{\alpha}{\cal E}^{\beta}
    {\cal E}^{\gamma}
    +
    iT{\cal E}^{\mu}B_{\mu\nu}\of{X}X^{\prime\nu}
  \right)
  \of{\sigma}
\end{eqnarray}
and hence, using \refer{T_F in terms of d_K}, to
\begin{eqnarray}
  \label{B field shift in delta T_F}
  \delta T_F\of{\sigma}
  &=&
  -\frac{i}{12\sqrt{T}}
    H_{\alpha\beta\gamma}\left(
       {\cal E}^{\dag \alpha}{\cal E}^{\dag \beta}{\cal E}^{\dag \gamma}
      +
      {\cal E}^{\alpha}{\cal E}^{\beta}{\cal E}^{\gamma}
    \right)
   -\frac{1}{\sqrt{8}}
   \Gamma_+^\mu
   B_{\mu\nu} \left({\cal P}_+^\nu - {\cal P}_-^\nu\right)
  \,.
\end{eqnarray}
In this special case $\delta T_F$ happens
to be the exact shift of $T_F$ (there are no higher order perturbations of $T_F$ in this background).
As has been noted already in \S\fullref{loop space and B-field background} 
the same result is obtained by canonically
quantizing the supersymmetric $2d$ $\sigma$-model \refer{susy action of string in b-field background} 
which describes superstrings in a Kalb-Ramond background.  

By means of \refer{first order delta T from anticommutators of T_F and delta T_F} the shift
$\delta T$ is easily found to be
\begin{eqnarray}
  \label{delta T for B field background}
  \delta T\of{\sigma}
  &=&
  \Bigg(
    -\frac{1}{12 T}
    \partial_\delta H_{\alpha\beta\gamma}
    \left(
       {\cal E}^{\delta}{\cal E}^{\dag \alpha}{\cal E}^{\dag \beta}{\cal E}^{\dag \gamma}
      +
      {\cal E}^{\dag \delta}{\cal E}^{\alpha}{\cal E}^{\beta}{\cal E}^{\gamma}
   \right)
   -
   i
   \frac{1}{2\sqrt{2T}}
   H_{\alpha\beta\gamma}
    \left(
       {\cal E}^{\dag \alpha}{\cal E}^{\dag \beta}
      +
      {\cal E}^{\alpha}{\cal E}^{\beta}
   \right)       
   {\cal P}_+^\gamma
   \nonumber\\
   &&
   +\frac{i}{\sqrt{4T}}
   \partial_\delta B_{\mu\nu}\left({\cal P}_+^\nu - {\cal P}_-^\nu\right)
   \Gamma_+^\delta\Gamma_+^\mu
   +
   B_{\mu\nu}
   {\cal P}_+^\mu
   {\cal P}_-^\nu
  \Bigg)
  \of{\sigma}
  \,.
\end{eqnarray}
This is of total weight $2$ and contains the weight (1,1)
vertex operator
\begin{eqnarray}
  V &=& B_{\mu\nu}{\cal P}_+^\mu {\cal P}_-^\nu
\end{eqnarray}
of the Kalb-Ramond field
(\cf eqs. (52),(53) in \cite{BaggerGiannakis:2001}). That $T + \delta T$
satisfies the Virasoro algebra to first order at the level of Poisson brackets
follows from the fact that it derives from a consistent deformation of the
form \refer{isomorphism of Virasoro algebra} (as well as from the fact that
it also derives from the respective $\sigma$-model Lagrangian).

\subsection{Relations between the various Superconformal Algebras}
\label{relations between susy algebras}

We have found classical deformations of the superconformal algebra associated
with several massless target space background fields. The special algebraic nature
of the form in which we obtain these superconformal algebras admits a convenient
treatment of gauge and duality transformations among the associated background fields.
This is discussed in the following.

\subsubsection{$\extd_K$-exact Deformation Operators}
\label{dK exact deformations}

Deformation operators $\bf W$ which are $\extd_K$-exact, i.e. which are of the form
\begin{eqnarray}
  {\bf W}
  &=&
  \superCommutator{\extd_K}{\bf w}
  \,,
\end{eqnarray}
(where $\superCommutator{\cdot}{\cdot}$ is the supercommutator) 
and which furthermore satisfy
\begin{eqnarray}
  \label{condition for extdK-exact W to be algebra preserving}
  {\bf W}
  &=&
  \superCommutator{\extd_{K,\xi}}{{\bf w}_{\xi^{-1}}}  
  \,,\hspace{.7cm}
  \forall\; \xi
\end{eqnarray}
(where ${\bf w}_{\xi} \defas \int d\sigma\; \xi\; w\of{\sigma}$)
are special 
because for them\footnote{
One way to see this is the following:
\begin{eqnarray}
  \superCommutator{\extd_{K,\xi}}{\superCommutator{\extd_K}{\bf w}}
  &=&
  \superCommutator
    {\extd_{K,\xi}}
    {
      \superCommutator
       { \extd_{K,\xi} }
       { {\bf w}_{\xi^{-1}} }
    }
  \nonumber\\
  &=&
  \superCommutator{{\cal L}_{K,\xi^2}}{{\bf w}_{\xi^{-1}}}  
  \nonumber\\
  &=&
  \int d\sigma\; \left(\xi^2 \xi^{-1} w^\prime  + \frac{1}{2}(\xi^2)^\prime\xi^{-1}w\right)\of{\sigma}
  \nonumber\\
  &=&
  \int d\sigma\; \left(\xi w\right)^\prime
  \,,
\end{eqnarray}
where we used that $w\of{\sigma}$ must be of weight $1/2$ in order that
$W\of{\sigma}$ satisfies condition \refer{weight condition on deformation operator}. 
}
\begin{eqnarray}
  \superCommutator{\extd_{K,\xi}}{\superCommutator{\extd_K}{\bf w}}
  &=& 0
\end{eqnarray}
and hence they
leave the generators of the algebra \refer{suVir algebra in almost usual form} 
invariant:
\begin{eqnarray}
  \extd_{K,\xi}^{\bf W}
  &=&
  \extd_{K,\xi}
  \,.
\end{eqnarray}

Two interesting choices for $\bf w$ are
\begin{eqnarray}
  \label{prepotential for AB gauge transformation}
  {\bf w} &=&  A_{(\mu,\sigma)}\of{X} {\cal E}^{\dag {(\mu,\sigma)}}
\end{eqnarray}
and
\begin{eqnarray}
  \label{prepotential for diffeomorphisms}
  {\bf w} &=& V^{(\mu,\sigma)}\of{X} {\cal E}_{{(\mu,\sigma)}}
  \,,
\end{eqnarray}
which both satisfy \refer{condition for extdK-exact W to be algebra preserving}.
They correspond to $B$-field gauge transformations and to diffeomorphisms, respectively:

\paragraph{$B$-field gauge transformations.}

For the choice \refer{prepotential for AB gauge transformation} one gets
\begin{eqnarray}
  {\bf W}
  &=&
  \antiCommutator{\extd_K}{A_{(\mu,\sigma)}{\cal E}^{\dag (\mu,\sigma)}}
  \nonumber\\
  &=&
  \frac{1}{2}(dA)_{(\mu,\sigma)(\nu,\sigma^\prime)}
  {\cal E}^{\dag (\mu,\sigma)}{\cal E}^{\dag (\nu,\sigma^\prime)}
  +
  i T A_{(\mu,\sigma)}X^{\prime (\mu,\sigma)}
  \,.
\end{eqnarray}
Comparison with \refer{B-field deformation operator} and 
\refer{A field deformation operator} shows that this $\bf W$ induces
a $B$-field background with $B = dA$ and a gauge field background
with $F = T \,dA$. According to \refer{combination of B and F} these
two backgrounds indeed precisely cancel.

This ties up a loose end from \S\fullref{loop space and B-field background}: 
A pure gauge transformation $B \to B + dA$ of the $B$-field does not affect
physics of the closed string and hence should manifest itself as an
algebra isomorphism. Indeed, this isomorphism is that induced by
${\bf W} = iT A_{(\mu,\sigma)} X^{\prime (\mu,\sigma)}$.

\paragraph{Target space diffeomorphisms.}

For the choice \refer{prepotential for diffeomorphisms} one gets
\begin{eqnarray}
  {\bf W}
  &=&
  \antiCommutator{\extd_K}{V^{(\mu,\sigma)}{\cal E}_{(\mu,\sigma)}}
  \nonumber\\
  &=&
  \int d\sigma\;
  \left(
    V^{\mu}\partial_{\mu}
    +
    (\partial_\mu V^\nu)
    {\cal E}^{\dag \mu}{\cal E}_{\nu}
  \right)
  \of{\sigma}
  \nonumber\\
  &=&
  {\cal L}_{V}
  \,,
\end{eqnarray}
where ${\cal L}_V$ is the operator inducing the Lie derivative along
$V$ on forms over loop space (\cf A.4 of \cite{Schreiber:2003a}).
According to \S\fullref{gravitational background by algebra isomorphism} 
the part involving
$
    (\partial_\mu V^\nu)
    {\cal E}^{\dag \mu}{\cal E}_{\nu}
$
changes the metric field at every point of target space, while the part involving
$V^\mu\partial_\mu$ translates the fields that enter in the superconformal generators.
This $\bf W$ apparently induces target space diffeomorphisms.

\subsubsection{T-duality}
\label{T-duality}

It is well known (\cite{LizziSzabo:1998} and references given there)
that in the context of the non-commutative-geometry description of
stringy spacetime physics T-duality corresponds to an inner 
automorphism
\begin{eqnarray}
  {\cal T}: {\cal A} &\to& e^{-{\bf W}}\; {\cal A}\, e^{\bf W} = {\cal A}
\hspace{.7cm}\mbox{with ${\bf W}^\dag = -{\bf W}$}
\end{eqnarray}
of the algebra ${\cal A}$ that enters the spectral triple.
This has been worked out in detail for the bosonic string in
\cite{EvansGiannakis:1996}. In the following this construction
is adapted to and rederived in the present context
for the superstring and then generalized to the various backgrounds
that we have found by deformations.

Following \cite{LizziSzabo:1998}
we first consider T-duality along all dimensions, or
equivalently, restrict attention to the field components along the
directions that are T-dualized. Then we show that 
the \emph{Buscher rules} (see \cite{BandosJulia:2003} for a recent reference) 
for factorized T-duality (i.e. for T-duality along only a single direction)
can very conveniently be derived in our framework, too.

\paragraph{Ordinary T-duality.}

Since T-dualizing along spacetime directions that are not characterized by
commuting isometries is a little subtle (\cf \S4 of \cite{EvansGiannakis:1996}),
assume that a background consisting of a non-trivial metric $g$ and Kalb-Ramond
field $b$ is given together with Killing vectors $\partial_{\mu_n}$
such that
\begin{eqnarray}
  \label{constancy of fields used for T-duality}
  \partial_{\mu_n} g_{\alpha\beta} &=& 0 
  \nonumber\\
  \partial_{\mu_n} b_{\alpha\beta} &=& 0 
  \,.
\end{eqnarray}
For convenience of notation we restrict attention in the following to 
the coordinates $x^{\mu_n}$, since all other coordinates are
mere spectators when T-dualizing. Furthermore we will suppress the
subindex $n$ altogether.

The inner automorphism ${\cal T}$ of the algebra of operators on
sections of the exterior bundle over loop space is defined by
its action on the canonical fields as follows:
\begin{eqnarray}
  \label{T-duality automorphism}
  {\cal T}\of{-i\partial_\mu} &=& X^{\prime \mu}
  \nonumber\\
  {\cal T}\of{X^{\prime \mu}} &=& -i\partial_\mu
  \nonumber\\
  \nonumber\\
  {\cal T}\of{{\cal E}^{\dag a}} &=& {\cal E}_a
  \nonumber\\
  {\cal T}\of{{\cal E}_a} &=& {\cal E}^{\dag a}
  \,.    
\end{eqnarray}
It is possible 
(see \cite{EvansGiannakis:1996}
and pp. 47 of \cite{LizziSzabo:1998}) to express this automorphism manifestly as a
similarity transformation 
\begin{eqnarray}
  {\cal T}\of{A} &=& e^{-\bf W} \,A\, e^{\bf W}
  \,.
\end{eqnarray}
This however requires taking into account normal ordering, 
which would lead us too far afield in the present discussion.
For our purposes it is fully sufficient to
note that ${\cal T}$ preserves the canonical brackets
\begin{eqnarray}
  \label{T duality preserves canonical brackets}
  \commutator{-i\partial_{(\mu,\sigma)}}{X^{\prime (\nu,\sigma^\prime)}} &=& 
   i\delta^\nu_\mu \delta^\prime\of{\sigma,\sigma^\prime}
  \nonumber\\
  &=&
  \commutator{{\cal T}\of{-i\partial_{(i,\sigma)}}}
   {{\cal T}\of{X^{\prime (j,\sigma^\prime)}}}
\end{eqnarray}
and
\begin{eqnarray}
  \antiCommutator{{\cal E}_{(i,\sigma)}}{{\cal E}^{\dag (j,\sigma)}}
  &=&
  \delta_i^j \delta\of{\sigma,\sigma^\prime}
  \nonumber\\
  &=&
  \antiCommutator{{\cal T}\of{{\cal E}_{(i,\sigma)}}}{
    {\cal T}\of{{\cal E}^{\dag (j,\sigma)}}}
\end{eqnarray}
(with the other transformed brackets vanishing)
and must therefore be an algebra automorphism. 

Acting on the $K$-deformed exterior (co)derivative on loop space the transformation
${\cal T}$ produces (we suppress the variable $\sigma$ and the mode functions $\xi$ for
convenience)
\begin{eqnarray}
  {\cal T}\of{
    \extd_{K}
  }
  &=&
  {\cal T}\of{
   {\cal E}^{\dag a}
   E_{a}{}^{\mu}
   \partial_{\mu}
   +
   iT
   {\cal E}_{a}
   E^{a}{}_{\mu}
   X^{\prime \mu}
  }
  \nonumber\\
  &=&
  i{\cal E}_{a}E_{a}{}^{\mu}X^{\prime \mu}
  +
  T {\cal E}^{\dag  a}E^{a}{}_{\mu}\partial_{\mu}  
  \nonumber\\
  &=&
   {\cal E}^{\dag a}
   \tilde E_{a}{}^{\mu}
   \partial_{\mu}
   +
   iT
   {\cal E}_{a}
   \tilde E^{a}{}_{\mu}
   X^{\prime \mu}
  \nonumber\\
  \nonumber\\
  {\cal T}\of{
    \coextd_{K}
  }
  &=&
   \left(
  {\cal T}\of{\extd_{K}}\right)^\dag
  \,,
\end{eqnarray}
where the T-dual vielbein $\tilde E$ is defined as
\begin{eqnarray}
  \tilde E^a{}_\mu &\defas& \frac{1}{T}E_a{}^\mu
  \,.
\end{eqnarray}
(This is obviously not a tensor equation but true in the special coordinates that have been chosen.)
Therefore T-duality sends the deformed exterior derivative associated with the metric defined
by the vielbein $E_a{}^\mu$ to that associated with the metric defined by the vielbein
$\tilde E_a{}^\mu$.
This yields the usual inversion of the spacetime radius $R \mapsto \alpha^\prime/R$:
\begin{eqnarray}
  E^a{}_{\mu} \;=\; \delta^a_\mu\, \sqrt{2\pi} R
  &\Rightarrow&
  \tilde E^a{}_\mu \;=\; \delta^a_\mu\, \frac{1}{T}\frac{1}{\sqrt{2\pi} R} \;=\; 
  \delta^a_\mu \sqrt{2\pi}\; \frac{\alpha^\prime}{R}
  \,.
\end{eqnarray}
Furthermore it is readily checked that the bosonic and fermionic
worldsheet oscillators  
transform as expected:
\begin{eqnarray}
  {\cal T}\of{{\cal P}_{\pm,a}}
  &=&
  {\cal T}\of{
  \frac{1}{\sqrt{2T}}
  \left(
    -i E_a{}^\mu \partial_\mu \pm T E_{a\mu}X^{\prime\mu}
  \right)
  }
  \nonumber\\
  &=&
  \frac{1}{\sqrt{2T}}
  \left(
     E_a{}^\mu X^{\prime\mu} \pm -iT E_{a\mu}\partial_\mu
  \right)
  \nonumber\\
  &=&
  \pm 
  \frac{1}{\sqrt{2T}}
  \left(-i
    \tilde E_a{}^\mu \partial_\mu
    \pm 
    T 
    E_{a\mu}X^{\prime\mu	}
  \right)
  \nonumber\\
  &=&
  \pm \tilde {\cal P}_{\pm,a}
\end{eqnarray}
and
\begin{eqnarray}
  {\cal T}\of{\Gamma^a_\pm}
  &=&
  \pm \Gamma^a_\pm
  \,.
\end{eqnarray}

More generally, when the Kalb-Ramond field is included one finds
\begin{eqnarray}
  {\cal T}\of{
  \extd_{K}^{(B)}
  \pm
  \coextd_{K}^{(B)}
  }
  &=&
  {\cal T}\of{
    \Gamma_\mp^a
    E_a{}^\mu
  \left(
    \partial_\mu
    \mp
    iT
    \left(
      G_{\mu\nu} \pm
      B_{\mu\nu}  
    \right)
   X^{\prime \nu}
  \right)
  }
  \nonumber\\
  &=&
  \mp
    \Gamma_\mp^a
    E_a{}^\mu
  \left(
    i X^{\prime\mu }
    \mp
    T
    \left(
      G_{\mu\nu}
      \pm
      B_{\mu\nu}  
    \right)
    \partial_\nu
  \right)
  \nonumber\\
  &=&
    \Gamma_\mp^a
    \tilde E_a{}^\mu
  \left(
    \partial_\mu
    \mp
    [T
    \left(
      G_{\mu\nu} \pm B_{\mu\nu}  
    \right)]^{-1}
    X^{\prime \nu}
  \right)
\end{eqnarray}
with
\begin{eqnarray}
  \tilde E_a{}^\mu
  &\defas&
  T
  E_a{}^\nu (G_{\nu\mu}\pm B_{\nu\mu})
  \,,
\end{eqnarray}
which reproduces the well known result 
(equation (2.4.39) of \cite{GiveonPorratiRabinovici:1994})
that the T-dual spacetime metric is given by
\begin{eqnarray}
  \tilde G^{\mu\nu}
  &=&
  T^2
  [(G\mp B)G^{-1}(G \pm B)]_{\mu\nu}
\end{eqnarray}
and that the T-dual Kalb-Ramond field is
\begin{eqnarray}
  \tilde B_{\mu\nu} &=& \pm[\frac{1}{T^2}(G\pm B)^{-1} -\tilde G]_{\mu\nu}
  \nonumber\\
  &=&
  \left[
    T^2(G\mp B)B^{-1}(G\pm B)
  \right]^{-1}_{\mu\nu}
  \,.
\end{eqnarray}
\begin{eqnarray}
\end{eqnarray}
It is also very easy in our framework to derive the Buscher rules 
for T-duality along a single direction $y$ (``factorized duality''): 
Let $\mathcal{T}_y$ be the transformation \refer{T-duality automorphism} restricted to
the $\partial_y$ direction, then 
from
\begin{eqnarray}
  \label{Buscher transformation}
  {\cal T}\of{
  \extd_{K}^{(B)}
  \pm
  \coextd_{K}^{(B)}
  }
  &=&
    \Gamma_\mp^a
    \left(
    E_a{}^i
    \partial_i
    \mp
    iT
    E_a{}^\mu
    \left(
      G_{\mu i} \pm
      B_{\mu i}  
    \right)
   X^{\prime i}
   \right)
  \nonumber\\
   &&+
  {\cal T}\of{
    \Gamma_\mp^a
    \left(
    E_a{}^y
    \partial_y
    \mp
    iT
    E_{}^\mu
    \left(
      G_{\mu y} \pm
      B_{\mu y}  
    \right)
   X^{\prime y}
  \right)
  }
  \nonumber\\
  &=&
    \Gamma_\mp^a
    \left(
    E_a{}^i
    \partial_i
    \mp
    iT
    E_a{}^\mu
    \left(
      G_{\mu i} \pm
      B_{\mu i}  
    \right)
   X^{\prime i}
   \right)
  \nonumber\\
   &&
  +
    \Gamma_\mp^a
    \left(
    T
    E_a{}^\mu
    \left(
      G_{\mu y} \pm
      B_{\mu y}  
    \right)
   \partial_y
   \mp
    i
    E_a{}^y
    e^{\Phi/2}
    X^{\prime y}
  \right)
\end{eqnarray}
one reads off the T-dual inverse vielbein
\begin{eqnarray}
  \tilde E_a{}^i &=& E_a{}^i
  \nonumber\\
  \tilde E_a{}^y &=& 
      T
    E_a{}^\mu
    \left(
      G_{\mu y} \pm
      B_{\mu y}  
    \right)
\end{eqnarray}
whose inverse $\tilde E_\mu{}^a$ is easily seen to be
\begin{eqnarray}
  \tilde E_{i}{}^a
  &=&
  E_i{}^a -\frac{G_{iy}\pm B_{iy}}{G_{yy}}E_y{}^a
  \nonumber\\
  \tilde E_y{}^a 
  &=&
  \frac{1}{T G_{yy}}E_y{}^a
  \,,
\end{eqnarray}
which gives the T-dual metric with minimal computational effort:
\begin{eqnarray}
  \tilde G_{yy}
  &=&
  \frac{1}{T G_{yy}}
  \nonumber\\
  \tilde G_{iy}
  &=&
  \mp \frac{B_{iy}}{TG_{yy}}
  \nonumber\\
  \tilde G_{ij}
  &=&
    G_{ij} - \frac{1}{G_{yy}}
    \left(
      G_{iy}G_{jy}
      -
      B_{iy}B_jy
    \right)
  \,.
\end{eqnarray}
Similarly the relations 
\begin{eqnarray}
  \tilde E_a{}^\mu(\tilde G_{\mu i}\pm \tilde B_{\mu i})
  &=&
  E_a{}^\mu(G_{\mu i} \pm B_{\mu i})
  \nonumber\\
  \tilde E_a{}^\mu(\tilde G_{\mu y}\pm \tilde B_{\mu y})
  &=&
  \frac{1}{T}
  E_a{}^y
\end{eqnarray}
for the T-dual $B$-field $\tilde B$ are read
off from \refer{Buscher transformation}. Solving them for $\tilde B$ is
straightforward and yields
\begin{eqnarray}
  \tilde B_{ij}
  &=&
  B_{ij}
  \mp
  \frac{1}{G_{yy}}
  \left(
    B_{jy}G_{iy} - B_{iy}G_{iy}
  \right)
  \nonumber\\
  \tilde B_{iy}
  &=&
  \frac{1}{T G_{yy}}G_{iy}
  \,.
\end{eqnarray}
These are the well known \emph{Buscher rules} for factorized T-duality
(see eq. (4.1.9) of \cite{GiveonPorratiRabinovici:1994}).

The constant dilaton can be formally absorbed into the string tension $T$ 
and is hence seen to be invariant under ${\cal T}_y$. This is
correct in the classical limit that we are working in. It
is well known (e.g. eq. (4.1.10) of \cite{GiveonPorratiRabinovici:1994}),
that there are higher loop corrections to the T-dual dilaton. These corrections are 
not visible with the methods discussed here.

$\,$\\

Using our representation for the superconformal generators
in various backgrounds it is now straightfroward to include more general background fields than
just $G$ and $B$ in the above construction:

\paragraph{T-duality for various backgrounds.}
\label{T-duality for various background fields}

When turning on all fields $G$, $B$, $A$, $C$ and $\Phi$, requiring them to be
constant in the sense of \refer{constancy of fields used for T-duality} and assuming
for convenience of notation that $B\inner C = 0$ the supercurrents read according
to the considerations in \S\ref{purely gravitational}-\S\ref{A-field background}
\begin{eqnarray}
  \label{general supercurrents for constant fields}
    \extd_K^{(\Phi)(A+B+C)} \pm \coextd_K^{(\Phi)(A+B+C)}
  &=&
    \Gamma_\mp^a E_a{}^\mu
    \left(
       e^{\Phi/2}(G_\mu{}^\nu \pm C_\mu{}^\nu)\partial_\nu
       \mp iT e^{-\Phi/2}(G_{\mu\nu}\pm (B_{\mu\nu}+\frac{1}{T}F_{\mu\nu})X^{\prime \nu}
    \right)
  \,.
  \nonumber\\
\end{eqnarray}

It is straightforward to apply $\cal T$ to this expression and read off the
new fields. However, since the resulting expressions are not too
enlightening we instead use a modification $\tilde {\cal T}$ of $\cal T$, which, too,
induces an algebra isomorphism, but which produces more
accessible field redefinitions. The operation $\tilde {\cal T}$ differs
from $\cal T$ in that index shifts are included:
\begin{eqnarray}
  \tilde {\cal T}\of{-i\partial_\mu}
  &\defas&
  T
  g_{\mu\nu}X^{\prime \nu}
  \nonumber\\
  \tilde {\cal T}\of{X^{\prime\mu}}
  &\defas&
  -\frac{i}{T}g^{\mu\nu}\partial_\nu
  \nonumber\\
  \nonumber\\
  \tilde {\cal T}\of{{\cal E}^{\dag a}}
  &\defas&
  {\cal E}^a
  \nonumber\\
  \tilde {\cal T}\of{{\cal E}^{a}}
  &\defas&
  {\cal E}^{\dag a}  
  \,.
\end{eqnarray}
Due to the constancy of $g_{\mu\nu}$ this preserves the canonical brackets
just as in \refer{T duality preserves canonical brackets} and hence is
indeed an algebra isomorphism.

Applying it to the supercurrents \refer{general supercurrents for constant fields} yields
\begin{eqnarray}
  \tilde {\cal T}
  \left[
    \extd_K^{(\Phi)(B+C)} \pm \extd_K^{(\Phi)(B+C)}
  \right]
  &=&
    \Gamma_\mp^a E_a{}^\mu
    \left(
        e^{-\Phi/2}(G_{\mu}{}^{\nu}\pm (B_{\mu}{}^{\nu} + \frac{1}{T}F_\mu{}^\nu)\partial_\nu
        \mp i Te^{\Phi/2}(G_{\mu\nu} \pm C_{\mu\nu}) X^{\prime \nu}
    \right)  
  \,.
  \nonumber\\
\end{eqnarray}
Comparison shows that under $\tilde {\cal T}$ the background fields transform as
\begin{eqnarray}
  \label{possible S-duality}
  B_{\mu\nu}+ \frac{1}{T}F_{\mu\nu} &\to& C_{\mu\nu}
  \nonumber\\
  C_{\mu\nu} &\to& B_{\mu\nu} + \frac{1}{T}F_{\mu\nu}
  \nonumber\\
  G_{\mu\nu} &\to& G_{\mu\nu}
  \nonumber\\
  \Phi &\to& -\Phi
  \,.
\end{eqnarray}
The fact that under this transformation 
the NS-NS 2-form is exchanged with what we interpreted as the 
R-R 2-form and that the
dilaton reverses its sign is reminiscent of S-duality. 
It is well known \cite{Duff:1995} that T-duality and S-duality are
themselves dual under the exchange of the fundamental F-string and the 
D-string. How exactly this applies to the constructions
presented here remains to be investigated. (For instance the
sign that distinguishes \refer{possible S-duality} from
the expected result would need to be explained, maybe by a change
of orientation of the string.)

\subsubsection{Hodge Duality on Loop Space}
\label{Hodge duality on loop space}

For the sake of completeness in the following the relation of
loop space Hodge duality to the above discussion is briefly indicated.
It is found that ordinary Hodge duality is at least superficially related to
the algebra isomorphisms discussed in \S\fullref{T-duality}. 
Furthermore a deformed version of Hodge duality is considered which
preserves the familiar relation $\extd^\dagger = \pm \star\,\extd\,\star^{-1}$.

\paragraph{Ordinary Hodge duality.}

On a finite dimensional pseudo-Riemannian manifold, let $\bar \star$ be the
phase-shifted Hodge star operator which is normalized so
as to satisfy
\begin{eqnarray}
  (\bar \star)^\dag &=& - \bar \star
  \nonumber\\
  (\bar \star)^2 &=& 1
  \,.
\end{eqnarray}
(For the precise relation of $\bar \star$ to the ordinary Hodge $\star$ see (A.18) of \cite{Schreiber:2003a}.)
The crucial property of this operator can be expressed as
\begin{eqnarray}
  \bar \star\, \onbCreator^{\mu} &=& \onbAnnihilator^\mu\, \bar \star
  \,,
\end{eqnarray}
where $\onbCreator^\mu$ is the operator of exterior multiplication by $dx^\mu$
and $\onbAnnihilator^\mu$ is its adjoint under the Hodge inner product. 

It has been pointed out in \cite{Witten:1982} 
that the notion of Hodge duality can be carried over to
infinite dimensional manifolds. This means in particular that on loop space there
is an idempotent operator $\bar\star$ so that
\begin{eqnarray}
  \bar \star\, {\cal E}^{\dag \mu} &=& {\cal E}^\mu\, \bar \star
\end{eqnarray}
and
\begin{eqnarray}
  \commutator{\bar \star}{X^{(\mu,\sigma)}}
  = 0 = 
  \commutator{\bar \star}{\gradOp_{(\mu,\sigma)}}
  \,. 
\end{eqnarray}
It follows in particular that the $K$-deformed exterior derivative is
related to its adjoint by
\begin{eqnarray}
  \coextd_K &=& - \bar \star\; \extd_K \; \bar \star
  \,.
\end{eqnarray}
In fact this holds for all the modes:
\begin{eqnarray}
  \coextd_{K,\xi}^\dag &=& - \bar \star\; \extd_{K,\xi} \; \bar \star
  \,.
\end{eqnarray}

In the spirit of the discussion of T-duality by algebra isomorphisms in 
\S\fullref{T-duality} one can equivalently say that $\bar \star$ induces
an algebra isomorphism $\cal H$ defined by
\begin{eqnarray}
  {\cal H}\of{A} &\defas& \bar \star\; A \bar \star
  \,,
\end{eqnarray}
i.e.
\begin{eqnarray}
  {\cal H }\of{-i\partial_\mu} &=& -i\partial_\mu
  \nonumber\\
  {\cal H}\of{X^\mu} &=& X^\mu
  \nonumber\\
  \nonumber\\
  {\cal H}\of{{\cal E}^{\dag a}} &=& {\cal E}^{a}
  \nonumber\\
  {\cal H}\of{{\cal E}^{a}} &=& {\cal E}^{\dag a}
  \,.  
\end{eqnarray}

It is somewhat interesting to consider the result of first applying $\cal H$ to
$\extd_K$ and then acting with the deformation operators $\exp\of{\bf W}$
considered before. This is equivalent to considering the deformation obtained by
$\bar \star\, e^{\bf W}$. This yields
\begin{eqnarray}
  \extd_K
  &\to&
  \left(e^{-\bf W}\bar \star\right)
  \extd_K
  \left(
  \bar \star\, e^{\bf W}\right)
  \;=\;
  -
  e^{-\bf W}\,\coextd_K \,e^{\bf W}
  \nonumber\\
  \coextd_K
  &\to&
  \left(e^{{\bf W}^\dag}\bar \star\right)
  \coextd_K
  \left(
  \bar \star\, e^{-{\bf W}^\dag}\right)
  \;=\;
  -
  e^{{\bf W}^\dag}\,\extd_K \,e^{-{\bf W}^\dag}
  \,.
\end{eqnarray}
Hence, except for a global and irrelevant sign, the deformations induced by
$e^{\bf W}$ and $\bar \star\, e^{\bf W}$ are related by
\begin{eqnarray}
  {\bf W} &\leftrightarrow& -{\bf W}^\dag
  \,.
\end{eqnarray}
Looking back at the above results for the backgrounds induced by various ${\bf W}$ 
this corresponds to
\begin{eqnarray}
  B &\leftrightarrow& C
  \nonumber\\
  A &\leftrightarrow& A
  \nonumber\\
  \Phi &\leftrightarrow& -\Phi
  \,.
\end{eqnarray}

It should be noted though, that unlike the similar correspondence 
\refer{possible S-duality} both sides of this relation are not unitarily
equivalent in the sense that the corresponding superconformal generators
$e^{-\bf W}\bar \star\, \extd_K \, \bar \star e^{\bf W}$
and
$e^{-\bf W} \, \extd_K \,  e^{\bf W}$
are not unitarily equivalent.

Nevertheless, it might be that the physics described by both generators
is somehow related. This remains to be investigated.

\paragraph{Deformed Hodge duality.}
\label{deformed Hodge duality}

The above shows that for general background fields (general deformations of the
superconformal algebra) the familiar equality of
$\coextd^{\bf W}_{K,\xi}$ with $- {\bar \star} \,\extd^{\bf W}_{K,\xi}\,{\bar \star}^{-1}$
is violated. It is however possible to consider a deformation ${\bar\star}^{\bf W}$ 
of $\bar \star$ itself
which restores this relation:
\begin{eqnarray}
  {\bar \star}^{\bf W}
  &\defas&
  e^{{\bf W}^\dag}\,\bar\star\, e^{\bf W}
  \,.
\end{eqnarray}
Obviously this operator satisfies
\begin{eqnarray}
  \coextd_{K,\xi}^{\bf W}
  &=&
  - {\bar \star}^{\bf W}
  \,
    \extd_{K,\xi}
  \,
  ({\bar \star}^{\bf W})^{-1}	
  \,.
\end{eqnarray}

The Hodge star remains invariant under this deformation when
$\bf W$ is anti-Hodge-dual:
\begin{eqnarray}
  \bar \star = {\bar \star}^{\bf W}
  &\Leftrightarrow&
  \bar \star \,{\bf W}\,\bar \star
  =
  - {\bf W}^\dagger
  \,.
\end{eqnarray} 
This is in particular true for the gravitational deformation of
\S\fullref{gravitational background by algebra isomorphism}.
It follows that ${\bar \star}^{(G)} = {\bar \star}$. This
can be understood in terms of the fact that the definition of the
Hodge star involves only the orthonormal metric on the tangent space
(\cf (A.14) of \cite{Schreiber:2003a}).

\subsubsection{Deformed inner Products on Loop Space.}
\label{deformed inner products on loop space}

The above discussion of deformed Hodge duality on loop space motivates the
following possibly interesting observation:

From the point of view of differential geometry the exterior derivative
$\extd$ on a manifold is a purely topological object which does not
depend in any way on the geometry, i.e. on the metric tensor. The geometric information
is instead contained in the Hodge star operator $\star$, the 
Hodge inner product  
$\langle \alpha |\beta\rangle = \int\limits_\manifold \alpha \wedge \star \beta$
on differential forms
and the adjoint $\coextd$ of $\extd$ with respect to $\langle\cdot|\cdot\rangle$.

We have seen in \S\fullref{deformed Hodge duality} that deformations of the
Hodge star operator on loop space may encode not only information about the
geometry of target space, but also about other background fields, like Kalb-Ramond
and dilaton fields. These deformations are accompanied by analogous deformations
\refer{isomorphism of Virasoro algebra} of $\extd$ and $\coextd$.

But from this point of view of differential geometry it appears unnatural to associate a 
deformation of both $\coextd$ as well as $\extd$ with a deformed Hodge star operator.
One would rather expect that $\extd$ remains unaffected by any background fields while
the information about these is contained in $\star$, $\langle\cdot|\cdot \rangle$ and 
$\coextd.$

Here we want to point out that both points of view are equivalent and indeed
related by a global similarity transformation ('duality') and that the change in 
point of view makes an interesting relation to noncommutative geometry transparent.

Namely consider deformed operators
\begin{eqnarray}
  \label{deformed ops once again}
  \extd^{({\bf W})}
  &=&
  e^{-{\bf W}}\extd e^{\bf W}
  \nonumber\\
  \coextd^{({\bf W})}
  &=&
  e^{{\bf W}^\dagger}\coextd e^{-{\bf W}^\dagger}
\end{eqnarray}
on an inner product space $\cal H$ with inner product $\langle\cdot | \cdot \rangle$
as in \refer{isomorphism of Virasoro algebra}.

By applying a global similarity transformation
\begin{eqnarray}
  \ket{\psi} &\to& \tilde {\ket{ \psi}} \defas e^{{\bf W}}\ket{\psi}
  \nonumber\\
  A &\to& \tilde A \defas e^{\bf W}Ae^{- {\bf W}}
\end{eqnarray}
to all elements $\ket{\psi} \in {\cal H}$ and all operators $A$ on $\cal H$
one of course finds
\begin{eqnarray}
  \left({\extd^{(\bf W)}}\right)^{\tilde{}}
  &=&
  \extd
  \nonumber\\
  \left({\coextd^{({\bf W})}}\right)^{\tilde {}}
  &=&
  e^{{\bf W} + {\bf W}^\dagger}\coextd e^{-{\bf W}- {\bf W}^\dagger}
  \,.
\end{eqnarray}

By construction, the algebra of $\left({\extd^{(\bf W)}}\right)^{\tilde{}}$
and $\left({\coextd^{(\bf W)}}\right)^{\tilde{}}$ is the same as that of
${\extd^{(\bf W)}}$ and ${\coextd^{(\bf W)}}$. But now all the information about the
deformation induced by $\bf W$ is contained in $\left({\coextd^{(\bf W)}}\right)^{\tilde{}}$
alone. This has the advantage that we can consider a deformed innner product
\begin{eqnarray}
  \label{deformed inner product}
  \langle\cdot | \cdot \rangle_{{}_{({ \bf W})}}
  &\defas&
  \langle\cdot | \,e^{-({\bf W}+{\bf W}^\dagger)}\, \cdot \rangle  
\end{eqnarray}
on ${\cal H}$ with respect to which 
\begin{eqnarray}
  \extd^{\dagger_{(\bf W)}}
  &=&
 \left({\coextd^{(\bf W)}}\right)^{\tilde{}}
  \,,
\end{eqnarray}
where $\langle A \cdot | \cdot \rangle_{{}_{({ \bf W})}} \defas 
\langle\cdot | A^{\dagger_{({\bf W})}} |\cdot \rangle_{{}_{({ \bf W})}}
$. If the original inner product came from a Hodge star this corresponds to
a deformation
\begin{eqnarray}
  \star &\to& \star \;e^{-({\bf W} + {\bf W}^\dagger)}
 \,.
\end{eqnarray}
This way now indeed the entire deformation comes from a deformation of the Hodge star
and the inner product.

That this is equivalent to the original notion \refer{deformed ops once again} 
of deformation can be checked again by noting that the deformed inner product
of the deformed states agrees with the original inner prodcut on the original states
\begin{eqnarray}
  \langle \tilde \psi | \tilde \phi \rangle_{{}_{({\bf W})}}
  &=&
  \langle \psi | \phi \rangle 
  \,.
\end{eqnarray}

These algebraic manipulations by themselves are rather trivial, but the interesting aspect
is that the form \refer{deformed inner product} of the deformation appears
in the context of noncommutative spectral geometry \cite{ForgySchreiber:2004}.
The picture that emerges is roughly that of a spectral triple 
$(\mathcal{A},\extd_K \pm \extd^{\dag_{({\mathbf W})}}_K, \mathcal{H})$,
where $\mathcal{A}$ is an algebra of functions on loop space 
(\cf \fullref{loop space definitions}), $\mathcal{H}$ is the inner product
space of differential forms over loop space equipped with a deformed Hodge
inner product \refer{deformed inner product} which encodes all the information
of the background fields on target space, and where two Dirac operators are given by
$\extd_K \pm \extd^{\dag_{({\mathbf W})}}_K$. There have once been attempts
\cite{Chamseddine:1997,Chamseddine:1997b,FroehlichGrandjeanRecknagel:1997,
FroehlichGrandjeanRecknagel:1996,FroehlichGawedzki:1993} 
to understand the superstring by regarding the worldsheet supercharges
as Dirac operators in a spectral triple. Maybe the insight that 
and how target space background fields manifest themselves as simple algebraic
deformations \refer{isomorphism of Virasoro algebra} of the Dirac operators, or, equivalently, 
\refer{deformed inner product} of the inner product on $\mathcal{H}$ can help to make progress with this approach.

\subsection{Appendix: Canonical Analysis of Bosonic D1 Brane Action}
\label{Canonical analysis of bosonic D1 brane action}

The bosonic part of the worldsheet action of the D-string is
\begin{eqnarray}
  {\cal S}
  &=&
    -
    T
    \int
    d^2 \sigma\;
    e^{-\Phi}\sqrt{-{\rm det} (G_{ab} + B_{ab} + \frac{1}{T}F_{ab})}
    +
    T
    \int
    \left(
      C_2 + C_0(B +\frac{1}{T}F)
   \right)
  \,,
\end{eqnarray}
where $G$, $B$, $C_0$ and $C_2$ are the respective background fields and
$F_{ab} = (dA)_{ab}$ is the gauge field on the worldsheet. Indices $a,b$ range
over the worldsheet dimensions and indices $\mu,\nu$ over target space dimensions.

Using \emph{Nambu-Brackets}
$\{X^\mu,X^\nu\} \defas \epsilon^{ab}\partial_a X^\mu \,\partial_b X^\nu$
(with $\epsilon^{01} = 1$, $\epsilon^{ab} = -\epsilon^{ba}$) the term in the
square root can be rewritten as
\begin{eqnarray}
  -{\rm det}\of{G_{ab} + B_{ab} + \frac{1}{T}F_{ab}}
  &=&
  -
  \frac{1}{2}
  \lbrace X^{\mu},X^{\nu}\rbrace
  G_{\mu\mu^\prime}
  G_{\nu\nu^\prime}
  \lbrace X^{\mu^\prime},X^{\nu^\prime}\rbrace
   - (B_{01} + \frac{1}{T}F_{01})^2
  \,.
\end{eqnarray}
The canonical momenta associated with the embedding coordinates $X^\mu$ are
\begin{eqnarray}
  P_\mu &=&
  \frac{\delta {\cal L}}{\delta \dot X^\mu}
  \nonumber\\
  &=&
  T
  \left(
  \frac{1}{e^{\Phi}\sqrt{-{\rm det}(G+B+F/T)}}
  \left(
  X^{\prime \nu}
  G_{\mu\mu^\prime}
  G_{\nu\nu^\prime}
  \lbrace X^{\mu^\prime},X^{\nu^\prime}\rbrace
  +
  B_{\mu\nu}X^{\prime\nu}(B_{01}+ \frac{1}{T}F_{01})
  \right)
  \right)+
  \nonumber\\
  &&
  \;+
  T
  (C_2 + C_0 B){}_{\mu\nu}X^{\prime\nu}
  \,.
  \nonumber\\
\end{eqnarray}
On the other hand the canonical momenta associated with the gauge field read
\begin{eqnarray}
  \label{D1 gauge field canonical momenta}
  E_0
  &\defas&
  \frac{\delta {\cal L}}{\delta \dot A_1}
  \;=\;
  0
  \nonumber\\
  E_1
  &\defas&
  \frac{\delta {\cal L}}{\delta \dot A_1}
  \;=\;
  \frac{1}{e^{\Phi}\sqrt{-{\rm det}(G+B+F/T)}}
  (B_{01}+ \frac{1}{T}F_{01})
  +
  C_0
  \,.
\end{eqnarray}
Since the gauge group is ${\rm U}\of{1}$, $A_\mu$ is a periodic variable and hence the 
eigenvalues of $E_1$ are discrete \cite{Witten:1995}:
\begin{eqnarray}
  E_1 &\defas& p \,\in \Z
  \,.
\end{eqnarray}
Inverting \refer{D1 gauge field canonical momenta} 
allows to rewrite the canonical momenta $P_\mu$ as
\begin{eqnarray}
  P_\mu &=&
  \frac{1}{\sqrt{-\mathrm{det}\of{G}}}
  \tilde T
  X^{\prime \nu}
  G_{\mu\mu^\prime}
  G_{\nu\nu^\prime}
  \lbrace X^{\mu^\prime},X^{\nu^\prime}\rbrace
  +
  T
  (C_2 + p B){}_{\mu\nu}X^{\prime\nu}
  \,,
\end{eqnarray}
where
\begin{eqnarray}
  \tilde T
  &\defas&
  T\sqrt{e^{-2\Phi} + (p-C_0)^2}
\end{eqnarray}
is the tension of a bound state of one D-string with p F-strings \cite{IgarishiItohMamimuraKuriki:1998}.
In this form it is easy to check that the following two constraints are satisfied:
\begin{eqnarray}
  \label{D-string constraints}
  &&(P-T(C_2 + p B)\inner X^\prime) \inner
  (P-T(C_2 + p B)\inner X^\prime)
  +
  \tilde T^2
  X^\prime \inner X^\prime
  = 0 
  \nonumber\\
  &&(P-T(C_2 + p B)\inner X^\prime)\inner X^\prime = 0
  \,,
\end{eqnarray}
which express temporal and spatial reparameterization invariance, respectively.
For constant $\tilde T$ this differes from the familiar constraints for the pure F-string 
only in a redefinition of the tension and the couplings to the background 2-forms.

For non-constant $\tilde T$, however, things are a little different. 
For the purpose of comparison with the results in \S\fullref{Dilaton background}
consider the case $B = C_0 = C_2 = p = 0$ and $\Phi$ possibly non-constant. In this case the constraints
\refer{D-string constraints} can be equivalently rewritten as
\begin{eqnarray}
  \mathcal{P_\pm}^2 &=& 0
\end{eqnarray}
with
\begin{eqnarray}
  \label{Born-Infeld current for pure dilaton background}
  {\cal P}_{\mu,\pm}
  &=&
  e^{\Phi/2}
  P_\mu 
  \pm
  T e^{-\Phi/2}
   G_{\mu\nu} 
   X^{\prime\nu}
  \,.
\end{eqnarray}
Up to fermionic terms this is the form found in \refer{Phi Dirac operators}.

\addtocontents{toc}{\vspace*{-.2em}}

\clearpage
\section{Worldsheet Invariants and Boundary States}
\label{Worldsheet Invariants and Boundary States}

The following is taken from
\cite{Schreiber:2004b}, which was mainly a reaction to attempts
\cite{Thiemann:2004,Pohlmeyer:2002,MeusburgerRehren:2002,Pohlmeyer:1998,
Pohlmeyer:1988} to find an alternative quantization of the
string in terms of non-standard invariants. 
Only after \cite{Schreiber:2004b} was published did I become aware that
essentially the same result had appeared long before in
\cite{BorodulinIsaev:1982,Isaev:1983}.

\subsection{DDF and Pohlmeyer invariants}

The classical string is known to be a completely integrable system with 
an infinite number of classical observables that Poisson-commute with all the
constraints. A concise and comprehensive review  of the work by Pohlmeyer,
Rehren, Bahns, et. al. 
\cite{Pohlmeyer:2002,MeusburgerRehren:2002,Pohlmeyer:1998,Pohlmeyer:1988}
on a particular manifestation of these \emph{gauge invariant observables}, known as 
\emph{Pohlmeyer invariants}, 
is given in \cite{Thiemann:2004}.

Since the Virasoro algebra is the direct sum of two copies
of $\mathrm{diff}(S^1)$, the diffeomorphism algebra of the circle, invariant
observables are simply those that are reparameterization invariant with respect
to these two algebras. Two common types of reparameterization invariant objects are
\begin{itemize}
  \item Wilson lines,
  \item integrals over densities of unit reparameterization weight .
\end{itemize} 
More precisely (see \S\fullref{Pohlmeyer invariants} 
for a detailed derivation), let $\mathcal{P}^\mu_\pm\of{\sigma}$ 
be the left- and right-moving classical fields 
on the free closed bosonic string with Poisson brackets of the form
\begin{eqnarray}
  \commutator{\mathcal{P}^\mu_\pm\of{\sigma}}{\mathcal{P}^\nu_\pm\of{\sigma^\prime}}_\mathrm{PB}
  &=&
  \pm \eta^{\mu\nu}\delta^\prime\of{\sigma,\sigma^\prime}
  \,.
\end{eqnarray}
These transform with unit weight under the action of the Virasoro algebra
and hence the \emph{Wilson line}
\begin{eqnarray}
  \mathrm{Tr}\mathbf{P}\exp\of{\int\limits_{S^1} \mathcal{P}^\mu_+  A_{+\mu}}
  \mathrm{Tr}\mathbf{P}\exp\of{\int\limits_{S^1} \mathcal{P}^\mu_-  A_{-\mu}}\,,
\end{eqnarray}
Poisson-commutes with all Virasoro constraints
(where $\mathbf{P}$ denotes path-ordering and 
$A_{\pm \mu}$ are two \emph{constant} Lie-algebra-valued 1-forms on target space). 
It is easy to see that also
the coefficients of $\mathrm{Tr}\of{A^n}$ in the Taylor expansion of this
object commute with the constraints. These coefficients are known as
the \emph{Pohlmeyer invariants}. The Poisson algebra of these
observables is rather convoluted. The problem of finding a quantum deformation of this
algebra turns out to be difficult and involved and has up to now remained
unsolved \cite{MeusburgerRehren:2002,Thiemann:2004,Bahns:2004}. 
Furthermore, by itself, it is not obvious how the above 
construction should generalize to the superstring.

For these reasons it seems worthwhile to consider the possibility
of alternatively using integrals over unit weight densities to construct a complete set
of classical invariant charges. A little reflection shows that the well-known
\emph{DDF operators} \cite{DelGiudiceDiVecchiaFubini:1972} for the 
covariantly quantized string,
which are operators that commute with all the quantum (super-)Virasoro constraints,
are built using essentially this principle:

From elementary CFT it follows that
for $\mathcal{O}(z)$ any primary CFT field of conformal weight $h=1$ we have
(after the usual introduction of a complex coordinate $z$ on the worldsheet)
\begin{eqnarray}
  \commutator{L_n}{\oint dz\, \mathcal{O}(z)} &=& 0\,, \hspace{.5cm}\forall\, n
  \,.
\end{eqnarray}
By choosing $\mathcal{O}(z) = \commutator{G_{-\nu}}{{\tilde {\mathcal{O}(z)}}}$,
(with $G_{-\nu}$ the $(\nu=0)$-mode (R sector) or $(\nu = -1/2)$-mode (NS sector) 
of the worldsheet supercurrent and 
$h\of{{\tilde {\mathcal{O}}}} = 1/2$) this generalizes to the superstring
\begin{eqnarray}
  \commutator{G_{n-\nu}}{\oint dz\, \mathcal{O}(z)} &=& 0 \,, \hspace{.5cm}\forall\, n
  \,,
\end{eqnarray}
as is reviewed below in \S\ref{DDF operators}. It hence only remains to find
$\mathcal{O}$ (or ${\tilde {\mathcal{O}}}$) of weight 1 (or 1/2) such that 
the resulting integrals have nice (super-)commutators and exhaust
the space of all invariant charges. Doing this in a natural way
yields the DDF operators. 

It is readily checked that this construction of the DDF operators can 
be mimicked in terms of the classical Poisson algebra to yield 
a complete set of classical invariants which we shall call
\emph{classical DDF invariants}. These inherit all the nice properties
of their quantum cousins. 

Most importantly, as is shown in \S\fullref{classical bosonic DDF invariants}, 
the Pohlmeyer invariants can be expressed in 
terms of the classical DDF invariants. Since it is known how the latter
have to be quantized (i.e. the crucial quantum corrections to these
charges is well known \cf \S\fullref{DDF operators}) 
this also tells us how the Pohlmeyer invariants
can consistently be quantized.

In particular this shows that any normal ordering in the quantization of the
Pohlmeyer invariants must be applied
only inside each DDF operator, while the DDF operators among themselves
need not be reordered. This clarifies the result of \cite{Bahns:2004},
where it was demonstrated that the Pohlmeyer invariants cannot be consistently
quantized by writing them in terms of worldsheet oscillators and
applying normal ordering with respect to these. Rather, as will be
shown here, one has to replace these oscillators with the corresponding
DDF observables, and the assertion is that the Pohlmeyer invariants, like
any other reparameterization invariant observable, are unaffected
by this replacement.

Because the DDF operators, together with the identity operator, form
a closed algebra, the quantization of the Pohlmeyer invariants in terms
of DDF operators, as demonstrated here, is manifestly consistent in the sense
that the quantum commutator of two such invariants is itself again an invariant.

It should be emphasized that, in contrast to what has been stated in 
\cite{Bahns:2004}, the construction of classical DDF invariants does \emph{not}
require that any worldsheet coordinate gauge has to be fixed, in particular
their construction has nothing to do with fixing conformal gauge. This
is obvious due to the fact that the classical DDF invariants are 
constructable (as the Pohlmeyer invariants, too) by proceeding from just the
Nambu-Goto action, which does not
even have an auxiliary worldsheet metric which could be gauge fixed. 
Furthermore the canonical data and the form of the Virasoro constraints as
obtained from the Nambu-Goto action are precisely the same as those
obtained from the Polyakov action with or without fixed worldsheet gauge.

Furthermore, our proof that the Pohlmeyer invariants can be equivalently
expressed in terms of DDF invariants (i.e. certain polynomials in DDF invariants are
equal to the Pohlmeyer invariants) constructively demonstrates that both are on the
same footing as far as requirements for their respective construction is concerned.\\

The organization of this paper is as follows:
  
We first review the construction of DDF operators in \S\fullref{DDF operators}
and then that of Pohlmeyer invariants in \S\fullref{Pohlmeyer invariants}.

Then in \S\fullref{classical bosonic DDF} we discuss the classical DDF invariants in detail, 
show how they can be used to 
express the Pohlmeyer invariants (\S\fullref{expressing Pohlmeyer invariants in terms of DDF invariants})
and how this generalizes to the superstring (\S\fullref{DDF and Pohlmeyer invariants for superstring}).

A brief summary of the results presented here will be published in 
\cite{Schreiber:2004c}.

\subsubsection{DDF operators and Pohlmeyer invariants}
\label{DDF operators and Pohlmeyer invariants}

We review first the DDF operators, then the Pohlmeyer invariants, then show
how both are related.

\paragraph{DDF operators.}
\label{DDF operators}

The construction of DDF operators \cite{DelGiudiceDiVecchiaFubini:1972}
is very well known, but to the best of our knowledge there is no
comprehensive review of all possible cases (transversal and longitudinal,
bosonic and fermionic)
available in the standard literature. The following section
tries to list and derive all the essential facts.

In the standard textbook literature one can find 
\begin{itemize}
  \item in \cite{GreenSchwarzWitten:1987} (in non-CFT language)
the construction of 
\begin{itemize}
  \item transversal bosonic (\S 2.3.2) 
  \item transversal supersymmetric (\S 4.3.2) 
  \item longitudinal bosonic (pp. 111),
\end{itemize}
\item
and in \cite{Polchinski:1998} (in
CFT language) the construction of 
\begin{itemize}
  \item transversal bosonic (eq. (8.2.29)) 
\end{itemize}
DDF states, which go back to \cite{DelGiudiceDiVecchiaFubini:1972}.
\end{itemize}

The following summarizes and derives (in CFT language) \emph{all}
\begin{itemize}
  \item
    transversal and longitudinal
  \item
    bosonic and fermionic
\end{itemize}
DDF operators (for a free supersymmetric worldsheet theory).

Using the standard normalization of the OPE
\begin{eqnarray}
  X^\mu\of{z} X^\nu\of{0} 
  &\sim&
  -\frac{\alpha^\prime}{2}\eta^{\mu\nu}
  \ln z
  \nonumber\\
  \psi^\mu\of{z}\psi^\nu\of{0} &\sim&  \frac{\eta^{\mu\nu}}{z}
\end{eqnarray}
for the bosonic and fermionic worldsheet fields, the (super-)Virasoro currents read
\begin{eqnarray}
  T\of{z} &=& 
  -\frac{1}{\alpha^\prime} \partial X \inner \partial X\of{z}
  -\frac{1}{2}\psi \inner \partial \psi
  \nonumber\\
  T_{\rm F}\of{z}
  &=&
  i \sqrt{\frac{\alpha^\prime}{2}}
  \psi \inner \partial X
  \,.
\end{eqnarray}
The \emph{DDF operators} are defined as a set of operators that commute with all 
modes of $T$ and $T_\mathrm{F}$ (are 'gauge invariant observables') 
and satisfy an algebra that mimics that of worldsheet oscillator creation/annihilation operators.

First of all one needs to single out two linearly independent lightlike Killing vectors $l$ and $k$
on target space, 
and in the context of this subsection we choose to normalize them as $l\inner k = 2$. 
The span of $l$ and $k$ is called the \emph{longitudinal} space and its orthogonal
complement is the \emph{tranverse} space.

For $\mathcal{O}\of{z} = \sum\limits_{-\infty}^{\infty}\mathcal{O}_m z^{-(m+h)}$
a primary field of weight $h$  we shall refer to the OPE
$T\of{z}\mathcal{O}\of{0}\sim \frac{h}{z^2}\mathcal{O}\of{0} 
+ \frac{1}{z}\partial\mathcal{O}\of{0}$ as the \emph{tensor law} in some
of the following formulas, instead of writing out all terms.
The modes of $T$ and $T_\mathrm{F}$ are denoted by $L_m$ and $G_{m-\nu}$ as
usual.

The elementary but crucial fact used for the construction of DDF operators
is that 0-modes of tensor operators of
weight $h=1$ commute with all $L_m$ generators
according to
\begin{eqnarray}
  \label{tensor law}
  \commutator{L_m}{{\cal O}_n} &=& \left((h-1)m-n\right) {\cal O}_{m+n}
  \,.
\end{eqnarray}

Therefore the task of finding DDF states is reduced to that of finding 
linearly independent $h=1$  fields that have the desired commutation relations 
and, in the case of the superstring, are closed with respect to $T_\mathrm{F}$  
(see below).

\subparagraph{Bosonic string.}

For the bosonic string the DDF operators $A^\mu_n$ are defined by
\begin{eqnarray}
  \label{bosonic string DDF operators}
  A^\mu_n
  &\propto&
  \oint \frac{dz}{2\pi i}
  \left(
    \partial X^\mu  
    + k^\mu\frac{\alpha^\prime}{8}in \partial \ln \left(k\inner \partial X\right) 
  \right)e^{in k\inner X}\of{z}
  \,.
\end{eqnarray}
(These are of course nothing but integrated vertex operators
of the massless fields.
Note that the logarithmic terms of $k \inner \partial X$, as well as the inverse powers
that will be used further below,
are well defined operators, as is discussed above equation (2.3.87) in \cite{GreenSchwarzWitten:1987}.)\\

It is straightforward to check that the operators \refer{bosonic string DDF operators} 
are really invariant:

First consider the transverse DDF operators.
For $v$ a transverse target space vector (such that in particular $v\inner k $ = 0 ) 
the operator $v\inner A_n$ is 
manifestly the 0-mode of an $h=1$ primary field 
(the exponential factor has $h=0$ due to $k\inner k = 0$) and hence is
invariant.

Furthermore 
$k \inner A_n \propto \delta_{n,0} k\inner \oint \partial X$ 
(for $n\neq 0$ the integrand is a total derivative)
also obvioulsy commutes with
the $L_m$. 

The only subtlety arises for the longitudinal $l\inner A_n$. Here,
the non-tensor behaviour of
\begin{eqnarray}
  T\of{z}  l\inner \partial X e^{in k\inner X}\of{w}
  &\sim&
  -\frac{\alpha^\prime}{2}\frac{in}{(z-w)^3}e^{in k\inner X}\of{w}
  +
  \mbox{$(h=1)$-tensor law}
\end{eqnarray}
is precisely canceled by the curious logarithmic correction term
$
  \partial \ln \left(k\inner \partial X\right) \of{z}
  =
  \frac{k\inner \partial^2 X}{k\inner \partial X}\of{z}
$.
Namely because of
\begin{eqnarray}
  T\of{z} \partial^2 X^\mu\of{w}
  &\sim&
  \frac{2 \partial X^\mu\of{w}}{(z-w)^3}
  +
  \mbox{$(h=2)$-tensor law}
\end{eqnarray}
one has
\begin{eqnarray}
  \Rightarrow
  T\of{z}
  \,
  \frac{k\inner \partial^2 X}{k\inner \partial X}e^{in k\inner X}\of{w}
  &\sim&
  \frac{2 e^{in k\inner X}}{(z-w)^3}
  +
  \mbox{$(h=1)$-tensor law}
  \,,
\end{eqnarray}
which hence makes the entire integrand of $l\inner A_m$ transform as an $h=1$  
primary, as desired.

\subparagraph{Superstring}
The analogous construction for the superstring has to ensure in addition that the DDF operators
commute with the supercharges $G_{m-\nu}$. 
This is simply achieved by `closing' the integral over 
a given weight $h=1/2$ primary field $D\of{z}$ to obtain the operator
\begin{eqnarray}
  \superCommutator{G_{-\nu}}{D_{\nu}}
  &=&
  \commutator{\oint \frac{dz}{2\pi i}T_F\of{z}}{\oint \frac{dz}{2\pi i}D\of{z}}
  \hspace{1cm}
  \left\lbrace
    \begin{array}{ll}
      \nu = 0 & \mbox{R sector}\\
      \nu = 1/2 & \mbox{NS sector}
    \end{array}
  \right.
  \,.
\end{eqnarray}
Here and in the remainder of this subsection the brackets denote supercommutators.

The resulting operator is manifestly the zero mode of a weight $h=1$ tensor and hence
commutes with all $L_n$. Furthermore it commutes with $G_{-\nu}$ because of
\begin{eqnarray}
  \superCommutator{G_{-\nu}}{\superCommutator{G_{-\nu}}{D_\nu}} 
  &=& 
  \superCommutator{L_{-2\nu}}{D_\nu}
  \;\stackrel{\refer{tensor law}}{=}\;
  0
  \,.
\end{eqnarray}
Since
$G_{-\nu}$ and $L_m, \forall\, m$ generate the entire algebra, the `closed' operator
$\commutator{G_{-\nu}}{D_\nu}$ indeed
commutes with all $L_m$ and $G_{m-\nu}\,,\forall\,m$.

It is therefore clear that the superstring DDF operators, which can be defined as
\begin{eqnarray}
  A_n^\mu
  &\defas&
  \commutator
  {G_\nu}
  {
     \oint \frac{dz}{2\pi i}
     \psi^\mu e^{i n k\inner X}\of{z} 
  }
  \nonumber\\
  B_n^\mu 
  &\defas&
  \commutator
  {G_\nu}
  {
    \oint \frac{dz}{2\pi i}
    \left(
      \psi^\mu \,k\inner \psi 
      -
      \frac{1}{4}
      k^\mu
      \partial \ln \left(k\inner \partial X\right)
    \right)
    \frac{e^{in k \inner X}}{\sqrt{k\inner \partial X}}
  }
\end{eqnarray}
commute with  the super-Virasoro generators, since the second arguments of the commutators
are integrals over weight 1/2 tensors. (And of course the latter are nothing but the 
integrated vertex operators
as they appear in the (-1) superghost picture). The nature and purpose of the logarithmic correction
term in the second line is just as discussed for the bosonic theory above: It cancels
the non-tensor term in
\begin{eqnarray}
  T\of{z}
  \,
  l\inner \psi\, k\inner\psi \frac{e^{i k\inner  X}}{\sqrt{k\inner \partial X}}\of{w}
  &\sim&
  \frac{1}{(z-w)^3}\frac{e^{i n k\inner X}}{\sqrt{k\inner \partial X}}
  +
  \mbox{$(h=1/2)$-tensor law}
  \,.
\end{eqnarray}

Evaluating the above supercommutators yields the explicit form for $A^\mu_n$ and $B^\mu_n$:
\begin{eqnarray}
  A^\mu_n &=&
  i \sqrt{\frac{2}{\alpha^\prime}}
  \oint \frac{dz}{2\pi i}
  \left(
    \partial X^\mu
    + 
    \frac{\alpha^\prime}{2}in \psi^\mu \, k\inner \psi
  \right)
  e^{ink\inner X}
  \of{z}
  \nonumber\\
  B^\mu_n
  &=&
  i \sqrt{\frac{2}{\alpha^\prime}}
  \oint \frac{dz}{2\pi i}
  \left(
    \partial X^\mu \, k\inner \psi
    -
    \psi^\mu k\inner \partial X
    +
    \frac{\alpha^\prime}{4}
    \psi^\mu \, k\inner \psi\, k\inner \partial \psi
    \frac{1}{k\inner \partial X}
  \right)
  \frac{e^{ink\inner X}}{\sqrt{k\inner \partial X}}
  \of{z}
  \nonumber\\
  &&+
  i \sqrt{\frac{2}{\alpha^\prime}}
  \oint \frac{dz}{2\pi i}
  k^\mu
  \left(
     k\inner \psi f_1\of{k\inner X,k\inner \partial X}
     +
     k\inner \partial \psi f_2\of{k\inner X, k\inner \partial X}
  \right)
  \of{z}
  \,,
\end{eqnarray}
where $f_1$ and $f_2$ are functions which we don not need to write out here.\\

The above discussion has focused on only a single chirality sector (left-moving, say).
It must be noted that the exponent $in k\inner X$ involved in the definition of
all the above DDF operators contains the 0-mode $k\inner x$ of the coordinate
field $k\inner X$. The existence of this 0-mode implies that the above DDF operators do
\emph{not} commute with the (super-)Virasoro generators of the opposite chirality.
In order to account for that one has to suitably multiply left- and right-moving DDF operators.
The details of this will be discussed in \S\fullref{classical bosonic DDF}.

\paragraph{Pohlmeyer invariants.}
\label{Pohlmeyer invariants}

We now turn to the classical bosonic string and discuss the
invartiants which have been studied by Pohlmeyer et al. 

In the literature the invariance of the Pohlmeyer charges is demonstrated
by the method of \emph{Lax pairs}. But the same fact follows already from the 
well-known reparameterization invariance property of Wilson loops.
To recall how this works for the classical bosonic string consider the following:

Denote the left- or rightmoving classical worldsheet fields in canonical language by
$\mathcal{P}^\mu\of{\sigma}$, which have the canonical Poisson bracket
\begin{eqnarray}
  \label{chiral CCR}
  \commutator{
    \mathcal{P}^\mu\of{\sigma}
  }
  {
    \mathcal{P}^\nu\of{\sigma^\prime}
  }_{\mathrm{PB}}
  &=&
  -
  \eta^{\mu\nu}
  \delta^\prime\of{\sigma-\sigma^\prime}
  \,.
\end{eqnarray}
The modes of the Virasoro constraints are
\begin{eqnarray}
  L_m &\defas&
  \frac{1}{2}
  \int d\sigma\,
  e^{-im\sigma}
  \eta_{\mu\nu}
  \mathcal{P}^\mu\of{\sigma}\mathcal{P}^\nu\of{\sigma}
\end{eqnarray}
and the $\mathcal{P}\of{\sigma}$ transform with unit weight under
their Poisson action:
\begin{eqnarray}
  \label{classical [Lm,Y]}
  \commutator{L_m}{\mathcal{P}^\mu\of{\sigma}}_{\mathrm{PB}}
  &=&
  \left(
    e^{-im\sigma}
    \mathcal{P}^\mu\of{\sigma}
  \right)^\prime
  \,.
\end{eqnarray}

This is all one needs to show that the \emph{Pohlmeyer invariants} 
$Z^{\mu_1 \cdots \mu_N}$ defined by
\begin{eqnarray}
  \label{definition Pohlmeyer invariants}
  Z^{\mu_1\cdots \mu_N}\of{\mathcal{P}}
  &:=&
  \frac{1}{N}
  \int\limits_0^{2\pi}
  d\sigma^1\,
  \int\limits_{\sigma^1}^{\sigma^1 + 2\pi}
  d\sigma^2\,  
  \cdots
  \int\limits_{\sigma^{N-1}}^{\sigma^1 + 2\pi}
  d\sigma^N\,  
    \mathcal{P}^{\mu_1}\of{\sigma^1}
    \mathcal{P}^{\mu_2}\of{\sigma^2}
    \cdots
    \mathcal{P}^{\mu_N}\of{\sigma^N}
  \nonumber\\
\end{eqnarray}
Poisson-commute with all the $L_m$.

The \emph{proof} involves just a little combinatorics and algebra:

First note that if 
$F\of{\sigma^1,\sigma^2,\cdots, \sigma^N}$
is any function which is periodic with period $2\pi$ in each
of its $N$ arguments, the cyclically permuted path-ordered
integral over $F$ is equal to the integral used in 
\refer{definition Pohlmeyer invariants}
\begin{eqnarray}
  &&
  \left[
    \int\limits_{0 < \sigma^1 < \sigma^2 < \cdots < \sigma^N < 2\pi}
    \!\!\!\!\!\!\!\!\!\!\!\!\!\!\!\!\!\!\!\!\!\!d^N \sigma
    \;\;\;\;\;\;\;+ 
    \int\limits_{0 < \sigma^N < \sigma^1 < \cdots < \sigma^{N-1} < 2\pi}
    \!\!\!\!\!\!\!\!\!\!\!\!\!\!\!\!\!\!\!\!\!\!d^N \sigma
    \;\;\;\;\;\;\;+
    \int\limits_{0 < \sigma^{N-1} < \sigma^N < \cdots < \sigma^{N-2} < 2\pi}
    \!\!\!\!\!\!\!\!\!\!\!\!\!\!\!\!\!\!\!\!\!\!d^N \sigma
  \;\;\;\;\;\;\;\right]  
  F\of{\sigma_1,\sigma_2,\cdots, \sigma_N}
  \nonumber\\
  &=&
  \int\limits_0^{2\pi}
  d\sigma^1\,
  \int\limits_{\sigma^1}^{\sigma^1 + 2\pi}
  d\sigma^2\,  
  \cdots
  \int\limits_{\sigma^{N-1}}^{\sigma^1 + 2\pi}
  d\sigma^N\,  
    F\of{\sigma_1,\sigma_2,\cdots ,\sigma_N}
  \,.
\end{eqnarray}
(This follows by noting that while, for instance, $\sigma^1$ runs from $0$ to $2\pi$
all other $\sigma^i$ can be taken to run from $\sigma^1$ to $\sigma^1 + 2\pi$ while
remaining in the correct order.)

This shows that the Pohlmeyer observables \refer{definition Pohlmeyer invariants} 
are 
invariant under cyclic permutation of their indices. It can also
be used to write their variation as
\begin{eqnarray}
  \delta Z^{\mu_1 \cdots \mu_N}
  &=&
  \frac{1}{N}
  \int\limits_0^{2\pi}
  d\sigma^1\,
  \int\limits_{\sigma^1}^{\sigma^1 + 2\pi}
  d\sigma^2\,  
  \cdots
  \int\limits_{\sigma^{N-1}}^{\sigma^1 + 2\pi}
  d\sigma^N\,  
  \Bigg(
    \mathcal{P}^{\mu_1}\of{\sigma^1}
    \mathcal{P}^{\mu_2}\of{\sigma^2}
    \cdots
    \delta\mathcal{P}^{\mu_N}\of{\sigma^N}  
   \nonumber\\
   &&
   \skiph{$
  \frac{1}{N}
  \int\limits_0^{2\pi}
  d\sigma^1\,
  \int\limits_{\sigma^1}^{\sigma^1 + 2\pi}
  d\sigma^2\,  
  \cdots
  \int\limits_{\sigma^{N-1}}^{\sigma^1 + 2\pi}
  d\sigma^N\,  
  \Bigg(\;
$}
  +
    \mathcal{P}^{\mu_N}\of{\sigma^1}
    \mathcal{P}^{\mu_1}\of{\sigma^2}
    \cdots
    \delta\mathcal{P}^{\mu_{N-1}}\of{\sigma^N}     
   \nonumber\\
   &&
   \skiph{$
  \frac{1}{N}
  \int\limits_0^{2\pi}
  d\sigma^1\,
  \int\limits_{\sigma^1}^{\sigma^1 + 2\pi}
  d\sigma^2\,  
  \cdots
  \int\limits_{\sigma^{N-1}}^{\sigma^1 + 2\pi}
  d\sigma^N\,  
  \Bigg(\;
$}
  + \cdots
   \Bigg)
  \,,
\end{eqnarray}
because we may cyclically permute the integration variables.
But if one now sets $\delta \mathcal{P}^\mu\of{\sigma} = 
\commutator{L_m}{\mathcal{P}^\mu\of{\sigma}}_\mathrm{PB}$
one gets, using \refer{classical [Lm,Y]},
\begin{eqnarray}
  &&
  \!\!\!\!\!\!\!\!\!\delta Z^{\mu_1 \cdots \mu_N} = 
  \nonumber\\
  &&
  \!\!\!\!\!\!\!
  \frac{1}{N}
  \int\limits_0^{2\pi}
  d\sigma^1\,
  \int\limits_{\sigma^1}^{\sigma^1 + 2\pi}
  d\sigma^2\,  
  \cdots
  \int\limits_{\sigma^{N-2}}^{\sigma^1 + 2\pi}
  d\sigma^{N-1}\,  
  \Bigg(
  \nonumber\\
  &&
    \xi\mathcal{P}^{\mu_N}
    \mathcal{P}^{\mu_1}\of{\sigma^1}
    \cdots
    \mathcal{P}^{\mu_{N-1}}\of{\sigma^{N-1}}
    -
    \mathcal{P}^{\mu_1}\of{\sigma^1}
    \cdots
    \xi
    \mathcal{P}^{\mu_{N-1}}
    \mathcal{P}^{\mu_N}\of{\sigma^{N-1}}
   \nonumber\\
   &&
  +
    \xi
    \mathcal{P}^{\mu_{N-1}}
    \mathcal{P}^{\mu_N}\of{\sigma^1}
    \cdots
    \mathcal{P}^{\mu_{N-2}}\of{\sigma^{N-1}}
    -
    \mathcal{P}^{\mu_N}\of{\sigma^1}
    \cdots
    \xi
    \mathcal{P}^{\mu_{N-2}}
    \mathcal{P}^{\mu_{N-1}}\of{\sigma^{N-1}}
   \nonumber\\
   &&
   \!\!\!\!\!\!\! 
   +
   \cdots
   \Bigg)
  \nonumber\\
  && 
  \!\!\!\!\!\!\!\!\! = 0
\end{eqnarray}
(where we have written $\xi\of{\sigma} = e^{-im\sigma}$ for brevity).
The contributions from the innermost integration cancel due to the cyclic
permutation of integrands and integration variables.
\endofproof

We note that the identity 
$\commutator{L_m}{Z^{\mu_1 \cdots \mu_N}\of{\mathcal{P}}}_\mathrm{PB} = 0$
is just the infinitesimal version of the fact that the Pohlmeyer observables
are invariant under \emph{finite reparameterizations}
\begin{eqnarray}
  \label{reparameterized cal P}
  \mathcal{P}\of{\sigma}
  &\mapsto&
  {\tilde {\mathcal{P}}}\of{\sigma}
  \defas
  R^\prime\of{\sigma}
  \mathcal{P}\of{R\of{\sigma}}
\end{eqnarray}
induced by the \emph{invertible} function $R$ which is assumed to satisfy
\begin{eqnarray}
  \label{cyclic property of R}
  R\of{\sigma+2\pi} = R\of{\sigma} + 2\pi
  \,.
\end{eqnarray}

Indeed, we have the important relation
\begin{eqnarray}
  \label{finite invariance}
  Z^{\mu_1 \cdots \mu_N}\of{\mathcal{P}}
  &=&
  Z^{\mu_1 \cdots \mu_N}\of{{\tilde {\mathcal{P}}}}
  \,,
\end{eqnarray}
which is at the heart of our derivation in \S\fullref{expressing Pohlmeyer invariants in terms of DDF invariants} 
that the Pohlmeyer invariants can be expressed in terms of DDF invariants

The \emph{proof} of this involves just a simple change of variables in the 
integral:
\begin{eqnarray}
  \label{proof of finite invariance}
  &&Z^{\mu_1 \cdots \mu_N}\of{{\tilde {\mathcal{P}}}}
  \nonumber\\
  &=&
  \frac{1}{N}
  \int\limits_0^{2\pi}
  d\sigma^1\,
  \int\limits_{\sigma^1}^{\sigma^1 + 2\pi}
  d\sigma^2\,  
  \cdots
  \int\limits_{\sigma^{N-1}}^{\sigma^1 + 2\pi}
  d\sigma^N\,  
    R^\prime\of{\sigma^1}R^\prime\of{\sigma^2}\cdots R^\prime\of{\sigma^N}
    \mathcal{P}^{\mu_1}\of{R\of{\sigma^1}}
    \cdots
    \mathcal{P}^{\mu_N}\of{R\of{\sigma^N}}  
  \nonumber\\
  &\stackrel{\tilde \sigma^i \defas R\of{\sigma^i}}{=}&
  \frac{1}{N}
  \int\limits_{R\of{0}}^{R\of{2\pi}}
  d\tilde \sigma^1\,
  \int\limits_{R\of{\sigma^1}}^{R\of{\sigma^1 + 2\pi}}
  d\tilde \sigma^2\,  
  \cdots
  \int\limits_{R\of{\sigma^{N-1}}}^{R\of{\sigma^1 + 2\pi}}
  d\tilde \sigma^N\,  
    \mathcal{P}^{\mu_1}\of{\tilde \sigma^1}
    \mathcal{P}^{\mu_2}\of{\tilde \sigma^2}
    \cdots
    \mathcal{P}^{\mu_N}\of{\tilde \sigma^N}  
  \nonumber\\
  &\equalby{cyclic property of R}&
  \frac{1}{N}
  \int\limits_{R\of{0}}^{R\of{0}+2\pi}
  d\tilde \sigma^1\,
  \int\limits_{\tilde \sigma^1}^{\tilde \sigma^1 + 2\pi}
  d\tilde \sigma^2\,  
  \cdots
  \int\limits_{\tilde\sigma^{N-1}}^{\tilde \sigma^1 + 2\pi}
  d\tilde \sigma^N\,  
    \mathcal{P}^{\mu_1}\of{\tilde \sigma^1}
    \mathcal{P}^{\mu_2}\of{\tilde \sigma^2}
    \cdots
    \mathcal{P}^{\mu_N}\of{\tilde \sigma^N}
  \nonumber\\
  &=&  
  Z^{\mu_1 \cdots \mu_N}\of{\mathcal{P}}
  \,.
\end{eqnarray}
\endofproof

\begin{eqnarray}
  \commutator{\mathcal{P}^\nu\of{\sigma}}{Z^{\mu_1 \cdots \mu_N}}_\mathrm{PB}
  &=&
  \frac{N-2}{N}
  \left(
  \mathcal{P}^{\mu_1}\of{\sigma} Z^{\mu_2 \cdots \mu_{N-1}}\eta^{\nu \mu_N}
  -
  Z^{\mu_1\cdots \mu_{N-2}} \mathcal{P}^{\mu_{N-1}}\of{\sigma} \eta^{\nu \mu_N}
  \right)
  + \mathrm{cycl.}
  \nonumber\\
\end{eqnarray}

(all indices transverse)
\begin{eqnarray}
  \commutator{A_m^\nu}{Z^{\mu_1 \cdots \mu_N}}_\mathrm{PB}
  &=&
  \frac{N-2}{N}
  \left(
  A_m^{\mu_1} Z^{\mu_2 \cdots \mu_{N-1}}\eta^{\nu \mu_N}
  -
  Z^{\mu_1\cdots \mu_{N-2}} A_m^{\mu_{N-1}} \eta^{\nu \mu_N}
  \right)
  + \mathrm{cycl.}
\end{eqnarray}

Finally, for the sake of completeness, 
we note the well-known fact that the Pohlmeyer invariants appear naturally as the 
Taylor-coefficients of
\emph{Wilson loops} along the string at constant worldsheet time.
Let $A_\mu$ be a  \emph{constant but otherwise arbitrary} $\mathrm{GL}\of{N,\C}$ 
connection on target space, then the Wilson loop around the string of this connection with respect
to $\mathcal{P}$ is
\begin{eqnarray}
  {\rm Tr}\,
  \mathbf{P}
  \exp\of{
    \int\limits_0^{2\pi}
    d\sigma\,
    A_\mu \mathcal{P}^\mu\of{\sigma}
  }
  &=&
  \sum\limits_{n=0}^\infty
  Z^{\mu_1 \cdots \mu_n}\of{\mathcal{P}}
  \,
  \mathrm{Tr}
  \of{A_{\mu_1}\cdots A_{\mu_2}}
  \,,
\end{eqnarray}
where $\mathbf{P}$ denotes path-ordering.

This way of getting string ``states'' by means of Wilson lines
of constant (but possibly large $N$) gauge connections 
along the string is intriguingly reminiscent of similar constructions 
used in the
IIB Matrix Model (IKKT model) \cite{AokiIsoKawaiYoshihisaKitazawaTsuchiyaTada:1999}.\\

In the next sections 
the classical DDF invariants are described and it is shown how the Pohlmeyer
invariants can be expressed in terms of these.

\subsubsection{Classical bosonic DDF invariants and their relation to the Pohlmeyer invariants}
\label{classical bosonic DDF invariants}

The construction of classical DDF-like invariants for the
\emph{super}string, which is the content of
\S\fullref{DDF and Pohlmeyer invariants for superstring},
is straightforward once the bosonic case is understood. 
The basic idea is very simple and shall 
therefore be given here first for
the bosonic string, in order to demonstrate how 
\S\fullref{DDF operators} and \S\fullref{Pohlmeyer invariants} 
fit together.

\paragraph{Classical bosonic DDF invariants.}
\label{classical bosonic DDF}

In order to establish our notation and sign conventions we 
briefly list some definitions and relations which are in principle 
well known from elementary CFT but are rarely written out in the 
canonical language which we will need here.

So let $X\of{\sigma}$ and $P\of{\sigma}$ be canonical coordinates and
momenta of the bosonic string with Poisson brackets
\begin{eqnarray}
  \commutator{X^\mu\of{\sigma}}{P_\nu\of{\kappa}}_\mathrm{PB}
  &=&
  \delta^\mu_\nu\, \delta\of{\sigma-\kappa}
  \,.
\end{eqnarray}

In close analogy to the CFT notation $\partial X$ and $\bar \partial X$ 
we define
\begin{eqnarray}
  \mathcal{P}_\pm^\mu(\sigma)
  &=&
  \frac{1}{\sqrt{2T}}\left(
    P^\mu\of{\sigma} \pm T X^{\prime \mu}\of{\sigma}
  \right)
  \,.
\end{eqnarray}
(Here $T = 1/2\pi \alpha^\prime$ is the string tension and we assume a trivial
Minkowski background and shift all spacetime indices with 
$\eta_{\mu\nu} = \mathrm{diag}(-1,1,\cdots,1)$.)

Their Poisson brackets are of course
\begin{eqnarray}
  \commutator{\mathcal{P}_\pm^\mu\of{\sigma}}{\mathcal{P}_\pm^\nu\of{\kappa}}_\mathrm{PB}
  &=&
  \pm \eta^{\mu\nu}\delta^\prime\of{\sigma-\kappa}
  \nonumber\\
  \commutator{\mathcal{P}_\pm^\mu\of{\sigma}}{\mathcal{P}_\mp^\nu\of{\kappa}}_\mathrm{PB}
  &=&
  0
  \,.
\end{eqnarray}
From the mode expansion
\begin{eqnarray}
  \label{oscillator expansion}
  \mathcal{P}^\mu_+\of{\sigma}
  &\defas&
  \frac{1}{\sqrt{2\pi}}\sum_m \tilde \alpha_m^\mu e^{-im\sigma}
  \nonumber\\
  \mathcal{P}^\mu_-\of{\sigma}
  &\defas&
  \frac{1}{\sqrt{2\pi}}\sum_m \alpha_m^\mu e^{+im\sigma}
\end{eqnarray}
one finds the string oscillator Poisson algebra
\begin{eqnarray}
  \commutator{\alpha_m^\mu}{\alpha_n^\nu}_\mathrm{PB}
  &=&
  -i\, m\, \eta^{\mu\nu}\delta_{m+n,0}
  \,,
\end{eqnarray}
as well as
\begin{eqnarray}
  \commutator{x^\mu}{p^\nu}_\mathrm{PB} 
  &=&
  \eta^{\mu\nu}
  \,,
\end{eqnarray}
where
\begin{eqnarray}
  x^\mu &\defas& \frac{1}{2\pi}\int X^\mu\of{\sigma}\,d\sigma
  \nonumber\\
  p^\mu &\defas& \int P^\mu\of{\sigma}\, d\sigma \;=\; \frac{1}{\sqrt{4\pi T}}\alpha_0 
    = \frac{1}{\sqrt{4\pi T}}\tilde \alpha_0
  \,.
\end{eqnarray}

In terms of these oscillators the field $X^\prime$ reads
\begin{eqnarray}
  X^{\prime\mu}\of{\sigma}
  &=&
  \frac{1}{\sqrt{2T}}
  \left(
    \mathcal{P}_+^\mu\of{\sigma}
    -
    \mathcal{P}_-^\mu\of{\sigma}
  \right)
  \nonumber\\
  &=&
  \frac{1}{\sqrt{4\pi T}}
  \sum\limits_{m=-\infty}^\infty
  \left(
    -\alpha^\mu_m + \tilde \alpha_{-m}^\mu
  \right)
  e^{+im\sigma}
\end{eqnarray}
and hence the canonical coordinate field itself is
\begin{eqnarray}
  X^\mu\of{\sigma}
  &=&
  x^\mu +
  \frac{i}{\sqrt{4\pi T}}
  \sum\limits_{m\neq 0}
  \frac{1}{m}
  \left(
    \alpha_m^\mu - \tilde \alpha_{-m}^\mu
  \right)
  e^{+im\sigma}
  \,.
\end{eqnarray}

Any field $A\of{\sigma}$ is said to have \emph{classical conformal weight} $w\of{A}$ iff
\begin{eqnarray}
  \commutator{L_m}{A\of{\sigma}}
  &=&
  e^{-im\sigma}A^\prime\of{\sigma}
  +
  w\of{A}(e^{-im\sigma})^\prime A\of{\sigma}
\end{eqnarray}
and is said to have classical conformal weight $\tilde w\of{A}$ iff
\begin{eqnarray}
  \commutator{\tilde L_m}{A\of{\sigma}}
  &=&
  -e^{+im\sigma}A^\prime\of{\sigma}
  -
  \tilde w\of{A}(e^{+im\sigma})^\prime A\of{\sigma}
  \,,
\end{eqnarray}
where 
\begin{eqnarray}
  L_m &\defas&
  \frac{1}{2}\int e^{-im\sigma} \mathcal{P}_-\of{\sigma} \inner \mathcal{P}_-\of{\sigma}
  \;=\;
  \frac{1}{2}
  \sum\limits_{k=-\infty}^\infty
  \alpha_{m-k}\inner \alpha_k
  \nonumber\\
  \tilde L_m &\defas&
  \frac{1}{2}\int e^{+im\sigma} \mathcal{P}_+\of{\sigma} \inner \mathcal{P}_+\of{\sigma}  
  \;=\;
  \frac{1}{2}
  \sum\limits_{k=-\infty}^\infty
  \tilde \alpha_{m-k}\inner \tilde \alpha_k  
\end{eqnarray}
are the usual modes of the Virasoro generators.

The parts of $X\of{\sigma}$ which have $w = 0$ and $\tilde w = 0$, respectively, are
\begin{eqnarray}
  X^\mu_-\of{\sigma}
  &\defas&
  x^\mu
  -
  \frac{\sigma}{4\pi T}p^\mu
  +
  \frac{i}{\sqrt{4\pi T}}\sum_{m\neq 0} \frac{1}{m}\alpha_m^\mu e^{+im\sigma}
\end{eqnarray}
and
\begin{eqnarray}
  X^\mu_+\of{\sigma}
  &\defas&
  x^\mu
  +
  \frac{\sigma}{4\pi T}p^\mu
  +
  \frac{i}{\sqrt{4\pi T}}\sum_{m\neq 0} \frac{1}{m}\tilde \alpha_m^\mu e^{-im\sigma}
  \,.
\end{eqnarray}
This is checked by noticing the crucial property
\begin{eqnarray}
  \left(X^{\mu}_-\right)^\prime\of{\sigma}
  &=&
  -\frac{1}{\sqrt{2T}}
  \mathcal{P}^\mu_-\of{\sigma}  
  \nonumber\\
  \left(X^{\mu}_+\right)^\prime\of{\sigma}
  &=&
  \frac{1}{\sqrt{2T}}
  \mathcal{P}^\mu_+\of{\sigma}
  \,.  
\end{eqnarray}

These weight 0 fields can now be used to construct ``invariant oscillators'', namely the 
classical DDF invariants:

To that end fix any lightlike vector field $k$ on target space and consider the fields
\begin{eqnarray}
  R_\pm\of{\sigma}
  &\defas&
  \pm
  \frac{4\pi T}{k\inner p}\,
    k\inner X_\pm\of{\sigma}
  \,.
\end{eqnarray}
The prefactor is an invariant and chosen so that
\begin{eqnarray}
  R_\pm\of{\sigma + 2\pi}
  &=&
  R_\pm\of{\sigma} + 2\pi
  \,.
\end{eqnarray}
Furthermore the derivative of $R_\pm$ is
\begin{eqnarray}
  R^\prime_\pm\of{\sigma}
  &=&
  \frac{2\pi\sqrt{2T}}{k\inner p}\, k\inner \mathcal{P}_\pm\of{\sigma}
  \,.
\end{eqnarray}
It has been observed \cite{Rehren:2004} that this derivative vanishes only on a
subset of phase space of vanishing measure. This can be seen as follows:

The classical Virasoro constraints $\mathcal{P}_\pm^2 = 0$ say that $\mathcal{P}_\pm\of{\sigma}$ 
is lightlike. Because $k$ is also lightlike this implies that $k\inner \mathcal{P}_\pm\of{\sigma}$
vanishes iff $\mathcal{P}_\pm\of{\sigma}$ is parallel to $k$. 

By writing $\mathcal{P}_\pm = \mathcal{P}_\pm^0\left[1,\mathcal{P}_\pm^i/\mathcal{P}_\pm^0\right]$
and noting that the spatial unit vector $e^i_\pm\of{\sigma} \defas
\mathcal{P}_\pm^i/\mathcal{P}_\pm^0$ is of weight $w = 0$ or $\tilde w = 0$
(while it Poisson commutes with the respective opposite Virasoro algebra), and
hence transforms under the action of the Virasoro generators (which includes time evolution)
as $e^i_\pm\of{\sigma} \to e^i_\pm\of{\sigma + f\of{\sigma}}$, one sees that this condition
is satisfied for some $\sigma$ at some instance of time if and only if it is satisfied for some $\sigma$
at any given time. In other words the time evolution of the string traces out trajectories
in phase space which either have $\mathcal{P}_\pm$ parallel to $k$ for some $\sigma$ 
at \emph{all times} or \emph{never}.

In summary this means that except on the subset of phase space (of vanishing measure) 
of those trajectories where there exists a $\sigma$ such that
$k\inner \mathcal{P}_\pm\of{\sigma} = 0$, the observables $R_\pm\of{\sigma}$ define invertible
reparameterizations of the interval $[0,2\pi)$, as considered in \refer{cyclic property of R}.

The above fact will be crucial below for expressing the Pohlmeyer invariants in terms of DDF invariants.
For later usage 
let us introduce the notation $\mathbf{P}_k$ for the total phase space minus that set of 
vanishing measure:
\begin{eqnarray}
  \label{nice part of phase space}
  \mathbf{P}_k
  &\defas&
  \set{
    (X\of{\sigma},P\of{\sigma})_{\sigma \in (0,2\pi)}
    |
    k\inner\mathcal{P}_\pm\of{\sigma} \neq 0\;\forall \sigma
  }
  \,.
\end{eqnarray}

Now the classical DDF observables $A_m^\mu$ and $\tilde A_m^\mu$ of the closed bosonic string are finally defined 
(adapting the construction of \refer{bosonic string DDF operators} but using slighly different
normalizations)
by
\begin{eqnarray}
  \label{definition classical DDF}
  A_m^\mu
  &\defas&
  \frac{1}{\sqrt{2\pi}}
  \int d \sigma\,
  \mathcal{P}_-^\mu\of{\sigma}
  e^{
    -im R_-\of{\sigma}
  }
  \nonumber\\
  \tilde A_m^\mu
  &\defas&
  \frac{1}{\sqrt{2\pi}}
  \int d \sigma\,
  \mathcal{P}_+^\mu\of{\sigma}
  e^{
    im R_+\of{\sigma}
  }
  \,.
\end{eqnarray} 
Note that the construction principle of these objects is 
essentially the same as that of the ordinary oscillators \refer{oscillator expansion}
except that the parameterization of the string used here differs from one point in phase space to 
the other.

Being integrals over fields of total weight $w=1$ and $\tilde w = 1$, respectively, the 
DDF observables obviously Poisson-commute with their associated \emph{half} of the Virasoro generators:
\begin{eqnarray}
  \label{invariance pure}
  \commutator{L_m}{A_n^\mu} &=& 0
  \nonumber\\
  \commutator{\tilde L_m}{\tilde A_n^\mu} &=& 0
  \,.
\end{eqnarray}

But due to the coordinate 0-mode 
$\frac{2T}{k\inner p}\,k\inner x$
that enters the definition of
$R_\pm$, the mixed Poisson-brackets do not vanish. In order to construct invariants one therefore
has to split off this 0-mode and define the truncated observables
\begin{eqnarray}
  a^\mu_m &\defas& A^\mu_m e^{-im \frac{2T}{k\inner p}\,k\inner x}
  \nonumber\\
  \tilde a^\mu_m &\defas& A^\mu_m e^{-im \frac{2T}{k\inner p}\,k\inner x}
  \,.
\end{eqnarray}
These now obviously have vanishing \emph{mixed} Poisson brackets:
\begin{eqnarray}
  \label{invariance mixed}
  \commutator{L_m}{\tilde a_n^\mu} &=& 0
  \nonumber\\
  \commutator{\tilde L_m}{a_n^\mu} &=& 0
  \,.  
\end{eqnarray}

Therefore classical DDF invariants which Poisson commute with \emph{all}
Virasoro constraints are obtained by forming products 
\begin{eqnarray}
  \label{DDF invariants}
  D_{\set{m_i,\tilde m_j}} 
  &\defas& 
  a^{\mu_1}_{m_1}\cdots a^{\mu_r}_{m_r}\,\tilde a^{\nu_1}_{\tilde n_1}\cdots a^{\nu_s}_{\tilde m_s}
  e^{i N \frac{2T}{k\inner p}\, k\inner x}
\end{eqnarray}
which satisfy the \emph{level matching condition}:
\begin{eqnarray}
  \label{level matching condition}
  \sum\limits_i m_i \;=\; N \;=\;  \sum\limits_j \tilde m_j
  \,.
\end{eqnarray}

In order to see this explicitly write
\begin{eqnarray}
  \commutator{L_n}{
  D_{\set{m_i,\tilde m_j}}
  }_\mathrm{PB}
  &=& 
  \underbrace{
  \commutator{L_n}
  {
  a^{\mu_1}_{m_1}\cdots a^{\mu_r}_{m_r}e^{i N \frac{2T}{k\inner p}\,k\inner x}
  }_\mathrm{PB}
  }_{\equalby{invariance pure} 0}
  \tilde a^{\nu_1}_{\tilde n_1}\cdots a^{\nu_s}_{\tilde m_s}
  +
  \nonumber\\
  &&+
  a^{\mu_1}_{m_1}\cdots a^{\mu_r}_{m_r}e^{i N \frac{2T}{k\inner p}\,k\inner x}
  \underbrace{
  \commutator{L_n}{
    \tilde a^{\nu_1}_{\tilde n_1}\cdots a^{\nu_s}_{\tilde m_s}
  }_\mathrm{PB}
  }_{\equalby{invariance mixed} 0}
  \nonumber\\
  \commutator{\tilde L_n}{
  D_{\set{m_i,\tilde m_j}}
  }_\mathrm{PB}
  &=& 
  \underbrace{
  \commutator{\tilde L_n}
  {
  a^{\mu_1}_{m_1}\cdots a^{\mu_r}_{m_r}
  }_\mathrm{PB}
  }_{\equalby{invariance mixed} 0}
  \tilde a^{\nu_1}_{\tilde n_1}\cdots a^{\nu_s}_{\tilde m_s}e^{i N \frac{2T}{k\inner p}\,k\inner x}
  +
  \nonumber\\
  &&+  a^{\mu_1}_{m_1}\cdots a^{\mu_r}_{m_r}
  \underbrace{
  \commutator{L_n}{
    \tilde a^{\nu_1}_{\tilde n_1}\cdots a^{\nu_s}_{\tilde m_s}e^{i N \frac{2T}{k\inner p}\,k\inner x}
  }_\mathrm{PB}
  }_{\equalby{invariance pure} 0}
  \,.
\end{eqnarray}

This establishes the classical invariance of the DDF observables $D_{\set{m_i,\tilde m_j}}$.
We next demonstrate how the Pohlmeyer invariants can be expressed in terms of DDF invariants.

\paragraph{Expressing Pohlmeyer invartiants in terms of DDF invariants.}
\label{expressing Pohlmeyer invariants in terms of DDF invariants}

From the Fourier mode-like objects $A_m^\mu$ and $\tilde A_m^\mu$ one reobtains
quasi-local fields\footnote{
  It is interesting to discuss these fields, and in particular their
  quantization, from the point of view of
  worldsheet (quantum) gravity:

    Clearly the $\mathcal{P}_\pm^\mu\of{\sigma}$ are `not physical'
  (do not Poisson commute with the constraints) because they evaluate the
string's momentum and tension energy at a given value of the parameter $\sigma$, which 
of course has no physical relevance whatsoever. Heuristically, a physical observable
may make recourse only to values of fields of the theory, not to values of
auxiliary unphysical parameters. That is precisely the role played by the fields $R_\pm$.
They allow to characterize a point of the string purely in terms of physical
fields (string oscillations). Instead of asking: ``What is the value of $\mathcal{P}_\pm$
at $\sigma = 3$?'', we may ask the physically meaningful question:
``What is the value of $\mathcal{P}_\pm$ at a
point on the string where its configuration is such that $R_+ = 3$?''	
Quasi-local observables like the $\mathcal{P}^R_\pm$, or rather their absence,
are related to old and well known issues of (quantum) gravity in higher dimensions, 
often referred to in the context of ``\emph{the problem of time}'' \protect\cite{Carlip:2001}.

It is maybe instructive to note how these issues are resolved here for the 
\emph{worldsheet} theory of the relativistic string,
a toy example for quantum gravity when regarded as a theory of 1+1 dimensional gravity.
(Of course the string is rather more than a toy example for quantum gravity from the
\emph{target space} perspective.)
} $\mathcal{P}^R_\pm$ by an inverse Fourier transformation:
\begin{eqnarray}
  \label{local pre-invariants}
  \mathcal{P}^R_-\of{\sigma}
  &\defas&
  \frac{1}{\sqrt{2\pi}}
  \sum\limits_m A_m^\mu e^{+im\sigma}
  \;=\;
  \left((R_-)^{-1}\right)^\prime\of{\sigma}
  \mathcal{P}^\mu\of{(R_-)^{-1}\of{\sigma}}
  \nonumber\\
  \mathcal{P}^R_+\of{\sigma}
  &\defas&
  \frac{1}{\sqrt{2\pi}}
  \sum\limits_m \tilde A_m^\mu e^{-im\sigma}
  \;=\;
  \left((R_+)^{-1}\right)^\prime\of{\sigma}
  \mathcal{P}^\mu\of{(R_+)^{-1}\of{\sigma}}
  \,.
\end{eqnarray}

This holds true on $\mathbf{P}_k$ \refer{nice part of phase space} where we can use the fact
that $R_\pm$ are invertible.

Comparison with \refer{reparameterized cal P} shows that these are just 
reparameterizations
of the original local worldsheet fields $\mathcal{P}_\pm^\mu$, albeit with
a reparameterization that varies from phase space point to phase space point,
which is crucial for their invariance. But because the proof
\refer{proof of finite invariance} of \refer{finite invariance} 
involves only data available at a
single point in phase space,
it follows that for \emph{every} invariant expression
$F\of{\mathcal{P}_-}$ of the worldsheet fields $\mathcal{P}_-^\mu$ with
$\commutator{L_m}{F\of{\mathcal{P}_-}} = 0\,,\forall m$ we have
\begin{eqnarray}
  F({\mathcal{P}_-}) &=& F({\mathcal{P}^R_-})
\end{eqnarray}
(on $\mathbf{P}_k$), and analogously for $\mathcal{P}_+$.

\emph{In summary} we therefore obtain the following result:

On the restricted phase space $\mathbf{P}_\mathrm{k}$
\refer{nice part of phase space}
the classical Pohlmeyer invariants \refer{definition Pohlmeyer invariants} 
can be expressed in terms of
the classical DDF invariants 
\refer{definition classical DDF}
and the relation is 
\begin{eqnarray}
  \label{relation Pohlmeyer-DDF}
  Z^{\mu_1\cdots \mu_N}\of{\mathcal{P}}
  &=&
  Z^{\mu_1\cdots \mu_N}(\mathcal{P}^R)
  \,,
\end{eqnarray}
where $\mathcal{P}$ is the ordinary worldsheet field \refer{chiral CCR},
and $\mathcal{P}^R$ is the linear combination 
\refer{local pre-invariants} of classical DDF observables.

This can be expressed in words also as follows: The Pohlmeyer invariants
are left intact when replacing oscillators by respective DDF observables
in their oscillator expansion ($\alpha_m^\mu \to A_m^\mu$
, $\tilde \alpha_m^\mu \to \tilde A_m^\mu$).
Note that the Pohlmeyer invariants are all of level 0 in the sense of
\refer{level matching condition} so that the level matching condition is
trivially satisfied.

Because every polynomial in the DDF observables 
is consistently quantized by replacing $A_m^\mu$ and $\tilde A_m^\mu$ by the
respective operators discussed in \S\fullref{DDF operators}, this 
yields a consistent quantization of the Pohlmeyer invariants.\\

Finally, by simply generalizing the DDF invariants to the superstring,
equation \refer{relation Pohlmeyer-DDF} defines the generalization of the
Pohlmeyer invariants to the superstring. This is discussed in the 
next subsection:

\paragraph{DDF and Pohlmeyer invariants for superstring.}
\label{DDF and Pohlmeyer invariants for superstring}

The additional fermionc fields on the classical superstring shall here be 
denoted by $\Gamma_\pm^\mu\of{\sigma}$, which are taken to be normalized
so that their fermionic Poisson bracket reads
\begin{eqnarray}
  \antiCommutator{\Gamma_\pm^\mu\of{\sigma}}{\Gamma^\nu_\pm\of{\kappa}}_\mathrm{PB}
  &=&
  \pm 2\eta^{\mu\nu}\delta\of{\sigma-\kappa}
  \nonumber\\
  \antiCommutator{\Gamma_\pm^\mu\of{\sigma}}{\Gamma^\nu_\mp\of{\kappa}}_\mathrm{PB}
  &=& 0
  \,.
\end{eqnarray}
The modes are of course
\begin{eqnarray}
  b_r^\mu 
  &\defas&
  \frac{i}{\sqrt{4\pi}}
  \int e^{-ir\sigma}
  \Gamma_-^\mu\of{\sigma}
  \nonumber\\
  \tilde b_r^\mu
  &=&
  \frac{1}{\sqrt{4\pi}}
  \int e^{+ir\sigma}
  \Gamma_+^\mu\of{\sigma}  
\end{eqnarray}
with non-vanishing brackets
\begin{eqnarray}
  \antiCommutator{b^\mu_r}{b^\nu_s}_\mathrm{PB}
  &=&
  -i
  \eta^{\mu\nu}
  \delta_{r+s,0}
  \nonumber\\
  \antiCommutator{\tilde b^\mu_r}{\tilde b^\nu_s}_\mathrm{PB}
  &=&
  -i
  \eta^{\mu\nu}
  \delta_{r+s,0}  
  \,,
\end{eqnarray}
and the fermionic part of the super Virasoro constraints are
\begin{eqnarray}
  G_r &\defas&
  \frac{i}{\sqrt{2}}
  \int e^{-ir\sigma}
  \Gamma_-\of{\sigma} \inner \mathcal{P}_-\of{\sigma}
  \;d\sigma
  \;=\;
  \sum\limits_{m=-\infty}^\infty
  b_{r+m}\inner \alpha_{-m}
  \nonumber\\
  \tilde G_r 
  &\defas&
  \frac{1}{\sqrt{2}}
  \int e^{+ir\sigma}
  \Gamma_+\of{\sigma} \inner \mathcal{P}_+\of{\sigma}
  \;d\sigma
  \;=\;
  \sum\limits_{m=-\infty}^\infty
  \tilde b_{r+m}\inner \tilde \alpha_{-m}
  \,.  
\end{eqnarray}
The point is that we can entirely follow the constructions discussed in \S\fullref{DDF operators} to
get classical DDF invariants $A_m^\mu$ and $B_m^\mu$ which Poisson-commute
with the full set of super Virasoro constraints. For instance in the R sector 
the DDF observable $A_m^\mu$ is
\begin{eqnarray}
  A_m^\mu
  &\defas&
  \antiCommutator{G_0}{\frac{i}{\sqrt{4\pi}}\int \Gamma_-^\mu\of{\sigma}e^{-im R_-\of{\sigma}}}
  \nonumber\\
  &=&
  \frac{1}{\sqrt{2\pi}}
  \int d\sigma\;
  \left(
    \mathcal{P}_-^\mu\of{\sigma}
    +
    im \frac{2\pi\sqrt{2T}}{k\inner p}\Gamma_-^\mu\of{\sigma} k\inner \Gamma_-\of{\sigma}
  \right)
  e^{-imR_-\of{\sigma}}
  \,.
\end{eqnarray}
By making the replacement $\alpha_m^\mu \to A_m^\mu$ in the ordinary Pohlmeyer
invariant $Z^{\mu_1\cdots \mu_N}\of{\mathcal{P}_-}$ one obtains an object
whose purely bosonic terms exactly coincide with the ordinary bosonic
Pohlmeyer invariant and which furthermore has fermionic terms such that it
super-Poisson-commutes with all super Virasoro constraints. This object
is therefore obviously the superstring generalization of the ordinary Pohlmeyer invariant
of the bosonic string.

\subsection{Boundary States for D-Branes with Nonabelian Gauge Fields}

In this subsection, which is taken from \cite{Schreiber:2004d},
we demonstrate a relation between two apparently unrelated 
aspects of superstrings: boundary states for
nonabelian gauge fields and (super-)Pohlmeyer invariants. 

On the one hand side superstring boundary states describing excitations of non-abelian
gauge fields on D-branes are still the subject of investigations 
\cite{MaedaNakatsuOonishi:2004,MurakaniNakatsu:2002,Schreiber:2004e} and
are of general interest for superstring theory, as they directly mediate between
string theory and gauge theory.

On the other hand, studies of string quantization focusing on non-standard 
worldsheet invariants, the so-called Pohlmeyer invariants, done in 
\cite{Pohlmeyer:2002,MeusburgerRehren:2002,Pohlmeyer:1998,Pohlmeyer:1988} and
recalled in \cite{Thiemann:2004}, were shown in \cite{Schreiber:2004b,Schreiber:2004c}
to be related to the standard quantization of the string by way of the well-known DDF invariants. 
This
raised the question whether the Pohlmeyer invariants are of any genuine interest in (super-)string theory
as commonly understood.

Here it shall be shown that the (super-)Pohlmeyer invariants do indeed play an interesting
role as boundary state deformation operators for non-abelian gauge fields, thus connecting
the above two topics and illuminating aspects of both them.\\

A boundary state is a state in the closed string's Hilbert space constructed in such a way
that inserting the vertex operator of that state in the path integral over the sphere reproduces
the disk amplitudes for certain boundary conditions (D-branes) of the open string. In accord with the
general fact that the worldsheet path integral insertions which describe background field excitations
are exponentiations of the corresponding vertex operators, it turns out that the boundary
states which describe gauge field excitations on the D-brane have the form
of (generalized) Wilson lines of the gauge field along the closed string 
\cite{Hashimoto:2000,Hashimoto:1999,MaedaNakatsuOonishi:2004,MurakaniNakatsu:2002,Schreiber:2004e}.

Long before these investigations, it was noted by Pohlmeyer \cite{Pohlmeyer:1988}, in the context of the classical 
string, that generalized Wilson lines along the closed string with respect to an auxiliary
gauge connection on spacetime provide a ``complete'' set of invariants of the theory, i.e.
a complete set of observables which (Poisson-)commute with all the Virasoro constraints.

Given these two developments it is natural to suspect that there is a relation between Pohlmeyer
invariants and boundary states. Just like the DDF invariants (introduced in \cite{DelGiudiceDiVecchiaFubini:1972}
and recently reviewed in \cite{Schreiber:2004b}), 
which are the more commonly considered
complete set of invariants of the string, commute with all the constraints and hence generate physical
states when acting on the worldsheet vacuum, a consistently quantized version of the
Pohlmeyer invariants should send boundary states of bare D-branes to those involving the
excitation of a gauge field.

Indeed, up to a certain condition on the gauge field, this turns out to be true and works as follows:\\

If $X^\mu(\sigma)$ and $P_\mu(\sigma)$ are the canonical coordinates and momenta of the
bosonic string, then
$\mathcal{P}_\pm^\mu\of{\sigma} \defas \frac{1}{\sqrt{2}T}(P_\mu\of{\sigma} \pm T \eta_{\mu\nu}X^{\prime\nu}\of{\sigma})$,
(where $T$ is the string's tension and a prime denotes the derivative with respect to $\sigma$)
are the left- and right-moving bosonic worldsheet fields for flat Minkowski background (in CFT context
denoted by $\partial X$ and $\bar \partial X$) and for any given constant gauge field $A$ on target space
the objects
\begin{eqnarray}
  \label{Pohlmeyer in introduction}
  W_\pm^{\mathcal{P}}[A]
  &\defas&
  \mathrm{Tr}\, \mathrm{P}
  \exp\of{\int_0^{2\pi}d\sigma\, A \inner \mathcal{P}_\pm\of{\sigma}}
\end{eqnarray}
(where $\mathrm{Tr}$ is the trace in the given representation of the 
gauge group's Lie algebra and $\mathrm{P}$ denotes path-ordering
along $\sigma$) Poisson-commute with all Virasoro constraints. In fact 
the coefficients of $\mathrm{Tr}(A^n)$ in these generalized Wilson lines do so seperately, and these
are usually addressed as the \emph{Pohlmeyer invariants}, even though we shall use this term 
for the full object \refer{Pohlmeyer in introduction}.

Fundamentally, the reason for this invariance is just the reparameterization invariance of the Wilson line,
which can be seen to imply that \refer{Pohlmeyer in introduction} remains unchanged under a substitution of
$\mathcal{P}$ with a reparameterized version of this field. In \cite{Schreiber:2004b} it was
observed that an interesting example for such a substitution is obtained by taking the ordinary DDF oscillators
\begin{eqnarray}
  \label{bosonic DDF in intro}
  A_m^\mu 
  &\propto&
  \int\limits_0^{2\pi}
  d\sigma\;
  \mathcal{P}_-^\mu\of{\sigma}e^{im\frac{4 \pi T}{k\inner p}\, k\inner X_-\of{\sigma}}
\end{eqnarray}
(where $k$ is a lightlike vector on target space,
$X_-$ is the left-moving component of $X$, $p$ is the center-of-mass momentum,  and an analogous expression
exists for $\mathcal{P}_+$)
and forming ``quasi-local'' invariants
\begin{eqnarray}
  \label{bosonic quasi local in intro}
 \mathcal{P}_-^{R\mu}\of{\sigma}
  &\defas&
  \frac{1}{\sqrt{2\pi}}
  \sum\limits_{m=0}^\infty
  A_m^\mu e^{im\sigma}
\end{eqnarray}
from them.\footnote{We dare to use the same symbol $A$ for the gauge field and for the DDF oscillators
in order to comply with established conventions. The DDF oscillators will always carry a mode index
$m$, however, and it should always be clear which object is meant.} 

One finds 
\begin{eqnarray}
  W^{\mathcal{P}}[A] &=& W^{\mathcal{P}^R}[A]
\end{eqnarray}
 and since the quantization of the $\mathcal{P}^R$
in terms of DDF oscillators is well known, this gives a consistent quantization of the Pohlmeyer invariants.
This is the quantization that we shall use here to study boundary states.

The above construction has a straightforward generalization to the superstring and this is the context
in which the relation between the Pohlmeyer invariants and boundary states turns out to have interesting aspects,
(while the bosonic case follows as a simple restriction, when all fermions are set to 0).

So we consider the supersymmetric extension of \refer{bosonic DDF in intro}, which, by 
convenient abuse of notation,
we shall also denote by $A_m^\mu$:
\begin{eqnarray}
  \label{susy DDF in intro}
  A_m^\mu
  &\propto&
  \int_0^{2\pi}
  d\sigma
  \left(
    \mathcal{P}_-\of{\sigma}
    +im \frac{\pi \sqrt{2T}}{k\inner p}
    k\inner\Gamma_-\of{\sigma}\, \Gamma_-^\mu\of{\sigma}
    \right)
    e^{im\frac{4 \pi T}{k\inner p}\, k\inner X_-\of{\sigma}}
  \,,
\end{eqnarray}
where $\Gamma_\pm\of{\sigma}$ denote the fermionic superpartners of $\mathcal{P}_\pm$. From
these we build again the objects \refer{bosonic quasi local in intro} and 
finally $W^{\mathcal{P}^R}[A]$, which we address as the \emph{super-Pohlmeyer} invariants. 

Being constructed from the supersymmetric invariants $\mathcal{P}^R$, which again are 
built from \refer{susy DDF in intro}, 
these manifestly commute with all of the super-Virasoro constraints. But in order to
relate them to boundary states they need to be re-expressed in terms of the plain objects
$\mathcal{P}$ and $\Gamma$. This turns out to be non-trivial and has some interesting aspects
to it. \\

After these peliminaries we can state 
the first result to be reported here, which is
\begin{enumerate}
\item
that on that subspace $\mathbf{P}_k$ of phase space where $k\inner X_-$
is invertible as a function of $\sigma$ (a condition that plays also a crucial role for the 
considerations of the bosonic DDF/Pohlmeyer relationship as discussed in \cite{Schreiber:2004b})
the super-Pohlmeyer invariants built from \refer{susy DDF in intro} are equal to 
\begin{eqnarray}
  \label{restricted super-Pohlmeyer invariant in intro}
  \left.
    W^{\mathcal{P}^R}[A]
  \right|_{\mathbf{P}_k}
  &=&
  \mathrm{Tr}\;
  \mathrm{P}
  \exp\of{
    \int_0^{2\pi}
    d\sigma\;
    \left(
      i A_\mu + 
      \commutator{A_\mu}{A_\nu}
      \frac{
k \inner \Gamma\;
      \Gamma^\nu
}{2k\inner \mathcal{P}}
    \right)
    \mathcal{P}^\mu
  }
  \,,
\end{eqnarray}

\item 
 that this expression extends to an invariant on all of phase space precisely if
the transversal components of $A$ mutually commute,

\item
  and that in this case the above is equal to
\begin{eqnarray}
  \label{nonlocal susy extension of Pohlmeyer intro}
  Y[A]
  &\defas&
  \mathrm{Tr}\,\mathrm{P}
  \exp\of{\int_0^{2\pi}d\sigma\;
    \left(
    i A_\mu \mathcal{P}^\mu
    +
    \frac{1}{4}\commutator{A_\mu}{A_\nu}
    \Gamma^\mu \Gamma^\nu
    \right)
  }
  \,.
\end{eqnarray}

\end{enumerate}

The second result concerns the application of the quantum version of these observables to 
the bare boundary state $\ket{\mathrm{D9}}$ of a space-filling D9-brane 
(see for instance appendix A of \cite{Schreiber:2004e} for a brief review of boundary state formalism and
further literature). Denoting by
$\mathcal{E}^\dagger\of{\sigma} = \frac{1}{2}\left(\Gamma_+\of{\sigma} + \Gamma_-\of{\sigma}\right)$
the differential forms on loop space (\cf section 2.3.1. of \cite{Schreiber:2004e} and
section 2.2 of \cite{Schreiber:2004} for the notation and nomenclature used here, and see
\cite{Schreiber:2004f} for a more general discussion of the loop space perspective)
we find
  that for the above case of commuting transversal $A$ the application of 
\refer{nonlocal susy extension of Pohlmeyer intro} to $\ket{\mathrm{D9}}$ yields
\begin{eqnarray}
  &&\mathrm{Tr}\, \mathrm{P}
  \exp\of{
    \int\limits_0^{2\pi}
     d\sigma\,
    \left(
      i A_\mu \mathcal{P}^\mu + \frac{1}{4}(F_A)_{\mu\nu}\Gamma^\mu\Gamma^\nu
    \right)
  }
  \ket{\mathrm{D9}}
  \nonumber\\
  &=&
  \mathrm{Tr}\, \mathrm{P}
  \exp\of{
    \int\limits_0^{2\pi}
     d\sigma\,
    \left(
      -i\sqrt{\frac{T}{2}} 
      A_\mu X^{\prime \mu} + \frac{1}{4}(F_A)_{\mu\nu}
      \mathcal{E}^{\dagger \mu}
      \mathcal{E}^{\dagger \nu}
    \right)
  }
  \ket{\mathrm{D9}}
  \,.
\end{eqnarray}
which is, on the right hand side, precisely the boundary state describing a non-abelian gauge field
on the D9 brane \cite{MaedaNakatsuOonishi:2004,Schreiber:2004e} (for comparison one should rescale
$A$ as discussed in \refer{rescaling of A} below).

In summary this shows that and under which conditions the application of a quantized super-Pohlmeyer
invariant to the boundary state of a bare D9 brane produces the boundary state describing a
non-abelian gauge field excitation. \\

The structure of this paper closely follows the above outline:

First of all \S\fullref{Super-Pohlmeyer invariants} 
is concerned with the classical super-Pohlmeyer invariants and
their expression in terms of local fields. Then \S\fullref{Super-Pohlmeyer invariants} 
discusses their cousins, the
invariants of the general form \refer{nonlocal susy extension of Pohlmeyer intro}.
Both are related in \S\fullref{Invariance of the extension of the restricted super-Pohlmeyer invariants}. 

Then the quantization of the super-Pohlmeyer invariants
is started in \S\fullref{Quantum super-Pohlmeyer invariants}. 
After an intermediate result concerning an operator ordering issue
is treated in \S\fullref{On an operator ordering issue in Wilson lines along the closed string} 
the quantum Pohlmeyer invariants are finally 
applied to the bare boundary state in \S\fullref{Super-Pohlmeyer and boundary states}.

\subsubsection{DDF operators, Pohlmeyer invariants and boundary states}

\paragraph{Super-Pohlmeyer invariants.}
\label{Super-Pohlmeyer invariants}

In \cite{Schreiber:2004b} it was shown how from the classical DDF oscillators of the bosonic string one can construct
\emph{quasilocal} fields $\mathcal{P}^R$, which (Poisson-)commute with all the constraints and
which, when used in place of $X^\prime$ in a Wilson line of a constant gauge field
along the string, reproduce the Pohlmeyer invariants. It was mentioned that using the DDF oscillators
of the superstring in this procedure leads to a generalization of the Pohlmeyer invariants
to the superstring. Here we will work out the explicit form of the \emph{super-Pohlmeyer invariants}
obtained this way and point out that they are interesting in their own right.

Using the notation of \cite{Schreiber:2004b} we denote by $\mathcal{P}^\mu(\sigma)$ the 
classical canonical left- or right-moving bosonic fields on the string, and by $\Gamma^\mu(\sigma)$
their fermionic partners, where the relation to the usual CFT notation is 
$\mathcal{P}^\mu \propto \partial X^\mu$
and $\Gamma^\mu \propto \psi^\mu$.

Our normalization is chosen such that the graded Poisson-brackets read
\begin{eqnarray}
  \label{classical CCM}
  &&\commutator{\Gamma^\mu\of{\sigma}}{\Gamma^\nu\of{\kappa}}
  =
  -2\eta^{\mu\nu}\delta\of{\sigma-\kappa}
  \nonumber\\
  &&
  \commutator{\mathcal{P}^\mu\of{\sigma}}{\mathcal{P}^\nu\of{\kappa}}
  =
  -\eta^{\mu\nu}\delta^\prime\of{\sigma-\kappa}
  \,.
\end{eqnarray}

The classical bosonic DDF oscillators $A_m^\mu$ of the superstring are obtained by acting with the supercharge
\begin{eqnarray}
  \label{the supercharge}
  G_0 
  &=&
  \frac{i}{\sqrt{2}}
  \int
  d\sigma\;
  \Gamma^\mu \mathcal{P}_\mu 
\end{eqnarray}
(we concentrate on the Ramond sector for notational simplicity) on integrals over weight 1/2 fields:
\begin{eqnarray}
  \label{bosonic susy DDF oscillator}
  A_m^\mu
  &\defas&
  \commutator{
    G_0  
  }{
    \frac{i}{\sqrt{4\pi}}
    \oint d\sigma\;
    \Gamma^\mu e^{-imR}
}
  \nonumber\\
  &=&
  \frac{1}{\sqrt{2\pi}}
  \oint d\sigma\;
  \left(
    {\cal P}^\mu
    +
    i
    m
    \frac{\pi\sqrt{2T}}{k \inner p}
    k \inner \Gamma
    \Gamma^\mu
  \right)
  e^{-im R}
  \,,
\end{eqnarray}
where
\begin{eqnarray}
    R\of{\sigma}
  &\defas&
  -
  \frac{4\pi T}{k\inner p}\,
    k\inner X_\pm\of{\sigma}
\end{eqnarray}
and $p^\mu = \int_0^{2\pi}P^\mu\of{\sigma}$.

By construction, the $A_m^\mu$ super-Poisson-commute with all the constraints.
From the $A_m^\mu$ quasi-local objects $\mathcal{P}^R$ are reobtained by Fourier transforming from the integral
mode index $m$ to the parameter $\sigma$:

\begin{eqnarray}
  \label{quasilocal observable}
  \mathcal{P}^{R}\of{\sigma}
  &\defas&
  \frac{1}{\sqrt{2\pi}}
  \sum\limits_{n = -\infty}^\infty
  A_n\,
  e^{in\sigma}
  \nonumber\\
  &=&
  \int_0^{2\pi}
  d\tilde \sigma\;
  \left(
    \mathcal{P}\of{\tilde \sigma}
    \delta\of{R\of{\tilde\sigma} - \sigma}
    +
    \frac{i\pi \sqrt{2T}}{k\inner p}
    k\inner \Gamma\of{\tilde \sigma}
    \Gamma\of{\tilde \sigma}
    \frac{\partial}{\partial \sigma}\delta\of{R\of{\tilde \sigma}-\sigma}
  \right)
  \,.
\end{eqnarray}

The role of the DDF oscillators played here is the derivation of this expression. Their invariance
was rather easy to enforce and check, but by taking combinations of them as in
\refer{quasilocal observable} and further constructions below, 
we can now build objects which are necessarily still invariants,
but whose invariance is much less obvious.

Since the DDF oscillators $A_m^\mu$ won't be explicitly needed anymore in the following, we
take the liberty to reserve the letter $A$ from now on to describe a gauge connection on target space.
We shall be interested in the Wilson line
\begin{eqnarray}
  \label{generalized Wilso line}
  W^{\mathcal{P}^R}[A]
  &\defas&
  \mathrm{Tr}\,
  \mathrm{P}
  \exp\of{
    i
    \int_0^{2\pi}
    d\sigma\;
    A\inner \mathcal{P}^{R}\of{\sigma}
  }
\end{eqnarray}
with respect to this gauge connection $A$, constructed using the ``generalized tangent vector'' $\mathcal{P}^{R}$
which plays the role of the true tangent vector $X^\prime$ found in ordinary Wilson lines.
Because this object follows in spirit closely the construction principle of the bosonic Pohlmeyer invariants,
and because its bosonic component coincides with the purely bosonic Pohlmeyer invariant, we shall here
address it as the \emph{super-Pohlmeyer invariant}. 	In the following a
form of this object in terms of the original local fields $\mathcal{P}$ and $\Gamma$ is derived, 
which will illuminate its relation to supersymmetric boundary states.

The integrand of \refer{generalized Wilso line} can be put in a more insightful form by means of 
a couple of manipulations:

Following the development in \cite{Schreiber:2004b} (\cf equation (2.43)) 
we now temporarily restrict attention to the subspace
$\mathbf{P}_k$
of phase space on which the function $R$ is invertible, in which case it is, by construction,
$2\pi$-periodic.  On that part of phase space (and only there)
the integral in \refer{quasilocal observable} can be evaluated to yield
\begin{eqnarray}
  \label{quasi local super P}
  \left.\mathcal{P}^R\of{\sigma}\right|_{\mathbf{P}_k}
  &=&
  \left(
    R^{-1}
  \right)^\prime
  \of{\sigma}
  \mathcal{P}
  \of{
    R^{-1}\of{\sigma}  
  }
  +
  \frac{\pi i \sqrt{2T}}{k \inner p}
  \frac{\partial}{\partial\sigma}
  \left(
  \left(
    R^{-1}
  \right)^\prime
  \of{\sigma}
  k \inner \Gamma\of{R^{-1}\of{\sigma}}
  \Gamma\of{R^{-1}\of{\sigma}}
  \right)
  \,.
  \nonumber\\
\end{eqnarray}
The first term is known from the bosonic theory (equation (2.51) in \cite{Schreiber:2004b}). The second
term involves the fermionic correction due to supersymmetry, and its remarkable property is
that it is a total $\sigma$-derivative. This means that when $\mathcal{P}^R$ is inserted in
a multi-integral as they appear in \refer{generalized Wilso line}, the fermionic term will produce boundary terms
and hence coalesce with neighbouring integrands. 

Before writing this down in more detail first note that due to $k$ being a null vector the
fermionic terms can never coalesce with themselves, because of
\begin{eqnarray}
  \label{some vanishing of fermionic combinations}
  &&
  \left(
    R^{-1}
  \right)^\prime
  k \inner \Gamma\of{R^{-1}}
  A \inner \Gamma\of{R^{-1}}
  \of{\sigma}
 \frac{\partial}{\partial\sigma}
  \left(
  \left(
    R^{-1}
  \right)^\prime
  k \inner \Gamma\of{R^{-1}}
  A \inner \Gamma\of{R^{-1}}
  \right)
  \of{\sigma}
  \nonumber\\
  &=&
  \frac{1}{2}
  \frac{\partial}{\partial \sigma}
  \underbrace{
  \left(
  \left(
    R^{-1}
  \right)^\prime
  k \inner \Gamma\of{R^{-1}}
  A_\mu\Gamma^\mu_-\of{R^{-1}}    
  \right)^2\of{\sigma}}_{ = 0}
  \nonumber\\
  &=&
  0
  \,.
\end{eqnarray}
This vanishing result depends on the Grassmann properties of the classical fermions
$\Gamma$, which we are dealing with here. 
The generalization of the present development to the quantum theory requires
more care and is dealt with below.

Using \refer{some vanishing of fermionic combinations} a little reflection shows that, when the total derivative 
terms in \refer{generalized Wilso line} are all integrated over and coalesced at the integration bounds
with the neighbouring terms $i A\inner \mathcal{P}$, this yields

\hspace{-1.5cm}\parbox{12cm}{
\begin{eqnarray}
  \label{super-Pohlmeyer invariant, ugly form}
  &&\left.\mathrm{Tr}\;
  \mathrm{P}
  \exp\of{i
    \int_0^{2\pi}
    d\sigma\;
    A \inner \mathcal{P}^R\of{\sigma}
  }\right|_{\mathbf{P}_k}
  \nonumber\\
  &=&
  \mathrm{Tr}\;
  \mathrm{P}
  \exp\of{
    \int_0^{2\pi}
    d\sigma\;
    \left(
      i A_\mu + 
      \commutator{A_\mu}{A_\nu}
      \frac{\pi \sqrt{2T}}{k\inner p}
      (R^{-1})^\prime\of{\sigma}
      k \inner \Gamma\of{R^{-1}\of{\sigma}}\;
      \Gamma^\nu\of{R^{-1}\of{\sigma}}
    \right)
    (R^{-1})^\prime\of{\sigma}
    \mathcal{P}^\mu\of{R^{-1}\of{\sigma}}
  }
  \,.
  \nonumber\\
\end{eqnarray}
}

This expression simplifies drastically when a change of variable $\tilde \sigma \defas R^{-1}\of{\sigma}$ is
performed in the integral, as in (2.23) of \cite{Schreiber:2004b}:
\begin{eqnarray}
  \cdots 
  &=&
  \mathrm{Tr}\;
  \mathrm{P}
  \exp\of{
    \int_0^{2\pi}
    d\tilde \sigma\;
    \left(
      i A_\mu + 
      \commutator{A_\mu}{A_\nu}
      \frac{\pi  \sqrt{2T}}{k\inner p}
      (R^{-1})^\prime\of{R\of{\tilde \sigma}}
      k \inner \Gamma\of{\tilde \sigma}\;
      \Gamma^\nu\of{\tilde \sigma}
    \right)
    \mathcal{P}^\mu\of{\tilde \sigma}
  }
  \,.
  \nonumber\\
\end{eqnarray}
The fermionic term further simplifies by using $(R^{-1})^\prime\of{R\of{\tilde \sigma}} = 
1/R^\prime\of{\tilde \sigma}$ and then equation (2.42) of \cite{Schreiber:2004b}, which gives
$(R^{-1})^\prime\of{R\of{\tilde \sigma}} = \frac{k\inner p}{2\pi \sqrt{2T}}\frac{1}{k\inner \mathcal{P}}$.
This way the above is finally rewritten as
\begin{eqnarray}
  \label{super-Pohlmeyer invariant}
  \left.
    W^{\mathcal{P}^R}[A]
  \right|_{\mathbf{P}_k}
  &=&
  \mathrm{Tr}\;
  \mathrm{P}
  \exp\of{
    \int_0^{2\pi}
    d\sigma\;
    \left(
      i A_\mu + 
      \commutator{A_\mu}{A_\nu}
      \frac{
k \inner \Gamma\;
      \Gamma^\nu
}{2k\inner \mathcal{P}}
    \right)
    \mathcal{P}^\mu
  }
  \,.
  \nonumber\\
\end{eqnarray}
This is the advertized explicit form of the super-Pohlmeyer invariant in terms of local fields,
when restricted to $\mathbf{P}_k$. 

The right hand side extends to an observable on all of phase space in the obvious way
and it is of interest to study if this extension is still an invariant. 
This is the content of the following subsections.

\subsubsection{Another supersymmetric extension of the bosonic Pohlmeyer invariants}
\label{Another supersymmetric extension of the bosonic Pohlmeyer invariants}

We address the objects
\refer{generalized Wilso line}
as super-Pohlmeyer invariants, because they are obtained from the bosonic Pohlmeyer invariants written in the
form 
$\mathrm{Tr}\,\mathrm{P}\exp\of{\int_0^{2\pi}d\sigma\, A\inner \mathcal{P}^R\of{\sigma}}$
of equation (2.52) of \cite{Schreiber:2004b} by replacing the bosonic \emph{quasi-local} invariants
$\mathcal{P}^R$ by their supersymmetric version \refer{quasilocal observable}. In this sense this
supersymmetric extension is \emph{local}, or rather ``quasi-local'', since the
$\mathcal{P}^R$ are. But it turns out that there is another fermionic extension of the
bosonic Pohlmeyer invariant $\mathrm{Tr}\, \mathrm{P}\exp\of{\int_0^{2\pi} d\sigma\;A \inner \mathcal{P}\of{\sigma}}$
which Poisson-commutes with all the super-Virasoro generators, and which is not local in this sense, namely 
\begin{eqnarray}
  \label{nonlocal susy extension of Pohlmeyer}
  Y[A]
  &\defas&
  \mathrm{Tr}\,\mathrm{P}
  \exp\of{\int_0^{2\pi}d\sigma\;
    \left(
    i A_\mu \mathcal{P}^\mu\of{\sigma}
    +
    \frac{1}{4}\commutator{A_\mu}{A_\nu}
    \Gamma^\mu\of{\sigma} \Gamma^\nu\of{\sigma}
    \right)
  }
  \,.
\end{eqnarray}
Here the integrand itself does not Poisson-commute with the supercharge $G_0$, but $Y[A]$ 
as a whole does.
(This can easily be generalized even to non-constant $A$, but we will here be content with writing down
all expression for the case of constant $A$. Non-constant $A$ will be discussed in the context of the
quantum theory further below.)

Invariance under the bosonic Virasoro generators is  immediate, because the integrand has unit weight.
All that remains to be checked is hence
\begin{eqnarray}
  \commutator{G_0}{Y[A]} &=& 0
  \,.
\end{eqnarray}

{\it Proof:}
This is best seen by following the logic involved in the derivation of equation (3.8) 
in \cite{Schreiber:2004e}:
There are terms coming from
$
  \commutator{G_0}{iA\inner \mathcal{P}\of{\sigma}}
  \propto
  i A\inner \Gamma^\prime\of{\sigma}  
$
which coalesce at the integration boundary with $iA\inner \mathcal{P}$ to give 
$
  -\commutator{A_\mu}{A_\nu} \Gamma^\mu \mathcal{P}^\nu
  \,.
$
This cancels with the contribution from
$
  \commutator{G_0}{\frac{1}{4}\commutator{A_\mu}{A_\nu}\Gamma^\mu\Gamma^\nu}
  \propto
  \commutator{A_\mu}{A_\nu}\Gamma^\mu\mathcal{P}^\nu
  .
$
(Here we write $\propto$ only as a means to ignore the irrelevant global prefactor $i/\sqrt{2}$ in
\refer{the supercharge}.)
Moreover, there is coalescence of $A\inner \Gamma^\prime$ with 
$\commutator{A_\mu}{A_\nu}\Gamma^\mu\Gamma^\nu$ which yields
$\commutator{A_\kappa}{\commutator{A_\mu}{A_\nu}}\Gamma^\kappa\Gamma^\mu\Gamma^\nu = 0$,
so that everything vanishes. This establishes the full invariance of $Y[A]$ under the
super-Virasoro algebra.
\endofproof

With this insight in hand, one can make a curious observation. Write $A_+ \defas k\inner A$
and consider the special case where all transversal components of $A$ 
together with the additional lightlike component $A_-$
mutually commute
\begin{eqnarray}
  \label{transversal A mutually commute}
  \commutator{A_i}{A_j} &=& 0\,,\;\; \forall\, i,j \neq +
  \,.
\end{eqnarray} 
Then
\begin{eqnarray}
  \commutator{A_\mu}{A_\nu}\frac{k\inner \Gamma \Gamma^\nu}{2 k\inner \mathcal{P}} \mathcal{P}^\mu
  &=&
  \frac{1}{2}\commutator{A_+}{A_i}\Gamma^+ \Gamma^i
  \nonumber\\
  &=&
  \frac{1}{4}
  \commutator{A_\mu}{A_\nu}\Gamma^\mu \Gamma^\nu
  \,.
\end{eqnarray}
Comparison of \refer{super-Pohlmeyer invariant} with \refer{nonlocal susy extension of Pohlmeyer} 
hence shows that in this case the super-Pohlmeyer invariant \refer{super-Pohlmeyer invariant} and the
invariant \refer{nonlocal susy extension of Pohlmeyer} coincide:
\begin{eqnarray}
  \label{two Pohlmeyer susy extensions coincide}
  \commutator{A_i}{A_j} = 0\,,\;\; \forall\, i,j \neq +
  &\Rightarrow&
  \left.
    W^{\mathcal{P}^R}[A]
  \right|_{\mathbf{P}_k}
  =
  Y[A]
  \,.
\end{eqnarray}
So in particular in the case \refer{transversal A mutually commute} the extension of the right hand
side of \refer{super-Pohlmeyer invariant} to all of phase space is still an invariant.

Comparing \refer{super-Pohlmeyer invariant} with equation (3.14) of \cite{Schreiber:2004e}
it is obvious, and will be discussed in more detail below, 
that $Y[A]$ must somehow be closely related to the boundary deformation operator describing
non-abelian $A$-field excitations. Together with \refer{two Pohlmeyer susy extensions coincide}
this gives a first indication of how super-Pohlmeyer invariants give insight into boundary states
of the superstring.

Before discussing this in more detail the next section investigates the most general condition
under which the extension of the right hand side of \refer{super-Pohlmeyer invariant} to all of phase space
is still an invariant. It turns out that there are other cases besides \refer{transversal A mutually commute}.

\subsubsection{Invariance of the extension of the restricted super-Pohlmeyer invariants}
\label{Invariance of the extension of the restricted super-Pohlmeyer invariants}

For the bosonic string the constraint $\mathcal{P}\inner \mathcal{P} = 0$, which
says that $\mathcal{P}$ is a null vector in target space, ensured that
the invertibility of $R$ was preserved by the evolution generated by the constraints
(\cf the discussion on p.12 of \cite{Schreiber:2004b}). 

The same is no longer true for the
superstring, where we schematically have $\mathcal{P}\inner \mathcal{P} + \Gamma\inner \Gamma^\prime = 0$,
instead. It follows that we cannot expect the extension of the right hand side of
\refer{super-Pohlmeyer invariant} to all of (super-)phase space to super-Poisson commute
with all the constraints, since the flow induced by the constraints will in general
leave the subspace $\mathbf{P}_k$. Only for the bosonic string does the flow induced by the
constraints respect $\mathbf{P}_k$.

Notice that this is not in contradiction to the above result that on $\mathbf{P}_k$ the
super-Pohlmeyer invariant \refer{generalized Wilso line} (which by construction super-Poisson commutes
with all the constraints) coincides with \refer{super-Pohlmeyer invariant}. Two functions
which conincide on a subset of their mutual domains need not have coinciding derivatives
at these points.

First of all one notes that the invariance under the action of the bosonic constraints
is still manifest in \refer{super-Pohlmeyer invariant}. Because the integrand still has unit weight
one checks this simply by using the same reasoning as in equation (2.19) of \cite{Schreiber:2004b}.

But the result of super-Poisson commuting with the supercharge $G_0$ is rather non-obvious.
A careful calculation shows that the result vanishes if and only if
\begin{eqnarray}
  \label{first condition  for invariance}
  \commutator{A_i}{A_j} &=& 0\,,\;\;\forall\, i,j\neq +
\end{eqnarray}
or
\begin{eqnarray}
  \label{second condition  for invariance}
  k\inner \Gamma^\prime = 0 = k\inner \mathcal{P}^\prime
  \,.
\end{eqnarray}

The first condition is that already discussed in 
\S\fullref{Another supersymmetric extension of the bosonic Pohlmeyer invariants}.
The second condition is nothing but the defining condition of \emph{lightcone gauge}
on the worldsheet.

Notice that these two conditions are very different in character. When the first
\refer{first condition  for invariance} is satisfied it means that the extension of the
right hand side of \refer{super-Pohlmeyer invariant} to all of phase space is indeed 
an honest invariant. When the first condition is not satisfied then the extenstion of the
right hand side of \refer{first condition  for invariance} to all of phase space
is simply not an invariant. Still, it is an object whose Poisson-commutator with the
super-Virasoro constraints vanishes on that part of phase space where
\refer{second condition  for invariance} holds.

We now conclude this subsection by giving the detailed {\it proof} for the above two conditions.

{\it Proof:}

First consider the
terms of fermionic grade 1. These are contributed by 
\begin{eqnarray}
  \label{an equation}
  \commutator{G_0}{A\inner \mathcal P} \propto A\inner \Gamma^\prime
\end{eqnarray}
as well as 
\begin{eqnarray}
  \label{another equation}
  \commutator{A_\mu}{A_\nu}
\frac{\commutator{G_0}{k\inner \Gamma}\Gamma^\nu \mathcal{P}^\mu}{2 k\inner \mathcal{P}}
  &\propto&
  -
  \commutator{A_\mu}{A_\nu}\Gamma^\nu\mathcal{P}^\mu
  \,.
\end{eqnarray} 
The other remaining
fermionic contraction does not contribute, due to
\begin{eqnarray}
  \commutator{A_\mu}{A_\nu}\commutator{G_0}{\Gamma^\nu}\mathcal{P}^\mu \propto
  -2\commutator{A_\mu}{A_\nu}\mathcal{P}^\nu\mathcal{P}^\mu = 0
  \,.
\end{eqnarray}
In the path ordered integral the terms \refer{an equation} appear as
\begin{eqnarray}
  &&
  \cdots iA\inner \mathcal{P}\of{\sigma_{i-1}}
  \int_{\sigma_{i-1}}^{\sigma_{i+1}}
  iA \inner \Gamma^\prime\of{\sigma_i}
  \; d\sigma_i\;
  iA \inner \mathcal{P}\of{\sigma_{i+1}}
  \cdots  
  \nonumber\\
  &=&
  \cdots
  \left( 
  \left(A\inner\mathcal{P}A\inner \Gamma\right)\of{\sigma_{i-1}}
  iA\inner \mathcal{P}\of{\sigma_{i+1}}
  -
  iA\inner \mathcal{P}\of{\sigma_{i-1}}
  \left(A\inner \Gamma  A\inner \mathcal{P}\right)\of{\sigma_{i+1}}
  \right)
  \cdots
  \,.
\end{eqnarray}
(This is really a special case of the general formula (3.8) in
\cite{Schreiber:2004e}.)
This way the term
\begin{eqnarray}
  \label{yet another equation}
  iA_\mu\Gamma^\mu\, iA_\nu \mathcal{P}^\nu - iA_\mu\mathcal{P}^\nu\,iA_\mu\Gamma^\mu
  &=&
  \commutator{A_\mu}{A_\nu}\Gamma^\nu\mathcal{P}^\mu
\end{eqnarray}
is produced, and it cancels precisely with 
\refer{another equation}.

This verifies that there are no terms of grade 1.

Now consider the remaining terms of grade 3. It is helpful to write
\begin{eqnarray}
  \label{helpful decopmposition}
  \commutator{A_\mu}{A_\nu}
  \frac{k\inner\Gamma \Gamma^\nu}{2 k\inner \Gamma}
   \mathcal{P^\mu}
  &=&
  \frac{1}{2}
  \commutator{A_+}{A_i} \Gamma^+ \Gamma^i 
  +
  \commutator{A_i}{A_j}
  \frac{k\inner\Gamma \Gamma^j}{2 k\inner \mathcal{P}}
   \mathcal{P}^i
  \,.  
\end{eqnarray}
The first term on the right hand side gives nothing of grade 3 when Poisson-commuted with 
$G_0$. The second term however gives rise to
\begin{eqnarray}
  \label{some intermediate caclulation}
  \commutator{G_0}{  \commutator{A_i}{A_j}\frac{k\inner\Gamma\, \Gamma^j}{2 k\inner \mathcal{P}}
   \mathcal{P}^i
}
  &=&
   \commutator{A_i}{A_j}
    \left(
    \frac{k\inner\Gamma\, \Gamma^j}{2 k\inner \mathcal{P}}
   \Gamma^{\prime i}
  -
    \frac{k\inner\Gamma\, \Gamma^j}{2 (k\inner \mathcal{P})^2}
   k\inner \Gamma^{\prime}
  \right)
  +
  \mbox{terms already considered}
  \nonumber\\
  &=&
   \commutator{A_i}{A_j}
    \left(
    \frac{k\inner\Gamma}{4 k\inner \mathcal{P}}
   (\Gamma^j\Gamma^{i})^\prime
  -
    \frac{k\inner\Gamma\, \Gamma^j}{2 (k\inner \mathcal{P})^2}
   k\inner \Gamma^{\prime}
  \right)
  +
  \mbox{terms already considered}  
  \nonumber\\
  &=&
   \commutator{A_i}{A_j}
    \left(
    \frac{k\inner\Gamma}{4 k\inner \mathcal{P}}
   \Gamma^j\Gamma^{i}
   \right)^\prime
  +
  \alpha
  +
  \mbox{terms already considered}
  \,,
  \nonumber\\  
\end{eqnarray}
where we have abbreviated with
\begin{eqnarray}
  \label{remaining terms}
  \alpha &\defas&
   -
   \commutator{A_i}{A_j}
   \left(
    \left(\frac{k\inner\Gamma}{2 k\inner \mathcal{P}}\right)^\prime
   \Gamma^j\Gamma^{i}   
    +
    \frac{k\inner\Gamma\, \Gamma^j}{2 (k\inner \mathcal{P})^2}
   k\inner \Gamma^{\prime}
  \right)
\end{eqnarray}
two terms which will \emph{not} cancel with anything in the following. (Notice that they are proportional 
to $\sigma$-derivatives of longitudinal objects (along $k$).)

The remaining first term on the right hand side of \refer{some intermediate caclulation}
coalesces with $i A_+ \mathcal{P}^+$ to yield
$\frac{i}{4}\commutator{A_+}{\commutator{A_i}{A_j}}\Gamma^+ \Gamma^i\Gamma^j$.
This cancels against the 
coalescence of \refer{an equation} with the first term on the right hand side of
\refer{helpful decopmposition}
which gives the term
$\frac{i}{2}\commutator{A_j}{\commutator{A_+}{A_i}}\Gamma^j \Gamma^+ \Gamma^i$, because
together they become the longitudinal component of the exterior covariant derivative of the
field strength of $A$, which vanishes. The transversal component of this exterior derivative
of the field strength appears in the remaining terms: 

First there is the remaining coalescence of \refer{an equation} with the second term
on the right hand side of \refer{helpful decopmposition}, which yields
$
  i\commutator{A_k}{\commutator{A_i}{A_j}}
  \frac{k\inner\Gamma}{2k\inner \mathcal{P}}
  \Gamma^k \Gamma^j \mathcal{P}^i
$. Together with the remaining coalescence of the first term on the right of
\refer{some intermediate caclulation} with the transversal $i A_j \mathcal{P}^j$ which
gives rise to
$
  i\commutator{A_k}{\commutator{A_i}{A_j}}
  \frac{k\inner \Gamma}{4 k\inner \mathcal{P}}
  \Gamma^i \Gamma^j\mathcal{P}^k
$
one gets something proportional to
$
  \Big[
    G_0,
  \underbrace{
  \commutator{A_k}{\commutator{A_i}{A_j}}\Gamma^k\Gamma^i\Gamma^j
  }_{=0}
  \Big]
  =0\,,
$
which vanishes because it involves the transversal part of the gauge covariant exterior
derivative of the field strength of $A$.

In summary, the only terms that remain are those of \refer{remaining terms}. When the $\sigma$-derivative
is written out this are three terms which have to vanish seperately, because they
contain different combinations of fermions. Clearly they vanish precisely if \refer{first condition for invariance}
or \refer{second condition for invariance} are satisfied. This completes the proof. \endofproof.

\subsubsection{Quantum super-Pohlmeyer invariants}
\label{Quantum super-Pohlmeyer invariants}

The DDF-invariants \refer{bosonic susy DDF oscillator} 
are, as discussed in equation (2.12) of \cite{Schreiber:2004b}, still invariants after quantization
in terms of DDF oscillators.
If we take the liberty to denote the quantized objects $\mathcal{P}$ and $\Gamma$ by the same symbols
as their classical counterparts, then the only thing that changes in the notation of the above sections is
that the canonical super-commutation relations \refer{classical CCM} pick up an imaginary factor
\begin{eqnarray}
  \commutator{\mathcal{P}^\mu\of{\sigma}}{\mathcal{P}\of{\kappa}} 
  &=& 
  -i \eta^{\mu\nu}\delta^\prime\of{\sigma-\kappa}
  \,.
\end{eqnarray}
This again introduces that same factor in the second term of \refer{bosonic susy DDF oscillator} and similarly
in the following expressions.

The quantization of the super-Pohlmeyer invariant \refer{generalized Wilso line} is a trivial consequence
of the quantization of the DDF invariants that it is built from, and, with that imaginary unit taken care of,
its restriction \refer{super-Pohlmeyer invariant} to the case where $R$ is invertible reads
\begin{eqnarray}
 \label{quantized restricted super-Pohlmeyer}
  \mathrm{Tr}\,\mathrm{P}
  \exp\of{
    \int\limits_0^{2\pi}
    d\sigma\;
    \left(
      i A_\mu + \frac{i}{2}\commutator{A_\mu}{A_\nu}\frac{k\inner \Gamma \Gamma^\nu}{k \inner \mathcal{P}}
    \right)
    \mathcal{P}^\mu
  }
  \,.
\end{eqnarray}
Noting that our $A$ is taken to be hermitian and that hence the gauge field strength is
\begin{eqnarray}
  F_A &=& -i(d + iA)^2
  \nonumber\\
  &=&
  dA + i A \wedge A
  \nonumber\\
  &=&
  \left(\partial_{[\mu} A_{\nu]} + \frac{i}{2}\commutator{A_\mu}{A_\nu}\right)dx^\mu \wedge dx^\nu
  \nonumber\\
  &=&
  \frac{1}{2}(F_A)_{\mu\nu}dx^\mu \wedge dx^\nu
\end{eqnarray}
the second term in the integrand is related to the field strength as in
\begin{eqnarray}
  \cdots 
  &=&
  \mathrm{Tr}\,\mathrm{P}
  \exp\of{
    \int\limits_0^{2\pi}
    d\sigma\;
    \left(
      i A_\mu + \frac{1}{2}(F_A)_{\mu\nu}\frac{k\inner \Gamma \Gamma^\nu}{k \inner \mathcal{P}}
    \right)
    \mathcal{P}^\mu
  }
  \,.
\end{eqnarray}
In the case $\commutator{A_i}{A_j} = 0$ \refer{transversal A mutually commute} we hence obtain the
quantized version of \refer{nonlocal susy extension of Pohlmeyer} in the form
\begin{eqnarray}
  \label{restricted and extended superPohl}
  Y[A] 
  &=&
  \mathrm{Tr}\,\mathrm{P}
  \exp\of{
    \int_0^{2\pi}
     d\sigma\;
     \left(
        i A_\mu \mathcal{P}^\mu + \frac{1}{4}(F_A)_{\mu\mu}\Gamma^\mu\Gamma^\nu
     \right)
  }
  \,.
\end{eqnarray}

While the quantized super-Pohlmeyer invariant, being constructed from invariant DDF operators,
is itself a quantum invariant in that it commutes with all the super-Virasoro constraints, the proof
in \S\fullref{Invariance of the extension of the restricted super-Pohlmeyer invariants} of the invariance 
condition of the restricted and then extended form \refer{super-Pohlmeyer invariant} receives quantum corrections. 
In its
classical version the proof makes use of the Grassmann property of the fermions $\Gamma$. Quantumly
there will be diverging contractions in products of $\Gamma$s which not only prevent the application of
the proof to the quantum theory but also make the expression \refer{quantized restricted super-Pohlmeyer}
ill defined without some regularization prescription. 

This application of \refer{restricted and extended superPohl} to a bare boundary states 
is the content of \S\fullref{Super-Pohlmeyer and boundary states}. But before coming to that
a technicality needs to be discussed, which is done in the next section.

\subsubsection{On an operator ordering issue in Wilson lines along the closed string}
\label{On an operator ordering issue in Wilson lines along the closed string}

For applying a generalized Wilson line of the kind discussed above to any string state, it 
is helpful to understand how the operators in the Wilson line can be commuted past each other 
to act on the state on the right. It turns out that under a certain condition, which is
fulfilled in the cases we are interested in, the operators can be freely commuted. This works
as follows:\\

A generalized Wilson line of the form
\begin{eqnarray}
  W^{\mathcal{P}}[A] 
   &=& 
  \mathrm{Tr}\,\mathrm{P}\exp\of{\int\limits_0^{2\pi} A\inner \mathcal{P}\of{\sigma}\, d\sigma}
\end{eqnarray}
with even graded $\mathcal{P}$ 
(which could be the $\mathcal{P}$ or $\mathcal{P}^R$ of the
previous sections but also more general objects) breaks up like
\begin{eqnarray}
  W^{\mathcal{P}}[A]
  &=&
  \sum\limits_{n=0}^\infty
  Z^{\mu_1 \cdots \mu_n}
  \mathrm{Tr}\of{A_{\mu_1}\cdots A_{\mu_2}}
\end{eqnarray}
into iterated integrals
\begin{eqnarray}
  \label{definition of Z}
  &&
  \!\!\!\!\!\!\!\!\!\!Z^{\mu_1 \cdots \mu_N}
  \nonumber\\
  &=&
  \frac{1}{N}
  \left[
    \int\limits_{0 < \sigma^1 < \sigma^2 < \cdots < \sigma^N < 2\pi}
    \!\!\!\!\!\!\!\!\!\!\!\!\!\!\!\!\!\!\!\!\!\!d^N \sigma
    \;\;\;\;\;\;\;+ 
    \int\limits_{0 < \sigma^N < \sigma^1 < \cdots < \sigma^{N-1} < 2\pi}
    \!\!\!\!\!\!\!\!\!\!\!\!\!\!\!\!\!\!\!\!\!\!d^N \sigma
    \;\;\;\;\;\;\;+
    \int\limits_{0 < \sigma^{N-1} < \sigma^N < \cdots < \sigma^{N-2} < 2\pi}
    \!\!\!\!\!\!\!\!\!\!\!\!\!\!\!\!\!\!\!\!\!\!d^N \sigma
  \;\;\;\;\;\;\;
  + \cdots\right]  
  \mathcal{P}^{\mu_1}\of{\sigma^1}
  \cdots
  \mathcal{P}^{\mu_N}\of{\sigma^N}
  \,.
  \nonumber\\
\end{eqnarray}
In equation (2.17) of \cite{Schreiber:2004b} if was noted that the integration domain
can equivalently be written as
\begin{eqnarray} 
  \label{alternative version of Z}
  Z^{\mu_1 \cdots \mu_N}
  &=&
  \frac{1}{N}
  \int\limits_0^{2\pi}
  d\sigma^1
  \;
  \int\limits_{\sigma^1}^{\sigma^1 + 2\pi}
  d\sigma^2
  \;
  \cdots
  \int\limits_{\sigma^{N-1}}^{\sigma^1 + 2\pi}
  d\sigma^N
  \;  
  \mathcal{P}^{\mu_1}\of{\sigma^1}
  \mathcal{P}^{\mu_2}\of{\sigma^2}
  \cdots
  \mathcal{P}^{\mu_N}\of{\sigma^N}
  \,.
\end{eqnarray}
This is seen by simply replacing all $\sigma^i < \sigma^1$ for $i>1$ by $\sigma^i + 2\pi$. Due to
the periodicity of $\mathcal{P}$ this does not change the value of the integral but yields the
integration bounds used in \refer{alternative version of Z}.

The reason why this is recalled here is that a slight generalization of this fact will be needed
in the following. Namely for any integer $M$ with $1 < M < N$ one can obviously more generally write
\begin{eqnarray}
  \label{yet another version of Z}
  &&Z^{\mu_1 \cdots \mu_N}
  \nonumber\\
  &=& 
  \int\limits_0^{2\pi}
  d\sigma^1
  \;
  \int\limits_{\sigma^1}^{\sigma^1 + 2\pi}
  d\sigma^M
  \;\;
  \int\limits_{\sigma^1}^{\sigma^M}
  d\sigma^2
  \;
  \cdots
  \int\limits_{\sigma^{M-2}}^{\sigma^M}
  d\sigma^{M-1}
  \;\;
  \int\limits_{\sigma^{M}}^{\sigma^1 + 2\pi}
  d\sigma^{M+1}
  \cdots
  \int\limits_{\sigma^{N-1}}^{\sigma^1 + 2\pi}
  d\sigma^N
  \;  
  \mathcal{P}^{\mu_1}\of{\sigma^1}
  \mathcal{P}^{\mu_2}\of{\sigma^2}
  \cdots
  \mathcal{P}^{\mu_N}\of{\sigma^N}
  \,.
  \nonumber\\
\end{eqnarray}
Equation \refer{alternative version of Z} follows as the special case with $M=2$.

The motivation for these considerations is the following:

Classically, the $\mathcal{P}$ commute among each other. Therefore the ordering of the
$\mathcal{P}$ in the integrand makes no difference, only the combination of spacetime
index $\mu_i$ with integration variable $\sigma^i$ does.

Here we want to note that this remains true at the quantum level \emph{if}
\begin{eqnarray}
  \label{condition for interchangability of objects in Wilson line}
  \commutator{\mathcal{P}\of{\sigma}}{\mathcal{P}\of{\kappa}} 
  &\propto& 
  \delta^\prime\of{\sigma - \kappa}
  \,.
\end{eqnarray}

This is readily seen by commuting $\mathcal{P}\of{\sigma^1}$ with $\mathcal{P}\of{\sigma^M}$
in \refer{yet another version of Z}. The result has the form
\begin{eqnarray}
  \int\limits_0^{2\pi}
  d\sigma^1
  \;
  \int\limits_{\sigma^1}^{\sigma^1 + 2\pi}
  d\sigma^M
  \;
  \delta^\prime\of{\sigma^1 - \sigma^M}
  F\of{\sigma^1, \cdots, \sigma^N}
  &=&
  \int\limits_0^{2\pi}
  d\sigma^1
  \;
  \int\limits_{\sigma^1}^{\sigma^1 + 2\pi}
  d\sigma^M
  \;
  \delta\of{\sigma^1 - \sigma^\kappa}
  \frac{\partial}{\partial \sigma^M}
  F\of{\sigma^1, \cdots, \sigma^N}
  \nonumber\\
  &=&
  \int\limits_{\sigma^1}^{\sigma^1 + 2\pi}
  d\sigma^M
  \;
  \delta\of{\sigma^1 - \sigma^\kappa}
  \frac{\partial}{\partial \sigma^M}
  F\of{\sigma^1, \cdots, \sigma^N}
  \nonumber\\
  &=&
  0
  \,,
\end{eqnarray}
so that all resulting commutator terms vanish. Every other commutator can be obtained by 
using the cyclic invariance in the integration variables.

More generally, any two (even graded, periodic) 
objects $A\of{\sigma}$, $B\of{\kappa}$ in the integrand of an iterated 
integral of the form \refer{definition of Z} whose commutator is proportional to 
$\commutator{A\of{\sigma}}{B\of{\kappa}} \propto \delta^\prime\of{\sigma-\kappa}$ can be 
commuted past each other in the Wilson line without affecting the value of the integral.

This simple but crucial observation will be needed below for the demonstration that
Pohlmeyer-invariants map the boundary state of a bare D-brane to that describing a brane
with a nonabelian gauge field turned on.

\subsubsection{Super-Pohlmeyer and boundary states}
\label{Super-Pohlmeyer and boundary states}

We now have all ingredients in place to apply the super-Pohlmeyer invariant to the boundary state
of a bare D9 brane. A brief review of the idea of boundary states adapted to the present context
is given in \cite{Schreiber:2004e}, but in fact only two simple relations are needed for the following:

If $\ket{\mathrm{D9}}$ is the boundary state of the space-filling BPS D9 brane, then
(due to equation (2.26) in \cite{Schreiber:2004b} and section 2.3.1 in \cite{Schreiber:2004e})
we have
\begin{eqnarray}
  \mathcal{P}^\mu\of{\sigma}\ket{\mathrm{D9}}
  &=&
  \sqrt{
    \frac{T}{2}
  }
    X^{\prime\mu}\of{\sigma}\ket{\mathrm{D9}}
\end{eqnarray}
and
\begin{eqnarray}
  \Gamma^\mu\of{\sigma}
  \ket{\mathrm{D9}}
  &=&
  {\cal E}^{\dagger \mu}\of{\sigma}
  \ket{\mathrm{D9}}
  \,.
\end{eqnarray}

Using the results of \S\fullref{On an operator ordering issue in Wilson lines along the closed string}
such a replacement extends to the full Wilson line made up from these objects:

Consider the extension \refer{restricted and extended superPohl} of the restricted super-Pohlmeyer invariant
with $A_+ \neq 0$ and furthermore only mutually commuting spatial components of $A$ nonvanishing. In this case
the fermionic terms in the integrand have trivial commutators so that the integrand as a whole 
satisfies condition \refer{condition for interchangability of objects in Wilson line}. Therefore,
according to the result of \S\fullref{On an operator ordering issue in Wilson lines along the closed string},
we can move all appearances of $\mathcal{P}^\mu  + \frac{1}{4}(F_A)_{\mu\nu}\Gamma^\mu\Gamma^\nu$
to the boundary state $\ket{\mathrm{D9}}$ on the right, change it there to 
$\sqrt{\frac{T}{2}}X^{\prime \mu} + \frac{1}{4}(F_A)_{\mu\nu}\mathcal{E}^{\dagger \mu}\mathcal{E}^{\dagger \nu}$
and then move this back to the original position (noting that still 
$\commutator{X^{\prime}\of{\sigma}}{\mathcal{P}\of{\kappa}} \propto \delta^\prime\of{\sigma-\kappa}$).
This way we have
\begin{eqnarray}
  \label{from Pohlmeyer to boundary}
  &&\mathrm{Tr}\, \mathrm{P}
  \exp\of{
    \int\limits_0^{2\pi}
     d\sigma\,
    \left(
      i A_\mu \mathcal{P}^\mu + \frac{1}{4}(F_A)_{\mu\nu}\Gamma^\mu\Gamma^\nu
    \right)
  }
  \ket{\mathrm{D9}}
  \nonumber\\
  &=&
  \mathrm{Tr}\, \mathrm{P}
  \exp\of{
    \int\limits_0^{2\pi}
     d\sigma\,
    \left(
      -i 
      \sqrt{\frac{T}{2}} A_\mu X^{\prime \mu} + \frac{1}{4}(F_A)_{\mu\nu}
      \mathcal{E}^{\dagger \mu}
      \mathcal{E}^{\dagger \nu}
    \right)
  }
  \ket{\mathrm{D9}}
  \,.
\end{eqnarray}

If we allowed ourself to regulate all the generalized Wilson lines considered here by a point-splitting
method as in \cite{MaedaNakatsuOonishi:2004}, i.e. by taking care that no local fields in the
Wilson line ever come closer than some samll distance $\sigma$, then the above step becomes a triviality.
Indeed, the result of \cite{MaedaNakatsuOonishi:2004} together with those of
\cite{Hashimoto:2000,Hashimoto:1999,Schreiber:2004e}
shows that this is a viable approach, because the condition for the $\epsilon$-regularized
Wilson line to be still an invariant is the same as that of the non-regularized Wilson line to be
free of divergences and hence well defined.

It will be convenient for our purposes to rescale $A$ as
\begin{eqnarray}
  \label{rescaling of A}
  A \mapsto -\sqrt{\frac{2}{T}}A
  \,, 
\end{eqnarray}
so that this becomes
\begin{eqnarray}
  \label{boundary state from super-Pohlmeyer}
  \cdots 
  &=&
  \mathrm{Tr}\, \mathrm{P}
  \exp\of{
    \int\limits_0^{2\pi}
     d\sigma\,
    \left(
      i 
      A_\mu X^{\prime \mu} 
      + 
      \frac{1}{2T}(F_A)_{\mu\nu}
      \mathcal{E}^{\dagger \mu}
      \mathcal{E}^{\dagger \nu}
    \right)
  }
  \ket{\mathrm{D9}}
  \,.
\end{eqnarray}
This is finally our main result, because this is precisely the boundary state of a nonabelian gauge field
as considered in 
equation (3.14) of \cite{Schreiber:2004e}, which is a generalization of the abelian case studied
in \cite{Hashimoto:2000,Hashimoto:1999}. The same form of the boundary state is obtained from
equations (3.7), (3.8)
in \cite{MaedaNakatsuOonishi:2004} 
when in the expression given there the integral over the Grassmann variables is performed
(following the computation described on pp. 236-237 of \cite{AndreevTseytlin:1988}).

The boundary state \refer{boundary state from super-Pohlmeyer} has two important properties:

\begin{enumerate}

\item

\paragraph{Super-Ishibashi property of the boundary state.}

The defining property of boundary states is that they are annihilated by the generators 
$\mathcal{L}_K$ of
$\sigma$-reparameterization as well as, in the superstring case, by their square root 
$\extd_K$, which is a deformed exterior derivative on loop space. $\mathcal{L}_K$ is a
linear combination of left- and right-moving bosonic super-Virasoro generators, while 
$\extd_K$ is a combination of fermionic super-Virasoro generators, as discussed in 
\cite{Schreiber:2004e}. 

It is noteworthy that the state \refer{boundary state from super-Pohlmeyer} 
indeed satisfies the Ishibashi conditions.
Naively this must be the case, because this state is obtained from the bare $\ket{\mathrm{D9}}$,
which does satisfy it by definition, by acting on it with a super-Pohlmeyer operator, that
commutes with all constraints and hence leaves the Ishibashi property of $\ket{\mathrm{D9}}$
intact. But above we mentioned that the restricted form \refer{quantized restricted super-Pohlmeyer}
of the quantized Pohlmeyer invariants that this state comes from has potential quantum anomalies
which would spoil this invariance. These are due to the non-Grassmann property of the
quantized fermions $\Gamma$. However, after application to the bare $\ket{\mathrm{D9}}$
which gives \refer{boundary state from super-Pohlmeyer}, the left- and right-moving 
fermions are replaced by their polar combination $\mathcal{E}^\dagger$, and these again
enjoy the Grassmann property (they are nothing but differential forms on loop space). For this
reason the final result can again enjoy the Ishibashi property, which means nothing but
super-reparameterization invariance with respect to $\sigma$.

{\it Proof:}

The invariance under reparameterizations induced by $\mathcal{L}_K$ is manifest, analogous
in all the cases considered here before, since \refer{boundary state from super-Pohlmeyer} is
the generalized Wilson line over an object of unit reparameterization weight. 

The only nontrivial
part that hence needs to be checked is the commutation with $\extd_K $ and here we only need to
know that
$\commutator{\extd_K}{X^\mu\of{\sigma}} = \mathcal{E}^{\dagger \mu}\of{\sigma}$.

Applying this to \refer{boundary state from super-Pohlmeyer}
we get, in the same manner as in the similar computations before,
from $\commutator{\extd_K}{i A_\mu X^{\prime \mu}} =
 i (\partial_\mu A_\nu - \partial_\nu A_\mu)\mathcal{E}^{\dagger \mu}X^{\prime \nu}
+ \left(iA_\mu \mathcal{E}^{\dagger \mu}\right)^\prime$
coalesced terms
$-\commutator{A_\mu}{A_\nu}\mathcal{E}^{\dagger \mu}X^{\prime \nu}$
and
$\frac{i}{2T}\commutator{A_\kappa}{(F_A)_{\mu\nu}}\mathcal{E}^{\dagger \kappa}\mathcal{E}^{\dagger \mu}\mathcal{E}^{\dagger \nu}$
at the integration boundaries.

These combine with the terms
$\commutator{\extd_K}{\frac{1}{2T}(F_A)_{\mu\nu}\mathcal{E}^{\dagger \mu}\mathcal{E}^{\dagger \nu}}
=
\frac{1}{2T} (\partial_{[\kappa} (F_A)_{\mu\nu]}
  \mathcal{E}^{\dagger \kappa}\mathcal{E}^{\dagger \mu}\mathcal{E}^{\dagger \nu})
+
-i (F_A)_{\mu\nu}\mathcal{E}^{\dagger \mu}X^{\prime \nu}
$
to
$
  \left(
   i (\partial_\mu A_\nu - \partial_\nu A_\mu) - \commutator{A_\mu}{A_\nu} -i (F_A)_{\mu\nu}
  \right)\mathcal{E}^{\dagger \mu}X^{\prime \nu}
  = 0
$
and\\
$
 \frac{1}{2T}
  \left(
    \partial_{[\kappa}(F_A)_{\mu\nu}
    +
    i \commutator{A_{[\kappa}}{(F_A)_{\mu\nu]}}
  \right)
  \mathcal{E}^{\dagger \kappa}\mathcal{E}^{\dagger \mu}\mathcal{E}^{\dagger \nu}
  = 0
  \,.
$
Hence all terms vanish and $\extd_K$ commutes with \refer{boundary state from super-Pohlmeyer}. \endofproof

\item

\paragraph{Nonlinear gauge invariance of the boundary state.}

A generic state constructed from gluon vertices for nonabelian $A$ will generically not
be invariant under a target space gauge transformation $A \to UAU^\dagger + U(dU^\dagger)$.
The generalized Wilson line in \refer{boundary state from super-Pohlmeyer} however does
have this invariance  - at least at the classical level. This follows from the general
invariance properties of Wilson lines (for details see appendix B of \cite{Schreiber:2004e})
and depends crucially on the appearance of the gauge covariant field strength $F_A$ in 
\refer{boundary state from super-Pohlmeyer}. 

\end{enumerate}

\addtocontents{toc}{\vspace*{-.2em}}

\clearpage
\section{Local Connections on Loop Space from Worldsheet Deformations}
\label{Connections on Loop Space from Worldsheet Deformations}

The following is taken from the preprint \cite{Schreiber:2004e}.
It leads over from the SCFT deformation theory in part
\ref{SQM on Loop Space} to the 2-bundle theory in part 
\ref{Nonabelian Strings}.

\subsection{SCFT deformations in loop space formalism}
\label{SCFT deformations in loop space formalism}

This introductory section discusses aspects of loop space formalism and
deformation theory that will be applied in \S\fullref{BSCFT deformation for nonabelian 2-form fields}
to the description of nonabelian 2-form background fields.

\subsubsection{SCFT deformations and backgrounds using Morse theory technique}
\label{Deformations and backgrounds using Morse theory technique}

The reasoning by which we intend to derive the worldsheet theory for superstrings in 
nonabelian 2-form backgrounds involves an interplay of deformation theory of 
superconformal field theories for closed strings, as described in \cite{Schreiber:2004},
as well as the generalization to boundary state deformations, which are disucssed
further below in \S\fullref{Boundary state deformations from unitary loop space deformations}.
The deformation method we use consists of adding deformation terms to the super Virasoro generators
and in this respect is in the tradition of similiar approaches as for instance described in
\cite{Giannakis:1999,OvrutRama:1992,EvansOvrut:1990,EvansOvrut:1989,FreericksHalpern:1988}
(as opposed to, say, deformations of the CFT correlators).
What is new here is the systematic use of similarity transformations on a certain combination
of the supercharges, as explained below.

In this section the SCFT deformation technique for the closed string is briefly reviewed
in a manner which should alleviate the change of perspective from the string's Fock
space to loop space.

Consider some realization of the superconformal generators $L_n, \bar L_n, G_r, \bar G_r$
(we follow the standard notation of \cite{Polchinski:1998})
of the type II superstring. We are looking for consistent deformations of these operators
to operators $L^\Phi_n, \bar L^\Phi_n, G^\Phi_r, \bar G^\Phi_r$ 
($\Phi$ indicates some unspecified background field consiguration which is associated with the deformation)
which still satisfy the
superconformal algebra and so that the generator of spatial worldsheet reparametrizations
remains invariant:
\begin{eqnarray}
  \label{invariance of spatial rep}
  L^\Phi_n - \bar L^\Phi_{-n} &\shallbe& L_n - \bar L_{-n}
  \,.
\end{eqnarray}
This condition follows from a canonical analysis of the worldsheet action, which
is nothing but 1+1 dimensional supergravity coupled to various matter fields. As for all
gravitational theories, their ADM constraints break up into spatial diffeomorphism
constraints as well as the Hamiltonian constraint, which alone encodes the dynamics.

The condition \refer{invariance of spatial rep} can also be understood in terms of
boundary state formalism, which is briefly reviewed in 
\S\fullref{Boundary state formalism}. As discussed below, the operator 
$\mathcal{B}$ related to a nontrivial bounday state $\ket{\mathcal{B}}$ can be interpreted as 
inducing a deformation $G_r^\Phi \defas \mathcal{B}^{-1}G_r \mathcal{B}$, etc. and the
condition \refer{invariance of spatial rep} is then equivalent to
\refer{Virasoro constraint on boundary state}.

In any case, we are looking for isomorphisms of the superconformal algebra
which satisfy \refer{invariance of spatial rep}:

To that end,
let $d_r$ and $d^\dagger_r$ be the modes of the polar combinations of the left- and right-moving
supercurrents 
\begin{eqnarray}
  \label{polar supercurrent modes}
  d_r 
  &\defas&
    G_r + i \bar G_{-r}
  \nonumber\\
  d^\dagger_r
  \defas 
  (d_r)^\dagger 
  &=&
    G_r - i \bar G_{-r}
  \,.
\end{eqnarray}
These are the 'square roots' of the reparametrization generator
\begin{eqnarray}
  \label{rep gen}
  \mathcal{L}_n 
  &\defas&
  -i
  \left(
  L_n - \bar L_{-n}
  \right)
  \,,
\end{eqnarray}
i.e.
\begin{eqnarray}
  \antiCommutator{d_r}{d_s}
  =
  \lbrace d^\dagger_r, d^\dagger_s\rbrace
  =
  2i\mathcal{L}_{r+s}
  \,.
\end{eqnarray}
Under a deformation the right hand side of this equation must stay invariant 
\refer{invariance of spatial rep} so that
\begin{eqnarray}
  d_r^{\Phi} &\defas& d_r + \Delta_\Phi  d_r
  \nonumber\\
  d^{\dagger\Phi}_r &\defas& d^\dagger_r + (\Delta_\Phi  d_r)^\dagger
\end{eqnarray}
implies that the shift $\Delta_\Phi d_r$ of $d_r$ has to satisfy
\begin{eqnarray}
  \label{shift in loop space extd}
  \antiCommutator{d_r}{\Delta_\Phi d_s}
  +
  \antiCommutator{d_s}{\Delta_\Phi d_r}
  +
  \antiCommutator{\Delta_\Phi d_r}{\Delta_\Phi d_s}
  &=&
  0
  \,.
\end{eqnarray}

One large class of solutions of this equation is
\begin{eqnarray}
  \label{similarity transform solutions of deformation condition}
  \Delta_\Phi d_r &=& A^{-1} \commutator{d_r}{A}\,,\hspace{.8cm} 
  \mbox{for $\commutator{\mathcal{L}_n}{A} = 0\;\, \forall\, n$}
  \,,
\end{eqnarray}
where $A$ is any even graded operator that is spatially reparametrization invariant, i.e. which 
commutes with \refer{rep gen}.

When this is rewritten as
\begin{eqnarray}
  \label{simtrafo on susy currents}
  d^\Phi_r &=& A^{-1} \circ d_r \circ A
  \nonumber\\
  d_r^{\dagger \Phi} &=& A^\dagger \circ d^\dagger_r \circ A^{\dagger -1}
\end{eqnarray}
one sees explicitly that the formal structure involved here is a direct generalization of
that used in \cite{Witten:1982} in the study of the relation of deformed generators in
supersymmetric quantum \emph{mechanics} to Morse theory. Here we are concerned with 
the direct generalization of this mechanism from $1+0$ to $1+1$ dimensional 
supersymmetric field theory.

In $1+0$ dimensional SQFT (i.e. supersymmetric quantum mechanics) relation 
\refer{simtrafo on susy currents} is sufficient for the deformation to be truly an isomorphism
of the algebra of generators. In 1+1 dimensions, on the superstring's worldsheet,
there is however one further necessary condition for this to be the case.
Namely the (modes of the) new worldsheet Hamiltonian constraint 
$H_n = L_n + \bar L_-n$ 
must clearly be defined as
\begin{eqnarray}
  \label{def deformed worldsheet Hamiltonian}
  H^\Phi_n
  &\defas&
  \frac{1}{2}
  \antiCommutator{d^{\Phi}_r}{d^{\dagger\Phi}_{n-r}}
  -
  \delta_{n,0}
  \frac{c}{12}\left(4r^2 - 1\right)
\end{eqnarray}
and \refer{shift in loop space extd} alone does not guarantee that this is \emph{unique} 
for all $r \neq n/2$. 
\emph{If} it is, however, then the Jacobi identity already implies that 
\begin{eqnarray}
  G^\Phi_r &\defas& \frac{1}{2}\left(d^\Phi_r + d^{\dagger \Phi}_r\right)
  \nonumber\\
  L^\Phi_n &\defas& \frac{1}{4}\left(
    \antiCommutator{d_r^{\Phi}}{d^{\dagger \Phi}_{n-r}}
    +
    \antiCommutator{d^\Phi_r}{d^\Phi_{n-r}}
   \right)
  -
  \delta_{r,n/2}
  \frac{c}{24}\left(4r^2 - 1\right)
  \nonumber\\
  \bar G^\Phi_{r} &\defas& -\frac{i}{2}\left(d^\Phi_{-r} - d^{\dagger \Phi}_{-r}\right)
  \nonumber\\
  L^\Phi_n &\defas& \frac{1}{2}\left(
    \antiCommutator{d_{-r}^{\Phi}}{d^{\dagger \Phi}_{r-n}}
    -
    \antiCommutator{d^\Phi_{-r}}{d^\Phi_{r-n}}
   \right)  
  \,,
  \hspace{.8cm}
  \forall\,r\neq n/2
\end{eqnarray}
generate two mutually commuting copies of the super Virasoro algebra.

In order to see this first note that the two copies of the unperturbed Virasoro
algebra in terms of the 'polar' generators $d_r, d^\dagger_r, i\mathcal{L}_m, H_m$
read
\begin{eqnarray}
  &&\antiCommutator{d_r}{d_s} = 2\, i\mathcal{L}_{r+s} = \antiCommutator{d^\dagger_r}{d^\dagger_s}
  \nonumber\\
  &&
  \commutator{i\mathcal{L}_m}{d_r} = \frac{m-2r}{2}d_{m+r}
  \nonumber\\
  &&
  \commutator{i \mathcal{L}_m}{d^\dagger_r} = \frac{m-2r}{2}d^\dagger_{m+r}
  \nonumber\\
  &&
  \commutator{i\mathcal{L}_m}{i\mathcal{L}_n} = (m-n)i\mathcal{L}_{m+n}
  \nonumber\\
  &&
  \commutator{i\mathcal{L}_m}{H_n} = (m-n)i\mathcal{H}_{m+n} + \frac{c}{6}(m^3 - m)\delta_{m,-n}
  \nonumber\\
  &&
  \commutator{H_m}{d_r} = \frac{m-2r}{2}d^\dagger_{m+r}
  \nonumber\\
  &&
  \commutator{H_m}{d^\dagger_r} = \frac{m-2r}{2}d_{m+r}
  \nonumber\\
  &&
  \commutator{H_m}{H_n} = (m-n)\,i\mathcal{L}_{m+n}
  \,.
\end{eqnarray}

Now check that these relations are obeyed also by the deformed generators
$d^\Phi_r$, $d^{\dagger\Phi}_r$, $i\mathcal{L}_m$, $H^\Phi_m$
using the
two conditions \refer{simtrafo on susy currents} and \refer{def deformed worldsheet Hamiltonian}:

First of all the relations
\begin{eqnarray}
  \label{bracket Lm dr}
  \commutator{i\mathcal{L}_m}{d^\Phi_r} &=& \frac{m-2r}{2}d^\Phi_{m+r}
  \nonumber\\
  \commutator{i\mathcal{L}_m}{d^{\dagger \Phi}_r} &=& \frac{m-2r}{2}d^{\dagger \Phi}_{m+r}  
\end{eqnarray}
follow simply from \refer{simtrafo on susy currents} and the original bracket 
$\commutator{L_m}{G_r} = \frac{m-2r}{2}G_{m+r}$ and immediately imply
\begin{eqnarray}
  \commutator{i\mathcal{L}_m}{i\mathcal{L}_n} &=& (m-n)i\mathcal{L}_{m+n}
\end{eqnarray}
(note that here the anomaly of the left-moving sector cancels that of the right-moving one).

Furthermore 
\begin{eqnarray}
  \commutator{i\mathcal{L}_m}{H^\Phi_n} &=& 
  \commutator{i\mathcal{L}_m}{\frac{1}{2}\antiCommutator{d^\Phi_r}{d^{\dagger \Phi}_{n-r}}
  }
  \nonumber\\
  &\equalby{bracket Lm dr}&
  \frac{m-2r}{4}\antiCommutator{d^\Phi_{m+r}}{d^{\dagger \Phi}_{n-r}}
  +
  \frac{m-2(n-r)}{4}\antiCommutator{d^\Phi_{r}}{d^{\dagger \Phi}_{m+n-r}}
  \nonumber\\
  &\equalby{def deformed worldsheet Hamiltonian}&
  (m-n)H^\Phi_{m+n}
  +
  \delta_{m,-n}
  \frac{c}{6}
  \left(
    \frac{m-2r}{4}
    (4(m+r)^2-1)
    +
    \frac{m-2(n-r)}{4}(4r^2-1)
  \right)
  \nonumber\\
  &=&
  (m-n)H^\Phi_{m+n}
  +
  \delta_{m,-n}\frac{c}{6}\left(m^3 - m\right)
  \,.
\end{eqnarray}
(Here the anomalies from both sectors add.)

The commutator of the Hamiltonian with the supercurrents is obtained for instance by first writing:
\begin{eqnarray}
  \commutator{H^\Phi_m}{d^\Phi_r}
  &=&
  \frac{1}{2}
  \commutator{\antiCommutator{d_r^\Phi}{d^{\dagger \Phi}_{m-r}}}{d^\Phi_r}
  \nonumber\\
  &=&
  -
  \frac{1}{2}
  \commutator{
    \antiCommutator{d^\Phi_r}{d^\Phi_r}
  }
  {d^{\dagger \Phi}_{m-r}}
  -
  \frac{1}{2}
  \commutator{\antiCommutator{d_r^\Phi}{d^{\dagger \Phi}_{m-r}}}{d^\Phi_r}  
  \nonumber\\
  &=&
  -\commutator{i\mathcal{L}_{2r}}{d^{\dagger \Phi}_{m-r}}
  -
  \commutator{H_m^\Phi}{d^\Phi_r}
  \nonumber\\
   &=&
   (m-2r)d^{\dagger \Phi}_{m+r}
  -
  \commutator{H_m^\Phi}{d^\Phi_r}   
  \,,
\end{eqnarray}
from which it follows that
\begin{eqnarray}
  \commutator{H^\Phi_m}{d^\Phi_r}
  &=&
  \frac{(m-2r)}{2}d^{\dagger \Phi}_{m+r}
\end{eqnarray}
and similarly
\begin{eqnarray}
  \commutator{H^\Phi_m}{d^{\dagger \Phi}_r}
  &=&
  \frac{(m-2r)}{2}d^{\Phi}_{m+r}
  \,.
\end{eqnarray}
This can finally be used to obtain
\begin{eqnarray}
  \commutator{H^\Phi_m}{H_n^\Phi}
  &=&
  (m-n)i \mathcal{L}_{m+n}
  \,.
\end{eqnarray}

In summary this shows that every operator $A$ which 
\begin{enumerate}
\item
commutes with $i\mathcal{L}_m$
\item
  is such that
  $\antiCommutator{A^{-1} d_r A}{A^\dagger d^\dagger_{n-r} A^{\dagger -1}} - \delta_{n,0}\frac{c}{12}(4r^2-1)$
  is \emph{independent} of $r$
\end{enumerate}
defines a consistent deformation of the super Virasoro generators and hence a string background which
satisfies the classical equations of motion of string field theory.

In \cite{Schreiber:2004} it was shown how at least all massless NS and NS-NS backgrounds
can be obtained by deformations $A$ of the form $A = e^{\mathbf{W}}$,
where $\mathbf{W}$ is related to the vertex operator of the respective background field.
For instance a Kalb-Ramond $B$-field background is induced by setting
\begin{eqnarray}
  \label{B field deformation}
  \mathbf{W}^{(B)}
  &=&
  \frac{1}{2}\int d\sigma\; \left(\frac{1}{T}dA + B\right)_{\mu\nu}{\cal E}^{\dagger \mu}{\cal E}^{\dagger \nu}
  \,,
\end{eqnarray}
where ${\cal E}^\dagger$ are operators of exterior multiplication with differential forms on loop space,
to be discussed in more detail below in \S\fullref{Differential geometry on loop space}, and
we have included the well known contribution of the 1-form gauge field $A$.

Moreover, it was demonstrated in \cite{Schreiber:2003a} that the structure 
\refer{simtrafo on susy currents}
of the SCFT deformations allows to handle superstring evolution in nontrivial backgrounds
as generalized Dirac-K{\"a}hler evolution in loop space.

In the special case where $A$ is \emph{unitary} the similarity transformations 
\refer{simtrafo on susy currents} of $d$ and $d^\dagger$ 
and hence of all other elements of the super-Virasoro
algebra are identical and the deformation is nothing but a unitary transformation. It was
discussed in \cite{Schreiber:2004} that gauge transformations of the background fields,
such as reparameterizations or gauge shifts of the Kalb-Ramond field, are described by such 
unitary transformation.

In particular, an \emph{abelian} 
gauge field background was shown to be induced by the Wilson line
\begin{eqnarray}
  \label{abelian gauge field deformation}
  \mathbf{W}^{(A)} &=& i \oint d\sigma\; A_\mu\of{X\of{\sigma}}X^{\prime \mu}\of{\sigma}
\end{eqnarray}
of the gauge field along the closed string.

While the above considerations apply to closed superstrings,
in this paper we shall be concerned with open superstrings, since these carry the
Chan-Paton factors that will transform under the nonabelian group that 
we are concerned with in the context on nonablian 2-form background fields.

It turns out that the above method for obtaining closed string backgrounds by
deformations of the differential geometry of loop space nicely generalizes to
open strings when boundary state formalism is used. This is the content of the
next section.

\subsubsection{Boundary state deformations from unitary loop space deformations}
\label{Boundary state deformations from unitary loop space deformations}

The tree-level diagram of an open string attached to a D-brane is a disk attached to
that brane with a certain boundary condition on the disk characterizing the presence of the
D-brane. In what is essentially a generalization of the method of image charges in electrostatics 
this can be equivalently
described by the original disk ``attached'' to an auxiliary disc, so that a sphere is formed,
and with the auxiliary disk describing incoming closed strings in just such a way, that the 
correct boundary condition is reproduced. 

Some details behind this heuristic picture are recalled in
\S\fullref{Boundary state formalism}. For our purposes it suffices to note that 
a deformation \refer{simtrafo on susy currents} of the superconformal generators for
closed strings with $A$ a \emph{unitary} operator (as for instance given by 
\refer{abelian gauge field deformation}) is of course equivalent to a corresponding unitary
transformation of the closed string states. But this means that the boundary state formalism implies that 
open string dynamics in a given background described by a unitary deformation operator $A$ on loop space is
described by a boundary state $A^\dagger\ket{D9}$, where $\ket{D9}$ is the boundary state
of a bare space-filling brane, which again, as discussed below in 
\refer{condition on constant 0-form on loop space}, is nothing but the
constant 0-form on loop space.

In this way boundary state formalism rather nicely generalizes the loop space formalism 
used here from closed to open strings.

In a completely different context, the above general picture has in fact been
verified for abelian gauge fields in \cite{Hashimoto:2000,Hashimoto:1999}. 
There it is shown that acting with \refer{abelian gauge field deformation} 
and the unitary part\footnote{
  When acting on $\ket{D9}$ 
  the non-unitary part of \refer{B field deformation} is projected out automatically.
} 
of \refer{B field deformation} for $B = 0$ on $\ket{D9}$, one obtains the correct boundary state
deformation operator
\begin{eqnarray}
  \label{deform op for abelian A field}
  \exp\of{\mathbf{W}^{(A)(B = \frac{1}{T}dA)}}
  \exp\of{
    \int\limits_0^{2\pi}
   \left(
     i A_\mu X^{\prime\mu}
     +
     \frac{1}{2T}(dA)_{\mu\nu}{\cal E}^{\dagger \mu}{\cal E}^{\dagger \nu}
   \right)
  }
\end{eqnarray}
which describes open strings on a D9 brane with the given gauge field turned on.

Here, we want to show how this construction directly generalizes to deformations
describing nonabelian 1- and 2-form backgrounds. 
It turns out that the loop space perspective together with boundary state formalism
allows to identify the relation between the
nonablian 2-form background and the corresponding connection on loop space, which 
again allows to get insight into the gauge invariances of gauge theories with nonabelian 2-forms.

One simple observation of the abelian theory proves to be crucial for the non-abelian generalization:
Since \refer{deform op for abelian A field} \emph{commutes} with $\extd_K$ the loop space connection
it induces (following the reasoning to be described in \S\fullref{connections on loop space})
\emph{vanishes}. This makes good sense, since the closed string does
not feel the background $A$ field.

But the generalization of a \emph{vanishing} loop space connection to something less trivial but still trivial
enough so that it can describe something which does not couple to the closed string is a
\emph{flat} loop space connection. Flatness in loop space means that every closed curve in loop space,
which is a torus worldsheet (for the space of oriented loops) in target space, is assigned
surface holonomy $g=1$, the identity element. This means that only open worldsheets with
boundary can feel the presence of a flat loopspace connection, just as it should be.

From this heuristic picture we expect that abelian but flat loop space connections play a special
role. Indeed, we shall find in \S\fullref{2-form gauge transformations} that only these
are apparently well behaved enough to avoid a couple of well known problems.\\\

The next section 
first demonstrates that the meaning of the above constructions become
rather transparent when the superconformal generators are identified as deformed
deRham operators on loop space.

\subsubsection{Superconformal generators as deformed deRham operators on loop space.}

Details of the representation of the super Virasoro generators on loop space have
been given in \cite{Schreiber:2004} and we here follow the notation introduced there.

\paragraph{Differential geometry on loop space.}
\label{Differential geometry on loop space}

Again, the loop space formulation can nicely be motivated from boundary state formalism:

The boundary state $\ket{\mathrm{b}}$ 
describing the space-filling brane in Minkowski space is, according to
\refer{constraint on boundary state}, given by the constraints
\begin{eqnarray}
  \label{boundary constraints of bare D9 brane}
  \left(
    \alpha_n^\mu + \bar \alpha^\mu_{-n} 
  \right)
  \ket{\mathrm{b}}
  = 0\,, \hspace{.8cm} \forall\, n,\mu
 \nonumber\\
  \left(
    \psi_r^\mu - i \bar \psi^\mu_{-r} 
  \right)
  \ket{\mathrm{b}}
  = 0\,, \hspace{.8cm} \forall\, r,\mu
\end{eqnarray}
(in the open string R sector).

We can think of the super-Virasoro constraints
as a Dirac-K{\"a}hler system on the exterior bundle over loop space 
$\mathcal{L}\of{\manifold}$ with coordinates
\begin{eqnarray}
  \label{X mode expansion}
  X^{(\mu,\sigma)} &=& \frac{1}{\sqrt{2\pi}}X_0^\mu
  +
  \frac{i}{\sqrt{4\pi T}}
  \sum\limits_{n\neq 0}
  \frac{1}{n}
  \left(
    \alpha^\mu_n - \tilde \alpha_{-n}^\mu
  \right)
  e^{in\sigma}\,,
\end{eqnarray}
holonomic vector fields
\begin{eqnarray}
  \frac{\delta}{\delta X^\mu\of{\sigma}}
  \defas
  \partial_{(\mu,\sigma)}
  &=&
  i \sqrt{\frac{T}{4\pi}}
  \sum\limits_{n = -\infty}^\infty
  \eta_{\mu\nu}\left(
   \alpha_n^\nu
   +
   \tilde \alpha_{-n}^\nu
  \right)
  e^{in\sigma}
\end{eqnarray}
differential form creators
\begin{eqnarray}
  \label{loop space differential form}
  {\cal E}^{\dagger (\mu,\sigma)}
  &=&
  \frac{1}{2}
  \left(
    \psi^\mu_+\of{\sigma}
    + 
    i
    \psi^\mu_-\of{\sigma}
  \right)
  \nonumber\\
  &=&
  \frac{1}{\sqrt{2\pi}}
  \sum\limits_r
  \left(
    \bar \psi_{-r} + i \psi_r
  \right)
  e^{ir\sigma}
\end{eqnarray}
and annihilators
\begin{eqnarray}
  {\cal E}^{(\mu,\sigma)}
  &=&
  \frac{1}{2}
  \left(
    \psi^\mu_+\of{\sigma}
    -
    i
    \psi^\mu_-\of{\sigma}
  \right)
  \nonumber\\
  &=&
  \frac{1}{\sqrt{2\pi}}
  \sum\limits_r
  \left(
    \bar \psi_{-r} - i \psi_r
  \right)
  e^{ir\sigma}
  \,.
\end{eqnarray}
In the polar form \refer{polar supercurrent modes} the fermionic super Virasoro constraints are identified with
the modes of the exterior derivative on loop space
\begin{eqnarray}
  \label{deformed extd on loop space}
  \extd_K
  &=&
  \int\limits_0^{2\pi}
  d\sigma\;
  \left(
  {\cal E}^{\dagger \mu}
  \partial_{\mu}
  \of{\sigma}
  +
  i T X^{\prime \mu}
  {\cal E}_{\mu}
  \of{\sigma}
  \right)
  \,,
\end{eqnarray}
deformed by the reparametrization Killing vector 
\begin{eqnarray}
  \label{rep Killing vector}
  K^{(\mu,\sigma)}
  &\defas&
  X^{\prime \mu}\of{\sigma}
  \,,
\end{eqnarray}
where $T = \frac{1}{2\pi \alpha^\prime}$ is the string tension.
The Fourier modes of this operator are the polar operators of \refer{polar supercurrent modes}
\begin{eqnarray}
  d_r &\propto& \oint d\sigma\; e^{-ir\sigma} \extd_K\of{\sigma}
  \,. 
\end{eqnarray}

Using this formulation of the super-Virasoro constraints it would seem natural to 
represent them on a Hilbert space whose 'vacuum' state $\ket{\mathrm{vac}}$ is the 
\emph{constant 0-form} on loop space, i.e.
\begin{eqnarray}
  \label{condition on constant 0-form on loop space}
  \partial_{(\mu,\sigma)}\ket{\mathrm{vac}} = 0 = {\cal E}_{(\mu,\sigma)}\ket{\mathrm{vac}}
  \hspace{.8cm}
  \forall\, \mu,\sigma
  \,.
\end{eqnarray}
While this is not the usual $\mathrm{SL}\of{2,\C}$ invariant vacuum of the closed
string, it is precisely the boundary state 
\refer{boundary constraints of bare D9 brane}
\begin{eqnarray}
  \label{D9 boundary state}
  \ket{\mathrm{vac}} &=& \ket{\mathrm{b}}
\end{eqnarray}
describing the D9 brane.

For the open string NS sector the last relation of \refer{boundary constraints of bare D9 brane}
changes the sign
\begin{eqnarray}
  \left(
    \psi_r^\mu + i \bar \psi^\mu_{-r} 
  \right)
  \ket{\mathrm{b}^\prime}
  = 0\,, \hspace{.8cm} \forall\, r,\mu  \;\;\mbox{NS sector}
\end{eqnarray}
and now implies that the vacuum is, from the loop space perspective, 
the formal \emph{volume form} instead of the
constant 0-form, i.e. that form annihilated by all differential form multiplication operators:
\begin{eqnarray}
  \partial_{(\mu,\sigma)}\ket{\mathrm{b}^\prime} = 0 = {\cal E}^{\dagger (\mu,\sigma)}\ket{\mathrm{b}^\prime}
  \hspace{.8cm}
  \forall\, \mu,\sigma
  \,.
\end{eqnarray}
In finite dimensional flat manifolds of course both are related simply by \emph{Hodge duality}:
\begin{eqnarray}
  \ket{\mathrm{b}^\prime} = \star \ket{\mathrm{b}}
  \,.
\end{eqnarray}
So Hodge duality on loop space translates to the $\mbox{NS} \leftrightarrow {R}$ transition on the
open string sectors.

\paragraph{Connections on loop space.}
\label{connections on loop space}

It is now straightforward to identify the relation between background fields induced by
deformations \refer{simtrafo on susy currents} and connections on loop space. A glance at
\refer{deformed extd on loop space} shows that we have to interpret the term of
differential form grade $+1$ in the polar supersymmetry generator as 
${\cal E}^{\dagger}\gradOp^{(\Phi)}_\mu$, where $\gradOp^\Phi_\mu$ is a loop space
connection (covariant derivative) induced by the target space background field $\Phi$.

Indeed, as was shown in \cite{Schreiber:2004}, one finds for instance that a gravitational background
$G_{\mu\nu}$ leads to $\gradOp^{(G)}$ which is just the Levi-Civita connection on loop
space with respect to the metric induced from target space. Furthermore, an \emph{abelian}
2-form field background is associated with a deformation operator 
\begin{eqnarray}
  \label{abelian B field deformation}
  \mathbf{W}^{(B)}
  &=&
  \oint d\sigma\
  B_{\mu\nu}{\cal E}^{\dagger \mu}{\cal E}^{\dagger \nu}
\end{eqnarray}
and leads to a connection
\begin{eqnarray}
  \label{abelian loop space connection}
  \gradOp^{(G)(B)}_\mu &=& \gradOp^{(G)}_\mu - i T B_{\mu\nu}X^{\prime \nu}
  \,,
\end{eqnarray}
just as expected for a string each of whose points carries $U(1)$ 
charge under $B$ proportional to the length element $X^\prime \,d\sigma$.

\subsection{BSCFT deformation for nonabelian 2-form fields}
\label{BSCFT deformation for nonabelian 2-form fields}

The above mentioned construction can now be used to examine deformations that involve
nonabelian 2-forms:

\subsubsection{Nonabelian Lie-algebra valued forms on loop space}

When gauge connections on loop space take values in nonabelian 
algebras deformation operators such as $\exp\of{\mathbf{W}^{(A)}}$ \refer{abelian gauge field deformation} and 
$\exp\of{\mathbf{W}^{(B)}}$
\refer{abelian B field deformation} obviously have to be replaced by path ordered exponentiated
integrals. The elementary properties of loop space differential forms involving such
path ordered integrals are easily derived, and were for instance given in 
\cite{GetzlerJonesPetrack:1991,Hofman:2002}.

So consider a differential $p+1$ form $\omega$ on \emph{target space}. It lifts
to a $p+1$-form $\Omega$ on loop space given by
\begin{eqnarray}
  \Omega
  &\defas&
  \frac{1}{(p+1)!}
  \int\limits_{S^1}
  \omega_{\mu_1\cdots \mu_{p+1}}\of{X}
   {\cal E}^{\dagger \mu_1}
  \cdots
   {\cal E}^{\dagger \mu_{p+1}} 
  \,. 
\end{eqnarray}

Let $\hat K = X^{\prime (\mu,\sigma)}{\cal E}_{(\mu,\sigma)}$ be the operator of interior
multiplication with the reparametrization Killing vector $K$ \refer{rep Killing vector}
on loop space. The above $p+1$-form
is sent to a $p$-form $\oint (\omega)$ on loop space by contracting with this Killing vector
(brackets will always denote the graded commutator):
\begin{eqnarray}
  \oint (\omega)
  &\defas&
  \commutator{\hat K}{\Omega}
  \nonumber\\
  &=&
  \frac{1}{p!}
  \int\limits_{S^1}
  d\sigma\;
  \omega_{\mu_1 \cdots \mu_{p+1}}
  X^{\prime \mu_{1}}
  {\cal E}^{\dagger \mu_2}\cdots
  {\cal E}^{\dagger \mu_{p+1}}
  \,.
\end{eqnarray}

The anticommutator of the loop space exterior derivative $\extd$ with $\hat K$
is just the reparametrization Killing Lie derivative
\begin{eqnarray}
  \commutator{\extd}{\hat K}
  &=&
  i{\cal L}_K
\end{eqnarray}
which commutes with 0-modes of fields of definite reparametrization weight, e.g.
\begin{eqnarray}
  \commutator{{\cal L}}{\Omega} &=& 0
  \,.
\end{eqnarray}
It follows that 
\begin{eqnarray}
  \label{an equation}
  \commutator{\extd }{\commutator{\hat K}{ \Omega}}
  &=&
  \commutator{\cal L}{\Omega}
  -
  \commutator{\hat K }{\commutator{\extd}{ \Omega}}
\end{eqnarray}
which implies that
\begin{eqnarray}
  \commutator{\extd}{\oint (\omega)}
  &=&
  \oint (-d\omega)
  \,.
\end{eqnarray}

The generalization to multiple path-ordered integrals
\begin{eqnarray}
  \label{path loop space integral}
  \oint (\omega_1, \cdots, \omega_n)
  &\defas&
  \int\limits_{0 < \sigma_{i-1} < \sigma_i < \sigma_{i+1} < \pi}
  \!\!\!\!\!\!\!
  d^n\sigma\;
  \commutator{\hat K}{\omega_1}\of{\sigma_1}\cdots
  \commutator{\hat K}{\omega_n}\of{\sigma_n}
\end{eqnarray}
is
\begin{eqnarray}
  \label{extd on path ordered forms}
  &&\commutator{\extd}{\oint (\omega_1, \cdots,\omega_n)} = 
  \nonumber\\
  &=&
  -\sum\limits_k (-1)^{\sum\limits_{i < k}p_i}
  \left(
    \oint (\omega_1, \cdots, d\omega_k, \cdots, \omega_n)
    +
    \oint (\omega_1,\cdots, \omega_{k-1}\wedge \omega_k,\cdots,\omega_n)
  \right)    
  \,.
  \nonumber\\
\end{eqnarray}
This is proposition 1.6 in \cite{GetzlerJonesPetrack:1991}.

In the light of \refer{simtrafo on susy currents} we are furthermore interested in expressions of the form
$  U_A\of{2\pi,0}
  \circ
  \extd_K
  \circ
  U_A\of{0,2\pi}$
where $U_A$ is the holonomy of $A$.

Using
\begin{eqnarray}
  \commutator{\extd}{U_A\of{0,2\pi}}
  &=&
  \commutator{\extd}{
    \sum\limits_{n=0}^\infty
    \oint \underbrace{(iA, \cdots,iA)}_{\mbox{$n$ times}}
  }
  \nonumber\\
  &=&
  -
  \sum\limits_{n=0}^\infty
  \sum\limits_{k}
  \oint 
  (iA, \cdots, iA,iF_A,iA,\cdots iA)_{\mbox{\tiny $n$ occurences of $iA$, $F_A$ at $k$}}
  \nonumber\\
  &=&
  \int\limits_0^{2\pi}
  d\sigma\;
  U_A\of{0,\sigma}
  \commutator{i F_A}{\hat K}\of{\sigma}
  U\of{\sigma,2\pi}\,
\end{eqnarray}
where
\begin{eqnarray}
  F_A &=& -i(d + iA)^2 
  \nonumber\\
  &=&
  dA + i A\wedge A
\end{eqnarray}
is the field strength of $A$ (which is taken to be hermitean),
one finds
\begin{eqnarray}
  \label{lextd deformed by Wilson line}
  U_A\of{2\pi,0}
  \circ
  \extd
  \circ
  U_A\of{0,2\pi}
  &=&
  \extd
  +
  \int\limits_0^{2\pi}
  d\sigma\,
  U_A\of{2\pi,\sigma}
  \commutator{i F_A}{\hat K}\of{\sigma}
  U_A\of{\sigma,2\pi}
  \,.
\end{eqnarray}

The point that will prove to be crucial in the following discussion is that there
is an $A$-holonomy on \emph{both} sides of the 1-form factor. The operator on the
right describes parallel transport with $A$ from $2\pi$ to $\sigma$, application of
$\commutator{\extd_A A}{\hat K}$ at $\sigma$ and then parallel transport back from $\sigma$
to $2\pi$. Following \cite{Hofman:2002} the abbreviating notation
\begin{eqnarray}
  \label{def ointA}
  \oint_A (\omega)
  &\defas&
  \int_0^{2\pi}
  d\sigma\;
  U_A\of{2\pi,\sigma}
  \commutator{\hat K}{\omega}
  U_A\of{\sigma,2\pi}
\end{eqnarray}
will prove convenient. (But notice that in \refer{def ointA} there is also a factor 
$U_A\of{\sigma,2\pi}$ on the  \emph{right}, which does not appear in \cite{Hofman:2002}.)
Using this notation \refer{lextd deformed by Wilson line} is rewritten as
\begin{eqnarray}
  \label{pure Wilson line deformation}
  U_A\of{2\pi,0}
  \circ
  \extd
  \circ
  U_A\of{0,2\pi}
  &=&
  \extd -i \oint_A (F_A)
  \,.  
\end{eqnarray}

This expression will prove to play a key role in the further development. In order
to see why this is the case we now turn to the computation and disucssion of the
connection on loop space which is induced by the nonabelian 2-form background.

\subsubsection{Nonabelian 2-form field deformation}
\label{Nonabelian 2-form field deformation}

With the above considerations it is now immediate how to incorporate a 
nonabelian 2-form in the target space of a boundary superconformal field
theory on the worldsheet. The direct generalization of 
\refer{abelian gauge field deformation} and 
\refer{abelian B field deformation} is obviously the deformation operator
\begin{eqnarray}
  \label{nonab B deformation}
  \exp\of{\bf W}^{(A)(B)_\mathrm{nonab}}
  &=&
  \mathrm{P}
  \exp\of{\int\limits_0^{2\pi}
    d\sigma\;
    \left(
      i A_\mu X^{\prime\mu}
      +
      \frac{1}{2}
      \left(\frac{1}{T}F_A + B\right)_{\mu\nu} {\cal E}^{\dagger \mu}{\cal E}^{\dagger \nu}
    \right)
  }
\end{eqnarray}
for non-abelian and hermitean $A$ and $B$. (P denotes path ordering)
Note that this is indeed reparametrization invariant on 
$\mathcal{L}\of{\manifold}$ and that
there is no trace in \refer{nonab B deformation}, so that
this must act on an appropriate bundle, 
which is naturally associated with a stack of $N$ branes (\cf pp. 3-4 of \cite{Zunger:2002}).

According to \S\fullref{connections on loop space} the loop space connection induced by
this deformation operator is given by the term of degree $+1$ in the deformation
of the superconformal generator \refer{deformed extd on loop space}. Using
\refer{lextd deformed by Wilson line} this is found to be
\begin{eqnarray}
  \label{non-abelian 2-form field deformation}
  \exp\of{-{\bf W}}^{(A)(B)_\mathrm{nonab}}
    \circ
    \extd_K
    \circ
  \exp\of{\bf W}^{(A)(B)_\mathrm{nonab}}
  &=&
  \extd
  +iT
  \oint_A (B)
  \nonumber\\
  &&\;\;
  +
  \mbox{(terms of grade $\neq 1$)}
  \,,
  \nonumber\\
\end{eqnarray}
where the notation \refer{def ointA} is used.

The second term $iT\oint_A (B)$ is the nonablian 1-form connection on loop space which is induced
by the target space 2-form $B$. Note that the terms involving
the $A$-field strength $d_A A$ coming from the $X^\prime$ term and 
those coming from the ${\cal E}^\dagger{\cal E}^\dagger$ 
term in \refer{non-abelian 2-form field deformation} mutually cancel. 

The connection \refer{non-abelian 2-form field deformation} is essentially that found,
by different means in different cóntexts, in 
\cite{Hofman:2002},\cite{AlvarezFerreiraSanchezGuillen:1998} and \cite{Chepelev:2002}.

\subsubsection{2-Form Gauge Transformations}
\label{2-form gauge transformations}

In a gauge theory with a nonabelian 2-form one expects the usual gauge invariance
\begin{eqnarray}
  \label{1st gauge trafo}
   A &\mapsto& U\,A\,U^\dagger + U(dU^\dagger)
  \nonumber\\
  B &\mapsto& U\,B\,U^\dagger
\end{eqnarray}
together with some nonabelian analogue of the infinitesimal shift
\begin{eqnarray}
  \label{2nd gauge trafo}
  A &\mapsto& A + \Lambda
  \nonumber\\
  B &\mapsto& B - d_A \Lambda + \cdots
\end{eqnarray}
familiar from the abelian theory.

With the above results, it should be possible to derive some properties of the gauge invariances of 
a nonabelian 2-form theory from loop space reasoning.
That's because on loop space 
\refer{non-abelian 2-form field deformation} is an ordinary
connection 1-form. The ordinary 1-form gauge transformations of that loop space connection should 
give rise to something like \refer{1st gauge trafo} and \refer{2nd gauge trafo} automatically.

Indeed, ``global'' gauge transformations 
(i.e. position-independent ones)
of 
\refer{non-abelian 2-form field deformation} on loop space give rise to 
\refer{1st gauge trafo}, while infinitesimal gauge transformations on loop space
give rise to \refer{2nd gauge trafo}, but with correction terms that only have
an interpretation on loop space.

More precisely, let $U\of{X} = U$ be any \emph{constant} group valued function on (a local patch of) loop space
and let $V(X) : \manifold \to G$ 
be such that $\lim\limits_{\epsilon \to 0} V(X)\of{X\of{\epsilon}} = V(X)\of{X\of{2\pi - \epsilon}} = U$, 
then
\begin{eqnarray}
  \label{1st gauge trafor from loop space}
  U \left(\oint_A (B)\right) U^\dagger + U(dU^\dagger)
  &=&
  \oint_{A^\prime}(B^\prime)
\end{eqnarray}
with
\begin{eqnarray}
  A^\prime &=& V\, A \, V^\dagger + V(dV^\dagger)
  \nonumber\\
  B^\prime &=& V\, B\, V^\dagger
  \,,
\end{eqnarray}
which reproduces \refer{1st gauge trafo}.

If, on the other hand, $U$ is taken to be a \emph{nonconstant} infinitesimal gauge transformation 
with a 1-form gauge parameter $\Lambda$ of the form
\begin{eqnarray}
  U\of{X} &=& 1 - i\oint_A (\Lambda)
\end{eqnarray}
then
\begin{eqnarray}
  U \left(\oint_A (B)\right)U^\dagger + U(dU^\dagger)
  &=&
  \oint_{A + \Lambda} (B + d_A B)
  +
  \cdots
  \,.
\end{eqnarray}

The first term reproduces \refer{2nd gauge trafo}, but there are further terms which 
do not have analogs on target space.

The reason for this problem can be understood from the boundary deformation operator point of view:

Let 
\begin{eqnarray}
  \mathrm{P} \exp\of{i\int\limits_0^{2\pi} R}
  =
  \lim\limits_{N=1/i\epsilon \to \infty} (1 + i\epsilon R(0))\cdots (1+i\epsilon R\of{\epsilon 2\pi})\cdots
  (1 + i\epsilon R\of{2\pi})
\end{eqnarray}
be the path ordered integral over some object $R$. Then a small shift $R \to R + \delta R$ amounts to
the ``gauge transformation''
\begin{eqnarray}
  \label{shift in boundary deformation operator}
  \mathrm{P} \exp\of{i\int\limits_0^{2\pi} R}
  &\to&
  \mathrm{P} \exp\of{i\int\limits_0^{2\pi} R}
  \left(
    1 + i\oint_R \delta R
  \right)
\end{eqnarray}
to first order in $\delta R$. 
Notice how, using the definition of $\oint_R$ given in \refer{def ointA}, the term $\oint_R \delta R$
inserts $\delta R$ successively at all $\sigma$ in the preceding Wilson line.

So it would seem that $U = 1 - i \oint_R \delta R$ is the
correct unitary operator, to that order, for the associated transformation. But the
problem ist that $R$ is in general not purely bosonic, but contains fermionic
contributions. These spoil the ordinary interpretation of the above $U$ as a gauge
transformation.

This means that an ordinary notion of gauge transormation is obtained if and only
if the fermionic contributions in 
\refer{nonab B deformation} disappear, which is the case when 
\begin{eqnarray}
  \label{fake flatness condition in BSCFT discussion}
  B &=& - \frac{1}{T}F_A
  \,.
\end{eqnarray}
Then 
\begin{eqnarray}
  \label{pure A Wilson line}
  \exp\of{\mathbf{W}}^{(A)(B = - \frac{1}{T}F_A)}
  &=&
  \mathrm{P}
  \exp\of{\int\limits_0^{2\pi} d\sigma\; i A_\mu X^{\prime\mu}\of{\sigma}}
\end{eqnarray}
is the pure $A$-Wilson line
and and the corresponding gauge covariant exterior derivative on loop space is
\begin{eqnarray}
  \label{flat loop space connection}
  \extd^{(A)(B)} &=& \extd - i \oint_A (F_A)
  \,,
\end{eqnarray}
as in \refer{pure Wilson line deformation}.
Now the transformation 
\begin{eqnarray}
  A &\mapsto& A + \Lambda
  \nonumber\\
  B &\mapsto& B - \frac{1}{T}\left( d\Lambda + i A \wedge \Lambda + i \Lambda \wedge A\right)
\end{eqnarray}
is correctly, to first order, induced by the loop space gauge transformation
\begin{eqnarray}
  U\of{X} &=& 1 - i \oint_A (\Lambda)
  \,.
\end{eqnarray}

The restriction 
\refer{fake flatness condition in BSCFT discussion}
was previously found in \cite{GirelliPfeiffer:2004}
in the context of categorified lattice gauge theory.
The discussion there uses the generalization of the above considerations
from the case where $A$ and $B$ take values in the same algebra
to that where they take values in a differential crossed module.
The above loop space considerations are generalized to this case in 
\S\fullref{section: Path Space}. See in particular
Prop. \refer{effect of infinitesimal gauge transformations on path space}
(p. \pageref{effect of infinitesimal gauge transformations on path space}).

\subsection{Gerstenhaber Brackets and Hochschild Cohomology}
\label{Gerstenhaber brackets and Hochschild Cohomology}

In \cite{Hofman:2002} C. Hofman had presented some considerations
which in some respects are closely related to the discussion above,
though different. Here we briefly review the key
observations of \cite{Hofman:2002} and discuss how they 
are related to the above. Interestingly, this involves 
considerations rather similar to but again different from
some aspects to be discussed in 
\S\fullref{The Differential Picture: Morphisms between p-Algebroids}. 
While at the moment this
appears like a coincidence, it could be that there is something
deeper going on which is not yet fully understood.

\vskip 2em

In \cite{Hofman:2002}
it was noted that equation \refer{extd on path ordered forms}
suggests that in differential calculus on pull-back forms on path space
\emph{graded multi-derivations} play an important role.

Clearly, if we consider the string of forms $\of{\omega_1,\dots,\omega_n}$
as an $n$-fold associative abstract product of some unspecified sort and 
if we associate a grade
\begin{eqnarray}
  \label{grade of a pulled back form}
  |\omega_i| &\defas& p_i = \mathrm{deg}\of{\omega_i} - 1
\end{eqnarray}
with each factor 
then the first term in \refer{extd on path ordered forms}
can be thought of as coming from the application of the 
\emph{unary} derivation
$\phi_d$ of odd grade $|\phi_d| = 1$ defined by\footnote{
  We have some (inessential) signs different from those in \cite{Hofman:2002}.
}
\begin{eqnarray}
  \phi_d\of{\omega_1,\dots,\omega_n}
  &\defas&
  \sum\limits_{k=0}^{n-1}
  (-1)^{\sum_{i=1}^k |\phi_d||\omega_i| }
  \of{
    \omega_1,\dots,\omega_k,-d\omega_{k+1},\omega_{k+2},\dots,\omega_n
   }
\end{eqnarray}
while the second term can be thought of as due to a \emph{binary} derivation $\phi_M$
of grade $|\phi_M| = 1$
\begin{eqnarray}
  \phi_M\of{\omega_1,\dots,\omega_n}
  &\defas&
  \sum\limits_{k=0}^{n-2}
  (-1)^{\sum_{i=1}^k |\phi_M||\omega_i| }
  \of{
    \omega_1,\dots,\omega_k,M\of{\omega_{k+1},\omega_{k+2}},\dots,\omega_n
   }
  \,,
\end{eqnarray}
where the binary map $M$ is, up to a sign, the (wedge) product operation
\begin{eqnarray}
  \label{product multi-derivation with special sign}
  M\of{\omega_1,\omega_2} &=& (-1)^{|\omega_1|}\omega_1 \wedge \omega_2
  \,.
\end{eqnarray}
For these two (multi-)derivations it so happens that their grade $|\phi|$ 
(defined by the signs in the above sums)
is related to the grade of their
image (defined by \refer{grade of a pulled back form}) simply by
\begin{eqnarray}
  \label{relation of grade to grade of image}
  |\phi\of{\omega_1,\dots,\omega_n}|
  =
  |\phi| + \sum\limits_{k=1}^n |\omega_k|
  \,.
\end{eqnarray}
An arbitrary graded multi-derivation has no reason to satisfy this relation. But those that do
have the nice property that they form a closed algebra under the graded Lie bracket 
given by the graded commutator 
\begin{eqnarray}
  \commutator{\phi_1}{\phi_2}
  &\defas&
  \phi_1 \circ \phi_2 - (-1)^{|\phi_1||\phi_2|}\;\phi_2\circ \phi_1
  \,.
\end{eqnarray}
A simple calculation shows that this bracket respects the grade
\begin{eqnarray}
  |\commutator{\phi_1}{\phi_2}| &=& |\phi_1| + |\phi_2|
\end{eqnarray}
and that if $\phi_1$ is $n_1$-ary and $\phi_2$ is $n_2$-ary the resulting derivation
is $(n_1+n_2-1)$-ary and given by
\begin{eqnarray}
  \label{graded bracket of two multi-derivations}
  &&\commutator{\phi_1}{\phi_2}\of{\omega_1,\dots,\omega_{n_1+n_2-1}}
  \nonumber\\
  &=&
  \sum\limits_{r=0}^{n_1-1}
  (-1)^{\sum_{i=1}^r |\phi_2||\omega_{i}|}
  \phi_1\of{
    \omega_1,\dots,\omega_r,
    \phi_2\of{\omega_{r+1},\dots,\omega_{r+n_2}},
    \omega_{r+n_2+1},\dots, \omega_{r+n_1+n_2 - 1}
  }
  \nonumber\\
  &&
  -
  (-1)^{|\phi_1||\phi_2|}
  \sum\limits_{r=0}^{n_2-1}
  (-1)^{\sum_{i=1}^r |\phi_1||\omega_{i}|}
  \phi_2\of{
    \omega_1,\dots,\omega_r,
    \phi_1\of{\omega_{r+1},\dots,\omega_{r+n_1}},
    \omega_{r+n_1+1},\dots, \omega_{r+n_1+n_2 - 1}
  }
  \,.
  \nonumber\\
\end{eqnarray}

As noted in \cite{Hofman:2002}, this operation is related to the so-called Gerstenhaber bracket
between multilinear maps. 

In order to see this, first note that for pull-back forms 
the grade $|\phi|$ of the multi-derivation
which satisfy \refer{relation of grade to grade of image} is the sum of the 
differential form
degree they carry and their arity reduced by one, i.e.
\begin{eqnarray}
  |\phi| &=& \mathrm{deg}\of{\phi} + n_\phi - 1
\end{eqnarray}
(because they remove $n$ contractions with $X^\prime$ and add one.)

Next it is instructive to first restrict attention to the case where all form degrees of 
the $\omega$s as well as of the
$\phi$ are \emph{even}, which is the purely 'bosonic' case. In this case the
grade $|\omega|$ of a form in \refer{grade of a pulled back form} is \emph{odd} and the
grade $|\phi|$ of our multi-derivations is $n-1$.

So in this case \refer{graded bracket of two multi-derivations} reduces to the ordinary
\emph{Gerstenhaber bracket}
\begin{eqnarray}
  \commutator{\phi_1}{\phi_2}
  &=&
  \phi_1\of{\phi_2\of{\dots},\dots}
  -
  (-1)^{(n_1-1)(n_2-1)}
  \phi_2\of{\phi_1\of{\dots},\dots}
 \nonumber\\
  &&+
  \left(
  (-1)^{(n_2-1)}
  \phi_1\of{\cdot,\phi_2\of{\dots},\dots}
  -
  (-1)^{(n_1-1)((n_2-1)+1)}
  \phi_2\of{\cdot,\phi_1\of{\dots},\dots}
  \right)
  \nonumber\\
  &&
  +
  \left(
  \phi_1\of{\cdot,\cdot,\phi_2\of{\dots},\dots}
  -
  (-1)^{(n_1-1)(n_2-1)}
  \phi_2\of{\cdot,\cdot,\phi_1\of{\dots},\dots}
  \right)
  \nonumber\\
  &&+
  \cdots
  \,,
\end{eqnarray}
which reproduces all the familiar algebraic relations between maps: 
For instance let $M$
be binary, $\mu$ unary and $\alpha$ $0$-ary (all of even form degree), then 
(for all their arguments of even form degree, too) the above says that

\begin{eqnarray}
  \commutator{\mu\of{\cdot}}{\alpha} &=& \mu\of{\alpha}
  \nonumber\\
  \commutator{\mu\of{\cdot}}{\nu\of{\cdot}} &=& \mu\of{\nu\of{\cdot}} - \nu\of{\mu\of{\cdot}}
  \nonumber\\
  \commutator{M\of{\cdot,\cdot}}{\alpha}
  &=&
  M\of{\alpha,\cdot} - M\of{\cdot,\alpha}
  \nonumber\\
  \commutator{M\of{\cdot,\cdot}}{\mu\of{\cdot}} 
  &=& 
  M\of{\mu\of{\cdot},\cdot} -
  M\of{\cdot,\mu\of{\cdot}}
  -
  \mu\of{M\of{\cdot,\cdot}}
 \nonumber\\
  \commutator{M\of{\cdot,\cdot}}{M\of{\cdot,\cdot}}
  &=&
  2 \left(
   \;
    M\of{M\of{\cdot,\cdot},\cdot} - M\of{\cdot,M\of{\cdot,\cdot}}
   \;
  \right)
  \,.
\end{eqnarray}
The expressions here are, respectively,
\begin{enumerate}
 \item application of a function to  its argument
 \item commutator of two operators
 \item commutator (a measure for the failure of $M$ to define a commutaive product)
 \item 'derivator' (a measure for the failure of $\mu$ to be a derivation of $M$)
 \item associator (a measure for the failure of $M$ to define an associative product).
\end{enumerate} 

Now it is easy to understand the general case where the form degrees may be odd:
The above commutators simply become graded commutators and the derivator becomes a graded derivator
in the familiar way. Note that the associator remains intact if for $M$ the product
\refer{product multi-derivation with special sign} is used, so that
\begin{eqnarray}
   \commutator{M}{M} &=& 0
\end{eqnarray}
(if the product on any internal degrees of freedom is associative).

With this in hand some interesting things about the exterior derivative on pull-back forms on loop space
can be said:

If from now on the letter $M$ is reserved for the special product derivation 
\refer{product multi-derivation with special sign} and if $d$ denotes the
obvious unary derivation, we can write
\begin{eqnarray}
  \extd \oint\of{\omega_1,\dots,\omega_n}
  &=&
  \oint (d + M)\of{\omega_1,\dots,\omega_n}
  \,.
\end{eqnarray}
The square is
\begin{eqnarray}
  \extd^2 \oint\of{\omega_1,\dots,\omega_n}
  &=&
  \oint 
  \left(
    \frac{1}{2}\commutator{d}{d} + \frac{1}{2}\commutator{M}{M} + \commutator{d}{M}
  \right)
  \of{\omega_1,\dots,\omega_n}
  \,.
\end{eqnarray}
Using the above insight we find that all three terms vanish by themselves:
\begin{eqnarray}
  \frac{1}{2}\commutator{d}{d}\of{\omega_1} &=& dd\omega_1 = 0
  \nonumber\\
  \frac{1}{2}
  \commutator{M}{M}\of{\omega_1,\omega_2,\omega_3} &=&
  (-1)^{|\omega_1||\omega_2|}
  \left(
    (-1)^{|\omega_1|}
    \omega_1\wedge \omega_2
  \right)
  \wedge
  \omega_3
  -
  (-1)^{2|\omega_1|}
  \omega_1
  \wedge
  \left(
    (-1)^{|\omega_2|}
    \omega_2 \wedge \omega_3
  \right)
  = 0
  \nonumber\\
  \commutator{d}{M}\of{\omega_1,\omega_2}
  &=&
  (-1)^{|\omega_1|}\,d(\omega_1\wedge\omega_2)
  +
  (-1)^{|\omega_1|+1}\, (d\omega_1)\wedge \omega_2 + (-1)^{2|\omega_1|}\,\omega_1\wedge (d\omega_2)
  \nonumber\\
  &=&
  (-1)^{|\omega_1|}
  \left(
    d(\omega_1\wedge \omega_2)
    -
    (d\omega_1)\wedge \omega_2
    -
    (-1)^{\mathrm{deg}\of{\omega_1}}
    \,
    \omega_1\wedge (d\omega_2)
  \right)
  = 0 
  \,,
\end{eqnarray}
due to the fact that the ordinary exterior derivative is nilpotent and a graded 
derivation of the ordinary wegde product, which is associative.

\vskip 2em

In \cite{Hofman:2002} it was argued that one should consider generalizations of
the multi-derivation $\phi_d + \phi_M$ using a unary 1-form $A$ and a 0-ary 2-form
$B$ multi-derivation (thus using all possible derivations of grade 1), 
to obtain the derivation $\phi_d + \phi_M + \phi_A + \phi_B$.
However, the motivation for this proposal used an argument which was a little shaky
(for instance according to that argument there should have really been a term proportional to the
field strength of $A$ in equation (15) of \cite{Hofman:2002}). But there is a way to 
derive this total derivation from an interesting loop space expression:

We know from the tensionful superstring  
that the natural modification of the exterior
derivative on loop space is a polar combination of the worldsheet supercharges,
namely the object \refer{k-defomred extds on loop space}
(p. \pageref{k-defomred extds on loop space})
\begin{eqnarray}
  \extd_K &\defas& \extd + iT\,\iota_K
  \,,
\end{eqnarray}
where
$\iota_K$ is the operator of inner multiplication with the loop space vector
$K^{(\mu,\sigma)} = X^{\prime\mu}\of{\sigma}$, which gerenrates rigid reparameterizations,
and we have kept the string tension $T$ (a constant) for later discussion of the limit
$T \to 0$.

Moreover,  from the above we know 
that the Wilson line of $A$ along the string
naturally generalizes to the multi-form
\begin{eqnarray}
  \label{intermediator}
  W[X]\of{\sigma,\sigma^\prime}
  &\defas&
  \mathrm{P}
  \exp\of{
    \int\limits_\sigma^{\sigma^\prime} d\sigma\; 
    \left(
      i A\mu \inner X^{\prime\mu}
      +
      \frac{1}{2}\left(
        \frac{1}{T}
        F_A
        +
        B
      \right)_{\mu\nu}
      dX^\mu\wedge dX^\nu \wedge
    \right)\of{\sigma}
  }
  \mathbf{1}
  \nonumber\\
  \,,
\end{eqnarray}
where $\mathbf{1}$ denotes the constant unit 0-form on loop space.
This was our object of interest in 
\S\fullref{BSCFT deformation for nonabelian 2-form fields}. 

But now let us try to generalize both this approach as well as the one
in \cite{Hofman:2002} and consider modified pull-back forms that have the above
generalized Wilson line between each factor:
\begin{eqnarray}
  \oint\limits_{(A,B)}
  \of\omega\of{
    \omega_1, \dots, \omega_n
  }
  &\defas&
  \int\limits_{0 < \sigma_i < \sigma_{i+1} < 1\,\forall\, i}
  W\of{0,\sigma_1}
  \iota_K\of{\omega_1}\of{\sigma_1}
  W\of{\sigma_1,\sigma_2}
  \iota_K\of{\omega_1}\of{\sigma_1}
  \cdots
  W\of{\sigma_n,1}  
    \,.
  \nonumber\\
\end{eqnarray}
The point of this definition is that the action of the modified exterior derivative
$\extd_K$ on this object in a certain scaling limit reproduces the action of the multi-derivations
proposed in \cite{Hofman:2002} up to an extra term:

If we let $B$ scale as $1/T$ then 
\begin{eqnarray}
  \extd_K \oint\limits_{(A,B)}\of{\omega_1,\dots,\omega_n}
  &=&
  \;\;\;
  \oint\limits_{(A,B)} (d+M+A+B)\of{\omega_1,\dots,\omega_n}
  +
  \order{1/T}
  \,,
\end{eqnarray}
where the remaining terms of order $1/T$ have no further contractions with $\iota_K$.
Hence there is a scaling limit of \emph{large} string tension $T \to \infty$ with
$T\,B$ fixed in which Hofman's multi-derivation are obtained from a proper loop space
differential.

Applying $\extd_K$ twice yields (\cf equation (23) in \cite{Hofman:2002})
\begin{eqnarray}
  (d+M+A+B)(d+M+A+B)
  &=&
  \mathcal{H} + \mathcal{F} + \mathcal{N} + \mathcal{K}
\end{eqnarray}
where
\begin{eqnarray}
  \mathcal{H} &=& 
  \commutator{d}{B} + \commutator{A}{B}
  \nonumber\\
  &=&
  dB + A\of{B}
  \nonumber\\
  \mathcal{F}\of{\omega} &=&
  \left(
    \commutator{d}{A} + \frac{1}{2}\commutator{A}{A} + \commutator{M}{B}
  \right)
  \of{\omega}
  \nonumber\\
  &=&
  d\of{A\of{\omega}} + A\of{A\of{\omega}} + M\of{B,\omega} + (-1)^{|\omega|}M\of{\omega,B}
  \nonumber\\
  &=&
  d\of{A\of{\omega}} + A\of{A\of{\omega}} - B\wedge\omega + \omega \wedge B
  \nonumber\\
  \mathcal{N}\of{\omega_1,\omega_2}
  &=&
  \left(\commutator{d}{M} + \commutator{A}{M}\right)\of{\omega_1,\omega_2} 
   \nonumber\\
  &=&
  (-1)^{|\omega_1|}
  \left(
    d_A\of{\omega_1\wedge \omega_2}
    -
  (d_A\omega_1)\wedge \omega_2
  -
  (-1)^{\mathrm{deg}\of{\omega_2}}
  \omega_1 \wedge (d_A\omega_2) 
  \right)
  \nonumber\\
  \mathcal{K}\of{\omega_1,\omega_2,\omega_3}
  &=&
  \frac{1}{2}\commutator{M}{M}\of{\omega_1,\omega_2,\omega_3}  
  \nonumber\\
  &=&
  (-1)^{|\omega_1||\omega_2|}
  \left(
    (-1)^{|\omega_1|}
    \omega_1\wedge \omega_2
  \right)
  \wedge
  \omega_3
  -
  (-1)^{2|\omega_1|}
  \omega_1
  \wedge
  \left(
    (-1)^{|\omega_2|}
    \omega_2 \wedge \omega_3
  \right)
  \,.
  \nonumber\\
\end{eqnarray}
The $\mathcal{H}$ here is the 3-form curvature and
$\mathcal{F}$ is the ``fake curvature'' of a 2-bundle with 
2-connection (or of a gerbe with connection and curving)
which we will re-encounter in part \ref{Nonabelian Strings}.

\subsection{Appendix: Boundary state formalism}
\label{Boundary state formalism}

As a background for \S\fullref{Boundary state deformations from unitary loop space deformations}
this section summarizes  basic aspects of boundary conformal field theory 
(as discussed for instance in \cite{RecknagelSchomerus:1999,Schomerus:2002,Baumgartl:2003}).

Given a conformal field theory on the complex plane (with coordinates
$z,\bar z$) we get an associated ('descendant') 
\emph{boundary conformal field theory} (BCFT) on the upper half plane (UHP),
$\Im\of{z} > 0$, by demanding suitable boundary condition on the real line. 
The only class of cases well understood so far is that where the chiral fields
$W\of{z}, \bar W\of{\bar z}$ can be analytically continued to the real line
$\Im\of{z} = 0$ and a local automorphism of the chiral algebra exists,
the \emph{gluing map} $\Omega$, such that on the boundary the left- and right-moving
fields are related by
\begin{eqnarray}
  \label{gluing condition}
  W\of{z} &=& \Omega \bar W\of{\bar z}\,,\hspace{1cm}\mbox{at $z = \bar z$}
  \,.
\end{eqnarray}
In particular $\Omega$ always acts trivially on the energy momentum current
\begin{eqnarray}
  \Omega \bar T\of{\bar z} &=& \bar T\of{\bar z}
\end{eqnarray}
so that
\begin{eqnarray}
  T\of{z} &=& \bar T\of{\bar z}\,,\hspace{1cm}\mbox{at $z = \bar z$}
  \,,
\end{eqnarray}
which ensures that no energy-momentum flows off the boundary.

This condition allows to introduce for every chiral $W, \bar W$ the single chiral field
\begin{eqnarray}
  \mathrm{W}\of{z}
  &=&
  \left\lbrace
    \begin{array}{ll}
       W\of{z} & \mbox{for $\Im\of{z} \geq 0$}\\
       \Omega\of{\bar W}\of{\bar z} & \mbox{for $\Im\of{z} < 0$}
    \end{array}
  \right.
\end{eqnarray}
defined in the entire plane. (This is known as the 'doubling trick'.)

Since it is relatively awkward to work with explicit constraints it is
desirable to find a framework where the boundary condition on fields at the real
line can be replaced by an operator insertion in a bulk theory without boundary.

Imagine an open string propagating with both ends attached to some D-brane. 
The worldsheet is topologically the disk (with appropriate operator insertions at the
boundary). This disk can equivalently be regarded as the half sphere glued to the
brane. But from this point of view it represents the worldsheet of a closed string
with a certain source at the brane. Therefore the open string disk correlator
on the brane is physically the same as a closed string emission from the brane with
a certain source term corresponding to the open string boundary condition.
The source term at the boundary of the half sphere can be represented 
by an operator insertion in the full sphere. The state corresponding to this vertex
insertion is the \emph{boundary state}.

In formal terms this heuristic picture translates to the following procedure:

First map the open string worldsheet to the sphere, in the above sense.
By stereographic projection, the sphere is mapped to the plane and the upper half sphere
which represents the open string worldsheet disk gets mapped to the complement of the
unit disk in the plane. Denote the complex coordinates on this complement by $\zeta, \bar \zeta$ 
and let the open string worldsheet time $\tau = -\infty$ be mappped to $\zeta=1$
and $\tau = +\infty$ mapped to $\zeta=-1$ (so that the open string propagates 'from right to left'
in these worldsheet coordinates). With $z,\bar z$ the coordinates on the UHP this corresponds to
$z = 0 \mapsto \zeta = 1$ and $z = \infty \mapsto \zeta = -1$.
The rest of the boundary of the string must get mapped to the unit circle, which is where the
string is glued to the brane. An invertible holomorphic map from the UHP to the complement of the unit disk
with these features\footnote{
  \begin{eqnarray}
    |\zeta|^2 &=& \frac{1 + |z|^2 + 2\Im\of{z}}{1 + |z|^2 - 2 \Im\of{z}} \geq 1\;\;\;
    \mbox{for $\Im\of{z} \geq 0$}
  \end{eqnarray}
} is
\begin{eqnarray}
  \zeta\of{z}
  &\defas&
  \frac{1- i z}{1 + iz}
  \,.
\end{eqnarray}

For a given boundary condition $\alpha$ the boundary state $\ket{\alpha}$ is now
defined as the state corresponding to the operator which, when inserted in the
sphere, makes the correlator of some open string field $\Phi$ 
on the sphere equal to that on the UHP with boundary condition $\alpha$:
\begin{eqnarray}
  \langle
    \Phi^{(\mathrm{H})}\of{z,\bar z}
  \rangle_\alpha
  &=&
  \left(
    \frac{\partial \zeta}{\partial z}
  \right)^{h}
  \left(
    \frac{\partial \bar \zeta}{\partial \bar z}
  \right)^{\bar h}
  \bra{0}
   \Phi^{(\mathrm{P})}\of{\zeta,\bar \zeta}
  \ket{\alpha}
  \,.
\end{eqnarray}
 
Noting that on the boundary we have
\begin{eqnarray}
  \frac{\partial \zeta}{\partial z}
  &=&
  -i \zeta
  \,,
  \hspace{.8cm}
  \mbox{at $z = \bar z \Leftrightarrow \zeta = 1/\bar \zeta$}
\end{eqnarray}
the gluing condition \refer{gluing condition} becomes
in the new coordinates
\begin{eqnarray}
  \label{gluing condition in zeta coords}
  &&\left(\frac{\partial \zeta}{\partial z}\right)^{h}
  W\of{\zeta}
  =
  \left(\frac{\partial \bar \zeta}{\partial \bar z}\right)^{h}
  \Omega \bar W\of{\bar \zeta}  
  \nonumber\\
  &\Leftrightarrow&
  W\of{\zeta}
  =
  (-1)^h \bar \zeta^{2h}\,\Omega\bar W\of{\bar \zeta}
  \,,
  \hspace{.8cm}\mbox{at $\zeta = 1/\bar \zeta$}\,.
\end{eqnarray}
In the theory living on the plane this condition translates into a constraint on the boundary state
$\ket{\alpha}$:
\begin{eqnarray}
  0 &\shallbe&
  \bra{0}
    \cdots
  \sum\limits_{n=-\infty}^\infty
  \left(
  W_n \zeta^{-n-h}
  -
  (-1)^h
  \zeta^{-2h}
  \Omega\bar W_n \zeta^{n+h}
  \right)
  \ket{\alpha}
  \nonumber\\
  &=&
  \bra{0}
    \cdots
  \sum\limits_{n=-\infty}^\infty
  \left(
  W_n \zeta^{-n-h}
  -
  (-1)^h
  \Omega\bar W_n \zeta^{n-h}
  \right)
  \ket{\alpha}
  \,,
  \hspace{.8cm}
  \forall\, \zeta = 1/\bar \zeta
  \,,
\end{eqnarray}
i.e.
\begin{eqnarray}
  \label{constraint on boundary state}
  \left(W_n - (-1)^h\Omega\bar W_{-n}\right)\ket{\alpha}
  &=& 0
  \,,
  \hspace{.8cm}
  \forall\, n\in \N
  \,.
\end{eqnarray}

Since $\Omega\bar T = \bar T$ holds for all BCFTs this implies in particular that
one always has
\begin{eqnarray}
  \label{Virasoro constraint on boundary state}
  \left(L_n - \bar L_{-n}\right)\ket{\alpha}  &=& 0\hspace{.8cm}\forall\, n
  \,,
\end{eqnarray}
which says 
that $\ket{\alpha}$ is invariant with respect to reparametrizations of the
spatial worldsheet variable $\sigma$ parameterizing the boundary
(\cf for instance section 3 of \cite{Schreiber:2004}).

\addtocontents{toc}{\protect\newpage}

\clearpage
\parbox{6cm}{
  \flushleft
  ``A picture says \\ more than a thousand \\ words.''
}

\part{Nonabelian Strings}
\label{Nonabelian Strings}

This part is concerned with categorified gauge theory.
The concept of a categorified principal bundle with 
connection and holonomy is defined and investigated
both in the ``integral picture'' and in the
``differential picure'' (\cf figure 
\ref{figure: 2-bundle as simpl map}, p. 
\pageref{figure: 2-bundle as simpl map}). The key notion
is that of global 2-holonomy (surface holonomy) which 
allows to construct ``nonabelian strings.''

\vskip 1em

Some basic concepts necessary for the following
discussions are introduced in
\S\fullref{nonabelian strings: preliminaries}
(see also the general introduction to categories in
\S\fullref{more details on category theory}). 
This material is
taken from the paper \cite{BaezSchreiber:2004}, written in 
collaboration with John Baez. 

Then Lie 2-groups, Lie
2-algebras and loop groups are discussed in 
\S\fullref{2-Groups, Loop Groups  and String}, with an
emphasis on how the group $\String(n)$ 
(\cf \S\fullref{more details on Spinning Strings})
is expressible in 
the language of 2-groups. This is taken from
\cite{BaezCransSchreiberStevenson:2005}, which is a collaboration
with John Baez, Alissa Crans and Danny Stevenson.

The analysis of 2-bundles and of 2-connections with 2-holonomy
is begun in \S\fullref{principal 2-Bundles}, which is
again taken from the work \cite{BaezSchreiber:2004}
with John Baez.

Buidling on that, globally defined 2-holonomy
(surface holonomy) is defined and
investigated in \S\fullref{the integral picture}. This is 
material from a paper in preparation \cite{Schreiber:2005}.

The differential version of these considerations,
which gives rise to generalized Deligne hypercohomology, is then 
discussed in 
\S\fullref{The Differential Picture: Morphisms between p-Algebroids}.

\clearpage
\section{Preliminaries}
\label{nonabelian strings: preliminaries}

To develop the theory of 2-connections,
we need some mathematical preliminaries on 
internalization (\S\fullref{section: Internalization}),
with special emphasis on Lie 2-groups and Lie 2-algebras
(\S\fullref{section: Lie 2-Groups}) and also 2-spaces
(\S\fullref{section: 2-Spaces}).  We also review the
theory of nonabelian gerbes (\S\fullref{section: Nonabelian Gerbes}).

\subsection{Internalization}
\label{section: Internalization}

To categorify concepts from differential geometry, we will use a procedure
called `internalization'.  Developed by Lawvere, 
Ehresmann \cite{Ehresmann:1966} and others, internalization is a method for 
generalizing concepts from ordinary set-based mathematics to other contexts 
--- or more precisely, to other \emph{categories}.  This method is simple
and elegant.  To internalize a concept, we merely have to describe it using 
commutative diagrams in the category of sets, and then interpret these diagrams 
in some other category $K$.  For example, if we internalize the concept of 
`group' in the category of topological spaces, we obtain the concept of 
`topological group'.

For categorification, the main concept we need to internalize is that
of a category.   To do this, we start by writing down the definition
of category using commutative diagrams.  We do this in terms of the functions
$s$ and $t$ assigning to any morphism $f \maps x \to y$ its 
source and target:
\[        s(f) = x, \qquad t(f) = y , \]
the function $\id$ assigning to any object its identity morphism:
\[      \id(x) = 1_x , \]
and the function $\circ$ assigning to any composable pair of morphisms
their composite:
\[      \circ (f, g) = f \circ g  \]
If we write $\Ob(C)$ for the set of objects and $\Mor(C)$ for the set of 
morphisms of a category $C$, the set of composable pairs of morphisms is 
denoted $\Mor(C) {{}_{s}\times_{t}} \Mor(C)$, since it consists of pairs 
$(f,g)$ with $s(f) = t(g)$.

In these terms, the definition of category looks like this:

\begin{quote}
A {\bf small category}, say $C$,
has a \underline{set} of objects $\Ob(C)$, a
\underline{set} of morphisms $\Mor(C)$, source and target 
\underline{functions}:
\[               s,t \maps \Mor(C) \to \Ob(C)  , \]
an identity-assigning \underline{function}:
\[      \id \maps \Ob(C) \to \Mor(C) \]
and a composition \underline{function}:
\[     \circ \maps \Mor(C) {{}_{s}\times_{t}} \Mor(C) \to \Mor(C) \]
making diagrams commute that express associativity of
composition, the left and right unit laws for identity
morphisms, and the behaviour of source and target under composition.
\end{quote}
We omit the actual diagrams because they 
are not very enlightening: the reader can find them elsewhere 
\cite{BaezCrans:2003} or reinvent them.  The main point here
is not so much what they are, as that they \emph{can} be written down.

To internalize this definition,
we replace the word `set' by `object of $K$' and replace 
the word `function' by `morphism of $K$':
\begin{quote}
A {\bf category in $K$}, say $C$,
has an \underline{object} $\Ob(C) \in K$, an
\underline{object} $\Mor(C) \in K$, source and target 
\underline{morphisms}:
\[               s,t \maps \Ob(C) \to \Mor(C)  , \]
an identity-assigning \underline{morphism}: 
\[      \id \maps \Ob(C) \to \Mor(C) ,\]
and a composition \underline{morphism}:
\[     \circ \maps \Mor(C) {{}_{s}\times_{t}} \Mor(C) \to \Mor(C) \]
making diagrams commute that express associativity of
composition, the left and right unit laws for identity
morphisms, and the behaviour of source and target under composition.
\end{quote}
Here we must define $\Mor(C) {{}_{s}\times_{t}} \Mor(C)$ using
a category-theoretic notion called a `pullback' \cite{MacLane}.
Luckily, in examples it is usually obvious what this pullback 
should be, since it consists of composable pairs of morphisms in $C$.

Using this method, we can instantly categorify various concepts used in 
gauge theory:
\begin{definition}\et
\label{Lie 2-group}
A {\bf Lie 2-group} is a category in $\Lie\Grp$, the category
whose objects are Lie groups and whose morphisms are smooth 
group homomorphisms.   
\end{definition}
\begin{definition}\et
\label{Lie 2-algebra}
A {\bf Lie 2-algebra} is a category in $\Lie\Alg$, the category
whose objects are Lie algebras and whose morphisms are Lie algebra
homomorphisms.
\end{definition}
(For the benefit of experts, we should admit that we are only
defining `strict' Lie 2-groups and Lie 2-algebras here.)

We could also define a `smooth 2-space' to be a category
in $\Diff$, the category whose objects are finite-dimensional smooth 
manifolds and whose morphisms are smooth maps.  However, this notion is
slightly awkward for two reasons.  First, unlike $\Lie\Grp$ and 
$\Lie\Alg$, $\Diff$ does not have pullbacks in general.  So, 
the subset of $\Mor(C) \times \Mor(C)$ consisting of composable 
pairs of morphisms may not be a submanifold.   Second, and more importantly,
we will also be interested in infinite-dimensional examples.

To solve these problems, we need a category of `smooth spaces' that has
pullbacks and includes a sufficiently large class of infinite-dimensional 
manifolds.  Various categories of this sort have been proposed.
It is unclear which one is best, but we shall use a slight variant of 
an idea proposed by Chen \cite{Chen:1975}.  We call this category 
$C^\infty$.   For the present 
purposes, all that really matters about this category is that it has
many nice features, including:

\begin{itemize}
\item  Every finite-dimensional smooth manifold is a smooth space,
with smooth maps between these being precisely those 
that are smooth in the usual sense.
\item  Every smooth space has a topology, and all smooth maps
between smooth spaces are continuous.
\item  Every subset of a smooth space is a smooth space.
\item  Every quotient of a smooth space by an equivalence relation
whose equivalence classes are closed subsets is a smooth space.
\item  If $\{X_\alpha\}_{\alpha \in A}$ are smooth
spaces, so is their product $\prod_{\alpha \in A} X_\alpha$.
\item   If $\{X_\alpha\}_{\alpha \in A}$ are smooth
spaces, so is their disjoint union $\coprod_{\alpha \in A} X_\alpha$.
\item  If $X$ and $Y$ are smooth spaces, so is the set
$C^\infty(X,Y)$ consisting of smooth maps from $X$ to $Y$. 
\item  There is an isomorphism of smooth spaces
\[    C^\infty(A \times X, Y)  \iso C^\infty(A, C^\infty(X,Y))  \]
sending any function $f \maps A \times X \to Y$ to the function
$\hat f \maps A \to C^\infty(X,Y)$ given by $\hat{f}(x)(a) = f(x,a)$.
\item  We can define vector fields and differential forms on smooth 
spaces, with many of the usual properties.
\end{itemize}

With the notion of smooth space in hand, we can make the following
definition:

\begin{definition}\et 
\label{smooth 2-space}
A {\bf (smooth) 2-space} is a category in $C^\infty$, the category whose 
objects are smooth spaces and whose morphisms are smooth maps.
\end{definition}

Not only can we categorify Lie groups, Lie algebras
and smooth spaces, we can also categorify the maps between them.  
The right sort of map between categories is a functor: a pair of
functions sending objects to objects and morphisms to morphisms,
preserving source, target, identities and composition.   
If we internalize this concept, we get the definition of
a `functor in $K$'.  We then say:
\begin{definition}\et
A {\bf homomorphism} between Lie 2-groups is a functor in $\Lie\Grp$.
\end{definition}
\begin{definition}\et
A {\bf homomorphism} between Lie 2-algebras is a functor in $\Lie\Alg$.
\end{definition}
\begin{definition} \label{smooth.map} \et
A {\bf (smooth) map} between 2-spaces is a functor in $C^\infty$.
\end{definition}

There are also natural transformations between functors, and internalizing
this notion we can make the following definitions:
\begin{definition}\et
A {\bf 2-homomorphism} between homomorphisms between 
Lie 2-groups is a natural transformation in $\Lie\Grp$.
\end{definition}
\begin{definition}\et
A {\bf 2-homomorphism} between homomorphisms between Lie 2-algebras is 
a natural transformation in $\Lie\Alg$.
\end{definition}
\begin{definition} \label{smooth.2-map} \et
A {\bf (smooth) 2-map} between maps between 2-spaces is 
a natural transformation in $C^\infty$.
\end{definition}

Writing down these definitions is quick and easy.  It takes longer to
understand them and apply them to higher gauge theory.
For this we must unpack them and look at examples.  We do this in the 
next two sections.

\subsection{2-Spaces}
\label{section: 2-Spaces}

Unraveling \refdef{smooth 2-space}, a smooth 2-space, or {\bf 2-space}
for short, is a category $X$ where:
\begin{itemize}
\item The set of objects, $\Ob(X)$, is a smooth space.
\item The set of morphisms, $\Mor(X)$, is a smooth space.
\item The functions mapping any morphism to its source and target,
$s,t \maps \Mor(X) \to \Ob(X)$, are smooth maps.
\item The function mapping any object to its identity morphism,
$\id \maps \Ob(X) \to \Mor(X)$, is a smooth map
\item The function mapping any composable pair of morphisms to their
composite, \hfill \break $\circ \maps \Mor(X) {{}_{s}\times_{t}} 
\Mor(X) \to \Mor(X)$, is a smooth map.
\end{itemize}
2-spaces are more common than one might at first guess.  One only
needs to know where to look. Here are some examples, working up to three
that arise naturally in string theory: the path groupoid of $M$, the loop 
groupoid of $M$, and the 2-space of infinitesimal loops in $M$.  

\begin{definition} \et \label{trivial-simple.2-space} 
A 2-space with only identity morphisms is called {\bf trivial}.  
A 2-space for which the source and target maps coincide is called 
{\bf simple}.
\end{definition}

\begin{example} \et \label{2space.1}
{\rm 
Any smooth space $M$ gives a trivial 2-space $X$ with
$\Ob(X) = M$.  This 2-space has $\Mor(X) = M$, with $s,t,i,\circ$ 
all being the identity map from $M$ to itself.   Every trivial 2-space 
is of this form.
}\end{example}

\begin{example} \et \label{2space.2}
{\rm A {\bf smooth monoid} is a smooth space with a smooth associative
product and an identity element.  Suppose that $E \stackto{p} B$
is a smooth bundle of smooth monoids, not necessarily locally trivial.
In other words, suppose that $E \stackto{p} B$ 
is a map of smooth spaces, each fiber $p^{-1}(b)$ is equipped with an
associative product and unit, and the fiberwise-defined product
$\circ \maps E {{}_{p}\times_{p}} E \to E$ and the map $i \maps B \to E$ 
sending each point $b \in B$ to the identity element in its fiber
are smooth.   Then there is a simple 2-space
$X$ with $\Ob(X) = B$, $\Mor(X) = E$, $s = t = p$, and $i,\circ$
as above.  Moreover, every simple 2-space is of this form.
}\end{example}

\begin{example} \et \label{2space.3}
{\rm Given a smooth space $M$, there is a smooth 2-space $\P_1(M)$,
the {\bf path groupoid of $M$}, such that:
\begin{itemize}
\item the objects of $\P_1(M)$ are points of $M$,
\item the morphisms of $\P_1(M)$ are thin homotopy classes of smooth paths
$\gamma \maps [0,1] \to M$ such that $\gamma(t)$ is constant near $t = 0$
and $t = 1$.
\end{itemize}
Here a {\bf thin homotopy} between smooth paths $\gamma_0,\gamma_1
\maps [0,1] \to M$ is a smooth map $F \maps [0,1]^2 \to M$ such that:
\begin{itemize}
\item 
$F(0,t) = \gamma_0(t)$ and 
$F(1,t) = \gamma_1(t)$, 
\item 
$F(s,t)$ is constant for $t$ near $0$ and constant for $t$ near $1$,
\item
$F(s,t)$ is independent of $s$ for $s$ near $0$ and for $s$ near $1$,
\item
the rank of the differential $dF(s,t)$ is $\le 1$ for all $(s,t) \in
[0,1]^2$. 
\end{itemize}
The last condition is what makes the homotopy `thin': it guarantees
that the homotopy sweeps out a surface of vanishing area.

To see how $\P_1(M)$ becomes a 2-space, first note that
the space of smooth maps $\gamma \maps [0,1] \to M$ becomes a
smooth space in a natural way, as does the subspace satisfying
the constancy conditions near $t = 0,1$, and finally the quotient
of this subspace by the thin homotopy relation.  
This guarantees
that $\Mor(\P_1(M))$ is a smooth space.  For short, we call this 
smooth space $PM$, the {\bf path space of $M$}.  $\Ob(\P_1(M)) = M$ is 
obviously a smooth space as well.  The source and target maps 
\[          s, t \maps \Mor(\P_1(M)) \to \Ob(\P_1(M))  \]
send any equivalence class of paths to its endpoints:
\[        s([\gamma]) = \gamma(0), \qquad t([\gamma]) = \gamma(1) .\]
The identity-assigning map sends any point $x \in M$ to the 
constant path at this point.  The composition map $\circ$ sends any
composable pair of morphisms $[\gamma], [\gamma']$ to $[\gamma \circ
\gamma']$, where 
\[    \gamma \circ \gamma'(t) = 
\left\{ 
\begin{array}{lcl}
\gamma(2t) &{\rm if}& 0 \le t \le \frac{1}{2} \\
\gamma'(2t - 1) &{\rm if} & \frac{1}{2} \le t \le 1 
\end{array}
\right.
\]
One can check that $\gamma \circ \gamma'$ is a smooth path
and that $[\gamma \circ \gamma']$ is well-defined and
independent of the choice of representatives for
$[\gamma]$ and $[\gamma']$.  One can also check that the maps
$s,t,i,\circ$ are smooth and that the usual rules of a category
hold.  It follows that $\P_1(M)$ is a 2-space.

In fact, $\P_1(M)$ is not just a category: it is also a {\bf groupoid}:
a category where every morphism has an inverse.  The inverse of 
$[\gamma]$ is just $[\overline{\gamma}]$, where $\overline{\gamma}$
is obtained by reversing the orientation of the path $\gamma$:
\[        \overline{\gamma}(t) = \gamma(1 - t)  .\]
Moreover, the map sending any morphism to its inverse is 
smooth.  Thus $\P_1(M)$ is a {\bf smooth groupoid}: a 2-space
where every morphism is invertible and the map sending every morphism
to its inverse is smooth.
}\end{example}  

\begin{example} \et \label{2space.4}
{\rm Given a 2-space $X$, any subcategory of $X$ becomes a
2-space in its own right.  Here a {\bf subcategory} is a category 
$Y$ with $\Ob(Y) \subseteq \Ob(X)$ and $\Mor(Y) \subseteq \Mor(X)$,
where the source, target, identity-assigning and composition maps
of $Y$ are restrictions of those for $X$.  The reason $Y$ becomes
a 2-space is that any subspace of a smooth space becomes a smooth 
space in a natural way 
and restrictions of smooth maps to subspaces are smooth.  We call
$Y$ a {\bf sub-2-space} of $X$.
}\end{example}

\begin{example} \et \label{2space.5}
{\rm Given a smooth space $M$, the path groupoid $\P_1(M)$ has
a sub-2-space ${\cal L}M$ whose objects are all the points of
$M$ and whose morphisms are those equivalence classes $[\gamma]$
where $\gamma$ is a loop: that is, a path with $\gamma(0) = \gamma(1)$.
We call ${\cal L}M$ the {\bf loop groupoid} of $M$.  Like
the path groupoid, the loop groupoid of $M$ is not just a 2-space, 
but a smooth groupoid.  
}\end{example}

\begin{example} \et \label{2space.6}
{\rm Given a smooth vector bundle $E \stackto{p} B$ over a smooth
space $B$, there is a simple 2-space $\cal E$ with $B$ as its space of 
objects, $E$ as its space of morphisms, and addition within each fiber 
as the operation of composing morphisms.  This is a special case of 
Example \ref{2space.2}.  Since every vector in each fiber of $E$ has
an additive inverse, $\cal E$ is actually a smooth groupoid.

In particular, if $E = \Lambda^2 TB$ is the bundle of antisymmetric 
rank $(2,0)$ tensors over the smooth space $B$, we call $\cal E$ the 
{\bf 2-space of infinitesimal loops in $B$}.  We can think of this as 
a kind of `limit' of the loop groupoid of $M$ in which the loops 
shrink to zero size.
}\end{example} 

For 2-spaces, and indeed for all categorified concepts, the usual
notion of `isomorphism' is less useful than that of `equivalence'.
For example, in categorified gauge theory what matters is not 
2-bundles whose fibers are all isomorphic to some standard fiber,
but those whose fibers are all {\it equivalent} to some standard fiber. 
We recall the concept of equivalence here:

\begin{definition} \et \label{equivalent.2-spaces}
Given 2-spaces $X$ and $Y$, an {\bf isomorphism} $f \maps X \to Y$ 
is a map equipped with a map $\bar{f} \maps Y \to X$ such that 
$\bar{f}f = 1_X$ and $f\bar{f} = 1_Y$.  An {\bf equivalence} 
$f \maps X \to Y$ is a map equipped with a map $\bar{f} \maps Y \to X$ 
and invertible 2-maps $\phi \maps \bar{f}f \To 1_X$ and 
$\bar{\phi} \maps f\bar{f} \To 1_Y$.  
\end{definition}  

\subsection{Lie 2-Groups and Lie 2-Algebras}
\label{section: Lie 2-Groups}

Unravelling \refdef{Lie 2-group}, we see that a 
Lie 2-group $\twogroup$ is a category where: 
\begin{itemize}
\item The set of objects, $\Ob(\twogroup)$, is a Lie group.
\item The set of morphisms, $\Mor(\twogroup)$, is a Lie group. 
\item The functions mapping any morphism to its source and target,
$s,t \maps \Mor(\twogroup) \to \Ob(\twogroup)$, are homomorphisms.
\item The function mapping any object to its identity morphism,
$\id \maps \Ob(\twogroup) \to \Mor(\twogroup)$, is a homomorphism.
\item The function mapping any composable pair of morphisms to their
composite, \hfill \break $\circ \maps \Mor(\twogroup) {{}_{s}\times_{t}} 
\Mor(\twogroup) \to \Mor(\twogroup)$, is a homomorphism.
\end{itemize}
For applications to higher gauge theory it is suggestive to draw 
objects of $\twogroup$ as arrows:
\[
\xymatrix{
   \bullet \ar@/^1pc/[rr]^{g}
&& \bullet
}
\]
and morphisms $f \maps g \to g'$ as surfaces of this sort:
\[
\xymatrix{
   \bullet \ar@/^1pc/[rr]^{g}_{}="0"
           \ar@/_1pc/[rr]_{g'}_{}="1"
           \ar@{=>}"0";"1"^{f}
&& \bullet
}
\]
This lets us can draw multiplication in $\Ob(\twogroup)$ as 
composition of arrows, 
multiplication in $\Mor(\twogroup)$ as `horizontal
composition' of surfaces, 
and composition of morphisms $f \maps g \to g'$ and
$f' \maps g' \to g''$ as `vertical composition' of surfaces.

In this notation, the fact that composition is a homomorphism says that
the `exchange law'
\[ (f_1 \circ f_1')(f_2 \circ f_2') =  (f_1 f_2) \circ (f_1' f_2') \]
holds whenever we have a situation of this sort:
\[
\xymatrix{
   \bullet \ar@/^2pc/[rr]^{g_1}_{}="0"
           \ar[rr]^<<<<<<{g_2}_{}="1"
           \ar@{=>}"0";"1"^{f_1}
           \ar@/_2pc/[rr]_{g_3}_{}="2"
           \ar@{=>}"1";"2"^{f_1'}
&& \bullet \ar@/^2pc/[rr]^{g_1'}_{}="3"
           \ar[rr]^<<<<<<{g_2'}_{}="4"
           \ar@{=>}"3";"4"^{f_2}
           \ar@/_2pc/[rr]_{g_3'}_{}="5"
           \ar@{=>}"4";"5"^{f_2'}
&& \bullet
}
\]
In other words, we can interpret this picture either as a horizontal
composite of vertical composites or a vertical composite of 
horizontal composites, without any ambiguity.

A Lie 2-group with only identity morphisms is the same thing as a
Lie group.  To get more interesting examples it is handy to think of a
Lie 2-group as special sort of `crossed module'.
To do this, start with a Lie 2-group $\twogroup$
and form the pair of Lie groups
\[   G = \Ob(\twogroup) , \qquad H = \ker s \subseteq \Mor(\twogroup).  \]
The target map restricts to a homomorphism
\[   t \maps H  \to G . \]
Besides the usual action of $G$ on itself by conjugation,
there is also an action of $G$ on $H$,
\[    \alpha \maps G \to \Aut(H) ,\]
given by
\[
\begin{array}{ccl}   \alpha(g)(h)  &=&  1_g \, h \, 1_{g^{-1}}  \\
&=& 
\xymatrix{
   \bullet \ar@/^1pc/[rr]^{g}_{}="0"
           \ar@/_1pc/[rr]_{g}_{}="1"
           \ar@{=>}"0";"1"^{1_g}
&& \bullet \ar@/^1pc/[rr]^{1}_{}="2"
           \ar@/_1pc/[rr]_{t(h)}_{}="3"
           \ar@{=>}"2";"3"^{h}
&& \bullet \ar@/^1pc/[rr]^{g^{-1}}_{}="4"
           \ar@/_1pc/[rr]_{g^{-1}}_{}="5"
           \ar@{=>}"4";"5"^{1_{g^{-1}}}
&& \bullet
} \; .
\end{array}
\]
The target map is equivariant with respect to this
action:
\[   t(\alpha(g)(h)) = g \, t(h)\, g^{-1} \]
and satisfies the so-called `Peiffer identity':
\[    \alpha(t(h))(h') = hh'h^{-1} . \]
A setup like this with groups rather than Lie
groups is called a `crossed module', so here we are getting a `Lie
crossed module':

\begin{definition}\et
A {\bf Lie crossed module} is a quadruple $(G,H,t,\alpha)$
consisting of Lie groups $G$ and $H$, a homomorphism $t \maps H \to G$,
and an action of $G$ on $H$ (that is, a homomorphism
$\alpha \maps G \to \Aut(H)$) such that $t$ is equivariant:
\[   t(\alpha(g)(h)) = g \, t(h)\, g^{-1} \]
and satisfies the Peiffer identity:
\[    \alpha(t(h))(h') = hh'h^{-1}  \]
for all $g \in G$ and $h,h' \in H$.
\end{definition}

\noindent This definition becomes a bit more memorable if
we abuse language and write $\alpha(g)(h)$ as $ghg^{-1}$; then the
equations above become
\[   t(ghg^{-1}) = g \, t(h)\, g^{-1} \]
and
\[    t(h) \, h' \, t(h)^{-1} = hh'h^{-1} . \]

As we shall see, Lie 2-groups are essentially the same as 
Lie crossed modules.  The same is true for the homomorphisms between
them.  We have already defined a homomorphism of Lie 2-groups as
a functor in $\Lie\Grp$.  We can also define a homomorphism of Lie
crossed modules:

\begin{definition}\et 
A {\bf homomorphism} from the Lie crossed module $(G,H,t,\alpha)$ to the
Lie crossed module $(G',H',t',\alpha')$ is a pair of homomorphisms
$f \maps G \to G'$, $\tilde{f} \maps H \to H'$ such that 
\[      t(\tilde{f}(h)) = f(t'(h))  \]
and 
\[   \tilde{f}(\alpha(g)(h)) = \alpha'(f(g)) (\tilde{f}(h)) \]
for all $g \in G$, $h \in H$.
\end{definition}

Not only does every Lie 2-group give a Lie crossed module; every
Lie crossed module gives a Lie 2-group.  In fact:

\begin{proposition}\et \label{liecrossedmodule} 
The category of Lie 2-groups is equivalent to the category of Lie
crossed modules. \end{proposition}
 
\Proof 
This follows easily from the well-known equivalence between
crossed modules and 2-groups \cite{BrownSpencer:1976}; details
can also be found in \cite{BaezLauda:2003}.
For the convenience of the reader, we sketch how to recover a Lie 2-group 
from a Lie crossed module.

Suppose we have a Lie crossed module $(G,H,t,\alpha)$.  Let
\[   \Ob(\twogroup) = G, \qquad \Mor(\twogroup) = G \ltimes H  \]
where the semidirect product is formed using the action of $G$
on $H$, so that multiplication in $\Mor(\twogroup)$ is given by:

\begin{equation}
\label{multiplication of morphisms}
    (g,h) (g',h') = (gg', h \alpha(g)(h')) .
\end{equation}
The inverse of an element of $\Mor(\twogroup)$ is given by:
\[  (g,h)^{-1} = (g^{-1}, \alpha\of{g^{-1}}\of{h^{-1}}) \,.
\]
We make this into a Lie 2-group where the source
and target maps $s,t \maps \Mor(\twogroup) \to \Ob(\twogroup)$ are given by
\begin{equation}
\label{source and target of 2-group element}
   s(g,h) = g , \qquad   t(g,h) = t(h)g  ,\
\end{equation}
the identity-assigning map $\id \maps \Ob(\twogroup) \to 
\Mor(\twogroup)$ is given by
\begin{equation}
 \id(g) = (g,1), 
\end{equation}
and the composite of the morphisms
\[
  (g,h) \maps g  \to g'  , \qquad \quad
  (g',h') \maps g' \to g'', 
\]
is
\begin{equation}
\label{composition of morphisms}
   (g,h) \circ (g',h') = (g,h'h) \maps g \to g'' .
\end{equation}
It is also worth noting that every morphism has an inverse
with respect to composition, which we denote by
\[ 
\overline{(g,h)} = \left(t\of{h} g, h^{-1}\right) \,.
\]
One can check that this construction indeed gives a Lie 2-group, and
that together with the previous construction it sets up an equivalence
between the categories of Lie 2-groups and Lie crossed modules.  \endofproof

Crossed modules are important in homotopy theory 
\cite{Brown:1999},
and the reader who is fonder of crossed modules than categories
is free to think of Lie 2-groups as a way of talking
about Lie crossed modules.  Both perspectives are useful, but one
advantage of Lie crossed modules is that they allow us to
quickly describe some examples:

\begin{example}\et %\label{crossed.1} 
{\rm Given any abelian group $H$, there is a Lie crossed module
where $G$ is the trivial group and $t, \alpha$ are trivial.  
This gives a Lie 2-group $\twogroup$ with one object
and $H$ as the group of morphisms.  Lie 2-groups of this
sort are important in the theory of \emph{abelian} gerbes.
}\end{example}

\begin{example}\et \label{crossed.2}  {\rm
More generally, given any Lie group $G$, abelian Lie group $H$, and
action $\alpha$ of $G$ as automorphisms of $H$, there is a Lie 
crossed module with $t \maps G \to H$ the trivial homomorphism.
For example, we can take $H$ to be a finite-dimensional vector space
and $\alpha$ to be a representation of $G$ on this vector space.

In particular, if $G$ is the Lorentz group and $\alpha$ is the defining
representation of this group on Minkowski spacetime, this construction
gives a Lie 2-group called the {\bf Poincar\'e 2-group}, because its group
of morphisms is the Poincar\'e group.  After its introduction in work
on higher gauge theory \cite{Baez:2002}, this 2-group has come to play
an important role in some recent work on quantum gravity by 
Crane, Sheppeard and Yetter \cite{CraneSheppeard:2003,CraneYetter:2003}.
}\end{example}

\begin{example}\et \label{crossed.3} {\rm
Given any Lie group $H$, there is a Lie
crossed module with $G = \Aut(H)$, $t \maps H \to G$ the homomorphism
assigning to each element of $H$ the corresponding inner automorphism,
and the obvious action of $G$ as automorphisms of $H$.  We call the
corresponding Lie 2-group the {\bf automorphism 2-group} of $H$, and
denote it by $\AUT(H)$.   This sort of 2-group is important in the theory 
of \emph{nonabelian} gerbes.  

In particular, if we take $H$ to be the multiplicative group of nonzero
quaternions, then $G = \SU(2)$ and we obtain a 2-group that plays a basic
role in Thompson's theory of quaternionic gerbes \cite{Thompson}.

We use the term `automorphism 2-group' because $\AUT(H)$ really is a
2-group of symmetries of $H$.  An object of $\AUT(H)$ is a symmetry of
the group $H$ in the usual sense: that is, an automorphism
$f \maps H \to H$.  On the other hand, a morphism $\theta \maps f \to f'$ 
in $\AUT(H)$ is a `symmetry between symmetries': that is, an
element $h \in H$ that sends $f$ to $f'$ in the following
sense: $hf(x)h^{-1} = f'(x)$ for all $x \in H$.
}\end{example}

\begin{example}\et \label{crossed.4} {\rm 
Suppose that $1 \to A \hookrightarrow H \stackto{t} G \to 1$ is a central
extension of the Lie group $G$ by the Lie group $H$.  Then there is
a Lie crossed module with this choice of $t \maps G \to H$.  To construct
$\alpha$ we pick any section $s$, that is, any function
$s \maps G \to H$ with $t(s(g)) = g$, and define 
\[      \alpha(g) h = s(g) h s(g)^{-1}  .\]
Since $A$ lies in the center of $H$, $\alpha$ independent of the choice of
$s$.  We do not need a global smooth section $s$ to show $\alpha(g)$ depends 
smoothly on $g$; it suffices that there exist a local smooth section in a
neighborhood of each $g \in G$.

It is easy to generalize this idea to infinite-dimensional examples, like
central extensions of loop groups, if we work not with Lie groups
but {\bf smooth groups}: that is, groups in the category of smooth spaces.  
The basic theory of smooth 2-groups and smooth crossed modules works just like
the finite-dimensional case, but with the category of smooth spaces replacing 
$\Diff$.

In particular, given a simply-connected connected simple Lie group 
$G$, the loop group $\Omega G$ is a smooth group. 
For each level $k \in \Z$, this group has a central extension
\[           1 \to \U(1) \hookrightarrow 
\widehat{\Omega_k G} \stackto{t} \Omega G \to 1   \]
as explained by Pressley and Segal \cite{PressleySegal:1986}.  The above 
diagram lives in the category of smooth groups, and there exist local smooth 
sections for $t \maps \tilde{L}G \to LG$, so we obtain a smooth crossed 
module $(\Omega G, \widehat{\Omega_k G}, t,\alpha)$ with 
$\alpha$ given as above.  This in turn gives a smooth 2-group which 
we call the {\bf loop 2-group} of $G$, ${\cal L}_k G$.
It has recently been shown 
\cite{BaezCransSchreiberStevenson:2005}
(see \S\fullref{2-Groups, Loop Groups  and String})
that this fits into an exact sequence of smooth 2-groups:
\[           1 \to {\cal L}_k G \hookrightarrow 
{\cal P}_k G \stackto{t} G \to 1   \]
where the middle term, the {\bf path 2-group} of $G$, has extremely
interesting properties.  In particular, it gives a
new construction of the group ${\rm String}(n)$ when $G = {\rm Spin}(n)$.
So, we expect that these 2-groups ${\cal P}_k G$ will be especially
interesting for applications of 2-bundles to string theory.
}\end{example}

Just as Lie groups give rise to Lie algebras, Lie 2-groups give
rise to Lie 2-algebras.  These can also be described using 
a differential version of crossed modules.  Recall that a Lie
2-algebra is a category $\lietwoalgebra$ where:
\begin{itemize}
\item The set of objects, $\Ob(\lietwoalgebra)$, is a Lie algebra.
\item The set of morphisms, $\Mor(\lietwoalgebra)$, is a Lie algebra.
\item The functions mapping any morphism to its source and target,
$s,t \maps \Mor(\lietwoalgebra) \to \Ob(\lietwoalgebra)$, are Lie
algebra homomorphisms,
\item The function mapping any object to its identity morphism,
$\id \maps \Ob(\lietwoalgebra) \to \Mor(\lietwoalgebra)$, is a Lie
algebra homomorphism.
\item The function mapping any composable pair of morphisms to their
composite, \hfill \break $\circ \maps \Mor(\lietwoalgebra) {{}_{s}\times_{t}} 
\Mor(\lietwoalgebra) \to \Mor(\lietwoalgebra)$,
is a Lie algebra homomorphism.
\end{itemize}

We can get a Lie 2-algebra by differentiating all the 
data in a Lie 2-group.  Similarly, we can get a `differential crossed 
module' by differentiating all the data in a Lie crossed module:

\begin{definition}\et 
\label{differential crossed module}
A {\bf differential crossed module} is a quadruple $\crossedmodule = 
(\g,\h,dt,d\alpha)$
consisting of Lie algebras $\g,\h$, a homomorphism $dt \maps \h \to \g$,
and an action $\alpha$ of $\g$ as derivations of $\h$
(that is, a homomorphism $\alpha \maps \g \to \Der(\h)$) satisfying
\begin{equation}
   dt(d\alpha(x)(y)) = [x, dt(y)] 
\end{equation}
and
\begin{equation}
  \label{notation for action of h on h'}
   d\alpha(dt(y)) (y') = [y, y'] 
\end{equation}
for all $x \in \g$ and $y,y' \in \h$.
\end{definition}

\noindent
This definition becomes easier to remember if we allow ourselves to
write $d\alpha(x)(y)$ as $[x,y]$.  Then the fact that $d\alpha$ is an
action of $\g$ as derivations of $\h$ simply means that $[x,y]$ is
linear in each argument and the following `Jacobi identities' hold:
\begin{equation}
 \label{jacobi1}
         [[x,x'],y] = [x,[x',y]] - [x',[x,y]] ,
\end{equation}
\begin{equation}
\label{jacobi2}
         [x,[y,y']] = [[x,y],y'] - [[x,y'],y]
\end{equation}
for all $x,x' \in \g$ and $y,y' \in \h$.
Furthermore, the two equations in the above definition become
\begin{equation} \label{cross1}
  t([x,y]) = [x, t(y)]
\end{equation}
and
\begin{equation}
 \label{cross2}
   [t(y), y'] = [y, y'] .
\end{equation}

\begin{proposition}\et \label{lie.2-algebra.2}
The category of Lie 2-algebras is equivalent to the
category of differential crossed modules.
\end{proposition}

\Proof 
The proof is just like that of Prop.\ \ref{liecrossedmodule}.
\endofproof

Since every Lie 2-group gives a Lie 2-algebra and a differential
crossed module, we get plenty of examples of the latter concepts from 
our example of Lie 2-groups.  Here is another interesting class of
examples:

\begin{example} \et {\rm 
Just as every Lie 2-group gives rise to a Lie 2-algebra, so does
every smooth 2-group.  The reason is that not only smooth manifolds
but also smooth spaces have tangent spaces, 
and the usual construction of Lie algebras from Lie groups
generalizes to smooth groups.  So, any smooth 2-group 
$\twogroup$ gives a Lie 2-algebra $\lietwoalgebra$ for which 
$\Ob(\lietwoalgebra)$ is the Lie algebra of $\Ob(\twogroup)$, 
$\Mor(\lietwoalgebra)$ is the Lie algebra of $\Mor(\twogroup)$, 
and the maps $s,t,i,\circ$ for $\lietwoalgebra$ are obtained by 
differentiating the corresponding maps for $\twogroup$.

In particular, suppose $G$ is a simply-connected compact simple Lie group
with Lie algebra $\g$.  Then the loop 2-group of $G$, as defined in Example 
\ref{crossed.4}, has a Lie 2-algebra.
This Lie 2-algebra has $L\g = C^\infty(S^1,\g)$ as its Lie algebra of
objects, and a certain central extension $\tilde{L}\g$ of $L\g$ as
its Lie algebra of morphisms.  We call this Lie 2-algebra 
the {\bf loop Lie 2-algebra of $\g$}.  It is another way of organizing
the data in the affine Lie algebra corresponding to $\g$.

Alternatively, we can take this central extension of smooth groups:
\[       1 \to \U(1) \hookrightarrow \tilde{L}G \stackto{t} LG \to 1   \]
and differentiate all the maps to obtain a central extension of Lie
algebras:
\[       1 \to \u(1) \hookrightarrow \tilde{L}\g \stackto{dt} L\g \to 1 .  \]
Just as every central extension of Lie groups gives a Lie 2-group,
every central extension of Lie algebras gives a Lie 2-algebra.  So,
we obtain a Lie 2-algebra, which is the loop Lie 2-algebra of $\g$.
}\end{example}

We conclude the preliminaries
with a brief review of nonabelian gerbes.

\subsection{Nonabelian Gerbes}
\label{section: Nonabelian Gerbes}

Given a bundle $E \stackto{p} M$, the sections of $E$ defined on all possible
open sets of $B$ are naturally organized into a structure called a `sheaf'.
This codifies the fact that we can restrict a section from an open set $U 
\subseteq B$ to a smaller open set $U' \subseteq U$, and also piece together 
sections on open sets $U_i$ covering $U$ to obtain a unique section on $U$, as 
long as the sections agree on the intersections $U_i \cap U_j$.  While 
mathematical physicists tend to be more familiar with bundles than sheaves, 
the greater generality of sheaves is important in algebraic geometry.  

While the 2-bundles to be discussed in the present paper arise from
categorifying the concept of `bundle', most previous work on this subject
starts by categorifying the concept of `sheaf' to obtain the concept of 
`stack', with `gerbes' as a key special case.  We suspect that just as
mathematical physicists are more comfortable with bundles than sheaves, they 
will eventually prefer 2-bundles to gerbes.  At present, however, it is crucial to
clarify the relation between 2-bundles and gerbes.  So, one of the goals of
this paper is to relate 2-connections on 2-bundles to the already established
notion of connections on gerbes.   We begin here by recalling the history of 
stacks and gerbes, and the concept of a gerbe with connection.

The idea of a stack goes back to Grothendieck \cite{Grothendieck}.  
Just as a sheaf over a space $M$ assigns a {\em set} of sections to any open
set $U \subseteq M$, a stack assigns a {\em category} of sections to any open 
set $U \subseteq M$.  Indeed, one may crudely define a stack as 
a `sheaf of categories'.  However, all the usual sheaf axioms need to be 
`weakened', meaning that instead of equations between objects, we must use 
isomorphisms satisfying suitable equations of their own.  For example, in a 
sheaf we can obtain a section $s$ over $U$ from sections $s_i$ over open sets 
$U_i$ covering $U$ when these sections are {\em equal} on double intersections:
\[      s_i|_{U_i \cap U_j} = s_j|_{U_i \cap U_j}  \]
For a stack, on the other hand, we can obtain a section $s$ over $U$ 
when the sections $s_i$ are {\it isomorphic} over double intersections:
\[      h_{ij} \maps  s_i|_{U_i \cap U_j} \stackto{\sim} s_j|_{U_i \cap U_j},  \]
as long as the isomorphisms satisfy the familiar `cocycle condition' on triple
intersections: $h_{ij} h_{jk} = h_{ik}$ on $U_i \cap U_j \cap U_k$.  

A good example is the stack of principal $H$-bundles
over $M$, where $H$ is any fixed Lie group.  This associates to each open set 
$U \subseteq M$ the category whose objects are principal $H$-bundles over $U$ and 
whose morphisms are $H$-bundle isomorphisms.  The above cocycle condition is very 
familiar in this case: it says when we can build a $H$-bundle $s$ over $U$ by 
gluing together $H$-bundles $s_i$ over open sets covering $U$, using 
$H$-valued transition functions $h_{ij}$ defined on double intersections.

This example also motivates the notion of a `gerbe', which is a special sort of 
stack introduced by Giraud \cite{Giraud:1971,Moerdijk:2002}.  
For a stack over $M$ to be a gerbe, it must satisfy three properties:
\begin{itemize}
\item 
Its category of sections over any open set must be a {\it groupoid}: that
is, a category where all the morphisms are invertible.
\item 
Each point of $M$ must
have a neighborhood over which the groupoid of sections is nonempty.
\item
Given two sections $s,s'$ over an open set $U \subseteq M$, each point of 
$U$ must have a neighborhood $V \subseteq U$ such that $s|_V \iso s'|_V$. 
\end{itemize}
It is easy to see that the stack of principal $H$-bundles satisfies all these 
conditions.  It satisfies another condition as well: for any section $s$ over 
an open set $U \subseteq M$, each point of $U$ has a neighborhood $V$ such 
that the automorphisms of $s|_V$ form a group isomorphic to the group of 
smooth $H$-valued functions on $V$.  A gerbe of this sort is called an 
`$H$-gerbe'.  Sometimes these are called `nonabelian gerbes', to distinguish 
them from another class of gerbes that only make sense when the group $H$ 
is abelian.

There is a precise sense in which the gerbe of principal $H$-bundles 
is the `trivial' $H$-gerbe.
Every $H$-gerbe is {\it locally} equivalent to this one, but not globally.
So, we can think of a $H$-gerbe as a thing whose sections look locally like 
principal $H$-bundles, but not globally.  This viewpoint is emphasized by the
concept of `bundle gerbe', defined first in the abelian case by Murray 
\cite{Murray:1994,Stevenson:2000} and more recently in the nonabelian case
that concerns us here by Aschieri, Cantini and Jur{\v c}o 
\cite{AschieriCantiniJurco:2004}. 

However, the most concrete way of getting our hands
on $H$-gerbes over $M$ is by gluing together trivial $H$-gerbes
defined on open sets $U_i$ that cover $M$.  This leads to a simple 
description of $H$-gerbes in terms of transition functions satisfying cocycle 
conditions.   Now the transition functions defined on double intersections 
take values not in $H$ but in $G = \Aut(H)$:
\[        g_{ij} \maps U_i \cap U_j \to G  \]
Moreover, they need not satisfy the usual cocycle condition for
triple intersections `on the nose', but only up to conjugation by 
certain functions
\[        h_{ijk} \maps U_i \cap U_j \cap U_k \to H  .\]
In other words, we demand:
\[         g_{ij} g_{jk} = t\of{h_{ijk}} g_{ik} \]
where $t \maps H \to G$ sends $h \in H$ to the operation of conjugating by $h$.
Finally, the functions $h_{ijk}$ should satisfy an cocycle condition on quadruple
intersections:
\[   \alpha\of{g_{ij}}\of{h_{jkl}} \, h_{ijl} = h_{ijk} h_{ikl} \]
where $\alpha$ is the natural action of $G = \Aut(H)$ on $H$.
All this can be formalized most clearly using the automorphism 2-group $\AUT(H)$
described in Example \ref{crossed.3}, since this Lie 2-group
has $(G,H,t,\alpha)$ as its corresponding Lie crossed module.  Indeed, one 
way that 2-bundles generalize gerbes is by letting an arbitrary Lie 2-group
play the role that $\AUT(H)$ plays here; we call this 2-group the
`structure 2-group' of the 2-bundle.

Given an $H$-gerbe, we can specify a `connection' on it 
by means of some additional local data.  We begin by choosing $\g$-valued 
1-forms $A_i$ on the open sets $U_i$, which describe parallel transport along 
paths.  But these 1-forms need not satisfy the usual consistency condition 
on double intersections!  Instead, they satisfy it only up to $\h$-valued 
1-forms $a_{ij}$:
\[        A_i + dt\of{a_{ij}} 
  = g_{ij} A_j g_{ij}^{-1} + g_{ij} \extd g_{ij}^{-1}.
\]
These, in turn, must satisfy a consistency condition on triple intersections:
\[      a_{ij} + \alpha\of{g_{ij}}\of{a_{jk}} = 
        h_{ijk} a_{ik} h_{ijk}^{-1} + h_{ijk}\, d\alpha(A_i)(h_{ijk}^{-1}) . \]
Next, we choose $\h$-valued 2-forms $B_i$ describing parallel transport along 
surfaces.  These satisfy a consistency condition on double intersections:
\[      
  \alpha(g_{ij})\of{B_j} 
  = 
  B_i
  -
  k_{ij}
  + 
  b_{ij}  
  \,,
\]
where the $\h$-valued 2-forms
$b_{ij}$
and
\[
  k_{ij}
  \defas
  \extd a_{ij} + a_{ij}\wedge a_{ij} - d\alpha(A_i)\wedge a_{ij}
\]
measure the failure of $B_i$ to transform covariantly. 
The 2-form $k_{ij}$ is essentially the curvature of $a_{ij}$,
while $b_{ij}$ is a new object which turns out to have a 
transition law of its own, this time on triple intersections:
\[
    b_{ij} + \alpha\of{g_{ij}}\of{d_{jk}}
    =
    h_{ijk} \,b_{ik}\, h_{ijk}^{-1}
    +
    h_{ijk}\, d\alpha\of{dt\of{B_i} + F_{A_i}}(h_{ijk}^{-1})
   \,.
\]

This description of connections on nonabelian gerbes was first given by Breen 
and Messing \cite{BreenMessing:2001}.  Aschieri, Cantini and Jur{\v c}o then 
gave a similar treatment using bundle gerbes 
\cite{AschieriCantiniJurco:2004}.

Later, Aschieri and Jur{\v c}o \cite{AschieriJurco:2004} introduced 
connections on so-called `twisted' nonabelian gerbes.  These are a 
categorified version of the somewhat more familiar `twisted bundles', 
so let us first recall the latter.  Suppose the group $H$ has a $\U(1)$ 
subgroup in its center, giving a central extension
\[   1 \to \U(1) \hookrightarrow H \to H/\U(1) \to 1   . \]
Then we can build a `twisted $H$-bundle' using $H$-valued transition functions
$h_{ij}$ that only satisfy the usual cocycle condition on triple intersections  
up to a phase: $h_{ij} h_{jk} = \lambda_{ijk} h_{ik}$.  These phases 
automatically satisfy a cocycle condition on quadruple intersections:
$\lambda_{ijk} \lambda_{ikl} = \lambda_{jkl} \lambda_{ijl}$.

Incidentally, this is just the cocycle condition for the transition
function in an abelian gerbe.
Since a bundle can be regarded as a 0-gerbe,
this is an example of a general pattern in which the twist
of a nonabelian twisted $p$-gerbe defines an abelian $(p+1)$-gerbe.
This is called the `lifting $(p+1)$-gerbe' because
precisely when it is trivial does the obstruction to lifting
the structure group of the $p$-gerbe to its central extension vanish.  
In fact, we can regard a principal $\U(1)$ bundle itself as an
abelian 0-gerbe.  This can be thought of as
measuring the `twist' of a (-1)-gerbe, which is just an
ordinary $H$-valued function when the twist vanishes.

Following these considerations we can
define a `twisted $H$-gerbe' by relaxing the aforementioned cocycle condition
on the functions $h_{ijk}$, requiring only that it hold up to a phase:
\[
\alpha\of{g_{ij}}\of{h_{jkl}} \, h_{ijl}  = \lambda_{ijkl} \, h_{ijk}\, h_{ikl} 
\]
where
\[     \lambda_{ijkl} \maps U_i \cap U_j \cap U_k \cap U_l \to \U(1) .\]
These phases automatically satisfy a cocycle condition on quintuple 
intersections,
\[     \lambda_{ijkl} \lambda_{ijlm} \lambda_{jklm} = 
        \lambda_{iklm} \lambda_{ijkm} 
  \,,
\]
and this is indeed precisely the cocycle condition for an abelian 
2-gerbe.

In an analogous way, the cocycle conditions for 
the connection $a_{ij}$ and curving $B_{i}$ of
a nonabelian gerbe can also pick up a twist. This amounts to
adding $\mathrm{Lie}(U(1))$-valued objects $\alpha_{ijk}$ 
and $\beta_{ij}$ to the previous equations:
\begin{eqnarray*}
  \alpha_{ijk}
   &=&     
   a_{ij} + \alpha\of{g_{ij}}\of{a_{jk}} 
   - 
   h_{ijk} \, a_{ik} \, h_{ijk}^{-1} 
   -
    h_{ijk} \, d\alpha(A_i)(h_{ijk}^{-1}) 
  \\
  \beta_{ij}
  &=&
    b_{ij} + \alpha\of{g_{ij}}\of{d_{jk}}
    -
    h_{ijk} \,b_{ik}\, h_{ijk}^{-1}
    -
    h_{ijk}\, d\alpha\of{dt\of{B_i} 
    -
    F_{A_i}}(h_{ijk}^{-1})
  \,. 
\end{eqnarray*}
From these equations it follows that the twists $\lambda_{ijkl}$,
$\alpha_{ijk}$ and $\beta_{ij}$ themselves satisfy cocycle
conditions that identify them as the transition functions,
connection and curving of a $\U(1)$ 2-gerbe, the `lifting 2-gerbe'.

The phenomenon of lifting gerbes has of course its analogue in the
language of 2-bundles, but this will not concern us here.

\vskip 1em

To summarize, we list the local data for a twisted nonabelian gerbe 
with connection.  For maximum generality, we start
with an arbitrary Lie 2-group $\twogroup$ and form its Lie crossed module 
$(G,H,\alpha,t)$.   The definition below reduces to that of Aschieri and 
Jur{\v c}o when $\twogroup = \AUT(H)$ and the `phases' $\lambda_{ijkl}$ 
lie in a chosen $\U(1)$ subgroup of the center of $H$.
In general, we merely require these `phases' to lie in the kernel of
$t \maps H \to G$.  This kernel always lies in the center of $H$, 
since if $h \in \ker t$ and $h' \in H$, the Peiffer identity gives
\[    hh'h^{-1} = \alpha(t(h))(h') = h'.  \]
So, all of Aschieri and Juc{\v c}o's calculations which require the 
phases to commute with other elements of $H$ still go through.

\begin{definition} \et 
\label{twisted nonabelian gerbe}
A {\bf twisted nonabelian gerbe with connection} consists of: 

\begin{itemize}
  \item
    a base space $\manifold$,
  \item
    an open cover $U$ of $\manifold$, with $U^{[n]}$ denoting the union of
    all $n$-fold intersections of patches in $U$,
  \item
    a Lie 2-group $\twogroup$ with Lie crossed module $(G,H,\alpha,t)$ 
    and differential crossed module $(\g,\h,d\alpha,dt)$,
  \item
    transition functions:
    \begin{eqnarray}
     g \maps U^{[2]} &\to& G  
      \nonumber\\
      (x,i,j) &\mapsto& g_{ij}\of{x} \in G
    \end{eqnarray}
   \item
    transition transformation functions:
    \begin{eqnarray}
      h \maps U^{[3]} &\to& H
      \nonumber\\
      (x,i,j,k) &\mapsto& h_{ijk}\of{x} \in H
    \end{eqnarray}
   \item
    connection 1-forms:
    \begin{eqnarray}
      A &\in& \Omega^1(U^{[1]},\g)
      \nonumber\\
        (x,i) &\mapsto& A_i\of{x} \in \g
    \end{eqnarray}
  \item
    curving 2-forms:
    \begin{eqnarray}
      B &\in&  \Omega^2(U^{[1]},\h)
      \nonumber\\
        (x,i) &\mapsto& B_i\of{x} \in \h
    \end{eqnarray}
   \item
    connection transformation 1-forms:
    \begin{eqnarray}
      a &\in&  \Omega^1(U^{[1]},\h)
      \nonumber\\
        (x,i,j) &\mapsto& a_{ij}\of{x}
    \end{eqnarray}
  \item
   curving transformation 2-forms:
    \begin{eqnarray}
      \label{curving transformation 2-forms of nonabelian gerbe}
       d &\in& \Omega^2(U^{[2]},\h)
      \nonumber\\
        (x,i,j) &\mapsto& d_{ij}\of{x}
    \end{eqnarray}
  \item
   phases twisting the cocycle condition for the $h_{ijk}$: 
    \begin{eqnarray}
      \label{phases}
      \lambda \maps U^{[4]} &\to& \ker\of{t} \subseteq H
      \nonumber\\
      (x,i,j,k,l) &\mapsto& \lambda_{ijkl}\of{x}
      \end{eqnarray}
   \item 
   phases twisting the cocycle condition for the $a_{ij}$
      \begin{eqnarray}
      \alpha &\in&  \Omega^1(U^{[3]},\ker\of{dt})
      \nonumber\\
      (x,i,j,k) &\mapsto& \alpha_{ijk}\of{x}
      \end{eqnarray}
    \item
   phases twisting the cocycle condition for the $b_{ij}$
      \begin{eqnarray}
      \beta &\in& \Omega^2(U^{[2]},\ker\of{dt})
      \nonumber\\
      (x,i,j) &\mapsto& \beta_{ij}\of{x}
      \end{eqnarray}
    \item
      \begin{eqnarray}
      \gamma &\in&  \Omega^3(U^{[1]},\ker\of{dt})
      \nonumber\\
      (x,i) &\mapsto& \gamma_{i}\of{x}
     \end{eqnarray}
\end{itemize}
such that the following cocycle conditions are satisfied:
\begin{itemize}

  \item cocycle condition for the $g_{ij}$:
\begin{eqnarray}
  \label{gerbe transition law for the transition functions}
  g_{ij}g_{jk} &=& t\of{h_{ijk}}g_{ik} 
\end{eqnarray}
  \item cocycle condition for the $A_i$:
\begin{eqnarray}
  \label{gerbe trans law for connection 1-form}
  A_i + dt\of{a_{ij}} 
  &=& 
  g_{ij} A_j g_{ij}^{-1} + 
  g_{ij} \extd g_{ij}^{-1}
\end{eqnarray}
  \item cocycle condition for the $B_i$:
\begin{eqnarray}
  \label{gerbe trans law for curving 2-forms}
  B_i
  &=& 
  \alpha\of{g_{ij}}\of{B_j}
  +
  k_{ij}
  -
  d_{ij}
  -
  \beta_{ij}
    \,.
\end{eqnarray}
where
\begin{eqnarray}
\label{curvature of a_{ij}}
  k_{ij}
  \defas
  \extd a_{ij} + a_{ij}\wedge a_{ij} - d\alpha(A_i)\wedge a_{ij}
\end{eqnarray}
 \item cocycle condition for the $d_{ij}$:
  \begin{eqnarray}
  \label{gerbe trans law for curving transformation 2-forms}
    d_{ij} + g_{ij}\of{d_{jk}}
    &=&
    h_{ijk} \,d_{ik}\, h_{ijk}^{-1}
    +
    h_{ijk}d\alpha\of{dt\of{B_i} + F_{A_i}}{h_{ijk}^{-1}}
  \end{eqnarray}
  \item cocycle condition for the $h_{ijk}$:
\begin{eqnarray}
  \label{gerbe coherence law for transformators of transition functions}
   \alpha\of{g_{ij}}\of{h_{jkl}} \, h_{ijl} 
  &=&
  \lambda_{ijkl} \, h_{ijk} \, h_{ikl}
\end{eqnarray}
  \item
  cocycle condition for the $a_{ij}$:
  \begin{eqnarray}
    \label{gerbe coherence law for transformators of connection forms}    
    \alpha_{ijk}
    &=&
    a_{ij} + g_{ij}\of{a_{jk}}
    -
    h_{ijk} \,a_{ik}\, h_{ijk}^{-1}
    -
    h_{ijk}\, \extd h^{-1}_{ijk}
    -
    h_{ijk}\, d\alpha\of{A_i}(h_{ijk}^{-1})
    \,.
  \end{eqnarray}
\end{itemize}

\end{definition}

Finally, the \emph{curvature 3-form} of the nonabelian gerbe is defined as
\begin{eqnarray}
  \label{gerbe curvature 3-form}
  H_i 
   &\defas&
 d_{A_i}B_i + \gamma_i ,
\end{eqnarray}
and its transformation law is:
\begin{eqnarray}
  \label{gerbe curvature transition law}
  H_i
  &=&
  \phi_{ij}\of{H_j}
  -
  \extd d_{ij}
  -
  \commutator{a_{ij}}{d_{ij}}
  -
  d\alpha\of{dt\of{B_i} + F_{A_i}}\of{a_{ij}}
  -
  d\alpha\of{A_i}\of{d_{ij}}
  \,.
\end{eqnarray}

\clearpage
\section{2-Groups, Loop Groups  and the $\mathrm{String}$-group}
\label{2-Groups, Loop Groups  and String}

The following is taken from 
\cite{BaezCransSchreiberStevenson:2005}, which is joint work
with John Baez, Alissa Crans and Danny Stevenson.

\subsection{Introduction} \label{introductionsection}

We describe an interesting relation between Lie 2-algebras, the
Kac--Moody central extensions of loop groups, and the group
$\String(n)$.  A Lie 2-algebra is a categorified version of a Lie
algebra where the Jacobi identity holds up to a natural isomorphism
called the `Jacobiator'.  Similarly, a Lie 2-group is a categorified
version of a Lie group.  If $G$ is a simply-connected compact simple
Lie group, there is a 1-parameter family of Lie 2-algebras $\g_k$ each
having $\g$ as its Lie algebra of objects, but with a Jacobiator built
from the canonical 3-form on $G$.  There appears to be no Lie 2-group
having $\g_k$ as its Lie 2-algebra, except when $k = 0$.  Here,
however, we construct for integral $k$ an infinite-dimensional Lie
2-group $\PG$ whose Lie 2-algebra is {\it equivalent} to $\g_k$.  The
objects of $\PG$ are based paths in $G$, while the automorphisms of
any object form the level-$k$ Kac--Moody central extension of the loop
group $\OG$.  This 2-group is closely related to the $k$th power of
the canonical gerbe over $G$.  Its nerve gives a topological group
$|\PG|$ that is an extension of $G$ by $K(\Z,2)$.  When $k = \pm 1$,
$|\PG|$ can also be obtained by killing the third homotopy group of
$G$.  Thus, when $G = \Spin(n)$, $|\PG|$ is none other than $\String(n)$.

The theory of simple Lie groups and Lie algebras has long played a 
central role in mathematics.  Starting in the 1980s, a wave of research
motivated by physics has revitalized this theory, expanding it to include 
structures such as quantum groups, affine Lie algebras, and central 
extensions of loop groups.  All these structures rely for their existence 
on the left-invariant closed 3-form $\nu$ naturally possessed by any compact 
simple Lie group $G$:
\[              \nu(x,y,z) = \langle x, [y,z] \rangle   \qquad 
x,y,z \in \g , \]
or its close relative, the left-invariant closed 2-form $\omega$ on the loop 
group $\OG$:
\[      \omega(f, g) = 2
\int_{S^1} \langle f(\theta), g'(\theta) \rangle \, d \theta  
\qquad f,g \in \Og .\]
Moreover, all these new structures fit together in a grand framework that 
can best be understood with ideas from physics --- in particular, the 
Wess--Zumino--Witten model and Chern--Simons theory.  Since these ideas
arose from work on string theory, which replaces point particles by 
higher-dimensional extended objects, it is not surprising that their study 
uses concepts from higher-dimensional algebra, such as gerbes 
\cite{Brylinski,BrylinskiMcLaughlin:1994,CareyJohnsonMurrayStevenson:2004}.

More recently, work on higher-dimensional algebra has focused attention
on Lie 2-groups \cite{BaezLauda:2003} and Lie 2-algebras 
\cite{BaezCrans:2003}.  A
`2-group' is a category equipped with operations analogous to those of
a group, where all the usual group axioms hold only up to specified 
natural isomorphisms satisfying certain coherence laws of their own.  
A `Lie 2-group' is a 2-group where the set of objects and the set of 
morphisms are smooth manifolds, and all the operations and natural 
isomorphisms are smooth.  Similarly, a `Lie 2-algebra' is a category 
equipped with operations analogous to those of a Lie algebra, satisfying 
the usual laws up to coherent natural isomorphisms.  Just as Lie
groups and Lie algebras are important in gauge theory, Lie 2-groups 
and Lie 2-algebras are important in `higher gauge theory', which 
describes the parallel transport of higher-dimensional extended objects 
\cite{BaezSchreiber:2004,Bartels:2004}.

The question naturally arises whether every finite-dimensional Lie
2-algebra comes from a Lie 2-group.  The answer is surprisingly subtle, as
illustrated by a class of Lie 2-algebras coming from simple Lie algebras. 
Suppose $G$ is a simply-connected compact simple Lie group $G$, and let
$\g$ be its Lie algebra.   For any real number $k$, there is a Lie 2-algebra 
$\g_k$ for which the space of objects is $\g$, the space of endomorphisms 
of any object is $\R$, and the `Jacobiator'
\[     J_{x,y,z} \maps [[x,y], z] \stackto{\sim} [x, [y,z]] + [[x,z],y]  \]
is given by
\[      J_{x,y,z} = k \, \nu(x,y,z)  \]
where $\nu$ is as above.   If we normalize the invariant inner 
product $\langle \cdot, \cdot \rangle$ on $\g$ so that the de Rham
cohomology class of the closed form $\nu/2\pi$ generates the third integral 
cohomology of $G$, then there is a 2-group $G_k$ corresponding 
to $\g_k$ in a certain sense explained below whenever $k$ is an integer.  
The construction of this 2-group is very interesting, because it uses
Chern--Simons theory in an essential way.  However, for $k \ne 0$
there is no good way to make this 2-group into a Lie 2-group!  The
set of objects is naturally a smooth manifold, and so is the set
of morphisms, and the group operations are smooth, but the
associator
\[           a_{x,y,z} \maps (xy)z \stackto{\sim} x(yz)   \]
cannot be made everywhere smooth, or even continuous.

It would be disappointing if such a fundamental Lie 2-algebra as 
$\g_k$ failed to come from a Lie 2-group even when $k$ was an integer.  
Here we resolve this dilemma by finding a Lie 2-algebra {\it
equivalent} to $\g_k$ that {\it does} come from a Lie 2-group --- albeit 
an {\it infinite-dimensional} one. 

The point is that the natural concept of `sameness' for categories is 
a bit subtle: not isomorphism, but equivalence.  Two categories 
are `equivalent' if there are functors going back and forth between them 
that are inverses {\it up to natural isomorphism}.  Categories that 
superficially look quite different can turn out to be equivalent.
The same is true for 2-groups and Lie 2-algebras.    Taking advantage
of this, we show that while the finite-dimensional Lie 2-algebra $\g_k$ 
has no corresponding Lie 2-group, it is equivalent to an infinite-dimensional 
Lie 2-algebra $\Pg$ which comes from an infinite-dimensional Lie 
2-group $\PG$.

The 2-group $\PG$ is easy to describe, in part because it is `strict':
all the usual group axioms hold as equations.   The basic idea is easiest 
to understand using some geometry.  Apart from some technical fine print, 
an object of $\PG$ is just a path in $G$ starting at the identity.  A 
morphism from the path $f_1$ to the path $f_2$ is an equivalence class 
of pairs $(D,z)$ consisting of a disk $D$ going from $f_1$ to $f_2$ 
together with a unit complex number $z$:

\vskip 1em
\[
 %%
 %% This diagram is made using pstricks and xypic.  The pstricks diagram 
 %% is placed as an object at the grid position (0,0) in the xypic 
 %% diagram.  The double arrow is drawn using xypic.  All labels are 
 %% placed using pstricks.  This diagram requires the macro packages
 %%
 %%         \usepackage{pstricks}
 %%         \usepackage{pst-grad}
 %%         \usepackage[all]{xy}
 \xy
  (0,0)*{
    \begin{pspicture}(3,3)
  %% Makes bottom part of the sphere with a slightly darker shading
  %% than the top part of the sphere
    \pscustom[fillcolor=lightgray,fillstyle=gradient,
      gradbegin=gray, gradend=white, gradmidpoint=1,gradangle=110]{
        \psarc(1.5,1.5){1.24}{180}{0}
        \psellipse[linestyle=none](1.5,1.5)(1.25,.5)
    }
  %% Makes the top part of the sphere with slightly lighter shading
    \pscustom[linestyle=dotted,fillcolor=lightgray,fillstyle=gradient,
            gradbegin=lightgray, gradend=white, 
		  gradmidpoint=1,gradangle=110]{
        \psellipse[linestyle=dotted](1.5,1.5)(1.25,.5)
        \psarc[linestyle=dotted](1.5,1.5){1.24}{0}{180}
    }
  %% Creates the dark\non-dotted part of the ellipse
    \begin{psclip}{\pswedge[linestyle=none](1.5,1.5){1.25}{180}{0}}
          \psellipse(1.5,1.5)(1.25,.5)
    \end{psclip}
  %% Makes one big continuous circle
    \pscircle(1.5,1.5){1.25}
  %% Bullets
    \psdots(1.5,2.75)(2,1.2)
  %% Right part of curve
    \psbezier[arrows=->](1.5,2.75)(1.4,2.4)(2.4,2.3)(2,1.7)
    \psbezier(2,1.7)(1.9,1.5)(2.1,1.4)(2,1.2)
  %% Left part of curve
    \psbezier[arrows=->](1.5,2.75)(1,2.4)(1,1.9)(1.1,1.7)
    \psbezier(1.1,1.7)(1.3,1.3)(1.7,1.3)(2,1.2)
  %% Label for the group G
     \rput(-.3,2){$G$}
  %% Labels
    \rput(1.5,2.95){$\scriptstyle 1$}
    \rput(.85,1.7){$\scriptstyle f_1$}
    \rput(2.2,1.7){$\scriptstyle f_2$}
   \end{pspicture}};
   %% Double arrow and label D
        {\ar@{=>}_<<{\scriptstyle D} (-1,3);(3.5,3) };
 \endxy
\]
\vskip 1em

\noindent
Given two such pairs $(D_1,z_1)$ and $(D_2,z_2)$, we can
always find a 3-ball $B$ whose boundary is $D_1 \cup D_2$, and we say
the pairs are equivalent when
\[   z_2/z_1 = e^{ik \int_B \nu} \]
where $\nu$ is the left-invariant closed 3-form on $G$ given as
above.  Note that $\exp(ik \int_B \nu)$ is independent of the choice 
of $B$, because the integral of $\nu$ over any 3-sphere is $2\pi$
times an integer.  There is an obvious way to compose morphisms in $\PG$, 
and the resulting category inherits a Lie 2-group structure from the Lie 
group structure of $G$.

The above description of $\PG$ is modeled after Murray's construction 
\cite{Murray:1994} of a gerbe from an integral closed 3-form on a manifold 
with a chosen basepoint.  Indeed, $\PG$ is just another way of talking 
about the $k$th power of the canonical gerbe on $G$, and the 2-group 
structure on $\PG$ is a reflection of the fact that this gerbe is 
`multiplicative' in the sense of Brylinski \cite{Brylinski:2000}.  The
3-form $k\nu$, which plays the role of the Jacobiator in $\g_k$, is
the 3-curvature of a connection on this gerbe.

In most of this paper we take a slightly 
different viewpoint.  Let $P_0 G$ be the space of smooth 
paths $f \maps [0,2\pi] \to G$ that start at the identity of $G$. 
This becomes an infinite-dimensional Lie group under pointwise 
multiplication.  The map $f \mapsto f(2\pi)$ is a homomorphism from 
$P_0 G$ to $G$ whose kernel is precisely $\Omega G$.  For any $k \in \Z$, 
the loop group $\OG$ has a central extension
\[  1 \stackto{\;} \U(1) \stackto{\;} \wOkG \stackto{p} \OG \stackto{\;} 1 \]
which at the Lie algebra level is determined by the 2-cocycle 
$ik\omega$, with $\omega$ defined as above.  This is called the
`level-$k$ Kac--Moody central extension' of $G$.
The infinite-dimensional Lie 2-group $\PG$ has $P_0 G$ as its group of
objects, and given $f_1, f_2 \in P_0 G$, a morphism $\hat{\ell}
\maps f_1 \to f_2$ is an element $\hat{\ell} \in \wOkG$ such that
\[  f_2 / f_1 = p(\hat{\ell}) . \] 
In this description, composition of morphisms in $\PG$ is multiplication 
in $\wOkG$, while again $\PG$ becomes a Lie 2-group using the Lie group
structure of $G$.

To better understand the significance of the Lie 2-algebra $\g_k$
and the 2-group $G_k$ it is helpful to recall the classification of
2-groups and Lie 2-algebras.  In \cite{BaezCrans:2003} it is shown that Lie 
2-algebras are classified up to equivalence by quadruples consisting of:
\begin{itemize}
\item a Lie algebra $\g$,
\item an abelian Lie algebra $\h$,
\item a representation $\rho$ of $\g$ on $\h$,
\item an element $[j] \in H^3(\g,\h)$ of the Lie algebra cohomology of
$\g$.
\end{itemize}
Given a Lie 2-algebra $\lietwoalg$, we obtain this data
by choosing a `skeleton' $\lietwoalg_0$ of $\lietwoalg$: that is,
an equivalent Lie 2-algebra in which any pair of isomorphic objects
are equal.   The objects in this skeleton form the
Lie algebra $\g$, while the endomorphisms of any object
form the abelian Lie algebra $\h$.
The representation of $\g$ on $\h$ comes from the bracket in
$\lietwoalg_0$, and the element $[j]$ comes from the Jacobiator.

Similarly, in \cite{BaezLauda:2003} we give a proof of the already known
fact that 2-groups are classified up to equivalence by quadruples
consisting of:
\begin{itemize}
\item a group $G$,
\item an abelian group $H$,
\item an action $\alpha$ of $G$ as automorphisms of $H$,
\item an element $[a] \in H^3(G,H)$ of group cohomology of $G$.
\end{itemize}
Given a 2-group $\lietwogrp$, we obtain this data by choosing a
skeleton $\lietwogrp_0$: that is, an equivalent 2-group in which
any pair of isomorphic objects are equal.
The objects in this skeleton form the group $G$, while the
automorphisms of any object form the abelian group $H$.
The action of $G$ on $H$ comes from conjugation in $\lietwogrp_0$, 
and the element $[a]$ comes from the associator.

These strikingly parallel classifications suggest that 2-groups
should behave like Lie 2-algebras to the extent that group cohomology
resembles Lie algebra cohomology.  But this is where the subtleties
begin!

Suppose $G$ is a simply-connected compact simple Lie group, and let
$\g$ be its Lie algebra.   If $\rho$ is the trivial representation of $\g$ 
on $\u(1)$, we have 
\[           H^3(\g,\u(1)) \iso \R  \]
because this cohomology group can be identified with the third de Rham
cohomology group of $G$, which has the class $[\nu]$ as a basis.  
Thus, for any $k \in \R$ we 
obtain a skeletal Lie 2-algebra $\g_k$ having $\g$ as its Lie algebra 
of objects and $\u(1)$ as the endomorphisms of any object, where the 
Jacobiator in $\g_k$ is given by
\[            J_{x,y,z} =  k \nu(x,y,z) .\]

To build a 2-group $G_k$ analogous to this
Lie 2-algebra $\g_k$, we need to understand the relation between
$H^3(G,\U(1))$ and $H^3(\g,\u(1))$.  They are not isomorphic.  However,
$H^3(\g,\u(1))$ contains a lattice $\Lambda$ consisting of the 
integer multiples of $[\nu]$.  The papers of Chern--Simons 
\cite{ChernSimons:1974} and Cheeger--Simons 
\cite{CheegerSimons:1985} construct an inclusion
\[          \iota \maps \Lambda \hookrightarrow H^3(G,\U(1))  .\]
Thus, when $k$ is an integer, we can build a skeletal 2-group
$G_k$ having $G$ as its group of objects, $\U(1)$ as the group
of automorphisms of any object, the trivial action of
$G$ on $\U(1)$, and $[a] \in H^3(G,\U(1))$ given by $k \, \iota[\nu]$.

The question naturally arises whether $G_k$ can be made into
a Lie 2-group.  The problem is that there is no continuous
representative of the cohomology class $k\, \iota[\nu]$
unless $k = 0$.  Thus, for $k$ nonzero, we cannot make
$G_k$ into a Lie 2-group in any reasonable way.
More precisely, we have this result \cite{BaezLauda:2003}:

\begin{theorem} \et Let $G$ be a simply-connected compact
simple Lie group.  Unless $k = 0$, there is no way to give the
2-group $G_k$ the structure of a Lie 2-group for which the group 
$G$ of objects and the group $\U(1)$ of endomorphisms of any 
object are given their usual topology.
\end{theorem}

\noindent
The goal of this paper is to sidestep this `no-go theorem' by
finding a Lie 2-algebra equivalent to $\g_k$ which does come
from an (infinite-dimensional) Lie group when $k \in \Z$.
We show:

\begin{theorem} \et Let $G$ be a simply-connected compact
simple Lie group.  For any $k \in \Z$, there is a Fr\'echet Lie
2-group $\PG$ whose Lie 2-algebra $\Pg$ is equivalent to $\g_k$.
\end{theorem}

We also study the relation between $\PG$ and the topological group
$\hat G$ obtained by killing the third homotopy group of $G$.  When $G
= \Spin(n)$, this topological group is famous under the name of
$\String(n)$, since it plays a role in string theory
\cite{MurrayStevenson:2001,StolzTeichner:2004,Witten:1988}.  
More generally, any compact simple
Lie group $G$ has $\pi_3(G) = \Z$, but after killing $\pi_1(G)$ by
passing to the universal cover of $G$, one can then kill $\pi_3(G)$ by
passing to $\hat G$, which is defined as the homotopy fiber of the
canonical map from $G$ to the Eilenberg--Mac Lane space $K(\Z,3)$.
This specifies $\hat G$ up to homotopy, but there is still the
interesting problem of finding nice geometrical models for $\hat G$.

Before presenting their solution to this problem, Stolz and Teichner
\cite{StolzTeichner:2004} wrote: ``To our best knowledge, there has yet not
been found a canonical construction for $\String(n)$ which has
reasonable `size' and a geometric interpretation.''  Here we present
another solution.  There is a way to turn any topological 2-group $C$
into a topological group $|C|$, which we explain in Section
\ref{topology.section}.  Applying this to $\PG$ when $k = \pm 1$, we
obtain $\hat G$:

\begin{theorem} \et  Let $G$ be a simply-connected compact
simple Lie group.  Then $|\PG|$ is an extension of $G$ by a
topological group that is homotopy equivalent to $K(\Z,2)$.
Moreover, $|\PG| \simeq \hat{G}$ when $k = \pm 1$.
\end{theorem}

\noindent
While this construction of $\hat G$ uses simplicial methods and
is thus arguably less `geometric' than that of Stolz and Teichner,
it avoids their use of type III$_1$ von Neumann algebras, and 
has a simple relation to the Kac--Moody central extension of $G$.

\subsection{2-Groups and 2-Algebras}

We begin with a review of Lie 2-algebras and Lie 2-groups.
More details can be found in our papers HDA5 \cite{BaezLauda:2003} and 
HDA6 \cite{BaezCrans:2003}.  
Our notation largely follows that of these papers,
but the reader should be warned that here we denote the composite of 
morphisms $f \maps x \rightarrow y$ and $g \maps y \rightarrow z$ as 
$g \circ f \maps x \rightarrow z.$  

\subsubsection{Lie 2-algebras}
\label{Lie 2-algebras}

The concept of `Lie 2-algebra' blends together the notion of a
Lie algebra with that of a category.  Just as a Lie algebra has an
underlying vector space, a Lie 2-algebra has an underlying
2-vector space: that is, a category where everything is {\it
linear}.  More precisely, a {\bf 2-vector space} $L$ is a
category for which:
\begin{itemize}
\item the set of objects $\Ob(L)$, 
\item the set of morphisms $\Mor(L)$
\end{itemize}
are both vector spaces, and:
\begin{itemize} 
\item 
the maps $s, t \maps \Mor(L) \to
\Ob(L)$ sending any morphism to its source and target,
\item the map $i \maps \Ob(L) \to \Mor(L)$ sending any object 
to its identity morphism, 
\item
the map $\circ$ sending any composable pair of morphisms 
to its composite
\end{itemize}
are all linear.  As usual, we write a
morphism as $f \maps x \to y$ when $s(f) = x$ and $t(f) = y$, and
we often write $i(x)$ as $1_x$.

To obtain a Lie $2$-algebra, we begin with a $2$-vector space
and equip it with a bracket functor, which satisfies the
Jacobi identity up to a natural isomorphism called the 
`Jacobiator'. Then we require that the Jacobiator satisfy a new
coherence law of its own: the `Jacobiator identity'.

\begin{definition} \et \label{defnlie2alg}
A {\bf Lie $2$-algebra} consists of:

\begin{itemize}
\item a $2$-vector space $L$
\end{itemize}
equipped with:
\begin{itemize}
\item a functor called the {\bf bracket}
\[   [\cdot, \cdot] \maps L \times L \to L ,\]
bilinear and skew-symmetric as a function of objects and morphisms,
\item a natural isomorphism called the {\bf Jacobiator},
\[ J_{x,y,z} \maps [[x,y],z] \to [x,[y,z]] + [[x,z],y],\]
trilinear and antisymmetric as a function of the objects $x,y,z \in L$,
\end{itemize}
such that: 
\begin{itemize}
\item the {\bf Jacobiator identity} holds: the following 
diagram commutes for all objects $w,x,y,z \in L$:
$$ \def\objectstyle{\scriptstyle}
    \def\labelstyle{\scriptstyle}
     \xy
     (0,35)*+{[[[w,x],y],z]}="1";
     (-40,20)*+{[[[w,y],x],z] + [[w,[x,y]],z]}="2";
     (40,20)*+{[[[w,x],z],y] + [[w,x],[y,z]]}="3";
     (-40,0)*+{[[[w,y],z],x] + [[w,y],[x,z]]}="4'";
     (-40,-4)*+{+ [w,[[x,y],z]] + [[w,z],[x,y]]}="4";
     (40,0)*+{[[w,[x,z]],y]}="5'";
     (40,-4)*+{+ [[w,x],[y,z]] + [[[w,z],x],y]}="5";
     (-35,-23)*+{[[[w,z],y],x] + [[w,[y,z]],x]}="6'";
     (-35,-28)*+{+ [[w,y], [x,z]] + [w,[[x,y],z]] + [[w,z],[x,y]]}="6";
     (35,-23)*+{[[[w,z],y],x] + [[w,z],[x,y]]  + [[w,y],[x,z]]}="7'";
     (35,-28)*+{+ [w,[[x,z],y]]  + [[w,[y,z]],x] + [w,[x,[y,z]]]}="7";
     %(0,-40)*+{[[[w,z],y],x] + [[w,z],[x,y]]  + [[w,y],[x,z]]}="8'";
     %(0,-44)*+{+ [w,[[x,z],y]]  + [[w,[y,z]],x] + [w,[x,[y,z]]]}="8";
              %(32,-31)*{J_{w,[x,z],y} }; %Label for 7-8 arrow
              %(32,-34.5)*{+  J_{[w,z],x,y} + J_{w,x,[y,z]}};
          {\ar_{[J_{w,x,y},z]}                   "1";"2"};
          {\ar^{J_{[w,x],y,z}}                               "1";"3"};
          {\ar_{J_{[w,y],x,z} + J_{w,[x,y],z}}   "2";"4'"};
          {\ar_{[J_{w,y,z},x]+1}                 "4";"6'"};
          {\ar^{[J_{w,x,z},y]+1}                   "3";"5'"};
          {\ar^{J_{w,[x,z],y}+  J_{[w,z],x,y} + J_{w,x,[y,z]}}                 "5";"7'"};
          {\ar_{[w,J_{x,y,z}]+1 \; \; }          "6";"7"};
          %{\ar^{}                                "7";"8'"};
\endxy
\\ \\
$$
\end{itemize}
\end{definition}

A homomorphism between Lie $2$-algebras is a linear functor
preserving the bracket, but only up to a specified natural isomorphism 
satisfying a suitable coherence law.  More precisely:

\begin{definition} \et \label{lie2algfunct} Given
Lie $2$-algebras $L$ and $L'$, a {\bf homomorphism} $F \maps L
\rightarrow L'$ consists of:

\begin{itemize}
    \item a functor $F$ from the underlying $2$-vector space of
        $L$ to that of $L'$, linear on objects and morphisms,
\item a natural isomorphism
        $$F_{2}(x,y)\maps [F(x), F(y)] \rightarrow F[x,y],$$
bilinear and skew-symmetric as a function of the objects $x, y \in L$,
\end{itemize}
such that:
\begin{itemize}
\item 
the following diagram commutes for all objects $x,y,z \in L$:
$$\xymatrix{
      [F(x), [F(y), F(z)]]
        \ar[rrrr]^<<<<<<<<<<<<<<<<<<<<<<{J_{F(x), F(y),
F(z)}}        \ar[dd]_{[1, F_{2}]}
         &&&& [[F(x), F(y)], F(z)] + [F(y), [F(x), F(z)]]
        \ar[dd]^{[F_{2}, 1] + [1, F_{2}]} \\ \\
         [F(x), F[y,z]]
        \ar[dd]_{F_{2}}
         &&&& [F[x,y], F(z)] + [F(y), F[x,z]]
        \ar[dd]^{F_{2} + F_{2}} \\ \\
         F[x,[y,z]]
        \ar[rrrr]^{F(J_{x,y,z})}
         &&&& F[[x,y],z] + F[y,[x,z]]}$$
\end{itemize}
\end{definition}

\noindent Here and elsewhere we omit the arguments of natural
transformations such as $F_2$ and $G_2$ when these are obvious
from context.

Similarly, a `2-homomorphism' is a linear natural isomorphism that is
compatible with the bracket structure:

\begin{definition} \et \label{lie2algnattrans} Let $F,G \maps
L \to L'$ be Lie 2-algebra homomorphisms.  A {\bf 2-homomorphism}
$\theta \maps F \To G$ is a natural transformation 
\[          \theta_x \maps F(x) \to G(x) , \]
linear as a function of the object $x \in L$, such that the following 
diagram commutes for all $x, y \in L$:
$$\xymatrix{
      [F(x), F(y)]
       \ar[rr]^{F_{2}}
       \ar[dd]_{[\theta_{x}, \theta_{y}]}
        && F[x,y]
       \ar[dd]^{\theta_{[x,y]}} \\ \\
        [G(x), G(y)]
       \ar[rr]^{G_{2}}
        && G[x,y] }$$

\end{definition}

In HDA6 we showed:

\begin{proposition} \et There is a strict 2-category {\bf Lie2Alg}
with Lie $2$-algebras as \break objects, 
homomorphisms between these as morphisms, and $2$-homomorphisms
between those as 2-morphisms.
\end{proposition}

\subsubsection{$L_\infty$-algebras}
\label{Linfty.section}

Just as the concept of Lie 2-algebra blends the notions of Lie
algebra and category, the concept of `$L_\infty$-algebra' blends
the notions of Lie algebra and chain complex.  More precisely, an
$L_\infty$-algebra is a chain complex equipped with a bilinear
skew-symmetric bracket operation that satisfies the Jacobi
identity up to a chain homotopy, which in turn satisfies a 
law of its own up to chain homotopy, and so on {\it ad infinitum}.   
In fact, $L_\infty$-algebras were defined long before Lie 2-algebras, 
going back to a 1985 paper by Schlessinger and Stasheff 
\cite{SchlessingerStasheff:1985}.  
They are also called `strongly homotopy Lie algebras', 
or `sh Lie algebras' for short.

Our conventions regarding $L_\infty$-algebras follow those of Lada 
and Markl \cite{LadaMarkl:1994}.  In particular, for graded objects $x_{1}, 
\ldots, x_{n}$ and a permutation $\sigma \in S_{n}$ we define the 
{\bf Koszul sign} $\epsilon(\sigma)$ by the 
equation
\[
x_{1} \wedge \cdots \wedge x_{n} = \epsilon(\sigma)
\, x_{\sigma(1)} \wedge \cdots \wedge
x_{\sigma(n)},
\]
which must be satisfied in the free
graded-commutative algebra on $x_{1}, \ldots, x_{n}.$  
Furthermore, we define
\[
\chi(\sigma) = \textrm{sgn} (\sigma) \,
\epsilon(\sigma; x_{1}, \dots, x_{n}).
\] 
Thus, $\chi(\sigma)$ takes into account the sign of the permutation in 
$S_{n}$ as well as the Koszul sign.
Finally, if $n$ is a natural number and $1 \leq j \leq n-1$ we 
say that $\sigma \in S_{n}$ is an $(j,n-j)${\bf -unshuffle} if
\[
\sigma(1) \leq\sigma(2) \leq \cdots \leq \sigma(j)
\hspace{.2in} \textrm{and} \hspace{.2in} \sigma(j+1) \leq
\sigma(j+2) \leq \cdots \leq \sigma(n).
\]
Readers familiar with shuffles will recognize unshuffles as their 
inverses. 

\begin{definition} \et \label{L-alg} An
\textbf{\textit{L}$_{\mathbf{\infty}}$-{\bf algebra}} 
is a graded vector space $V$
equipped with a system $\{l_{k}| 1 \leq k < \infty\}$ of linear
maps $l_{k} \maps V^{\otimes k} \rightarrow V$ with $\deg(l_{k}) =
k-2$ which are totally antisymmetric in the sense that
\begin{eqnarray}
   l_{k}(x_{\sigma(1)}, \dots,x_{\sigma(k)}) =
   \chi(\sigma)l_{k}(x_{1}, \dots, x_{n})
\label{antisymmetry}
\end{eqnarray}
for all $\sigma \in S_{n}$ and $x_{1}, \dots, x_{n} \in V,$ and,
moreover, the following generalized form of the Jacobi identity
holds for $0 \le n < \infty :$
\begin{eqnarray}
   \displaystyle{\sum_{i+j = n+1}
   \sum_{\sigma}
   \chi(\sigma)(-1)^{i(j-1)} l_{j}
   (l_{i}(x_{\sigma(1)}, \dots, x_{\sigma(i)}), x_{\sigma(i+1)},
   \ldots, x_{\sigma(n)}) =0,}
\label{megajacobi}
\end{eqnarray}
where the summation is taken over all $(i,n-i)$-unshuffles with $i
\geq 1.$
\end{definition}

In this definition 
the map $l_1$ makes $V$ into a chain complex, since this map has
degree $-1$ and Equation (\ref{megajacobi}) says its square is
zero.  In what follows, we denote $l_1$ as $d$.  
The map $l_{2}$ resembles a Lie bracket, since it
is skew-symmetric in the graded sense by Equation
(\ref{antisymmetry}). The higher $l_k$ maps are related to the 
Jacobiator and the Jacobiator identity.

To make this more precise, we make the following definition:

\begin{definition} \et A
\textbf{k-{\bf term} \textit{L}$_{\mathbf{\infty}}$-{\bf algebra}} 
is an $L_{\infty}$-algebra $V$ with $V_{n} = 0$ for $n \geq k$. 
\end{definition}

A $1$-term $L_{\infty}$-algebra is simply an ordinary Lie algebra, 
where $l_{3} =0$ gives the Jacobi identity.  However, in a $2$-term
$L_{\infty}$-algebra, we no longer have $l_3 = 0$.
Instead, Equation (\ref{megajacobi}) says that the Jacobi identity
for $x,y,z \in V_0$ holds up to a term of the form
$dl_3(x,y,z)$.  We do, however, have $l_{4} = 0$, which provides
us with the coherence law that $l_{3}$ must satisfy.
It follows that a $2$-term $L_{\infty}$-algebra consists of:
\begin{itemize}
    \item vector spaces $V_{0}$ and
     $V_{1}$,

    \item a linear map $d\maps V_{1} \rightarrow V_{0},$

    \item bilinear maps $l_{2}\maps V_{i} \times V_{j}
     \rightarrow V_{i+j},$ where $0 \le i+j \le 1$,

    \item a trilinear map $l_{3}\maps V_{0} \times V_{0} \times
     V_{0} \rightarrow V_{1}$

\end{itemize}
satisfying a list of equations coming from Equations (\ref{antisymmetry})
and (\ref{megajacobi}) and the fact that $l_4 = 0$.
This list can be found in HDA6, but we will not need it here.

In fact, $2$-vector spaces are 
equivalent to 2-term chain complexes of vector spaces: that
is, chain complexes of the form 
\[                     V_1 \stackto{d} V_0 . \] 
To obtain such a chain complex from a 2-vector space $L$, we let 
$V_0$ be the space of objects of $L$.   
However, $V_1$ is not the space of morphisms.  Instead, we define 
the {\bf arrow part} $\vec{f}$ of a morphism $f \maps x \to y$ by
$$\vec{f} = f - i(s(f)), $$
and let $V_1$ be the space of these arrow parts.
The map $d \maps V_1 \to V_0$ is then just the target map $t
\maps \Mor(L) \to \Ob(L)$ restricted to $V_1 \subseteq \Mor(L)$.

To understand this construction a bit better, note 
that given any morphism $f \maps x \to y$, its arrow
part is a morphism $\vec{f} \maps 0 \rightarrow y-x.$   
Thus, taking the arrow part has the effect of `translating $f$ 
to the origin'.  We can always recover any morphism from its source together 
with its arrow part, since $f = \vec{f} + i(s(f))$.   It follows that
any morphism $f \maps x \to y$ can be identified 
with the ordered pair $(x, \vec{f})$ consisting of 
its source and arrow part.  So, we have $\Mor(L) \iso V_0 \oplus V_1$.

We can actually recover the whole 2-vector space structure of $L$
from just the chain complex $d \maps V_1 \to V_0$.
To do this, we take:
\begin{eqnarray*}
  \Ob(L) & = & V_{0} \\
  \Mor(L) & = & V_{0} \oplus V_{1},
\end{eqnarray*}
with source, target and identity-assigning maps defined by:
\[
\begin{array}{ccl}
  s(x, \vec{f}) &=& x \\
  t(x, \vec{f}) &=& x + d\vec{f} \\
  i(x) &=& (x, 0) 
\end{array}
\]
and with the composite of $f \maps x \to y$ and $g \maps y \to z$
defined by: 
\[  g \circ f = (x, \vec{f} + \vec{g}).
\]
So, 2-vector spaces are equivalent to 2-term chain complexes.

Given this, it should not be surprising that Lie $2$-algebras 
are equivalent to 2-term $L_\infty$-algebras.  Since we make
frequent use of this fact in the calculations to come, we recall
the details here.

Suppose $V$ is a 2-term $L_\infty$-algebra.   We obtain a 2-vector
space $L$ from the underlying chain complex of $V$ as above.
We continue by equipping $L$ with additional structure that
makes it a Lie $2$-algebra. It is sufficient to define the bracket
functor $[\cdot, \cdot] \maps L \times L \rightarrow L$ on a pair
of objects and on a pair of morphisms where one is an identity
morphism.  So, we set:
\[  
\begin{array}{ccl}
[x,y] &=& l_2(x,y)       , \cr
[1_z,f] &=& (l_2(z,x), l_2(z,\vec{f})) , \cr
[f,1_z] &=& (l_2(x,z), l_2(\vec{f}, z))  , 
\end{array}
\]
where $f \maps x \to y$ is a morphism in $L$ and $z$ is an object.
Finally, we define the Jacobiator for $L$ in terms of its source and arrow
part as follows:
$$J_{x,y,z} = ([[x,y],z], l_{3}(x,y,z)).  $$
For a proof that $L$ defined this way is actually a Lie 2-algebra, see
HDA6.

In our calculations we shall often describe Lie 2-algebra homomorphisms
as homomorphisms between the corresponding $2$-term $L_{\infty}$-algebras:

\begin{definition} \et \label{Linftyhomo}
Let $V$ and $V'$ be $2$-term $L_{\infty}$-algebras. An
\textbf{\textit{L}$_{\mathbf{\infty}}$-{\bf homomorphism}} 
$\phi \maps V \rightarrow V'$ consists of:
\begin{itemize}
\item a chain map $\phi \maps V \to V'$ consisting of linear maps
$\phi_0 \maps V_0 \to V'_0$  and
$\phi_1 \maps V_1 \to V'_1$, 

\item a skew-symmetric bilinear map 
$\phi_{2} \maps V_{0} \times V_{0} \to V_{1}'$,
\end{itemize}

such that the following equations hold for all $x,y,z \in V_0$
and $h \in V_{1}:$
\begin{eqnarray}
d (\phi_{2}(x,y)) = \phi_{0}(l_{2} (x,y)) - l_{2}(\phi_{0}(x), \phi_{0}(y)) 
\label{homo1}
\end{eqnarray}
\begin{eqnarray}
\phi_{2}(x,dh) = \phi_{1}(l_{2}(x,h)) - l_{2}(\phi_{0}(x), \phi_{1}(h)) 
\label{homo2}
\end{eqnarray}
\begin{equation}
\begin{array}{l}
l_3(\phi_0(x),\phi_0(y), \phi_0(z))
  -\phi_1(l_3(x,y,z)) = \\  
   \phi_2(x,l_2(y,z))
   + \phi_2(y,l_2(z,x))
   + \phi_2(z,l_2(x,y)) \; +  \\
  {}   l_2(\phi_0(x),\phi_2(y,z))
  +    l_2(\phi_0(y),\phi_2(z,x))
  +    l_2(\phi_0(z),\phi_2(x,y))
\label{homo3}
\end{array}
\end{equation}
\end{definition}

\noindent 
Equations (\ref{homo1}) and (\ref{homo2})
say that $\phi_{2}$ defines a chain homotopy from
$l_{2}(\phi(\cdot), \phi(\cdot))$ to $\phi(l_{2}(\cdot, \cdot))$,
where these are regarded as chain maps from $V \otimes V$ to $V'$.
Equation (\ref{homo3}) is just a chain complex
version of the commutative diagram in Definition
\ref{lie2algfunct}.

Without providing too many details, let us sketch how to obtain
the Lie 2-algebra homomorphism $F$ corresponding to a given
$L_{\infty}$-homomorphism $\phi \maps V \to V'$. We define
the chain map $F \maps L \to L'$ in terms of $\phi$ using the 
fact that objects of a 2-vector space are 0-chains in the corresponding 
chain complex, while morphisms are pairs consisting of a 0-chain
and a 1-chain.  To make $F$ into a Lie 2-algebra homomorphism
we must equip it with a skew-symmetric bilinear natural
transformation $F_{2}$ satisfying the conditions in Definition
\ref{lie2algfunct}. We do this using the skew-symmetric bilinear
map $\phi_{2} \maps V_{0} \times V_{0} \to V_{1}'$.  In terms of
its source and arrow parts, we let
$$F_{2}(x,y) = (l_{2}(\phi_{0}(x), \phi_{0}(y)), \phi_{2}(x,y)).$$

We should also know how to compose $L_\infty$-homomorphisms.
We compose a pair of $L_{\infty}$-homomorphisms $\phi
\maps V \rightarrow V'$ and $\psi \maps V' \rightarrow V''$ by
letting the chain map $\psi \circ \phi \maps V \to V''$ be the
usual composite:
$$
V \stackto{\phi} V' \stackto{\psi} V''
$$
while defining $(\psi \circ \phi)_{2}$ as follows:
\begin{equation}
 (\psi \circ \phi)_2(x,y) = \psi_2(\phi_0(x),\phi_0(y)) +
                                \psi_1(\phi_2(x,y)).
\label{composite.homo}
\end{equation}
This is just a chain complex version of how we compose
homomorphisms between Lie 2-algebras.  Note that the identity homomorphism
$1_V \maps V \to V$ has the identity chain map as its underlying
map, together with $(1_V)_2 = 0$.

We also have `2-homomorphisms' between homomorphisms:

\begin{definition} \et \label{Linfty2homo} Let $V$ and $V'$ be
2-term $L_\infty$-algebras and let $\phi, \psi \maps V \to V'$ be
$L_{\infty}$-homomorphisms.  An 
\textbf{\textit{L}$_{\mathbf{\infty}}$-{\bf 2-homomorphism}} 
$\tau \maps \phi \To \psi$ is a chain
homotopy $\tau$ from $\phi$ to $\psi$
such that the following equation holds for all $x,y \in
V_0$: 
\begin{equation}
\phi_2(x,y) - \psi_2(x,y) =  
l_{2}(\phi_{0}(x), \tau(y)) + l_{2}(\tau(x), \psi_{0}(y)) -
\tau ( l_{2}(x,y)) 
\label{2homo}
\end{equation}
\end{definition}

Given an $L_{\infty}$-2-homomorphism $\tau \maps \phi \To \psi$ between
$L_\infty$-homomorphisms $\phi, \psi \maps V \to V'$, there is a 
corresponding Lie 2-algebra 2-homomorphism $\theta$ 
whose source and arrow part are:
\[
\theta(x) = (\phi_{0} (x), \tau(x))
\]
for any object $x$.  
Checking that this really is a Lie 2-algebra 2-homomorphism is routine. 
In particular, Equation (\ref{2homo}) is just a chain complex version of
the commutative diagram in the Definition \ref{lie2algnattrans}.

In HDA6, we showed:

\begin{proposition} \et There is a strict $2$-category
{\bf 2TermL$_\mathbf\infty$} with $2$-term $L_{\infty}$-algebras
as objects, $L_\infty$-homomorphisms as morphisms, and
$L_\infty$-$2$-homomorphisms as $2$-morphisms.
\end{proposition}

Using the equivalence between $2$-vector spaces and $2$-term chain
complexes, we established the equivalence between Lie $2$-algebras
and $2$-term $L_{\infty}$-algebras:

\begin{theorem} \label{1-1'} \et The $2$-categories {\rm Lie$2$Alg}
and {\rm 2TermL$_{\infty}$} are $2$-equivalent.
\end{theorem}

We use this result extensively in Section \ref{equivalence.section}.
Instead of working in Lie2Alg, we do calculations in 2TermL$_\infty$.
The reason is that defining Lie 2-algebra homomorphisms and
2-homomorphisms would require specifying both source and arrow parts
of morphisms, while defining the corresponding $L_\infty$-morphisms
and 2-morphisms only requires us to specify the arrow parts.
Manipulating the arrow parts rather than the full-fledged morphisms
leads to less complicated computations.

\subsubsection{The Lie 2-Algebra $\g_k$}
\label{ghbar.section}

Another benefit of the
equivalence between Lie 2-algebras and $L_{\infty}$-algebras is
that it gives some important examples of Lie
$2$-algebras.  Instead of thinking of a Lie $2$-algebra as a
category equipped with extra structure, we may work with a
$2$-term chain complex endowed with the structure described in
Definition \ref{L-alg}.   This is especially simple when the
differential $d$ vanishes.  Thanks to the formula
\[ 
  d\vec{f} =  t(f) - s(f)   ,
\]
this implies that
the source of any morphism in the Lie $2$-algebra equals its
target.  In other words, the corresponding Lie $2$-algebra is 
`skeletal':

\begin{definition} \et
A category is {\bf skeletal} if isomorphic objects are always equal.
\end{definition}

Every category is equivalent to a skeletal one formed by choosing one
representative of each isomorphism class of objects 
\cite{MacLane}.
As shown in HDA6, the same sort of thing is true in the context of 
Lie $2$-algebras:

\begin{proposition} \et \label{skeletal}
Every Lie $2$-algebra is equivalent, as an object of {\rm
Lie2Alg}, to a skeletal one.
\end{proposition}

This result helps us classify Lie 2-algebras up to 
equivalence.  We begin by reminding the reader of the relationship between
$L_{\infty}$-algebras and Lie algebra cohomology described in HDA6:

\begin{theorem} \et \label{trivd}
There is a one-to-one correspondence between isomorphism classes
of $L_{\infty}$-algebras
consisting of only two nonzero terms $V_{0}$ and $V_{n}$ with
$d=0,$ and isomorphism classes of quadruples $(\g, V, \rho, [l_{n+2}])$
where $\g$ is a Lie algebra, $V$ is a vector space, $\rho$ is
a representation of $\g$ on $V$, and $[l_{n+2}]$ is an element 
of the Lie algebra cohomology group $H^{n+2}(\g,V)$.
\end{theorem}

\noindent
Here the representation $\rho$ comes from $\ell_2 \maps V_0 \times V_n 
\to V_n$.  

Because $L_{\infty}$-algebras are equivalent to Lie
2-algebras, which all have equivalent skeletal versions,
Theorem \ref{trivd} implies:

\begin{corollary} \et \label{class} Up to equivalence, Lie 2-algebras
are classified by isomorphism classes of quadruples $(\g,\rho,V,[\ell_3])$
where: 
\begin{itemize}
\item $\g$ is a Lie algebra, 
\item $V$ is a vector space,
\item $\rho$ is a representation of $\g$ on $V$,
\item $[\ell_3]$ is an element of $H^3(\g,V)$. 
\end{itemize}
\end{corollary}

\noindent
This classification of Lie 2-algebras is just another way of stating 
the result mentioned in the Introduction.  And, as mentioned there,
this classification lets us construct a 1-parameter family of
Lie 2-algebras $\g_k$ for any simple real Lie algebra $\g$:

\begin{example} \label{ghbar} \et Suppose $\g$ is a simple real
Lie algebra and $k \in \R$.  Then there is a skeletal Lie $2$-algebra
$\g_k$ given by taking $V_{0} = \g$, $V_{1} = \R$, $\rho$ the trivial 
representation,  and $l_3(x,y,z) = k\langle x,[y,z]\rangle $. 
\end{example}

\noindent
Here $\langle \cdot, \cdot \rangle$ is a suitably rescaled version
of the Killing form $\tr(\ad(\cdot)\ad(\cdot))$.
The precise rescaling factor will only
become important in Section \ref{loop.basic.section}.  
The equation saying that $l_3$ is a $3$-cocycle is equivalent to the 
equation saying that the left-invariant 3-form $\nu$ on $G$ with 
$\nu(x,y,z) = \langle x,[y,z] \rangle$ is {\it closed}.

\subsubsection{The Lie 2-Algebra of a Fr\'echet Lie 2-Group}
\label{frechet.section}

Just as Lie groups have Lie algebras, `strict Lie 2-groups' have
`strict Lie 2-algebras'.  Strict Lie $2$-groups and Lie
$2$-algebras are categorified versions of Lie groups and Lie
algebras in which all laws hold `on the nose' as equations, rather
than up to isomorphism.    All the Lie 2-groups discussed in this
paper are strict.  However, most of them are infinite-dimensional
`Fr\'echet' Lie 2-groups.  

Since the concept of a Fr\'echet Lie group is easy to explain but 
perhaps not familiar to all readers, we begin by recalling this.
For more details we refer the interested reader to the survey article 
by Milnor \cite{Milnor:1984}, or Pressley and Segal's book on loop groups 
\cite{PressleySegal:1986}. 

A {\bf Fr\'echet space} is a vector space equipped with a topology 
given by a countable family of seminorms $\| \cdot \|_n$, or 
equivalently by the metric
\[ d(x,y) = \sum_n 2^{-n}  \frac{\|x - y\|_n}{\|x - y\|_n + 1} \, ,\]
where we require that this metric be complete.
A classic example is the space of smooth maps from the 
interval or circle to a finite-dimensional normed vector space,
where $\|f\|_n$ is the supremum of the norm of the $n$th derivative
of $f$.  In particular, the space of smooth paths or loops in a 
finite-dimensional simple Lie algebra is a Fr\'echet space.  This 
is the sort of example we shall need.

The theory of manifolds generalizes from the finite-dimensional
case to the infinite-dimensional case by replacing $\R^n$ with 
a Fr\'echet space \cite{Hamilton:1982}.  In particular, there is a 
concept of the `Fr\'echet derivative' of a map between 
Fr\'echet spaces, and higher derivatives of such maps can also be 
defined.  If $V, W$ are Fr\'echet spaces and $U \subset V$ is an 
open set, a map 
$\phi \maps U \to W$ is called {\bf smooth} if its $n$th derivative 
exists for all $n$.  A {\bf Fr\'echet manifold} modeled on the 
Fr\'echet space $V$ is a paracompact Hausdorff space $M$ that can be 
covered with open sets $U_\alpha$ equipped with homeomorphisms 
$\phi_\alpha \maps U_\alpha \to V$ called {\bf charts}
such that the maps $\phi_\alpha \circ \phi_\beta^{-1}$ are 
smooth where defined.  In particular, the space of smooth paths 
or loops in a compact simple Lie group $G$ is naturally a Fr\'echet 
manifold modeled on the Fr\'echet space of smooth paths or loops 
in the Lie algebra $\g$.  

A map between Fr\'echet manifolds is {\bf smooth} if composing it
with charts and their inverses in the usual way, we get 
functions between Fr\'echet spaces that are smooth where defined.
A {\bf Fr\'echet Lie group} is a Fr\'echet manifold $G$ such that
the multiplication map $m \maps G \times G \to G$ and the inverse
map $\inv \maps G \to G$ are smooth.   A {\bf homomorphism} of
Fr\'echet Lie groups is a group homomorphism that is also smooth.

Finally:

\begin{definition} \label{frechetlietwogroup}  \et
A {\bf strict Fr\'echet Lie $2$-group} $C$ is a
category such that:
\begin{itemize}
\item the set of objects $\Ob(C)$ and
\item the set of morphisms $\Mor(C)$
\end{itemize}
are both Fr\'echet Lie groups, and:
\begin{itemize} 
\item
the maps $s, t \maps \Mor(C) \to
\Ob(L)$ sending any morphism to its source and target,
\item the map $i \maps \Ob(C) \to \Mor(C)$ sending any object 
to its identity morphism, 
\item
the map $\circ \maps \Mor(C) \times_{\Ob(C)} \Mor(C) \to \Mor(C)$
sending any composable pair of morphisms to its composite
\end{itemize}
are all Fr\'echet Lie group homomorphisms.
\end{definition}

\noindent
Here $\Mor(C) \times_{\Ob(C)} \Mor(C)$ is the set of composable
pairs of morphisms, which we require to be a Fr\'echet Lie group.  

Just as for ordinary Lie groups, taking the tangent space at the 
identity of a Fr\'echet Lie group gives a Lie algebra.
Using this, it is not hard to see that strict Fr\'echet Lie 2-groups
give rise to Lie 2-algebras.  These Lie 2-algebras are actually
`strict':

\begin{definition} \et
A Lie 2-algebra is {\bf strict} if its Jacobiator is the identity.
\end{definition}  

\noindent 
This means that the map $l_3$ vanishes in the corresponding 
$L_\infty$-algebra.  Alternatively:

\begin{proposition} \et
A strict Lie 2-algebra is the same as a 2-vector space 
$L$ such that:
\begin{itemize}
\item $\Ob(L)$ is equipped with the structure of a Lie algebra,
\item $\Mor(L)$ is equipped with the structure of a Lie algebra, 
\end{itemize}
and: 
\begin{itemize}
\item
the source and target maps 
$s,t \maps \Mor(L) \to \Ob(L)$, 
\item
the identity-assigning map
$i \maps \Ob(L) \to \Mor(L)$, and 
\item
the composition map
$\circ \maps \Mor(L) \times_{\Ob(L)} \Mor(L) \to \Mor(L)$
\end{itemize}
are Lie algebra homomorphisms.
\end{proposition}

\Proof - 
A straightforward verification; see also HDA6.  
\endofproof

\begin{proposition} \label{lietwoalg.of.lietwogrp} \et
Given a strict Fr\'echet Lie $2$-group $\lietwogrp$, there is 
a strict Lie $2$-algebra $\lietwoalg$ for which:
\begin{itemize} 
\item $\Ob(\lietwoalg)$ is the Lie algebra
of the Fr\'echet Lie group $\Ob(\lietwogrp)$, 
\item $\Mor(\lietwoalg)$ is the Lie
algebra of the Fr\'echet Lie group $\Mor(\lietwogrp)$,
\end{itemize}
and the maps:
\begin{itemize}
\item
$s,t \maps \Mor(\lietwoalg) \to \Ob(\lietwoalg)$, 
\item
$i \maps \Ob(\lietwoalg) \to \Mor(\lietwoalg)$, and 
\item
$\circ \maps \Mor(\lietwoalg) \times_{\Ob(\lietwoalg)} \Mor(\lietwoalg) 
\to \Mor(\lietwoalg)$
\end{itemize}
are the differentials of the corresponding maps for $\lietwogrp$.  
\end{proposition}

\Proof  This is a generalization
of a result in HDA6 for ordinary Lie 2-groups,
which is straightforward to show directly.  
\endofproof

In what follows all Fr\'echet Lie
2-groups are strict, so we omit the term `strict'.

\subsection{Loop Groups}
\label{loop.section}

Next we give a brief review of loop groups and their central
extensions.  More details can be found in the canonical text
on the subject, written by Pressley and Segal \cite{PressleySegal:1986}.

\subsubsection{Definitions and Basic Properties}
\label{loop.basic.section}

Let $G$ be a simply-connected compact simple Lie group.  We shall be
interested in the {\bf loop group} $\OG$ consisting of all smooth maps
from $[0,2\pi]$ to $G$ with $f(0) = f(2\pi) = 1$.  We make $\OG$ into
a group by pointwise multiplication of loops: $(fg)(\theta) =
f(\theta) g(\theta)$.  Equipped with its $C^\infty$ topology, $\OG$
naturally becomes an infinite-dimensional Fr\'echet manifold.  In fact
$\OG$ is a Fr\'echet Lie group, as defined in Section
\ref{frechet.section}.

As remarked by Pressley and Segal, the behaviour of the group
$\OG$ is ``untypical in its simplicity,'' since it turns out to
behave remarkably
like a compact Lie group.  For example, it has an exponential
map that is locally one-to-one and onto, and it has a
well-understood highest weight theory of representations.  One
striking difference between $\OG$ and $G$, though, is the
existence of nontrivial central extensions of $\OG$ by the circle
$\TT$:
\begin{equation}
\label{eq: Kac--Moody extn}
1\to \TT \to \widehat{\Omega G}\stackrel{p}{\to} \Omega G\to 1 .
\end{equation}
It is important to understand that these extensions are nontrivial, 
not merely in that they are classified by a nonzero $2$-cocycle, but 
also \emph{topologically}.  In other words, $\wOG$ is a nontrivial
principal $\TT$-bundle over $\OG$ with the property that $\wOG$ is a
Fr\'echet Lie group, and $\TT$ sits inside $\widehat{\Omega G}$ as a 
central
subgroup in such a way that the quotient $\widehat{\Omega G}/\TT$
can be identified with $\Omega G$.  Perhaps the
best analogy is with the double cover of $\SO(3)$: there
$\SU(2)$ fibers over $\SO(3)$ as a $2$-sheeted covering and
$\SU(2)$ is not homeomorphic to $SO(3)\times \Z/2\Z$.  $\wOG$ is called
the \textbf{Kac--Moody group}.  

Associated to the central extension~\eqref{eq: Kac--Moody extn}
there is a central extension of Lie algebras:
\begin{equation}
\label{eq: Kac--Moody alg extn}
0 \to \u(1) \to \wOg \stackto{p} \Og \to 0
\end{equation}
Here $\Og$ is the Lie algebra of $\OG$,
consisting of all smooth maps $f \maps S^1\to \g$ such
that $f(0) = 0$.  The
bracket operation on $\Omega \mathfrak{g}$ is given by the
pointwise bracket of functions: thus $[f,g](\theta) =
[f(\theta),g(\theta)]$ if $f,g\in \Omega \mathfrak{g}$.
$\wOg$ is the simplest example of an affine Lie algebra.

The Lie algebra extension~\eqref{eq: Kac--Moody alg extn}
is simpler than the group extension~\eqref{eq: Kac--Moody extn}
in that it is determined up to isomorphism by a Lie algebra
$2$-cocycle $\omega(f,g)$, i.e.\ a skew bilinear map
$\omega\colon \Og\times \Og \to \R$ satisfying the
{\bf 2-cocycle condition}
\begin{equation}
\label{eq: 2-cocycle eqn}
\omega([f,g],h) + \omega([g,h],f) + \omega([h,f],g) = 0 .
\end{equation}
For $G$ as above we may assume the cocycle $\omega$ equal, up
to a scalar multiple, to the {\bf Kac--Moody $2$-cocycle}
\begin{equation}
\label{eq: Kac--Moody cocycle}
\omega(f,g) = 2\int^{2\pi}_0
\langle f(\theta),g'(\theta)\rangle\, d\theta
\end{equation}
Here $\langle \cdot, \cdot \rangle$ is an invariant symmetric bilinear
form on $\g$.
Thus, as a vector space $\wOg$ is isomorphic to $\Og\oplus \R$,
but the bracket is given by
$$
[(f,\a),(g,\b)] = ([f,g],\omega(f,g))
$$
Since $\omega$ is a skew form on $\Og$, it defines a left-invariant
2-form $\omega$ on $\OG$.  The cocycle 
condition, Equation~\eqref{eq: 2-cocycle eqn},
says precisely that $\omega$ is closed.  We quote the following 
theorem from Pressley and Segal, slightly corrected:

\begin{theorem} \label{extension.thm} \et
Suppose $G$ is a simply-connected compact simple Lie group.  Then:

\begin{enumerate}

\item The central extension of Lie algebras
\[
0\to \u(1) \to \wOg \to \Og \to 0
\]
defined by the cocycle $\omega$ above corresponds to a
central extension of Fr\'echet Lie groups
\[
   1 \to \U(1) \to \wOG \to \OG \to 1
\]
in the sense that $i \omega$ is the curvature
of a left-invariant connection on the principal $\TT$-bundle
$\wOG$ iff the $2$-form $\omega/2\pi$ on $\OG$ has integral periods.

\item The 2-form $\omega/2\pi$ has integral periods iff the
invariant symmetric bilinear form $\langle \cdot, \cdot \rangle$
on $\g$ satisfies this integrality condition: $\langle
h_\theta ,h_\theta \rangle \in \frac{1}{2\pi} \Z$ for the 
coroot $h_{\theta}$ associated to the highest root $\theta$ of $G$.

\end{enumerate}
\end{theorem}

Since $G$ is simple, all invariant symmetric bilinear forms on its Lie
algebra are proportional, so there is a unique invariant
inner product $(\cdot, \cdot)$ with
$(h_{\theta},h_{\theta}) = 2$.
Pressley and Segal \cite{PressleySegal:1986} call this inner
product the {\bf basic inner product} on $\g$.
In what follows, we always use $\langle \cdot, \cdot \rangle$ to stand
for this basic inner product divided by $4 \pi$.  This is the smallest
inner product to satisfy the integrality condition in the above
theorem.

More generally, for any integer $k$, the inner product 
$k\langle \cdot, \cdot \rangle$ satisfies the integrality condition 
in Theorem \ref{extension.thm}.
It thus gives rise to a central extension
\[
   1 \to \TT \to \wOkG \to \OG \to 1
\]
of $\OG$.  The integer $k$ is called the {\bf level} of
the central extension $\wOkG$.

\subsubsection{The Kac--Moody group $\wOkG$}
\label{Kac-Moody.section}

In this section we begin by recalling an explicit construction of $\wOkG$
due to Murray and Stevenson \cite{MurrayStevenson:2001}, 
inspired by the work of
Mickelsson \cite{Mickelsson:1987}.  We then use this
to prove a result, Proposition \ref{conjugation}, that will be
crucial for constructing the 2-group $\PG$.

First, suppose that $\cG$ is any Fr\'echet Lie group.
Let $P_0\cG$ denote the space of smooth based paths in $\cG$: 
\[   P_0\cG = \{ f \in C^\infty([0,2\pi], \cG) \colon \; f(0) = 1  \}
\]
$P_0\cG$ is a Fr\'echet Lie group under pointwise multiplication of 
paths, whose Lie algebra is
\[   P_0L = \{ f \in C^\infty([0,2\pi], L) \colon \; f(0) = 0  \}
\]
where $L$ is the Lie algebra of $\cG$.  Furthermore, the map $\pi\colon
P_0\cG\to \cG$ which evaluates a path at its endpoint is a 
homomorphism of Fr\'echet Lie groups.  The kernel 
of $\pi$ is equal to 
\[   \Omega\cG = \{ f \in C^\infty([0,2\pi], \cG) \colon \;
f(0) = f(1) = 1  \, \}
\]
Thus, $\Omega \cG$ is a normal subgroup of $P_0 \cG$.  Note
that we are defining $\Omega\cG$ in a somewhat nonstandard way,
since its elements can be thought of as loops $f \maps S^1 \to \cG$
that are smooth everywhere except at the basepoint, where both left
and right derivatives exist to all orders, but need not agree.
However, we need this for the sequence
\[     1 \stackto{} \Omega\cG \stackto{} P_0 \cG \stackto{\pi} \cG 
         \stackto{} 1 \]
to be exact, and our $\Omega\cG$ is homotopy equivalent
to the standard one.

At present we are most interested in the case where $\cG = \OG$.
Then a point in $P_0\cG$ gives a map $f\maps [0,2\pi] \times
S^1 \to G$ with $f(0,\theta) = 1$ for all $\theta\in S^1$,
$f(t,0) = 1$ for all $t \in [0,2\pi]$.  It is an easy calculation 
\cite{MurrayStevenson:2001} to show that the
map $\kappa \colon P_0\OG\times P_0\OG \to \TT$ defined by
\begin{equation}
\label{eq: loop cocycle} \kappa(f,g) =
\exp\left(2ik\int^{2\pi}_0\int^{2\pi}_0 \langle f(t)^{-1}f'(t),
g'(\theta) g(\theta)^{-1}\rangle\, d\theta \, dt\right)
\end{equation}
is a group $2$-cocycle.  This 2-cocycle $\kappa$ makes
$P_0\OG \times \TT$ into a group with the following product:
\[
  (f_1,z_1)\cdot (f_2,z_2) =
  \left(f_1f_2,z_1 z_2 \,\kappa(f_1,f_2)\right) .
\]
Let $N$ be the subset of $P_0\OG\times \TT$
consisting of pairs $(\c,z)$ such that $\c\colon [0,2\pi] \to \OG$
is a loop based at $1\in \OG$ and
\[
z = \exp\left(-ik \int_{D_\c} \omega \right)
\]
where $D_\c$ is any disk in $\OG$ with $\gamma$ as its boundary.
It is easy to check that $N$
is a normal subgroup of the group $P_0\OG\times \TT$ with the product
defined as above.  To construct
$\wOkG$ we form the quotient group $(P_0\OG\times
\TT)/N$.  In \cite{MurrayStevenson:2001} it is shown
that the resulting central extension is isomorphic to the
central extension of $\OG$ at level $k$.
So we have the commutative diagram
\begin{equation}
\label{centext}
\xymatrix{
P_0\OG \times \TT \ar[d] \ar[r] & \wOkG \ar[d] \\
P_0\OG \ar[r]^-{\pi} & \OG                                 }
\end{equation}
where the horizontal maps are quotient maps, the upper horizontal
map corresponding to the normal subgroup $N$, and the lower horizontal
map corresponding to the normal subgroup $\Omega^2G$ of $P_0\OG$.

Notice that the group of based paths $P_0G$ acts on $\OG$ by
conjugation.  The next proposition shows that this action 
lifts to an action on $\wOkG$:

\begin{proposition} \label{conjugation} \et
The action of $P_0G$ on $\OG$ by conjugation
lifts to a smooth action $\alpha$ of $P_0G$ on $\wOkG$, whose
differential gives an action $d\alpha$ of the Lie
algebra $P_0\g$ on the Lie algebra $\wOkg$ with
\[           d\alpha(p)(\ell,c) = 
\big([p,\ell], \; 
2k \int_0^{2\pi} \langle p(\theta), \ell'(\theta)\rangle \, d\theta \,) .
\]
for all $p \in P_0\g$ and all $(\ell,c) \in \Og \oplus \R
\iso \wOkg$.
\end{proposition}

\Proof 
To construct $\alpha$ it suffices to construct a smooth action of
$P_0G$ on $P_0\OG\times \TT$ that preserves the product
on this group and also preserves the normal subgroup $N$.
Let $p\colon [0,2\pi]\to G$ be an element of $P_0G$, so that
$p(0) = 1$.  Define the action of $p$ on a point $(f,z)\in
P_0\OG\times \TT$ to be
\[ p\cdot (f,z) =
\big( pfp^{-1}, \; z \exp(ik \int^{2\pi}_0 \beta_p(f(t)^{-1}f'(t))\, 
dt)\; \big) \]
where $\beta_p$ is the left-invariant
$1$-form on $\OG$ corresponding to the linear map
$\beta_p\colon \Og\to \R$ given by:
$$
\beta_p(\xi) =
-2 \int^{2\pi}_0 \langle \xi(\theta), p(\theta)^{-1} p'(\theta)\rangle \, 
d\theta .
$$
for $\xi\in \Og$.
To check that this action preserves the product on
$P_0\OG\times \TT$, we have to show that
\[
\big(pf_1p^{-1}, \, z_1 
\exp(ik\int^{2\pi}_0\beta_p(f_1(t)^{-1}f_1'(t))\, dt) \, \big)
\cdot
\big(pf_2p^{-1}, \, z_2 
\exp(ik\int^{2\pi}_0\beta_p(f_2(t)^{-1}f_2'(t))\, dt) \, \big)  \]
\[
= \big(pf_1f_2p^{-1}, \,
z_1z_2 \kappa(f_1,f_2)
\exp(ik\int^{2\pi}_0 \beta_p((f_1f_2)(t)^{-1}(f_1f_2)'(t))\, dt) \, 
\big).
\]
It therefore suffices to establish the identity
\begin{multline*}
\kappa(pf_1p^{-1},pf_2p^{-1}) =
\kappa(f_1,f_2)\exp \Big(ik\int^{2\pi}_0
\left(\beta_p((f_1f_2)(t)^{-1}(f_1f_2)'(t)) - \right. \\
\left.\beta_p(f_1(t)^{-1}f_1'(t)) - \beta_p(f_2(t)^{-1}f_2'(t))\right)
\, dt\Big) .
\end{multline*}
This is a straightforward computation that can safely
be left to the reader.

Next we check that the normal subgroup $N$ is preserved by the action
of $P_0G$.  For this we must show that if $(f,z)\in N$ then
$$
\big(pfp^{-1},\,z\exp(ik\int_0^{2\pi}\beta_p(f^{-1}f')\, dt)\big) \in N .
$$ 
Recall that $N$ consists
of pairs $(\c,z)$ such that $\c\in \Omega^2G$ and $z =
\exp(-ik\int_{D_\c} \omega)$ where $D_\c$ is a disk in $\OG$ with
boundary $\c$.  Therefore we need to show that
$$
\exp\left(ik\int_{D_{p^{-1}\c p}}\omega \right) =
\exp\left(ik\int_{D_{\c}}\omega \right)
\exp\left(-ik\int_0^{2\pi}\beta_p(\c^{-1}\c')dt \right) .
$$
This follows immediately from the identity
$$
\mathrm{Ad}(p)^*\omega = \omega - d\beta_p,
$$
which is easily established by direct computation.

Finally, we have to check the formula for $d\alpha$.
On passing to Lie algebras, diagram (\ref{centext})
gives rise to the following
commutative diagram of Lie algebras:
$$
\xymatrix{
P_0\Omega \mathfrak{g} \oplus \R \ar[r]^-{\overline{\mathrm{ev}}}
\ar[d] & \Omega \mathfrak{g} \oplus \R \ar[d] \\
P_0\Omega \mathfrak{g} \ar[r]^-{\mathrm{ev}} &
\Omega \mathfrak{g} }
$$
where $\overline{\mathrm{ev}}$ is the homomorphism
$(f,c)\mapsto (f(2\pi),c)$ for $f\in P_0\Omega \g$ and
$c\in \R$.  To calculate $d\alpha(p)(\ell,c)$ we compute
$\overline{\mathrm{ev}}(d\alpha(p)(\tilde{\ell},c))$ where
$\tilde{\ell}$ satisfies $\mathrm{ev}(\tilde{\ell}) = \ell$ (take,
for example, $\tilde{\ell}(t) = t\ell/2\pi$).  It is then 
straightforward to compute that
$$
\overline{\mathrm{ev}}(d\alpha(p)(\tilde{\ell},c)) = 
\big([p,\ell],2k\int^{2\pi}_0
\langle p(\theta),\ell'(\theta)\rangle\, d\theta\big).
$$
\hbox{\hskip 60ex} \endofproof

\subsection{The 2-Group $\mathcal{P}_k G$ and $\mathrm{String}\of{n}$}
\label{PG.section}

Having completed our review of Lie 2-algebras and loop groups,
we now study a Lie 2-group $\PG$ whose Lie 
2-algebra $\Pg$ is equivalent to $\g_k$.   We begin in 
Section \ref{PG.construction.section} by giving a construction of 
$\PG$ in terms of the central extension $\wOkG$ of the loop group 
of $G$.  This yields a description of $\Pg$ which we use later 
to prove that this Lie 2-algebra is equivalent to $\g_k$.  

Section \ref{topology.section} gives another viewpoint on $\PG$,
which goes a long way toward explaining the significance of this 2-group.  
For this, we study the topological group $|\PG|$ formed by taking the 
geometric realization of the nerve of $\PG$.

\subsubsection{Constructing $\PG$}
\label{PG.construction.section}

In Proposition \ref{conjugation} we saw that the action of
the path group $P_0G$ on the loop group $\OG$ 
by conjugation lifts to an action 
$\alpha$ of $P_0G$ on the central extension $\wOkG$.  
This allows us to define a Fr\'echet Lie group $P_0G 
\ltimes \wOkG$ in which multiplication is given by:
\[
(p_1,\hat{\ell}_1)\cdot (p_2,\hat{\ell}_2) =
\big(\, p_1p_2, \, \hat{\ell_1} \alpha(p_1)(\hat{\ell}_2) \,\big) .
\]
This, in turn, allows us to construct the 
2-group $\PG$ which plays the starring role in this paper:

\begin{proposition} \label{PG.construction} \et
Suppose $G$ is a simply-connected compact simple Lie 
group and $k \in \Z$.  Then there is a Fr\'echet Lie 2-group 
$\PG$ for which:
\begin{itemize}
\item The Fr\'echet Lie group of objects $\Ob(\PG)$ is $P_0G$.
\item The Fr\'echet Lie group of morphisms $\Mor(\PG)$ is 
$P_0G \ltimes \wOkG$.
\end{itemize}
\begin{itemize}
\item  The source and target maps $s,t \maps \Mor(\PG) \to \Ob(\PG)$
are given by:
\[   
\begin{array}{ccl}
s(p,\hat{\ell}) &=& p  \\
t(p,\hat{\ell}) &=&  \partial(\hat{\ell}) p
\end{array}
\]
where $p \in P_0G$, $\hat{\ell} \in \wOkG$, and 
$\partial \maps \wOkG \to P_0 G$ is the composite:
\[     \wOkG \to \OG \hookrightarrow P_0 G \, .\]
\item The identity-assigning map $i \maps \Ob(\PG) \to \Mor(\PG)$
is given by:
\[    i(p) = (p,1) . \]
\item  The composition map $\circ \maps \Mor(\PG) \times_{\Ob(\PG)} 
\Mor(\PG) \to \Mor(\PG)$ is given by:
\[      (p_1,\hat{\ell}_1) \circ (p_2,\hat{\ell}_2) = 
        (p_2, \hat{\ell}_1 \hat{\ell}_2)  \]
whenever $(p_1, \hat{\ell}_1), (p_2, \hat{\ell}_2)$ are composable
morphisms in $\PG$.
\end{itemize}
\end{proposition}

\Proof  One can check directly that $s,t,i,\circ$ are Fr\'echet Lie
group homomorphisms and that these operations 
make $\PG$ into a category.   Alternatively, one can check that
$(P_0 G, \wOkG, \alpha, \partial)$ is a crossed module
in the category of Fr\'echet manifolds.
This merely requires checking that
\begin{equation} %\label{crossed.1}
   \partial( \alpha(p) (\hat{\ell})) = 
p \, \partial(\hat{\ell})\, p^{-1} 
\end{equation}
and 
\begin{equation} %\label{crossed.2}
    \alpha(\partial(\hat{\ell}_1)) (\hat{\ell_2}) = 
\hat{\ell}_1 \hat{\ell}_2 {\hat{\ell}_1}^{-1} . 
\end{equation}
Then one can use the fact that crossed modules in the
category of Fr\'echet manifolds are the same as Fr\'echet
Lie 2-groups (see for example HDA6).  \endofproof

We denote the Lie 2-algebra of $\PG$ by $\Pg$.  To prove
this Lie 2-algebra is equivalent to $\g_k$ in Section
\ref{equivalence.section}, we will use 
an explicit description of its corresponding 
$L_\infty$-algebra:

\begin{proposition} \label{Pg} \et
The 2-term $L_\infty$-algebra $V$ corresponding to the 
Lie 2-algebra $\Pg$ has:
\begin{itemize}
\item $V_{0} = P_0 \g$ and $V_{1} = \wOkg \iso \Og \oplus \R$,

\item $d\maps V_{1} \rightarrow V_{0}$ equal to
      the composite 
     \[     \wOkg \to \Og \hookrightarrow P_0 \g \, ,\]

\item $l_2 \maps V_0 \times V_0 \to V_1$ given by
      the bracket in $P_0 \g$:
      \[       l_2(p_1,p_2) = [p_1,p_2],     \]
      and $l_2 \maps V_0 \times V_1 \to V_1$ given by the action
      $d\alpha$ of $P_0 \g$ on $\wOkg$, or explicitly:
      \[  
      l_2(p, (\ell, c)) = 
      \big([p,\ell],  \;
      2k\int_0^{2\pi} \langle p(\theta),
      \ell'(\theta) \rangle \, d\theta \; \big) 
      \]
for all $p \in P_0\g$, $\ell \in \OG$ and $c \in \R$.

    \item $l_{3}\maps V_{0} \times V_{0} \times
     V_{0} \rightarrow V_{1}$ equal to zero.
\end{itemize}
\end{proposition}

\Proof 
This is a straightforward application of the correspondence
described in Section \ref{Linfty.section}.  The formula for
$l_2 \maps V_0 \times V_1 \to V_1$ comes from Proposition
\ref{conjugation}, while $\ell_3 = 0$
because the Lie 2-algebra $\Pg$ is strict.  
\endofproof

\subsubsection{The Topology of $|\PG|$}
\label{topology.section}

In this section we construct an exact sequence of Fr\'echet Lie
2-groups: 
\[
1\to \LG \stackto{\iota} \PG \stackto{\pi} G \to 1 \, ,
\]
where $G$ is considered as a Fr\'echet Lie $2$-group with only 
identity morphisms.  Applying a certain procedure for turning
topological 2-groups into topological groups, described below,
we obtain this exact sequence of topological groups: 
\[
1 \to |\LG| \stackto{|\iota|} |\PG| \stackto{|\pi|} G \to 1  \, .
\]
Note that $|G| = G$.  We then show that the topological group $|\LG|$ 
has the homotopy type of the Eilenberg--Mac Lane space $K(\Z,2)$.
Since $K(\Z,2)$ is also the classifying space $B\U(1)$, the above exact 
sequence is a topological analogue of the exact sequence of Lie 2-algebras 
describing how $\g_k$ is built from $\g$ and $\u(1)$:
\[   0 \to {\rm b}\u(1) \to \g_k \to \g \to 0 \, , \]
where ${\rm b}\u(1)$ is the Lie 2-algebra with a 0-dimensional
space of objects and $\u(1)$ as its space of morphisms.  

The above exact sequence of topological groups exhibits $|\PG|$ as
the total space of a principal $K(\Z,2)$ bundle over $G$.  Bundles
of this sort are classified by their `Dixmier--Douady class',
which is an element of the integral third cohomology group of the base
space.  In the case at hand, this cohomology group is $H^3(G) \iso
\Z$, generated by the element we called $[\nu/2\pi]$ in the
Introduction.  We shall show that the Dixmier--Douady class of the
bundle $|\PG| \to G$ equals $k [\nu/2\pi]$.  Using this, we show that
for $k = \pm 1$, $|\PG|$ is a version of $\hat G$ --- the topological group 
obtained from $G$ by killing its third homotopy group.

We start by defining a map $\pi \maps \PG \to G$ as follows.  
We define $\pi$ on objects $p \in \PG$ as follows:
\[       \pi(p) = p(2 \pi) \in G . \]
In other words, $\pi$ applied to a based path in $G$ gives the
endpoint of this path.  We define $\pi$ on morphisms in the only
way possible, sending any morphism $(p,\hat{\ell}) \maps p \to 
\partial(\hat{\ell}) p$
to the identity morphism on $\pi(p)$.  It is easy to see that
$\pi$ is a {\bf strict homomorphism} of Fr\'echet Lie 2-groups: 
in other words, a map that strictly preserves all the Fr\'echet 
Lie 2-group structure.  Moreover, it is easy to see that
$\pi$ is onto both for objects and morphisms.   

Next, we define the Fr\'echet Lie 2-group $\LG$ to be the
{\bf strict kernel} of $\pi$.   In other words, the objects of $\LG$ are 
objects of $\PG$ that are mapped to $1$ by $\pi$, 
and similarly for the morphisms of $\LG$, while the source,
target, identity-assigning and composition maps for $\LG$ are
just restrictions of those for $\PG$.  So:
\begin{itemize}
\item the Fr\'echet Lie group of objects $\Ob(\LG)$ is $\OG$,
\item the Fr\'echet Lie group of morphisms $\Mor(\LG)$ is 
$\OG \ltimes \wOkG$,
\end{itemize}
where the semidirect product is formed using the action 
$\alpha$ restricted to $\OG$.  Moreover, the formulas for
$s,t,i,\circ$ are just as in Proposition \ref{PG.construction},
but with loops replacing paths.  

It is easy to see that the inclusion $\iota \maps \LG \to \PG$
is a strict homomorphism of Fr\'echet Lie 2-groups.
We thus obtain:

\begin{proposition} \label{exact1} \et
The sequence of strict Fr\'echet 2-group homomorphisms
\[
1\to \LG \stackto{\iota} \PG \stackto{\pi} G \to 1
\] 
is {\bf strictly exact}, meaning that the image of each arrow
is equal to the kernel of the next, both on objects and on morphisms.
\end{proposition} 

Any Fr\'echet Lie 2-group $C$ is, among other things, a {\bf
topological category}: a category where the sets $\Ob(C)$ and
$\Mor(C)$ are topological spaces and the source, target,
identity-assigning and composition maps are continuous.  Homotopy
theorists have a standard procedure for taking the `nerve' of a
topological category and obtaining a simplicial space.  They also know
how to take the `geometric realization' of any simplicial space,
obtaining a topological space.  We use $|C|$ to denote the geometric
realization of the nerve of a topological category $C$.  If $C$ is in
fact a topological 2-group --- for example a Fr\'echet Lie 2-group ---
then $|C|$ naturally becomes a topological group.

For readers unfamiliar with these constructions, let us give a more
hands-on description of how to build $|C|$.  First for any $n \in \N$
we construct a space $|C|_n$.  A point in $|C|_n$ consists of a string
of $n$ composable morphisms in $C$:
\[    x_0 \stackto{f_1} x_1 \stackto{f_2} \cdots
 \stackto{f_{n-1}} x_{n-1} \stackto{f_n} x_n \]
together with a point in the standard $n$-simplex: 
\[    a \in \Delta_n = \{ (a_0, \dots, a_n) \in [0,1] \colon \; 
                       a_0 + \cdots + a_n = 1  \}.  \]
Since $|C|_n$ is a subset of $\Mor(C)^n \times \Delta_n$, we
give it the subspace topology.
There are well-known face maps $d_i \maps \Delta_n \to
\Delta_{n+1}$ and degeneracies $s_i \maps \Delta_n \to
\Delta_{n-1}$.  We use these to build $|C|$ by gluing together 
all the spaces $|C|_n$ via the following identifications:
\[   \left(x_0 \stackto{f_1} \cdots \stackto{f_n} x_n, a \right) 
\sim
     \left(x_0 \stackto{f_1} \cdots \stackto{f_i} x_i \stackto{1} x_i
      \stackto{f_{i+1}} \cdots \stackto{f_n} x_n, d_i(a)\right)
\]
for $0 \le i \le n$, and
\[   \left(x_0 \stackto{f_1} \cdots \stackto{f_n} x_n, a\right) 
\sim
     \left(x_0 \stackto{f_1} \cdots \stackto{f_{i-2}} 
       x_{i-1} \stackto{f_{i}f_{i+1}} x_{i+1} 
      \stackto{f_{i+2}} \cdots \stackto{f_n} x_n, s_i(a)\right) 
\]
for $0 < i < n$, together with
\[   \left(x_0 \stackto{f_1} \cdots \stackto{f_n} x_n, a\right) 
\sim
     \left(x_1 \stackto{f_2} x_2 \stackto{f_3}
     \cdots \stackto{f_n} x_n, s_0(a)\right) \]
and
\[   \left(x_0 \stackto{f_1} \cdots \stackto{f_n} x_n, a\right) 
\sim
    \left(x_0 \stackto{f_1} \cdots \stackto{f_{n-2}} x_{n-2} 
\stackto{f_{n-1}} x_{n-1}, s_{n}(a)\right)  
\]
This defines $|C|$ as a topological space, but when $C$ is a
topological 2-group the multiplication in $C$ makes $|C|$ into a
topological group.  Moreover, if $G$ is a topological group viewed as
a topological 2-group with only identity morphisms, we have $|G| \iso
G$.

Applying the functor $|\cdot|$ to the exact sequence in
Proposition \ref{exact1}, we obtain this result, which
implies Theorem 3:

\begin{theorem} \label{exact2} \et
The sequence of topological groups
\[
1\to |\LG| \stackto{|\iota|} |\PG| \stackto{|\pi|} G \to 1
\] 
is exact, and $|\LG|$ has the homotopy type of $K(\Z,2)$.
Thus, $|\PG|$ is the total space of a $K(\Z,2)$ bundle over $G$.
The Dixmier--Douady class of this bundle is $k [\nu/2\pi] \in
H^3(G)$.   Moreover, $\PG$ is $\hat G$ when $k = \pm 1$.  
\end{theorem} 

\Proof 
It is easy to see directly that the functor $|\cdot|$ carries strictly
exact sequences of topological 2-groups to exact sequences of
topological groups.  To show that $|\LG|$ is a $K(\Z,2)$, we prove
there is a strictly exact sequence of Fr\'echet Lie 2-groups
\begin{equation}
\label{exact.sequence}
1 \to \U(1) \to \T\wOkG \to \LG \to 1 \, .
\end{equation}
Here $\U(1)$ is regarded as a Fr\'echet Lie 2-group with only identity
morphisms, while $\T\wOkG$ is the Fr\'echet Lie 2-group with $\wOkG$
as its Fr\'echet Lie group of objects and precisely one morphism from
any object to any other.  In general:

\begin{lemma} \label{T} \et
For any Fr\'echet Lie group
$\cG$, there is a Fr\'echet Lie 2-group $\T\cG$ with:
\begin{itemize}
\item $\cG$ as its Fr\'echet Lie group of objects,
\item $\cG \ltimes \cG$ as its Fr\'echet Lie group of morphisms,
where the semidirect product is defined using the conjugation 
action of $\cG$ on itself,
\end{itemize}
and:
\begin{itemize}
\item source and target maps given by $s(g,g') = g$, $t(g,g') = gg'$,
\item identity-assigning map given by $i(g) = (g,1)$,
\item composition map given by $(g_1,g_1') \circ (g_2,g_2') = 
(g_2,g_1'g_2')$ whenever $(g_1,g_1')$, $(g_2,g_2')$ are composable
morphisms in $\T\cG$.
\end{itemize}
\end{lemma}

\Proof 
It is straightforward to check that
this gives a Fr\'echet Lie 2-group.  Note that
$\T\cG$ has $\cG$ as objects and 
one morphism from any object to any other.  
\hbox{\hskip 60ex} \endofproof

In fact, Segal \cite{Segal:1968} has already introduced $\T\cG$ under the 
name $\overline{\cG}$, treating it as a topological category.  He 
proved that $|\T\cG|$ is contractible.  In fact, he exhibited 
$|\T\cG|$ as a model of $E\cG$, the total space of the universal bundle
over the classifying space $B\cG$ of $\cG$.  
Therefore, applying the functor $| \cdot |$ to the exact
sequence (\ref{exact.sequence}), we obtain this short exact sequence of 
topological groups:
\[
1 \to \U(1) \to E\wOkG \to |\LG| \to 1 \, .
\]
Since $E\wOkG$ is contractible, it follows
that $|\LG| \iso E\wOG/\U(1)$ has the homotopy
type of $B\U(1) \simeq K(\Z,2)$.   

One can check that $|\pi| \maps |\PG| \to G$ is a locally trivial
fiber bundle, so it defines a principal $K(\Z,2)$ bundle over $G$.
Like any such bundle, this is the pullback of the universal
principal $K(\Z,2)$ bundle $p \maps EK(\Z,2) \to BK(\Z,2)$ along some 
map $f \maps G \to BK(\Z,2)$, giving a commutative diagram of spaces:
\[
\begin{diagram}  
\node{|\LG|} \arrow{e,t}{|\iota|} \arrow{s,l}{\sim}    
\node{|\PG|} \arrow{e,t}{|\pi|} \arrow{s,r}{p^\ast f}
\node{G} \arrow{s,r}{f}                   \\
\node{K(\Z,2)} \arrow{e,t}{i} 
\node{EK(\Z,2)} \arrow{e,t}{p} 
\node{BK(\Z,2)}                  
\end{diagram}
\]
Indeed, such bundles are classified up to isomorphism by the homotopy class 
of $f$.  Since $BK(\Z,2) \simeq K(\Z,3)$, this homotopy class is determined
by the Dixmier--Douady class $f^\ast \kappa$, where $\kappa$ is the
generator of $H^3(K(\Z,3)) \iso \Z$.  The next order of business is
to show that $f^\ast \kappa = k[\nu/2\pi]$.  

For this, it suffices to show that $f$ maps the generator of
$\pi_3(G) \iso \Z$ to $k$ times the generator of $\pi_3(K(\Z,3))
\iso \Z$.   Consider this bit of the long exact sequences of homotopy
groups coming from the above diagram:
\[
\begin{diagram}  
\node{\pi_3(G)} \arrow{e,t}{\partial} \arrow{s,l}{\pi_3(f)}             
\node{\pi_2(|\LG|)} \arrow{s,r}{\iso}   \\ 
\node{\pi_3(K(\Z,3))} \arrow{e,t}{\partial'} \node{\pi_2(K(\Z,2))} 
\end{diagram}
\]
Since the connecting homomorphism $\partial'$ and the 
map from $\pi_2(|\LG|)$ to $\pi_2(K(\Z,2))$ are isomorphisms,
we can treat these as the identity by a suitable choice of
generators.  Thus, to show that $\pi_3(f)$ is multiplication by $k$ 
it suffices to show this for the connecting homomorphism $\partial$.

To do so, consider this commuting diagram of Frech\'et Lie 2-groups:
\[
\begin{diagram}  
\node{\OG} \arrow{e,t}{\iota} \arrow{s,l}{i}
\node{P_0 G} \arrow{e,t}{\pi} \arrow{s,l}{i'}
\node{G} \arrow{s,r}{1} 
              \\
\node{\LG} \arrow{e,t}{\iota} 
\node{\PG}  \arrow{e,t}{\pi}
\node{G}  
\end{diagram}
\]
Here we regard the groups on top as 2-groups with only identity
morphisms; the downwards-pointing arrows include these in
the 2-groups on the bottom row.  Applying the functor $|\cdot|$, 
we obtain a diagram where each row is a principal bundle:
\[
\begin{diagram}  
\node{\OG} \arrow{e,t}{|\iota|} \arrow{s,l}{|i|} 
\node{P_0 G} \arrow{e,t}{|\pi|} \arrow{s,l}{|i'|} 
\node{G} \arrow{s,r}{1} 
              \\
\node{|\LG|} \arrow{e,t}{|\iota|} 
\node{|\PG|}  \arrow{e,t}{|\pi|}
\node{G}  
\end{diagram}
\]
Taking long exact sequences of homotopy groups, this gives:
\[
\begin{diagram}  
\node{\pi_3(G)} \arrow{e,t}{1} \arrow{s,l}{1} 
\node{\pi_2(\OG)} \arrow{s,r}{\pi_2(|i|)} 
              \\
\node{\pi_3(G)} \arrow{e,t}{\partial}
\node{\pi_2(|\LG|)}  
\end{diagram}
\]
Thus, to show that $\partial$ is multiplication by $k$ 
it suffices to show this for $\pi_2(|i|)$.  

For this, we consider yet another commuting diagram of
Frech\'et Lie 2-groups:
\[
\begin{diagram}  
\node{\U(1)} \arrow{e} \arrow{s}
\node{\wOkG} \arrow{e} \arrow{s}
\node{\OG}             \arrow{s,l}{i}
                          \\
\node{\U(1)}   \arrow{e}
\node{\T\wOkG} \arrow{e}
\node{\LG}  
\end{diagram}
\]
Applying $|\cdot|$, we obtain a diagram where each row is a 
principal $\U(1)$ bundle:
\[
\begin{diagram}  
\node{\U(1)} \arrow{e} \arrow{s}
\node{\wOkG} \arrow{e} \arrow{s}
\node{\OG}             \arrow{s,r}{|i|}
                          \\
\node{\U(1)}   \arrow{e}
\node{|\T\wOkG|} \arrow{e}
\node{|\LG| \simeq K(\Z,2)}    
\end{diagram}
\]
Recall that the bottom row is the universal principal $\U(1)$ bundle.
The arrow $|i|$ is the classifying map for the $\U(1)$ bundle $\wOkG \to \OG$.  
By Theorem \ref{extension.thm}, the Chern class of this bundle
is $k$ times the generator of $H^2(\OG)$, so $\pi_2(|i|)$ must map the 
generator of $\pi_2(\OG)$ to $k$ times the generator of $\pi_2(K(\Z,2))$.  

Finally, let us show that $|\PG|$ is $\hat{G}$ when $k = \pm 1$.
For this, it suffices to show that when $k = \pm 1$, the map
$|\pi| \maps |\PG| \to G$ induces isomorphisms on all homotopy
groups except the third, and that $\pi_3(|\PG|) = 0$.  For this
we examine the long exact sequence:
\[        \cdots \stackto{} 
    \pi_n(|\LG|) \stackto{} \pi_n(|\PG|)  
                 \stackto{} \pi_n(G) 
                 \stackto{\partial} \pi_{n-1}(|\LG|)
          \stackto{} \cdots  \, .
\] 
Since $|\LG| \simeq K(\Z,2)$, its homotopy groups vanish
except for $\pi_2(|\LG|) \iso \Z$, so $|\pi|$ induces an
isomorphism on $\pi_n$ except possibly for $n = 2,3$.  
In this portion of the long exact sequence we have
\[ 
                    0  
                \stackto{} \pi_3(|\PG|)  
                 \stackto{} \Z 
                 \stackto{k} \Z
                 \stackto{} \pi_2(|\PG|)
                 \stackto{} 0
\] 
so $\pi_3(|\PG|) \iso 0$ unless $k = 0$, and 
$\pi_2(|\PG|) \iso \Z/k\Z$, so $\pi_2(|\PG|) \iso \pi_2(G) \iso 0$
when $k = \pm 1$.
\endofproof

\subsection{The Equivalence between $\Pg$ and $\g_k$}
\label{equivalence.section}

In this section we prove our main result, which implies
Theorem 2:

\begin{theorem} \et \label{bigthm}
There is a strictly exact sequence of Lie 2-algebra homomorphisms
\[
    0 \to \T\Og \stackto{\lambda}
     \Pg \stackto{\phi}  \g_k  \to 0  \]
where $\T\Og$ is equivalent to the trivial Lie 2-algebra
and $\phi$ is an equivalence of Lie 2-algebras.  
\end{theorem}

\noindent
Recall that by `strictly exact' we mean that both on the vector
spaces of objects and the vector spaces of morphisms,
the image of each map is the kernel of the next.  

We prove this result in a series of lemmas.  We begin by describing
$\T{\Og}$ and showing that it is equivalent to the trivial Lie 
2-algebra.  Recall that in Lemma \ref{T} we constructed for any 
Fr\'echet Lie group $\cG$ a Fr\'echet Lie 2-group $\T\cG$ with $\cG$ 
as its group of objects and precisely one morphism from any object
to any other.  We saw that the space $|\T\cG|$ is contractible; this is 
a topological reflection of the fact that $\T\cG$ is equivalent to the 
trivial Lie 2-group.  Now we need the Lie algebra analogue of this 
construction:

\begin{lemma} \label{TL} \et Given a Lie algebra $L$, there is
a 2-term $L_{\infty}$-algebra $V$ for which:

\begin{itemize}
    \item $V_{0} = L$ and $V_{1} = L$,

    \item $d\maps V_{1} \rightarrow V_{0}$ is the identity,

    \item $l_2 \maps V_0 \times V_0 \to V_1$ and 
    $l_2 \maps V_0 \times V_1 \to V_1$
    are given by the bracket in $L$,

    \item $l_{3}\maps V_{0} \times V_{0} \times
     V_{0} \rightarrow V_{1}$ is equal to zero.
\end{itemize}
We call the corresponding strict Lie 2-algebra $\T{L}$.
\end{lemma}

\Proof 
Straightforward. 
\endofproof

\begin{lemma} \label{trivial} \et 
For any Lie algebra $L$, the Lie 2-algebra $\T{L}$ is equivalent to the 
trivial Lie 2-algebra.  That is, $\T{L} \simeq 0$.
\end{lemma}

\Proof There is a unique homomorphism $\beta \maps \T{L} \to 0$
and a unique homomorphism $\gamma \maps 0 \to \T{L}$.  Clearly
$\beta \circ \gamma$ equals the identity.  The composite $\gamma \circ
\beta$ has:
\[
\begin{array}{lccl}
    (\gamma \circ \beta)_0 \maps &x &\mapsto& 0
    \\
    (\gamma \circ \beta)_1 \maps &x &\mapsto& 0
    \\
    (\gamma \circ \beta)_2 \maps &(x_1,x_2) &\mapsto& 0 \, ,
\end{array}
\]
while the identity homomorphism from $\T{L}$ to itself has:
\[
\begin{array}{lccl}
    \mathrm{id}_0 \maps &x &\mapsto& x
    \\
    \mathrm{id}_1 \maps &x &\mapsto& x
    \\
    \mathrm{id}_2 \maps &(x_1,x_2) &\mapsto& 0
    \,.
\end{array}
\]
There is a 2-isomorphism
\[
    \tau \maps \gamma \circ \beta \stackTo{\sim} \mathrm {id} 
\]
given by
\[
    \tau\of{x} = x
    \,,
\]
where the $x$ on the left is in $V_0$ and that on the right in
$V_1$, but of course $V_0 = V_1$ here. \endofproof

We continue by defining the Lie 2-algebra homomorphism $\Pg
\stackto{\phi} \g_k$.

\begin{lemma} \label{phi} \et 
There exists a Lie 2-algebra homomorphism
\[
    \phi \maps \Pg \to \g_k
\]
which we describe in terms of its corresponding $L_{\infty}$-homomorphism:
\[
\begin{array}{ccl}
\phi_0 (p) &=& p(2\pi) \\
           &&          \\
\phi_1 (\ell, c) &=& c \\
           &&          \\
\phi_2 (p_1, p_2) &=& 
       k \displaystyle{\int_0^{2\pi}} \left( \bracket{p_1}{p_2'} 
      - \bracket{p_1'}{p_2} \right) \, d\theta 
\end{array}
\]
where $p, p_1, p_2 \in P_0 \g$ and 
$(\ell, c) \in \Og \oplus \R \iso \wOkg$. 
\end{lemma}

\noindent Before beginning, note that the quantity
\[
    \displaystyle{\int_0^{2\pi} \! \left( \bracket{p_1}{p_2'} - 
     \bracket{p_1'}{p_2} \right) \, d\theta
     = 
    2  \int_0^{2\pi} \! \bracket{p_1}{p_2'} \; d\theta}
    \; - \; 
     \bracket{p_1(2\pi)}{p_2 (2\pi)}
\]
is skew-symmetric, but not in general equal to
\[
    \displaystyle{2\int_0^{2\pi}
       \bracket{p_1}{p_2'} \; d\theta}
\]
due to the boundary term.  However, these quantities are equal
when either $p_1$ or $p_2$ is a loop.

\vskip 1em

\Proof We must check that $\phi$ satisfies the conditions of Definition
\ref{Linftyhomo}.  First we show that $\phi$ is a chain map.  
That is, we show that $\phi_0$ and $\phi_1$ preserve the differentials:
$$\xymatrix{
    \wOkg \ar[rr]^{d}
    \ar[dd]_{\phi_{1}} &&
    P_0 \g  \ar[dd]^{\phi_{0}} \\ \\
    \R \ar[rr]^{d'} &&
    \g
}$$
where $d$ is the composite given in Proposition \ref{Pg}, and $d'=0$ 
since $\g_k$ is skeletal.  This square commutes since $\phi_0$ is
also zero.

We continue by verifying conditions (\ref{homo1}) - (\ref{homo3}) of 
Definition \ref{Linftyhomo}.  The bracket on objects is preserved
on the nose, which implies that the right-hand side of (\ref{homo1}) is 
zero.  This is consistent with the fact that the differential in 
the $L_\infty$-algebra for $\g_k$ is zero, which implies that the 
left-hand side of (\ref{homo1}) is also zero.

The right-hand side of (\ref{homo2}) is given by:
\begin{eqnarray*}
      \phi_1( l_2(p,(\ell,c))
      -
      l_2 (
        \phi_0 (p),
        \phi_1 (\ell,c)
      ) &=&
      \phi_1 \Big( [p,\ell],
      2k\int \bracket{p}{\ell'} \; d\theta \Big) \
      -
      \underbrace{l_2(p(2\pi),c)}_{=0} \cr
      &=&
      2k \int \bracket{p}{\ell'}
      \; d\theta .
    \end{eqnarray*}
This matches the left-hand side of (\ref{homo2}), namely:
\begin{eqnarray*}
    \phi_2 (p, d (\ell,c)  )
    &=&
    \phi_2 (p, \ell) \\
    &=&
    k
    \int
      (\bracket{p}{\ell'}
      -
      \bracket{p'}{\ell} ) \; d\theta
\nonumber \cr
    &=&
    2k \int
      \bracket{p}{\ell'}
    \; d \theta
\end{eqnarray*}
Note that no boundary term appears here 
since one of the arguments is a loop.

Finally, we check condition (\ref{homo3}).  Four terms in
this equation vanish because $l_3 = 0$ in $\Pg$ and 
$l_2 = 0$ in $\g_k$.  We are left needing to show
\[
l_3 (\phi_0 (p_1),\phi_0 (p_2), \phi_0 (p_3)) = 
\phi_2 (p_1,l_2 (p_2,p_3)) + \phi_2 (p_2,l_2 (p_3,p_1)) + 
\phi_2 (p_3,l_2 (p_1,p_2)) \, . 
\]
The left-hand side here equals
$k \langle p_1(2\pi), [p_2(2\pi), p_3(2\pi)]\rangle $. 
The right-hand side equals:
\[
\begin{array}{ccl}
&& \phi_2(p_1, l_2(p_2,p_3)) \;\; + \;\; {\rm cyclic\; permutations}   \\
\\
&=&
k \displaystyle{\int \left( \bracket{p_1}{[p_2,p_3]'}
-\bracket{p_1'}{[p_2,p_3]} \right) \, d\theta \;\; + \; {\rm cyclic\; perms.} }
\\
\\
&=&
k \displaystyle{ \int \left( \bracket{p_1}{[p'_2,p_3]} 
             +\bracket{p_1}{[p_2,p'_3]}
             -\bracket{p'_1}{[p_2,p_3]} \right) \, d\theta \;\; + \;\; 
{\rm cyclic \; perms.} } \\
\end{array}
\]
Using the antisymmetry of $\langle \cdot, [\cdot, \cdot] \rangle$, this
becomes:
\[ k \int \left( \bracket{p'_2}{[p_3,p_1]} 
                +\bracket{p'_3}{[p_1,p_2]}
                -\bracket{p'_1}{[p_2,p_3]} \right) 
                 \, d\theta \;\; + \;\; {\rm cyclic \; perms.} \\
\]
The last two terms cancel when we add all their cyclic permutations,
so we are left with all three cyclic permutations of the first term:
\[ k \int \left( \bracket{p'_1}{[p_2,p_3]} 
                +\bracket{p'_2}{[p_3,p_1]} 
                +\bracket{p'_3}{[p_1,p_2]} \right) \, d\theta \, . \]
If we apply integration by parts to the first term, we get:
\[ k \int \left(-\bracket{p_1}{[p'_2,p_3]} - \bracket{p_1}{[p_2,p'_3]}
                +\bracket{p'_2}{[p_3,p_1]} 
                +\bracket{p'_3}{[p_1,p_2]} \right) \; d\theta \; + \;  \]
\[ k \bracket{p_1(2\pi)}{[p_2(2\pi),p_3(2\pi)]} \, .\]
By the antisymmetry of $\langle \cdot, [\cdot, \cdot] \rangle$, 
the four terms in the integral cancel, leaving just
$k \langle p_1(2\pi), [p_2(2\pi), p_3(2\pi)] \rangle$, as desired. 
\endofproof

Next we show that the strict kernel of 
$\phi \maps \Pg \to \g_k$ is $\T\Omega\g$:

\begin{lemma} \et There is a Lie 2-algebra homomorphism
\[   \lambda \maps \T{\Omega\g} \to \Pg, \] 
that is one-to-one both on objects and on morphisms, and
whose range is precisely the kernel of 
$\phi \maps \Pg \to \g_k$, both on objects and on morphisms.
\end{lemma}

\Proof 
Glancing at the formula for $\phi$ in Lemma \ref{phi}, we
see that the kernel of $\phi_0$ and the kernel of
$\phi_1$ are both $\Og$.  We see from Lemma \ref{TL} that
these are precisely the spaces $V_0$ and $V_1$ in the 
2-term $L_\infty$-algebra $V$ corresponding to $\T{\Og}$.  
The differential $d \maps \ker(\phi_1) \to \ker(\phi_0)$
inherited from $\T\Og$ also matches that in $V$: it is 
the identity map on $\Og$.

Thus, we obtain an inclusion of 2-vector spaces
$\lambda \maps  \T{\Omega\g} \to \Pg$.
This uniquely extends to a Lie $2$-algebra homomorphism, which we 
describe in terms of its corresponding $L_{\infty}$-homomorphism:
{
\begin{eqnarray*}
\lambda_0 (\ell) &=& \ell \cr 
&&                         \cr
\lambda_1 (\ell) &=& (\ell, 0) \cr
\lambda_2 (\ell_1, \ell_2) &=&
\displaystyle{\big(0, - 2k 
             \int_0^{2\pi} \bracket{\ell_1}{\ell_2'} \, d\theta \big)}
\end{eqnarray*}
}
where $\ell, \ell_1, \ell_2 \in \Omega\g$, and the zero in the last
line denotes the zero loop.

To prove this, we must show that the conditions of Definition \ref{Linftyhomo}
are satisfied.  We first check that $\lambda$ is a chain map, i.e., this 
square commutes:
$$\xymatrix{
    \Omega\g \ar[rr]^{d}
    \ar[dd]_{\lambda_{1}} &&
    \Omega\g   \ar[dd]^{\lambda_{0}} \\ \\
     \wOkg \ar[rr]^{d'} &&
    P_0 \g
}$$
where $d$ is the identity and $d'$ is the composite given in
Proposition \ref{Pg}.  To see this, note that
$d'(\lambda_1 (\ell)) = d'(\ell,0) = \ell$ and
$\lambda_0 (d(\ell)) = \lambda_0 (\ell) = \ell$.

We continue by verifying conditions (\ref{homo1}) - (\ref{homo3}) of 
Definition \ref{Linftyhomo}.
The bracket on the space $V_0$ is strictly preserved 
by $\lambda_0$, which implies that the right-hand side of 
(\ref{homo1}) is zero.  It remains to show that the left-hand side, 
$d'(\lambda_2 (\ell_1, \ell_2))$, is also zero.
Indeed, we have:
\[
d'(\lambda_2 (\ell_1, \ell_2)) =
d'\left(0, - 2k \int \bracket{\ell_1}{\ell_2'} \; d\theta \right) =
0 \, .
\]

Next we check property (\ref{homo2}).  On the
right-hand side, we have:
\begin{eqnarray*}
        \lambda_1 ( l_2 (\ell_1,\ell_2) )
        -
        l_2 (\lambda_0 (\ell_1) , \lambda_1 (\ell_2))
        &=&
        ([\ell_1, \ell_2],0)
        -
        \big( [\ell_1, \ell_2],
        2k \int \bracket{\ell_1}{\ell_2 '} \, d\theta \big) \\
        &=&
        \big(0, -2k  \int \bracket{\ell_1}{\ell_2 '} \, d\theta \big) \, .
\end{eqnarray*}
On the left-hand side, we have:
$$\lambda_2(\ell_1, d(\ell_2)) = \lambda_2 (\ell_1, \ell_2)
= \big(0, - 2k \int \bracket{\ell_1}{\ell_2 '} \; d\theta \big)$$
Note that this also shows that given the chain map defined by 
$\lambda_0$ and $\lambda_1$, the function $\lambda_2$ that extends
this chain map to an $L_\infty$-homomorphisms is unquely
fixed by condition (\ref{homo2}).

Finally, we show that $\lambda_2$ satisfies condition (\ref{homo3}).
The two terms involving $l_3$ vanish since $\lambda$ is a map between two
strict Lie $2$-algebras.  The three terms of the form
$l_2 ( \lambda_0 (\cdot), \lambda_2 (\cdot, \cdot))$ vanish because
the image of $\lambda_2$ lies in the center of $\wOkg$.
It thus remains to show that
$$
\lambda_2(\ell_1, l_2(\ell_2, \ell_3)) + 
\lambda_2 (\ell_2, l_2 (\ell_3, \ell_1))
+ \lambda_2(\ell_3, l_2(\ell_1, \ell_2)) = 0. $$
This is just the cocycle property of the Kac--Moody cocycle,
Equation (\ref{eq: 2-cocycle eqn}).
\hbox{\hskip 60ex} \endofproof

Next we check the exactness of the sequence
\[    0 \to \T{\Omega\g} \stackto{\lambda}
            \Pg \stackto{\phi} \g_k  \to 0  \]
at the middle point.  Before doing so, we recall the formulas for
the $L_\infty$-homomorphisms corresponding to $\lambda$ and $\phi$.  The 
$L_\infty$-homomorphism corresponding to $\lambda \maps \T{\Omega\g} 
\to \Pg$ is given by
\begin{eqnarray*}
\lambda_0 (\ell) &=& \ell \cr
\lambda_1 (\ell) &=& (\ell, 0) \cr
\lambda_2 (\ell_1, \ell_2) &=&
\displaystyle{\big( 0,\, - 2k \int_0^{2\pi} 
\bracket{\ell_1}{\ell_2'} \, d\theta \big)}
\end{eqnarray*}
where $\ell, \ell_1, \ell_2 \in \Omega\g$, and that corresponding
to $\phi \maps \Pg \to \g_k$ is given by:
\begin{eqnarray*}
\phi_0 (p) &=& p(2\pi) \cr
\phi_1 (\ell, c) &=& c \cr
\phi_2 (p_1, p_2) &=& \displaystyle{k
       \int_0^{2\pi} \big( \bracket{p_1}{p_2'} - 
                   \bracket{p_1'}{p_2} \big) \; d\theta }
\end{eqnarray*} where $p, p_1, p_2 \in P_0 \g$, 
$\ell \in \Og$, and $c \in \R$.

\begin{lemma} \et The composite
\[
    \T{\Omega\g}
    \stackto{\lambda}
    \Pg
    \stackto{\phi}
    \g_k
\]
is the zero homomorphism, and the kernel of $\phi$ is precisely the image
of $\lambda$, both on objects and on morphisms.
\end{lemma}

\Proof
The composites $(\phi \circ \lambda)_0$ and 
$(\phi \circ \lambda)_1$ clearly vanish.  Moreover 
$(\phi \circ \lambda)_2$ vanishes since:
\begin{eqnarray*}
    (\phi \circ \lambda)_2 (\ell_1,\ell_2)
    &=&
    \phi_2 (\lambda_0 (\ell_1) , \lambda_0 (\ell_2) )
    +
    \phi_1 ( \lambda_2 ( \ell_1,\ell_2 )) \qquad 
    \textrm{by} \; (\ref{composite.homo})
    \\
    &=&
    \phi_2 ( \ell_1,\ell_2 )
    +
    \phi_1 \big(0, - 2k \int \bracket{\ell_1}{\ell_2'} \, d\theta \big)
    \\
    &=&
    k \int ( \bracket{\ell_1}{\ell_2'}
    - \bracket{\ell_1 '}{\ell_2} ) \, d\theta
    - 2k \int \bracket{\ell_1}{\ell_2'} \, d\theta  \\
    &=& 0
\end{eqnarray*}
with the help of integration by parts.
The image of $\lambda$ is precisely the kernel of $\phi$ by construction.
\endofproof

Note that $\phi$ is obviously onto, both for objects and
morphisms, so we have an exact sequence
\[    0 \to \T\Omega\g \stackto{\lambda}
            \Pg \stackto{\phi} \g_k  \to 0 \, . \]
Next we construct a family of splittings
$\psi \maps \g_k \to \Pg$ for this exact sequence:

\begin{lemma} \et \label{psi}
Suppose
\[
    f \maps [0,2\pi] \to \mathbb{R}
\]
is a smooth function with $f(0) = 0$ and $f(2\pi) = 1$. 
Then there is a Lie 2-algebra homomorphism
\[
   \psi \maps \g_k \to \Pg
\]
whose corresponding $L_{\infty}$-homomorphism is given by:
\begin{eqnarray*}
\psi_0 (x) &=& xf \cr
\psi_1 (c) &=& (0, c) \cr
\psi_2 (x_1, x_2) &=& ([x_1, x_2](f-f^2), 0)
\end{eqnarray*}
where $x, x_1, x_2 \in \g$ and $c \in \R$.
\end{lemma}

\Proof 
We show that $\psi$ satisfies the conditions of Definition \ref{Linftyhomo}.
We begin by showing that $\psi$ is a chain map, meaning that the following
square commutes:
$$\xymatrix{
    \R \ar[rr]^{d}
    \ar[dd]_{\psi_{1}} &&
    \g  \ar[dd]^{\psi_{0}} \\ \\
    \wOkg \ar[rr]^{d'} &&
    P_0 \g
}$$
where $d =0 $ since $\g_k$ is skeletal and $d'$ is the composite given
in Proposition \ref{Pg}.  This square commutes because
$\psi_0(d(c)) = \psi_0 (0) = 0$ and $d'(\psi_1(c)) =
d'(0,c) = 0$.  

We continue by verifying conditions (\ref{homo1}) - (\ref{homo3})
of Definition \ref{Linftyhomo}.
The right-hand side of (\ref{homo1}) equals:
$$\psi_0 ( l_2(x_1,x_2) ) - l_2 ( \psi_0 (x_1) , \psi_0 (x_2) )
=  [x_1,x_2] (f-f^2) \, . $$
This equals the left-hand side $d'(\psi_2 (x_1,x_2)) $ by 
construction.

The right-hand side of (\ref{homo2}) equals:
\[ 
  \psi_1( l_2(x,c) ) - l_2( \psi_0(x), \psi_1(c) ) \; =  \;
  \psi_1( 0 ) - l_2( xf, (0,c) ) \; = \; 0 
\]
since both terms vanish separately. Since the left-hand side is
$\psi_2(x,dc) = \psi_2(x,0) = 0$, this shows that $\psi$ satisfies
condition (\ref{homo2}).

Finally we verify condition (\ref{homo3}).  The term
$l_3 ( \psi_0 (\cdot), \psi_0 (\cdot), \psi_0 (\cdot))$ vanishes
because $\Pg$ is strict. The sum of three other terms vanishes 
thanks to the Jacobi identity in $\g$:
\begin{eqnarray*}
&& \psi_2(x_1, l_2(x_2,x_3)) 
 + \psi_2(x_2, l_2(x_3,x_1)) 
 + \psi_2(x_3, l_2(x_1,x_2))  \\
    &=&
    \Big(([x_1,[x_2,x_3]] +  
          [x_2,[x_3,x_1]] +  
          [x_3,[x_1,x_2]]) \, (f - f^2),\, 0 \Big)  \\
    &=&
    (0,0) \, .
\end{eqnarray*}
Thus, it remains to show that:
$$
 -\psi_1 (l_3(x_1,x_2,x_3)) =  $$
$$  l_2(\psi_0 (x_1), \psi_2 (x_2, x_3))
+ l_2(\psi_0 (x_2), \psi_2 (x_3, x_1)) 
+ l_2(\psi_0 (x_3), \psi_2 (x_1, x_2)) \, .
$$
This goes as follows:
\begin{eqnarray*}
    &&   l_2(\psi_0 (x_1), \psi_2 (x_2, x_3))
       + l_2(\psi_0 (x_2), \psi_2 (x_3, x_1))
       + l_2(\psi_0 (x_3), \psi_2 (x_1, x_2)) \\
    &=&
    \Big(0, 3 \cdot 2k \int_0^{2\pi}
      \bracket{x_1}{[x_2,x_3]} \, f(f-f^2)' \, d\theta \; \Big) \\
    &=&
    \left(0,-k\bracket{x_1}{[x_2,x_3]} \right) \qquad 
\textrm{by the calculation below} \\
    &=&
    -\psi_1 (l_3(x_1, x_2, x_3)) \, .
\end{eqnarray*}
The value of the integral here is \emph{universal},
independent of the choice of $f$:
\begin{eqnarray*}
\int_0^{2\pi}  f (f-f^2)' \; d \theta
&=&
\int_0^{2\pi} \left(f(\theta) f'(\theta)
- 2 f^2 (\theta) f'(\theta) \right) \; d\theta
    \\
    &=&
    \frac{1}{2} - \frac{2}{3}
    =
    -\frac{1}{6}
    \,.
\end{eqnarray*}
\hbox{\hskip 50ex} \endofproof

The final step in proving Theorem \ref{bigthm}
is to show that $\phi \circ \psi$ is the
identity on $\g_k$, while $\psi \circ \phi$ is isomorphic to the
identity on $\Pg$.
For convenience, we recall the definitions first:
$\phi \maps \Pg \to \g_k$ is given by:
\begin{eqnarray*}
\phi_0 (p) &=& p(2\pi) \cr
\phi_1 (\ell, c) &=& c \cr
\phi_2 (p_1, p_2) &=& \displaystyle{k
     \int_0^{2\pi} 
\left( \bracket{p_1}{p_2'} - \bracket{p_1'}{p_2} \right) \; d\theta }
\end{eqnarray*} 
where $p, p_1, p_2 \in P_0 \g$, $\ell \in \wOkg$, and $c \in \R$, 
while $\psi \maps \g_k \to \Pg$ is given by:
\begin{eqnarray*}
\psi_0 (x) &=& xf \cr
\psi_1 (c) &=& (0, c) \cr
\psi_2 (x_1, x_2) &=& ([x_1, x_2] (f-f^2), 0)
\end{eqnarray*}
where $x, x_1, x_2 \in \g$, $c \in \R$, and
$f \maps [0,2\pi] \to \R$ satisfies the conditions of Lemma \ref{psi}.

\begin{lemma} \et
      With the above definitions we have:
    \begin{itemize}
      \item $\phi \circ \psi$ is the identity Lie 2-algebra homomorphism 
            on $\g_k$;
      \item $\psi \circ \phi$ is isomorphic, as a Lie 2-algebra
            homomorphism, to the identity on $\Pg$.
    \end{itemize}
\end{lemma}

\Proof 
We begin by demonstrating that $\phi \circ \psi$
is the identity on $\g_k$.
First,
$$(\phi \circ \psi)_0 (x) = \phi_0 (\psi_0 (x))
= \phi_0(xf) = xf(2\pi) = x,$$
since $f(2\pi) = 1$ by the definition of $f$ in Lemma \ref{psi}.
Second,
$$(\phi \circ \psi)_1 (c) = \phi_1 (\psi_1 (c))
= \phi_1 ((0,c)) = c$$
Finally,
\begin{eqnarray*}
(\phi \circ \psi)_2 (x_1, x_2)
&=&
\phi_2 (\psi_0 (x_1), \psi_0(x_2)) + \phi_1 (\psi_2 (x_1, x_2))
\qquad \textrm{by} \; (\ref{composite.homo}) \\
&=&
\phi_2(x_1 f, x_2 f) \; + \; \phi_1([x_1, x_2](f-f^2),0) \\
&=&
k \int (
\bracket{x_1 f}{x_2 f'} - \bracket{x_1 f'}{x_2 f} ) \, d\theta \; 
+ \; 0 \\
&=&
k \bracket{x_1}{x_2} \int (ff' - f'f)\; d\theta
\\
&=& 0  \, .
\end{eqnarray*}

Next we consider the composite
\[
    \psi \circ \phi \maps \Pg \to \Pg \, .
\]
The corresponding $L_{\infty}$-algebra homomorphism is given by:
\begin{eqnarray*}
(\psi \circ \phi)_0 (p) &=& p(2\pi)f \\
(\psi \circ \phi)_1 (\ell,c) &=& (0,c) \\
(\psi \circ \phi)_2 (p_1,p_2) &=&
\left( [p_1 (2\pi) , p_2 (2\pi)](f-f^2), \;
k \int \left( \bracket{ p_1}{ p_2'} - \bracket{ p_1'}{ p_2}
\right) \; d\theta \right)
\end{eqnarray*}
where again we use equation (\ref{composite.homo}) to obtain
the formula for $(\psi \circ \phi)_2$.

For this to be isomorphic to the identity there must
exist a Lie 2-algebra 2-isomorphism 
$$\tau \maps \psi \circ \phi \To \id$$
where $\id$ is the identity on $\Pg$.
We define this in terms of its corresponding 
$L_{\infty}$-2-homomorphism by setting:
$$\tau(p) = (p-p(2\pi)f,0) \,.$$
Thus, $\tau$ turns a path $p$ into the loop $p-p(2\pi)f$.

We must show that $\tau$ is a chain homotopy satisfying
condition (\ref{2homo}) of Definition \ref{Linfty2homo}.
We begin by showing that $\tau$ is a chain homotopy.  We have
\begin{eqnarray*}
d(\tau(p)) \; = \; d(p-p(2\pi)f,0) &=& p -p(2\pi)f \\
&=& \id_0(p) - (\psi \circ \phi)_0 (p)
\end{eqnarray*}
and
\begin{eqnarray*}
\tau (d(\ell,c)) \; = \; \tau (\ell) &=& (\ell,0) \\
&=& \id_1(\ell, c) - (\psi \circ \phi)_1 (\ell, c)
\end{eqnarray*}
so $\tau$ is indeed a chain homotopy.

We conclude by showing that $\tau$ satisfies 
condition (\ref{2homo}):
$$(\psi \circ \phi)_2 (p_1,p_2) =
l_2((\psi \circ \phi)_0 (p_1), \tau (p_2)) +
l_2 (\tau (p_1), p_2) - \tau ( l_2 (p_1, p_2))$$
In order to verify this equation, we write out the right-hand 
side more explicitly by inserting the formulas for 
$(\psi\circ\phi)_2$ and for $\tau$, obtaining:
\[
  l_2\big(p_1(2\pi)f,\, (p_2-p_2(2\pi)f,0)\big)
  +
  l_2\big((p_1-p_1(2\pi)f,0),\, p_2\big)
  -
  ([p_1,p_2] - [p_1(2\pi),p_2(2\pi)]f,\, 0)  
\]
This is an ordered pair consisting of a loop in $\g$ and a real number.
By collecting summands, the loop itself turns out to be:
\[
  [p_1(2\pi),p_2(2\pi)] (f-f^2)
  \,.
\]
Similarly, after some integration by parts the real number is found to be:
\[ 
 k  \int_0^{2\pi}
    \left(
       \bracket{p_1}{p_2'}
       -
       \bracket{p_1'}{p_2}
    \right)
    \, d\theta
  \,.
\]
Comparing these results with the value of 
$(\psi\circ \phi)_2(p_1,p_2)$ given
above, one sees that $\tau$ indeed satisfies (\ref{2homo}).
\endofproof

\subsection{Conclusions}
\label{conclusions.section}

We have seen that the Lie 2-algebra $\g_k$ is equivalent
to an infinite-dimensional Lie 2-algebra $\Pg$, and that when
$k$ is an integer, $\Pg$ comes from an infinite-dimensional
Lie 2-group $\PG$.  Just as the Lie 2-algebra $\g_k$ is built from
the simple Lie algebra $\g$ and a shifted version of $\u(1)$:
\[ 0 \stackto{\;} {\rm b}\u(1) \stackto{\;} \g_k \stackto{\;} 
\g \stackto{\;} 0\, , \]
the Lie 2-group $\PG$ is built from $G$ and another Lie 2-group:
\[ 1 \stackto{\;} \LG \stackto{\;} \PG \stackto{\;} G \stackto{\;} 1  \]
whose geometric realization is a shifted version of $\U(1)$:
\[ 1 \stackto{\;} B\U(1) \stackto{\;} 
|\PG| \stackto{\;} G \stackto{\;} 1\, . \]
None of these exact sequences split; in every case an interesting
cocycle plays a role in defining the middle term.  In the first case, 
the Jacobiator of $\g_k$ is $k\nu \maps \Lambda^3 \g \to \R$.
In the second case, composition of morphisms is defined using
multiplication in the level-$k$ Kac--Moody central extension of
$\OG$, which relies on the Kac--Moody cocycle $k\omega \maps 
\Lambda^2 \Og \to \R$.  In the third case, $|\PG|$ is the total
space of a twisted $B\U(1)$-bundle over $G$ whose Dixmier--Douady
class is $k[\nu/2\pi] \in H^3(G)$.  Of course, all these cocycles
are different manifestations of the fact that every simply-connected 
compact simple Lie algebra has $H^3(G) = \Z$. 

We conclude with some remarks of a more speculative nature.  
There is a theory of `2-bundles' in which a Lie 2-group plays 
the role of structure group \cite{BaezSchreiber:2004,Bartels:2004}.  
Connections on
2-bundles describe parallel transport of 1-dimensional
extended objects, e.g.\ strings.  Given the importance of the 
Kac--Moody extensions of loop groups in string theory, it is 
natural to guess that connections on 2-bundles with structure 
group $\PG$ will play a role in this theory.

The case when $G = \Spin(n)$ and $k = 1$ is particularly interesting,
since then $|\PG| = \String(n)$.  In this case we suspect that
$2$-bundles on a spin manifold $M$ with structure $2$-group $\PG$ 
can be thought as substitutes for principal $\String(n)$-bundles on 
$M$.  It is interesting to think about `string structures'
\cite{MurrayStevenson:2001} on $M$ from this perspective: given a principal $G$-bundle 
$P$ on $M$ (thought of as a $2$-bundle with only identity morphisms) 
one can consider the obstruction problem of trying to lift the structure 
$2$-group from $G$ to $\PG$.  There should be a single topological
obstruction in $H^4(M;\Z)$ to finding a lift, namely the characteristic 
class $p_1/2$.  When this characteristic class vanishes, every principal 
$G$-bundle on $M$ should have a lift to a $2$-bundle $\mathcal{P}$ on $M$ 
with structure $2$-group $\PG$.  It is tempting to conjecture that the 
geometry of these $2$-bundles is closely related to the enriched 
elliptic objects of Stolz and Teichner \cite{StolzTeichner:2004}.

\newpage

\subsection{Strict 3-Groups}
\label{Strict 3-Groups}

We end this section on 2-groups and the 2-group
$\mathcal{P}_k G$ with a brief remark on strict 3-groups and
a 3-group extension of $\mathcal{P}_k G$ (which is not
from \cite{BaezCransSchreiberStevenson:2005}). This thread
will then be taken up again in a remark on 3-bundles
in \S\fullref{Global 3-Holonomy in 3-Bundles} 
which serves to substantiate the statement
from the last section \S\fullref{conclusions.section}, 
that $\mathcal{P}_k G$-2-bundles should be obstructed by the
first Pontryagin class.

\subsubsection{Strict 3-Groups and 2-Crossed Modules}
\label{Strict 3-Groups and 2-Crossed Modules}

Let $\mathcal{C}$ be any 2-category. Denote composition of 
1-morphisms by $\circ_1$ and composition of 2-morphisms by
$\circ_2$.
A strict 2-functor on $\mathcal{C}$ respects both of these
compositions strictly.

A {\bf strict 3-group} 
$\mathcal{G}$
is a strict 2-category with a strict 2-functor
\[
  \cdot : \mathcal{G} \times \mathcal{G} \to \mathcal{G}
\]
which satisfies the axioms of a group product strictly.

There are other, equivalent, definitions of strict 3-groups.
For instance a strict 3-group is (I believe) also the same
as a strict 3-category with a single object and all 
1-, 2-, and 3-morphisms invertible.

In any case, there are several exchange laws for strict 3-groups.
In the formulation using a product 2-functor, which is one
one I will use here, two exchange laws
comes from the 2-functoriality of `$\cdot$'. Another one
is the exchange between $\circ_1$ and $\circ_2$ in a strict
2-category. But only two of these three laws turn out to be 
independent.

\vskip 1em

When we forget about the 2-morphisms in $\mathcal{G}$ we 
obtain a strict 2-group with objects and 1-morphisms those
of $\mathcal{G}$. 
This 2-group is given by a crossed module 
$(G,H,\alpha_1,t_1)$ where $G$ and $H$ are any groups and
\begin{eqnarray*}
  \alpha_1 &:& G \to \mathrm{Aut}(H)\\
  t_1 &:& H \to G
\end{eqnarray*}
are homomorphisms.

Recall that under the product operation `$\cdot$' the 1-morphisms 
of $\mathcal{G}$ form the group $G \semidir H$ with
\[
  (g,h)(g',h') = (gg', h \, \alpha\of{g}\of{h'})
  \,.
\] 

Alternatively, when we forget about the
objects in $\mathcal{G}$ we get another strict 2-group
$(G',J,\alpha_2,t_2)$
\begin{eqnarray*}
  \alpha_2 &:& G' \to \mathrm{Aut}(J)\\
  t_2 &:& J \to G'
\end{eqnarray*}
whose objects are the 1-morphisms and whose 1-morphisms 
are the 2-morphisms of $\mathcal{G}$. Hence $G'$ must be
the group of 1-morphisms under $\cdot$, i.e.
\[
  G' = G \semidir H
  \,.
\]
Collecting all this data we can describe the 3-group 
$\mathcal{G}$ by a tuple
\[
  (G,H,J,\alpha_1,\alpha_2,t_1,t_2)
\]
There are certain conditions imposed on 
this data coming from the two versions of the exchange law
in $\mathcal{G}$:

{\bf Claim}: 
\begin{enumerate}
\item
  $\mathrm{Im}\of{t_2} = (1,H) \simeq H \subset G\semidir H$

\item
$t_1,t_2$ form a sequence (not necessarily exact)
\[
  J \stackto{t_2} H \stackto{t_1} G
  \,,
\]
i.e.
\[
  t_1\of{t_2\of{j}} = 1
  \,.
\]

\item
  $J$ is necessarily an \emph{abelian} group.
\end{enumerate}

When all conditions are satisfied we should call
the tuple $(G,H,J,\alpha_i,t_i)$ a 
{\bf 2-crossed module}.

So, to summarize, a 2-crossed module is a tuple
\[
  (G,H,J,\alpha_1,\alpha_2,t_1,t_2)
\]
where $G$ and $H$ are any groups and $J$ is an abelian group,
and where
\begin{eqnarray*}
  \alpha_1 &:& G \to \mathrm{Aut}\of{H}
  \\
  \alpha_2 &:& G \semidir H \to \mathrm{Aut}\of{J}
  \\
  t_1 &:& H \to G
  \\
  t_2 &:& J \to H \subset G\semidir H
\end{eqnarray*}
are homomorphisms such that we have a sequence
\begin{eqnarray*}
  J \stackto{t_2} H \stackto{t_1} G
  \,.
\end{eqnarray*}
and such that
$(G,H,\alpha_1,t_1)$ and $(G\semidir H,J,\alpha_2,t_2)$
are two ordinary crossed modules.

\Proof

First some notation: 

We label 1-morphisms in the 3-group as usual by their
source object $g\in G$ and an element of $h\in H$ as
$(g,h)$. Similarly, 2-morphisms are labeled by their
source $(g,h) \in G\semidir H$ and an element $j\in J$
as 
\[
  ((g,h),j) : (g,h) \stackto{j} \left( (g,h') = t_2\of{j} (g,h) \right)
  \,.
\]

Here source and target are 1-morphisms
\begin{eqnarray*}
  (g,h) : g \stackto{h} \left(g' = t_1\of{h}g\right)
  \\
  (g,h') : g \stackto{h'} \left(g' = t_1\of{h'}g\right)  
\end{eqnarray*}
which must share the same source and target objects $g,g'\in G$.

It follows that for $h$ and $h'$ here we have
\begin{eqnarray}
  \label{t1(h) = t1(h')}
  t_1(h) = t_1\of{h'}
  \,.
\end{eqnarray}
When plugged into the equation 
\[
  (g,h') = t_2\of{j}(g,h)
\]
for the target of $((g,h),j)$ one gets
\begin{eqnarray*}
  t_2\of{j} 
  &=& 
  (g,h')(g,h)^{-1}
  \\
  &=&
  (g,h')(g^{-1},\alpha_1\of{g^{-1}}\of{h^{-1}})  
  \\
  &=&
  (1,h' h^{-1})
  \,.
\end{eqnarray*}
This proves first of all that the image of $t_2$ is 
$(1,H) \subset G\semidir H$.

So if we identify this image with $H$
\[
  (1,h'h^{-1}) \simeq h'h^{-1}
\]
and apply $t_1 : H \to G$ to it, we get
\begin{eqnarray*}
  t_1\of{t_2\of{j}} &=& t_1\of{h'h^{-1}}
  \\
  &=&
  t_1\of{h'}t_1\of{h}^{-1}
  \\
  &\equalby{t1(h) = t1(h')}&
  1
  \,.
\end{eqnarray*}

This shows that we have a sequence $J \stackto{t_2} H \stackto{t_1} G$.

The above facts were consequences of the exchange laws in 
the 2-groups $(G,H,\alpha_1,\alpha_2)$ and
$(G',J,\alpha_2,t_2)$ that sit inside the 3-group $\mathcal{G}$.
The exchange law in $(G',J,\alpha_2,t_2)$ comes from the
functoriality of `$\cdot$' with respect to $\circ_2$. 
The exchange law in $(G,H,\alpha_1,\alpha_2)$ comes from
functoriality of `$\cdot$' with respect to $\circ_1$ restricted
to identity 2-morphisms. When this is generalized to
non-identity 2-morphisms one obtains a further condition:

Consider two composable 1-morphisms $(g,h)$ and $(t_1\of{h}g,h')$
with composition
\[
  (g,h) \circ_1 (t_1\of{h}g,h')
  =
  (g,h'h)
  \,.
\]
Now let $((g,h),j)$ and $((t_1\of{h}g,h'),j')$ be two 2-morphisms 
with source $(g,h)$ and $(t_1\of{h}g,h')$, respectively. 

{\bf Lemma}: Under the composition $\circ_1$ the 2-morphisms 
behave as
\[
  \left(
    (g,h),j
  \right) 
  \circ_1 
 \left(
   (t_1\of{h}g,h'),j'
  \right)
  =
  \left(
    (g,h'h), \; j'\, \alpha_2\of{1,h'}\of{j}
     \,.
  \right)
\]

\emph{Proof of the lemma}:

First of all we can identically write:
\begin{eqnarray*}
  \left(
    (g,h),j
  \right) 
  \circ_1 
 \left(
   (t_1\of{h}g,h'),j'
  \right)
  &=&
  \left[ ((1,1),1) \cdot ((g,h),j) \right]
  \circ_1
  \left[
    ((1,h'),j') \cdot ((t_1\of{h}g,1),1)
  \right]
\end{eqnarray*}

Using the exchange law (2-functorality of `$\cdot$') this
becomes
\begin{eqnarray*}
  \cdots 
  &=&
 \left[
   ((1,1),1) 
    \circ_1
    ((1,h'),j')
  \right]
  \cdot
  \left[
    ((g,h),j)
    \circ_1
    ((t_1\of{h}g,1),1)
  \right]
  \,.
\end{eqnarray*}
Note how now the two $\circ_1$-compositions that appear each
involve an \emph{identity} 2-morphism.
The axioms for 2-categories say that the identites under 
$\circ_2$-composition are also identites for $\circ_1$-composition. 
It follows that
\begin{eqnarray*}
  \cdots
  &=&
  ((1,h'),j') \cdot ((g,h),j)
  \\
  &=&
  ((g,h'h),j'\, \alpha_2\of{1,h'}\of{j})
  \,.
\end{eqnarray*}
This poves the above lemma.
\endofproof
Noting that the $\circ_1$-product is hence once again a kind of
semidirect product one sees that the lemma can equivalently be
obtained by studying the exchange law between $\circ_1$ and $\circ_2$.

With this product rule in hand we can now do a variation of the
Eckmann-Hilton argument:

Consider the expression
\[
  \left[
    ((1,1),j) \circ_1 ((1,1),1)
  \right]
  \cdot
  \left[
    ((1,1),1) \circ_1 ((1,1),j')
  \right]
  \,.
\]
Evaluating this as indicated yields
\[
  \cdots = ((1,1),jj')
  \,.
\]
Using the exchange law first gives
\[
  \cdots = 
  \left[
    ((1,1),j)\cdot ((1,1),1)
  \right]
  \circ_1
  \left[
    ((1,1),1) \cdot ((1,1),j')
  \right]
\]
and using the above lemma this becomes
\[
  \cdots = 
  ((1,1),j'j)
  \,.
\]
This shows that $J$ must be an abelian group.
\endofproof

\subsubsection{A 3-group extension of $\mathcal{P}_k G$}
\label{a 3-group extension of PkG}

Given any crossed module $(G,H,\alpha_1,t_1)$ we may ask
what extensions $(G,H,J,\alpha_i,t_i)$ to a
2-crossed module there are.

By the above result this amounts to finding an abelian group $J$
and homomorphisms 
$\alpha_2 : G \semidir H \to \mathrm{Aut}\of{J}$
and $t_2 : J \to G \semidir H$ such that
\begin{eqnarray*}
  &&t_1\of{t_2\of{j}} = 1\,, \hspace{1cm} \forall\, j\in J
  \,.
\end{eqnarray*}

Consider the strict 2-group
\[
  \mathcal{P}_k G = (P_0 G, \widehat{\Omega_k G}, \alpha_1,t_1)
\]
from above. There is an obvious choice for the above extension:

Let $J = U\of{1}$ be the abelian group and identify this by
\begin{eqnarray*}
  t_2 : J &\to& \widehat{\Omega_k G}\\
        j &\mapsto& (1,j)
\end{eqnarray*}
with the center of $H = \widehat{\Omega_k G}$. This gives an
exact sequence
\[
  J = U\of{1} \stackto{t_2} \widehat{\Omega_k G} \stackto{t_1} P_0 G
\]
and hence this $t_2$ is admissible.

\newpage

\clearpage
\section{2-Connections with 2-Holonomy on 2-Bundles}
\label{principal 2-Bundles}

Bartels \cite{Bartels:2004} defined a concept of `2-bundle'
by categorifying the usual concept of bundle.  Originally his 
definition only treated 2-bundles over an ordinary space.   However, to handle 
twisted nonabelian gerbes, we need 2-bundles over a 2-space of infinitesimal 
loops.   So, we define locally trivial 2-bundles over an arbitrary simple
2-space in \S\fullref{section: Locally Trivial 2-Bundles}, describe them
using local data in \S\fullref{section: 2-Transitions in Terms of Local Data}, 
and specialize to the case where the base space consists of infinitesimal 
loops
in \S\fullref{Restriction to the case of trivial base 2-space}.

\subsection{Locally Trivial 2-Bundles}
\label{section: Locally Trivial 2-Bundles}

In differential geometry an
ordinary bundle consists of two smooth spaces, the {\bf total space}
$E$ and the {\bf base space} $B$, together with a {\bf projection map} 
\[
  E \stackto{p} B
  \,.
\]
To categorify the theory of bundles, we start by replacing smooth spaces by 
smooth 2-spaces:

\begin{definition}\et
  \label{2-bundle}
  A {\bf 2-bundle} consists of
  \begin{itemize}
    \item
      a 2-space $\twoP$ (the {\bf total 2-space})
    \item
      a 2-space $\twoB$ (the {\bf base 2-space})
    \item
      a smooth map $p \maps \twoP \to \twoB$ 
      (the {\bf projection})
    \,.
  \end{itemize}
\end{definition} 

In gauge theory we are interested in \emph{locally trivial} 2-bundles. 
Ordinarily, a locally trivial
bundle with fiber $F$ is a bundle $E \stackto{p} B$ together with an open
cover $U_i$ of $B$, such that the restriction of $E$ to any of the $U_i$ 
is equipped with an isomorphism to the trivial bundle $U_i \times F \to U_i$. 
To categorify this, we need to define a `2-cover' of the base 2-space $B$.

\begin{definition}\et
  \label{sub-2-spaces}
  A 2-space $S$ is called a {\bf sub-2-space} of the 2-space $B$ if 
  $\Ob(S) \subseteq \Ob(B)$, $\Mor(S) \subseteq \Mor(B)$, and the source, target, identity and 
  composition maps of $S$ are the restriction of these maps on $B$ to these 
  subsets.  In this case we write $S \subseteq B$.  
\end{definition}

\begin{definition}\et
  \label{open.sub-2-spaces}
  If $S$ is a sub-2-space of $B$, $\Ob(S)$ is open
  in $\Ob(B)$ and $\Mor(S)$ is open in $\Mor(B)$, we say $S$ is an {\bf open}
  sub-2-space of $B$.
\end{definition} 

\begin{definition}\et
  \label{union of 2-spaces}
  Given a simple 2-space $B$ (\refdef{trivial-simple.2-space}) and a
  collection of sub-2-spaces 
  $\{U_i\}_{i \in I}$ with $U_i \subseteq B$, their {\bf union}
  $\bigcup\limits_{i\in I} U_i$ is defined to be the
  sub-2-space of $B$ whose space of objects is 
  $\bigcup\limits_{i \in I} \Ob(U_i)$ and whose space of morphisms is
  $\bigcup\limits_{i \in I} \Mor(U_i)$.
  Similarly, the {\bf intersection} $\bigcap\limits_{i\in I} U_i$
  is defined to be the sub-2-space of $B$ whose space of objects is 
  $\bigcap\limits_{i \in I} \Ob(U_i)$ and whose space of morphisms is
  $\bigcap\limits_{i \in I} \Mor(U_i)$.
\end{definition}

\noindent
We restrict this definition of `union' to simple 2-spaces
$B$ to avoid situations where one can compose a morphism 
$f \maps x \to y$ in $U_{i_1}$ with a morphism $g \maps y \to z$ in $U_{i_2}$ 
to obtain a morphism that lies in none of the $U_i$.  This happens, for
example, when $B$ is the path groupoid of some space (\refdef{2space.3}).  
Such 2-spaces require a more general concept of `2-cover' than
we provide here.

\begin{definition}\et
  \label{2-cover}
  A {\bf 2-cover} of a simple 2-space B is a collection $\{U_i\}_{i \in I}$
  of open sub-2-spaces of $B$ with $\bigcup\limits_{i\in I} U_i = B$.
\end{definition}

Given a 2-cover $\{U_i\}_{i \in I}$ of $B$, we can form a 2-space 
called their {\bf disjoint union} $U = \bigdisunion\limits_{i\in I} U_i$ 
in an obvious way, and every 2-cover is equipped with a 2-map
\begin{eqnarray}
 \label{cover injection map}
  U \stackto{j} B
\end{eqnarray}
that restricts on each $U_i$ to the inclusion $U_i \hookrightarrow B$.
We often refer to the 2-cover $\{U_i\}_{i \in I}$ simply as $U$ for 
short.

\begin{definition}\et
Given a 2-cover $U$ of a simple 2-space $B$,
we define the {\bf 2-space of $n$-fold intersections} by:
\[
  U^{[n]}
  =
  \bigsqcup\limits_{i_1, i_2,\dots,i_n \in I}
    U_{i_1} \cap U_{i_2} \cap \dots \cap U_{i_n}
  \,.
\]
\end{definition}

\noindent
This 2-space comes with maps
\begin{eqnarray}
  \label{cover projection maps}
  U^{[n]} & \stackto{j_{01\cdots(k-1)(k+1)\cdots n}} & U^{[n-1]}
  \nonumber\\
  (i_1,i_2,\dots,i_n,x \stackto{\gamma} y)
  &\mapsto&
  (i_1,\dots,i_{k-1},i_{k+1},\dots,i_n,x \stackto{\gamma} y)  
\end{eqnarray}
that forget about the $k$th member of the multiple intersection.

With the notion of 2-cover in hand, we can now state the definition of a
locally trivial 2-bundle.  First note that we can restrict a
2-bundle $E \stackto{p} B$ to any sub-2-space $U \subseteq B$
to obtain a 2-bundle which we denote by $P|_U \stackto{p} U$.  Then:

\begin{definition}\et
\label{2-bundle with local trivialization}
Given a 2-space $F$, we define a {\bf locally trivial 2-bundle with fiber $F$} 
to be a 2-bundle $E \stackto{p} B$ (\refdef{2-bundle}) and a 2-cover 
$U$ of the base 2-space $B$ 
equipped with equivalences (\refdef{equivalent.2-spaces})
  \[
    P|_{\twoU_i} \stackto{t_i} \twoU_i \times \twoF
  \]
such that the diagrams
\[ 
\xymatrix @!0
{ P|_{\twoU_i}
 \ar [dddrr]_{p}
  \ar[rrrr]^{t_i} 
  & &  & &
  U_i \times F
  \ar[dddll]^{}
  \\ \\ &  \\ & &
   U_i 
 }
\]
commute up to invertible 2-maps for all $i\in I$.
\end{definition}

This definition is concise and elegant, but rather abstract. In 
\S\fullref{section: 2-Transitions in Terms of Local Data}
we translate its meaning into transition laws for local data
specifying the 2-bundle. In order to do so, we first need to 
extract \emph{transition functions} from a local trivialization:

By composing the local trivializations and their weak inverses 
on double intersections $U_{ij}$
one gets autoequivalences of $U_{ij}\times F$ 
of the form
\[
  U_{ij}\times \twoF \stackto{\bar t_i \circ t_j} U_{ij}\times \twoF
\]
and similarly for other index combinations.

\begin{definition}\et
\label{principal 2-bundle}
Given a 2-bundle equipped with a local trivialization 
such that
\begin{itemize}
\item 
all autoequivalences 
$U_{ij} \times F \stackto{\bar t_i \circ t_j} U_{ij} \times F$ 
act trivially on the $U_{ij}$ factor, so that
\[
  \bar t_i \circ t_j = \mathrm{id}_{U_{ij}} \times g_{ij}
  \,,
\]
\item
$F$ is a 2-group $\twogroup$,
\item
the $g_{ij}$ act by left horizontal 2-group multiplication on $\twoF$
\end{itemize}
we say that our 2-bundle is a {\bf principal $\twogroup$-2-bundle} and
that 
\begin{eqnarray}
  U^{[2]} & \stackto{g} & \twogroup
  \nonumber\\
  U_{ij} &\mapsto& g_{ij}
  \nonumber 
\end{eqnarray}
is the {\bf transition function}. 
\end{definition}

\subsection{2-Transitions in Terms of Local Data}
\label{section: 2-Transitions in Terms of Local Data}

Consider a triple intersection $U_{ijk} = U_i \cap U_j \cap U_k$
in a principal $\twogroup$-2-bundle (\refdef{principal 2-bundle}).
The existence of the local trivialization implies that the following diagram
2-commutes (all morphisms here are 2-maps and all 
2-morphisms are natural isomorphisms between these):
\begin{center}
\begin{picture}(300,340)
\includegraphics{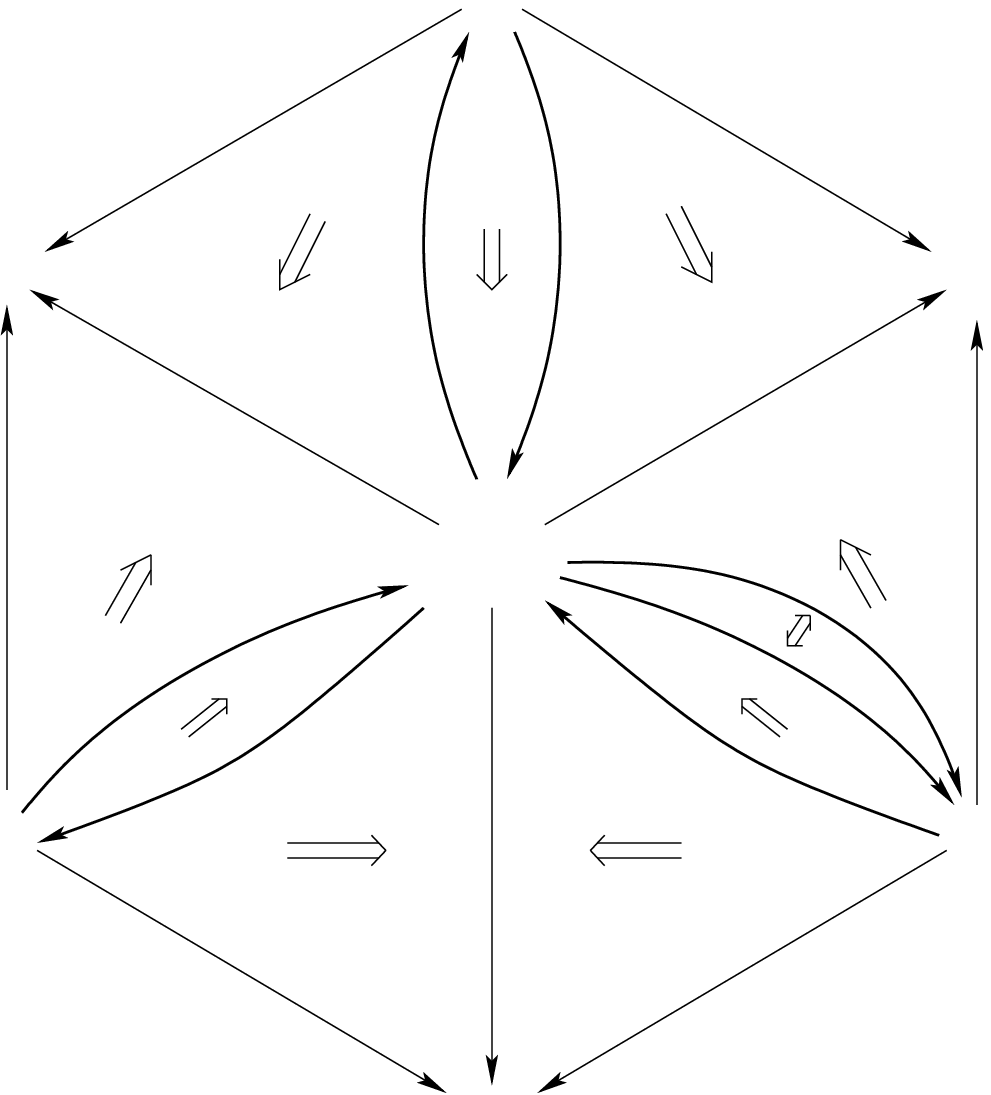}
\put(-162,152){$p^{-1}{U_{ijk}}$}
\put(-155,80){$p$}
\put(-65,190){$p$}
\put(-224,190){$p$}
\put(-150,-7){$U_{ijk}$}
\put(-10,235){$U_{ijk}$}
\put(-293,236){$U_{ijk}$}
\put(-160,320){$U_{ijk}\times F$}
\put(-5,70){$U_{ijk} \times F$}
\put(-320,70){$U_{ijk} \times F$}
\put(-80,87){$\bar t_i$}
\put(-45,120){$t_i$}
\put(-18,122){$t^\prime_i$}
\put(-215,87){$t_j$}
\put(-240,128){$\bar t_j$}
\put(-176,245){$t_k$}
\put(-120,245){$\bar t_k$}
\end{picture}
\end{center}

\vskip 2em

Compared to the analogous diagram for an uncategorified bundle two important 
new aspects are
that the barred morphisms are inverses-up-to-isomorphism of the local
trivializations and that the local trivialization itself is unique only up to
natural isomorphisms (\refdef{2-bundle with local trivialization}). 
The latter is indicated by the presence of an arrow
denoting a trivialization
$t^\prime_i$ naturally isomorphic to $t_i$.

 From the diagram it is clear that the usual transition law 
$g_{ij}g_{jk} = g_{ik}$
here becomes a natural isomorphism called a 2-transition, 
which was first
considered in \cite{Bartels:2004} for the special case of
trivial base 2-spaces, but which directly
generalizes to arbitrary base 2-spaces:

\begin{definition}\et
  \label{2-transition}
  Given a base 2-space $B$ with cover $U \stackrel{j}{\longrightarrow} B$
  a {\bf 2-transition} is 
  \begin{itemize}
  \item
 a 2-map
\[
  U^{[2]}
  \stackrel{g}{\longrightarrow}
  \twogroup
\]
called the {\bf transition function}, 
\item
and a natural isomorphism
  $f$ 
 \begin{eqnarray}
    &&U^{[3]} 
    \stackrel{j_{02}}{\longrightarrow}
    U^{[2]}
    \stackrel{g}{\longrightarrow}
    \twogroup
    \nonumber\\
    & \stackTo{f} &
    \nonumber\\
    &&
    U^{[3]} \stackrel{\stackrel{\vee}{U}{}^{[3]}}{\longrightarrow}
    U^{[3]} \times U^{[3]}
    \stackrel{j_{01}\times j_{12}}{\longrightarrow}
    U^{[2]} \times U^{[2]}
    \stackrel{g \times g}{\longrightarrow}
    \twogroup \times \twogroup
    \stackrel{m}{\longrightarrow}
    \twogroup
    \,,
    \nonumber
 \end{eqnarray}
(which expressed the categorification of the ordinary transition law
$g_{ij}g_{jk} = g_{ik}$),
  together with the coherence law for $f$ enforcing the
  associativity of the product $g_{ij}g_{jk}g_{kl}$,
  \item
  and a natural isomorphism
  \begin{eqnarray}
    && U \stackto{j_{00}} U^{[2]} \stackto{g} \twogroup
    \nonumber\\
    & \stackTo{\eta} &
    \nonumber\\
    &&
    U \stackto{\hat U} 1 \stackto{i} \twogroup \,.    
    \nonumber
  \end{eqnarray}
  (expressing the categorification of the ordinary $g_{ii} = 1$) together
  with its coherence laws.
\end{itemize}

\end{definition}

In the above $U^{[3]} \stackto{\stackrel{\vee}{U}{}^{[3]}}  U^{[3]}\times U^{[3]}$ 
denotes the
diagonal embedding of $U^{[3]}$ in its second tensor power and $m$ denotes the
horizontal multiplication (functor) in the 2-group $\twogroup$. The maps
$j_{\cdots}$ have been defined in \refer{cover projection maps}.

In terms of local functions this means the following:

\begin{proposition}\et
  \label{proposition on local meaning of 2-transition}
  A 2-transition 
  (\refdef{2-transition})
 on a $\twogroup$-2-bundle with base 2-space being 
a simple 2-space (\refdef{trivial-simple.2-space})
and $\twogroup$ a 2-group induces the
transition law \refer{gerbe transition law for the transition functions}
of a nonabelian gerbe.
\end{proposition}

\Proof

The existence of the natural isomorphism
means that there is a map
\begin{eqnarray}
  (U^{[3]})^1 &\stackrel{f}{\longrightarrow}& \twogroup^2
  \nonumber\\
  (x,i,j,k) &\mapsto& h_{ijk}\of{x}
  \,,
  \nonumber
\end{eqnarray}
with the property
\begin{eqnarray}
   \label{a relation in a 2-transition}
   g^2_{ik}\of{x} \circ h_{ijk}\of{x}
   &=&
   h_{ijk}\of{x}
   \circ
   (g^2_{ij}\of{x} \cdot g^2_{jk}\of{x})
   \,,
   \hspace{1cm}
   \forall\, (x,i,j) \in U^{[2]}
   \,.
\end{eqnarray}
(Here $\circ$ denotes the vertical and $\cdot$ the horizontal product in 
the 2-group,
see Prop.\ \ref{liecrossedmodule})

For $\twogroup$ the source/target matching condition implies that
(again Prop. \ref{liecrossedmodule})
\begin{eqnarray}
  \label{source/target matching in proof for 2-transition}
  t\of{f^2_{ijk}} g^1_{ik}
  &=&
  g^1_{ij}g^1_{jk}
  \,,
\end{eqnarray}
where we have decomposed the 2-group element
\[
  h_{ijk}\of{x} = (f^1_{ijk}\of{x},f^2_{ijk}\of{x})
\]
into its source label $f^1_{ijk}\of{x} \in G$ 
and its morphism label $f^2_{ijk}\of{x}\in H$.

Identifying $g_{ij} = \phi_{ij}$ this is the gerbe transition law 
\refer{gerbe transition law for the transition functions}.
\endofproof

\subsubsection{2-Transitions and Cocycles}

The functor that glues different local
trivializations of the 2-bundles is
\[
 \bar t_i \circ t_j : U_{ij} \times \twogroup \to U_{ij} \times \twogroup
 \,.
\]
Let us concentrate on the case where
these transitions act trivially on the $U_{ij}$ factor and act by 
left multiplication on $\twogroup$. 

A little thought shows that this means that the $\bar t_i \circ t_j$
must come from a functor
\[
  g_{ij} : U_{ij} \to \twogroup
  \,,
\]
which assigns a morphisms 
$g_{ij}\of{x} \stackto{g_{ij}\of{\gamma}} g'_{ij}\of{x}$ in the 
2-group $\twogroup$ to every
morphism $x \stackto{\gamma} y$ in $U_{ij}$, as follows:
\[
  \bar t_i \circ t_j\of{
    x \stackto{\gamma} y, f
  }
  =
  \left(
     x \stackto{\gamma} y,
     \; g_{ij}\of{\gamma}\cdot f
  \right)
  \,.
\]
Here $f\in \twogroup$. 

The usual equation 
\[
  g_{ij}g_{jk} = g_{ik}
\]
describing the consistency of transitions in an ordinary 
principal 2-bundle now becomes a natural isomorphism between 
functors, whose naturality square is the following:

\[
  \begin{array}{c|ccc}
     U_{ijk} & \twogroup &    & \twogroup  \\
     \hline
     x & g_{ik}\of{x}  & \stackto{f_{ijk}\of{x}} & g_{ij}\of{x}\cdot g_{jk}\of{x}\\
     \gamma \Bigg\downarrow  & 
            g_{ik}\of{\gamma}\Bigg\downarrow \skiph{$g_{ik}\of{\gamma}$} & &
            g_{ij}\of{\gamma}\cdot g_{jk}\of{\gamma}\Bigg\downarrow 
            \skiph{$g_{ij}\of{\gamma}g_{jk}\of{\gamma}$} 
            \\
     y & g_{ik}\of{y} & \stackto{f_{ijk}\of{y}} & g_{ij}\of{y}\cdot g_{jk}\of{y}
  \end{array}
\]
The morphisms
\[
   g_{ik}\of{x} \stackto{f_{ijk}\of{x}} g_{ij}\of{x}\cdot g_{jk}\of{x}   
\]
in $\twogroup$ define the natural transformation. When we write
the morphism $f_{ijk}\of{x}$ in terms of elements of $G \ltimes H$ as
\[
  h_{ijk}\of{x}
  =
  \left(
    g_{ij}\of{x}g_{jk}\of{x}, \; f_{ijk}\of{x}
  \right)
\]
the existence of these arrows is expressed by the equation
\[
  g_{ij}\of{x}g_{jk}\of{x}
  =
  t\of{f_{ijk}\of{x}}g_{ik}\of{x}
  \,.
\]
This is the first cocylce relation.

It is accompanied by another equation, which is not called a 
cocycle relation, but which appears here notwithstanding. It is 
the one expressing the commutativity of the above naturality
square and reads
\[
  f_{ijk}\of{y} g_{ik}\of{\gamma}f_{ijk}^{-1}\of{x}
  =
  h_{ij}\of{\gamma} \alpha\of{g_{ij}\of{x}}\of{h_{jk}\of{\gamma}}
  \,,
\]
where we have written $g\of{\gamma}$ in terms of $(g,h) \in G \ltimes H$.

\subsubsection{Interpretation in terms of transition bundles}

In order to grasp the meaning of this equation it may be instructive
to compare the above to the analogous step in the context of bundle 
gerbes, as detailed in \cite{AschieriCantiniJurco:2004} (where the last
equation above corresponds to equation (106)). 

In that context one declares that instead of transition \emph{functions}
one is to use transition \emph{bundles} $E$ which are
principal $H$-bibundles.

So on the fibers of such a bundle the group $H$ acts
freely and transitively from the left 
\begin{eqnarray*}
  H \ni h : E &\to& E\\
  e &\mapsto& h e
\end{eqnarray*}
and from the 
right $e \mapsto e h$. Being both free and transitive, 
these two actions must be related by an automorphism
$\phi_e \in \Aut\of{H}$ as
\[
   eh \defas \phi_e\of{h} e
  \,.
\]
A local trivializaton of such a bundle consists of choosing a good cover
$\set{O_\alpha}$ and local sections $\sigma^\alpha : O_\alpha \to E$ .
These give rise to the local maps
\[
  \phi^\alpha \defas \sigma^\alpha \circ \phi
\]
and on overlapping patches they are related by transition functions
$h^­{\alpha\beta} : O_{\alpha\beta} \to H$ as
\[
  \sigma_\alpha = h^{\alpha\beta}\sigma_\beta
  \,.
\]
For bibundles $E_1$ and $E_2$ one can form the product bundle
$E_1 E_2$ defined fiberwise over $H$. So if $E_1$ is trivialized by
$\set{\sigma_\alpha^1}$ with local data $\set{h^{\alpha\beta}_1,\phi^\alpha_1}$
and
$E_2$ by $\set{\sigma^\alpha_2}$ with local data $\set{h^{\alpha\beta}_2,
\phi^\alpha_2}$,
then the product bundle is trivialized by sections 
$\set{(\sigma^\alpha_1,\sigma^\alpha_2)}$
such that the transitions are given by
\begin{eqnarray*}
  (\sigma^\alpha_1,\sigma^\alpha_2) 
   &=& 
  (h^{\alpha\beta}_1 \sigma^\beta_1, h^{\alpha\beta}_2\sigma^\beta_2)
  \\
  &=&
  h^{\alpha\beta}_1 (\sigma^\beta_1 h^{\alpha\beta}_2, \sigma^\beta_2)
  \\
  &=&
  h^{\alpha\beta}_1 \phi^\beta_1\of{h^{\alpha\beta}_2}(\sigma^\beta_1,
  \sigma^\beta_2)
  \,.
\end{eqnarray*}
The maps $\phi$ of the product bundle are similarly seen to be given by
$\phi^i_1 \phi^i_2$.
Hence the product bundle is given by the local data
\[
  \set{
    h^{\alpha\beta}_1 \phi^\beta_1\of{h^{\alpha\beta}_2}
    \,,
    \phi^\alpha_1 \phi^\beta_2
    \,.
  }
\]
If a bibundle is trivial, it admits a global section represented by
$f^\alpha : O_\alpha \to H $ and its local data in the above sense reads
$\set{f^\alpha (f^\beta)^{-1}, \mathrm{Ad}\of{f^\alpha}}$.

 If $E_1$ is described by
$\set{\sigma_i^1}$ with $\set{h_{ij}^1,\phi_i^1}$ and 
$\set{\sigma_i^2}$ with $E_2$ by $\set{h_{ij}^2,\phi^2_i}$, then
their product, which is locally trivialized by the sections
$(\sigma_i^1,\sigma_i^2)$ is described by
\[
  \set{h_{ij}^1,\phi_i^1} 
  \cdot
  \set{h_{ij}^2,\phi^2_i}
  \defas
  \set{h^1_{ij} }
  \,.
\]

Now, in the context of bundle gerbes one chooses a good cover 
$\set{U_i}$ of base space $M$ and gets \emph{transition bi-bbundles}
$\set{E_{ij} \to U_{ij}}_{i,j}$ over the double intersections.
The transition law now says that on triple intersections 
\[
  E_{ij}E_{jk} = T_{ijk}E_{ik}
  \,,
\]
where $T_{ijk}$ is a bundle of the above form with 
\[
  T_{ijk} \simeq \set{f_{ijk}^{\alpha} (f_{ijk}^{\beta})^{-1}, 
   \mathrm{Ad}_{f_{ijk}^\alpha}}
  \,.
\]
By the above formulas this yields the following two conditions on the
local data of these bundles:
\[
  \phi_{ij}^\alpha \phi_{ij}^\beta
  =
  \mathrm{Ad}_{f_{ijk}^\alpha} \phi_{ik}^\alpha 
\]
and
\[
  h_{ij}^{\alpha\beta} \phi_{ij}^{\beta}\of{h_{ij}^{\alpha\beta}}
  =
  f^\alpha_{ijk}h_{ik}^{\alpha\beta}(f_{ijk}^{\beta})^{-1}
  \,.
\]
The first of these corresponds to the existence of the arrows in the
natural transformation discussed above. The second corresponds to the
commutativity of the naturality square.

The full discussion of these nonabelian bundle gerbes requires the
consideration of bundles over fiver products of fibrations of base space.
The interested reader is referred to \cite{AschieriCantiniJurco:2004}
for further details.

\subsubsection{The coherence law for the 2-transition}

The natural transformation $f$ which weakens the ordinary transition law
$g_{ij}g_{jk} = g_{ik}$ has to satisfy a coherence law which makes its application
on multiple products $g_{ij}g_{jk}g_{kl}$ well defined. 

Note that first of all that Prop.\ \refer{liecrossedmodule} 
implies a certain relation among the $h_{ijk}$:
By using the relation $g^1_{ij}g^1_{jk} = t\of{h_{ijk}}g^1_{ik}$ 
in the expression $g_{ij}g_{jk}g_{kl}$ in two
different ways one obtains
\[
  t\of{h_{ijk}}
  t\of{h_{ikl}}
  =
  g_{ij}t\of{h_{jkl}}g_{ij}^{-1}
  t\of{h_{ijl}}
  \,.
\]
This equation implies that
\begin{eqnarray}
  \label{2-transition coherence law}
  f^{-1}_{ikl}f^{-1}_{ijk}\alpha\of{g_{ij}}\of{h_{jkl}}h_{ijl}
  =
  \lambda_{ijkl}
\end{eqnarray}
with 
\[
  \lambda_{ijkl} \maps U^1_{ijkl} \to  \ker\of{t} \subset H
  \,.
\]

This is the gerbe transition law 
\refer{gerbe coherence law for transformators of transition functions}.
The function $\lambda_{ijkl}$ is the `twist' 0-form 
\refer{phases}.

 From the perspective of 2-bundles
the twist can be understood as coming from a nontrivial natural
transformation between 2-maps from $U^{[4]}$ to $U^{[2]}$:

First assume that the natural transformation
\begin{eqnarray}
  \label{natural transformation between j023j02 and j013j02}
  &&U^{[4]} \stackto{j_{023} \circ j_{02}} U^{[2]}
  \nonumber\\
  &\stackrel{\omega_{03}}{\Rightarrow}&
  \nonumber\\
  &&
  U^{[4]} \stackto{
  j_{013} \circ j_{02}
  }
  U^{[2]}
  \,.
\end{eqnarray}
is trivial, which means 
that sending a based loop $\gamma_{(x,i,j,k,l)}$ in $(U^{[4]})^2$
first to the based loop $\gamma_{(x,i,k,l)}$ in $(U^{[3]})^2$
and then to $\gamma_{(x,i,l)}$ in $(U^{[2]})^2$
yields the same result as first sending it to
$\gamma_{(x,i,j,l)}$ in $(U^{[3]})^2$ and then to 
$\gamma_{(x,i,l)}$ in $(U^{[2]})^2$.

Using \refer{a relation in a 2-transition} we have

\hspace{-3cm}\parbox{20cm}{
\begin{eqnarray}
  \label{coherence law for transition transformation}
  &&
  (g_{ij}^2 \cdot g^2_{jk}) \cdot g^2_{kl} = 
  g^2_{ij} \cdot (g^2_{jk} \cdot g^2_{kl})
  \nonumber\\
  &\stackrel{\refer{a relation in a 2-transition}}{\Leftrightarrow}&
  \left(
    (h_{ijk})^{r} \circ g^2_{ik} \circ h_{ijk}
  \right) 
  \cdot 
  \left(
    1_{g^1_{kl}} \circ g^2_{kl} \circ 1_{g^1_{kl}}
  \right)
  =
  \left(
    1_{g^1_{ij}}
    \circ
    g^2_{ij}
    \circ
    1_{g^1_{ij}}
  \right)
  \cdot
  \left(
    (h_{jkl})^{r} \circ g^2_{jl} \circ h_{jkl}
  \right)
  \nonumber\\
  &\Leftrightarrow&
  \left(
    (h_{ijk})^{r} \cdot 1_{g^1_{kl}}
  \right)
  \circ
  \left(
    g^2_{ik}\cdot g^2_{kl}
  \right)
  \circ
  \left(
    h_{ijk} \cdot 1_{g^1_{kl}}
  \right)
  =
  \left(
    1_{g^1_{ij}} \cdot (h_{jkl})^{r} 
  \right)
  \circ
  \left(
    g_{ij}^2 \cdot g_{jl}^2
  \right)
  \circ
  \left(
    1_{g^1_{ij}} \cdot h_{jkl}
  \right)
  \nonumber\\
  &\stackrel{\refer{a relation in a 2-transition}}{\Leftrightarrow}&
  \left(
    (h_{ijk})^{r} \cdot 1_{g^1_{kl}}
  \right)
  \circ
  \left(
    (h_{ikl})^{r} \circ g^2_{il} \circ h_{ikl}
  \right)
  \circ
  \left(
    h_{ijk} \cdot 1_{g^1_{kl}}
  \right)
  =
  \left(
    1_{g^1_{ij}} \cdot (h_{jkl})^{r} 
  \right)
  \circ
  \left(
    (h_{ijl})^{r} \circ g_{il}^2 \circ h_{ijl}^2
  \right)
  \circ
  \left(
    1_{g^1_{ij}} \cdot h_{jkl}
  \right) 
  \,.
  \nonumber\\
\end{eqnarray}
}
This has the form
\begin{eqnarray}
  A^r \circ g^2_{il} \circ A
  &=&
  B^r \circ g^2_{il} \circ B
  \nonumber
\end{eqnarray}
with
\begin{eqnarray}
  \label{conjugation elements in discussion of transition coherence law}
  A &=& h_{ikl} \circ (h_{ijk} \cdot 1_{g^1_{kl}})
  \nonumber\\
  B &=& h_{ijl} \circ (1_{g^1_{ij}} \cdot h_{jkl})
  \,.
\end{eqnarray}
If we identify both `conjugations' we obtain
\begin{eqnarray}
  \label{equality of conjugation elements in discussion of transition coherence law}
  A = B \;&\Leftrightarrow&\;
  (f^2_{ikl})^{-1}(f^2_{ijk})^{-1}\alpha\of{g^1_{ij}}\of{f^2_{jkl}}f^2_{ijl}
  = 1
  \,.
\end{eqnarray}
This reproduces \refer{2-transition coherence law} without the twist.

Now generalize to nontrivial
natural transformations
\refer{natural transformation between j023j02 and j013j02}.
This implies the existence of a function
\[
  (U^{[4]})^1 
  \stackto{\ell}
  (U^{[2]})^2  
\]
that assigns loops based in double overlaps to points in 
quadruple overlaps. Applying the `transition function' $g$ 
to these loops implies that
\begin{eqnarray}
  \label{natural transformation inducind transition twist}
  g^2\of{\ell} \circ g^2\of{j_{013}\circ j_{02}}
  &=&
  g^2\of{j_{023} \circ j_{02}}\circ g^2\of{\ell}
  \,.
\end{eqnarray}
The 2-group element $g^2\of{\ell}$ is specified by a function
\begin{eqnarray}
  (U^{[4]})^1 \stackto{\lambda} \ker\of{t} \subset H
   \nonumber
\end{eqnarray}
as
\begin{eqnarray}
  (U^{[4]})^1 & \stackto{\ell \circ g} & \twogroup^2
  \nonumber\\
  (x,i,j,k,l) & \mapsto & (g^1_{il}, \lambda_{ijkl}\of{x})
  \,.
  \nonumber
\end{eqnarray}

All this applies to \refer{coherence law for transition transformation}
by noting that there on the left hand side the $g_{il}$
in general is $g\of{j_{023}\circ j_{02}}$ while that on the right hand
is $g\of{j_{013}\circ j_{02}}$.

Hence in the case of nontrivial arrow base space we have to
replace the $g^2_{il}$ in the last line on 
the left with $g^2\of{j_{013}\circ j_{02}}$
and that on the right with
$g^2\of{\ell} \circ g^2\of{j_{013}\circ j_{02}} \circ (g^2\of{\ell})^{-1}$.

When doing so the 2-group elements $A$ and $B$ of 
\refer{conjugation elements in discussion of transition coherence law}
become
\begin{eqnarray}
  A &=& (g_{il}^1,\lambda^{-1}_{ijkl}) \circ h_{ikl} \circ (h_{ijk} \cdot 1_{g^1_{kl}})
  \nonumber\\
  B &=& h_{ijl} \circ (1_{g^1_{ij}} \cdot h_{jkl})
  \,.
  \nonumber
\end{eqnarray}
Equating these generalizes 
\refer{equality of conjugation elements in discussion of transition coherence law}
to
\begin{eqnarray}
  A = B \;&\Leftrightarrow&\;
  (f^2_{ikl})^{-1}(f^2_{ijk})^{-1}\alpha\of{g^1_{ij}}\of{f^2_{jkl}}f^2_{ijl}
  = \lambda_{ijkl}
  \,.
  \nonumber
\end{eqnarray}

\subsubsection{Restriction to the case of trivial base 2-space}
\label{Restriction to the case of trivial base 2-space}

It is instructive to restrict the above general disucssion to the
case where the base 2-space is trivial:

In that case the 2-transition specifies the following data:
\begin{itemize}
  \item
    smooth maps
    \[
       g_{ij} \maps U_i \cap U_j \to \twogroup^1
    \] 
   \item
     smooth maps
    \[
      h_{ijk} \maps U_i \cap U_j \cap U_k \to \twogroup^2
    \]
    with
    \[
       h_{ijk}\of{x} \maps  g_{ik}\of{x} \to g_{ij}\of{x}g_{jk}\of{x}
    \]
   \item
     smooth maps
    \[
      k_i : U_i \to \twogroup^2
    \]
    with
    \[
      k_i \maps g_{ii} \to 1 \in \twogroup
      \,.
    \]
\end{itemize}

The coherence law 
\refer{2-transition coherence law} 
says that on quadruple intersections
$U_i \cap U_j \cap U_k \cap U_l$
the following 2-morphisms in $\twogroup$ are identical:

\begin{center}
\begin{picture}(300,150)
\includegraphics{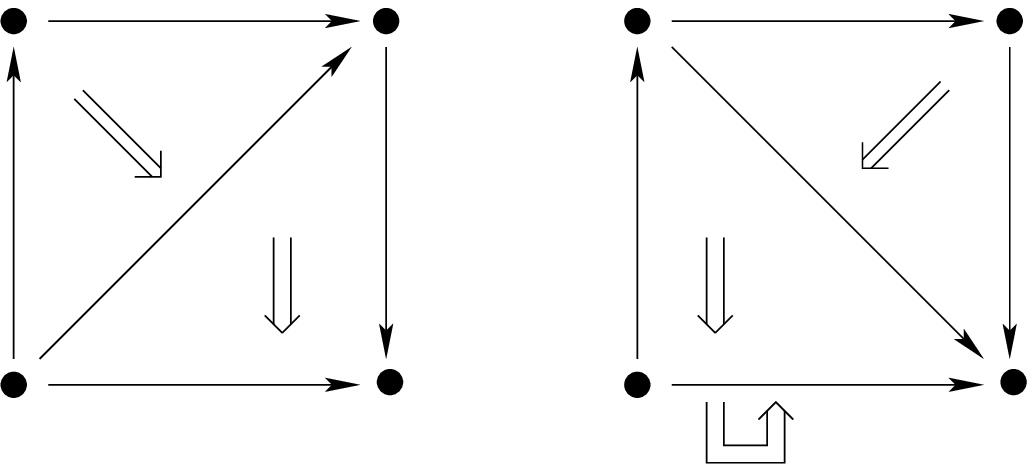}
\put(-5,20){
\begin{picture}(300,150)
\put(-245,115){$g_{jk}$}
\put(-65,115){$g_{jk}$}
\put(-245,-5){$g_{il}$}
\put(-45,-5){$g_{il}$}
\put(-90,-30){$\lambda_{ijkl}$}
\put(-307,55){$g_{ij}$}
\put(-127,55){$g_{ij}$}
\put(-180,55){$g_{kl}$}
\put(-1,55){$g_{kl}$}
\put(-152,55){$=$}
\put(-280,30){$g_{ik}$}
\put(-40,21){$g_{jl}$}
\put(-60,80){$\bar f_{jkl}$}
\put(-84,30){$\bar f_{ijl}$}
\put(-234,30){$\bar f_{ikl}$}
\put(-254,77){$\bar f_{ijk}$}
\end{picture}
}
\end{picture}
\end{center}

This diagram gives a nice visualization of the different ways to
go from the upper arc $g_{ij}g_{jk}g_{kl}$ of the square to the
bottom edge $g_{il}$.

There are also coherence laws for $k_i$,
the {\bf left unit law} and {\bf right unit law},
which express the relation of $k$
to $f$ when two of the indices of the latter coincide:
\begin{center}
\begin{picture}(230,200)
\includegraphics{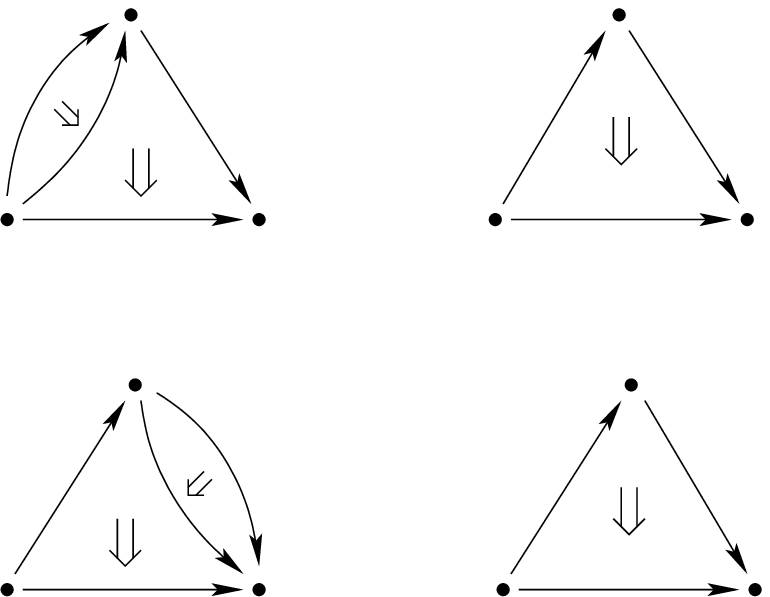}
\put(0,0){X}
\put(-45,-6){$g_{ij}$}
\put(-190,-6){$g_{ij}$}
\put(-45,100){$g_{ij}$}
\put(-190,100){$g_{ij}$}
\put(-20,145){$g_{ij}$}
\put(-161,145){$g_{ij}$}
\put(-71,35){$g_{ij}$}
\put(-215,35){$g_{ij}$}
\put(-75,145){$g_{ii}$}
\put(-224,145){$g_{ii}$}
\put(-200,144){$k_i$}
\put(-193,129){$1$}
\put(-176,122){$1$}
\put(-117,130){$=$}
\put(-55,120){$f^r_{iij}$}
\put(-117,20){$=$}
\put(-16,35){$g_{jj}$}
\put(-149,35){$g_{jj}$}
\put(-172,39){$k_j$}
\put(-175,20){$1$}
\put(-194,15){$1$}
\put(-52,13){$f^r_{ijj}$}
\end{picture}
\end{center}

\vskip 2em

The freedom of having nontrivial $k_i$ is special to 2-bundles and not
known in (nonabelian) gerbe theory. Gerbe cocycles involve {\v C}ech cohomology
and hence \emph{antisymmetry} in indices $i,j,k,\dots$ in the sense that
group valued functions go into their inverse on an odd permutation of their
cover indices.

Whenever we derive nonabelian gerbe cocycles from 2-bundles with 2-connection
we will hence have to restrict to $k_i = 1$ for all $i$.

\subsubsection{Weak Principal 2-Bundles}
\label{Weak Principal 2-Bundles}

The above discussion applied to principal 2-bundles whose structure
group is a \emph{strict} 2-group. It can however easily be
generalized to coherent weak structure 2-groups:

For weak 2-groups the above coherence law for the transition functions
has to take into account that the product of the edge labels is not
associative, but that instead there is the {\bf associator}, a
morphism
\[
  (g_{ij} \cdot g_{jk}) \cdot g_{kl}
  \stackTo{\alpha}
  g_{ij} (g_{jk} \cdot g_{kl})
  \,.
\]
Therfore the coherence law more generally looks like
\begin{center}
\begin{picture}(370,170)
\includegraphics{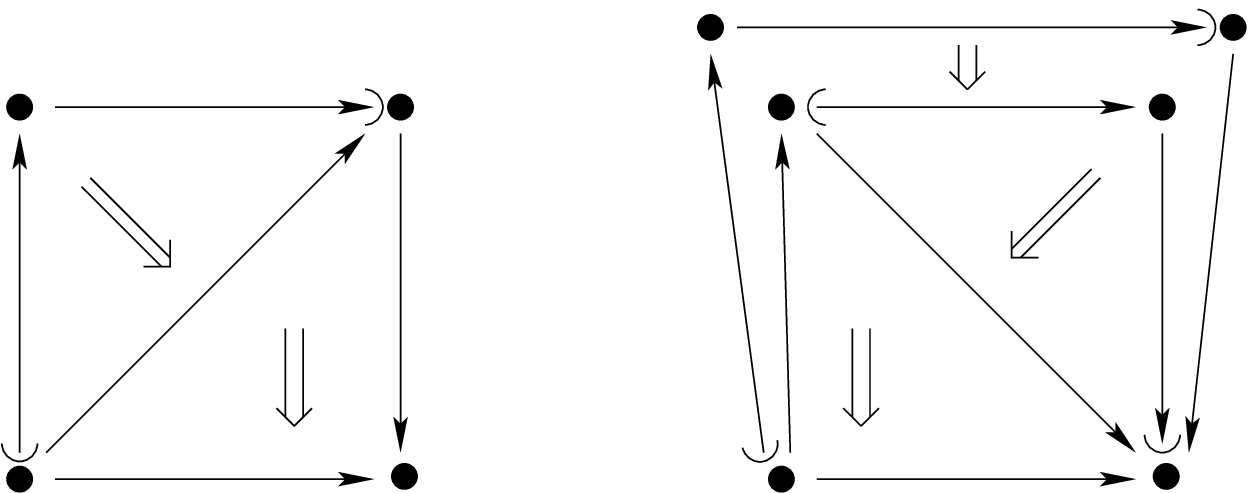}
\put(-66,2){
\begin{picture}(300,150)
\put(-307,55){$g_{ij}$}
\put(-100,55){$g_{ij}$}
\put(-180,55){$g_{kl}$}
\put(53,55){$g_{kl}$}
\put(-142,55){$=$}
\put(-280,30){$g_{ik}$}
\put(-8,27){$g_{jl}$}
\put(-20,80){$\bar f_{jkl}$}
\put(-44,30){$\bar f_{ijl}$}
\put(-234,30){$\bar f_{ikl}$}
\put(-254,77){$\bar f_{ijk}$}
\put(-245,115){$g_{jk}$}
\put(-28,140){$g_{jk}$}
\put(-32,120){$\alpha$}
\put(-245,-5){$g_{il}$}
\put(-45,-5){$g_{il}$}
\end{picture}
}
\end{picture}
\end{center}

This however no longer has a simple translation into 
a formula for group elements.

\vskip 2em

The above diagrams applied to trivial base 2-spaces.
We can also consider nontrivial base 2-spaces, i.e. those
with nontrivial base arrow-space. It turns out however that
the local description of a 2-bundle coincides with the cocycle data
of a nonabelian gerbe only in the limit that the base 2-space morphisms
differ `infinitesimally' from identity morphisms. What this means is
explained in the next subsection.

\subsubsection{Summary}

By internalizing the concept of an ordinary bundle in the 2-category of
2-spaces (which again are categories internalized in $\Diff$, the category
of smooth spaces) one obtains a categorified notion
of the fiber bundle concept, called 2-bundle, which differs from an 
ordinary bundle essentially in that what used to be ordinary maps between sets
(like the projection map of the bundle)
now become (smooth) functors between categories. This adds to the original 
bundle (at the `point level' of the 2-space) a dimensional generalization 
(at the `arrow level' of the 2-space)
of all concepts involved. In addition to providing new `degrees of freedom' the
categorification weakens former notions of equality. 

By re-expressing the abstract arrow-theoretic construction of a 2-bundle 
in terms of concrete local group- and algebra-valued $p$-forms,
we find a generalization of the ordinary transition laws for such local data
in an ordinary bundle. Under certain conditions these generalized transition
laws coincide with the cocycle data of nonabelian gerbes.

So far all of this pertained to 2-bundles (and nonabelian gerbes) without
a notion of connection. For constructing a categorified connection and
hence a notion of nonabelian surface holonomy, it is helpful to 
consider ordinary connections on spaces of paths in a manifold.
This is the content of the next section.

\subsection{Local 2-Holonomy and Transitions}
\label{Local 2-Holonomy and Transitions}

We now define what a {\bf 2-connection} in a 2-bundle 
over a categorically trivial base 2-space
is supposed to be. 
Then, in \S\fullref{section 2-connections in terms of local p-forms} 
we list the results concerning 
the expression of such 2-connections in terms of
local $p$-form data and sketch the proofs. 
Making these precise requires some technology, which is
developed in \S\fullref{section: Path Space}.

Locally a $p$-holonomy is nothing but a $p$-functor from the
$p$-groupoid of $p$-paths to the structure $p$-group. 
For $p=1,2$ these $p$-groupoids are defined in the next
subsection \S\ref{p-path p-groupoids}. The definition of 
global 2-holonomy is then given in \S\fullref{p-holonomy p-functors}.

\subsubsection{$p$-Path $p$-Groupoids}
\label{p-path p-groupoids}

\begin{definition}\et
  \label{groupoid of paths}

  The {\bf path groupoid} $\P_1\of{U}$ 
  of a manifold $U$ is the groupoid for which
  \begin{itemize}
    \item
      objects are points $x \in U$
    \item
      morphisms with source $s \in U$ and target $t\in U$
      are thin homotopy equivalence classes $[\gamma]$
      of parametrized paths $\gamma \in \paths_s^t\of{U}$
      \refdef{based path space} (p. \pageref{based path space})
     \[
       \xymatrix{
         x \ar@/^1pc/[rr]^{[\gamma]}
         && y
       }
     \]
    \item
      composition is given by
   \[
     \xymatrix{
     x \ar@/^1pc/[rr]^{[\gamma_1]}
     && y \ar@/^1pc/[rr]^{[\gamma_2]}
     && z
     }
     =
     \xymatrix{
       x \ar@/^1pc/[rr]^{[\gamma_1 \circ \gamma_2]}
       && z
      }
   \]
    where
    \begin{eqnarray}
       \circ \maps \paths_x^y\of{U} \times \paths_y^z\of{U}
       &\to&
       \paths_x^z\of{U}
       \nonumber\\
       (\gamma_1,\gamma_2) &\mapsto&
       \gamma_{1,2}
       \nonumber
    \end{eqnarray}
    with
    \begin{eqnarray}
      \gamma_{1,2}\of{\sigma}
      &\defas&
      \left\lbrace
        \begin{array}{cc}
           \gamma_1\of{2\sigma} & \mbox{for $0 \leq \sigma \leq 1/2$} \\
           \gamma_2\of{2\sigma-1} & \mbox{for $1/2 \leq \sigma \leq 1$}
        \end{array}
      \right.
      \,.
      \nonumber
    \end{eqnarray}
  \end{itemize}
\end{definition}

Note that taking thin homotopy equivalence classes makes this composition
asscociative and invertible.

Just like ordinary holonomy locally is a functor from the groupoid of paths
to an ordinary group, 2-holonomy
locally is a 2-functor from some 2-groupoid to a 2-group. This 2-groupoid
is roughly that consisting of bounded surfaces in $U$ whose horizontal
and vertical composition corresponds to the ordinary gluing of bounded
surfaces. This heuristic idea is made precise 
by the following defition of 
$\P_2\of{U}$, the {\bf 2-groupoid of bigons}.

\vskip 1em

First of all a bigon is a `surface with two corners'. More precisely:

\begin{definition}\et
  \label{parametrized bigon}
  Given any manifold $U$ a {\bf parametrized bigon} in $U$
  is a smooth map
  \begin{eqnarray}
    \Sigma \maps [0,1]^2 &\to& U
    \nonumber\\
    (\sigma,\tau) &\mapsto& \Sigma\of{\sigma,\tau}
  \end{eqnarray}
  with
  \begin{eqnarray}
    \Sigma\of{0,\tau} &=& s \in U
    \nonumber\\
    \Sigma\of{1,\tau} &=& t \in U 
    \nonumber  
  \end{eqnarray}
  for given $s,t\in U$,
  which is constant in a neighborhood of 
  $\sigma = 0,1$
  and independent of $\tau$ near $\tau = 0,1$.

  Equivalently, a parametrized bigon is a path in path space
  $\paths_s^t\of{U}$ \refdef{based path space}
  \begin{eqnarray}
    \Sigma \maps [0,1] &\to& \paths_s^t\of{U}
   \nonumber\\
      \tau &\mapsto& \Sigma\of{\cdot,\tau}   
    \,,
      \nonumber
  \end{eqnarray} 
  which is constant in a neighborhood of $\tau = 0,1$.
  We call $s$ the {\bf source vertex} of the bigon, $t$ the 
  {\bf target vertex}, $\Sigma\of{\cdot,0}$ the {\bf source edge}
  and $\Sigma\of{\cdot,1}$ the {\bf target edge}.
\end{definition}

As with paths, the parametrization involved here is ultimately
not of interest and should be divided out:

\begin{definition}\et
  \label{bigon}
  An {\bf unparametrized bigon} or simply a {\bf bigon}
  is a thin homotopy equivalence class $[\Sigma]$ of 
  parametrized bigons
  $\Sigma$ \refdef{parametrized bigon}.
\end{definition}

More in detail, this means (\cf for instance 
\cite{MackaayPicken:2000} p.26 and \cite{BaezLauda:2003} p.50) 
that two parametrized bigons 
$\Sigma_1, \Sigma_2 \maps [0,1]^2 \to U$ 
are taken to be equivalent
\[
  \Sigma_1 \sim \Sigma_2
\]
precisely if there exists a smooth map
\[
  H \maps [0,1]^3 \to U
\]
which takes one bigon smoothly into the other while preserving their boundary,
i.e. such that
\begin{eqnarray}
  H\of{\sigma,\tau,0} &=& \Sigma_1\of{\sigma,\tau}
  \nonumber\\
  H\of{\sigma,\tau,1} &=& \Sigma_2\of{\sigma,\tau}
  \nonumber\\
  H\of{\sigma,0,\nu} &=& \Sigma_1\of{\sigma,0} = \Sigma_2\of{\sigma,0}
  \nonumber\\
  H\of{\sigma,1,\nu} &=& \Sigma_1\of{\sigma,1} = \Sigma_2\of{\sigma,1}
  \nonumber\\
  H\of{0,\tau,\nu} &=& \Sigma_1\of{0,\tau} = \Sigma_2\of{0,\tau}
  \nonumber\\
  H\of{1,\tau,\nu} &=& \Sigma_1\of{1,\tau} = \Sigma_2\of{1,\tau}
  \,,
  \nonumber
\end{eqnarray}
but which does so in a degenerate fashion, meaning that
\[
  \mathrm{rank}\of{dH}\of{\sigma,\nu,\tau} < 3
\]
for all $\sigma,\tau,\nu \in [0,1]$.

These bigons naturally form a coherent 2-groupoid:

\begin{definition}\et
  \label{2-groupoid of bigons}
  The {\bf path 2-groupoid} 
  $\P_2\of{U}$
  of a manifold $U$ is the 2-groupoid whose
  \begin{itemize}
    \item 
      objects are points $x\in U$
    \item 
      morphisms are 
      paths $\gamma \in \paths_x^y\of{U}$
     \[
      \xymatrix{
       x \ar@/^1pc/[rr]^{\gamma} 
       && y 
      }
      \]
    \item
      2-morphisms       are bigons 
      \refdef{bigon} with source edge $\gamma_1$ and target
      edge $\gamma_2$
 \[
\xymatrix{
   x \ar@/^1pc/[rr]^{\gamma_1}_{}="0"
           \ar@/_1pc/[rr]_{\gamma_2}_{}="1"
           \ar@{=>}"0";"1"^{{}_{[\Sigma_1]}}
&& y
}
\]
  \end{itemize}
  and whose composition operations are defined as
  \begin{itemize}
   \item 
\[
\xymatrix{
   x \ar@/^1pc/[rr]^{\gamma_1}
&& y \ar@/^1pc/[rr]^{\gamma_2}
&& z
}
  =
\xymatrix{
   x \ar@/^1pc/[rr]^{\gamma_1 \circ \gamma_2}
&& z 
}
\]

\item
\[
\xymatrix{
   x \ar@/^2pc/[rr]^{\gamma_1}_{}="0"
           \ar[rr]^<<<<<<{\gamma_2}_{}="1"
           \ar@{=>}"0";"1"^{[\Sigma_1]}
           \ar@/_2pc/[rr]_{\gamma_3}_{}="2"
           \ar@{=>}"1";"2"^{[\Sigma_2]}
&& y
}
=
\xymatrix{
   x \ar@/^1pc/[rr]^{\gamma_1}_{}="0"
           \ar@/_1pc/[rr]_{\gamma_3}_{}="1"
           \ar@{=>}"0";"1"^{{}_{[\Sigma_1\circ \Sigma_2]}}
&& y
}
\]

\item
\[
\xymatrix{
   x \ar@/^1pc/[rr]^{\gamma_1}_{}="0"
           \ar@/_1pc/[rr]_{\gamma_1^\prime}_{}="1"
           \ar@{=>}"0";"1"_{[\Sigma_1]}
&& y \ar@/^1pc/[rr]^{\gamma_2}_{}="2"
           \ar@/_1pc/[rr]_{\gamma_2^\prime}_{}="3"
           \ar@{=>}"2";"3"_{[\Sigma_2]}
&& z
}
 =
\xymatrix{
   x \ar@/^1pc/[rr]^{\gamma_1\circ \gamma_2}_{}="0"
           \ar@/_1pc/[rr]_{\gamma_1^\prime\circ\gamma_2^\prime}_{}="1"
           \ar@{=>}"0";"1"_{{}_{[\Sigma_1\cdot \Sigma_2]}}
&& z 
}
\]
  \end{itemize}
  where
  \begin{eqnarray}
    \left(\gamma_1\circ\gamma_2\right)\of{\sigma}
    &\defas&
    \left\lbrace
      \begin{array}{cl}
         \gamma_1\of{2\sigma} & \mbox{for $0 \leq \sigma \leq 1/2$}\\
         \gamma_2\of{2\sigma-1} & \mbox{for $1/2\leq \sigma \leq 1$}
      \end{array}
    \right.
  \nonumber\\
    \left(\Sigma_1\circ\Sigma_2\right)\of{\sigma,\tau}
    &\defas&
    \left\lbrace
      \begin{array}{cl}
         \Sigma_1\of{\sigma,2\tau} & \mbox{for $0 \leq \tau \leq 1/2$}\\
         \Sigma_2\of{\sigma,2\tau-1} & \mbox{for $1/2\leq \tau \leq 1$}
      \end{array}
    \right.
  \nonumber\\
    \left(\Sigma_1\cdot\Sigma_2\right)\of{\sigma,\tau}
    &\defas&
    \left\lbrace
      \begin{array}{cl}
         \Sigma_1\of{2\sigma,\tau} & \mbox{for $0 \leq \sigma \leq 1/2$}\\
         \Sigma_2\of{2\sigma-1,\tau} & \mbox{for $1/2\leq \sigma \leq 1$}
      \end{array}
    \right.  
    \,.
    \nonumber
  \end{eqnarray}
\end{definition}

Note that in this definition we did \emph{not} divide out by thin
homotopy of parametrized \emph{paths} 
but only by thin homotopy of parametrized bigons. This implies that
the horizontal composition in this 2-groupoid is \emph{not}
associative. But one can check that 
the above indeed is a coherent 2-groupoid where associativity is \emph{weakly} 
preserved in a \emph{coherent} fashion, as described in 
\cite{BaezLauda:2003}.

Namely there are degenerate bigons for which $\mathrm{rank}\of{d\Sigma} \leq 1$,
whose vertical composition with any other bigon 
has the effect of applying a thin homotopy to that bigon's source or
target edges. Therefore associativity of horizontal composition of bigons holds
up to vertical composition with such degenerate bigons and hence up to natural
isomorphism.

\vskip 2 em

\subsubsection{$p$-Holonomy $p$-Functors}
\label{p-holonomy p-functors}

Our definition of 2-connection arises from categorifying the following
definition of a 1-connection in a principal 1-bundle, which is equivalent
to any of the other familiar definitions.

\begin{definition}
\label{definition of 1-connection in 1-bundle in terms of local holonomy}
   A {\bf (1-)connection} in a locally trivializable
   principal $G$-(1-)bundle $E \to B$ is the collection of the 
   following data:
   \begin{itemize}
     \item 
        A good covering $\covering = \bigsqcup\limits_{i \in I} U_i$ of $B$,
     \item 
       for each $i \in I$ a smooth functor
\[
  \begin{array}{rcccc}
    \hol_i &\maps& \P_1\of{U_i} &\to& G
    \\
    &&
\xymatrix{
   x \ar@/^1pc/[rr]^{\gamma}="0"
&& y
}
&\mapsto& 
\xymatrix{
   \bullet \ar@/^1pc/[rr]^{W_i[\gamma]}="0"
&& \bullet
}
  \end{array}
 \,,
\]
  called the {\bf local holonomy functor},
 from the groupoid $\P_1\of{U_i}$ of paths in $U_i$ 
  \refdef{groupoid of paths}, 
  to the structure
  group $G$ (regarded as a category with a single object),

\item
  for each $i,j\in I$ a natural isomorphism
  \[
    \hol_i|_{U_{ij}} \stackto{g_{ij}} \hol_j|_{U_{ij}}
    \,,
  \]
  i.e. a commuting diagram
\begin{center}
\begin{picture}(100,100)
  \includegraphics{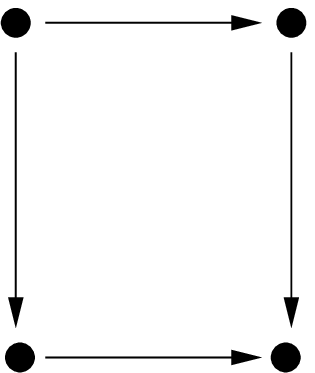}
  \put(-60,108){$g_{ij}\of{x}$}
  \put(-60,-5){$g_{ij}\of{y}$}
  \put(-115,54){$W_i[\gamma]$}
  \put(-3,54){$W_j[\gamma]$}
\end{picture}
\vskip 1em
\end{center}
(for $x \stackto{\gamma} y$ a path in $U_{ij}$)
  called the {\bf transition natural isomorphisms}
  between the local holonomy functors $\hol_i$ and $\hol_j$
  restricted to $U_{ij}$,

\item
  and for each $i,j,k \in I$ a commuting diagram
\begin{center}
 \begin{picture}(140,120)
  \includegraphics{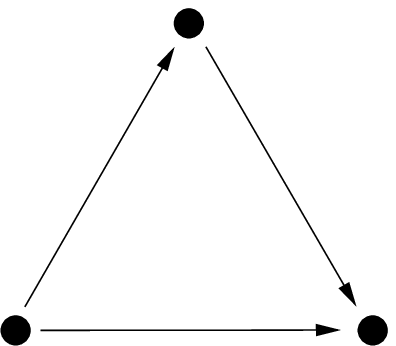}
  \put(-134,0){$\hol_i$}
  \put(-65,105){$\hol_j$}
  \put(4,0){$\hol_k$}
  \put(-100,50){$g_{ij}$}
  \put(-25,50){$g_{jk}$}
  \put(-70,-5){$g_{ik}$}
  \end{picture}
\vskip 1em
\end{center}
  equating the transition isomorphisms 
  $g_{ik}$ and $g_{ij}\circ g_{jk}$
  restricted to $U_{ijk}$
  \,.
\end{itemize}

\end{definition}

From this one obtains the familiar result that

  \begin{enumerate}
    \item
      The local holonomy functors $\hol_i$ are specified by
      1-forms 
      \[
        A_i \in \Omega^1\of{U_i,\mathrm{Lie}\of{G}}\,.
      \]
    \item
      The transition isomorphisms $g_{ij}$ 
      are specified by the transition functions
      \[
        g_{ij} \maps U_{ij} \to G\,,
      \] 
      satisfying the equation
      \[
        A_i = g_{ij}A_j g_{ij}^{-1} + g_{ij} \extd g_{ij}^{-1}
        \,, \hspace{1cm} \mbox{on $U_{ij}$}
        \,.
      \]
   \item
    The identity between these natural isomorphisms on triple overlaps
    $U_{ijk}$ is
    equivalent to the equation
    \[
      g_{ik} = g_{ij}g_{jk}\,, \hspace{1cm}\mbox{on $U_{ijk}$}
      \,.
    \]
  \end{enumerate}

What we are after is a categorification of this situation. 
This leads to the following definition.

\begin{definition}
  \label{def: 2-connection in a 2-bundle}  

   A {\bf 2-connection}
   in a locally trivializable
   principal $\twogroup$-2-bundle $E \to M$ 
   which admits a {\bf 2-holonomy}
   is the collection of the 
   following data:
   \begin{itemize}
     \item 
        A good covering $\covering = \bigsqcup\limits_{i \in I} U_i$ of $M$,
     \item 
       for each $i \in I$ a smooth 2-functor
\begin{eqnarray}
  \label{local holonomy 2-functor}
  \begin{array}{rcccc}
    \hol_i &\maps& \P_2\of{U_i} &\to& G_2
    \\
    &&
\xymatrix{
   x \ar@/^1pc/[rr]^{\gamma}="0"
           \ar@/_1pc/[rr]_{}_{\gamma'}="1"
           \ar@{=>}"0";"1"^{\Sigma}
&& y
}
&\mapsto& 
\xymatrix{
   \bullet \ar@/^1pc/[rr]^{\hol_i\of{\gamma}}="0"
           \ar@/_1pc/[rr]_{}_{\hol_i\of{\gamma'}}="1"
           \ar@{=>}"0";"1"^{\hol_i\of{\Sigma}}
&& \bullet
}
  \end{array}
  \,,
\end{eqnarray}
  called the {\bf local holonomy 2-functor},
 from the 2-groupoid $\P_2\of{U_i}$ of 2-paths in $U_i$ 
 (def. \ref{2-groupoid of bigons}, p. \pageref{2-groupoid of bigons})
 to the structure
  2-group $\twogroup$ (regarded as a 2-category with a single object),

\item
  for each $i,j\in I$ a pseudo-natural isomorphism
  \[
    \hol_i|_{U_{ij}} \stackto{g_{ij}} \hol_j|_{U_{ij}}
    \,,
  \]
  i.e. a 2-commuting diagram
\begin{eqnarray}
  \label{transition pseudo-natural iso}
\begin{picture}(240,140)
 \includegraphics{tincan.eps}
 \put(-3,-17){	
 \begin{picture}(0,0)
 \put(122,0){
  \begin{picture}(0,0)
   \put(-385,70){$\hol_i\of{\gamma_2}$}
   \put(-293,70){${}_{\hol_i\of{\gamma_1}}$}
   \put(-333,100){$\hol_i\of{\Sigma}$}
   \put(-290,106){$a_{ij}\of{\gamma_2}$}
   \put(-250,118){${}_{a_{ij}\of{\gamma_1}}$}
   \put(-250,163){$g_{ij}\of{x}$}
   \put(-250,10){$g_{ij}\of{y}$}
  \end{picture}
 }
 \put(288,0){
  \begin{picture}(0,0)
   \put(-385,70){$\hol_j\of{\gamma_2}$}
   \put(-293,70){${}_{\hol_j\of{\gamma_1}}$}
   \put(-333,100){$\hol_j\of{\Sigma}$}
  \end{picture}
 }
 \end{picture}
}
\end{picture}
\end{eqnarray}

  called the {\bf transition pseudo-natural isomorphisms}
  between the local holonomy 2-functors $\hol_i$ and $\hol_j$
  restricted to $U_{ij}$,

\item
  for each $i,j,k \in I$ a modification of pseudonatural 
  transformations
\begin{center}
 \begin{picture}(140,110)
  \includegraphics{triangle.eps}
  \put(-139,0){$\hol_i$}
  \put(-70,105){$\hol_j$}
  \put(2,0){$\hol_k$}
  \put(-100,50){$g_{ij}$}
  \put(-25,50){$g_{jk}$}
  \put(-70,-5){$g_{ik}$}
  \put(-70,24){$f_{ijk}$}
  \end{picture}
  \vskip 1em
\end{center}
i.e. a 2-commuting diagram
\begin{eqnarray}
  \label{transition modification}
\begin{picture}(240,190)
 \includegraphics{modification.eps}
 \put(0,20){\begin{picture}(0,0)
 \put(-3,-17){	
 \begin{picture}(0,0)
 \put(122,0){
  \begin{picture}(0,0)
   \put(-387,70){$\hol_i\of{\gamma_2}$}
   \put(-293,70){${}_{\hol_i\of{\gamma_1}}$}
   \put(-336,100){$\hol_i\of{\Sigma}$}
   \put(-290,106){$a_{ik}\of{\gamma_2}$}
   \put(-250,118){${}_{a_{ij}\cdot a_{jk}\of{\gamma_1}}$}
   \put(-284,149){$g_{ik}\of{x}$}
   \put(-256,183){${}_{g_{ij}\circ g_{jk}\of{x}}$}
   \put(-230,158){${}_{f_{ijk}\of{x}}$}
  \end{picture}
 }
 \put(125,-139){\begin{picture}(0,0)
   \put(-284,149){$g_{ik}\of{y}$}
   \put(-256,183){${}_{g_{ij}\circ g_{jk}\of{y}}$}
   \put(-230,158){${}_{f_{ijk}\of{y}}$}
 \end{picture}}
 \put(288,0){
  \begin{picture}(0,0)
   \put(-388,70){$\hol_k\of{\gamma_2}$}
   \put(-293,70){${}_{\hol_k\of{\gamma_1}}$}
   \put(-336,100){$\hol_k\of{\Sigma}$}
  \end{picture}
 }
 \end{picture}
}
\end{picture}}
\end{picture}
\end{eqnarray}

  between the transition pseudonatural isomorphisms 
  ${g}_{ik}$ and ${g}_{ij}\circ {g}_{jk}$
  restricted to $U_{ijk}$,
 
 \item
   for each $i,j,k,l \in I$ an identity
\begin{eqnarray}
  \label{transition tetrahedron}
\begin{picture}(300,130)
\includegraphics{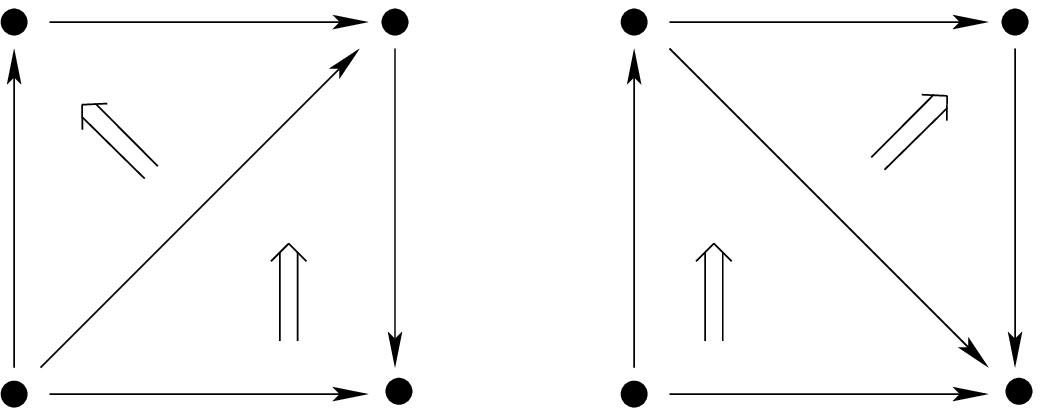}
\put(-7,2){
\begin{picture}(300,150)
\put(-245,115){${g}_{jk}$}
\put(-65,115){${g}_{jk}$}
\put(-245,-5){${g}_{il}$}
\put(-65,-5){${g}_{il}$}
\put(-307,55){${g}_{ij}$}
\put(-127,55){${g}_{ij}$}
\put(-180,55){${g}_{kl}$}
\put(-1,55){${g}_{kl}$}
\put(-152,55){$=$}
\put(-280,30){${g}_{ik}$}
\put(-40,21){${g}_{jl}$}
\put(-58,80){$f_{jkl}$}
\put(-82,30){$f_{ijl}$}
\put(-234,30){$f_{ikl}$}
\put(-254,77){$f_{ijk}$}
\put(-313,-9){${}_{\hol_i}$}
\put(-313,118){${}_{\hol_j}$}
\put(-178,-9){${}_{\hol_l}$}
\put(-178,118){${}_{\hol_k}$}
\put(183,0){\begin{picture}(0,0)
\put(-313,-9){${}_{\hol_i}$}
\put(-313,118){${}_{\hol_j}$}
\put(-178,-9){${}_{\hol_l}$}
\put(-178,118){${}_{\hol_k}$}
\end{picture}}
\end{picture}
}
\end{picture}
\end{eqnarray}
between these modifications,
expressing the 2-commutativity of tetrahedra of the form
\begin{eqnarray*}
\begin{picture}(180,150)
  \includegraphics{3dtetrahedron.eps}
  \put(-117,56){${}_{g_{ij}}$}
  \put(-63,56){${}_{g_{jk}}$}
  \put(-90,-3){$g_{ik}$}
  \put(-100,22){$f_{ijk}$}
  \put(-137,80){$g_{il}$}
  \put(-41,80){$g_{kl}$}
  \put(-82,95){${}_{g_{jl}}$}
  \put(-193,-3){$\hol_i$}
  \put(3,-3){$\hol_k$}
  \put(-82,67){${}_{\hol_j}$}
  \put(-82,150){${}_{\hol_l}$}
\end{picture}
\end{eqnarray*}

\end{itemize}

\end{definition}

In analogy to the situation for 1-connections in 1-bundles,
one would like to have an equivalent expression of this definition of
a 2-connection in a 2-bundle in terms of differential forms.
This is what we will work out in what follows. The result is this:

\begin{proposition}
\label{central proposition concerning 2-connections}

A 2-connection in a 2-bundle as defined in def. 
\refer{def: 2-connection in a 2-bundle} is 
expressible in terms of differential forms as follows:

\begin{enumerate}

\item
  
  The local holonomy 2-funtor $\hol_i$ is specified by 
  two differential forms
  \begin{eqnarray*}
     A_i \in \Omega^1\of{U_i,\g}
     \\
     B_i \in \Omega^2\of{U_i,\h}
  \end{eqnarray*}
  satisfying
  \begin{eqnarray}
    \label{vanishing of fake curvature in central prop conc 2-connections}
    F_{A_i} + dt\of{B_i} = 0
    \,,
  \end{eqnarray}
  where $F_{A_i}$ is the curvature 2-form of $A_i$.

  The simple idea behind the proof for this is sketched in
  \S\fullref{subsubsec: Definition of Local 2-Holonomy on Single Overlaps}.
  The full proof is the content of 
  \S\fullref{section: Local 2-Holonomy from local Path Space Holonomy}.

 \item
   The transition pseudo-isomorphism  
  $\hol_i \stackto{g_{ij}} \hol_j$
  is specified by
   \begin{eqnarray*}
     g_{ij} &\in& \Omega^0\of{U_{ij}, G}
     \\
     a_{ij} &\in& \Omega^1\of{U_{ij}, \h}
   \end{eqnarray*}
   satisfying the equations
  \begin{eqnarray}
    \label{equation: the transition cocycle condition}
    A_i &=& g_{ij}A_jg_{ij}^{-1} + g_{ij} \extd g_{ij}^{-1}
    \nonumber\\
    B_i &=& \alpha\of{g_{ij}}\of{B_i} + k_{ij}
  \end{eqnarray}
  on $U_{ij}$,
  where
  \[
    k_{ij} = \extd a_{ij} + a_{ij}\wedge a_{ij} + d\alpha\of{A_i}\wedge a_{ij}
    \,.
  \]
  The proof for this is
  given in \S\fullref{subsubsec: Transition Law on Double Overlaps}.
  Again, the idea is quite simple, but the proof has to make use of
  some facts only developed in \S\fullref{section: Path Space}.

 \item
  The modification 
  $g_{ik} \stackto{f_{ijk}} g_{ij}\circ g_{jk}$ 
  is specified by
  \[
    f_{ijk} \in \Omega^0\of{U_{ijk}, H}
  \]
  satisfying the equation
  \begin{eqnarray}
   \label{equation: the cocycle condition on triple overlaps}
      a_{ij} + g_{ij}\of{a_{jk}}
    -
    f_{ijk} \,a_{ik}\, f_{ijk}^{-1}
    -
    f_{ijk}\, \extd h^{-1}_{ijk}
    -
    f_{ijk}\, d\alpha\of{A_i}(f_{ijk}^{-1})
    = 0
\end{eqnarray}
on $U_{ijk}$.

 This is proven in \S\fullref{subsubsec: Transition Law on Triple Overlaps}.

\item
  The equation between four of these modifications
  on quadruple overlaps is the already familiar tetrahedron law
  \[
   \alpha\of{g_{ij}}\of{f_{jkl}} \, f_{ijl} 
  =
  f_{ijk} \, f_{ikl}
  \,.
  \]
  This was discussed before in 
\S\fullref{section: 2-Transitions in Terms of Local Data}.

\end{enumerate}

\end{proposition}

\newpage
\subsection{2-Holonomy in Terms of Local $p$-Forms}
\label{section 2-connections in terms of local p-forms}

In this subsection the proof of the central proposition
\ref{central proposition concerning 2-connections}
(p. \pageref{central proposition concerning 2-connections}) 
is sketched in a way that is supposed to
clearly point out the underlying mechanisms in a concise way. Several
technicalities that these proofs rely on are then discussed in
detail in \S\fullref{section: Path Space}.

\subsubsection{Definition on Single Overlaps}
\label{subsubsec: Definition of Local 2-Holonomy on Single Overlaps}

Consider any bigon $\Sigma$ in a patch $U_i$, 
i.e. a 2-morphism in $\P_2\of{U_i}$
\refdef{2-groupoid of bigons}, and consider a local 2-holonomy functor
$\hol_i \maps \P_2\of{U_i} \to \twogroup$
\refdef{def: 2-connection in a 2-bundle}. 
Since $\hol_i$ is a functor,
the 2-group 2-morphism 
which it associates to $\Sigma$ can be computed by dividing $\Sigma$
into many small sub-bigons, evaluating $\hol_i$ on each of these and
composing the result in $\twogroup$. This is illustrated in the following
sketchy figure.

\begin{center}
\begin{picture}(430,220)
\includegraphics{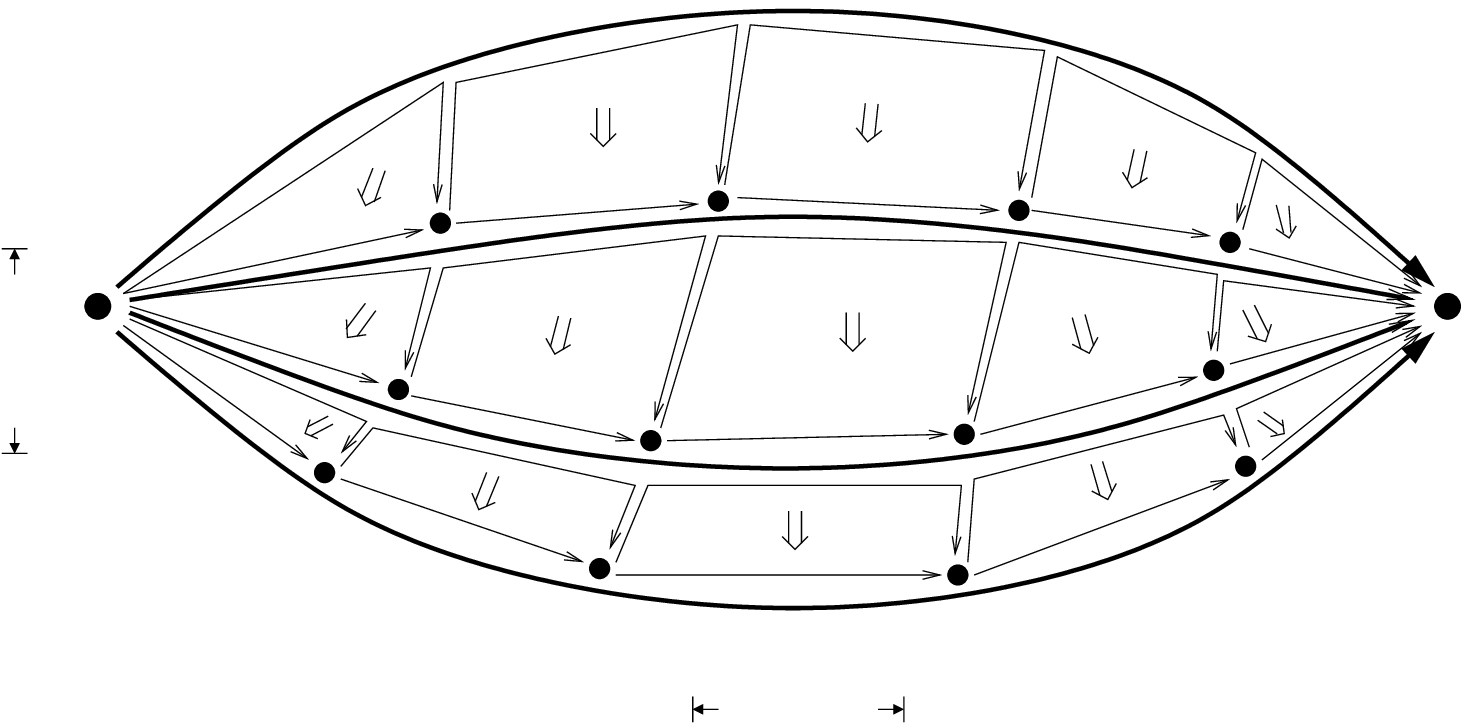}
 \put(0,33){\begin{picture}(0,0)
\put(-309,95){${}_{g_1}$}
\put(-332,83){${}_{h_1}$}
\put(-270,98){${}_{g_2}$}
\put(-274,79){${}_{h_2}$}
\put(-190,104){${}_{g_3}$}
\put(-190,79){${}_{h_3}$}
\put(-110,99){${}_{g_4}$}
\put(-122,79){${}_{h_4}$}
\end{picture}}
\put(-425,106){$\approx \epsilon$}
\put(-197,2){$\approx \epsilon$}
\end{picture}
\end{center}

Here the $j$-th 2-morphism is supposed to be given by 
\[
  \hol\of{\Sigma_j} = (g_j,h_j) \in \twogroup
\]
with $g \in G$ and $h\in H$.
By the rules of 2-group multiplication 
(Prop.\ \ref{liecrossedmodule})
the total horizontal product 
\[
  (g^\mathrm{tot},h^{\mathrm{tot}})
  \defas
  (g_1,h_1)\cdot (g_2,h_2) \cdot (g_3,h_3) \cdots
\]
of all these 2-morphisms
is given by
\begin{eqnarray}
  g^\mathrm{tot} &=& g_1\, g_2\, g_3 \cdots g_N
  \nonumber\\
  h^\mathrm{tot} &=& h_1 \,\alpha\of{g_1}\of{h_2}\, \alpha\of{g_1g_2}\of{h_3}
  \cdots \alpha\of{g_1g_2g_3\cdots g_{N-1}}\of{h_N}
  \,.
  \nonumber
\end{eqnarray}
The products of the $g_j$ can be addressed as a 
\emph{path holonomy} along the 
upper edges, which, for reasons to become clear shortly, we shall write as
\[
  g_1 \, g_2 \cdots g_{j} \defas (W_{j+1})^{-1}
  \,.
\]
Now suppose the group elements come from algebra elements $A_j \in \g$ and $B_j \in \h$
as
\begin{eqnarray}
  \label{connection 1- and 2-form motivated from discretization}
  g_j &\defas& \exp\of{\epsilon A_j}
  \nonumber\\
  h_j &\defas& \exp\of{\epsilon^2 B_j}
\end{eqnarray}
where
\[
  \epsilon \defas 1/N
  \,,
\]
then
\[
  h^\mathrm{tot}
  =
  1
  +
  \epsilon^2
  \sum\limits_{j=1}^N
  \alpha\of{W^{-1}_j}\of{B_j}
  +
  \order{\epsilon^4}
  \,.
\]
Using the notation
\begin{eqnarray}
  W_j &\defas& W\of{1-j\epsilon,1}
  \nonumber\\
  B_j &\defas& B\of{1-\epsilon j}
  \nonumber
\end{eqnarray}
we have
\begin{eqnarray}
  h^\mathrm{tot}
  &=&
  1+
  \epsilon \,
  \int\limits_0^1 d\sigma\;
  \alpha\of{W^{-1}\of{\sigma,1}}\of{B\of{\sigma}}
  + 
  \order{\epsilon^3}
  \,.
  \nonumber
\end{eqnarray}
Finally, imagine that the $\twogroup$-labels $h^{\mathrm{tot}}_k$ of 
many such thin horizontal rows of `surface elements' 
are composed \emph{vertically}. Each of them comes from algebra elements
\[
  B_k\of{\sigma} \defas B\of{\sigma,k\epsilon}
\]
and holonomies
\[
  W_k\of{\sigma,1} \defas W_{k\epsilon}\of{\sigma,1}
\]
as
\[
  h_k^\mathrm{tot}
  \defas
  1+
  \epsilon \,
  \int\limits_0^1 d\sigma\;
  \alpha\of{W_{k\epsilon}^{-1}\of{\sigma,1}}\of{B\of{\sigma,k\epsilon}}
  + 
  \order{\epsilon^3}  
  \,.
\]
In the limit of vanishing $\epsilon$ their total vertical product is
\begin{eqnarray}
  \lim\limits_{\epsilon = 1/N \to 0}
  h^\mathrm{tot}_0 h^\mathrm{tot}_\epsilon h^\mathrm{tot}_{2\epsilon}
  \cdots h^\mathrm{tot}_1
  &=&
  \mathrm{P}
  \exp\of{
    \int\limits_0^1
      d\tau\;
      \mathcal{A}\of{\tau}
  }
  \nonumber
\end{eqnarray}
for
\begin{eqnarray}
  \label{motivation for standard path space connection}
  \mathcal{A}\of{\tau} 
  &=& 
  \int\limits_0^1 d\sigma\; \alpha\of{W^{-1}_\tau\of{\sigma,1}}\of{B\of{\sigma,\tau}}
  \,,
\end{eqnarray}
where $\mathrm{P}$ denotes path ordering with respect to $\tau$.

Thinking of each of these vertical rows of surface elements as 
paths (in the limit $\epsilon \to 0$), this shows roughly how the computation
of total 2-group elements from vertical and horizontal products of many
`small' 2-group elements can be reformulated as the holonomy of a connection
on path space of the form \refer{motivation for standard path space connection}. 
This way the local 2-holonomy functor $\hol_i$ comes from a 1-form
$A_i \in  \Omega^1\of{U_i,\g}$ and a 2-form
$B_i \in \Omega^2\of{U_i,\h}$ that arise as the continuum limit of the
construction in 
\refer{connection 1- and 2-form motivated from discretization}.
This is made precise in 
\S\fullref{section: Local 2-Holonomy from local Path Space Holonomy}.

There it is discussed that given a bigon 
$\gamma \stackto{\Sigma} \tilde \gamma$ the 2-group morphism
\begin{eqnarray}
  \label{first definition of components of hol(Sigma)}
  \hol_i\of{\Sigma} = (W_i[\gamma_1] \in G, \; 
  \mathcal{W}_i^{-1}[\Sigma] \in H)
\end{eqnarray}
is obtained from the holonomy $W_i[\gamma]$ of $A_i$ along 
$\gamma$ and the
inverse of the path space holonomy 
$\mathcal{W}_i^{-1}[\Sigma]$
of $\mathcal{A}_{(A_i,B_i)}$
along a path in path space that maps to $\Sigma$.

But not every pair $(A,B)$ corresponds to a local holonomy-functor.
As first noticed in \cite{GirelliPfeiffer:2004} there is a consistency
condition which can be understood as follows: 

Let $g_j \stackto{h_j} g'_j$ be the $j$th 2-group 2-morphism
in the above figure.
The nature of 2-groups (prop. \ref{liecrossedmodule}) 
requires that
\[
  t\of{h_j} = g'_j g_j^{-1}
  \,.
\]
But, in the above sense, the left hand side is given by 
$\exp\of{\epsilon^2 dt\of{B}_j}$, while the right hand side
is $\approx \exp\of{-\epsilon F_{A_j}}$, where $F_{A_j}$ denotes the
curvature 2-form of $A$ evaluated on a 2-vector tangent to $\Sigma_j$.
Hence we get the condition
\[
  dt\of{B} + F_A = 0 
  \,.
\]
This is the content of 
prop. \ref{pre-2-holonomy asscoiates 2-group elements} 
(p. \pageref{pre-2-holonomy asscoiates 2-group elements}).
See also prop. 
\ref{effect of infinitesimal gauge transformations on path space}
(p. \pageref{effect of infinitesimal gauge transformations on path space}).

\subsubsection{Transition Law on Double Overlaps}
\label{subsubsec: Transition Law on Double Overlaps}

\begin{proposition}\et
\label{proposition stated again: transition of 2-holonomy}
  The 2-commutativity of the diagram
  \refer{transition pseudo-natural iso} 
  (p. \pageref{transition pseudo-natural iso})
  is equivalent to the equations
  \refer{equation: the transition cocycle condition}
    \begin{eqnarray*}
    A_i &=& g_{ij}A_jg_{ij}^{-1} + g_{ij} \extd g_{ij}^{-1} 
           - dt\of{a_{ij}}
    \\
    B_i &=& \alpha\of{g_{ij}}\of{B_i} + k_{ij}
    \,.
  \end{eqnarray*}

\end{proposition}

\Proof

The 2-commutativity of the diagram is equivalent to the equality
of the 2-morphism on its left face with the composition of the 
2-morphisms on the front, back and right faces:
\begin{eqnarray}
\begin{picture}(400,130)
\includegraphics{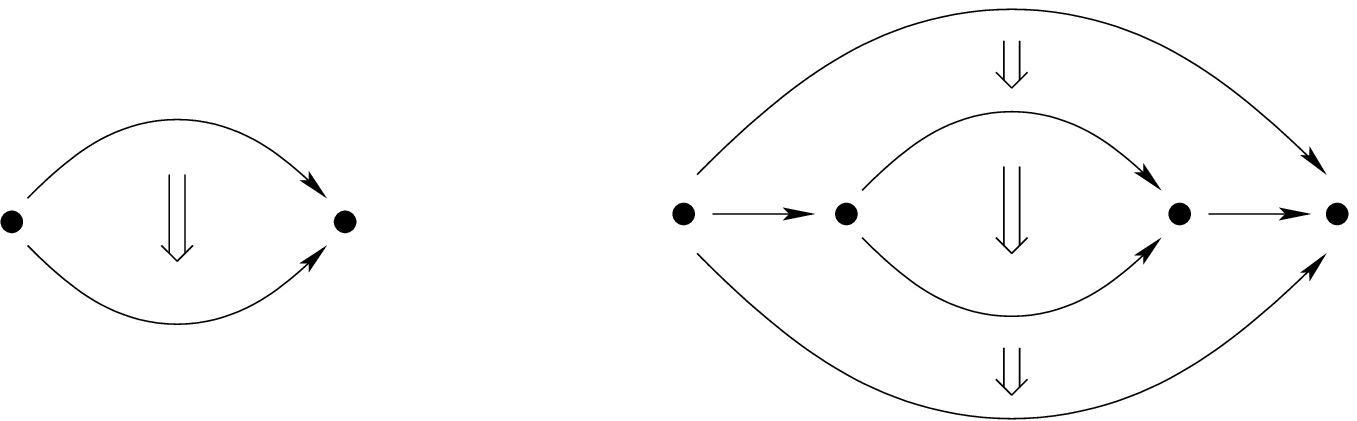}
\put(-334,58){${}_{\hol_i\of{\Sigma}}$}
\put(-94,58){${}_{\hol_j\of{\Sigma}}$}
\put(-348,94){${}_{\hol_i\of{\gamma}}$}
\put(-348,20){${}_{\hol_i\of{\tilde \gamma}}$}
\put(-149,86){${}_{\hol_j\of{\gamma}}$}
\put(-143,32){${}_{\hol_j\of{\tilde \gamma}}$}
\put(-180,67){${}_{g_{ij}\of{x}}$}
\put(-42,68){${}_{g_{ij}^{-1}\of{y}}$}
\put(-92,15){${}_{\bar a_{ij}\of{\tilde \gamma}}$}
\put(-92,104){${}_{a_{ij}\of{\gamma}}$}
\put(-250,55){$=$}
\put(-110,124){${}_{\hol_i\of{\gamma}}$}
\put(-110,-8){${}_{\hol_i\of{\tilde \gamma}}$}
\end{picture}
  \label{diagram version of 2-connection transition equation}
\end{eqnarray}

Recall from \refer{first definition of components of hol(Sigma)}
that $\hol_i\of{\Sigma}$ has source $\hol_i\of{\gamma} = W_i[\gamma]$.
So we write 
\[
  a_{ij}\of{\gamma}
  \defas
  \left(
    W_i[\gamma]\in G,\; E\of{a_{ij}}[\gamma] \in H
  \right)
  \,,
\]
where
\[
  a_{ij} \in \Omega^1\of{U_{ij},\h}
\]
is a 1-form (which we find convenient to denote by the same symbol
as the 2-morphism $a_{ij}\of{\gamma}$ that it is associated with) 
and where $E$ is a
function whose nature is to be determined by the source/target matching
condition. This says that
\begin{eqnarray}
  t\of{E\of{a_{ij}[\gamma]}}
  W_{i}[\gamma]
  &=&
  g_{ij}\of{x} W_{j}[\gamma] g_{ij}^{-1}\of{y}
  \,.
\end{eqnarray}
Expressions like this are handled by 
prop.\ 
\ref{difference in holonomies wrt different connections}
(p. \pageref{difference in holonomies wrt different connections}).
In order to apply it conveniently we take the inverse on both sides
to get
\begin{eqnarray}
  W_{i}[\gamma^{-1}]
  t\of{E\of{a_{ij}}[\gamma]}^{-1}
  &=&
  g_{ij} W_{j}[\gamma^{-1}] g_{ij}^{-1}
\end{eqnarray}
(using $W[\gamma^{-1}] = W^{-1}[\gamma]$).
Then the proposition tells us that
$t\of{E\of{a_{ij}}[\gamma]}^{-1}$ is of the form
\begin{eqnarray}
  t\of{E\of{a_{ij}}[\gamma]}^{-1}
  &=&
  \lim\limits_{\epsilon = 1/N \to 0}
  \left(
    1 + \epsilon\oint\limits_{A_i}\of{\alpha}
  \right)
  \left(
    1 + \epsilon\oint\limits_{A_i + \epsilon \alpha}\of{\alpha}
  \right)  
  \cdots
  \left(
    1 + \epsilon\oint\limits_{A_i + (1-\epsilon)a^1_{ij}}\of{\alpha}
  \right)_{|\gamma^{-1}}
  \,,
  \nonumber\\
\end{eqnarray}
where the right hand side is evaluated at $\gamma^{-1}$,
and where $\alpha \in \Omega^1\of{U_{ij},\g}$ is given by
\[
  \alpha 
   \defas
  g^1_{ij}(\extd+A_j)(g^1_{ij})^{-1}
  -
  A_i
  \,.
\]
The 1-form $\alpha$ must take values in the image of $dt$, 
and it is the
corresponding pre-image which we denote by $a_{ij}$, so that
$dt\of{a_{ij}} = \alpha$:
\begin{eqnarray}
  \label{A transition law from 2-transition}
  dt\of{a_{ij}} 
  &=&
  g^1_{ij}(d+A_j)(g^1_{ij})^{-1}
  -
  A_i
  \,.
\end{eqnarray}
This is the first of the two equations to be derived. 

It follows that $E\of{a_{ij}}[\gamma]$ itself is given by
\begin{eqnarray}
  (E\of{a_{ij}}[\gamma])^{-1}
  &=&
  \lim\limits_{\epsilon = 1/N \to 0}
  \left(
    1 + \epsilon\oint\limits_{A_i}\of{a_{ij}}
  \right)
  \left(
    1 + \epsilon\oint\limits_{A_i + \epsilon dt\of{a_{ij}}}\of{a_{ij}}
  \right)  
  \cdots
  \left(
    1 + \epsilon\oint\limits_{A_i + (1-\epsilon)dt\of{a_{ij}}}\of{a_{ij}}
  \right)_{|\gamma^{-1}}
  \,.
  \nonumber
\end{eqnarray}

Now that we have determined the 2-morphism $E\of{a_{ij}}[\gamma]$,
we can evaluate the diagrams in equation 
\refer{diagram version of 2-connection transition equation}. 
Recalling again
equation \refer{first definition of components of hol(Sigma)},
one sees that
the equality of the 
2-morphism on the left hand with that on the right means that
\[
  \mathcal{W}^{-1}_i\of{\Sigma}
  =
  (E\of{a_{ij}}[\tilde \gamma])^{-1} 
  \mathcal{W}^{-1}_j\of{\Sigma}
  E\of{a_{ij}}[\gamma] 
  \,.
\]
This is nothing but a gauge transformation of path space
holonomy. Using
Prop.\ \ref{effect of finite path space gauge transformations}
it implies the second of two equations to be proven.
\endofproof

Note that in the case that the 2-group in question is the abelian
one given by the crossed module 
$(G=1, H = U\of{1},\alpha = \mathrm{trivial}, t = \mathrm{trivial})$
the formula for $E\of{a_{ij}}[\gamma]$ 
reduces simply to the line holonomy of
$a_{ij}$ along $\gamma$:
\begin{eqnarray}
  \label{aij in the abelian case}
  E\of{a_{ij}}[\gamma]
  =
  \exp\of{\int_\gamma a_{ij}}
  \,.
\end{eqnarray}

\subsubsection{Transition Law on Triple Overlaps}
\label{subsubsec: Transition Law on Triple Overlaps}

\begin{proposition}

  The 2-commutativity of the diagram
  \refer{transition modification}
  (p. \pageref{transition modification})
  is equivalent to the equation
  \refer{equation: the cocycle condition on triple overlaps}
  (p. \pageref{equation: the cocycle condition on triple overlaps})
    \[
      a_{ij} + g_{ij}\of{a_{jk}}
    -
    f_{ijk} \,a_{ik}\, f_{ijk}^{-1}
    -
    f_{ijk}\, \extd f^{-1}_{ijk}
    -
    f_{ijk}\, d\alpha\of{A_i}(f_{ijk}^{-1})
    = 0
   \,.
\]

\end{proposition}

\Proof
Since our target category $\twogroup$ is a strict 2-group, so that
(when regarded as a 2-category with a single object)
all of its 1- and 2-morphisms 
are invertible, the diagram 
\refer{transition modification}
expressing the modifications on $U_{ijk}$ 
can be simplified. Using the transition diagram
\refer{transition pseudo-natural iso} we can equate the
composition of the 2-morphisms $\hol_i\of{\Sigma}$ and 
$\hol_k\of{\Sigma}$ as well as the 2-morphisms 
$a_{ik}[\tilde \gamma]$ on the front of this diagram with the
single 2-morphism $a_{ik}[\gamma]$ and hence get rid
of the dependency on $\tilde \gamma$ and $\Sigma$:
\begin{center}
\begin{picture}(210,185)
 \includegraphics{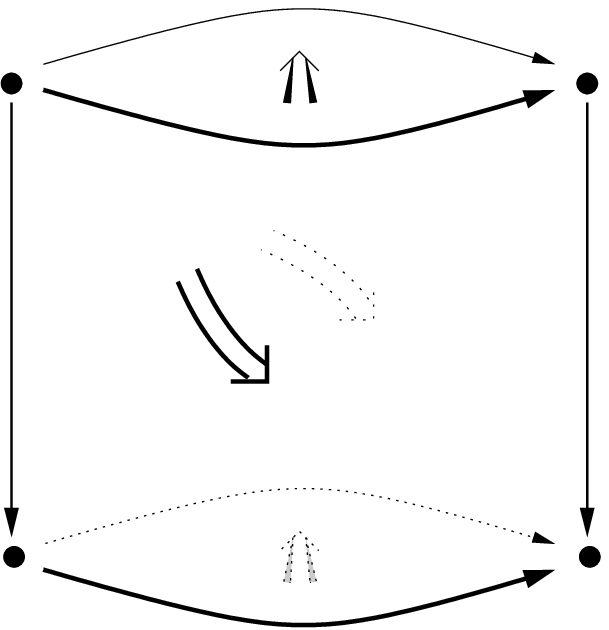}
 \put(23,20){\begin{picture}(0,0)
 \put(-3,-17){	
 \begin{picture}(0,0)
 \put(122,0){
  \begin{picture}(0,0)
   \put(-355,80){$\hol_i\of{\gamma}$}
   \put(-290,106){$a_{ik}\of{\gamma}$}
   \put(-250,118){${}_{a_{ij}\cdot a_{jk}\of{\gamma}}$}
   \put(-284,149){$g_{ik}\of{x}$}
   \put(-256,183){${}_{g_{ij}\circ g_{jk}\of{x}}$}
   \put(-230,158){${}_{f_{ijk}\of{x}}$}
  \end{picture}
 }
 \put(125,-139){\begin{picture}(0,0)
   \put(-284,149){$g_{ik}\of{y}$}
   \put(-256,183){${}_{g_{ij}\circ g_{jk}\of{y}}$}
   \put(-230,158){${}_{f_{ijk}\of{y}}$}
 \end{picture}}
 \put(288,0){
  \begin{picture}(0,0)
   \put(-318,80){$\hol_k\of{\gamma}$}
  \end{picture}
 }
 \end{picture}
}
\end{picture}}
\end{picture}
\end{center}
In order to emphasize the structure of this diagram it is useful
to make the triangular shape of the top and bottom explicit:
\begin{center}
\begin{picture}(200,240)
\includegraphics{2concoherence.eps}
  \put(-3,100){$\hol_k\of{\gamma}$}
  \put(-196,100){$\hol_i\of{\gamma}$}
  \put(-82,160){${}_{\hol_j\of{\gamma}}$}
  \put(-100,-7){$g_{ik}\of{y}$}
  \put(-100,187){$g_{ik}\of{x}$}
  \put(-138,43){${}_{g_{ij}\of{y}}$}
  \put(-138,237){${}_{g_{ij}\of{x}}$}
  \put(-47,43){${}_{g_{jk}\of{y}}$}
  \put(-47,237){${}_{g_{jk}\of{x}}$}
  \put(-95,23){${}_{f_{ijk}\of{y}}$}
  \put(-95,215){${}_{f_{ijk}\of{x}}$}
  \put(-100,90){$a_{ik}\of{\gamma}$}
  \put(-150,124){${}_{a_{ij}\of{\gamma}}$}
  \put(-50,117){${}_{a_{jk}\of{\gamma}}$}
\end{picture}
\end{center}
\begin{eqnarray}
   \label{simplified transition isomorphism}
\end{eqnarray}
The 2-commutativity of this diagram 
is equivalent to the following equality between the 2-morphism
obtained from its top, bottom and front face and the 2-morphism
obtained from the two faces on the back:
\begin{center}
\begin{picture}(300,300)
  \includegraphics{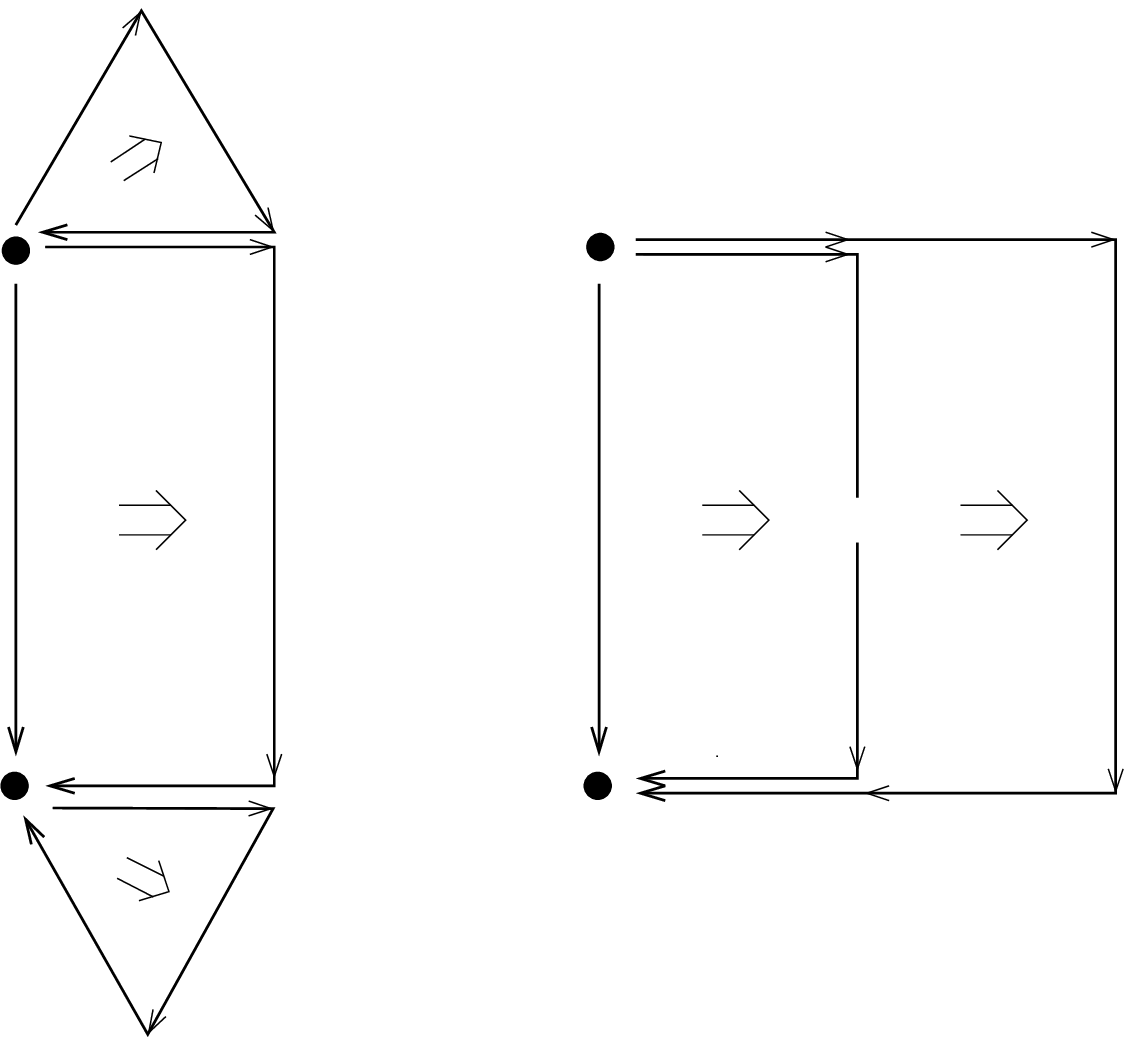}
  \put(-295,133){${}_{a_{ik}\of{\gamma}}$}
  \put(-345,150){${}_{W_i[\gamma]}$}
  \put(-244,150){${}_{W_k[\gamma]}$}
  \put(-300,245){${}_{f_{ijk}\of{x}}$}
  \put(-328,265){${}_{g_{ij}\of{x}}$}
  \put(-262,265){${}_{g_{jk}\of{x}}$}
  \put(-293,222){${}_{g_{ik}\of{x}}$}
  \put(-293,82){${}_{g_{ik}^{-1}\of{y}}$}
  \put(-262,38){${}_{g_{jk}^{-1}\of{y}}$}
  \put(-330,38){${}_{g_{ij}^{-1}\of{y}}$}
  \put(-300,58){${}_{f_{ijk}^{-1}\of{y}}$}
  \put(-205,147){$=$}
  \put(-123,238){${}_{g_{ij}\of{x}}$}
  \put(-56,238){${}_{g_{jk}\of{x}}$}
  \put(-123,65){${}_{g_{ij}^{-1}\of{y}}$}
  \put(-56,65){${}_{g_{jk}^{-1}\of{y}}$}
  \put(-175,150){${}_{W_i[\gamma]}$}
  \put(-90,150){${}_{W_j[\gamma]}$}
  \put(-2,150){${}_{W_k\of{\gamma}}$}
  \put(-126,135){${}_{a_{ij}\of{\gamma}}$}
  \put(-50,135){${}_{a_{jk}\of{\gamma}}$}
\end{picture}
\end{center}

In terms of group elements
this means that
\[
  f_{ijk}\of{x}
  \; 
  E\of{a_{ik}}[\gamma]
  \; 
  \alpha\of{W_i[\gamma]}\of{f_{ijk}^{-1}\of{y}}
  =
  \alpha\of{g_{ij}\of{x}}\of{E\of{a_{jk}}[\gamma]} 
  \;
  E\of{a_{ij}}[\gamma]  
  \,.
\]
Now expand around the point $x$ to get the differential version of this
statemment:
\begin{eqnarray*}
  W_i[\gamma]
  &\approx&
  1
  +
  \epsilon A_i\of{\gamma'}
  \\
  E\of{a_{ij}}[\gamma]
  &\approx&
  1
  +
  \epsilon a_{ij}\of{\gamma'}
  \\
  f_{ijk}^{-1}\of{y}
  &\approx&
  f_{ijk}^{-1}\of{x} 
  +
  \epsilon (\extd (f_{ijk}^2)^{-1})\of{\gamma'}\of{x} 
  \,.
\end{eqnarray*}
(Here $\gamma = \frac{d}{d\sigma}\gamma\of{0}$ 
is the tangent vector to $\gamma$ at $x= \gamma\of{0}$.)
Substituting this into the above equation and collecting terms of
first order in $\epsilon$ yields the promised equation.
\endofproof

\subsection{Path Space}
\label{section: Path Space}

As we have seen, the space of all paths in a manifold or more
general smooth spaces constitutes a smooth space
it itself.  In particular,
we study the notion of holonomy for curves in path space.
A curve in path space over $U$ maps to a (possibly degenerate) 
surface in $U$ and hence its path space holonomy gives rise to
a notion of surface holonomy in $U$.

In this section we first discuss
basic concepts of differential geometry on path spaces
and then apply them to define path space holonomy.
Using that,
a 2-functor $\hol_i$ from the 2-groupoid of bigons in $U_i$
(to be defined below)
to the structure 2-group is defined and shown to be consistent.

Throughout the following, various $p$-forms taking values in 
Lie algebras $\g$ and $\h$ are used, where $\g$ and $\h$ are
part of a differential crossed module
$\crossedmodule$ \refdef{differential crossed module}.

Elements of a basis of $\g$ will be denoted by $T_a$
with $a \in (1,\dots,\mathrm{dim}\of{\g})$
and those of a basis of $\h$ by 
$S_a$ with $a \in (1,\dots,\mathrm{dim}\of{\h})$. Arbitrary
elements will be expanded as $A = A^a T_a$.	

  Given a $\g$-valued 1-form $A$ its {\bf gauge covariant exterior derivative}
  is
  \begin{eqnarray}
    \extd_A \omega &\defas& 
    \commutator{\extd + A}{\omega}
    \nonumber\\
    &\defas&
    \extd \omega + A^a \wedge d\alpha\of{T_a}\of{\omega}
    \nonumber
  \end{eqnarray}
  and its {\bf curvature} is
  \begin{eqnarray}
    F_A 
    &\defas&
    (\extd + A)^2
    \nonumber\\
    &\defas&
    \extd A + \frac{1}{2}A^a \wedge A^b \commutator{T_a}{T_b}
    \,.
    \nonumber
  \end{eqnarray}
  By
  a {\bf $\crossedmodule$-valued (1,2)-form} on a manifold 
  $U$ we shall mean a pair $(A,B)$ with
  \begin{eqnarray}
    \label{definition 1,2 form}
    A &\in&  \Omega^1\of{U,\g}
    \nonumber\\
    B &\in&  \Omega^2\of{U,\h}  
   \,.
  \end{eqnarray}

Differential calculus on spaces of \emph{parametrized} paths can 
be handled rather easily.  We start by establishing some basic facts on
parametrized paths and then define the \emph{groupoid of paths} by
considering thin homotopy equivalence classes of parametrized paths. 

\begin{definition}\et
  \label{based path space}
  Given a manifold $U$,
  the {\bf based parametrized path space}
  $\paths_s^t\of{U}$
  over $U$ with source $s \in U$ and target $t \in U$
is the space of smooth maps  
\begin{eqnarray}
  X : [0,1] &\to& U
  \nonumber\\
  \sigma &\mapsto& X\of{\sigma}
\end{eqnarray} 
which are constant in a neighborhood of $\sigma = 0$ and in a neighborhood of
 $\sigma = 1$.
  When source and target coincide 
  \[
  \Omega_x\of{U}
  \defas
  \paths_x^x\of{U} 
  \]
  is the {\bf based loop space} over $U$ based at $x$.
\end{definition}

The constancy condition at the boundary is known as the property of having
{\bf sitting instant}, compare for instance \cite{CaetanoPicken:1993}.
It serves in def. \ref{groupoid of paths} to ensure that the
composition of two smooth parametrized paths is again a smooth
parametrized path.

  In the study of differential forms on 
  parametrized path space the following notions play an
  important role (\cf \cite{GetzlerJonesPetrack:1991}, section 2):

\begin{definition}\et
  \label{notions on path space}
  \begin{enumerate}
  \item
  Given any path space $\paths_s^t\of{U}$ \refdef{based path space}, 
  the 1-parameter family of maps
  \begin{eqnarray}
    e_\sigma : \paths_s^t\of{U} &\to& U
    \hspace{1cm}(\sigma \in (0,1))
    \nonumber\\
     \gamma &\mapsto& \gamma\of{\sigma}
     \nonumber
  \end{eqnarray}
  maps each path to its position in $U$ at parameter value $\sigma$.

  \item
  Given any differential $p$-form $\omega \in \Omega^p\of{U}$
  the pullback to $\paths_s^t\of{U}$ by $e_\sigma$ shall be denoted simply  by
  \begin{eqnarray}
    \omega\of{\sigma}
    &\defas&
    e^\ast_\sigma\of{\omega}
    \,.
    \nonumber
  \end{eqnarray}

  \item
  The contraction of $\omega\of{\sigma}$ with the vector
  \[
    \gamma' \defas \frac{d}{d\sigma}\gamma
  \]
   is denoted by $\iota_{\gamma'} \omega\of{\sigma}$.
\end{enumerate} 
\end{definition}

A special class of differential forms on path space play a major role:

\begin{definition}\et
  \label{Chen forms}
  Given a familiy $\set{\omega_i}_{i=1}^N$ of differential forms on a manifold $U$
  with degree
  \begin{eqnarray}
    \mathrm{deg}\of{\omega_i} &\defas& p_i +1
    \nonumber
  \end{eqnarray}
  one gets a differential form (\cf \refdef{notions on path space})
  \begin{eqnarray}
    \Omega_{\set{\omega_i},(\alpha,\beta)}\of{\gamma}
    &\defas&
    \oint\limits_{X|_{\alpha}^{\beta}} (\omega_1, \dots, \omega_n)
    \;\defas\;
    \int\limits_{\alpha < \sigma_i < \sigma_{i+1} < \beta}
    \!\!\!\!\!\!
    \iota_{\gamma'} \omega_1\of{\sigma^1}
    \wedge
    \cdots
    \wedge
    \iota_{\gamma'} \omega_N\of{\sigma^N}
    \nonumber
  \end{eqnarray}
  of degree
  \begin{eqnarray}
    \mathrm{deg}\of{\Omega_{\set{\omega_i}}} &=& \sum\limits_{i=1}^N p_i
    \,,
    \nonumber
  \end{eqnarray}
  on any based parametrized path space $\paths_s^t\of{U}$
  \refdef{based path space}.

  For $\alpha=0$ and $\beta = 1$ we write
  \begin{eqnarray}
    \Omega_{\set{\omega_i}}
    &\defas&
    \Omega_{\set{\omega_i},(0,1)}
    \,.
    \nonumber
  \end{eqnarray}
  These path space forms are known as
  {\bf multi integrals} or {\bf iterated integrals}
  or {\bf Chen forms} (\cf \cite{GetzlerJonesPetrack:1991,Hofman:2002}).
\end{definition}

It turns out that the exterior derivative on path space maps
Chen forms \refdef{Chen forms} to Chen forms in a nice way:

\begin{proposition}\et
  \label{action of extd on Chen forms} 
  The action of the path space exterior derivative 
  on Chen forms \refdef{Chen forms} is
\begin{eqnarray}
  \label{extd on path space Chen forms}
  \extd \oint (\omega_1, \cdots,\omega_n)
  &=&
  (\tilde \extd + \tilde M)
  \oint (\omega_1, \cdots,\omega_n)
  \,,
\end{eqnarray}
where
\begin{eqnarray}
  \tilde \extd \oint (\omega_1, \cdots,\omega_n)
  &\defas&
  -\sum\limits_k (-1)^{\sum\limits_{i < k}p_i}
    \oint (\omega_1, \cdots, \extd\omega_k, \cdots, \omega_n)
  \nonumber\\
  \tilde M \oint (\omega_1, \cdots,\omega_n)
  &\defas&
  -\sum\limits_k (-1)^{\sum\limits_{i < k}p_i}
    \oint (\omega_1,\cdots, \omega_{k-1}\wedge \omega_k,\cdots,\omega_n)
  \,,
  \nonumber
\end{eqnarray}
satisfying
\begin{eqnarray}
  \tilde \extd^2 &=& 0
  \nonumber\\
  \tilde M^2 &=& 0
  \nonumber\\
  \antiCommutator{\tilde \extd^2}{\tilde M^2} &=& 0
  \,. 
\end{eqnarray}
\end{proposition}
(\cf \cite{GetzlerJonesPetrack:1991,Hofman:2002})

\subsubsection{The Standard Connection 1-Form on Path Space}
\label{section: The Standard Connection 1-Form on Path Space}

There are many 1-forms on path space that one could consider as
local connection 1-forms in order to define a local holonomy on
path space. Here we restrict attention to a special class, to be called
the \emph{standard connection 1-forms} 
\refdef{standard path space 1-form}, because, as is shown 
in \S\fullref{section: Local 2-Holonomy from local Path Space Holonomy},
these turn out to be the ones which compute local 2-group holonomy.
(This same `standard connection 1-form' can however also be motivated
from other points of view, as done in 
\cite{AlvarezFerreiraSanchezGuillen:1998,Schreiber:2004e}.)

\paragraph{Holonomy and parallel transport.}

In order to set up some notation and conventions and for later references, 
the following gives a list of well-known 
definitions and facts that are crucial for the further developments:

\begin{definition}\et
  \label{line holonomy and parallel transport}
  Given a path space $\paths_s^t\of{U}$ \refdef{based path space}
  and a 
  $\crossedmodule$-valued (1,2)-form $(A,B)$ 
  \refer{definition 1,2 form}
  on $U$, the following objects are of interest:
  \begin{enumerate}
    \item
      The {\bf line holonomy} of $A$ along a given path $\gamma$ is denoted by
      \begin{eqnarray}
        \label{def line holonomy}
        W_A[\gamma]{\of{\sigma^1,\sigma^2}} 
        &\defas&
        \mathrm{P}\exp\of{\int\limits_{\gamma|_{\sigma^1}^{\sigma^2}} A}
        \nonumber\\
        &\defas&
        \sum\limits_{n=0}^{\infty}
        \oint\limits_{\gamma|_{\sigma^1}^{\sigma^2}}
        (A^{a_1},\dots ,A^{a_n})
        T_{a_1}\cdots T_{a_n}
        \,.
      \end{eqnarray}

   \item
    The {\bf parallel transport} of elements in $T \in \g$ and 
    $S\in \h$
   is written
\begin{eqnarray}
  \label{notions of parallel tranport}
  T^{W_A[\gamma]}\of{\sigma}
  &\defas&
  W^{-1}_A[\gamma|_\sigma^1]T\of{\sigma}W_A[\gamma]\of{\sigma,1}
  \nonumber\\
  &=&
  \sum_{n=0}^\infty
   \oint_{\gamma|_\sigma^1} (-A^{a_1},\cdots,-A^{a_n})
   \;
   \commutator{T_{a_n}}{\cdots \commutator{T_{a_1}}{T\of{\sigma}}\cdots}     
  \,,
  \nonumber\\
  S^{W_A[\gamma]}\of{\sigma}
  &\defas&
  \alpha\of{W_A[\gamma|_\sigma^1]}\of{S\of{\sigma}}
  \nonumber\\
  &\defas&
  \sum_{n=0}^\infty
   \oint_{\gamma|_\sigma^1} (-A^{a_1},\cdots,-A^{a_n})
   \;
   d\alpha(T_{a_n})\circ \cdots \circ d\alpha\of{T_{a_1}}\of{S\of{\sigma}}    
  \,.
  \nonumber\\
\end{eqnarray}
  \end{enumerate}
\end{definition}

For convenience the dependency $[\gamma]$ on the path $\gamma$ will often be omitted.

\begin{proposition}\et
  \label{properties of parallel transport}
  Parallel transport \refdef{line holonomy and parallel transport} has the following
  properties:
  \begin{enumerate}
   \item 
   Let $\sigma_1 \leq \sigma_2 \leq \sigma_3$
   then
   \begin{eqnarray}
     W_A[\gamma]\of{\sigma_1,\sigma_2} \circ W_A[\gamma]\of{\sigma_2,\sigma_3}  
     &=&
     W_A[\gamma]\of{\sigma_1,\sigma_3}
     \,.
     \nonumber
   \end{eqnarray}
   \item
  Conjugation of elements in $\g$ with parallel tranport of elements
in $\h$ yields
\begin{eqnarray}
  \label{conjugation of g by holonomies}
  W_A\of{\sigma,1}\of{
    d\alpha\of{T}\of{\sigma}
    \of{
      W_A^{-1}\of{\sigma,1}
      \of{
        S
      }
    }
  }
  &=&
  d\alpha\of{T^{W_A}\of{\sigma}}\of{S}
  \,.
\end{eqnarray}

 \item
  Given a $G$-valued 0-form
  $
    g \in \Omega^0\of{U,G} 
  $
  and a path $\gamma \in \paths_x^y\of{U}$
  we have
  \begin{eqnarray}
    \label{gauge trafo of parallel transport}
    g\of{x}W_A[\gamma](g\of{y})^{-1}
    &=&
    W_{(gAg^{-1} + g^{-1}\extd g)}[\gamma]
    \,.
  \end{eqnarray}

  \item
  Given a $G$-valued 0-form
  $
    g \in  \Omega^0\of{U,G} 
  $
  and a based loop $\gamma \in \paths_x^x\of{U}$
  we have
  \begin{eqnarray}
     \label{action of phi on parallel transport}
     \alpha\of{\phi\of{x}}\of{W_A[\gamma]\of{\sigma,1}\of{S\of{\sigma}}}
     &=&
     W_{A^\prime}[\gamma]\of{\sigma,1}\of{\alpha\of{\phi\of{\gamma\of{\sigma}}}\of{S\of{\sigma}}}
  \end{eqnarray}
  with
  \begin{eqnarray}
    A^\prime &\defas& \phi A \phi^{-1} + \phi (d\phi^{-1})
    \,.
    \nonumber
  \end{eqnarray}
  \end{enumerate}
\end{proposition}

Integrals over $p$-forms pulled back to a path and 
parallel tranported to some base point
play an important role for path space holonomy. Following
\cite{Hofman:2002,Schreiber:2004e} 
we introduce special notation to take care
of that automatically:

\begin{definition}\et
  \label{notation Chen form with parallel transport}
  A natural addition to the notation \refer{Chen forms}
  for iterated integrals in the presence of a $\g$-valued
  1-form $A$ is the abbreviation
  \begin{eqnarray}
    \oint_A\of{
      \omega_1,\dots\omega_N
    }
    &\defas&
    \oint\of{
      \omega_1^{W_A},\dots \omega_N^{W_A}
    }
    \,,
    \nonumber
  \end{eqnarray}
  where $(\cdot)^{W_A}$ is defined in def \ref{line holonomy and parallel transport}.
  When Lie algebra indices are displayed on the left they are defined to 
pertain to the parallel tranported object:
  \begin{eqnarray}
    \label{special notation in Chen form with parallel transport}
    \oint_A (\dots,\omega^a,\dots)
    &\defas&
    \oint (\dots,(\omega^{W_A})^a,\dots)
    \,.
  \end{eqnarray}
\end{definition}

Using this notation first of all the following fact can be conveniently 
stated, which plays a central role 
in the analysis of the 
transition law for the 2-holonomy in 
\S\fullref{subsubsec: Transition Law on Double Overlaps}:

\begin{proposition}\et
  \label{difference in holonomies wrt different connections}
  The difference in line holonomy \refdef{line holonomy and parallel transport}
  along a given loop with respect to two different 1-forms $A$ and $A^\prime$
  can be expressed as
  
  \hspace{-3cm}\parbox{20cm}{
  \begin{eqnarray}
    (W_{A}[\gamma])^{-1}W_{A^\prime}[\gamma] 
    &=&
    \lim\limits_{\epsilon = 1/N \to 0}
    \left(
      1 + \epsilon\oint\limits_{A} (\alpha)
    \right)
    \left(
      1 + \epsilon\oint\limits_{A+\epsilon(\alpha)} (\alpha)
    \right)
    \cdots
    \left(
      1 + \epsilon\oint\limits_{A^\prime- \epsilon (\alpha)} (\alpha)
    \right)_\gamma
    \,,
    \nonumber
  \end{eqnarray}
}
with $\alpha \defas A' - A$.

\end{proposition}
\Proof

First note that from def. \ref{line holonomy and parallel transport} 
it follows that
\begin{eqnarray}
  \oint\limits_A \of{\alpha}
  &=&
  \int\limits_{0}^1
  d\sigma
  (W_A[\gamma](\sigma,1))^{-1} \iota_{\gamma'}\alpha\of{\sigma}
  W_A[\gamma](\sigma,1)
  \,.
  \nonumber
\end{eqnarray}
This implies that
\begin{eqnarray}
  W_A[\gamma]\left(1 + \epsilon \oint_{A}\of{\alpha}\right)_\gamma
  &=&
  W_{A+\epsilon (\alpha)}[\gamma] + \order{\epsilon^2}
  \,.
  \nonumber
\end{eqnarray}
The proposition follows by iterating this.
\endofproof

\paragraph{Exterior derivative and curvature for Chen forms.}

The exterior derivative on path space maps Chen forms to Chen forms
(Prop.\ \ref{action of extd on Chen forms}). Since we shall be interested 
in Chen forms involving parallel transport 
\refdef{notation Chen form with parallel transport}, it is important
to know also the particular action of the exterior derivative on these:

\begin{proposition}\et
  \label{extd on A-Chen forms of one form}
  The action of the path space exterior derivative on
  $\oint_A\of{\omega}$ is
  \begin{eqnarray}
    \label{path space extd on Chen forms with parallel transport}
    \extd \oint_A (\omega)
    &=&
    -\oint_A (\extd_A \omega)
    -
    (-1)^{\mathrm{deg}\of{\omega}}
    \oint_A\of{
      d\alpha\of{T_a}\of{\omega}, F_A^a
    }
    \,.
  \end{eqnarray}
\end{proposition}
(Recall the convention \refer{special notation in Chen form with parallel transport}).

\Proof

This is a straightforward, though somewhat tedious, computation
using prop \ref{action of extd on Chen forms}.
\endofproof

We have restricted attention here to just a single insertion, i.e.
$\oint_A\of{\omega}$ instead of $\oint_A \of{\omega_1,\dots,\omega_n}$,
because this is the form that the
\emph{standard connection 1-form} has:

\begin{definition}\et
  \label{standard path space 1-form}
  Given a $\crossedmodule$-valued $(1,2)$-form \refer{definition 1,2 form} 
  the path space 1-form
\begin{eqnarray}
   \Omega^1\of{\paths_s^t\of{U},\h}
  \ni
  \;
  \mathcal{A}_{(A,B)}
  &\defas&
  \oint_A (B)
  \,.
  \nonumber
\end{eqnarray}
  is here called the
  {\bf standard local connection 1-form on path space}.
\end{definition}
(\cf \cite{AlvarezFerreiraSanchezGuillen:1998,Chepelev:2002,Schreiber:2004e})

Given a connection, one wants to know its curvature:

\begin{corollary}
  \label{corollary: curvature of standard path space 1-form}
  The curvature of the standard path space 1-form $\mathcal{A}_{(A,B)}$
  \refdef{standard path space 1-form}
  is
  \begin{eqnarray}
    \label{curvature of standard path space 1-form}
    \mathcal{F}_\mathcal{A}
   &=&
    -
    \oint_A \of{\extd_A B}
    -
    \oint_A\of{d\alpha\of{T_a}\of{B},(F_A+ dt\of{B})^a}
    \,.
  \end{eqnarray}
\end{corollary}

\Proof 
 Use Prop.\ \ref{extd on A-Chen forms of one form}. \endofproof

\begin{definition}\et
  \label{curvature and fake curvature}
  Given a standard path space connection 1-form $\mathcal{A}_{(A,B)}$
  \refdef{standard path space 1-form} coming from a $\g$-valued
  1-form $A$ and an $\h$-valued 2-form $B$
\begin{itemize}
  \item
     the 3-form 
     \begin{eqnarray}
        \label{definition curvature 3-form}
        H \defas \extd_A B
     \end{eqnarray}
     is called the {\bf curvature 3-form},
   \item
     the 2-form
     \begin{eqnarray}
       \label{definition fake curvature}
       \tilde F \defas F_A + dt\of{B}
     \end{eqnarray}
     is called the {\bf fake curvature 2-form}.
\end{itemize}
\end{definition}
The term `fake curvature' has been introduced in \cite{BreenMessing:2001}.
The notation $\tilde F$ follows \cite{GirelliPfeiffer:2004}. The 
curvature 3-form was used in \cite{Baez:2002}.

Using this notation the local path space curvature reads
\begin{eqnarray}
  \label{path space curvature in fake flat case}
  \mathcal{F}_\mathcal{A}
  &=&
  - \oint_A \of{H} - \oint_A \of{ d\alpha(T_a)\of{B},\tilde F^a}
  \,.
\end{eqnarray}

\subsubsection{Path Space Line Holonomy and Gauge Transformations}

With the usual tools of differential geometry available for
path space 
the holonomy on path space is defined as usual:

\begin{definition}\et
  \label{def path space holonomy}
  Given a path space 1-form $\mathcal{A}$ and a curve $\Sigma$ in path space the 
  {\bf path space line holonomy} of $\mathcal{A}$ along $\Sigma$ is
  \begin{eqnarray}
    \mathcal{W}_{\mathcal{A}}\of{\Sigma}
    &\defas&
    \mathrm{P} \exp\of{
      \int_\Sigma \mathcal{A}
    }
    \,.
    \nonumber
  \end{eqnarray}
\end{definition}

Note that by definition P here indicates path ordering with objects at 
higher parameter value to the 
\emph{right} of those with lower parameter value, just as in the definition of
ordinary line holonomy in \refdef{line holonomy and parallel transport}.

Path space line holonomy has a richer set of gauge transformations
than holonomy on base space. In fact, ordinary gauge transformations
on base space correspond to \emph{constant} (`global') gauge transformations
on path space in the following sense:

\begin{proposition}\et
  \label{target space gauge trafos on path space holonomy}
  Given a path space line holonomy
  \refdef{def path space holonomy}
  coming from a standard path space connection 1-form \refdef{standard path space 1-form}
  $\mathcal{A}_{(A,B)}$
  in a based loop space $\paths_x^x\of{U}$ 
  as well as a 
  $G$-valued 0-form $\phi \in \Omega^0\of{U,G}$
  we have
  \begin{eqnarray}
    \alpha\of{\phi\of{x}}\of{
      \mathcal{W}_{\mathcal{A}_{(A,B)}}\of{\Sigma}
    }
    &=&
      \mathcal{W}_{\mathcal{A}_{(A^\prime,B^\prime)}}\of{\Sigma}    
    \nonumber
  \end{eqnarray}
  with
  \begin{eqnarray}
    A^\prime &=& \phi A \phi^{-1} + \phi (d\phi^{-1}) 
    \nonumber\\
    B^\prime &=& \alpha\of{\phi}\of{B}
    \,.
    \nonumber
  \end{eqnarray}
\end{proposition}

\Proof
Write out the path space holonomy in infinitesimal steps and apply
\refer{action of phi on parallel transport} on each of them.
\endofproof

The usual notion of gauge transformation is obtained by conjugation:

\begin{definition}\et
  \label{infinitesimal path space holonomy gauge transformation}
  Given the path space holonomy $\mathcal{W}_{\mathcal{A}_{(A,B)}}\of{\Sigma|_{\gamma_0}^{\gamma_1}}$ 
  \refdef{def path space holonomy}
  of a standard local path space connection 1-form $\mathcal{A}_{(A,B)}$
  \refdef{standard path space 1-form} along a curve $\Sigma$ in
  $\paths_s^t\of{U}$ with endpaths
  $\gamma_0$ and $\gamma_1$,
  an {\bf infinitesimal path space holonomy gauge transformation}
 is a 1-parameter familiy maps
  \begin{eqnarray*}
    \mathcal{W}_{\mathcal{A}_{(A,B)}}\of{\Sigma|_{\gamma_0}^{\gamma_1}}
    &\mapsto&
    \left(
      1 - \epsilon \oint_A \of{a}
    \right)_{\gamma_0}
    \mathcal{W}_{\mathcal{A}_{(A,B)}}\of{\Sigma|_{\gamma_0}^{\gamma_1}}
    \left(
      1 + \epsilon \oint_A \of{a}
    \right)_{\gamma_1}
    \\
    &&\defas
    \mathrm{Ad}_{\gamma_0}^{\gamma_1}
    \left(
      1 - \epsilon \oint_A \of{a}
    \right)
    \of{
    \mathcal{W}_{\mathcal{A}_{(A,B)}}\of{\Sigma|_{\gamma_0}^{\gamma_1}}
    }
    \,,
  \end{eqnarray*}  
  for $\epsilon \in \R$ and for $a$ any 1-form 
  \begin{eqnarray}
    a &\in& \Omega^1\of{U,\h}
    \,.
    \nonumber
  \end{eqnarray}
\end{definition}

This 
yields a new sort of gauge transformation in terms of the 
target space (1,2) form $(A,B)$:

\begin{proposition}\et
  \label{effect of infinitesimal gauge transformations on path space}
  Infinitesimal path space holonomy gauge transformations
  \refdef{infinitesimal path space holonomy gauge transformation}
  for the holonomy of a standard path space connection 1-form
  $\mathcal{A}_{(A,B)}$ and arbitrary transformation parameter $a$
  yields to first order in the parameter $\epsilon$ the path space holonomy
  of a transformed standard path space connection 1-form
  $\mathcal{A}_{(A^\prime,B^\prime)}$ with
  \begin{eqnarray}
    \label{liner transformation of A and B under path space gauge trafos}
    A^\prime &=& A + dt\of{a}
    \nonumber\\
    B^\prime &=& B - \extd_A a
  \end{eqnarray}
  if and only if the fake curvature 
  \refdef{curvature and fake curvature}
  vanishes. 
\end{proposition}
(This was originally considered in \cite{Schreiber:2004e} for the special case $G=H$, 
$t = \mathrm{id},\, \alpha = \mathrm{Ad}$.)

\Proof

As for any holonomy, the gauge transformation induces a transformation of the
connection 1-form $\mathcal{A} \to \mathcal{A}^\prime$ given by
\begin{eqnarray}
  \mathcal{A}^\prime
  &=&
  \left(
    1- \epsilon\oint_A \of{a}
  \right)
  \left(\extd + \mathcal{A}\right)
  \left(
    1 + \epsilon\oint_A \of{a}
  \right)
  \nonumber\\
  &=&
  \mathcal{A} 
  + 
  \epsilon\; \extd_\mathcal{A}\oint_A\of{a}
  +
  \order{\epsilon^2}
  \,.
\end{eqnarray}
Using \refer{path space extd on Chen forms with parallel transport} one finds
(using the notation \refer{special notation in Chen form with parallel transport})
\begin{eqnarray}
  \mathcal{A} + 
  \epsilon\;\extd_{\mathcal{A}}\oint_A (a)
  &=&
  \oint_{A^\prime}\of{B^\prime}
    +
  \epsilon
    \oint_A
    \of{
      d\alpha\of{T_a}\of{a},
      (dt\of{B} + F)^a
    }
  + \order{\epsilon^2}
  \,.
  \nonumber
\end{eqnarray}

Since $a$ is by assumption arbitrary, the last line is equal to
a standard connection 1-form to order $\epsilon$  if and only if $dt\of{B} + F = 0$.
\endofproof

The above infinitesimal gauge transformation is easily integrated to a
finite gauge transformation:

\begin{definition}\et
  \label{finite path space gauge transformations}
  A {\bf finite path space holonomy gauge transformation}
  is the integration of infinitesimal  path space holonomy gauge transformations
  \refdef{infinitesimal path space holonomy gauge transformation}, i.e. it is 
  a map for any $a \in  \Omega^1\of{U,\h}$
  given by
  \begin{eqnarray*}
    &&\mathcal{W}_{\mathcal{A}_{(A,B)}}\of{\Sigma|_{\gamma_0}^{\gamma_1}}
    \mapsto
    \lim\limits_{\epsilon = 1/N \to 0}
    \underbrace{
    \mathrm{Ad}_{\gamma_0}^{\gamma_1}
    \left(
      1 - \epsilon \oint_{A + dt\of{a}} \of{a}
    \right)
    \cdots
    \mathrm{Ad}_{\gamma_0}^{\gamma_1}
    \left(
      1 - \epsilon \oint_{A} \of{a}
    \right)}_{\mbox{$N$ factors}}
    \of{
    \mathcal{W}_{\mathcal{A}_{(A,B)}}\of{\Sigma|_{\gamma_0}^{\gamma_1}}
    }
    \,.
  \end{eqnarray*}
\end{definition}

\begin{proposition}\et
  \label{effect of finite path space gauge transformations}
  A finite path space holonomy gauge transformation \refdef{finite path space gauge transformations}
  of the holonomy of a standard path space connection 1-form $\mathcal{A}_{(A,B)}$
  is equivalent to a transformation
  \begin{eqnarray}
    \mathcal{A}_{(A,B)} &\mapsto& \mathcal{A}_{(A^\prime,B^\prime)}
    \nonumber
  \end{eqnarray}
  where
  \begin{eqnarray}
    \label{finite path space gauge transformation on 1,2 form}
    A &\mapsto& A + dt\of{a}
    \nonumber\\
    B &\mapsto& B - \underbrace{(d_A a + a\wedge a)}_{\defas k_a} 
  \end{eqnarray}
  is the transformed $(1,2)$-form $(A,B)$.
\end{proposition}

\Proof
This is a standard computation.
\endofproof

  In summary the above yields two different notions of gauge transformations on path space:
  \begin{enumerate}
    \item If the path space in question is a based loop space then 
       according to Prop.\ \ref{target space gauge trafos on path space 
       holonomy} a gauge transformation on
       target space  
       yields an ordinary gauge transformation of the $(1,2)$-form $(A,B)$:
       \begin{eqnarray}
          \label{first kind trafo}
          A &\mapsto& \phi A \phi^{-1} + \phi (d\phi^{-1})
          \nonumber\\
          B &\mapsto& \alpha\of{\phi}\of{B}
          \,.
          \nonumber 
       \end{eqnarray}
      We shall call this a {\bf 2-gauge transformation of the first kind}.
    \item
     A gauge transformation in path space itself yields,
according to prop. \ref{effect of finite path space gauge transformations},
 a transformation
     \begin{eqnarray}
        \label{second kind trafo}
        A &\mapsto& A + dt\of{a}
        \nonumber\\
        B &\mapsto& B - (d_A a + a\wedge a)
        \,. 
        \nonumber
     \end{eqnarray}
     We shall call this a {\bf 2-gauge transformation of the second kind.}
  \end{enumerate}
  Recall that according to Prop.\ 
  \ref{effect of infinitesimal gauge transformations on path space}
  this works precisely when $(A,B)$ defines a standard connection 1-form 
   \refdef{standard path space 1-form} 
   on path space for which the `fake curvature' \refdef{curvature and fake curvature}
  vanishes
  $\tilde F = dt\of{B} + F_A = 0$.

In the context of loop space these two transformations and the conditions on them
were discussed for the special case $G=H$ and $t = \mathrm{id},\, \alpha = \mathrm{Ad}$ 
in \cite{Schreiber:2004e}. In the context of 2-groups and higher lattice gauge theory 
they were
found in section 3.4 of \cite{GirelliPfeiffer:2004}. They also appear in the
transition laws for nonabelian gerbes 
\cite{BreenMessing:2001,AschieriCantiniJurco:2004,AschieriJurco:2004},
as is discussed in detail in \S\fullref{section: Nonabelian Gerbes}.
The same transformation for the  special case where all groups are abelian 
is well known from abelian gerbe theory \cite{Chatterjee:1998} but also
for instance from string theory (e.g. section 8.7 of \cite{Polchinski:1998}).

\vskip 2em

With holonomy on path space understood, it is now possible to use the fact
that every curve in path space maps to a (possibly degenerate) surface in
target space in order to get a notion of (local) surface holonomy. That is
the content of the next subsection.

\subsubsection{The local 2-Holonomy Functor}
\label{section: Local 2-Holonomy from local Path Space Holonomy}

\begin{definition}\et
  \label{local 2-holonomy}
  Given a patch $U$ and a 2-group $\twogroup$
  a {\bf local 2-holonomy} 
  is a strict 2-functor 
  \[
    \hol \maps \P_2\of{U} \to \twogroup
  \]
  from the path 2-groupoid
  $\P_2\of{U}$ \refdef{2-groupoid of bigons}
  to the 2-group $\twogroup$.
\end{definition}

(The fact that this functor is strict means that it ignores the 
parametrization of the bigons' source and target edges. Eventually
one may want to replace the structure 2-group here with the more
general `coherent' 2-group dicussed in \cite{BaezLauda:2003}, and the 
strict 2-functor with a more general sort of 2-functor.)

We want to construct a local 2-holonomy from a standard path space
connection 1-form \refdef{standard path space 1-form}. In order
to do so we first construct a `pre-2-holonomy' for any
standard path space connection 1-form and then determine under
which conditions this actually gives a true 2-holonomy. It turns
out that the necessary and sufficient conditions for this is the
vanishing of the fake curvature \refdef{curvature and fake curvature}.

\begin{definition}\et
  \label{local pre-2-holonomy}
  Given a standard path space connection 1-form 
  \refdef{standard path space 1-form}
  and given any parametrized bigon \refdef{parametrized bigon}
  $
    \Sigma : [0,1]^2 \to U
  $
  with 
  source edge
  $
    \gamma_1 \defas \Sigma\of{\cdot,0}
  $
  and
  target edge
  $
    \gamma_2 \defas \Sigma\of{\cdot,1}\,,
  $
  the triple
  $
    (g_1, h, g_2 ) \in G\times H \times G
  $
  with
  \begin{eqnarray}
    \label{assignments of pre-2-holonomy}
    g_i &\defas& W_A\of{\gamma_i}
    \nonumber\\
    h &\defas& \mathcal{W}^{-1}_{\mathcal{A}}\of{\Sigma\of{1-\cdot,\cdot}}
  \end{eqnarray}
  is called the {\bf local pre-2-holonomy} of $\Sigma$ associated with 
  $\mathcal{A}$.
\end{definition}

In order for a pre-2-holonomy to give rise to a true 2-holonomy
two conditions have to be satisfied:
\begin{enumerate}
   \item 
     The triple $(g_1,h,g_2)$ has to specify a 2-group element.
      By Prop.\ \ref{liecrossedmodule} 
      this is the case precisely if
      $g_2 = t\of{h}g_1$ \refer{source and target of 2-group element}.
   \item
     The pre-2-holonomy has to be invariant under thin homotopy in
     order to be well defined on bigons.
\end{enumerate}

The solution of this is the content of Prop.\ 
\ref{from pre-2-holonomy to 2-holonomy}
below. In order to get there the following considerations are necessary:

In order to analyze the first of the above two points consider
the behaviour of the pre-2-holonomy under changes of the
target edge.

  Given a path space $\paths_s^t\of{U}$
  and a $\g$-valued 1-form with line 
  holonomy holonomy $W_A[\gamma]$ on $\gamma \in \paths_s^t$ 
  \refdef{line holonomy and parallel transport} 
  the {\bf change in holonomy} of $W_A$ as one changes $\gamma$ is well known
  to be given by the following:

\begin{proposition}\et
  \label{variation of holonomy under variation of the path}
  Let $\rho : \tau \mapsto \gamma\of{\tau}$ be the flow generated by the vector field $D$ on
  $\paths_s^t$, then
  \begin{eqnarray}
   \label{variation of holonomy}
    \left.
    \frac{d}{d\tau}
    W_A^{-1}[\gamma\of{0}]W_A[\gamma\of{\tau}]
    \right|_{\tau = 0}
    &=&
    -
    \left(
      \oint\limits_A (F_A)
    \right)\of{D}
    \,.
  \end{eqnarray}  
\end{proposition}
(Note that the right hand side denotes evaluation of the path space 1-form
$\oint_A(F_A)$ on the path space vector field $D$.)

\Proof The proof is standard. The only subtlety is to take care of the
various conventions for signs and orientations which give rise to the minus 
sign in \refer{variation of holonomy}. \endofproof

\begin{proposition}\et
  \label{pre-2-holonomy asscoiates 2-group elements}
  For the pre-2-holonomy \refdef{local pre-2-holonomy} 
  of parametrized bigons $\Sigma$ 
  associated with the standard connection 1-form
  $\mathcal{A}_{(A,B)}$
  to specify 2-group elements,
  i.e. for the triples $(g_1,h,g_2)$ to satisfy $g_2 = t\of{h}g_1$,
  we must have
  \[
    dt\of{B} + F_A = 0
    \,.
  \]
\end{proposition}

\Proof
According to def. \ref{local pre-2-holonomy}
the condition $g_2 = t\of{h}g_1$ translates into
\begin{eqnarray}
  \label{an equation}
  t\of{
    h
  }
  &=&
  W_A\of{\gamma_2}
  W^{-1}_A\of{\gamma_1}
  \nonumber\\
  &=&
  W_A^{-1}\of{\gamma_2^{-1}}
  W_A\of{\gamma_1^{-1}}
  \,.
  \nonumber
\end{eqnarray}
Now let 
there be a flow $\tau \mapsto \gamma_\tau$ on $\paths_s^t\of{U}$
generated by a vector field $D$ and choose $\gamma_2^{-1} = \gamma_\tau$
and $\gamma_1^{-1} = \gamma_0$. Then according to
Prop.\ \ref{variation of holonomy under variation of the path}
we have
\begin{eqnarray}
  \frac{d}{d\tau}
  W_A^{-1}\of{\gamma_2^{-1}}
  W_A\of{\gamma_1^{-1}}
  &=&
  + \left(\oint\limits_A\of{F_A}\right)_{\gamma_0}\of{D}
  \,,
  \nonumber
\end{eqnarray}
where the plus sign is due to the fact that $D$ here points opposite to the $D$
in Prop.\ \ref{variation of holonomy under variation of the path}.

Applying the same $\tau$-derivative on the left hand side 
of \refer{an equation} yields
\begin{eqnarray}
  - \left(\oint_A\of{dt\of{B}}\right)\of{D} &=& 
    \left(\oint_A \of{F_A}\right)\of{D}
  \,.
  \nonumber
\end{eqnarray}
(Here the minus sign on the left hand side comes from the fact that 
we have identified $t\of{h}$ with the \emph{inverse} path space holonomy
$\mathcal{W}^{-1}_{\mathcal{A}}$. This is necessary because the ordinary
path space holonomy is path-ordered to the right, while we need $t\of{h}$
to be path ordered to the left.)

This can be true for all $D$ only if $-dt\of{B} = F_A$.
\endofproof

This is nothing but the {\bf nonabelian Stokes theorem}.
(Compare for instance  \cite{KarpMansouriRno:1999} and references given there.)

\vskip 2em

Next it needs to be shown that a pre-2-holonomy with 
$dt\of{B} + F_A = 0$ is invariant under thin-homotopy:

\begin{proposition}\et
  \label{local condition for r-flatness}
  The standard path space connection 1-form 
  $\mathcal{A}_{(A,B)}$ \refdef{standard path space 1-form}
  is inavriant under thin homotopy  precisely if
  the path space 2-form
  \begin{eqnarray}
    \label{the r-flatness condition}
    \oint_A\of{ d\alpha\of{T_a}\of{B}, (F_A + dt\of{B})^a }
  \end{eqnarray}
  vanishes on all pairs path space vector fields 
  that generate thin homotopy flows.
 \end{proposition} 

\Proof
For the special case $G=H$ and $t = \mathrm{id},\, \alpha = \mathrm{Ad}$ this was proven 
by \cite{Alvarez:2004}.
The full proof is a straightforward generalization of this special case:

Consider a path $\Sigma$ in path space with tangent vector $T$ and
let $D$ be any vector field on $\paths_s^t\of{U}$. By a standard result
the path space holonomy $\mathcal{W}\of{\Sigma}$ is invariant under the
flow generated by $D$ iff the curvature of $\mathcal{A}$ vanishes on 
$T$ and $D$, $\mathcal{F}\of{T,D} = 0$.

But from corollary 
\ref{corollary: curvature of standard path space 1-form} we know
that $\mathcal{F} = - \oint \of{\extd_A B} -
 \oint (d\alpha\of{T_a}\of{B}, (F_A + dt\of{B})^a)$. 
It is easy to see that $\oint \of{\extd_A B}$ vanishes on all pairs
of tangent vectors that generate thin homotopy transformations of $\Sigma$
and that the remaining term vanishes on $(T,D)$ for all $D$ if it vanishes
on all pairs of tangent vector that generate thin homotopy transformations.
\endofproof
Now we can finally prove the following:
\begin{proposition}\et
  \label{from pre-2-holonomy to 2-holonomy}
  The pre-2-holonomy \refdef{local pre-2-holonomy}
  induces a true local 2-holonomy \refdef{local 2-holonomy}
  \[
  \begin{array}{rcccc}
    \hol_i &\maps& \P_2\of{U_i} &\to& G_2
    \\
    &&
\xymatrix{
   x \ar@/^1pc/[rr]^{\gamma}="0"
           \ar@/_1pc/[rr]_{}_{\gamma'}="1"
           \ar@{=>}"0";"1"^{\Sigma}
&& y
}
&\mapsto& 
\xymatrix{
   \bullet \ar@/^1pc/[rr]^{W_i[\gamma]}="0"
           \ar@/_1pc/[rr]_{}_{W_i[\gamma']}="1"
           \ar@{=>}"0";"1"^{\mathcal{W}^{-1}_i\of{\Sigma}}
&& \bullet
}
  \end{array}
  \]
  precisely if the fake curvature 
  \refdef{curvature and fake curvature} vanishes.
\end{proposition}
\Proof

We have already shown that for $dt\of{B} + F_A = 0$ the
pre-2-holonomy indeed maps into a 2-group 
(Prop.\ \ref{pre-2-holonomy asscoiates 2-group elements})
and that its values are well defined on bigons
(Prop.\ \ref{local condition for r-flatness}). What remains
to be shown is functoriality, i.e. that the pre-2-holonomy
respects the composition of bigons and 2-group elements.

First of all it is immediate that composition of paths is
respected, due to the properties of ordinary holonomy.
Vertical composition of 2-holonomy
(being composition of ordinary holonomy in path space)
is completely analogous. The fact that pre-2-holonomy involves
the \emph{inverse} path space holonomy takes care of the 
nature of the vertical product in the 2-group, which reverses
the order of factors:
In the diagram
\[
\begin{array}{c|ccc}
\twogroup &
\xymatrix{
   \bullet \ar@/^2pc/[rr]^{W_A[\gamma_1]}_{}="0"
           \ar[rr]^<<<<<<{{}_{W_A[\gamma_2]}}_{}="1"
           \ar@{=>}"0";"1"^{{}_{\mathcal{W}^{-1}_\mathcal{A}[\Sigma_1]}}
           \ar@/_2pc/[rr]_{W_A[\gamma_3]}_{}="2"
           \ar@{=>}"1";"2"^{{}_{\mathcal{W}^{-1}_\mathcal{A}[\Sigma_2]}}
&& \bullet
}
&=&
\xymatrix{
   \bullet \ar@/^1pc/[rr]^{W_A[\gamma_1]}_{}="0"
           \ar@/_1pc/[rr]_{W_A[\gamma_3]}_{}="1"
           \ar@{=>}"0";"1"^{{}_{{}_{\mathcal{W}^{-1}_\mathcal{A}[\Sigma_1\circ \Sigma_2]}}}
&& 
}
\\
\;\;\;\;\Big\Uparrow \hol
&
\Big\Uparrow 
&
&
\Big\Uparrow
\\
\P_2\of{U}
&
\xymatrix{
   x \ar@/^2pc/[rr]^{\gamma_1}_{}="0"
           \ar[rr]^<<<<<<{\gamma_2}_{}="1"
           \ar@{=>}"0";"1"^{[\Sigma_1]}
           \ar@/_2pc/[rr]_{\gamma_3}_{}="2"
           \ar@{=>}"1";"2"^{[\Sigma_2]}
&& y
}
&=&
\xymatrix{
   x \ar@/^1pc/[rr]^{\gamma_1}_{}="0"
           \ar@/_1pc/[rr]_{\gamma_3}_{}="1"
           \ar@{=>}"0";"1"^{{}_{[\Sigma_1\circ \Sigma_2]}}
&& y
}
\end{array}
\]
the top right bigon must be labeled (according to the properties
of 2-groups described in Prop.\ \ref{liecrossedmodule})
by
\begin{eqnarray}
  (W_A[\gamma_1], \mathcal{W}^{-1}_{\mathcal{A}}[\Sigma_1])
  \circ
  (W_A[\gamma_2], \mathcal{W}^{-1}_{\mathcal{A}}[\Sigma_2])
  &=&
  (W_A[\gamma_1], \mathcal{W}^{-1}_{\mathcal{A}}[\Sigma_2]\mathcal{W}^{-1}_{\mathcal{A}}[\Sigma_1])  
  \nonumber\\
  &=&
  (W_A[\gamma_1], \mathcal{W}^{-1}_{\mathcal{A}}[\Sigma_1\circ\Sigma_2])
  \nonumber\,,
\end{eqnarray}
which indeed is the label associated by the $\hol$-functor
in the right column of the diagram.

So far we have suppressed in these formulas the 
 reversal \refer{assignments of pre-2-holonomy}
in the first coordinate of $\Sigma$, since it plays no role for the above.
This reversal however is essential in order for the $\hol$-functor
to respect horizontal composition.

In order to see this it is sufficient to consider \emph{whiskering}, i.e. horizontal
composition with identity 2-morphisms.

When whiskering from the left we have
\[
\begin{array}{c|ccc}
\twogroup
&
\xymatrix{
   \bullet \ar[rr]^{W_A[\gamma_1]}_{}
   &&\bullet \ar@/^1pc/[rr]^{W_A[\gamma_2]}_{}="0"
           \ar@/_1pc/[rr]_{W_A[\gamma_2^\prime]}_{}="1"
           \ar@{=>}"0";"1"^{{}_{\mathcal{W}_\mathcal{A}^{-1}[\Sigma]}}
&& \bullet
}
&=&
\hspace{28pt}
\xymatrix{
   \bullet \ar@/^1pc/[rr]^{W_A[\gamma_1\circ\gamma_2]}_{}="0"
           \ar@/_1pc/[rr]_{W_A[\gamma_1\circ\gamma_2^\prime]}_{}="1"
           \ar@{=>}"0";"1"^{{}_{\alpha\of{W_A[\gamma_1]}\of{\mathcal{W}_\mathcal{A}^{-1}[\Sigma]}}}
&& 
}
\\
\;\;\;\;\;\Big\Uparrow \hol
&
\Big\Uparrow
&&
\Big\Uparrow 
\\
\P_2\of{U}
&
\xymatrix{
   x \ar[rr]^{\gamma_1}_{}
   &&y \ar@/^1pc/[rr]^{\gamma_2}_{}="0"
           \ar@/_1pc/[rr]_{\gamma_2^\prime}_{}="1"
           \ar@{=>}"0";"1"^{[\Sigma]}
&& z
}
&=&
\xymatrix{
   x \ar@/^1pc/[rr]^{\gamma_1\circ\gamma_2}_{}="0"
           \ar@/_1pc/[rr]_{\gamma_1\circ \gamma_2^\prime}_{}="1"
           \ar@{=>}"0";"1"^{[\Sigma]}
&& z
}
\end{array}
\]
Evaluating the 
line holonomy in path space for this situation
involves taking the path ordered
exponential of (\cf \refer{notions of parallel tranport})
\[
  \int\limits_{(\gamma_1\circ\gamma_2)^{-1}}
  d\sigma\;
    \alpha\of{W^{-1}_A[(\gamma_1\circ\gamma_2)^{-1}|_\sigma^1]}
    \of{B\of{\sigma}}
\]
evaluated on the tangent vector to the whiskered $\Sigma$.
Since this vanishes on $\gamma_1$ 
and using the reparameterization invariance of $W_A$ the above equals
\[
  \cdots
  =
  \alpha\of{W_A[\gamma_1]}
  \of{
    \int\limits_{\gamma_2^{-1}}
    d\sigma\;
      \alpha\of{W^{-1}_A[\gamma_2^{-1}|_\sigma^1]}\of{B\of{\sigma}}
  }
  \,.
\]
Hence the above diagram does commute. In this
computation the path reversal is essential, which of course is related
to our convention that parallel transport be to the point with parameter 
$\sigma = 1$. A simple plausibility argument for this was given at
the beginning of \S\fullref{section: The Standard Connection 1-Form on Path Space}.

Finally, whiskering to the right is trivial, since we can simply use reparametrization
invariance to obtain
\begin{eqnarray}
  \int\limits_{(\gamma_1\circ\gamma_2)^{-1}}
  d\sigma\;
    \alpha\of{W_A[(\gamma_1\circ\gamma_2)^{-1}|_\sigma^1]}\of{B\of{\sigma}}
  &=&
  \int\limits_{\gamma_1^{-1}}
  d\sigma\;
    \alpha\of{W_A[\gamma_1^{-1}|_{\sigma^1}]}\of{B\of{\sigma}}
  \,,
  \nonumber
\end{eqnarray}
because for right whiskers the integrand vanishes on $\gamma_2$.

Since general horizontal composition is obtained by first whiskering and then
composing vertically, this also proves that the $\hol$-functor respects 
general horizontal composition.

In summary, this shows that a pre-2-holonomy with 
vanishing fake curvature \refdef{curvature and fake curvature} 
$dt\of{B} + F_A = 0$ defines a 2-functor $\hol \maps \P_2\of{U} \to 
\twogroup$ and hence a local strict 2-holonomy.
\endofproof

\subsection{2-Curvature}
\label{2-Transition of Curvature}

Since curvature is the first order term in the holonomy around a small
loop, the 2-transition Prop.\ 
\ref{proposition stated again: transition of 2-holonomy}
of 2-holonomy immediately implies a transition law for the
path space curvature 2-form $\mathcal{F}_\mathcal{A} = - \oint_A (H)$
\refer{path space curvature in fake flat case} and hence of the
curvature 3-form $H = \extd_A B$ \refdef{curvature and fake curvature}.

First of all one notes the following:

\begin{proposition}\et
 The curvature 3-form \refdef{curvature and fake curvature}
 $H = \extd_A B$ transforms \emph{covariantly}
  under gauge transformations of the first kind \refer{first kind trafo}.
  Moreover, it is \emph{invariant} under gauge transformations of the second kind
  \refer{second kind trafo} if and only if the fake curvature vanishes.
\end{proposition}

\Proof

The covariant transformation under gauge transformations of the first kind
follows from simple standard reasoning. The invariance under 
infinitesimal transformations of the second kind
with $A \to A + \epsilon dt\of{a}$ and $B \to B - \epsilon \extd_A a$
follows from noting the invariance under `infinitesimal' shifts:
\begin{eqnarray}
  H = \extd_A B &\to& \extd_{A + \epsilon dt\of{a}}(B - \epsilon \extd_A a)
  \nonumber\\
  &=&
  \extd_A B 
   - \epsilon 
  \left(
     \extd_A \extd_A a
     -
     d\alpha\of{dt\of{a}}\of{B}
  \right)
  +
  \order{\epsilon^2}
  \nonumber\\
  &\equalby{notation for action of h on h'}&
  H -\epsilon
  \left(
    d\alpha\of{F_A}\of{a}
    +
    d\alpha\of{dt\of{B}}\of{a}
  \right)
  +
  \order{\epsilon^2}
  \nonumber\\
  &\equalby{definition fake curvature}&
  H - \epsilon d\alpha\of{\tilde F}\of{B} + \order{\epsilon^2}
  \nonumber\\
  &\stackrel{\tilde F = 0}{=}&
  H + \order{\epsilon^2}
  \,.
\end{eqnarray}
\endofproof
(\cf equation (3.43) of \cite{GirelliPfeiffer:2004}).

Note that the invariance of $H$ under transformations of the second
kind does imply invariance of the path space curvature
2-form $\mathcal{F}_\mathcal{A}$. Naively, this transforms as
\begin{eqnarray}
  \mathcal{F}_\mathcal{A}
  =
  -\oint\limits_A (H) 
  &\to&
  -\oint\limits_{A + dt\of{a}}(H)
  \,.
\nonumber
\end{eqnarray}
but since $\image\of{dt}$ acts trivially on $\ker\of{dt}$ 
(this is shown in  prop. \ref{im t acts trivially on ker t} 
right below)
we have
\begin{eqnarray*}
  -\oint\limits_{A + dt\of{a}}(H)
  &=&
    -\oint\limits_{A}(H)
  \,.
\end{eqnarray*}

\begin{proposition}
  \label{im t acts trivially on ker t}
  In a crossed module $\image\of{t}$ acts trivially on
  $\ker\of{t}$. Equivalently,
  in a differential crossed module
  $\image\of{dt}$ acts trivially on $\ker\of{dt}$.
\end{proposition}

\Proof This is a consequence of the property
\begin{eqnarray}
  \alpha\of{t\of{h_1}}\of{h_2} &=& h_1 h_2 h_1^{-1}
  \nonumber\\
  d\alpha\of{dt\of{S_1}}\of{S_2} &=& \commutator{S_1}{S_2}
\end{eqnarray}
of a crossed module, with
$h_i \in H$ and $S_i \in \h$:

Let $h\in H$ and $k \in \ker\of{t} \subset H$.
Then
\begin{eqnarray*}
  \alpha\of{t\of{h}}\of{k}
  &=&
  h k h^{-1}
  \\
  &=&
  k k^{-1} h k h^{-1}
  \\
  &=&
  k \underbrace{\alpha\of{t\of{k^{-1}}}\of{h}}_{= h} h^{-1}
  \\
  &=&
  k
  \,.
\end{eqnarray*} 

Similarly for differential crossed modules with
$S \in \h$ and $S_0\in \ker\of{dt} \subset \h$:

\begin{eqnarray}
  d\alpha\of{dt\of{S}}\of{S_0}
  &=&
  \commutator{S}{S_0}
  \nonumber\\
  &=&
  -\commutator{S_0}{S}
  \nonumber\\
  &=&
  -\underbrace{d\alpha\of{dt\of{S_0}}}_{=0}\of{S}
  \nonumber\\
  &=&
  0
  \,.
\end{eqnarray}
\endofproof

The transition law for $H_i \defas \extd_{A_i} B_i$ is now a simple corollary:

\begin{corollary}
  The local curvature 3-form $H_i = \extd_{A_i} B_i$ of the local standard path
  space connection of a 2-bundle with 2-connection has the transition law
  \[
    H_i = \alpha\of{g^1_{ij}}\of{H_j}
  \]
  on double intersections $U_{ij}$.
\end{corollary}
This is the transition law \refer{gerbe curvature transition law}
of the curvature 3-form of a nonabelian gerbe for vanishing
fake curvature and $d_{ij} = 0$.

One should note that also the fake curvature 
\refdef{curvature and fake curvature}
transforms covariantly
and can therefore indeed consistently be chosen to vanish:
The transition law for $F_{A_i}$ following from 
\refer{A transition law from 2-transition} is
\[
  F_{A_i} = g_{ij}F_{A_j}g_{ij}^{-1} - dt\of{k_{ij}}
\]
and that of $dt\of{B}$ following from 
Prop. \ref{proposition stated again: transition of 2-holonomy}
(p. \pageref{proposition stated again: transition of 2-holonomy})
\[
  dt\of{B_i} = g_{ij}\,dt\of{B_j} g_{ij}^{-1} + dt\of{k_{ij}}
  \,,
\]
so that
\[
  \tilde F_i = g_{ij} \tilde F_j g_{ij}^{-1}
  \,.
\]

The {\bf Bianchi-identity} on path space says that
\begin{eqnarray*}
  0 
  &=&
  \extd \oint\limits_A \of{H}
  +
  \underbrace{
  \oint\limits_A\of{B} 
   \wedge 
  \oint\limits_A \of{H}
  -
  \oint\limits_A \of{H}
  \wedge
  \oint\limits_A\of{B}
  }_{\mbox{$= 0$ by prop. \ref{im t acts trivially on ker t}}}
  \\
  &=&
  \extd \oint\limits_A \of{H}
  \\
  &=&
  - \oint\limits_A \of{\extd_A H}
  - \oint\limits_A\of{d\alpha\of{T_a}\of{H},F^a}
  \\
  &=&
  - \oint\limits_A \of{\extd_A H}
  + \oint\limits_A\of{d\alpha\of{T_a}\of{H},dt\of{B}^a}
  \\
  &=&
  \oint\limits_A \of{\extd_A H}
  \,.
\end{eqnarray*}
Hence the vanishing of the fake curvature ensures that the
3-form field strength is covariantly closed:
\[
  \extd_A H = 0
  \,.
\]
Since $\extd_A H = (\extd_A)^2 B = F_A \wedge B$
this can be seen more explicitly also as follows:
\begin{proposition}
  The vanishing of the fake curvature implies that
  \[
    F_A\wedge B = 0
    \,,
  \]
  which is shorthand for
  \[
    (F_A^a \wedge B^b) \; d\alpha\of{T_a}\of{S_b} = 0
    \,.
  \]
\end{proposition}
\Proof
Use $F_A = -dt\of{B}$ to get
\begin{eqnarray*}
  (F_A^a \wedge B^b) \; d\alpha\of{T_a}\of{S_b}
  &=&
  -(B^a \wedge B^b) \; d\alpha\of{dt\of{S_a}}\of{S_b}
  \\
  &=&
  -(B^a \wedge B^b) \; \commutator{S_a}{S_b}
  \\
  &=& 
  0  
  \,.
\end{eqnarray*}
This vanishes because 
$B^a \wedge B^b = B^b \wedge B^a$ 
(since $B$ is a 2-form)
while
$\commutator{S_a}{S_b} = -\commutator{S_b}{S_a}$.
\endofproof

This again ensures that self-duality
of the field strength, i.e. 
\[
  H = \pm \star H
\]
is sufficient to imply equations of motion of the form
\[
  \extd_A \star H = 0
  \,.
\]
In the abelian case this ensures that 
the 6-dimensional self-dual theory compactifies to
an ordinary gauge theory (\cf \S\fullref{Literature on M2M5}). 
Vanishing fake curvature also
ensures that this is gauge invariant.\footnote{
These observations 
concerning the equation $\extd_A H = 0$
arose in a discussion with Jens Fjelstad.}

\clearpage
\section{Global 2-Holonomy}
\label{the integral picture}

In \S\fullref{principal 2-Bundles} \cite{BaezSchreiber:2004}
2-bundles with 2-connection admitting 2-holonomy have been
described, and the transition laws for the 2-connection have
been worked out. With these ingredients it is now possible to
write down a definition of global surface holonomy obtained by
gluing together local surface holonomies. This is discussed in the
following.

The result is a diagrammatic explanation and generalization to
nonabelian and, with slight modifications, weakened cases of the
procedure that was apparently first stated by Alvarez
\cite{Alvarez:1985} in the context of 2D topological field theories and
later used by Gawedzki and Reis \cite{GawedzkiReis:2002}
in the context of the WZW model. 

\vskip 1em

{\bf Note on Notation:}

We will display many 2-commuting diagrams in the following. 
In order not to overburden the labeling of the 1- and 2-morphism
we will frequently leave ``re-whiskering'' of 2-morphisms implicit.
This means that whenever a displayed 2-morphism does not go between
its genuine source and target 1-morphisms we have implicitly 
composed it horizontally with appropriate identity 2-morphisms so 
that it goes between source and target as indicated in the
respective diagram. This should never lead to any ambiguities
but instead make the diagrams more easily readable.

\subsection{1-Bundles with 1-Connection}

We start by considering the situation that we are interested in
for the case of ordinary (1-)bundles with (1-)connection. 

Recall the definition of a principal 1-bundle with 1-connection
from def 
\ref{definition of 1-connection in 1-bundle in terms of local holonomy}
(p. \pageref{definition of 1-connection in 1-bundle in terms of local holonomy})
(see also
\S\fullref{introduction: 1-connections in 1-bundles}).  
This was given in terms of local holonomy functors which were
related on double overlaps by natural transformations.
One can easily see that this can equivalently
be described by a \emph{single} functor, called the
{\bf global (1-)holonomy (1-)functor} from what we call the
``\Cech-extended'' path (1-)groupoid to the structure group.

\subsubsection{The global 1-Holonomy 1-Functor}

There is a groupoid, called the \Cech-groupoid, whose morphisms
are the ``transitions'' of any given point from one patch $U_i$
of a good covering $\covering$ to another patch $U_j$:
\begin{definition}
  \label{def Cech 1-groupoid}

  The {\bf \Cech (1-)groupoid} $C\of{\covering}$ of a
  good covering $\covering \to M$
has as objects all points  in $\covering$ 
\[
  \Ob\of{C\of{U}} \defas \set{(x,i) |i\in I\,, x\in U_i}
\]
and its morphisms are formal arrows
\[
  \Mor\of{C\of{U}} \defas \set{(x,i)\to(x,j) 
    |i,j\in I\,, x\in U_{ij}}
\]
such that there is at most one morphism between 
any two objects, i.e. such that every triangle
\begin{center}
\begin{picture}(140,120)
  \includegraphics{baretriangle.eps}
  \put(-139,0){$(x,i)$}
  \put(-70,105){$(x,j)$}
  \put(2,0){$(x,k)$}
\end{picture}
\end{center}
commutes.

\end{definition}

We can combine this \Cech-groupoid with the path groupoids 
$\P_1\of{U_i}$ of all the patches $U_i$
\refdef{groupoid of paths} into a single groupoid which we call the 
{\bf \Cech-extended path groupoid} $\P_1^C\of{\covering}$.

\begin{definition}
\label{def Cech-extended 1-path 1-groupoid}
The {\bf \Cech-extended path (1-)groupoid} $\P_1^C\of{\covering}$
of a good covering $\covering \to M$
has objects
\[
  \Ob\of{\P_1^C\of{\covering}} \defas \set{(x,i) |i\in I\,, x\in U_i}
\]
and its set of morphisms is defined to be the set of morphism generated
by formal compositions of the elements in $\Mor\of{\P_1\of{U_i}}\,, 
\forall\, i\in I$
and $\Mor\of{C\of{\covering}}$, 
divided out by all commuting diagrams of the form
\begin{center}
\begin{picture}(60,200)
  \includegraphics{pathtransition.eps}
  \put(2,0){$(y,j)$}
  \put(2,162){$(x,j)$}
  \put(-116,0){$(y,i)$}
  \put(-116,162){$(x,i)$}
  \put(-103,80){$[\gamma_i]$}
  \put(-3,80){$[\gamma_j]$}
\end{picture}
\end{center}
whenever $[\gamma_i]|_{U_{ij}} = [\gamma_j]|_{U_{ij}}$
\end{definition}

Heuristically, this diagram expresses how a path in a double overlap
of two patches
can be regarded as a path in either of the two patches.
More technically, we can regard this diagram as expressing a 
certain natural isomorphism between two functors that map a given
pre-image path to $[\gamma_i]$ and to $[\gamma_j]$, respectively.

We can now define a {\bf global holonomy functor} $\hol$ to be any functor
$\hol$ 
from the \Cech-extended path groupoid to the structure group $G$:
\[
  \hol : \P_1^C\of{U} \to G
  \,.
\]
Given any such functor, denote its image on $[\gamma_i]$ by
\[
  \hol\of{[\gamma_i]} \defas \hol_i\of{\gamma}
\]
and denote its image of $(x,i)\to (x,j)$ by
\[
  \hol\of{
    (x,i) \to (x,j)
  }
  \defas
  g_{ij}\of{x}
  \,.
\]
Then the application of $\hol$ to the commuting diagram from
\refdef{def Cech 1-groupoid} yields the cocycle condition on the 
transition
function
\begin{eqnarray*}
  \hol\of{
   \raisebox{-40pt}{
  \hspace{20pt}
\begin{picture}(140,100)
  \includegraphics{baretriangle.eps}
  \put(-139,0){$(x,i)$}
  \put(-70,105){$(x,j)$}
  \put(2,0){$(x,k)$}
\end{picture}
}
}
=
 \raisebox{-40pt}{
  \hspace{20pt}
 \begin{picture}(140,120)
  \includegraphics{baretriangle.eps}
  \put(-100,50){$g_{ij}$}
  \put(-25,50){$g_{jk}$}
  \put(-70,-5){$g_{ik}$}
  \end{picture}
  }
  \,,
\end{eqnarray*}
\vskip 1em
while application of $\hol$ to the commuting diagram
from def. \ref{def Cech 1-groupoid} yields
\vskip 1em
\[
\hol\of{
\raisebox{-80pt}{
\hspace{23pt}
\begin{picture}(120,160)
  \includegraphics{pathtransition.eps}
  \put(2,0){$(y,j)$}
  \put(2,162){$(x,j)$}
  \put(-116,0){$(y,i)$}
  \put(-116,162){$(x,i)$}
  \put(-103,80){$[\gamma_i]$}
  \put(0,80){$[\gamma_j]$}
\end{picture}
}
}
=
\raisebox{-80pt}{
\hspace{33pt}
\begin{picture}(120,160)
  \includegraphics{pathtransition.eps}
  \put(-60,165){$g_{ij}\of{x}$}
  \put(-60,-4){$g_{ij}\of{y}$}
  \put(-125,80){$\hol_i\of{[\gamma]}$}
  \put(-2,80){$\hol_j\of{[\gamma]}$}
\end{picture}
}
\,,
\]
\vskip 1em
which is the transition law for the local holonomy functor
$\hol_i$.

So the global functor $\hol$ does capture all the information about
the local holonomy functors $\hol_i$ together with their gluing
conditions. 

Suppose we want to compute some holonomy
of a (class of a) path $[\gamma] \in \Mor\of{\P_1\of{M}}$
in the base manifold $M$. This path will in general not sit inside a
single patch $U_i$. We can however ``lift'' it to a morphism
in 
the \Cech-extended path groupoid
$\P_1^C\of{\covering}$ by dividing it into $N$ sub-paths
$[\gamma_n]\,, n \in \set{1,2,\dots, N}$ that all do sit in a
single $U_i$,
\[
  [\gamma_n] \in \Mor\of{\paths\of{U_{i_{n}}}}
  \,,
\]
and composing these with morphisms in $C\of{\covering}$:
\[
  \hspace{-2cm}
  (\gamma_0\of{0},i_{0}) 
    \stackto{[\gamma_0]} 
  (\gamma_0\of{1},i_{0})
  \to
  (\gamma_1\of{0},i_{1}) 
    \stackto{[\gamma_1]} 
  (\gamma_1\of{1},i_{1})
  \to 
  \cdots
  \to
  (\gamma_N\of{0},i_{N}) 
    \stackto{[\gamma_N]} 
  (\gamma_N\of{1},i_{N})
  \,.
\]
On this morphism we can apply $\hol$ and regard the result as the 
holonomy of the original path $[\gamma]$ (with respect to the given 
choice of trivialization at the endpoints of $[\gamma]$). 
\begin{center}
\begin{picture}(200,217)
\includegraphics{globallinehol.eps}
\put(-200,112){$U_i$}
\put(-126,114){$U_j$}
\put(-203,170){$\gamma_1$}
\put(-121,168){$\gamma_2$}
\put(-235,10){$\hol_i\of{\gamma_1}$}
\put(-117,10){$\hol_j\of{\gamma_2}$}
\put(-173,10){$g_{ij}\of{x}$}
\put(-167,147){$x$}
\put(-154,60){$\hol$}
\end{picture}
%\caption{\label{globallinehol eps}
%{\bf Computing global line holonomy}
%}
\end{center}

The result reproduces the familiar law for how to compute global
line holonomy. 
This is well defined since a gauge transformation amounts to
\begin{center}
\begin{picture}(290,66)
\includegraphics{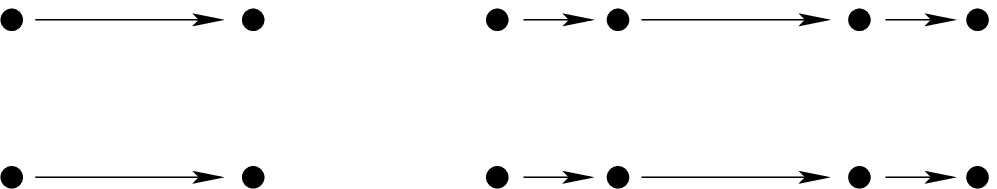}
\put(-183,47){$=$}
\put(-257,55){$\tilde g_{ij}$}
\put(-80,55){$g_{ij}$}
\put(-130,55){$h_i$}
\put(-30,56){$h_j^{-1}$}
\put(-3,-48){
\begin{picture}(0,0)
\put(-183,47){$=$}
\put(-257,55){$\tilde \hol_{i}$}
\put(-80,55){$\hol_i$}
\put(-130,55){$h_i$}
\put(-30,56){$h_i^{-1}$}
\end{picture}
}
\end{picture}
\end{center}
and since these two contributions cancel:
%\begin{figure}[h]
\begin{center}
\begin{picture}(250,260)
\includegraphics{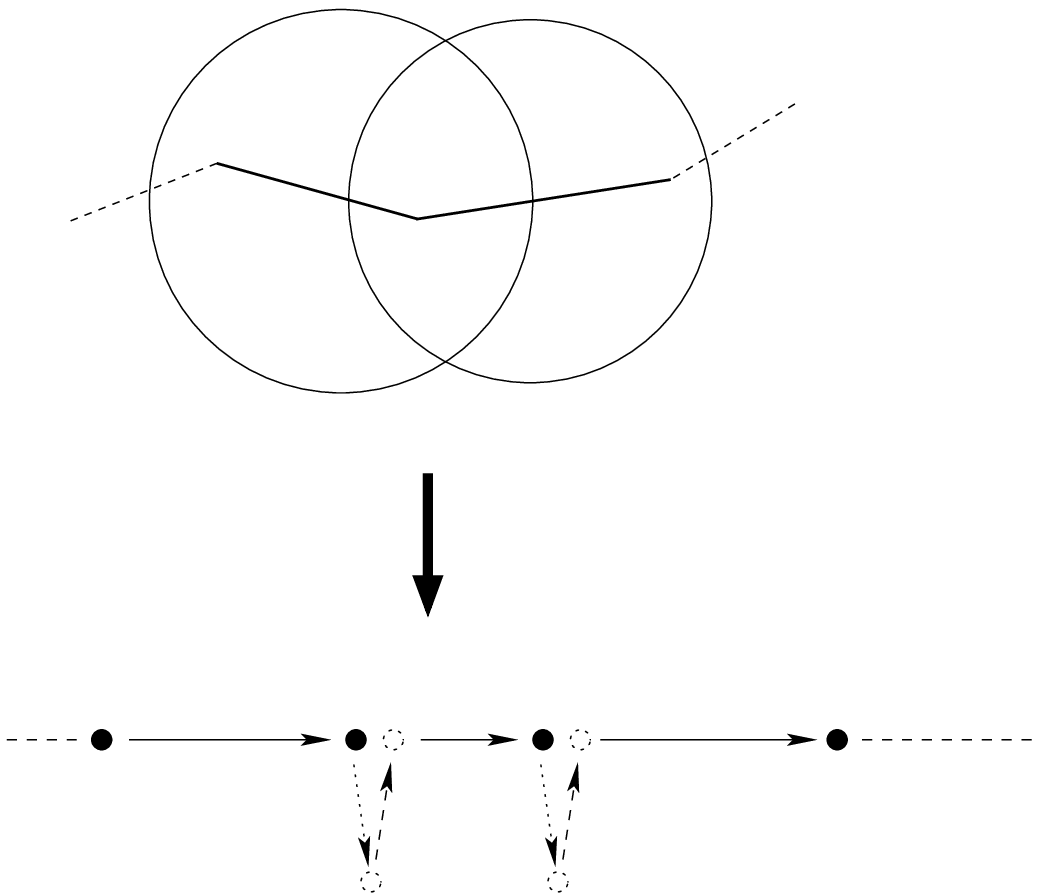}
\put(-20,42){
\begin{picture}(200,240)
\put(-200,112){$U_i$}
\put(-126,114){$U_j$}
\put(-203,170){$\gamma_1$}
\put(-121,168){$\gamma_2$}
\put(-235,10){$\hol_i\of{\gamma_1}$}
\put(-100,10){$\hol_j\of{\gamma_2}$}
\put(-165,10){$g_{ij}\of{x}$}
\put(-167,147){$x$}
\put(-154,60){$\hol$}
\put(-173,-25){${}_{h_i\of{x}}$}
\put(-208,-25){${}_{h_i^{-1}\of{x}}$}
\put(-118,-25){${}_{h_j\of{x}}$}
\put(-154,-11){${}_{h_j^{-1}\of{x}}$}
\end{picture}
}
\end{picture}
%\caption{\label{globallineholgauged eps}
%{\bf Gauge invariance of global line holonomy} corresponds
%  to `zig-zag moves' in the gauge group.
%}
\end{center}
%\end{figure}
From the point of view of the global holonomy functor this fact becomes
a tautology, becuase gauge transformations 
of the local trivialization of the bundle with connection
are nothing but natural transformations of the global holonomy functor,
as can be seen from the following naturality squares:
\begin{eqnarray}
\label{gauge trafo of Cech-extended holonomy 1-functor}
\begin{picture}(350,300)
  \includegraphics{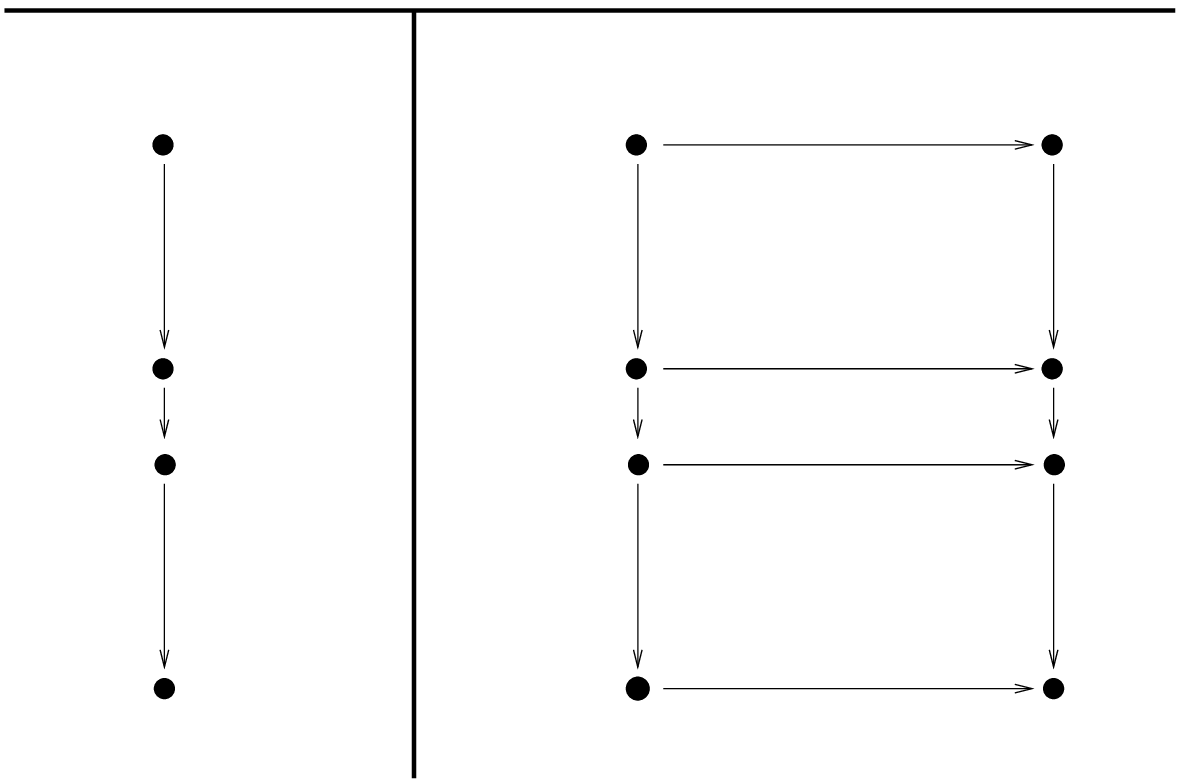}
  \put(-287,184){$(x,i)$}
  \put(-289,154){$\gamma_1$}
  \put(-287,118){$(y,i)$}
  \put(-287,91){$(y,j)$}
  \put(-289,59){$\gamma_2$}
  \put(-287,25){$(z,j)$}
  \put(92,-3){
  \begin{picture}(0,0)
  \put(-289,154){$\hol_i\of{\gamma_1}$}
  \put(-283,107){$g_{ij}\of{y}$}
  \put(-289,59){$\hol_j\of{\gamma_2}$}
  \end{picture}
  }
  \put(253,-3){
  \begin{picture}(0,0)
  \put(-289,154){$\tilde \hol_i\of{\gamma_1}$}
  \put(-289,107){$\tilde g_{ij}\of{y}$}
  \put(-289,59){$\tilde \hol_j\of{\gamma_2}$}
  \end{picture}
  }
  \put(-105,190){$h_i\of{x}$}
  \put(-105,125){$h_i\of{y}$}
  \put(-105,78){$h_j\of{y}$}
  \put(-105,15){$h_j\of{y}$}
  \put(-300,232){$\P_1^C\of{\covering}$}
  \put(-100,232){$G$}
\end{picture}
\end{eqnarray}
Here
\[
  \hol \stackto{h} \tilde \hol
\]
is a natural transformation between global holonomy functors
$\hol$ and $\tilde \hol$, given by
\begin{eqnarray*}
  h\of{(x,i)} \defas h_i\of{x} \in G
  \,.
\end{eqnarray*}

\newpage
\subsection{2-Bundles with 2-Connection}
\label{Global 2-Holonomy in 2-Bundles}

The above considerations for 1-connections with 1-holonomy
in 1-bundles
are straightforwardly generalized to 2-connection with 2-holonomy
in 2-bundles.

\subsubsection{The \Cech-extended 2-Path 2-Groupoid}

We are again interested in merging the 
2-path 2-groupoids 
$\P_2\of{U_i}$
\refdef{2-groupoid of bigons}
with 
the \Cech 2-groupoid $C_2\of{\covering}$ of the
covering $\covering = \bigsqcup\limits_{i \in I}U_i$, which
is the generalization of the \Cech-1-groupoid from
def. \ref{def Cech 1-groupoid}.

\begin{definition}

  Given a good covering 
$\covering \to M$, the {\bf \Cech 2-groupoid}
  $C_2\of{\covering}$ is defined as follows:
Its objects are all the elements
\[
  \Ob\of{C_2\of{\covering}} 
  = 
  \set{(x,i)| i \in I\,, x \in U_i}
\]
of the covering $\covering$ 
and its 1-morphisms are those \emph{generated}
from the 1-morphisms present in $C\of{\covering}$,
\[
  \Mor_1\of{C_2\of{U}} = 
  \left\langle
  \set{(x,i) \to (x,j)| i \in I\,, x \in U_i}
  \right\rangle
  \,,
\]
but
we no longer demand to have commuting triangle diagrams.
In particular, for every object $(x,i)$ there is now precisely one morphisms
\[
  (x,i) \to (x,i)
\]
not equal to the identity morphism (which we denote by $\mathrm{Id}_{(x,i)}$).

In addition, there is in $C_2\of{\covering}$
precisely one 2-morphism between any two
1-morphisms with coinciding endpoints:
\[
  \Mor_2\of{C_2\of{U}} = 
  \set{
    \begin{array}{c}
      (x,i) \to (x,j_1)\to \cdots (x,j_n)\to (x,k) \\
       \Big\Downarrow \\
      (x,i) \to (x,j'_1)\to \cdots (x,j'_m)\to (x,k)
    \end{array}
   }
  \,.
\]
Hence instead of 1-commuting triangles we have traingle 2-morphisms
\begin{center}
\begin{picture}(140,120)
  \includegraphics{triangle.eps}
  \put(-139,0){$(x,i)$}
  \put(-70,105){$(x,j)$}
  \put(2,0){$(x,k)$}
\end{picture}
\end{center}
and every tetrahedron of the form
\begin{eqnarray}
\label{tetrahedron in Cech 2-groupoid}
\begin{picture}(180,140)
  \includegraphics{3dtetrahedron.eps}
  \put(-190,-3){$(x,i)$}
  \put(3,-3){$(x,k)$}
  \put(-82,69){${}_{(x,j)}$}
  \put(-82,153){${}_{(x,l)}$}
\end{picture}
\end{eqnarray}
2-commutes.

\end{definition}

As before, the 2-groupoids $C_2\of{\covering}$ 
and $\P_2\of{U_i},\, i \in I$,  can be merged to what we shall call
the \emph{\Cech-extended 2-path 2-groupoid} $\P_2^C\of{\covering}$ of the
covering $\covering \to M$.

\begin{definition}

The {\bf \Cech-extended 2-path 2-groupoid} of a 
covering $\covering \to M$
is defined as follows:

The set of objects is the same as before, the set
of 1-morphisms is that generated by formally composing those of
$\P_2\of{U_i}, i\in I$,  and $C_2\of{\covering}$. The set of 2-morphisms
are the formal compositions generated by the
2-morphisms in  $\P_2\of{U_i}, i\in I$,  and $C_2\of{\covering}$
and in addition we throw in precisely one 2-morphism 
\begin{center}
\begin{picture}(60,200)
  \includegraphics{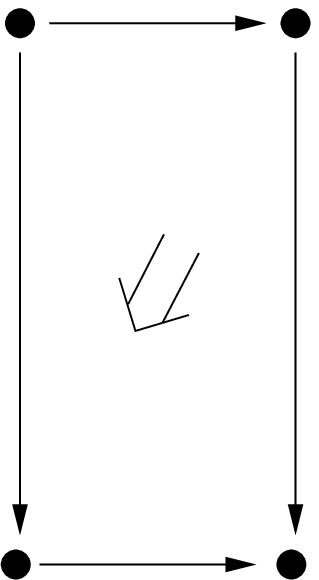}
  \put(2,0){$(y,j)$}
  \put(2,162){$(x,j)$}
  \put(-116,0){$(y,i)$}
  \put(-116,162){$(x,i)$}
  \put(-103,80){$\gamma_i$}
  \put(0,80){$\gamma_j$}
\end{picture}
\end{center}
whenever $\gamma_i|_{U_{ij}} = \gamma_j|_{U_{ij}}$.
This expresses how a path in
a double overlap of two patches can be regarded as a path in either of 
the two patches.

It follows that in $\P_2^C\of{\covering}$ we have 2-commuting
diagrams of the following form
\begin{eqnarray}
\label{triangular cyclinder diagram in Cech-extended 2-groupoid}
\begin{picture}(200,280)
\includegraphics{2concoherence.eps}
  \put(0,0){$(y,k)$}
  \put(-190,0){$(y,i)$}
  \put(0,200){$(x,k)$}
  \put(-190,200){$(x,i)$}
  \put(-80,66){${}_{(y,j)}$}
  \put(-80,260){${}_{(x,j)}$}
  \put(-3,100){$\gamma_k$}
  \put(-175,100){$\gamma_i$}
  \put(-82,150){${}_{\gamma_j}$}
\end{picture}
\vspace{20pt}
\end{eqnarray}
This is because the front side has the same boundary as the rest and since 
by the above there is precisely one 2-morphism for any boundary of this
form. 

But now there are also
surfaces (bigons) sitting in double overlaps. For these we similarly need
to postulate a 2-commuting (``tin can''-) diagram expressing how they can be 
realized as bigons in either of the two patches:
\begin{eqnarray}
\label{tincan diagram in Cech extended 2-path 2-groupoid}
\begin{picture}(240,180)
 \includegraphics{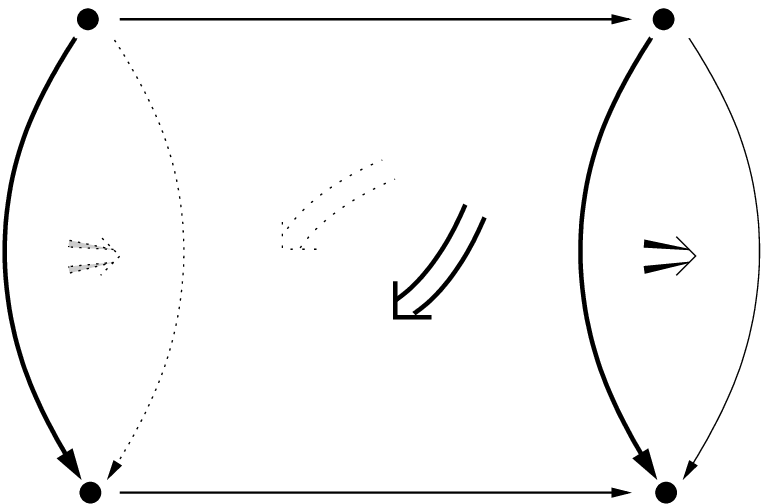}
 \put(-3,-17){
 \begin{picture}(0,0)
 \put(122,0){
  \begin{picture}(0,0)
   \put(-328,165){$(x,i)$}
   \put(-328,5){$(y,i)$}
   \put(-358,70){$\gamma_i$}
   \put(-293,70){${}_{\tilde \gamma_i}$}
   \put(-333,105){$[\Sigma_i]$}
  \end{picture}
 }
 \put(288,0){
  \begin{picture}(0,0)
   \put(-328,165){$(x,j)$}
   \put(-328,5){$(y,j)$}
   \put(-358,70){$\gamma_j$}
   \put(-293,70){${}_{\tilde \gamma_j}$}
   \put(-333,105){$[\Sigma_j]$}
  \end{picture}
 }
 \end{picture}
}
\end{picture}
\end{eqnarray}

\vskip 2em

\end{definition}

\subsubsection{The global 2-Holonomy 2-Functor}
\label{section: The global 2-Holonomy 2-Functor}

We can now define a \emph{global holonomy 2-functor}
to be any 2-functor $\hol$ from the \Cech-extended 2-path 2-groupoid to 
the structure 2-group:
\[
  \hol : \P_2^C\of{\covering} \to G_2
  \,.
\]
As before, it is a matter of introducing notation to define
the following quantities:
\begin{eqnarray*}
\hol\of{
\xymatrix{
   (x,i) \ar@/^1pc/[rr]^{}_{}="0"
&& (x,j)
}
}
&\defas&
\xymatrix{
   \bullet \ar@/^1pc/[rr]^{g_{ij}\of{x}}_{}="0"
&& \bullet
}
\end{eqnarray*}

\begin{eqnarray*}
\hol\of{
\xymatrix{
   (x,i) \ar@/^1pc/[rr]^{}_{}="0"
           \ar@/_1pc/[rr]_{}_{\mathrm{Id}}="1"
           \ar@{=>}"0";"1"^{}
&& (x,i)
}
}
&\defas&
\xymatrix{
   \bullet \ar@/^1pc/[rr]^{g_{ii}\of{x}}_{}="0"
           \ar@/_1pc/[rr]_{}_{1}="1"
           \ar@{=>}"0";"1"^{k_i\of{x}}
&& \bullet
}
\end{eqnarray*}

\begin{eqnarray*}
  \hol\of{
   \raisebox{-40pt}{
  \hspace{20pt}
\begin{picture}(140,100)
  \includegraphics{triangle.eps}
  \put(-139,0){$(x,i)$}
  \put(-70,105){$(x,j)$}
  \put(2,0){$(x,k)$}
\end{picture}
}
}
&\defas&
   \raisebox{-40pt}{
  \hspace{20pt}
 \begin{picture}(140,110)
  \includegraphics{triangle.eps}
  \put(-112,50){$g_{ij}\of{x}$}
  \put(-27,50){$g_{jk}\of{x}$}
  \put(-70,-5){$g_{ik}\of{x}$}
  \put(-70,24){$f_{ijk}\of{x}$}
  \end{picture}
 }
\end{eqnarray*}
\vskip 1em

\begin{eqnarray*}
\hol\of{
\raisebox{-80pt}{
\hspace{23pt}
\begin{picture}(120,160)
  \includegraphics{weakpathtransition.eps}
  \put(2,0){$(y,j)$}
  \put(2,162){$(x,j)$}
  \put(-116,0){$(y,i)$}
  \put(-116,162){$(x,i)$}
  \put(-103,80){$\gamma_i$}
  \put(0,80){$\gamma_j$}
\end{picture}
}
}
&\defas&
\raisebox{-80pt}{
\hspace{33pt}
\begin{picture}(120,160)
  \includegraphics{weakpathtransition.eps}
  \put(-118,80){$\hol_i\of{\gamma}$}
  \put(-3,80){$\hol_j\of{\gamma}$}
  \put(-50,104){$a_{ij}\of{\gamma}$}
  \put(-58,167){$g_{ij}\of{x}$}
  \put(-58,-6){$g_{ij}\of{y}$}
\end{picture}
}
\end{eqnarray*}

\begin{eqnarray*}
\hol\of{
\xymatrix{
   (x,i) \ar@/^1pc/[rr]^{\gamma_1}_{}="0"
           \ar@/_1pc/[rr]_{}_{\gamma_2}="1"
           \ar@{=>}"0";"1"^{[\Sigma]}
&& (y,i)
}
}
&\defas&
\xymatrix{
   \bullet \ar@/^1pc/[rr]^{\hol_i\of{\gamma_1}_{}}="0"
           \ar@/_1pc/[rr]_{}_{\hol_i\of{\gamma_2}}="1"
           \ar@{=>}"0";"1"^{\hol\of{\Sigma}}
&& \bullet
}
\end{eqnarray*}
\vskip 1em

Applying this global 2-holonomy 2-functor $\hol$ 
\begin{itemize}
\item
to the 2-commuting diagram
\refer{tincan diagram in Cech extended 2-path 2-groupoid}
in $\P_2^C\of{\covering}$ yields the transition law
\refer{transition pseudo-natural iso}
\item
to the
2-commuting diagram
\refer{triangular cyclinder diagram in Cech-extended 2-groupoid} 
yields the modification 
\refer{transition modification}
further
discussed in \S\fullref{subsubsec: Transition Law on Triple Overlaps}
\item
to the tetrahedron \refer{tetrahedron in Cech 2-groupoid} in
$C_2\of{\covering}$ yields the tetrahedron law
\refer{transition tetrahedron}.
\end{itemize}
Hence the global 2-holonomy 2-functor encodes precisely the
information of a 2-bundle with 2-connection and 2-holonomy
as defined in \S\fullref{p-holonomy p-functors}.

In complete analogy to how we proceeded before for ordinary
bundles, the 2-functor $\hol$ allows to compute surface
holonomy by ``lifting'' a given surface in $M$ to 	
a 2-morphism in $\P_2\of{\covering}$ and applying $\hol$ on that. 

This procedure is indicated by the following figure.
\clearpage
\begin{center}
\begin{picture}(350,530)
\includegraphics{globalsurfacehol.eps}
\put(-14,16){
\begin{picture}(0,0)
\put(-250,360){$U_i$}
\put(-105,360){$U_k$}
\put(-180,520){$U_j$}
\put(-210,405){$\Sigma_i$}
\put(-140,405){$\Sigma_k$}
\put(-175,470){$\Sigma_j$}
\put(-173,439){$x$}
\put(-182,395){$\gamma_3$}
\put(-140,442){$\gamma_2$}
\put(-200,454){$\gamma_1$}
\put(-168,300){$\hol$}
\put(-186,124){$g_{ik}\of{x}$}
\put(-180,43){$g_{ik}^{-1}$}
\put(-214,160){$g_{ij}\of{x}$}
\put(-160,160){$g_{jk}\of{x}$}
\put(-190,140){$f_{ijk}\of{x}$}
\put(-186,70){$a_{ik}\of{\gamma_3}$}
\put(-238,95){$\hol_i\of{\gamma_3}$}
\put(-147,95){$\hol_k\of{\gamma_3}$}
\put(-260,140){$\hol_i\of{\gamma_1}$}
\put(-110,144){$\hol_k\of{\gamma_2}$}
\put(-220,210){$\hol_j\of{\gamma_1}$}
\put(-160,213){$\hol_j\of{\gamma_2}$}
\put(-267,185){$a_{ij}\of{\gamma_1}$}
\put(-106,185){$a_{jk}\of{\gamma_2}$}
\put(-290,213){$g_{ij}^{-1}$}
\put(-70,213){$g_{jk}^{-1}$}
\put(-295,90){$\hol_i\of{\Sigma_i}$}
\put(-84,90){$\hol_k\of{\Sigma_k}$}
\put(-186,239){$\hol_j\of{\Sigma_j}$}
\end{picture}
}
\end{picture}
\end{center}

That the global holonomy constructed this way is indeed well defined
(invariant under gauge transformations) follows again, as for the
case of 1-bundles discussed before, from the fact that gauge transformations
are nothing but (pseudo-)natural transformations of the global 
2-holonomy 2-functor. This is discussed in the following
\S\ref{gauge transformations in terms of global 2-holonomy 2-functor}.
An analysis of the gauge invariance of the global 2-holony depicted in
this figure from a slightly different
point of view is given in \S\fullref{more details on 2-holonomy}.

It is easy to see that in the abelian case the above figure encodes
precisely the well-known concept of surface holononmy in abelian gerbes
(and hence the proper action functional for strings in 
Kalb-Ramond backgrounds)
as described in 
\cite{Alvarez:1985,MackaayPicken:2000,GawedzkiReis:2002}
and summarized for instance in \cite{AschieriJurco:2004}.

For consider the case where the strict structure 2-group is given by the 
crossed module
$\twogroup = (G=1,H=U\of{1},\alpha = \mathrm{trivial}, t= \mathrm{trivial})$.
Then $\hol_i\of{\Sigma_i}$ is simply the
exponetiated integral of $B_i$ over $\Sigma_i$, $a_{ij}\of{\gamma}$ 
is (according to \refer{aij in the abelian case},
p. \pageref{aij in the abelian case})
simply the line holonomy of $a_{ij}$ along $\gamma$ and the composition 
of all the 2-group elements in the above figure simply amounts to 
multiplying all these elements of $U\of{1}$. This is precisely the
procedure discussed in the above mentioned references.

\subsubsection{Gauge Transformations}
\label{gauge transformations in terms of global 2-holonomy 2-functor}

What is very convenient about the abive formulation, where
all the information about a 2-bundle with 2-connection is 
encoded in a single 2-holonomy 2-functor on a \Cech-extended
path 2-groupoid, is that, as was the case analogously for
ordinary bundles before, the gauge transformations of the
2-bundle arise simply as natural pseudo-isomorphisms between two
such 2-functors.

The effect of a gauge transformation 
\[
  \hol \to \tilde \hol
\]
on the transition functions
$g_{ij}$ and $f_{ijk}$ is given by the following 
naturality diagram:
\begin{eqnarray}
  \label{gauge transformation of fijk}
\begin{picture}(360,220)
  \includegraphics{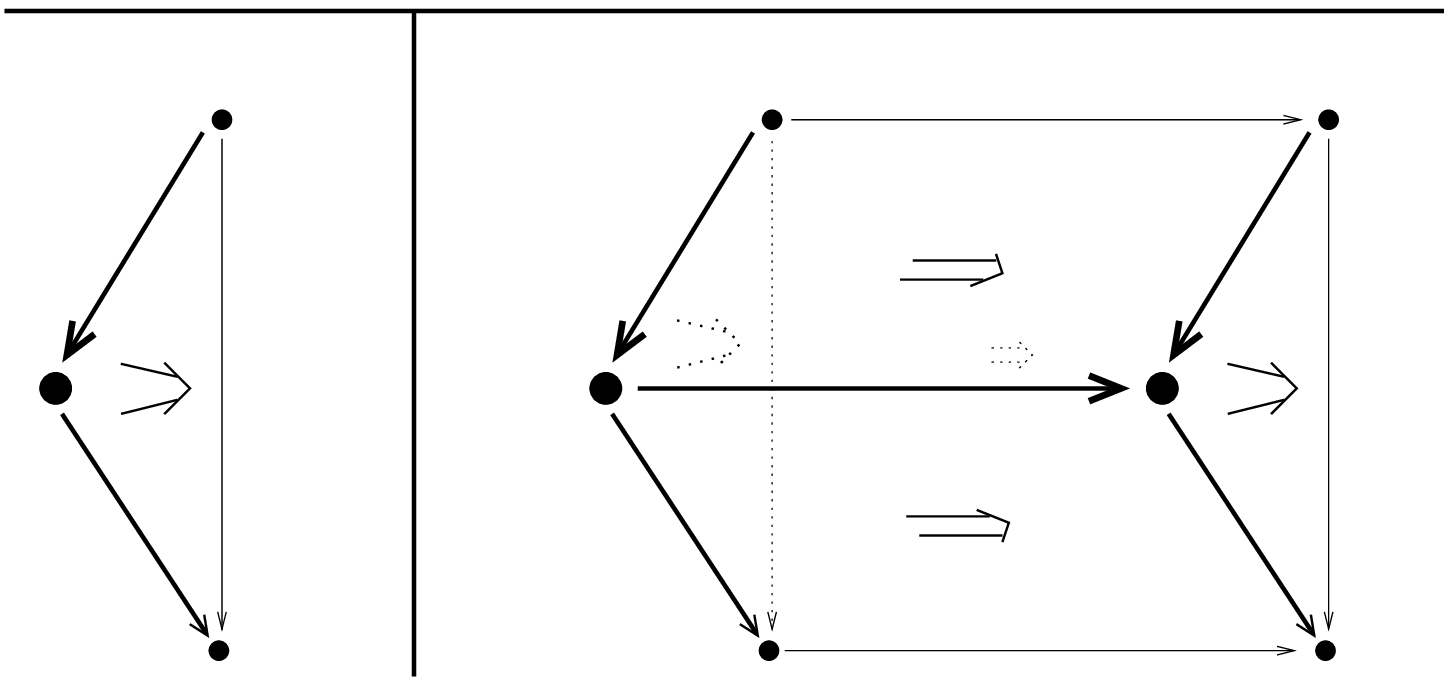}
 \put(-247,135){$g_{ij}\of{x}$}
 \put(-250,35){$g_{jk}\of{x}$}
 \put(-195,60){${}_{g_{ik}\of{x}}$}
 \put(-225,110){${}_{f_{ijk}\of{x}}$}
  \put(157,-2){
  \begin{picture}(0,0)
    \put(-247,135){$\tilde g_{ij}\of{x}$} 
    \put(-250,35){$\tilde g_{jk}\of{x}$}
    \put(-195,60){${}_{\tilde g_{ik}\of{x}}$}
    \put(-225,100){${}_{\tilde f_{ijk}\of{x}}$}
  \end{picture}
  }
  \put(-363,173){${}_{(x,i)}$}
  \put(-363,-2){${}_{(x,k)}$}
  \put(-437,80){$(x,j)$}
 \put(-140,170){${}_{h_i\of{x}}$}
 \put(-174,89){$h_j\of{x}$}
 \put(-140,01){${}_{h_k\of{x}}$}
 \put(-152,128){${}_{p_{ij}\of{x}}$}
 \put(-132,103){${}_{p_{ik}\of{x}}$}
 \put(-156,36){${}_{p_{jk}\of{x}}$}
 \put(-380,203){$\P_2^C\of{U}$}
 \put(-140,203){$G_2$}
\end{picture}
\nonumber\\
\end{eqnarray}
In terms of group elements this implies
\[
  t\of{p_{ij}}g_{ij} = h_i \tilde g_{ij}h_j^{-1} 
\]
as well as a more unwieldy transformation equation for $f_{ijk}$.

The effect of the same gauge transformation on the 2-holonomy
itself is given by
another naturality diagram:
\begin{eqnarray}
\label{diagram: gauge transformation of 2-holonomy 2-functor}
\begin{picture}(440,220)
 \includegraphics{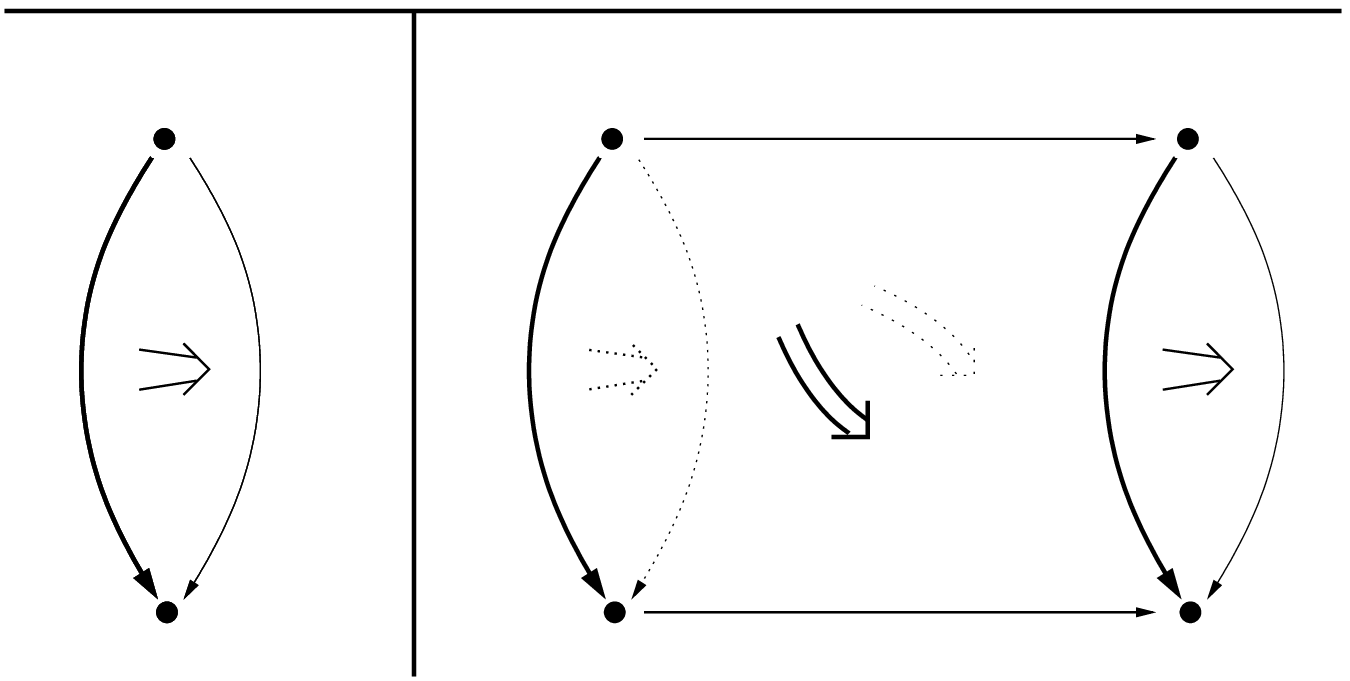}
 \put(-19,-2){
 \begin{picture}(0,0)
 \put(-10,0){
  \begin{picture}(0,0)
   \put(-356,70){$\gamma_1$}
   \put(-293,70){${}_{\gamma_2}$}
   \put(-325,105){$[\Sigma]$}
   \put(-328,168){$(x,i)$}
   \put(-328,7){$(y,i)$}
  \end{picture}
 }
 \put(122,0){
  \begin{picture}(0,0)
   \put(-378,50){$\hol_i\of{\gamma_1}$}
   \put(-293,70){${}_{\hol_i\of{\gamma_2}}$}
   \put(-333,105){$\hol_i\of{\Sigma}$}
   \put(-290,109){$\alpha_{i}\of{\gamma_1}$}
   \put(-250,120){${}_{\alpha_{i}\of{\gamma_2}}$}
   \put(-250,163){$h_i\of{x}$}
   \put(-250,10){$h_i\of{y}$}
  \end{picture}
 }
 \put(288,0){
  \begin{picture}(0,0)
   \put(-378,50){${\tilde \hol_i}\of{\gamma_1}$}
   \put(-293,70){${}_{{\tilde \hol_i}\of{\gamma_2}}$}
   \put(-333,105){${\tilde {\hol_i}}\of{\Sigma}$}
  \end{picture}
 }
 \end{picture}
}
  \put(-360,203){$\P_2^C\of{U}$}
  \put(-135,203){$G_2$}
\end{picture}
\end{eqnarray}
\vskip 2em
whose translation into formulas follows from 
proposition in \S\fullref{subsubsec: Transition Law on Double Overlaps} 
and reads:
\begin{eqnarray*}
  A_i &=& h_i \tilde A_i h_i^{-1} + h_i\extd h_i^{-1} - dt\of{\alpha_i}
  \\
  B_i &=& \alpha\of{h_i}\of{\tilde B_i} + \extd_{A_i}\alpha_i 
            + \alpha_i \wedge \alpha_i
  \,,
\end{eqnarray*}
Here we have again used the same symbol to denote the 
2-group morphism $\alpha_i$ and the 1-form it comes from.

Finally, the effect of the same gauge transformation on the 
transitions $a_{ij}$ of the connection is given by this naturality
diagram:
\begin{eqnarray}
  \label{natural transformation for aij}
  \begin{picture}(400,230)
    \includegraphics{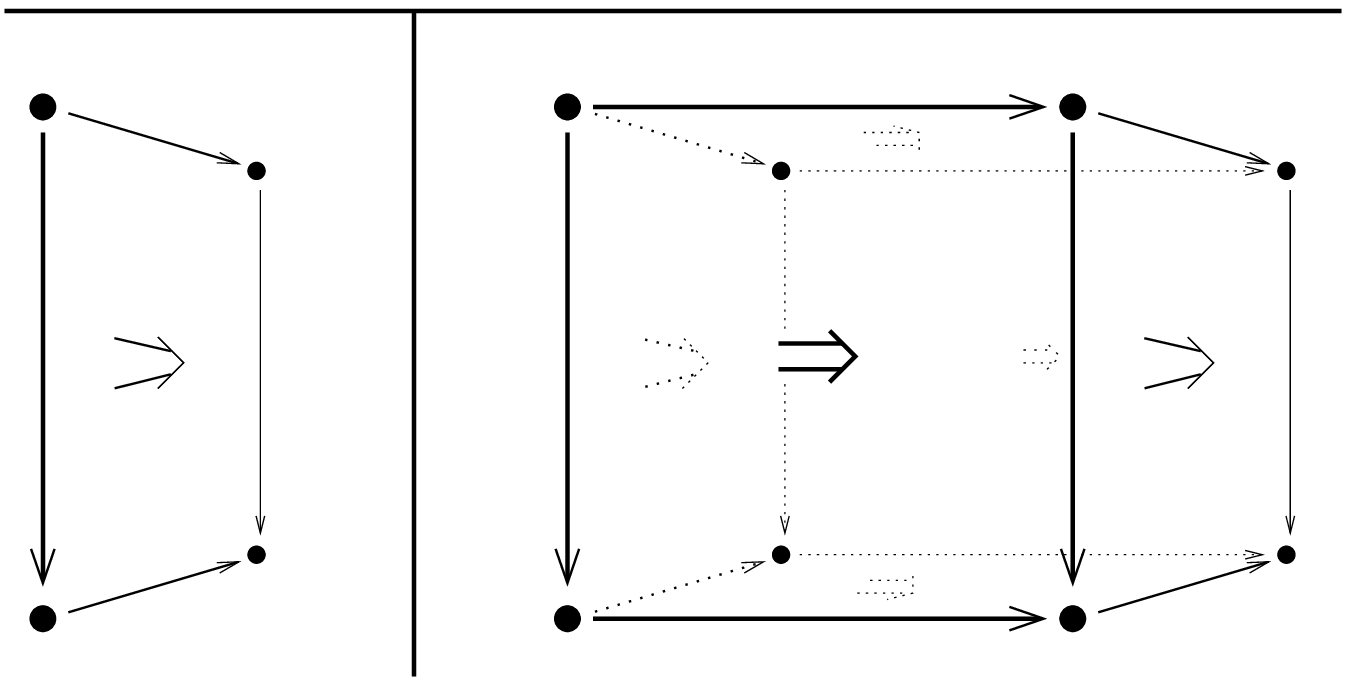}
  \put(-358,200){$\paths^C_2\of{U}$}
  \put(-140,200){$G_2$}
  \put(-390,174){$(x,i)$}
  \put(-390,3){$(y,i)$}
  \put(-390,90){$\gamma_i$}
  \put(-310,149){${}_{(x,j)}$}
  \put(-310,35){${}_{(y,j)}$}
  \put(-312,90){${}_{\gamma_j}$}
  \put(-256,120){$W\of{\gamma_i}$}
  \put(-187,120){${}_{W\of{\gamma_j}}$}  
  \put(-218,93){${}_{a_{ij}\of{\gamma}}$}  
  \put(-205,150){${}_{g_{ij}\of{x}}$}
  \put(-205,34){${}_{g_{ij}\of{y}}$}
  \put(170,0){
    \begin{picture}(0,0)
     \put(-250,120){${\tilde W}\of{\gamma_i}$}
     \put(-187,120){${}_{{\tilde W}\of{\gamma_j}}$}  
     \put(-245,93){${}_{{\tilde a}_{ij}\of{\gamma}}$}  
     \put(-224,163){${}_{{\tilde g}_{ij}\of{x}}$}
     \put(-224,21){${}_{{\tilde g}_{ij}\of{y}}$}
    \end{picture}
   }
   \put(-158,172){$h_i\of{x}$}
   \put(-158,7){$h_i\of{y}$}
   \put(-114,141){${}_{h_j\of{x}}$}
   \put(-114,43){${}_{h_j\of{y}}$}
   \put(-162,157){${}_{p_{ij}\of{x}}$}
   \put(-162,28){${}_{p_{ij}\of{y}}$}
  \put(-160,80){$\alpha_{i}\of{\gamma}$}
  \put(-107,87){${}_{\alpha_{j}\of{\gamma}}$}
  \end{picture}
\end{eqnarray}

These are the (pseudo-)natural transformations of the global 2-holonomy
2-functor.
Note that this can alternatively be understood as a 
(pseudo-)natural transformation of the simplicial map $\Omega$
defining a 2-bundle with 2-holonomy (as depicted in
figure \ref{figure: 2-bundle as simpl map}, 
p. \pageref{figure: 2-bundle as simpl map} and 
detailed in \S\fullref{p-holonomy p-functors}), which assigns
2-holonomy 2-functors $\hol_i$ to patches $U_i$, assigns
pseudo-natural transformations $\hol_i \stackto{g_{ij}} \hol_j$ 
between these to double overlaps $U_{ij}$ and assigns
modifications $g_{ik} \stackto{f_{ijk}} g_{ij} \circ g_{jk}$
to triple overlaps $U_{ijk}$. 
A transformation of such a map $\Omega$ looks like this:
 \begin{center}
\begin{picture}(420,220)
  \includegraphics{2nattraftriangle.eps}
 \put(-234,135){$g_{ij}$}
 \put(-234,35){$g_{jk}$}
 \put(-195,60){${}_{g_{ik}}$}
 \put(-225,110){${}_{f_{ijk}}$}
 \put(-214,167){${}_{\hol_i}$}
 \put(-214,2){${}_{\hol_k}$}
 \put(-269,83){$\hol_j$}
  \put(157,-2){
  \begin{picture}(0,0)
    \put(-233,135){$\tilde g_{ij}$} 
    \put(-233,35){$\tilde g_{jk}$}
    \put(-193,60){${}_{\tilde g_{ik}}$}
    \put(-225,102){${}_{\tilde f_{ijk}}$}
    \put(-214,170){${}_{\tilde \hol_i}$}
    \put(-214,4){${}_{\tilde \hol_k}$}
    \put(-264,73){$\tilde \hol_j$}
  \end{picture}
  }
  \put(-357,170){${}_{i}$}
  \put(-357,-2){${}_{k}$}
  \put(-417,80){$j$}
 \put(-140,168){${}_{h_i}$}
 \put(-174,92){$h_j$}
 \put(-140,01){${}_{h_k}$}
 \put(-152,128){${}_{p_{ij}}$}
 \put(-132,103){${}_{p_{ik}}$}
 \put(-156,36){${}_{p_{jk}}$}
 \put(-380,203){$$}
 \put(-140,203){$G_2^{\P_2\of{U_{ijk}}}$}
\end{picture}
\end{center}

Here $G_2^{\P_2\of{U_{ijk}}}$ is the (2-)functor
(2-)category 
(\cf
\S\fullref{elements of cat theory: 2-functors and pseudo-nat trafos})
of 2-holonomy 2-functors form the 2-path 2-groupoid $\P_2\of{U_{ijk}}$
of surfaces in the triple overlap $U_{ijk}$ to the structure 2-group
$G_2$.

The existence of 
\[
  \hol_i \stackto{h_i} \tilde \hol_i
\]
is equivalent to diagram 
\refer{diagram: gauge transformation of 2-holonomy 2-functor},
the existence of 
\[
  g_{ij} \stackto{p_{ij}}  h_i \circ \tilde g_{ij} \circ h_j^{-1}
\]
is given by diagram 
\refer{natural transformation for aij} and the 2-commutativity 
is given by diagram 
\refer{gauge transformation of fijk}.

\subsubsection{More Details}
\label{more details on 2-holonomy}

The reader might complain that the introduction of 
2-connections with 2-holonomy in 2-bundles in 
\S\fullref{p-holonomy p-functors} does not obviously 
follow the categorification dictionary 
advertized in \S\fullref{Mathematical Motivations}. 
But in fact it does. Spelling this out in a little detail
helps elucidate the nature of the more concise definitions
in terms of pseudo-natural transformations of 2-functors
that we have discussed above.

To begin with, consider the \emph{equation} which describes the
transition of an ordinary holonomy 
of a path $x \stackto{\gamma} y$
in an ordinary (1-)bundle from patch $U_i$ to patch $U_j$:
\[
  \hol_i\of{\gamma} = 
  g_{ij}\of{x}
  \cdot
  \hol_j\of{\gamma}
  \cdot
  g_{ij}^{-1}\of{y}
  \,.
\]
This is an equation between group element valued functions, where
$\hol_i$ is regarded as a function on path space $P\of{U_i}$ and
where $\cdot$ denotes the ordinary group product operation.
Following the dictionary in \S\fullref{Mathematical Motivations}
we are to replace this by a natural isomorphism between 2-group-valued
functors, where on the right the product is to become the product
functor in the 2-group. Restricting to the special case that
the categorification of $g_{ij}\of{x}$ are identity morphisms as in 
\refer{nat diagram for transition on catdisc base space}
(p. \pageref{nat diagram for transition on catdisc base space}),
this natural isomorphism is expressed as follows:
\begin{eqnarray}
\label{categorification of holonomy transition law}
\begin{picture}(400,150)
\includegraphics{2gauge1.eps}
\put(-334,55){$\hol_i\of{\Sigma}$}
\put(-94,58){$\hol_j\of{\Sigma}$}
\put(-365,94){$\hol_i\of{x \stackto{\gamma_1} y}$}
\put(-365,16){$\hol_i\of{x \stackto{\gamma_2} y}$}
\put(-149,88){$\hol_j\of{\gamma_1}$}
\put(-149,26){$\hol_j\of{\gamma_2}$}
\put(-180,65){$g_{ij}\of{x}$}
\put(-42,65){$g_{ij}^{-1}\of{y}$}
\put(-92,15){$\bar a_{ij}\of{\gamma_2} $}
\put(-92,101){$a_{ij}\of{\gamma_1}$}
\put(-250,55){$=$}
\put(-110,124){$\hol_i\of{\gamma_1}$}
\put(-110,-12){$\hol_i\of{\gamma_2}$}
\end{picture}
\end{eqnarray}
Here $\bar a_{ij}$ is the inverse of the morphism $a_{ij}$, which
encodes the natural transformation. Note how the 
``horizontal conjugation'' by $g_{ij}$ is accompanied now by 
a ``vertical conjugation'' by $a_{ij}$. 

But this is nothing but the 2-commutativity of the diagram
\refer{transition pseudo-natural iso}
(p. \pageref{transition pseudo-natural iso}), which expressed the
existence of a pseuo-natural transformation between 
2-holonomy 2-functors.

\paragraph{Transition on Triple Overlaps.}

From this perspective, the diagram
\refer{transition modification}
(p. \pageref{transition modification})
arises as a \emph{coherence law} for the above natural isomorphism.
Namely consider a transition
\[
  U_i \to U_j \to U_k \to U_i
\]
from $U_i$ to \emph{itself}
using $a_{ij}$, $a_{jk}$ and $a_{ki}$, and demand that the result
be the identity transformation. By the above, this amounts to
demanding that this diagram
\begin{center}
  \begin{picture}(600,150)
    \put(-60,0){\includegraphics{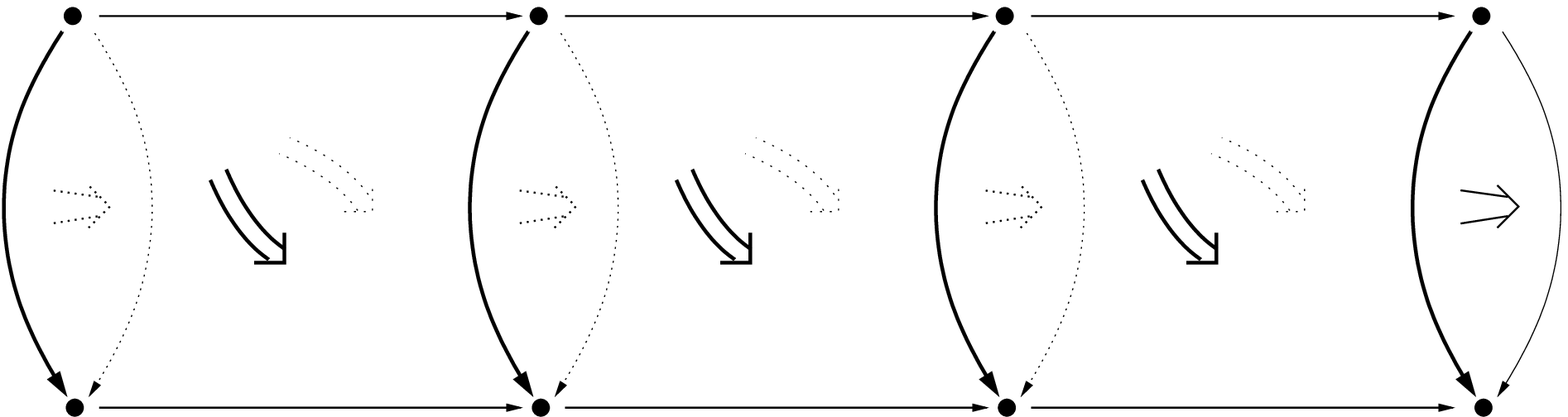}}
 \put(160,-17){\begin{picture}(0,0)
 \put(122,0){
  \begin{picture}(0,0)
   \put(-338,60){$\hol_i\of{\gamma_1}$}
   \put(-293,70){${}_{\hol_i\of{\gamma_2}}$}
   \put(-333,105){$\hol_i\of{\Sigma}$}
   \put(-290,106){$a_{ij}\of{\gamma_1}$}
   \put(-250,118){${}_{a_{ij}\of{\gamma_2}}$}
   \put(-250,163){$g_{ij}\of{x}$}
   \put(-250,10){$g_{ij}\of{y}$}
  \end{picture}
 }
 \put(288,0){
  \begin{picture}(0,0)
   \put(-382,50){$\hol_j\of{\gamma_1}$}
   \put(-293,70){${}_{\hol_j\of{\gamma_2}}$}
   \put(-333,105){$\hol_j\of{\Sigma}$}
  \end{picture}
 }
 \end{picture}}
 \put(320,-17){\begin{picture}(0,0)
 \put(122,0){
  \begin{picture}(0,0)
   \put(-290,106){$a_{jk}\of{\gamma_1}$}
   \put(-250,118){${}_{a_{jk}\of{\gamma_2}}$}
   \put(-250,163){$g_{jk}\of{x}$}
   \put(-250,10){$g_{jk}\of{y}$}
  \end{picture}
 }
 \put(288,0){
  \begin{picture}(0,0)
   \put(-382,50){$\hol_k\of{\gamma_1}$}
   \put(-293,70){${}_{\hol_k\of{\gamma_2}}$}
   \put(-333,105){$\hol_k\of{\Sigma}$}
  \end{picture}
 }
 \end{picture}}
 \put(487,-17){\begin{picture}(0,0)
 \put(122,0){
  \begin{picture}(0,0)
   \put(-290,106){$a_{ki}\of{\gamma_1}$}
   \put(-250,118){${}_{a_{ki}\of{\gamma_2}}$}
   \put(-250,163){$g_{ki}\of{x}$}
   \put(-250,10){$g_{ki}\of{y}$}
  \end{picture}
 }
 \put(288,0){
  \begin{picture}(0,0)
   \put(-382,50){$\hol_i\of{\gamma_1}$}
   \put(-328,60){${}_{\hol_i\of{\gamma_2}}$}
   \put(-333,105){$\hol_i\of{\Sigma}$}
  \end{picture}
 }
 \end{picture}}
  \end{picture}
\end{center}
equals this diagram:
\begin{center}
\begin{picture}(240,160)
 \includegraphics{tincan.eps}
 \put(-3,-17){
 \begin{picture}(0,0)
 \put(122,0){
  \begin{picture}(0,0)
   \put(-385,70){$\hol_i\of{\gamma_1}$}
   \put(-293,70){${}_{\hol_i\of{\gamma_2}}$}
   \put(-333,105){$\hol_i\of{\Sigma}$}
   \put(-282,104){$\mathrm{Id}$}
   \put(-255,113){${}_{\mathrm{Id}}$}
   \put(-250,163){$\mathrm{Id}$}
   \put(-250,10){$\mathrm{Id}$}
  \end{picture}
 }
 \put(288,0){
  \begin{picture}(0,0)
   \put(-385,70){$\hol_i\of{\gamma_1}$}
   \put(-293,70){${}_{\hol_i\of{\gamma_2}}$}
   \put(-333,105){$\hol_i\of{\Sigma}$}
  \end{picture}
 }
 \end{picture}
}
\end{picture}
\end{center}

We want to show that this equality gives the 2-commutativity of the
diagram 
\refer{transition modification}
(p. \pageref{transition modification}).

\newpage

For this purpose it is convenient to first redraw these cyclinder-like
diagrams in planar form as follows:

\hspace{-1cm}
\begin{picture}(500,420)
\includegraphics{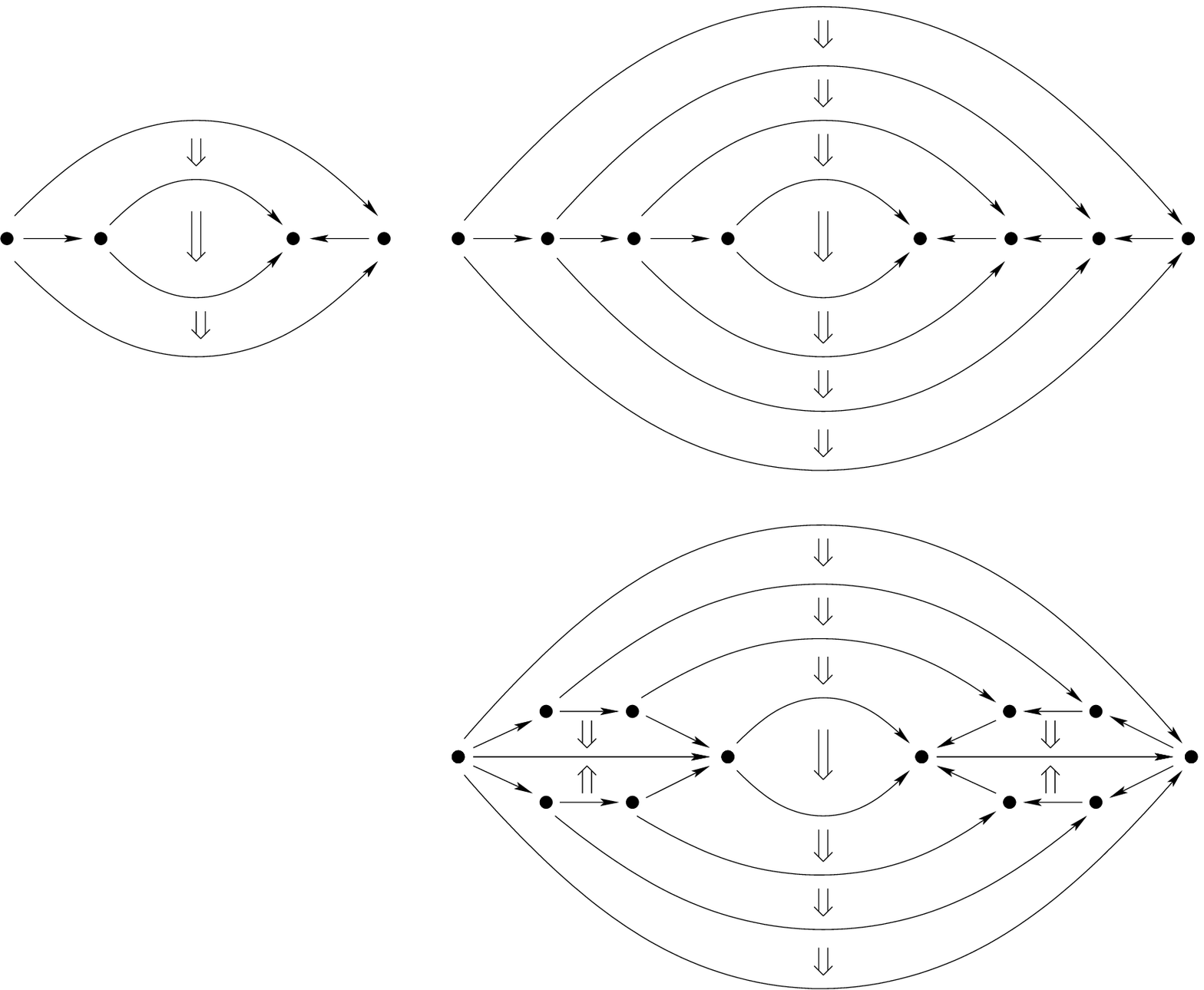}
\put(-322,303){$=$}
\put(-322,94){$=$}
\put(-470,309){${}_{\mathrm{Id}}$}
\put(-420,340){${}_{\mathrm{Id}}$}
\put(-420,270){${}_{\mathrm{Id}}$}
\put(-438,305){${}_{ \hol_i\of{\Sigma} }$}
\put(-358,309){${}_{\mathrm{Id}}$}
\put(-420,360){${}_{\hol_i\of{\gamma_1}}$}
\put(-397,333){${}_{\hol_i\of{\gamma_1}}$}
\put(-397,280){${}_{\hol_i\of{\gamma_2}}$}
\put(-420,250){${}_{\hol_i\of{\gamma_2}}$}
\put(250,0){
\begin{picture}(0,0)
\put(-473,311){${}_{g_{ki}\of{x}}$}
\put(-543,311){${}_{g_{ij}\of{x}}$}
\put(-509,311){${}_{g_{jk}\of{x}}$}
\put(-320,311){${}_{g_{jk}\of{y}}$}
\put(-433,338){${}_{a_{ki}\of{\gamma_1}}$}
\put(-433,390){${}_{a_{ij}\of{\gamma_1}}$}
\put(-435,365){${}_{a_{jk}\of{\gamma_1}}$}
\put(-435,250){${}_{\bar a_{jk}\of{\gamma_2}}$}
\put(-433,223){${}_{\bar a_{ij}\of{\gamma_2}}$}
\put(-433,274){${}_{\bar a_{ki}\of{\gamma_2}}$}
\put(-438,305){${}_{{{\hol_i}}\of{\Sigma}}$}
\put(-358,311){${}_{g_{ij}\of{y}}$}
\put(-285,311){${}_{g_{ki}\of{y}}$}
\put(-348,362){${}_{\hol_j\of{\gamma_1}}$}
\put(-363,343){${}_{\hol_k\of{\gamma_1}}$}
\put(-363,267){${}_{\hol_k\of{\gamma_2}}$}
\put(-378,290){${}_{\hol_i\of{\gamma_2}}$}
\put(-382,323){${}_{\hol_i\of{\gamma_1}}$}
\put(-350,248){${}_{\hol_j\of{\gamma_1}}$}
\put(-330,380){${}_{\hol_i\of{\gamma_1}}$}
\put(-330,232){${}_{\hol_i\of{\gamma_2}}$}
\end{picture}
}
\put(250,-211){
\begin{picture}(0,0)
\put(-468,323){${}_{g_{ki}\of{x}}$}
\put(-541,322){${}_{g_{ij}\of{x}}$}
\put(-506,329){${}_{g_{jk}\of{x}}$}
\put(-468,293){${}_{g_{ki}\of{x}}$}
\put(-541,293){${}_{g_{ij}\of{x}}$}
\put(-506,284){${}_{g_{jk}\of{x}}$}
\put(-498,315){${}_{f_{ijk}\of{x}}$}
\put(-498,298){${}_{f_{ijk}\of{x}}$}
\put(-520,310){${}_{\mathrm{Id}}$}
\put(182,0){
\begin{picture}(0,0)
\put(-468,323){${}_{g_{ki}\of{y}}$}
\put(-541,322){${}_{g_{ij}\of{y}}$}
\put(-506,329){${}_{g_{jk}\of{y}}$}
\put(-468,293){${}_{g_{ki}\of{y}}$}
\put(-541,293){${}_{g_{ij}\of{y}}$}
\put(-506,284){${}_{g_{jk}\of{y}}$}
\put(-498,315){${}_{f_{ijk}\of{y}}$}
\put(-498,298){${}_{f_{ijk}\of{y}}$}
\put(-520,310){${}_{\mathrm{Id}}$}
\end{picture}
}
\put(-433,338){${}_{a_{ki}\of{\gamma_1}}$}
\put(-433,390){${}_{a_{ij}\of{\gamma_1}}$}
\put(-435,365){${}_{a_{jk}\of{\gamma_1}}$}
\put(-435,250){${}_{\bar a_{jk}\of{\gamma_2}}$}
\put(-433,223){${}_{\bar a_{ij}\of{\gamma_2}}$}
\put(-433,274){${}_{\bar a_{ki}\of{\gamma_2}}$}
\put(-438,305){${}_{{{\hol_i}}\of{\Sigma}}$}
\put(-347,363){${}_{\hol_j\of{\gamma_1}}$}
\put(-360,346){${}_{\hol_k\of{\gamma_1}}$}
\put(-359,263){${}_{\hol_k\of{\gamma_2}}$}
\put(-388,282){${}_{\hol_i\of{\gamma_2}}$}
\put(-388,328){${}_{\hol_i\of{\gamma_1}}$}
\put(-347,245){${}_{\hol_j\of{\gamma_2}}$}
\put(-330,380){${}_{\hol_i\of{\gamma_1}}$}
\put(-330,232){${}_{\hol_i\of{\gamma_2}}$}
\end{picture}
}
\end{picture}

\vskip 2em

The first equality sign here expresses the above 
cylindrical diagram.
The point here is the step after the second equality sign, where
two pairs of mutually cancelling $f_{ijk}$ 2-morphisms have
been inserted in order to obtain the two horizontal identity
1-morphisms. (Note that these are really defined only up to a 
2-morphism $\mathrm{Id} \to \mathrm{Id}$, given by an element
$h \in \ker{t} \subset H$. A nontrivial $h$ here would give the
transition law for a \emph{3-bundle}.)

By comparison one sees that the upper half and the
lower half of the diagram after the second equality sign are 
just a planar version of the diagram discussed in 
\S\fullref{subsubsec: Transition Law on Triple Overlaps}. 
The equality of the
above two cyclindrical diagrams is equivalent to the 2-commutativity of 
this diagram. 
Hence we can interpret the existence of the modification
\refer{transition modification} in a 2-bundle with 2-holonomy
as a coherence law on the transition on double overlaps.

As a byproduct, this tells us that equation 
\refer{equation: the cocycle condition on triple overlaps}
(p. \pageref{equation: the cocycle condition on triple overlaps})
between the differential forms encoding these transitions
can be understood as equating the triple application of 
equations \refer{equation: the transition cocycle condition}
with the identity transformation.

In order to see this,
begin by writing down the transition law for the connection 1-form from
$U_i$ to $U_j$:
\[
  A_i =
  g_{ij} A_j (g_{ij})^{-1}
  +
  g_{ij}(\extd (g_{ij})^{-1})
  -
  dt\of{a_{ij}}
\]
This expresses $A_i$ in terms of $A_j$. Now 
use the same transition law but from $U_j$ to $U_k$
in order to express the
$A_j$ in this formula
in terms of $A_k$:
\[
  \cdots = 
  g_{ij}
  \underbrace{
    \left(
    g_{jk}A_k g_{jk}^{-1}
    +
    g_{jk}(\extd g_{jk}^{-1})
    -
    a_{jk}
  \right)
  }_{= A_j}
   (g_{ij})^{-1}
  +
  g_{ij}(\extd (g_{ij})^{-1})
  -
  dt\of{a_{ij}}  
\]
Finally, express $A_k$ in terms of the original $A_i$:
\begin{eqnarray*}
  \cdots
  &=&
  g_{ij}
  \underbrace{
  \left(
    g_{jk}
    \underbrace{
    \left(
       g_{ki}A_i g_{ki}^{-1}
       +
       g_{ki}(\extd g_{ki}^{-1})
       -
       dt\of{a_{ki}}
    \right)
     }_{A_k}
    (g_{jk})^{-1}
    g_{jk}(\extd g_{jk}^{-1})
    -
    dt\of{a_{jk}}
  \right)
  }_{= A_j}
   (g_{ij})^{-1}
    \\
   && +
  g_{ij}(\extd (g_{ij})^{-1})
  -
  dt\of{a_{ij}}  
\end{eqnarray*}
After multiplying out the brackets this reads
\begin{eqnarray*}
  A_i 
  &=&
  (g_{ij}g_{jk}g_{ki})
  A_i
  (g_{ij}g_{jk}g_{ki})^{-1}
  +
  (g_{ij}g_{jk}g_{ki})
  \extd
  (g_{ij}g_{jk}g_{ki})^{-1}  
  \\
  &&
  -
  dt\of{a_{ij}}
  -
  g_{ij}dt\of{a_{jk}}(g_{ij})^{-1}
  -
  g_{ij}g_{jk}dt\of{a_{ki}}(g_{ij}g_{jk})^{-1}
\end{eqnarray*}
Using the relation 
\refer{source/target matching in proof for 2-transition}
\[
  t\of{f_{ijk}}g_{ik} = g_{ij}g_{jk}
\]
this is simplified to
\begin{eqnarray*}
  A_i
  &=&
  A_i 
  +
  t\of{h_{ijk}}
  \commutator
  {A_i}
  {t\of{h_{ijk}}^{-1}}
  +
  t\of{h_{ijk}}
  \extd
  \left(t\of{h_{ijk}}^{-1}\right)  
  \\
  &&
  -
  dt\of{a_{ij}}
  -
  g_{ij}dt\of{a_{jk}}(g_{ij})^{-1}
  -
  t\of{h_{ijk}}g_{ik} \,dt\of{a_{ki}}\,g_{ik} (t\of{h_{ijk}})^{-1}  
  \,.
\end{eqnarray*}
Finally we can factor out the action of $dt$:
\begin{eqnarray*}
  A_i 
  &=&
  A_i
  + 
  dt\of{
    h_{ijk}d\alpha\of{A_i}\of{h_{ijk}^{-1}}
    +
    h_{ijk}\extd h_{ijk}^{-1}
    -
    a_{ij}
    -
    \alpha\of{g_{ij}}\of{a_{jk}}
    -
    h_{ijk}\,\alpha\of{g^1_{ik}}\of{a_{ki}}\, h_{ijk}^{-1}
  }
  \,.
  \nonumber
\end{eqnarray*}
It follows that the term in brackets has to be in the kernel of 
$dt$, i.e.
\begin{eqnarray}
    h_{ijk}d\alpha\of{A_i}\of{h_{ijk}^{-1}}
    +
    h_{ijk}\extd h_{ijk}^{-1}
    -
    a_{ij}
    -
    g_{ij}\of{a_{jk}}
    -
    h_{ijk}\,d\alpha\of{g^1_{ik}}\of{a_{ki}}\, h_{ijk}^{-1}
   &=& 
   -\alpha_{ijk}
  \nonumber
\end{eqnarray}
with $\alpha_{ijk} \in \ker\of{dt}$.
This can be simplified a little further: For
$j=k$ this equation reduces to
\begin{eqnarray*}
    a_{ik}
    +
    \alpha\of{g_{ik}}\of{a_{ki}}
   &=& 
   0
   \,.
\end{eqnarray*}
Reinserting this result yields
\begin{eqnarray*}
    h_{ijk}d\alpha\of{A_i}\of{h_{ijk}^{-1}}
    +
    h_{ijk}\extd h_{ijk}^{-1}
    -
    a_{ij}
    -
    g^1_{ij}\of{a_{jk}}
    +
    h_{ijk} a_{ik} h_{ijk}^{-1}
   &=& 
   -\alpha_{ijk}
  \,.
\end{eqnarray*}
Here $\alpha_{ijk}$ comes from the freedom to insert morphisms
$\mathrm{Id}\to \mathrm{Id}$, mentioned above, which corresponds to
freedom present in 3-bundles. Setting $\alpha_{ijk} = 0$ yields the
promised equation \refer{equation: the cocycle condition on triple overlaps}.

\paragraph{Gauge Transformations}
\label{2-Gauge Transformations}

In the same vein we can also understand the gauge transformations
of 2-holonomy in 2-bundles that were discussed in
\S\fullref{gauge transformations in terms of global 2-holonomy 2-functor}.

All we need is a categorification of the \emph{equation}
\[
  \tilde g_{ij} = h_i \cdot g_{ij} \cdot h_j^{-1}
  \,,
\]
which expresses the gauge transformed transition function
$\tilde g_{ij}$ in terms of the original one for ordinary bundles.
As in \refer{nat diagram for transition on catdisc base space}
(p. \pageref{nat diagram for transition on catdisc base space})
this gives a natural isomorphism
\[
\xymatrix{
   \bullet \ar@/^1pc/[rr]^{h_i g_{ij} h_j^{-1}}_{}="0"
           \ar@/_1pc/[rr]_{\tilde g_{ij}}_{}="1"
           \ar@{=>}"0";"1"^{p_{ij}}
&& \bullet
}
\]
such that we have this naturality diagram:
\begin{eqnarray}
  \label{nat diagram for transition on catdisc base space}
  \begin{picture}(370,150)
    \includegraphics{nattraf.eps}
    \put(-296,142){$U_{ij}$}
    \put(-100,142){$\twogroup$}
    \put(-296,107){$x$}
    \put(-296,20){$x$}
    \put(-305,65){$\mathrm{Id}$}
    \put(-180,107){$g_{ij}\of{x}$}
    \put(-180,20){$g_{ij}\of{x}$}
    \put(-170,65){$\mathrm{Id}$}
    \put(-50,107){$h_i \cdot g_{ij} \cdot h_j^{-1}\of{x}$}
    \put(-50,20){$h_i \cdot g_{ij} \cdot h_j^{-1}\of{x}$}
    \put(-33,65){$\mathrm{Id}$}
    \put(-120,104){$p_{ij}\of{x}$}
    \put(-120,20){$p_{ij}\of{x}$}
  \end{picture}
\end{eqnarray}

This implies for $f_{ijk}$ the gauge transformation law
\begin{center}
\begin{picture}(320,150)
\includegraphics{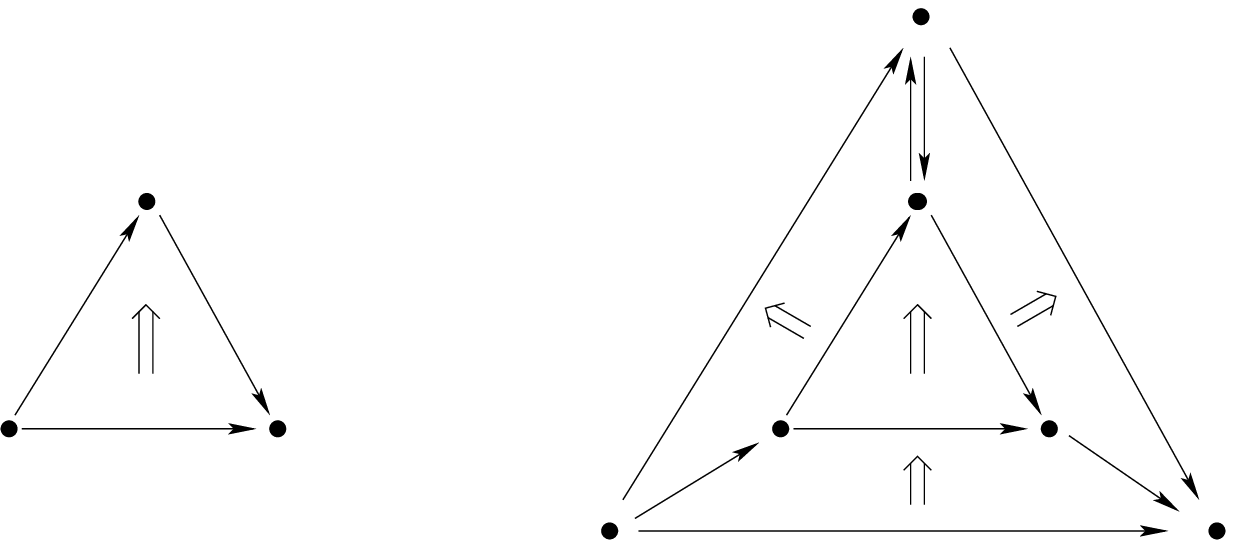}
\put(-210,70){$=$}
\put(-352,67){$\tilde g_{ij}$}
\put(-289,67){$\tilde g_{jk}$}
\put(-315,22){$\tilde g_{ik}$}
\put(-334,50){${\tilde f}_{ijk}$}
\put(237,0){
  \begin{picture}(30,30)
\put(-352,87){$g_{ij}$}
\put(-320,87){$g_{jk}$}
\put(-322,25){$g_{ik}$}
\put(-327,50){${f}_{ijk}$}
\put(-378,14){$h_i$}
\put(-290,11){$h_k^{-1}$}
\put(-334,-7){$\tilde g_{ik}$}
\put(-383,92){$\tilde g_{ij}$}
\put(-285,92){$\tilde g_{jk}$}
\put(-350,110){$h_j^{-1}$}
\put(-325,115){$h_j$}
\put(-347,15){$\bar p_{ik}$}
\put(-377,54){$p_{ij}$}
\put(-294,57){$p_{jk}$}
\end{picture}
}
\end{picture}
\end{center}
The equality here expresses precisely the 2-commutativity of the
naturality diagram \refer{gauge transformation of fijk}
(p. \pageref{gauge transformation of fijk}).

Furthermore, the gauge transformation equation for the holonomy
\[
  \tilde \hol_i\of{\gamma}
  =
  h_i\of{x} \cdot \hol_i\of{\gamma} \cdot h_i^{-1}\of{x}
\]
is categorified precisely as in 
\refer{categorification of holonomy transition law}
(p. \pageref{categorification of holonomy transition law}, recall the
discussion there) and yields
\begin{center}
\begin{picture}(400,150)
\includegraphics{2gauge1.eps}
\put(-334,55){${\tilde {\hol_i}}\of{\Sigma}$}
\put(-94,58){$\hol_i\of{\Sigma}$}
\put(-349,94){$\tilde \hol_i\of{\gamma_1}$}
\put(-349,16){$\tilde \hol_i\of{\gamma_2}$}
\put(-149,88){$\hol_i\of{\gamma_1}$}
\put(-149,26){$\hol_i\of{\gamma_2}$}
\put(-177,65){$h_i\of{x}$}
\put(-42,65){$h_i^{-1}\of{y}$}
\put(-92,15){$\alpha_i\of{\gamma_2} $}
\put(-92,101){$\alpha_i\of{\gamma_1}$}
\put(-250,55){$=$}
\put(-110,124){$\tilde \hol_i\of{\gamma_1}$}
\put(-110,-12){$\tilde \hol_i\of{\gamma_2}$}
\end{picture}
\end{center}
\vskip 1em
This yields the 2-commutativity of 
\refer{diagram: gauge transformation of 2-holonomy 2-functor}
(p. \pageref{diagram: gauge transformation of 2-holonomy 2-functor}).

What requires, once again, a little more work is the analogous
discussion for the $a_{ij}$
\begin{eqnarray}
\begin{picture}(180,80)
\includegraphics{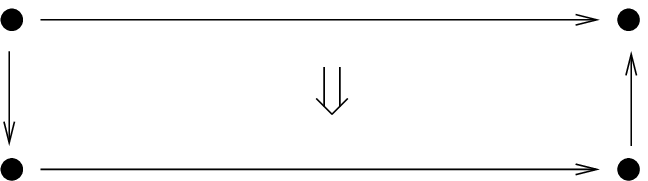}
\put(-80,25){$a_{ij}\of{x \stackto{\gamma} y}$}
\put(-120,57){$\hol_i\of{x \stackto{\gamma} y}$}
\put(-120,-12){$\hol_j\of{x \stackto{\gamma} y}$}
\put(-213,25){$g_{ij}\of{x}$}
\put(2,25){$g_{ij}^{-1}\of{y}$}
\end{picture}
  \label{aij transformation diagram}
\end{eqnarray}
giving the derivation of 
\refer{natural transformation for aij}
(p. \pageref{natural transformation for aij}).

This is obtained by considering a transition first in one
gauge and then in the other:
\hspace{-1cm}
\begin{picture}(500,420)
\includegraphics{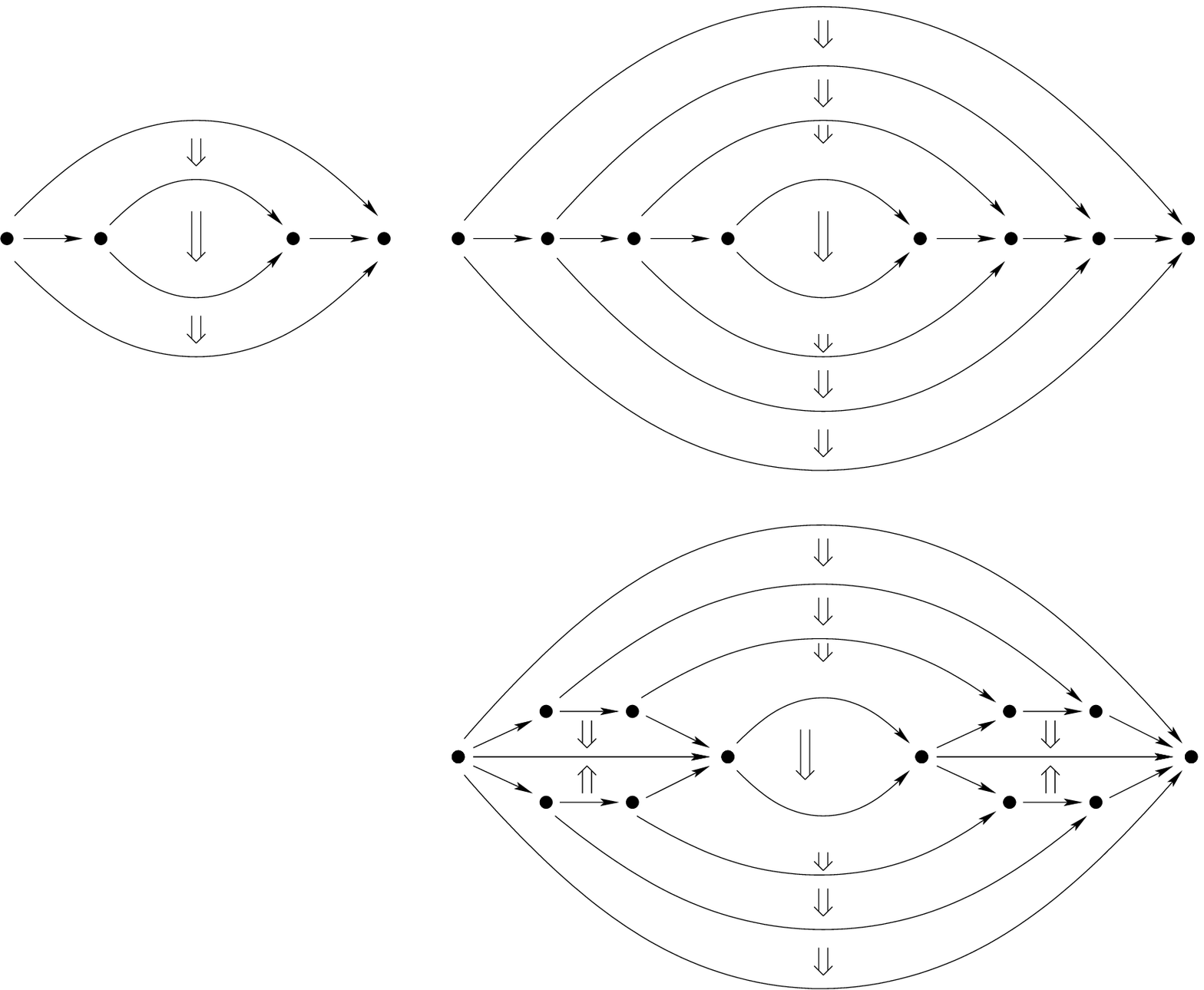}
\put(-322,303){$=$}
\put(-322,94){$=$}
\put(-470,312){${}_{\tilde g_{ij}}$}
\put(-436,340){${}_{\tilde a_{ij}\of{\gamma_1}}$}
\put(-436,270){${}_{\tilde a_{ij}\of{\gamma_2}}$}
\put(-438,305){${}_{ \tilde {\hol_j}\of{\Sigma}}$}
\put(-358,313){${}_{\tilde g_{ij}^{-1}}$}
\put(-420,360){${}_{\tilde \hol_i\of{\gamma_1}}$}
\put(-397,333){${}_{\tilde \hol_j\of{\gamma_1}}$}
\put(-397,280){${}_{\tilde \hol_j\of{\gamma_2}}$}
\put(-420,250){${}_{\tilde \hol_i\of{\gamma_2}}$}
\put(250,0){
\begin{picture}(0,0)
\put(-470,315){${}_{h_j^{-1}}$}
\put(-540,310){${}_{h_i}$}
\put(-506,312){${}_{g_{ij}}$}
\put(-320,315){${}_{g_{ij}^{-1}}$}
\put(-433,338){${}_{1_{h_j^{-1}}\cdot\bar \alpha_j\of{\gamma_1}\cdot 1_{h_j}}$}
\put(-433,390){${}_{\alpha_i\of{\gamma_1}}$}
\put(-435,365){${}_{a_{ij}\of{\gamma_1}}$}
\put(-435,250){${}_{\bar a_{ij}\of{\gamma_2}}$}
\put(-433,223){${}_{\bar \alpha_i\of{\gamma_2}}$}
\put(-433,274){${}_{1_{h_j^{-1}}\cdot\alpha_j\of{\gamma_2}\cdot 1_{h_j}}$}
\put(-436,305){${}_{{\tilde {\hol_j}}\of{\Sigma}}$}
\put(-358,311){${}_{h_j}$}
\put(-285,313){${}_{h_i^{-1}}$}
\put(-348,362){${}_{\hol_i\of{\gamma_1}}$}
\put(-363,343){${}_{\hol_j\of{\gamma_1}}$}
\put(-363,267){${}_{\hol_j\of{\gamma_2}}$}
\put(-378,290){${}_{\tilde \hol_j\of{\gamma_2}}$}
\put(-380,323){${}_{\tilde \hol_j\of{\gamma_1}}$}
\put(-350,248){${}_{\hol_i\of{\gamma_2}}$}
\put(-330,380){${}_{\tilde \hol_i\of{\gamma_1}}$}
\put(-330,232){${}_{\tilde \hol_i\of{\gamma_2}}$}
\end{picture}
}
\put(250,-211){
\begin{picture}(0,0)
\put(-468,323){${}_{h_j^{-1}}$}
\put(-541,322){${}_{h_i}$}
\put(-506,329){${}_{g_{ij}}$}
\put(-468,293){${}_{h_j^{-1}}$}
\put(-541,293){${}_{h_i}$}
\put(-506,284){${}_{g_{ij}}$}
\put(-498,315){${}_{p_{ij}}$}
\put(-498,298){${}_{p_{ij}}$}
\put(182,0){
\begin{picture}(0,0)
\put(-468,323){${}_{h_j}$}
\put(-543,322){${}_{h_i^{-1}}$}
\put(-510,331){${}_{g_{ij}^{-1}}$}
\put(-468,290){${}_{h_j}$}
\put(-545,292){${}_{h_i^{-1}}$}
\put(-510,283){${}_{g_{ij}^{-1}}$}
\put(-494,316){${}_{p_{ij}^{-1}}$}
\put(-494,298){${}_{p_{ij}^{-1}}$}
\end{picture}
}
\put(-433,338){${}_{1_{h_j^{-1}}\cdot\bar \alpha_j\of{\gamma_1}\cdot 1_{h_j}}$}
\put(-433,390){${}_{\alpha_i\of{\gamma_1}}$}
\put(-435,365){${}_{a_{ij}\of{\gamma_1}}$}
\put(-435,250){${}_{\bar a_{ij}\of{\gamma_2}}$}
\put(-433,223){${}_{\bar \alpha_i\of{\gamma_2}}$}
\put(-433,274){${}_{1_{h_j^{-1}}\cdot\alpha_j\of{\gamma_2}\cdot 1_{h_j}}$}
\put(-442,305){${}_{{\tilde {\hol_j}}\of{\Sigma}}$}
\put(-347,363){${}_{\hol_i\of{\gamma_1}}$}
\put(-360,346){${}_{\hol_j\of{\gamma_1}}$}
\put(-360,264){${}_{\hol_j\of{\gamma_2}}$}
\put(-406,300){${}_{\tilde \hol_j\of{\gamma_2}}$}
\put(-406,315){${}_{\tilde \hol_j\of{\gamma_1}}$}
\put(-347,245){${}_{\hol_i\of{\gamma_2}}$}
\put(-330,380){${}_{\tilde \hol_i\of{\gamma_1}}$}
\put(-330,232){${}_{\tilde \hol_i\of{\gamma_2}}$}
\end{picture}
}
\end{picture}
From this one reads off the gauge transformation law of the $a_{ij}$
as
\begin{center}
\begin{picture}(400,140)
\includegraphics{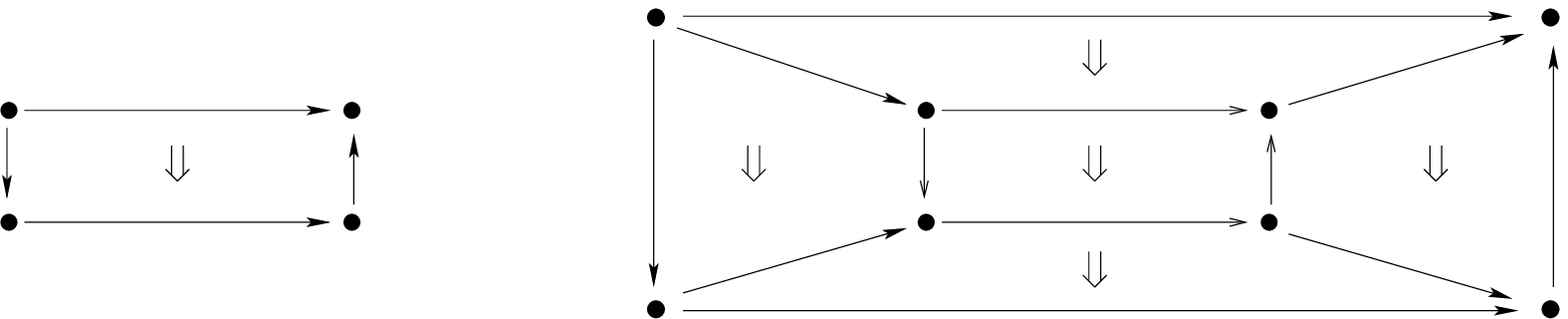}
\put(-310,40){$=$}
\put(-345,43){$\tilde g_{ij}^{-1}$}
\put(-468,43){$\tilde g_{ij}$}
\put(-420,43){$\tilde a_{ij}$}
\put(-410,70){$\tilde \hol_i$}
\put(-410,15){$\tilde \hol_j$}
\put(263,0){
\begin{picture}(0,0)
\put(-345,43){$g_{ij}^{-1}$}
\put(-265,43){$\tilde g_{ij}^{-1}$}
\put(-468,43){$g_{ij}$}
\put(-520,53){$p_{ij}\cdot 1_{h_j}$}
\put(-325,53){$(p_{ij}\cdot 1_{h_j})^{-1}$}
\put(-548,43){$\tilde g_{ij}$}
\put(-420,43){$a_{ij}$}
\put(-410,94){$\tilde \hol_i$}
\put(-410,-10){$\tilde \hol_j$}
\put(-381,65){$\hol_i$}
\put(-381,18){$\hol_j$}
\put(-473,71){$h_i$}
\put(-417,71){$\alpha_i$}
\put(-417,10){$\bar \alpha_j$}
\put(-473,13){$h_j$}
\put(-347,70){$h_i^{-1}$}
\put(-346,9){$h_j^{-1}$}
\end{picture}
}
\end{picture}
\end{center}
And indeed, this equation expresses the 2-commutativity 
of \refer{natural transformation for aij}.

This way, all the gauge transformation laws discussed in
\S\fullref{gauge transformations in terms of global 2-holonomy 2-functor}
are reobtained.

\paragraph{Gauge Invariance of Global 2-Holonomy.}

In the same pedestrian way we can now analyze the gauge invariance
of global 2-holonomy, which in the formulation of 
\S\fullref{Global 2-Holonomy in 2-Bundles} is nothing but 2-functoriality
of the global 2-holonomy 2-functor $\hol$.

\vskip 1em

Given any closed surface in base space whose 2-holonomy
is to be computed, we can triangulate it such 
a way that
each face comes to lie in an element of the cover,
each edge in a double overlap and each vertex in a
triple overlap. We can always assume the graph of the
triangulation to be trivalent. (If it is not
we replace the problematic vertices by small circles
of edges.)

The task is to assign 2-group elements to 
faces, edges and vertices of the triangularization
such that the result of gluing them all together is
independent of the choice of gauge (trivialization)
as well as of the choice of cover and the choice
of triangularization. For now
we restrict attention on independence of the 
gauge choice.

It is clear that local 2-holonomies $\hol_i\of{\Sigma}$
must be assigned to faces $\Sigma$. The only candidate 
2-group elements to be assigned to edges $\gamma$ 
are $a_{ij}\of{\gamma}$ and the only candidate 2-group 
elements to be assigned to vertices $x$ are $f_{ijk}\of{x}$.

There is only one way to glue all these pieces consistently,
and this is the way depicted in 
figure \ref{figure: global 2-holonomy}
(p. \pageref{figure: global 2-holonomy}).

Before looking at the gauge invariance of this definition
notice how ``2-conjugations'' (horizontal and vertical 
conjugation) respects the composition in the 2-group in the
following sense:
\newpage 
\hspace{-3.3cm}
\begin{picture}(400,350)
\includegraphics{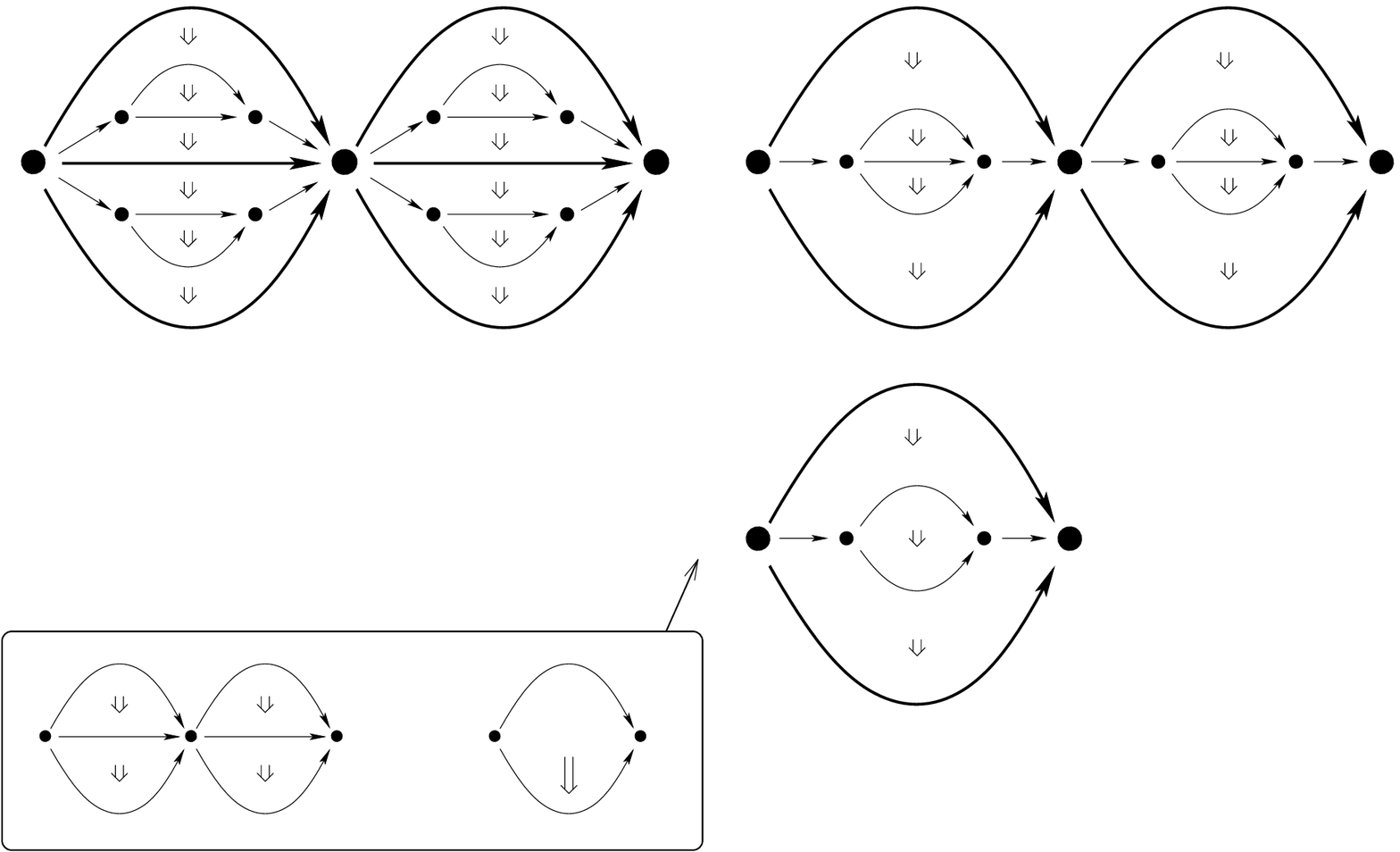}
\put(-269,118){$=$}
\put(-269,264){$=$}
\put(-384,43){$\defas$}
\put(-495,78){$g_1$}
\put(-510,50){$g_2$}
\put(-490,53){$f_1$}
\put(-490,28){$f_2$}
\put(-495,8){$g_3$}
\put(52,0){
\begin{picture}(0,0)
\put(-485,74){$g_1'$}
\put(-510,50){$g_2'$}
\put(-490,53){$f_1'$}
\put(-490,28){$f_2'$}
\put(-485,9){$g_3'$}
\end{picture}
}
\put(153,0){
\begin{picture}(0,0)
\put(-485,76){$g_1 g_1'$}
\put(-484,45){
  $f$}
\put(-485,7){$g_3 g_3'$}
\end{picture}
}
\put(-477,292){${}_{f_1}$}
\put(-477,240){${}_{f_2}$}
\put(-478,315){${}_{\alpha_1}$}
\put(-480,275){${}_{\bar \alpha_2}$}
\put(-508,285){${}_{h_x}$}
\put(-435,286){${}_{h_y^{-1}}$}
\put(-510,247){${}_{h_x}$}
\put(-435,246){${}_{h_y^{-1}}$}
\put(-480,255){${}_{\alpha_2}$}
\put(-480,215){${}_{\bar \alpha_3}$}
\put(117,0){
\begin{picture}(0,0)
\put(-477,292){${}_{f_1}$}
\put(-477,240){${}_{f_2}$}
\put(-478,315){${}_{\alpha_1'}$}
\put(-480,275){${}_{\bar \alpha_2'}$}
\put(-508,285){${}_{h_y}$}
\put(-435,286){${}_{h_z^{-1}}$}
\put(-510,247){${}_{h_y}$}
\put(-435,246){${}_{h_z^{-1}}$}
\put(-480,255){${}_{\alpha_2'}$}
\put(-480,215){${}_{\bar \alpha_3'}$}
\end{picture}
}
\put(276,0){
\begin{picture}(0,0)
\put(-478,305){${}_{\alpha_1}$}
\put(-510,273){${}_{h_x}$}
\put(-432,274){${}_{h_y^{-1}}$}
\put(-480,225){${}_{\bar \alpha_3}$}
\put(-475,275){${}_{f_1}$}
\put(-475,257){${}_{f_2}$}
\put(117,0){
\begin{picture}(0,0)
\put(-478,305){${}_{\alpha_1'}$}
\put(-508,274){${}_{h_y}$}
\put(-435,274){${}_{h_z^{-1}}$}
\put(-480,225){${}_{\bar \alpha_3'}$}
\put(-475,275){${}_{f_1'}$}
\put(-475,257){${}_{f_2'}$}
\end{picture}
}
\end{picture}
}
\put(276,-147){
\begin{picture}(0,0)
\begin{picture}(0,0)
\put(-476,300){${}_{\alpha_1\cdot\alpha_1'}$}
\put(-508,274){${}_{h_x}$}
\put(-435,274){${}_{h_z^{-1}}$}
\put(-476,237){${}_{\bar(\alpha_3\cdot\alpha_3')}$}
\put(-474,268){${}_{f}$}
\end{picture}
\end{picture}
}
\end{picture}

\vskip 1em

We display this rather simple fact in such a detail
because it gives a good illustration of the way 
how in a composite diagram of gauge transformed
quantities 2-conjugation operations mutually cancel
and leave the diagrams in the original gauge behind.
The demonstration of the gauge invariance of global 2-holonomy
further below 
proceeds in this fashion.

Fix the gauge $\tilde G$ and let the surface holonomy in 
the vicinity of some vertex $x$ be given by
\newpage
\hspace{-2cm}
%\begin{figure}[h]
\begin{picture}(500,450)
\includegraphics{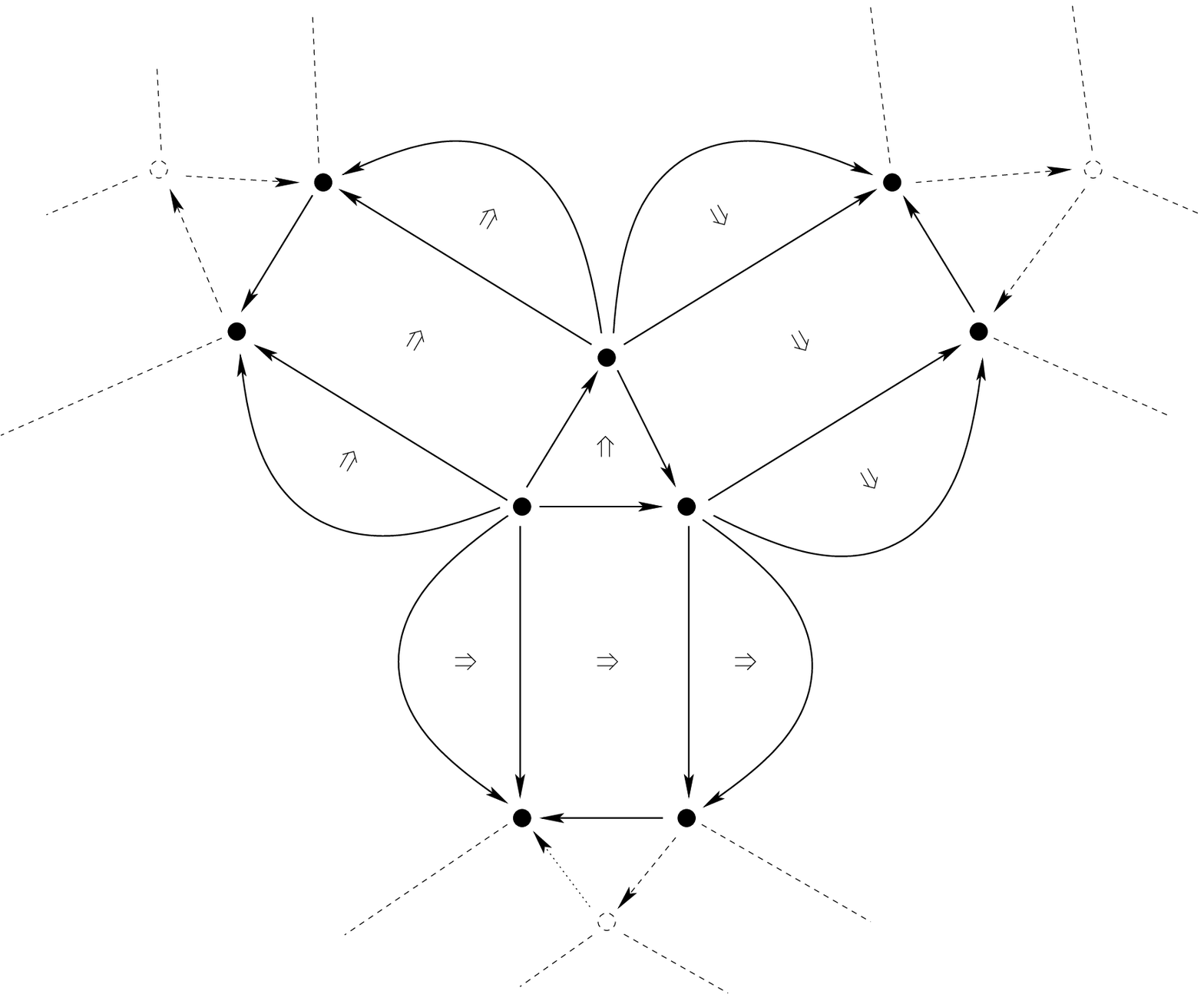}
\put(-267,123){$\tilde a_{ik}\of{\gamma_3}$}
\put(-328,263){$\tilde a_{ij}\of{\gamma_1}$}
\put(-190,263){$\tilde a_{jk}\of{\gamma_2}$}
\put(-267,220){$\tilde f_{ijk}\of{x}$}
\put(-267,195){$\tilde g_{ik}\of{x}$}
\put(-260,63){$\tilde g_{ik}^{-1}$}
\put(-303,230){$\tilde g_{ij}\of{x}$}
\put(-228,230){$\tilde g_{jk}\of{x}$}
\put(-405,318){$\tilde g_{ij}^{-1}$}
\put(-108,312){$\tilde g_{jk}^{-1}$}
\put(-326,115){$\tilde \hol_i\of{\gamma_3}$}
\put(-392,240){$\tilde \hol_i\of{\gamma_1}$}
\put(-185,334){$\tilde \hol_j\of{\gamma_2}$}
\put(-216,115){$\tilde \hol_k\of{\gamma_3}$}
\put(-350,332){$\tilde \hol_j\of{\gamma_1}$}
\put(-143,239){$\tilde \hol_k\of{\gamma_2}$}
\put(-327,147){${\tilde {\hol_i}}\of{\Sigma_1^{a}}$}
\put(-210,147){${\tilde {\hol_k}}\of{\Sigma_3^{a}}$}
\put(-376,210){${\tilde {\hol_i}}\of{\Sigma_1^b}$}
\put(-183,205){${\tilde {\hol_k}}\of{\Sigma_3^b}$}
\put(-310,311){${\tilde {\hol_j}}\of{\Sigma_2^a}$}
\put(-232,312){${\tilde {\hol_j}}\of{\Sigma_2^b}$}
\end{picture}
%\caption{
%  {\bf Surface holonomy} in the gauge $\tilde G$
%}
%\end{figure}

In this diagram the full local surface holonomies
${\tilde {\hol_i}}\of{\Sigma_1}$,
${\tilde {\hol_k}}\of{\Sigma_2}$,
${\tilde {\hol_j}}\of{\Sigma_3}$
are depicted only in terms of two surface sub-elements
$\Sigma_i^a,\Sigma_i^b \subset \Sigma_i$ etc., 
respectively,
adjacent to the edges meeting at the given vertex.

Now we insert into this diagram the equalities discussed
in \S\fullref{2-Gauge Transformations}, 
which re-express the diagrams in the
gauge $\tilde G$ in terms of those in the gauge $G$:

\newpage
\hspace{-2cm}
%\begin{figure}[h]
\begin{picture}(500,450)
\includegraphics{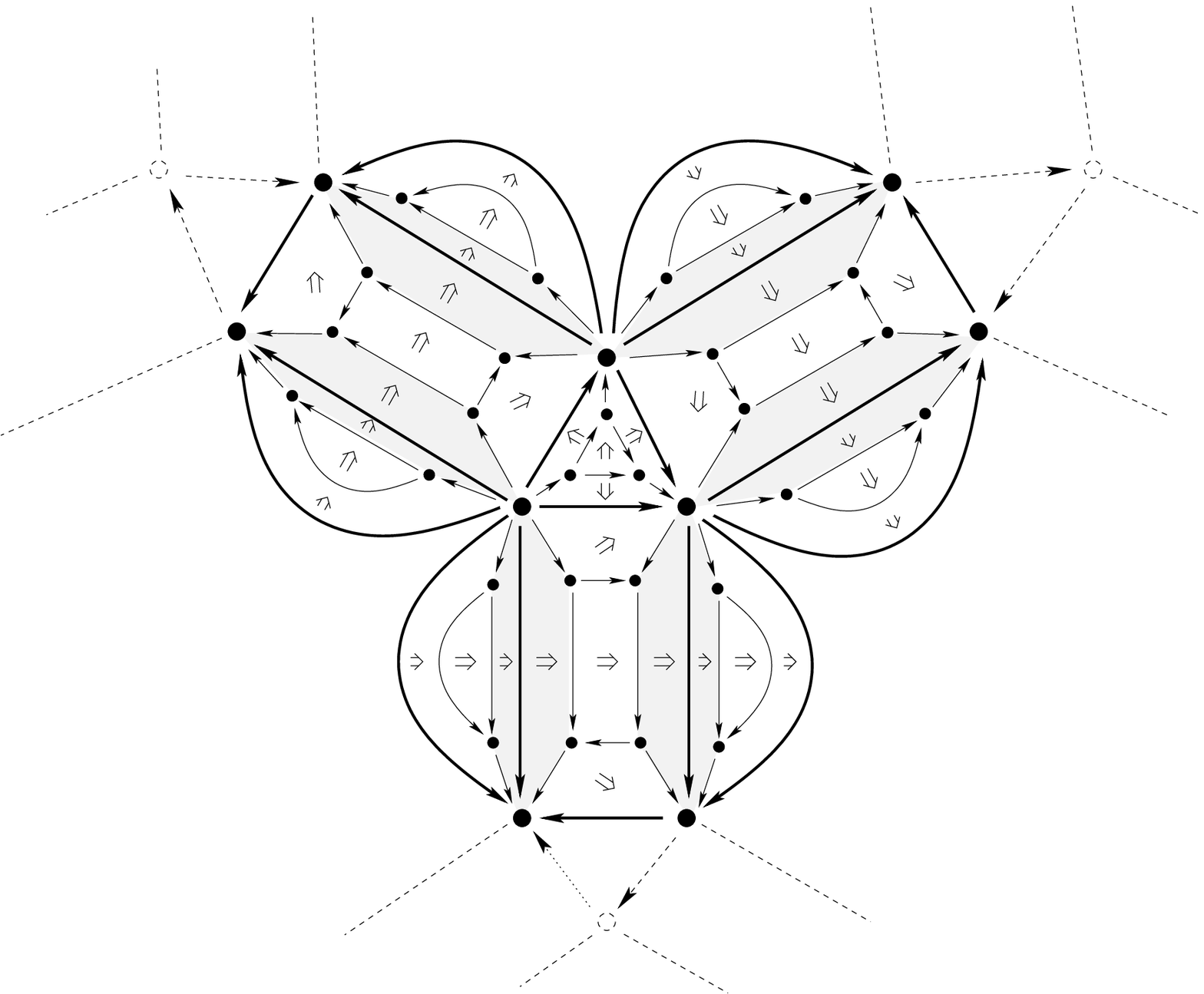}
\put(-258,131){${}_{ a_{ik}}$}
\put(-328,267){${}_{ a_{ij}}$}
\put(-185,270){${}_{ a_{jk}}$}
\put(-232,131){${}_{ \bar\alpha_{k}}$}
\put(-216,131){${}_{ \alpha_{k}}$}
\put(-280,131){${}_{ \alpha_{i}}$}
\put(-298,131){${}_{ \bar \alpha_{i}}$}
\put(-314,290){${}_{ \bar a_{j}}$}
\put(-306,307){${}_{ a_{j}}$}
\put(-354,254){${}_{ a_{i}}$}
\put(-364,242){${}_{ \bar a_{i}}$}
\put(-180,300){${}_{ a_{j}}$}
\put(-192,317){${}_{ \bar a_{j}}$}
\put(-154,255){${}_{ \bar a_{k}}$}
\put(-146,237){${}_{ a_{k}}$}
\put(-262,223){${}_{ f_{ijk}}$}
\put(-260,200){${}_{\tilde g_{ik}}$}
\put(-258,68){${}_{ \tilde g_{ik}^{-1}}$}
\put(-283,240){${}_{ \tilde g_{ij}}$}
\put(-231,239){${}_{ \tilde g_{jk}}$}
\put(-402,318){${}_{ \tilde g_{ij}^{-1}}$}
\put(-108,312){${}_{\tilde g_{jk}^{-1}}$}
\put(-317,147){${}_{\hol_i}$}
\put(-202,148){${}_{\hol_k}$}
\put(-376,233){${}_{\hol_i}$}
\put(-156,212){${}_{\hol_k}$}
\put(-300,320){${}_{\hol_j}$}
\put(-217,320){${}_{\hol_j}$}
\end{picture}
%\caption{
%  {\bf Surface holonomy} in the gauge $\tilde G$
%  expressed in terms of objects in the gauge $G$
%}
%\end{figure}

The result is that several 2-group elements are now adjacent
which mutually cancel to unity. First of all we can
cancel $a_{ij}$ against $\bar a_{ij}$ and analogously
for $jk$ and $ij$. The respective identity 
2-morphisms have been shaded in the diagram.

After removing them we are left with the following diagram:

\newpage
\hspace{-2cm}
%\begin{figure}[h]
\begin{picture}(500,450)
\includegraphics{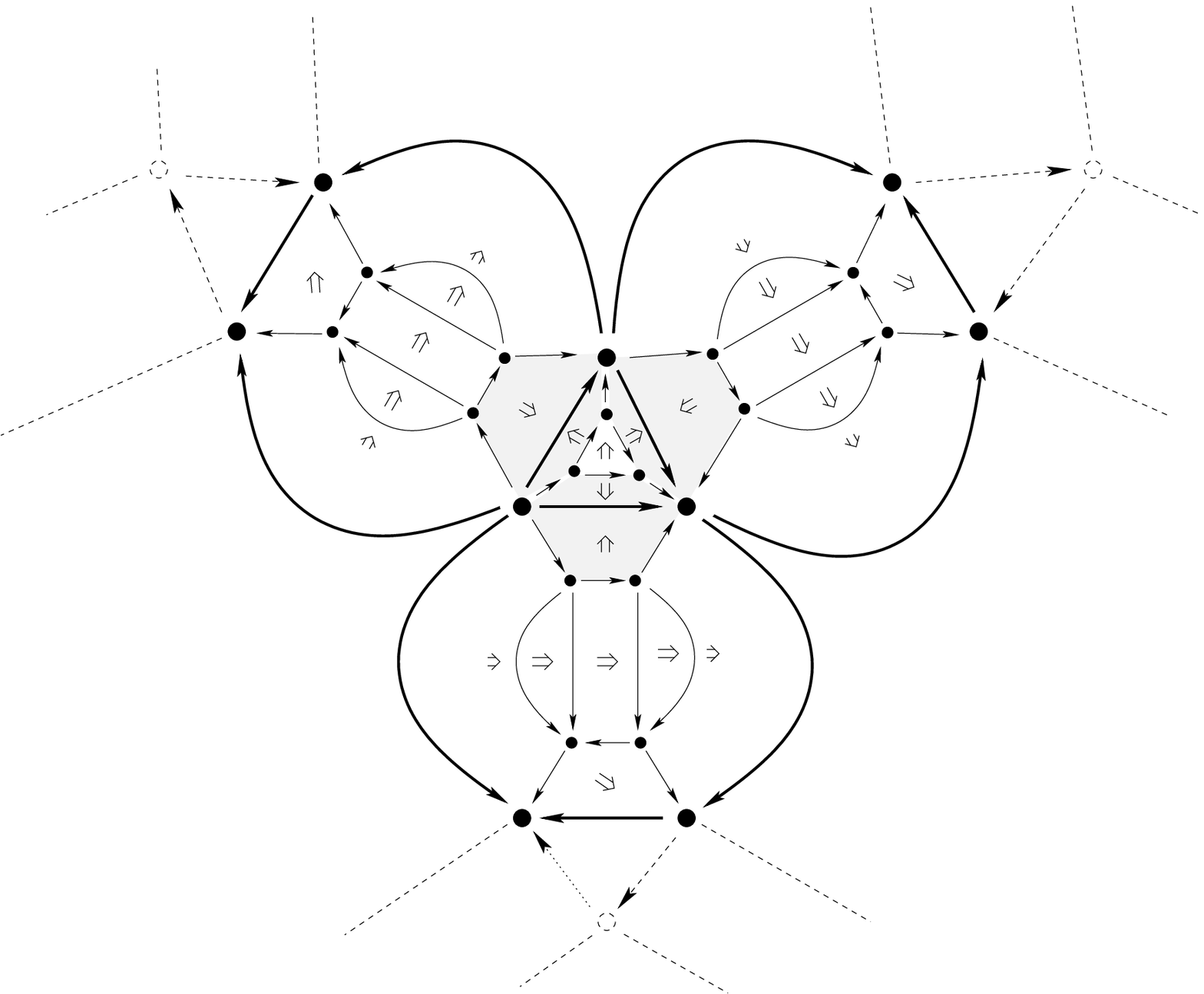}
\put(-258,131){${}_{ a_{ik}}$}
\put(-328,267){${}_{ a_{ij}}$}
\put(-185,270){${}_{ a_{jk}}$}
\put(-235,133){${}_{ \hol_k}$}
\put(-210,134){${}_{ 1_{h_k^{-1}} \cdot  \bar a_{k}}$}
\put(-283,131){${}_{ \hol_{i}}$}
\put(-303,131){${}_{ \tilde a_{i}}$}
\put(-314,287){${}_{ \hol_j}$}
\put(-302,307){${}_{ 1_{h_j^{-1}}\cdot \bar a_j}$}
\put(-356,258){${}_{\hol_i}$}
\put(-364,235){${}_{ \alpha_{i}}$}
\put(-180,300){${}_{ \hol_j}$}
\put(-190,317){${}_{ a_{j}}$}
\put(-158,257){${}_{\hol_k}$}
\put(-146,237){${}_{ 1_{h_k^{-1}}\cdot \bar a_{k}}$}
\put(-262,223){${}_{ f_{ijk}}$}
\put(-270,200){${}_{\tilde g_{ik}}$}
\put(-258,68){${}_{ \tilde g_{ik}^{-1}}$}
\put(-287,232){${}_{ \tilde g_{ij}}$}
\put(-227,232){${}_{ \tilde g_{jk}}$}
\put(-402,318){${}_{ \tilde g_{ij}^{-1}}$}
\put(-108,312){${}_{\tilde g_{jk}^{-1}}$}
\put(-287,254){${}_{ p_{ij}}$}
\put(-227,254){${}_{ p_{jk}}$}
\put(-248,190){${}_{p_{ik}}$}
\put(-248,212){${}_{p_{ik}}$}
\put(-248,241){${}_{p_{jk}}$}
\put(-268,241){${}_{p_{ij}}$}
\end{picture}
%\caption{
%  {\bf First step} in reducing the gauge transformed
%   surface holonomy to that in the original gauge
%}
%\end{figure}

We have also reversed the direction of some edges
by whiskering, so that we can now cancel
$p_{ij}$ against $\bar p_{ij}$. When the shaded identity
2-morphisms are removed one obtains the following 
diagram

\newpage
%\begin{figure}[h]
\begin{picture}(500,450)
\includegraphics{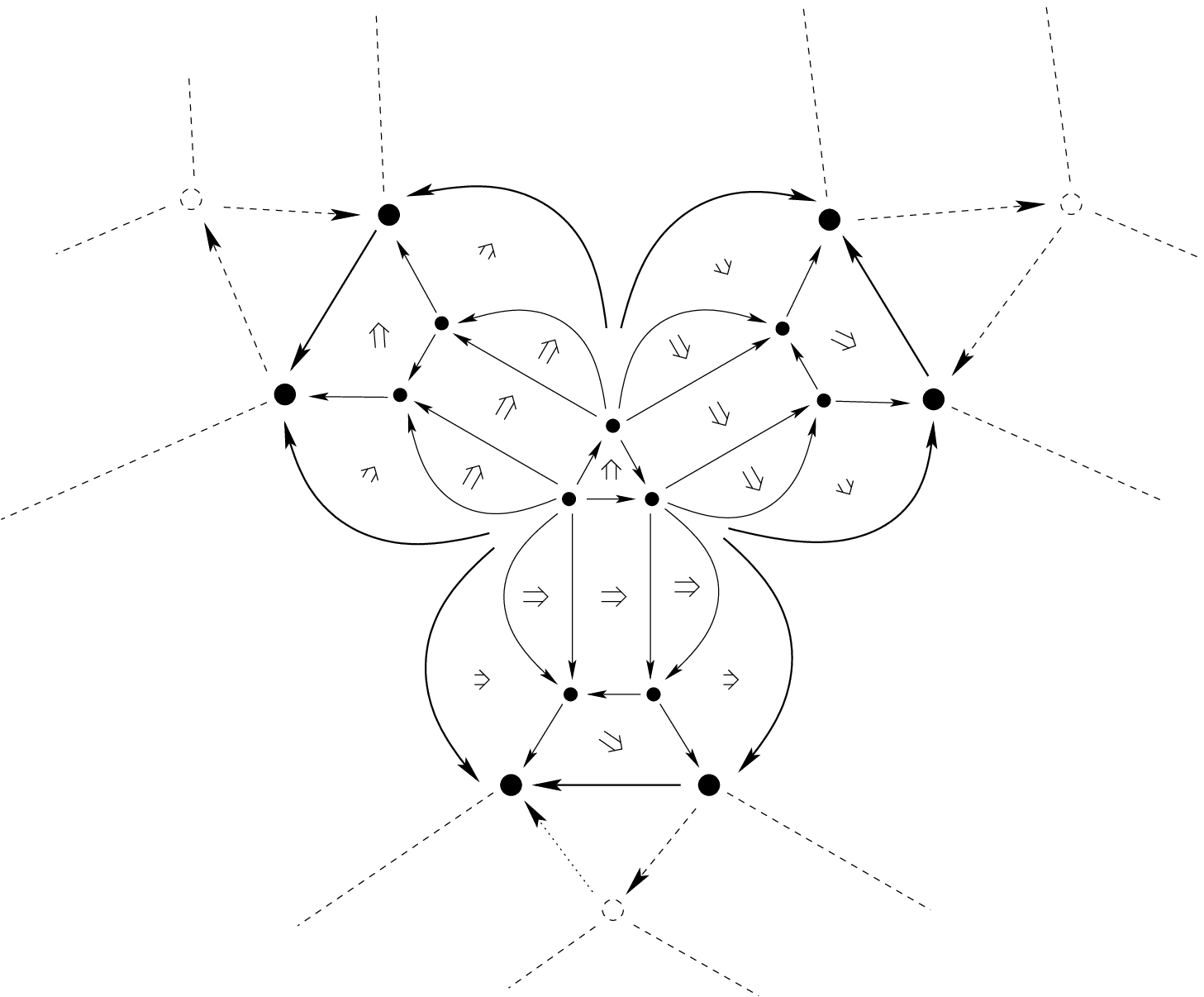}
\put(42,-43){
\begin{picture}(0,0)
\put(-258,175){${}_{ a_{ik}}$}
\put(-305,250){${}_{ a_{ij}}$}
\put(-212,250){${}_{ a_{jk}}$}
\put(-235,175){${}_{ \hol_k}$}
\put(-283,175){${}_{ \hol_{i}}$}
\put(-295,270){${}_{ \hol_j}$}
\put(-317,232){${}_{ \hol_i}$}
\put(-225,272){${}_{ \hol_j}$}
\put(-205,230){${}_{ \hol_k}$}
\put(-262,223){${}_{ f_{ijk}}$}
\end{picture}
}
\end{picture}
%\caption{
%  {\bf Second step} in reducing the gauge transformed
%   surface holonomy to that in the original gauge
%}
%\end{figure}

In the center of this diagram the surface holonomy
in the gauge $G$ has appeared. It is surrounded by
2-conjugations which cancel against the contributions 
from the other vertices.

This demonstrates the gauge invariance of our global 2-holonomy.
In order to demonstrate the invariance under different choices of
triangulations and of good covering, it is advisable to 
adopt a more sophisticated approach towards 2-holonomy, namely
a more intrinsic one that makes use of \emph{torsors} and does
not require local trivializations. This is the content of 
\S\fullref{p-Functors from p-Paths to p-Torsors}.

\newpage
\subsection{3-Bundles with 3-Connection}
\label{Global 3-Holonomy in 3-Bundles}

It is relatively straightforward to repeat the step from 1-bundles with
1-connection to 2-bundles with 2-connection again and again. The details
will get more and more involved, but the general principle remains the same. 

We will not enter a full discussion of 3-bundles here, but will 
want to emphasize the
 following easily accessible fact about 3-bundles which will have
relevance for a good understanding of 2-bundles:

\subsubsection{The \Cech-extended 3-Path 3-Groupoid}

From the above discussion it is clear that
the \Cech 3-Groupoid $C_3\of{\covering}$ and hence the \Cech-extended 
path 3-groupoid $\P_3^C\of{U}$ 
of a locally trivialized 3-bundle contains
3-morphisms of the following kind:
\begin{center}
\begin{picture}(300,150)
\includegraphics{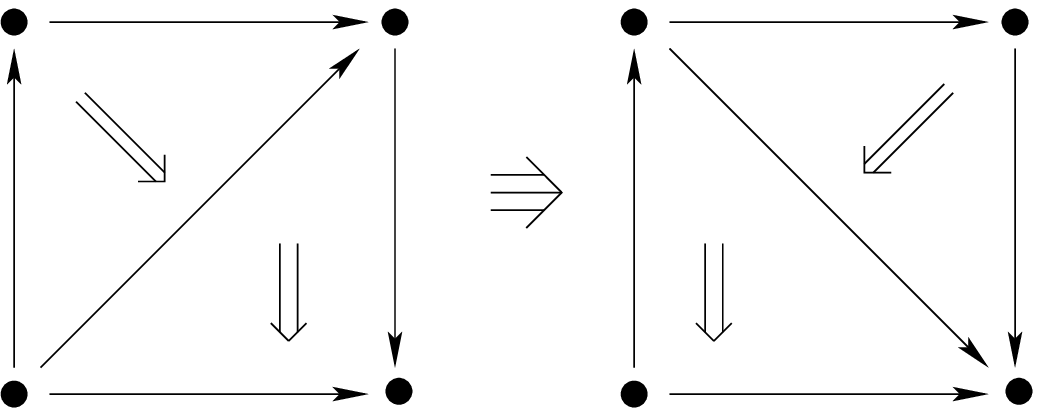}
  \put(-320,122){$(x,j)$}
  \put(-320,-12){$(x,i)$}
  \put(-183,122){$(x,k)$}
  \put(-183,-12){$(x,l)$}
  \put(175,0){
    \begin{picture}(0,0)
     \put(-320,122){$(x,j)$}
     \put(-320,-12){$(x,i)$}
     \put(-183,122){$(x,k)$}
     \put(-183,-12){$(x,l)$}
    \end{picture}
  }
\end{picture}
\end{center}
\vskip 2em
These are 3-morphisms inside of each \Cech-tetrahedron.

\subsubsection{The global 3-Holonomy 3-Functor}

Applying a holonomy 3-functor to these produces a 3-morphism
$\lambda_{ijkl}$ in a 3-group (\cf \S\fullref{Strict 3-Groups}):

\begin{center}
\begin{picture}(300,150)
\includegraphics{weaktetrahedron2.eps}
\put(-7,3){
\begin{picture}(300,150)
\put(-245,115){$g_{jk}$}
\put(-65,115){$g_{jk}$}
\put(-245,-5){$g_{il}$}
\put(-45,-5){$g_{il}$}
\put(-307,55){$g_{ij}$}
\put(-127,55){$g_{ij}$}
\put(-180,55){$g_{kl}$}
\put(-1,55){$g_{kl}$}
\put(-155,73){$\lambda_{ijkl}$}
\put(-280,30){$g_{ik}$}
\put(-40,21){$g_{jl}$}
\put(-60,80){$f_{jkl}$}
\put(-84,30){$f_{ijl}$}
\put(-234,30){$f_{ikl}$}
\put(-254,77){$f_{ijk}$}
\end{picture}
}
\end{picture}
\end{center}

In terms of the 2-crossed module $(G,H,J,\alpha_1,\alpha_2,t_1,t_2)
\simeq G_3$ of the structure 3-group 
(see \S\fullref{Strict 3-Groups and 2-Crossed Modules})
this says that
\[
  t_2\of{\lambda_{ijkl}}f_{ijk}f_{ikl}
  =
  \alpha_1\of{g_{ij}}\of{f_{jkl}}f_{ijl}
  \,.
\]
It is well known that $t_2\of{\lambda_{ijkl}}$ satisfying this
equation defines a class in \Cech cohomology
\[
  [t_2\of{\lambda_{ijkl}}] \in \check{H}^3\of{M} = H^4\of{M;\mathbb{Z}}
  \,.
\]

For the special case that the 3-group 
$G_3$
is an extension of the 2-group
$\mathcal{P}_k G$ as described in 
\S\fullref{a 3-group extension of PkG} it is a theorem in
\cite{BrylinskiMcLaughlin:1994} that this class is the 
characteristic class $p_1/2$ of a $G$-bundle
$E \to M$.\footnote{
I am grateful to Branislav Jur{\v c}o and 
Danny Stevenson for making me aware of this theorem.}

This should mean that the obstruction to having a 
$\mathcal{P}_k G$-2- bundle on $M$ 
lifting a principal $G$ bundle on $M$ is precisely 
this class. 
According to the discussion in 
\S\fullref{more details on Spinning Strings}
this would mean that $\mathcal{P}_1 \Spin\of{n}$-2-bundles
have the right properties to describe the parallel transport
of spinning strings.

\vskip 1em

Finally we note one fact about 3-connections with 3-holonomy in
3-bundles that generalizes the condition of vanishing fake curvature
for 2-holonomy and which will be rederived in linearized form
using nonabelian Deligne hypercohomology in 
\S\fullref{Strict Infinitesimal 3-Bundles with 3-Connection}:

\begin{proposition}
  \label{prop: conscond for local 3-holonomy}
  For $G_3$ a strict 3-group (\cf \fullref{Strict 3-Groups}),
  $G_3 = (G,H,J,\alpha_{1,2},t_{1,2})$,
  the 3-Holonomy functor over a patch $U_i$ 
  is specified by a set $\set{A_i,B_i,C_i}$
  with 
  \begin{eqnarray*}
    A_i \in \Omega^1\of{U_i,\g}
    \\
    B_i \in \Omega^2\of{U_i,\h}
    \\
    C_i \in \Omega^3\of{U_i,\j}
  \end{eqnarray*}
  which has to satisfy the two consistency conditions
  \begin{eqnarray*}
    &&dt_1\of{B_i} + F_{A_i} = 0
    \\
    &&dt_2\of{C_i} + \extd_{A_i} B_i = 0
    \,.
  \end{eqnarray*}
\end{proposition}

This follows from similar consideration as in one dimension lower,
using the fact \cite{BaezSchreiber:2004} that for vanishing
fake curvature the curvature on path
space given by
\[
  \mathcal{F}_i = \oint\limits_A \of{\extd_{A_i} B_i}
  \,.
\]

\subsection{$p$-Functors from $p$-Paths to $p$-Torsors}
\label{p-Functors from p-Paths to p-Torsors}

So far we have constructed $p$-holonomy making use of good coverings
of the base manifold. This has the advantage that 
this way $p$-holonomy can locally 
be defined as nothing but a $p$-functor to the structure $p$-group. 
But the
resulting global $p$-holonomy $p$-functor defined by gluing these
local $p$-functors by $p$-functor $n$-morphisms on $(n+1)$-fold
overlaps is not
\emph{manifestly} insensitive to the choice of good covering.

The reason for this is that a given fiber of a principal 
$G_p$-$p$-bundle is, while isomorphic to $G_p$, not
\emph{canonically} isomorphic to $G_p$. It is a 
$G_p$-$p$-\emph{torsor} rather than $G_p$ itself.

A nice pedagogical introduction to torsors can be found in 
\cite{Baez:2004}. The precise definition of (1-)torsors
and 2-torsors is stated for instance in Toby Bartels' paper 
on 2-bundles \cite{Bartels:2004}.

Hence a holonomy functor should really associate to any
path in the base manifold a \emph{torsor morphism} between the
fibers above the endpoints of the path. More precisely, a
$p$-holonomy $p$-functor should be a $p$-functor from $p$-paths
in the base manifold to $p$-morphisms of $p$-torsors.
\[
  \hol \maps \P_p\of{M} \to G_p\!-\!p\mathrm{Tor}
  \,.
\]
Even more precisely, since we would want $\hol$ to be a smooth
$p$-functor between smooth $p$-categories, while $G_p-p\mathrm{Tor}$
does not have a smooth structure, we should have a $p$-functor
\[
  \hol \maps \P_p\of{M} \to \trans_p\of{E}
  \,.
\]
Here $\trans_p\of{E}$ is the $p$-category of
{\bf $p$-transporters} in a principal $G_p$-$p$-bundle $E\to M$
over a categorically trivial base manifold $M$. Objects of 
$\trans_p\of{E}$ are the fibers $E_x$ of $E$ for all $x \in M$,
regarded as $G_p$-$p$-torsors,
and $n$-morphisms in $\trans_p\of{E}$ are their $p$-torsor 
$n$-morphisms.

Given any such $p$-functor to $\trans_p\of{M}$ we can always forget
about the smooth structure and regard it as a $p$-functor to 
$G_p\!-\!p\mathrm{Tor}$. For notational convenience, this is 
what we shall do in the following. 

Indeed, this more intrinsic description of 
global $p$-holonomy is \emph{equivalent} to the one we have been 
concentrating on up to this point. 
In \S\fullref{1-Torsors and 1-Bundles with Connection and Holonomy}
we recall the well-known way how this 
equivalence works for 1-bundles. Then
in \S\fullref{2-Torsors and 2-Bundles with 2-Holonomy} the
generalization to 2-bundles is spelled out.

\subsubsection{1-Torsors and 1-Bundles with Connection and Holonomy}
\label{1-Torsors and 1-Bundles with Connection and Holonomy}

A principal $G$-bundle with connection over a base manifold $M$
is specified by a functor
\[
  \hol \maps \P_1\of{M} \to \GTor
  \,.
\]
We shall now choose any good covering $\covering \to M$ of $M$,
$\covering = \bigsqcup\limits_{i\in I} U_i$, and
demonstrate how this single functor encodes local holonomy
functors 
\[
  \hol_i \maps \P_1\of{U_i} \to G
\]
on each $U_i$, which are related on double overlaps by natural
transformations $\hol_i \stackto{g_{ij}} \hol_j$, as described in 
\S\fullref{introduction: 1-connections in 1-bundles}. In the process
of doing so, the procedure for computing global 1-holonomy in terms of
the $\hol_i$, as described at the end of 
\S\fullref{introduction: 1-connections in 1-bundles}, drops out
automatically.

For 1-bundles this is all rather simple and very well known. We
find it worthwhile to restate these facts in order to make
the generalization to 2-bundles 
in \S\fullref{2-Torsors and 2-Bundles with 2-Holonomy}
more accessible.

\vskip 1em

So let $\covering \to M$ be a good covering of $M$,
$\covering = \bigsqcup\limits_{i\in I} U_i$. When restricted to
any of the $U_i$, the functor $\hol$ is naturally isomorphic to
a local holonomy functor
\[
  \hol_i \maps \P_2\of{U_i} \to G
  \,.
\]
Fix any such natural isomorphism
\[
  \hol|_{U_i} \stackto{t_i} \hol_i
  \,.
\]
It is specified by a naturality square
\begin{center}
  \begin{picture}(370,170)
    \includegraphics{nattraf.eps}
    \put(-305,142){$\P_1\of{U_i}$}
    \put(-110,142){$\GTor$}
    \put(-296,107){$x$}
    \put(-296,20){$y$}
    \put(-307,65){$[\gamma]$}
    \put(-170,104){$E_x$}
    \put(-170,20){$E_y$}
    \put(-200,65){$\hol|_{U_i}\of{\gamma}$}
    \put(-40,104){$G$}
    \put(-40,20){$G$}
    \put(-33,65){$\hol_i\of{\gamma}$}
    \put(-110,104){$t_i\of{x}$}
    \put(-110,20){$t_i\of{y}$}
  \end{picture}
\end{center}
Here we use the fact that $G$ itself is a $G$-torsor and that its 
automorphisms in $\GTor$ correspond to (right)-multiplication with 
elements in $G$. 

It follows that on double overlaps we have natural isomorphisms
\[
  \hol_i \stackto{g_{ij}} \hol_j
\]
given by this commuting diagram
\begin{center}
 \begin{picture}(140,110)
  \includegraphics{baretriangle.eps}
  \put(-107,50){$\bar t_i\of{x}$}
  \put(-27,50){$t_j\of{x}$}
  \put(-68,-6){$g_{ij}\of{x}$}
  \put(-125,-5){$\hol_i$}
  \put(-0,-5){$\hol_j$}
  \put(-65,102){$\hol|_{U_{ij}}$}
  \end{picture}
  \vskip 1em
  \,.
\end{center}
Here $\bar t_i$ denotes the inverse of $t_i$.
We write $\bar t_i$ because later on, when we categorify, $t_i$ will be
only weakly invertible 
(up to equivalence) and $\bar t_i$ will denote any of its weak 
inverses.

\paragraph{Line Holonomy in Terms of local Trivializations.}

\begin{figure}
\begin{center}
\begin{picture}(200,450)
\includegraphics{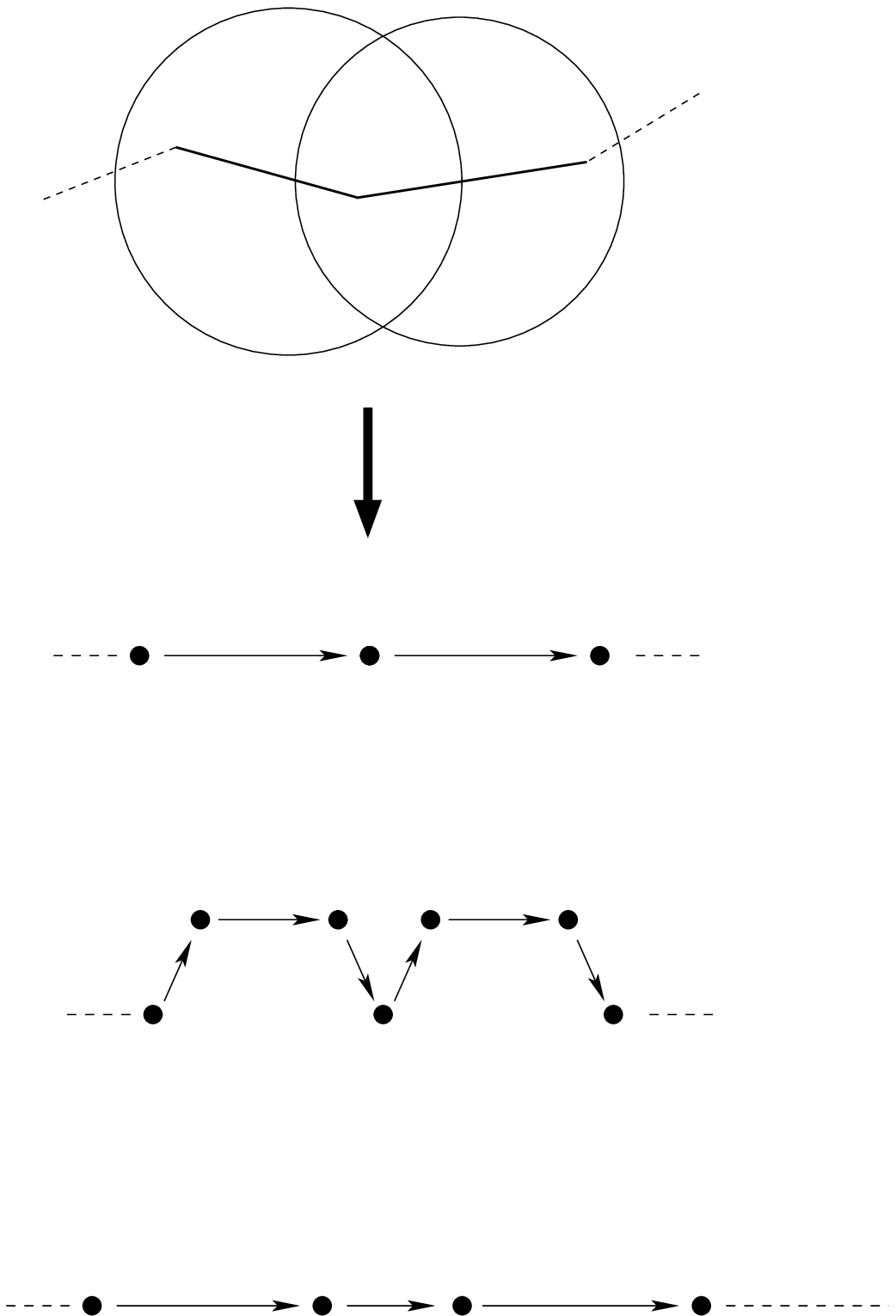}
\put(-220,393){$[\gamma_i]$}
\put(-215,420){$U_i$}
\put(-134,391){$[\gamma_j]$}
\put(-141,419){$U_j$}
\put(-243,383){$x$}
\put(-183,368){$y$}
\put(-105,380){$z$}
\put(-350,368){$\P_1\of{M}$}
\put(-350,220){$\GTor$}
\put(-350,97){$\GTor$}
\put(-342,130){$G$}
\put(-342,0){$G$}
\put(-200,283){$\hol$}
\put(-180,170){$=$}
\put(-180,50){$=$}
\put(-258,230){$E_x$}
\put(-182,230){$E_y$}
\put(-104,230){$E_z$}
\put(-230,227){$\hol\of{\gamma_i}$}
\put(-153,227){$\hol\of{\gamma_j}$}
\put(-227,140){$\hol_i\of{\gamma_i}$}
\put(-148,140){$\hol_j\of{\gamma_j}$}
\put(-263,114){$t_i\of{x}$}
\put(-255,88){$E_x$}
\put(-209,114){$t_i^{-1}\of{y}$}
\put(-177,88){$E_y$}
\put(-158,114){$t_j\of{y}$}
\put(-97,114){$t_j^{-1}\of{z}$}
\put(-100,88){$E_z$}
\put(-183,10){$g_{ij}\of{y}$}
\put(-250,10){$\hol_i\of{\gamma_i}$}
\put(-120,10){$\hol_j\of{\gamma_j}$}
\end{picture}
\end{center}
\caption{
\label{torsor 1-holonomy}
{\bf Global (1-)holonomy in terms of (1-)torsor (1-)morphisms}.
 The functor $\hol$ associates torsor morphisms between fibers
 to paths in the base manifold. Using trivializations $t_i$ on patches
  $U_i$ these torsor morphisms can be identified with elements of the
  structure group. The step $t_i^{-1} \circ t_j$ from one trivialization
  to another one on double overlaps $U_{ij}$ gives rise to multiplication
  by the transition function $g_{ij}$.}
\end{figure}

Now consider the application of $\hol$ to any morphism 
$[\gamma] \in \P_1\of{M}$ 
that does not necessarily sit in a single patch 
$U_i$
\[
  \hol\of{
    \xymatrix{
    x \ar@/^1pc/[rr]^{[\gamma]}
    && y
    }
  }
  =
    \xymatrix{
    E_x \ar@/^1pc/[rr]^{\hol\of{\gamma}}
    && E_y
    }  
  \,.
\]
We can always split up $[\gamma]$ into pieces
\[
    \xymatrix{
    x \ar@/^1pc/[rr]^{[\gamma]}
    && y
    }
  =
      \xymatrix{
       x  \ar@/^1pc/[rr]^{[\gamma_{i_1}]}
    && y_{i_1} \ar@/^1pc/[rr]^{[\gamma_{i_2}]}
    && y_{i_2} 
    }
    \cdots
      \xymatrix{
         \ar@/^1pc/[rr]^{[\gamma_{i_n}]}
    && y
    }
\]
such that $\gamma_{i_m} \in \P_2\of{U_{i_m}}\,,\;\forall\; m$.
Applying $\hol$ to that similarly yields
\[
\hol\of{
    \xymatrix{
    x \ar@/^1pc/[rr]^{[\gamma]}
    && y
    }
 }
  =
      \xymatrix{
       E_x  \ar@/^1pc/[rr]^{\hol\of{\gamma_{i_1}}}
    && P_{y_{i_1}} \ar@/^1pc/[rr]^{\hol\of{\gamma_{i_2}}}
    && P_{y_{i_2}} 
    }
    \cdots
      \xymatrix{
         \ar@/^1pc/[rr]^{\hol\of{\gamma_{i_n}}}
    && E_y
    }
  \,.
\]
But at this point we can apply the commutativity of the 
above naturality square and set
\[
  \hol\of{\gamma_i}
  =
  t_i\of{x} \circ \hol_i\of{\gamma_i} \circ t_i^{-1}\of{y}
  \,. 
\]
This procedure is indicated in figure \ref{torsor 1-holonomy}.

\clearpage

\paragraph{Gauge Transformations.}
\label{1-gauge transformations in torsor language}

Changing from one local trivialization
\[
  i \mapsto \left(\hol|_{U_i} \stackto{t_i} \hol_i\right)
\]
to another one
\[
  i \mapsto \left(\hol|_{U_i} \stackto{\tilde t_i} \tilde \hol_i\right)
\]
is called a {\bf gauge transformation}.

This corresponds to replacing the transition 
\begin{center}
 \begin{picture}(140,110)
  \includegraphics{baretriangle.eps}
  \put(-94,50){$\bar t_i$}
  \put(-27,50){$t_j$}
  \put(-68,-6){$g_{ij}$}
  \put(-125,-5){$\hol_i$}
  \put(-0,-5){$\hol_j$}
  \put(-65,102){$\hol|_{U_{ij}}$}
  \end{picture}
  \vskip 1em
\end{center}
by another transition $\tilde g_{ij}$ which is given by 
the commutativity of this diagram:
\begin{center}
 \begin{picture}(140,110)
  \includegraphics{baretriangle.eps}
  \put(-94,50){$\tilde {\bar t}_i$}
  \put(-27,50){$\tilde t_j$}
  \put(-68,-6){$\tilde g_{ij}$}
  \put(-122,-5){$\hol_i$}
  \put(-0,-5){$\hol_j$}
  \put(-65,102){$\hol|_{U_{ij}}$}
  \end{picture}
  \vskip 1em
\end{center}
This situation is depicted by the following commuting diagram:
\begin{eqnarray}
\begin{picture}(170,140)
  \includegraphics{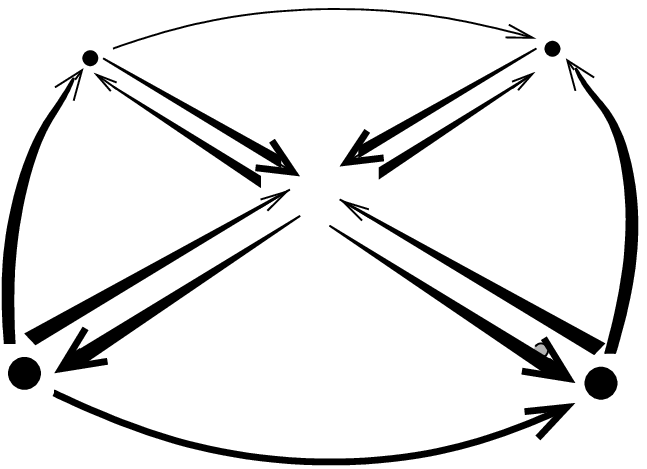}
  \put(0,-106){\begin{picture}(0,0)
  \put(-100,187){${}_{\hol|_{U_{ij}}}$}
  \put(-124,159){${}_{t_i}$}
  \put(-78,159){${}_{\tilde t_i}$}
  \put(-149,168){${}_{\bar t_i}$}
  \put(-63,175){${}_{{\bar {\tilde t}}_i}$}
  \put(2,180){${}_{\tilde g_{ij}}$}
  \put(-195,180){${}_{g_{ij}}$}
  \put(-142,202){${}_{{}_{t_j}}$}
  \put(-130,215){${}_{{}_{\bar t_j}}$}
  \put(-60,198){${}_{{}_{\tilde t_j}}$}
  \put(-68,217){${}_{{}_{\bar{\tilde t}_j}}$}
  \put(-104,98){$h_i$}
  \put(-104,245){${}_{{}_{h_j}}$}
  \put(-198,123){$\hol_i$}
  \put(-06,123){$\tilde \hol_i$}
  \put(-173,231){${}_{{}_{\hol_j}}$}
  \put(-23,233){${}_{{}_{\tilde \hol_j}}$}
  \end{picture}}
\end{picture}
\nonumber\\
\nonumber
\end{eqnarray}

In components this means that the following diagram commutes, too:
\begin{eqnarray}
\begin{picture}(170,270)
  \includegraphics{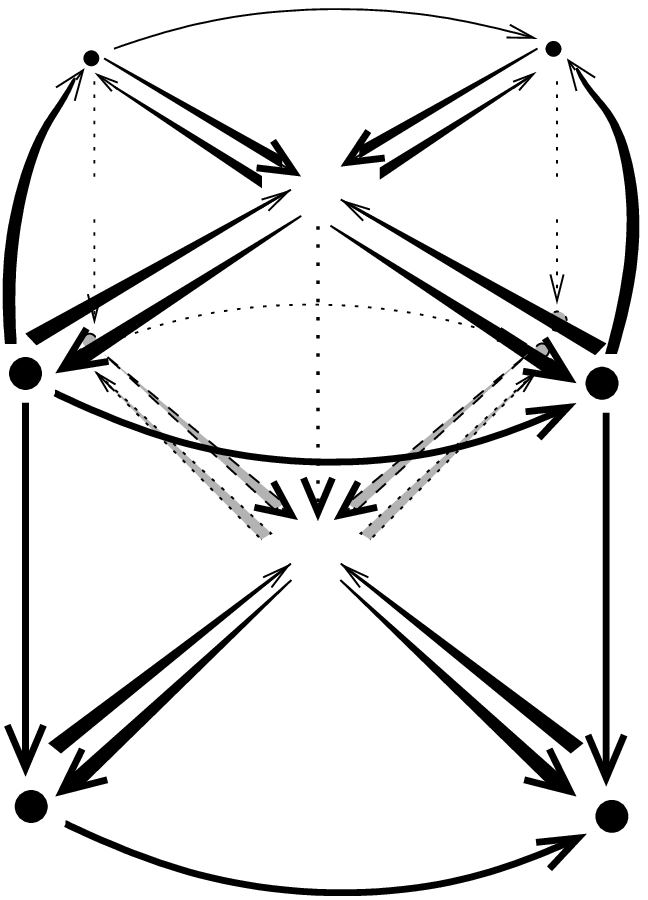}
  \put(0,19){\begin{picture}(0,0)
  \put(-100,83){${}_{P_y}$}
  \put(-100,184){${}_{P_x}$}
  \put(-92,135){${}_{\hol\of{\gamma}}$}
  \put(-124,159){${}_{t_i\of{x}}$}
  \put(-84,159){${}_{\tilde t_i\of{x}}$}
  \put(-149,173){${}_{\bar t_i\of{x}}$}
  \put(-130,41){${}_{t_i\of{y}}$}
  \put(-149,58){${}_{\bar t_i\of{y}}$}
  \put(-55,59){${}_{{\bar {\tilde t}}_i\of{y}}$}
  \put(-69,39){${}_{{{\tilde t}}_i\of{y}}$}
  \put(-63,175){${}_{{\bar {\tilde t}}_i\of{x}}$}
  \put(2,180){${}_{\tilde g_{ij}\of{x}}$}
  \put(-204,180){${}_{g_{ij}\of{x}}$}
  \put(-8,80){$\tilde \hol_i\of{\gamma}$}
  \put(-210,80){$\hol_i\of{\gamma}$}
  \put(-150,202){${}_{{}_{t_j\of{x}}}$}
  \put(-130,215){${}_{{}_{\bar t_j\of{x}}}$}
  \put(-60,198){${}_{{}_{\tilde t_j\of{x}}}$}
  \put(-68,220){${}_{{}_{\bar{\tilde t}_j\of{x}}}$}
  \put(-63,100){${}_{{}_{\tilde t_j\of{y}}}$}
  \put(-63,130){${}_{{}_{\bar {\tilde t}_j\of{y}}}$}
  \put(-140,100){${}_{{}_{t_j\of{y}}}$}
  \put(-136,127){${}_{{}_{\bar t_j\of{y}}}$}
  \put(-168,187){${}_{{}_{\hol_j\of{\gamma}}}$}
  \put(-34,187){${}_{{}_{\tilde \hol_j\of{\gamma}}}$}
  \put(-169,109){$h_i\of{x}$}
  \put(-104,-28){$h_i\of{y}$}
  \put(-104,245){${}_{{}_{h_j\of{x}}}$}
  \end{picture}}
\end{picture}
\nonumber\\
\nonumber
\end{eqnarray}

From this diagram one reads off that
\[
  \tilde g_{ij} = h_i^{-1} \, g_{ij} \, h_j
\]
and that
\[
  \tilde \hol_i = h_i^{-1} \, \hol_i \, h_i
  \,,
\]
where 
\[
  h_i = {\bar t}_i \circ \tilde t_i
\]
and
\[
  h_j = {\bar t}_j \circ \tilde t_j
  \,.
\]
This is the well-known gauge transformation law 
for 1-bundles with connection.

In fact, when read from the left to the right one sees that this
diagram is equivalent to the diagram 
\refer{gauge trafo of Cech-extended holonomy 1-functor}
(p. \pageref{gauge trafo of Cech-extended holonomy 1-functor})
which depicts a natural isomorphism of the global holonomy 1-functor
regarded as a functor
\[
  \hol \maps \P_1^C\of{\covering} \to G
\]
from the {\v C}ech-extended 1-path 1-groupoid 
(def. \ref{def Cech-extended 1-path 1-groupoid}, 
p. \pageref{def Cech-extended 1-path 1-groupoid})
of the good covering
$\covering$ to the structure group $G$. This justifies our use of the
same symbol ``$\hol$'' for both functors.

\vskip 2em

In the next subsection all this is categorified.

\newpage

\subsubsection{2-Torsors and 2-Bundles with 2-Holonomy}
\label{2-Torsors and 2-Bundles with 2-Holonomy}

Let $G_2$ be any 2-group and let $\GTwoTor$ be the 2-category of
$G_2$-2-torsors. 

A principal $G_2$-2-bundle with connection and holonomy is
specified by a 2-functor
\[
  \hol \maps \mathcal{P}_2\of{M} \to \GTwoTor
  \,.
\]

When restricted to any of the $U_i$, the 2-functor $\hol$
is pseudonaturally isomorphic to a local 2-holonomy 2-functor
\[
  \hol_i \maps \P_2\of{U_i} \to G_2
  \,.
\]
Fix any such natural isomorphism
\[
  \hol|_{U_i} \stackto{t_i} \hol_i
  \,.
\]
It is specified by a naturality tincan diagram

\begin{eqnarray}
\label{local trivialization pseudonat traf}
\begin{picture}(440,200)
 \includegraphics{2nattraf2.eps}
 \put(-19,-2){
 \begin{picture}(0,0)
 \put(-10,0){
  \begin{picture}(0,0)
   \put(-358,70){$\gamma_1$}
   \put(-290,70){${}_{\gamma_2}$}
   \put(-323,105){$[\Sigma]$}
   \put(-322,168){$y$}
   \put(-322,7){$x$}
  \end{picture}
 }
 \put(122,0){
  \begin{picture}(0,0)
   \put(-376,50){$\hol\of{\gamma_1}$}
   \put(-293,70){${}_{\hol\of{\gamma_2}}$}
   \put(-333,105){$\hol\of{\Sigma}$}
   \put(-330,167){$E_y$}
   \put(-330,10){$E_x$}
   \put(-287,108){$t_i\of{\gamma_1}$}
   \put(-250,120){${}_{t_i\of{\gamma_2}}$}
   \put(-250,165){$t_i\of{y}$}
   \put(-250,10){$t_i\of{x}$}
  \end{picture}
 }
 \put(288,0){
  \begin{picture}(0,0)
   \put(-330,167){$G_2$}
   \put(-327,8){$G_2$}
   \put(-378,50){$\hol_i\of{\gamma_1}$}
   \put(-293,70){${}_{\hol_i\of{\gamma_2}}$}
   \put(-333,105){$\hol_i\of{\Sigma}$}
   \put(-333,165){$$}
   \put(-333,10){$$}
  \end{picture}
 }
 \end{picture}
}
  \put(-360,203){$\P_2\of{U_i}$}
  \put(-145,203){$\GTwoTor$}
\end{picture}
\end{eqnarray}

Here we use the fact that $G_2$ itself is a $G_2$-torsor and that its 
automorphisms in $\GTwoTor$ correspond to (right)-multiplication with 
identity morphism in $G_2$ and that its 2-morphisms correspond to
2-torsor 2-morphisms between these.

It follows that on double overlaps we have pseudonatural isomorphisms
\[
  \hol_i \stackto{g_{ij}} \hol_j
\]
given by diagrams like this:
\begin{eqnarray}
\label{modification between gij and trivialization}
 \begin{picture}(140,110)
  \includegraphics{triangle.eps}
  \put(-94,50){$\bar t_i$}
  \put(-27,50){$t_j$}
  \put(-68,-6){$g_{ij}$}
  \put(-127,-5){$\hol_i$}
  \put(-0,-5){$\hol_j$}
  \put(-63,102){$\hol|_{U_{ij}}$}
  \put(-63,24){$\phi_{ij}$}
  \end{picture}
  \,.
 \nonumber\\
\end{eqnarray}
Here $\bar t_i$ is any one weak inverse of $t_i$ and 
\[
  g_{ij} \stackto{\phi_{ij}} \bar t_i \circ t_j
\] 
is a modification of pseudonatural transformations.

The pseudonatural
transformation $g_{ij}$ is given by a tincan diagram of this form:
\begin{eqnarray}
\label{pseudo nattraf on double overlap in torsor section}
\begin{picture}(240,160)
 \includegraphics{tincan.eps}
 \put(-3,-17){
 \begin{picture}(0,0)
 \put(122,0){
  \begin{picture}(0,0)
   \put(-378,70){$\hol_i\of{\gamma}$}
   \put(-293,70){${}_{\hol_i\of{\tilde \gamma}}$}
   \put(-333,105){$\hol_i\of{\Sigma}$}
   \put(-290,106){$a_{ij}\of{\gamma}$}
   \put(-250,118){${}_{a_{ij}\of{\tilde\gamma}}$}
   \put(-250,163){$g_{ij}\of{x}$}
   \put(-250,10){$g_{ij}\of{y}$}
  \end{picture}
 }
 \put(288,0){
  \begin{picture}(0,0)
   \put(-378,70){$\hol_j\of{\gamma}$}
   \put(-293,70){${}_{\hol_j\of{\tilde \gamma}}$}
   \put(-333,105){$\hol_j\of{\Sigma}$}
  \end{picture}
 }
 \end{picture}
}
\end{picture}
\,.
\end{eqnarray}
\vskip 2em

On triple overlaps $g_{ik}$ and $g_{ij}\circ g_{jk}$ are
related by a modification $f_{ijk}$ which is a composite of three of the
$\phi$ from \refer{modification between gij and trivialization}
and of a $\rho_j \maps \mathrm{Id}\to t_j \circ \bar t_j$:
\begin{center}
 \begin{picture}(300,160)
  \includegraphics{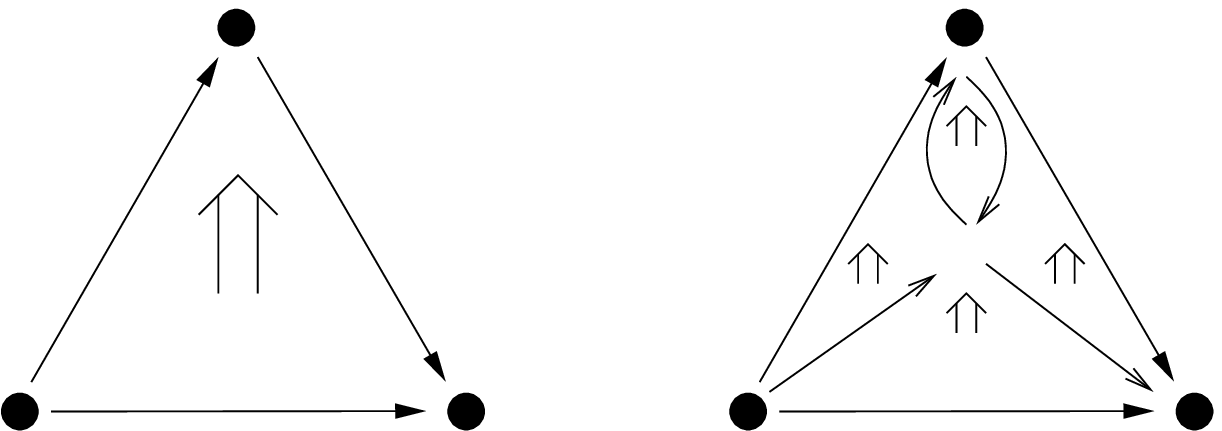}
\put(-230,0){\begin{picture}(0,0)
  \put(-103,55){$g_{ij}$}
  \put(-12,55){$g_{jk}$}
  \put(-64,-2){$g_{ik}$}
  \put(-134,-8){$\hol_i$}
  \put(13,-8){$\hol_k$}
  \put(-60,126){$\hol_j$}
  \put(-65,27){$f_{ijk}$}
\end{picture}}
\put(-20,0){\begin{picture}(0,0)
  \put(-103,55){$g_{ij}$}
  \put(-12,55){$g_{jk}$}
  \put(-64,-2){$g_{ik}$}
  \put(-134,-8){$\hol_i$}
  \put(13,-8){$\hol_k$}
  \put(-60,126){$\hol_j$}
  \put(-65,54){${}_{\hol|_{U_{ijk}}}$}
  \put(-57,74){$\rho_j$}
  \put(-86,18){$\bar t_i$}
  \put(-69,67){$t_j$}
  \put(-40,67){$\bar t_j$}
  \put(-28,20){$t_k$}
  \put(-58,20){$\phi_{ik}$}
  \put(-90,40){${}_{\phi_{ij}}$}
  \put(-28,39){${}_{\phi_{jk}}$}
\end{picture}}
\put(-180,60){$\defas$}
  \end{picture}
  \vskip 1em
  \,.
\end{center}

\paragraph{The various 2-morphisms involved.}
\label{The various 2-morphisms involved}

In the above diagrams we had, as usual, left all re-whiskering
implicit. But for the following considerations it turns out that
we need to take care of these details, lest a crucial point about
the final result remains invisble.

\vskip 1em

Since we are working in the weak 2-category $\GTwoTor$, there are
several different things one could want to mean by ``reversion'' of 
a 2-morphism, depending on what we want to do to the source and 
target 1-mophisms.
We shall define the following notation:

The 2-morphism in $\GTwoTor$ which we want to call $t_i$ is
precisely the following one:

\begin{eqnarray*}
\begin{picture}(100,100)
\includegraphics{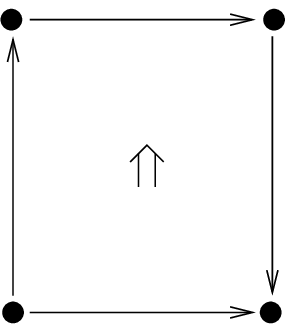}
\put(-63,-7){$\hol|_{U_i}\of{\gamma}$}
\put(-60,93){$\hol_i\of{\gamma}$}
\put(-68,42){$t_i\of{\gamma}$}
\put(-102,42){$t_i\of{x}$}
\put(-2,42){$\bar t_i\of{y}$}
\put(-93,-7){$E_x$}
\put(-2,-7){$E_y$}
\put(-93,93){$G_2$}
\put(-5,93){$G_2$}
\end{picture}
\,.
\end{eqnarray*}

Its inverse 2-morphism is this one:

\begin{eqnarray*}
\begin{picture}(100,100)
\includegraphics{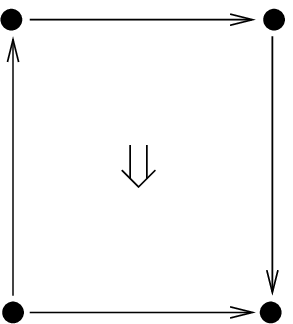}
\put(-63,-7){$\hol|_{U_i}\of{\gamma}$}
\put(-60,93){$\hol_i\of{\gamma}$}
\put(-75,50){$(t_i\of{\gamma})^{-1}$}
\put(-102,42){$t_i\of{x}$}
\put(-2,42){$\bar t_i\of{y}$}
\put(-93,-7){$E_x$}
\put(-2,-7){$E_y$}
\put(-93,93){$G_2$}
\put(-5,93){$G_2$}
\end{picture}
\,.
\end{eqnarray*}

We shall however be interested in the 2-morphism obtained from this
one by reversing the 1-morphism on the left and the right. 
To that end we rewhisker, i.e. we horizontally compose
$(t_i\of{\gamma})^{-1}$ with the identity 2-morphisms on
$\bar t_i\of{x}$ and on $t_i\of{y}$:

\begin{eqnarray*}
\begin{picture}(170,140)
\includegraphics{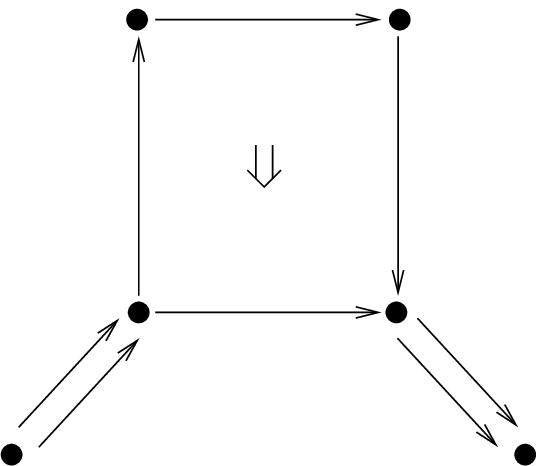}
\put(-35,41){\begin{picture}(0,0)
\put(-63,-7){$\hol|_{U_i}\of{\gamma}$}
\put(-60,93){$\hol_i\of{\gamma}$}
\put(-75,50){$(t_i\of{\gamma})^{-1}$}
\put(-102,42){$t_i\of{x}$}
\put(-2,42){$\bar t_i\of{y}$}
\put(-78,7){$E_x$}
\put(-19,8){$E_y$}
\put(-93,93){$G_2$}
\put(-5,93){$G_2$}
\put(-130,-47){$G_2$}
\put(32,-49){$G_2$}
\put(-93,-27){$\bar t_i\of{x}$}
\put(-120,-11){$\bar t_i\of{x}$}
\put(-11,-27){$t_i\of{y}$}
\put(14,-11){$t_i\of{y}$}
\end{picture}}
\end{picture}
\,.
\end{eqnarray*}

Since $\bar t_i$ is only the weak inverse of $t_i$, we furthermore need
to compose with $\eta_i \defas \mathrm{Id} \to \bar t_i \circ t_i$:

\begin{eqnarray*}
\begin{picture}(170,140)
\includegraphics{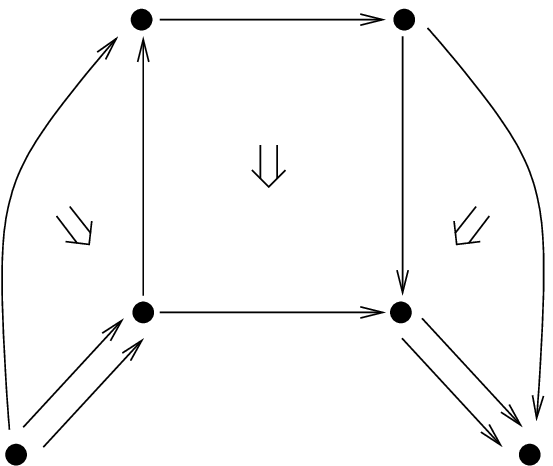}
\put(-37,41){\begin{picture}(0,0)
\put(-63,-7){$\hol|_{U_i}\of{\gamma}$}
\put(-60,93){$\hol_i\of{\gamma}$}
\put(-75,50){$(t_i\of{\gamma})^{-1}$}
\put(-96,49){${}_{t_i\of{x}}$}
\put(-4,49){${}_{\bar t_i\of{y}}$}
\put(-78,7){$E_x$}
\put(-19,8){$E_y$}
\put(-93,93){$G_2$}
\put(-5,93){$G_2$}
\put(-130,-47){$G_2$}
\put(32,-49){$G_2$}
\put(-93,-27){$\bar t_i\of{x}$}
\put(-116,-11){${}_{\bar t_i\of{x}}$}
\put(-11,-27){$t_i\of{y}$}
\put(14,-11){${}_{t_i\of{y}}$}
\put(-110,14){$\eta_i\of{x}$}
\put(3,14){$\eta_i\of{y}$}
\put(-132,24){$\mathrm{Id}$}
\put(37,24){$\mathrm{Id}$}
\end{picture}}
\end{picture}
\,.
\end{eqnarray*}

The resulting 2-morphism is the one we want to call $\bar t_i\of{\gamma}$:

\begin{eqnarray}
\raisebox{-50pt}{
\begin{picture}(100,100)
\includegraphics{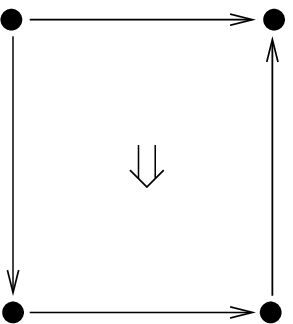}
\put(-63,-7){$\hol|_{U_i}\of{\gamma}$}
\put(-60,93){$\hol_i\of{\gamma}$}
\put(-68,42){$\bar t_i\of{\gamma}$}
\put(-102,42){$\bar t_i\of{x}$}
\put(-2,42){$t_i\of{y}$}
\put(-93,-7){$E_x$}
\put(-2,-7){$E_y$}
\put(-93,93){$G_2$}
\put(-5,93){$G_2$}
\end{picture}
}
\hspace{10pt}
&\defas&
\hspace{20pt}
\raisebox{-75pt}{
\begin{picture}(170,140)
\includegraphics{square2morphrevwhisconj.eps}
\put(-37,41){\begin{picture}(0,0)
\put(-63,-7){$\hol|_{U_i}\of{\gamma}$}
\put(-60,93){$\hol_i\of{\gamma}$}
\put(-75,50){$(t_i\of{\gamma})^{-1}$}
\put(-96,49){${}_{t_i\of{x}}$}
\put(-4,49){${}_{\bar t_i\of{y}}$}
\put(-78,7){$E_x$}
\put(-19,8){$E_y$}
\put(-93,93){$G_2$}
\put(-5,93){$G_2$}
\put(-130,-47){$G_2$}
\put(32,-49){$G_2$}
\put(-93,-27){$\bar t_i\of{x}$}
\put(-116,-11){${}_{\bar t_i\of{x}}$}
\put(-11,-27){$t_i\of{y}$}
\put(14,-11){${}_{t_i\of{y}}$}
\put(-110,14){$\eta_i\of{x}$}
\put(3,14){$\eta_i\of{y}$}
\put(-132,24){$\mathrm{Id}$}
\put(37,24){$\mathrm{Id}$}
\end{picture}}
\end{picture}
}
\,.
\end{eqnarray}

\vskip 1em

Below, it turns out that we need to apply this process of
going from $(t_i\of{\gamma})^{-1}$ to $\bar t_i\of{\gamma}$
the other way around. Since this is an important step, we
go through it again:

\vskip 1em

So start this time with the above 2-morphism

\begin{eqnarray*}
\begin{picture}(100,100)
\includegraphics{square2morph.eps}
\put(-60,-7){$\hol_i\of{\gamma}$}
\put(-63,93){$\hol|_{U_i}\of{\gamma}$}
\put(-68,42){$\bar t_i\of{\gamma}$}
\put(-102,42){$\bar t_i\of{x}$}
\put(-2,42){$t_i\of{y}$}
\put(-93,-7){$G_2$}
\put(-2,-7){$G_2$}
\put(-93,93){$E_x$}
\put(-5,93){$E_y$}
\end{picture}
\,,
\end{eqnarray*}

and take its inverse

\begin{eqnarray*}
\begin{picture}(100,100)
\includegraphics{square2morphrev.eps}
\put(-60,-7){$\hol_i\of{\gamma}$}
\put(-63,93){$\hol|_{U_i}\of{\gamma}$}
\put(-75,50){$(\bar t_i\of{\gamma})^{-1}$}
\put(-102,42){$\bar t_i\of{x}$}
\put(-2,42){$t_i\of{y}$}
\put(-93,-7){$G_2$}
\put(-2,-7){$G_2$}
\put(-93,93){$E_x$}
\put(-5,93){$E_y$}
\end{picture}
\,,
\end{eqnarray*}

then re-whisker

\begin{eqnarray*}
\begin{picture}(170,140)
\includegraphics{square2morphrevwhis.eps}
\put(-35,41){\begin{picture}(0,0)
\put(-60,-7){$\hol_i\of{\gamma}$}
\put(-63,93){$\hol|_{U_i}\of{\gamma}$}
\put(-75,50){$(\bar t_i\of{\gamma})^{-1}$}
\put(-102,42){$\bar t_i\of{x}$}
\put(-2,42){$t_i\of{y}$}
\put(-78,7){$G_2$}
\put(-19,8){$G_2$}
\put(-93,93){$E_x$}
\put(-5,93){$E_y$}
\put(-130,-47){$E_x$}
\put(32,-49){$E_x$}
\put(-93,-27){$t_i\of{x}$}
\put(-120,-11){$t_i\of{x}$}
\put(-11,-27){$\bar t_i\of{y}$}
\put(14,-11){$\bar t_i\of{y}$}
\end{picture}}
\end{picture}
\,,
\end{eqnarray*}

and compose with $\rho_i \defas \mathrm{Id} \to t_i \circ \bar t_i$

\begin{eqnarray*}
\begin{picture}(170,140)
\includegraphics{square2morphrevwhisconj.eps}
\put(-37,41){\begin{picture}(0,0)
\put(-60,-7){$\hol_i\of{\gamma}$}
\put(-63,93){$\hol|_{U_i}\of{\gamma}$}
\put(-75,50){$(\bar t_i\of{\gamma})^{-1}$}
\put(-96,49){${}_{\bar t_i\of{x}}$}
\put(-4,49){${}_{t_i\of{y}}$}
\put(-78,7){$G_2$}
\put(-19,8){$G_2$}
\put(-93,93){$E_x$}
\put(-5,93){$E_y$}
\put(-130,-47){$E_x$}
\put(32,-49){$E_y$}
\put(-93,-27){$t_i\of{x}$}
\put(-116,-11){${}_{t_i\of{x}}$}
\put(-11,-27){$\bar t_i\of{y}$}
\put(14,-11){${}_{\bar t_i\of{y}}$}
\put(-110,14){$\rho_i\of{x}$}
\put(3,14){$\rho_i\of{y}$}
\put(-132,24){$\mathrm{Id}$}
\put(37,24){$\mathrm{Id}$}
\end{picture}}
\end{picture}
\,.
\end{eqnarray*}

The resulting 2-morphism must be $t_i\of{\gamma}$:

\begin{eqnarray}
\label{expression tigamma in terms of inv bar tigamma}
\raisebox{-50pt}{
\begin{picture}(100,100)
\includegraphics{square2morphtotrev.eps}
\put(-60,-7){$\hol_i\of{\gamma}$}
\put(-63,93){$\hol|_{U_i}\of{\gamma}$}
\put(-68,42){$t_i\of{\gamma}$}
\put(-102,42){$t_i\of{x}$}
\put(-2,42){$\bar t_i\of{y}$}
\put(-93,-7){$G_2$}
\put(-2,-7){$G_2$}
\put(-93,93){$E_x$}
\put(-5,93){$E_y$}
\end{picture}
}
\hspace{10pt}
&=&
\hspace{20pt}
\raisebox{-75pt}{
\begin{picture}(170,140)
\includegraphics{square2morphrevwhisconj.eps}
\put(-37,41){\begin{picture}(0,0)
\put(-60,-7){$\hol_i\of{\gamma}$}
\put(-63,93){$\hol|_{U_i}\of{\gamma}$}
\put(-75,50){$(\bar t_i\of{\gamma})^{-1}$}
\put(-96,49){${}_{\bar t_i\of{x}}$}
\put(-4,49){${}_{t_i\of{y}}$}
\put(-78,7){$G_2$}
\put(-19,8){$G_2$}
\put(-93,93){$E_x$}
\put(-5,93){$E_y$}
\put(-130,-47){$E_x$}
\put(32,-49){$E_y$}
\put(-93,-27){$t_i\of{x}$}
\put(-116,-11){${}_{t_i\of{x}}$}
\put(-11,-27){$\bar t_i\of{y}$}
\put(14,-11){${}_{\bar t_i\of{y}}$}
\put(-110,14){$\rho_i\of{x}$}
\put(3,14){$\rho_i\of{y}$}
\put(-132,24){$\mathrm{Id}$}
\put(37,24){$\mathrm{Id}$}
\end{picture}}
\end{picture}
}
\,.
\end{eqnarray}

\vskip 1em

We can define the $a_{ij}$ as the composition of $\bar t_i\of{\gamma}$
with $t_j\of{\gamma}$ up to modification $\phi_{ij}$
\[
  a_{ij}\of{\gamma} 
  \stackto{\phi_{ij}} \bar t_i\of{\gamma} \circ t_j\of{\gamma}
\] as the composition $\bar t_i\of{\gamma} \circ t_j\of{\gamma}$
with boundary
\[
  g_{ij}\of{x} \stackto{\phi_{ij}\of{x}} \bar t_i\of{x} \circ t_j\of{x}
  \,.
\]
In order to perform this compositon we need to rewhisker $t_j$
from the left by the identity 2-morphism $\mathrm{Id}_{\bar t_i}$
and from the right by the identity 2-morphism $\mathrm{Id}_{t_i}$:

\hspace{-1cm}\parbox{10cm}{
\begin{eqnarray}
\label{precise definition of aij 2-morphism in terms of trivializations}
\begin{picture}(350,180)
\includegraphics{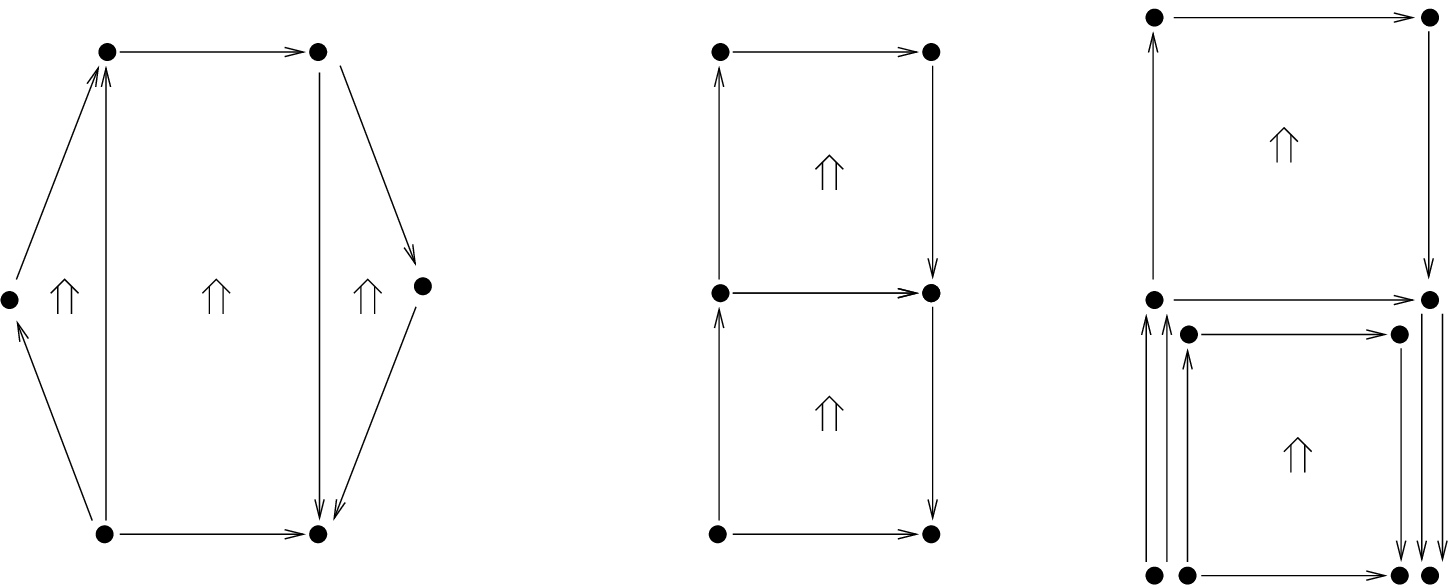}
\put(-60,-8){$\hol_i\of{\gamma}$}
\put(-196,4){$\hol_i\of{\gamma}$}
\put(-56,0){\begin{picture}(0,0)
  \put(-315,4){$\hol_i\of{\gamma}$}
\end{picture}}
\put(-60,170){$\hol_j\of{\gamma}$}
\put(-196,160){$\hol_j\of{\gamma}$}
\put(-56,0){\begin{picture}(0,0)
  \put(-315,160){$\hol_j\of{\gamma}$}
\end{picture}}
\put(-56,0){\begin{picture}(0,0)
  \put(-333,50){${}_{g_{ij}\of{x}}$}
  \put(-354,75){${}_{\phi_{ij}\of{x}}$}
  \put(-268,75){${}_{\phi_{ij}\of{y}}$}
  \put(-294,50){${}_{g_{ij}^{-1}\of{y}}$}
  \put(-310,72){$a_{ij}\of{\gamma}$}
\end{picture}}
\put(-267,80){$\defas$}
\put(-122,80){$\defas$}
\put(-234,40){$\bar t_i\of{x}$}
\put(-190,35){$\bar t_i\of{\gamma}$}
\put(-235,115){$t_j\of{x}$}
\put(-147,40){$t_i\of{y}$}
\put(-424,40){$\bar t_i\of{x}$}
\put(-425,115){$t_j\of{x}$}
\put(-310,40){$t_i\of{y}$}
\put(-306,115){$\bar t_j\of{y}$}
\put(-200,74){$\hol|_{U_{ij}}\of{\gamma}$}
\put(-147,115){$\bar t_j\of{y}$}
\put(-190,106){$t_j\of{\gamma}$}
\put(130,0){\begin{picture}(0,0)
\put(-240,40){$\bar t_i\of{x}$}
\put(-205,40){${}_{\bar t_i\of{x}}$}
\put(-160,40){${}_{t_i\of{y}}$}
\put(-184,23){$\bar t_i\of{\gamma}$}
\put(-238,115){$t_j\of{x}$}
\put(-130,40){$t_i\of{y}$}
\put(-199,87){$\hol|_{U_{ij}}\of{\gamma}$}
\put(-192,67){${}_{\hol|_{U_{ij}}\of{\gamma}}$}
\put(-134,115){$\bar t_j\of{y}$}
\put(-190,113){$t_j\of{\gamma}$}
\end{picture}}
\put(-56,0){\begin{picture}(0,0)
  \put(-343,4){$G_2$}
  \put(-272,4){$G_2$}
  \put(-343,160){$G_2$}
  \put(-272,160){$G_2$}
\end{picture}}
\put(120,0){\begin{picture}(0,0)
\put(-343,4){$G_2$}
\put(-272,4){$G_2$}
\put(-343,160){$G_2$}
\put(-272,160){$G_2$}
\put(-347,80){$E_x$}
\put(-267,80){$E_y$}
\end{picture}}
\put(240,0){\begin{picture}(0,0)
\put(-325,-9){$G_2$}
\put(-255,-9){$G_2$}
\put(-337,169){$G_2$}
\put(-247,169){$G_2$}
\put(-342,81){$E_x$}
\put(-240,81){$E_y$}
\end{picture}}
\end{picture}
\nonumber\\
\nonumber\\
\end{eqnarray}
}

\newpage
\paragraph{Surface Holonomy in Terms of Local Trivializations.}

Now consider the application of $\hol$ to any 2-morphism 
$[\Sigma] \in \P_2\of{M}$ 
that does not necessarily sit in a single patch.

\begin{center}
\begin{picture}(240,500)
\includegraphics{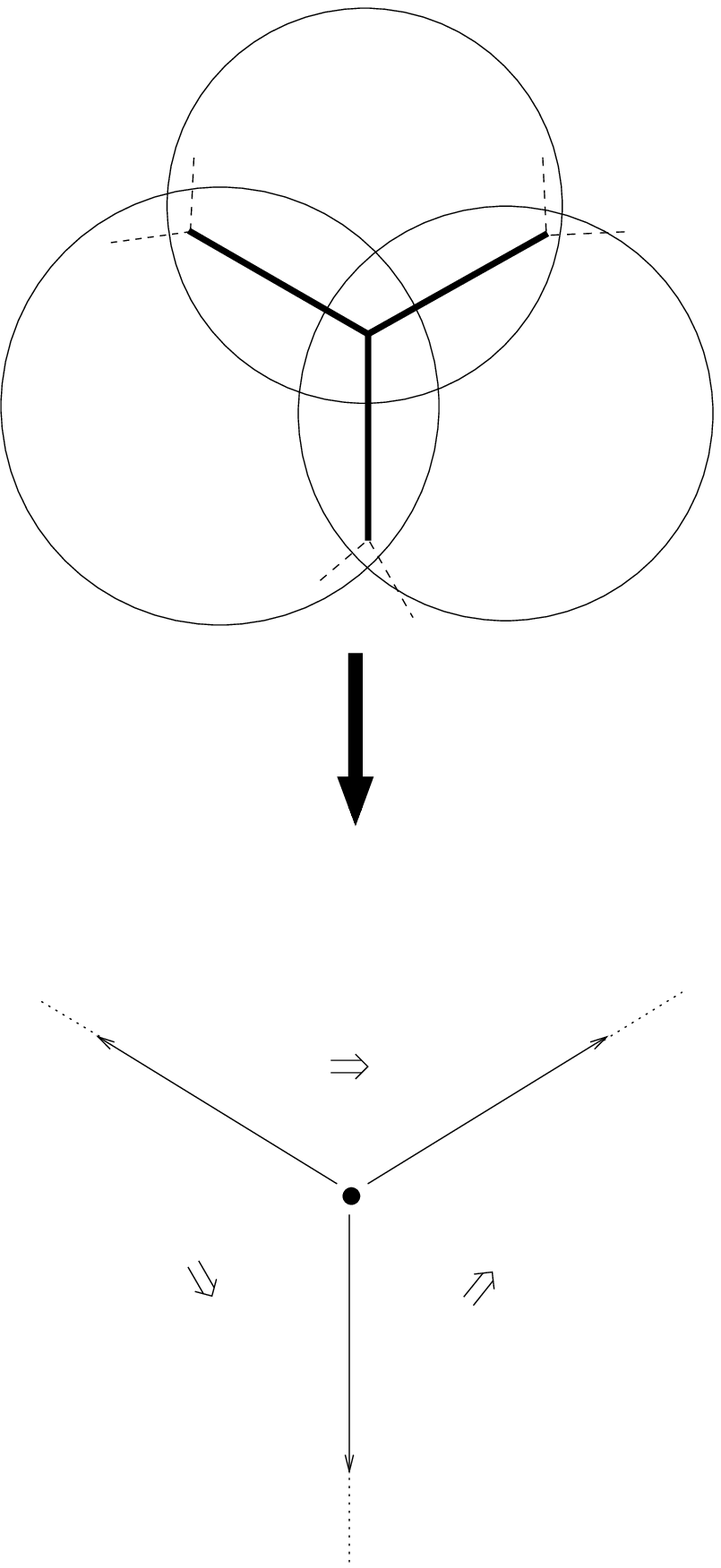}
\put(55,-35){
\begin{picture}(0,0)
\put(-250,360){$U_i$}
\put(-105,360){$U_k$}
\put(-180,520){$U_j$}
\put(-215,405){$[\Sigma_i]$}
\put(-140,405){$[\Sigma_k]$}
\put(-175,477){$[\Sigma_j]$}
\put(-173,439){$x$}
\put(-182,395){$\gamma_3$}
\put(-140,442){$\gamma_2$}
\put(-200,454){$\gamma_1$}
\put(-168,300){$\hol$}
\put(-350,420){$\P_2\of{M}$}
\put(-350,120){$\GTwoTor$}
\put(-260,115){$\hol\of{\Sigma_i}$}
\put(-190,208){$\hol\of{\Sigma_j}$}
\put(-128,115){$\hol\of{\Sigma_k}$}
\put(-255,175){$\hol\of{\gamma_1}$}
\put(-113,185){$\hol\of{\gamma_2}$}
\put(-211,95){$\hol\of{\gamma_3}$}
\put(-184,163){$E_x$}
\end{picture}
}
\end{picture}
\end{center}
 
We can decompose $[\Sigma]$ into several 2-morphisms that all do
sit inside a single patch of the covering. Their images under $\hol$
can be regarded as the ``bottom'' (leftmost surface) of the 
tincan diagram \refer{local trivialization pseudonat traf}. 
Since this tincan diagram 2-commutes, we can replace
each $\hol\of{\Sigma_i}$ by the the respective tincan with its
leftmost surface cut out.
This is shown in the following figure.

\begin{center}
\begin{picture}(240,500)
\includegraphics{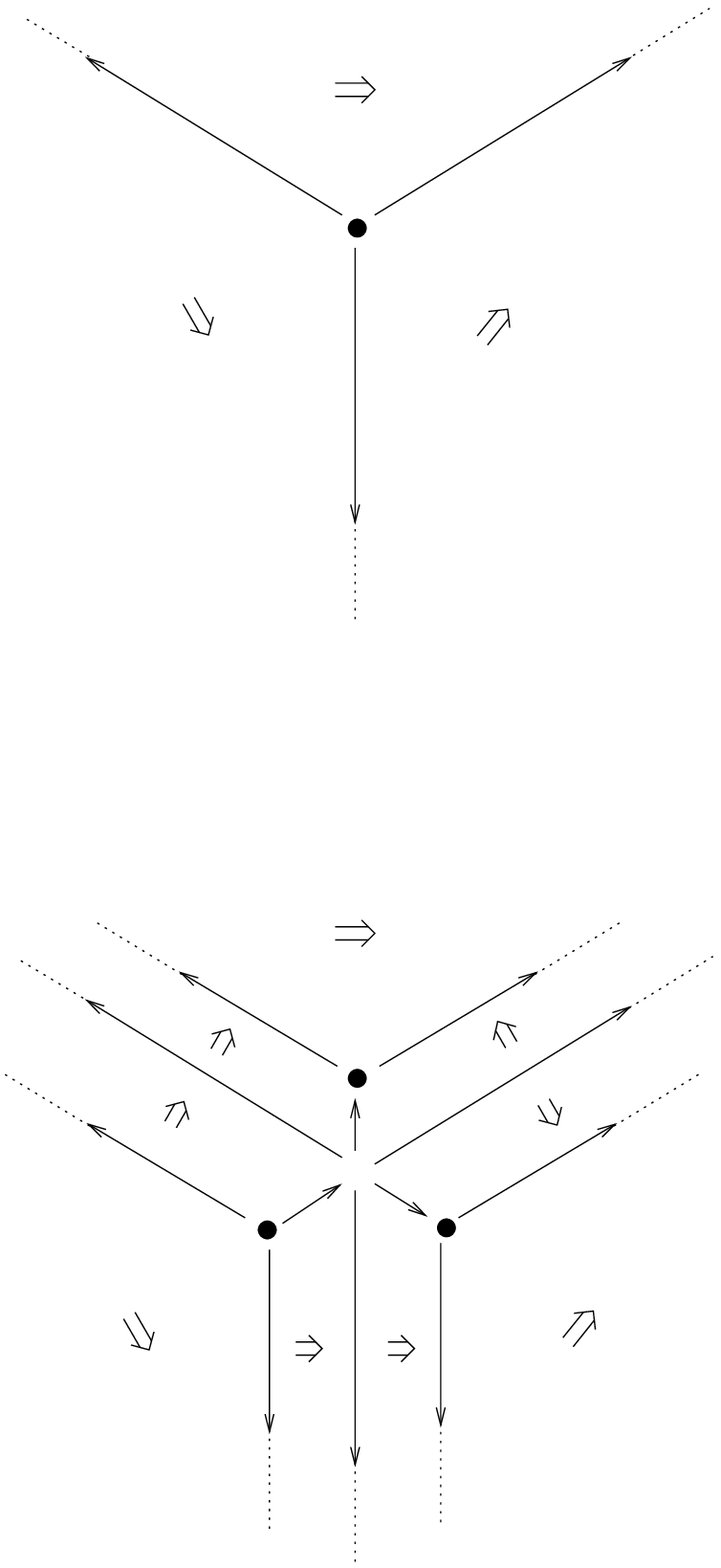}
\put(65,250){
\begin{picture}(0,0)
\put(-260,115){$\hol\of{\Sigma_i}$}
\put(-190,208){$\hol\of{\Sigma_j}$}
\put(-128,115){$\hol\of{\Sigma_k}$}
\put(-185,163){$E_x$}
\put(-255,175){$\hol\of{\gamma_1}$}
\put(-113,185){$\hol\of{\gamma_2}$}
\put(-212,95){$\hol\of{\gamma_3}$}
\end{picture}
}
\put(-115,255){$=$}
\put(65,-40){
\begin{picture}(0,0)
\put(-278,95){$\hol_i\of{\Sigma_i}$}
\put(-195,241){$\hol_j\of{\Sigma_j}$}
\put(-110,97){$\hol_k\of{\Sigma_k}$}
\put(-207,156){${}_{\bar t_i\of{x}}$}
\put(-165,156){${}_{t_k\of{x}}$}
\put(-176,176){${}_{t_j\of{x}}$}
\put(-182,159){${}_{E_x}$}
\put(-200,116){${}_{\bar t_i\of{\gamma_3}}$}
\put(-175,116){${}_{t_k\of{\gamma_3}}$}
\put(-255,178){${}_{\bar t_i\of{\gamma_1}}$}
\put(-241,205){${}_{t_k\of{\gamma_1}}$}
\put(-116,182){${}_{t_k\of{\gamma_2}}$}
\put(-129,205){${}_{t_j\of{\gamma_2}}$}
\put(-255,144){$\hol_i\of{\gamma_1}$}
\put(-220,169){${}_{\hol\of{\gamma_1}}$}
\put(-153,169){${}_{\hol\of{\gamma_2}}$}
\put(-115,155){$\hol_k\of{\gamma_2}$}
\put(-239,70){$\hol_i\of{\gamma_3}$}
\put(-202,60){${}_{\hol\of{\gamma_3}}$}
\put(-150,70){$\hol_k\of{\gamma_3}$}
\put(-219,215){$\hol_j\of{\gamma_3}$}
\put(-170,215){$\hol_j\of{\gamma_3}$}
\end{picture}
}
\end{picture}
\end{center}

We can express $t_j\of{\gamma_2}$ in terms of 
$\bar t_j\of{\gamma_2}$, using equation
\refer{expression tigamma in terms of inv bar tigamma}. This
introduces the 2-morphism $\mathrm{Id} \stackto{\rho_j} t_j \circ \bar t_j$
into the diagram:

\begin{center}
\begin{picture}(240,200)
\includegraphics{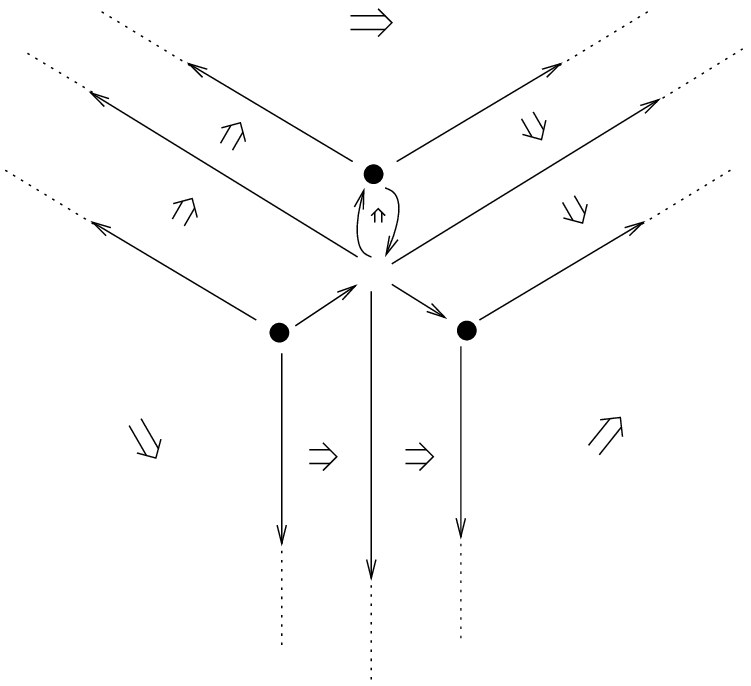}
\put(66,-40){
\begin{picture}(0,0)
\put(-278,95){$\hol_i\of{\Sigma_i}$}
\put(-195,241){$\hol_j\of{\Sigma_j}$}
\put(-110,97){$\hol_k\of{\Sigma_k}$}
\put(-207,156){${}_{\bar t_i\of{x}}$}
\put(-165,156){${}_{t_k\of{x}}$}
\put(-200,179){${}_{t_j\of{x}}$}
\put(-169,179){${}_{\bar t_j\of{x}}$}
\put(-182,159){${}_{E_x}$}
\put(-200,116){${}_{\bar t_i\of{\gamma_3}}$}
\put(-175,116){${}_{t_k\of{\gamma_3}}$}
\put(-255,178){${}_{\bar t_i\of{\gamma_1}}$}
\put(-241,205){${}_{t_k\of{\gamma_1}}$}
\put(-116,182){${}_{t_k\of{\gamma_2}}$}
\put(-129,205){${}_{\bar t_j\of{\gamma_2}}$}
\put(-255,144){$\hol_i\of{\gamma_1}$}
\put(-220,169){${}_{\hol\of{\gamma_1}}$}
\put(-153,169){${}_{\hol\of{\gamma_2}}$}
\put(-115,155){$\hol_k\of{\gamma_2}$}
\put(-239,70){$\hol_i\of{\gamma_3}$}
\put(-202,60){${}_{\hol\of{\gamma_3}}$}
\put(-150,70){$\hol_k\of{\gamma_3}$}
\put(-219,215){$\hol_j\of{\gamma_3}$}
\put(-170,215){$\hol_j\of{\gamma_3}$}
\end{picture}
}
\end{picture}
\end{center}

Then we can compose the $t$ and $\bar t$ pairwise, using
\refer{precise definition of aij 2-morphism in terms of trivializations}.
Compare this to the discussion in 
\S\fullref{section: Locally Trivial 2-Bundles}.
The result is the diagram already familiar from 
\S\fullref{section: The global 2-Holonomy 2-Functor}:

\begin{center}
\begin{picture}(350,310)
\includegraphics{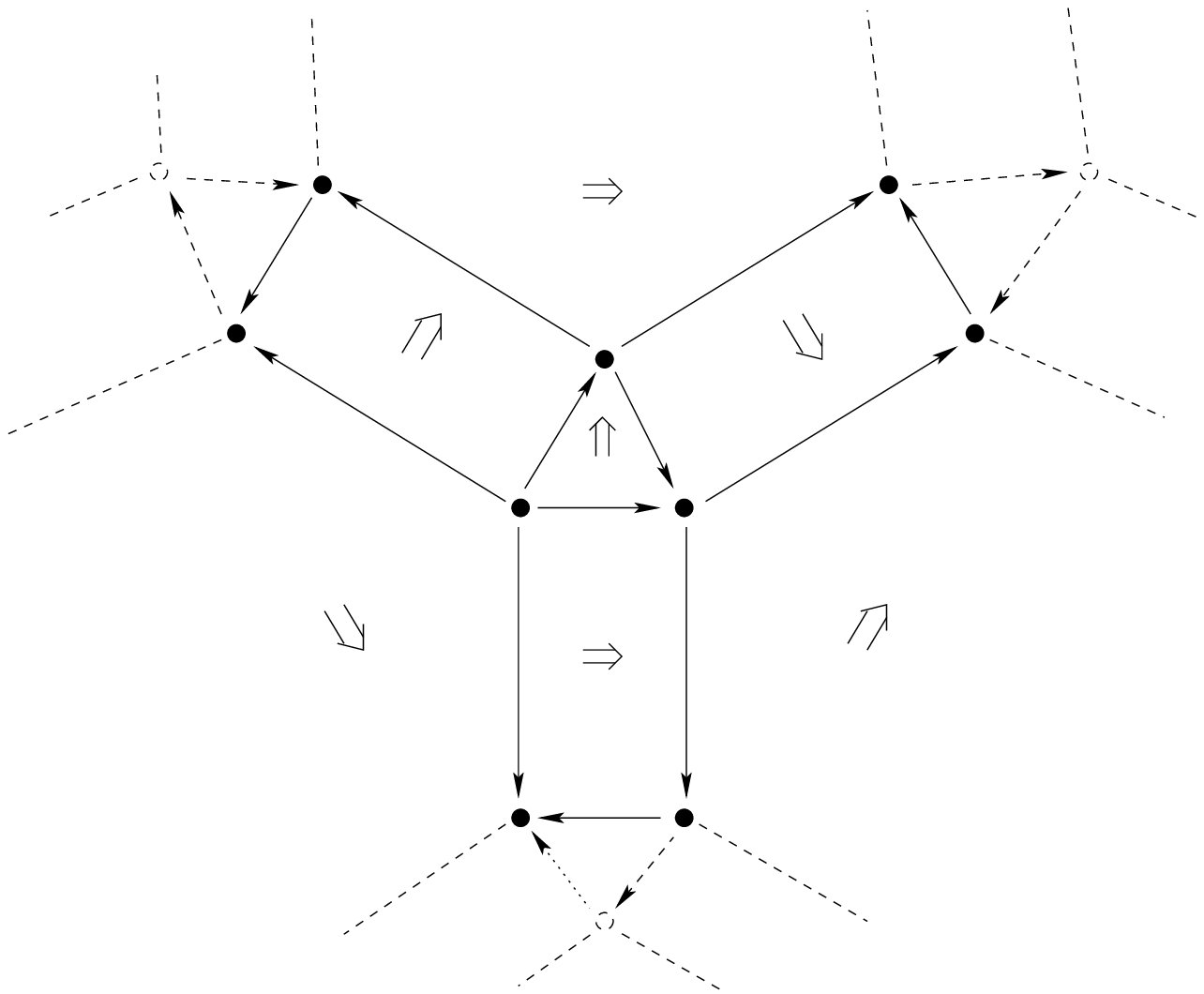}
\put(-14,16){
\begin{picture}(0,0)
\put(-186,124){$g_{ik}\of{x}$}
\put(-180,43){$g_{ik}^{-1}$}
\put(-214,160){$g_{ij}\of{x}$}
\put(-160,160){$g_{jk}\of{x}$}
\put(-190,140){$f_{ijk}\of{x}$}
\put(-186,70){$a_{ik}\of{\gamma_3}$}
\put(-238,95){$\hol_i\of{\gamma_3}$}
\put(-147,95){$\hol_k\of{\gamma_3}$}
\put(-260,140){$\hol_i\of{\gamma_1}$}
\put(-110,144){$\hol_k\of{\gamma_2}$}
\put(-220,210){$\hol_j\of{\gamma_1}$}
\put(-160,213){$\hol_j\of{\gamma_2}$}
\put(-267,185){$a_{ij}\of{\gamma_1}$}
\put(-106,185){$a_{jk}\of{\gamma_2}$}
\put(-290,213){$g_{ij}^{-1}$}
\put(-70,213){$g_{jk}^{-1}$}
\put(-295,90){$\hol_i\of{\Sigma_i}$}
\put(-84,90){$\hol_k\of{\Sigma_k}$}
\put(-186,239){$\hol_j\of{\Sigma_j}$}
\end{picture}
}
\end{picture}
\end{center}

This shows how $\hol \maps \P_2\of{M} \to \GTwoTor$ can be reexpressed 
in terms of $\hol_i \maps \P_2\of{U_i} \to G$ on each $U_i$.

\paragraph{Consistency conditions.}

There are two consistency conditions on the $t_i$.

One comes from the condition that degenerate surfaces do not contribute.
Consider moving $x \in U_{ijk}$ to $x' \in U_{ijk}$ by extending
all of $\gamma_1$, $\gamma_2$, $\gamma_3$ by the \emph{same} 
path $x \to x'$. 

\begin{center}
\begin{picture}(240,220)
\includegraphics{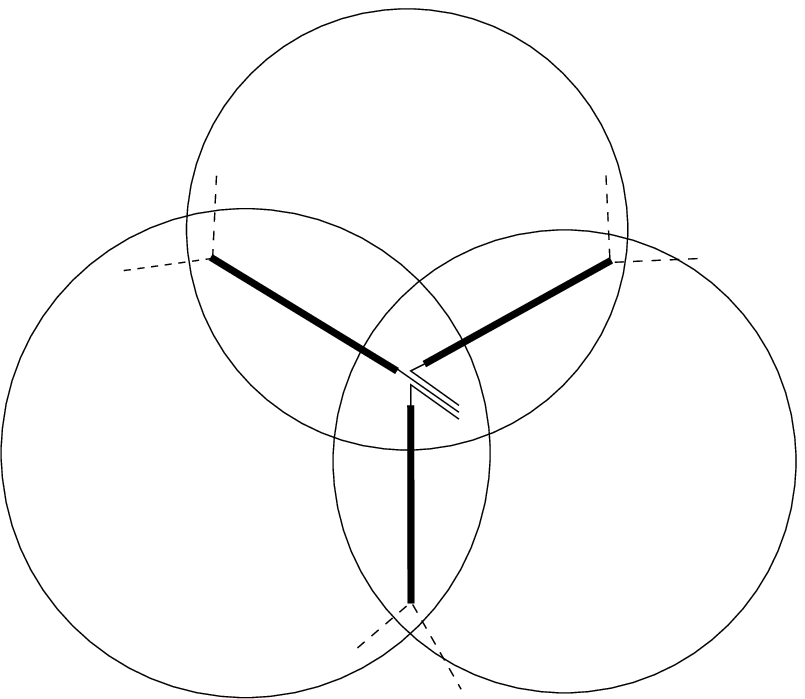}
\put(55,-338){
\begin{picture}(0,0)
\put(-250,360){$U_i$}
\put(-105,360){$U_k$}
\put(-180,520){$U_j$}
\put(-215,405){$[\Sigma_i]$}
\put(-140,405){$[\Sigma_k]$}
\put(-175,477){$[\Sigma_j]$}
\put(-173,436){${}_{x}$}
\put(-156,418){${}_{x'}$}
\put(-182,395){$\gamma_3$}
\put(-140,442){$\gamma_2$}
\put(-200,454){$\gamma_1$}
\end{picture}
}
\end{picture}
\end{center}

This removes $f_{ijk}\of{x}$ in the above diagram and replaces it by
a diagram of the above form around $x'$ but
with all $\Sigma_i$ vanishing. Since this must not contribute, this
diagram has to equal the 2-morphism $f_{ijk}\of{x}$ that was replaced.
In other words, the $a_{ij}$ must be such that the following diagram
2-commutes:

\begin{center}
\begin{picture}(200,280)
\includegraphics{2concoherence.eps}
  \put(-3,100){$\hol_k\of{\gamma}$}
  \put(-194,100){$\hol_i\of{\gamma}$}
  \put(-82,160){${}_{\hol_j\of{\gamma}}$}
  \put(-100,-7){$g_{ik}\of{x}$}
  \put(-100,187){$g_{ik}\of{x'}$}
  \put(-138,43){${}_{g_{ij}\of{x}}$}
  \put(-138,237){${}_{g_{ij}\of{x'}}$}
  \put(-47,43){${}_{g_{jk}\of{x}}$}
  \put(-47,237){${}_{g_{jk}\of{x'}}$}
  \put(-95,23){${}_{f_{ijk}\of{x}}$}
  \put(-95,215){${}_{f_{ijk}\of{x'}}$}
  \put(-100,90){$a_{ik}\of{\gamma}$}
  \put(-150,124){${}_{a_{ij}\of{\gamma}}$}
  \put(-50,117){${}_{a_{jk}\of{\gamma}}$}
\end{picture}
\end{center}
\vskip 2em

This is the transition law on triple overlaps discussed in
\S\fullref{subsubsec: Transition Law on Triple Overlaps}.

The other consistency condition is obtained by considering
vertices at which more than three edges meet. Whenever this is the
case, we can insert constant paths until only trivalent vertices are
left. But these constant paths can be inserted in more than one way.
Four a 4-valent vertex this is indicated by the following figure.

\hspace{-2.3cm}\parbox{10cm}{
\begin{picture}(500,460)
\includegraphics{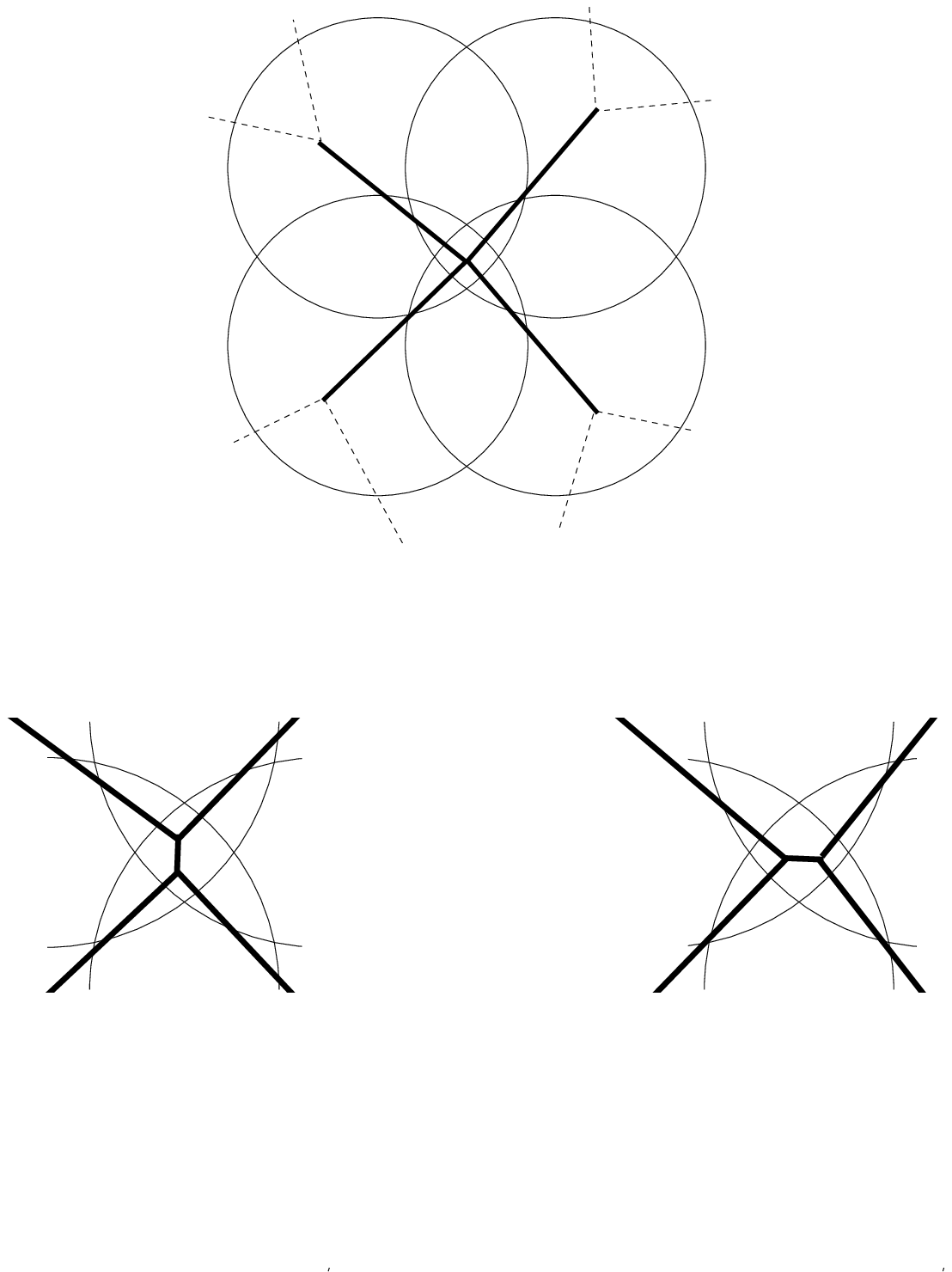}
\put(0,-9){\begin{picture}(0,0)
\put(-330,315){$U_i$}
\put(-330,370){$U_j$}
\put(-190,315){$U_l$}
\put(-190,370){$U_k$}
\put(-290,310){${}_{\gamma_1}$}
\put(-290,381){${}_{\gamma_2}$}
\put(-230,310){${}_{\gamma_4}$}
\put(-222,381){${}_{\gamma_3}$}
\put(-110,-190){\begin{picture}(0,0)
\put(-290,300){${}_{\gamma_1}$}
\put(-290,384){${}_{\gamma_2}$}
\put(-225,305){${}_{\gamma_4}$}
\put(-222,378){${}_{\gamma_3}$}
\put(-238,340){${}_{\gamma_0}$}
\end{picture}}
\put(100,-190){\begin{picture}(0,0)
\put(-290,304){${}_{\gamma_1}$}
\put(-290,379){${}_{\gamma_2}$}
\put(-220,305){${}_{\gamma_4}$}
\put(-198,378){${}_{\gamma_3}$}
\put(-245,344){${}_{\gamma_0}$}
\end{picture}}
\put(-250,150){$=$}
\end{picture}}
\end{picture}
}

\vspace{-2cm}
Here $\gamma_0$ denotes a constant path, which has been drawn with 
a spatial extension just for convenience. It is a well known theorem
(see for instance \cite{FukumaHosonoKawai:1992})
that all triangulations can be obtained from any given one by a series
of moves of two types, one of which is the
one from one of the lower two pictures to the other.
The other is the ``bubble move'' which was
called the ``left and right
unit law'' in \S\fullref{Restriction to the case of trivial base 2-space}.

Invariance of $\hol$ under this move is expressed by the
equation
\begin{eqnarray*}
\begin{picture}(300,130)
\includegraphics{reversedtetrahedron.eps}
\put(-7,2){
\begin{picture}(300,150)
\put(-245,115){${g}_{jk}$}
\put(-65,115){${g}_{jk}$}
\put(-245,-5){${g}_{il}$}
\put(-65,-5){${g}_{il}$}
\put(-307,55){${g}_{ij}$}
\put(-127,55){${g}_{ij}$}
\put(-180,55){${g}_{kl}$}
\put(-1,55){${g}_{kl}$}
\put(-152,55){$=$}
\put(-280,30){${g}_{ik}$}
\put(-40,21){${g}_{jl}$}
\put(-58,80){$f_{jkl}$}
\put(-82,30){$f_{ijl}$}
\put(-234,30){$f_{ikl}$}
\put(-254,77){$f_{ijk}$}
\put(-313,-9){${}_{\hol_i}$}
\put(-313,118){${}_{\hol_j}$}
\put(-178,-9){${}_{\hol_l}$}
\put(-178,118){${}_{\hol_k}$}
\put(183,0){\begin{picture}(0,0)
\put(-313,-9){${}_{\hol_i}$}
\put(-313,118){${}_{\hol_j}$}
\put(-178,-9){${}_{\hol_l}$}
\put(-178,118){${}_{\hol_k}$}
\end{picture}}
\end{picture}
}
\end{picture}
\,,
\end{eqnarray*}
which is equivalent to the 
2-commutativity of this tetrahedron:
\begin{center}
\begin{picture}(180,170)
  \includegraphics{3dtetrahedron.eps}
  \put(-117,56){${}_{g_{ij}}$}
  \put(-63,56){${}_{g_{jk}}$}
  \put(-90,-3){$g_{ik}$}
  \put(-100,22){$f_{ijk}$}
  \put(-137,80){$g_{il}$}
  \put(-41,80){$g_{kl}$}
  \put(-82,95){${}_{g_{jl}}$}
\end{picture}
\end{center} 

This is the tetrahedron transition law on quadruple overlaps
known from equation \refer{transition tetrahedron}
(p. \pageref{transition tetrahedron}) and from 
\S\fullref{Restriction to the case of trivial base 2-space}.

\paragraph{Gauge Transformations.}

The discussion of gauge transformations completely parallels
that in \S\fullref{1-gauge transformations in torsor language},
only that now nontrivial 2-morphisms appear where previously only
identity 2-morphisms were present.

Switching from one local trivialization to another corresponds to
replacing the transition
\begin{eqnarray}
 \begin{picture}(140,110)
  \includegraphics{triangle.eps}
  \put(-94,50){$\bar t_i$}
  \put(-27,50){$t_j$}
  \put(-68,-6){$g_{ij}$}
  \put(-127,-5){$\hol_i$}
  \put(-0,-5){$\hol_j$}
  \put(-63,102){$\hol|_{U_{ij}}$}
  \put(-63,24){$\phi_{ij}$}
  \end{picture}
  \,.
 \nonumber\\
\end{eqnarray}
by another transition
\begin{eqnarray}
 \begin{picture}(140,110)
  \includegraphics{triangle.eps}
  \put(-94,50){$\bar {\tilde t}_i$}
  \put(-27,50){$\tilde t_j$}
  \put(-68,-6){$\tilde g_{ij}$}
  \put(-127,-5){$\tilde \hol_i$}
  \put(-0,-5){$\tilde \hol_j$}
  \put(-63,102){$\hol|_{U_{ij}}$}
  \put(-63,24){$\tilde\phi_{ij}$}
  \end{picture}
  \,.
 \nonumber\\
\end{eqnarray}
Hence we get the following diagram
\begin{eqnarray}
\begin{picture}(170,140)
  \includegraphics{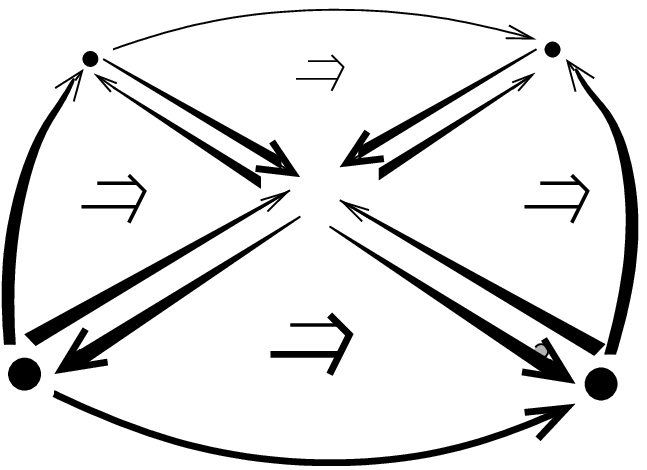}
  \put(0,-106){\begin{picture}(0,0)
  \put(-100,187){${}_{\hol|_{U_{ij}}}$}
  \put(-124,159){${}_{t_i}$}
  \put(-78,159){${}_{\tilde t_i}$}
  \put(-149,168){${}_{\bar t_i}$}
  \put(-63,175){${}_{{\bar {\tilde t}}_i}$}
  \put(2,180){${}_{\tilde g_{ij}}$}
  \put(-195,180){${}_{g_{ij}}$}
  \put(-171,188){${}_{\phi_{ij}}$}
  \put(-43,190){${}_{\tilde \phi_{ij}}$}
  \put(-142,202){${}_{{}_{t_j}}$}
  \put(-130,215){${}_{{}_{\bar t_j}}$}
  \put(-60,198){${}_{{}_{\tilde t_j}}$}
  \put(-68,217){${}_{{}_{\bar{\tilde t}_j}}$}
  \put(-104,98){$h_i$}
  \put(-104,245){${}_{{}_{h_j}}$}
  \put(-198,123){$\hol_i$}
  \put(-06,123){$\tilde \hol_i$}
  \put(-173,231){${}_{{}_{\hol_j}}$}
  \put(-23,233){${}_{{}_{\tilde \hol_j}}$}
  \end{picture}}
\end{picture}
\nonumber\\
\nonumber
\end{eqnarray}

The existence of these modifications of pseudonatural transformations
implies that a diagram as indicated in the following figure
2-commutes:
\begin{eqnarray}
\begin{picture}(170,270)
  \includegraphics{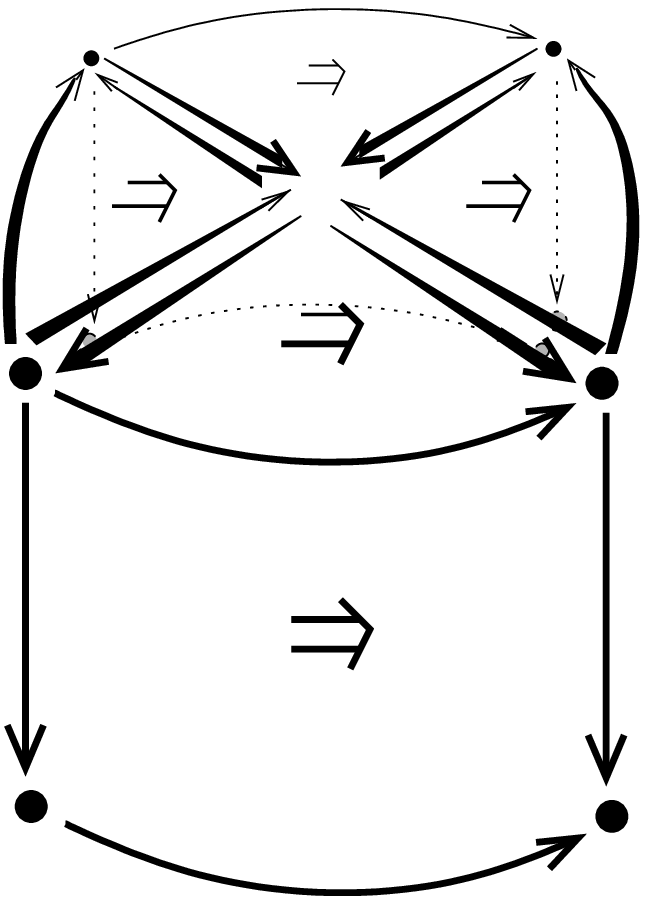}
  \put(0,19){\begin{picture}(0,0)
  \put(-153,195){${}_{\phi_{ij}\of{x}}$}
  \put(-71,187){${}_{\tilde \phi_{ij}\of{x}}$}
  \put(-100,71){$\alpha_i\of{\gamma}$}
  \put(-100,184){${}_{P_x}$}
  \put(2,180){${}_{\tilde g_{ij}\of{x}}$}
  \put(-204,180){${}_{g_{ij}\of{x}}$}
  \put(-8,80){$\tilde \hol_i\of{\gamma}$}
  \put(-210,80){$\hol_i\of{\gamma}$}
  \put(-169,109){$h_i\of{x}$}
  \put(-104,-28){$h_i\of{y}$}
  \put(-104,245){${}_{{}_{h_j\of{x}}}$}
  \end{picture}}
\end{picture}
\nonumber\\
\nonumber
\end{eqnarray}

(For readability, not all 2-morphism are shown.) 
But this is nothing but the naturality tincan diagram describing the
gauge transformations discussed in
\S\fullref{gauge transformations in terms of global 2-holonomy 2-functor}
This means that 
our 2-holonomy 2-functor
\[
  \hol \maps \P_2\of{M} \to \trans_2\of{E}
\]
encodes the same information as 
gauge equivalence classes of 2-functors
\[
  \hol \maps \P_2^C\of{\covering} \to G_2
\]
from the {\v C}ech-extended 2-path 2-groupoid to the structure
2-group, defined in \S\fullref{Global 2-Holonomy in 2-Bundles}.

\clearpage
\section{The Differential Picture: Nonabelian Deligne Hypercohomology}
\label{The Differential Picture: Morphisms between p-Algebroids}

We now study the ``differential version'' of the considerations in 
\S\fullref{principal 2-Bundles} and \S\fullref{the integral picture}.

The differential version of a (Lie-)$p$-groupoid is called a 
(Lie-)$p$-algebroid \cite{BaezCrans:2003}.
The differential version of a
holonomy functor between $p$-groupoids should be a
morphisms between $p$-algebroids. Similarly, the differential
version of a natural transformation 
between
two such $p$-functors should be a 2-isomorphism between morphisms
of $p$-algebroids, and so on. This is indicated on the 
right of figures 
\ref{figure: local 2-holonomy and local 2-connection},
p. \pageref{figure: local 2-holonomy and local 2-connection}
and \ref{figure: 2-bundle as simpl map}, p.
\pageref{figure: 2-bundle as simpl map}.

It is known that Lie $p$-algebras (and Lie $p$-algebroids)
are equivalently described in
terms of $L_\infty$ algebras \cite{BaezCrans:2003}
which again are known to be described
dually in terms of differential graded algebras (dg-algebras)
\cite{LadaStasheff:1992,LadaMarkl:1994,BarnichFulpLadaStasheff:1997}. 
Explicit 
translations from $p$-algebroids
to their dual dg-algebras
are, for $p=1$ and $p=2$, spelled out
in \cite{BojowaldKotovStrobl:2004,KotovSchallerStrobl:2004}.

In the following we first review aspects of the formalism 
of expresssing $p$-algebroid morphisms in terms
of dg-algebra morphisms,
following \cite{BojowaldKotovStrobl:2004,KotovSchallerStrobl:2004,
Mackenzie:1995}.
Then we discuss how the
differential picture of what was done in 
\S\fullref{the integral picture}
has an analogue in this formalism, thereby finding
a generalized form of Deligne hypercohomology.

Finally we check for the special case of (strict) nonabelian 
1- and 2-bundles that the cocycle relations
generated by nonabelian Deligne cohomology do indeed
reproduce the infinitesimal version  of the full
cocycle relations known from the integral picture.

\subsection{Introduction}

  Abelian gerbes with connection and curving 
  are well known to be given by classes in Deligne
  hypercohomology  \cite{Brylinski,Stevenson:2000,Chatterjee:1998,MackaayPicken:2000}, 
  which is the combination of {\v C}ech and
  de Rham cohomology.

  More recently, nonabelian (bundle-)gerbes have been studied in more
  detail \cite{BreenMessing:2001,AschieriCantiniJurco:2004},
  but without an appropriate 
  nonabelian generalization of
  abelian Deligne cohomology available.

  But abelian and nonabelian gerbes may equivalently be described in 
  terms of categorified
  principal fiber bundles \cite{Bartels:2004,BaezSchreiber:2004}. 
  These can be described in terms of
  $p$-functors $\hol_i$ from certain $p$-groupoids 
  $\P_p\of{U_i}$ of paths to the
  structure $p$-group(oid) $G_p$. (Here $p=1$ gives an ordinary bundle
  and $p=2$ a 2-bundle, or gerbe.)

  Similar to how a Lie goup has a differential description in terms 
  of its Lie algebra, there should be a differential description of these
  $p$-functors in terms of morphisms between 
  $p$-algebroids.
  
  Such morphisms have been studied in detail in the context
  of certain topological $\sigma$-models
  \cite{BojowaldKotovStrobl:2004,KotovSchallerStrobl:2004},
  and aspects of their relation to gerbes have been indicated in 
  \cite{Strobl:2004,GruetzmannStrobl:2005}.
  A powerful tool used in these studies is the
  description of $p$-algebroids in terms of their
  \emph{duals}, which are nothing but 
  differential graded algebras (dg-algebras).
  In this language a morphism of $p$-algebroids 
  corresponds to a chain map between dg-algebras,
  a 2-morphisms between two such morphisms corresponds to
  a chain homotopy, and so on.
  There is a natural operator, $Q$, which makes the space of
  dg-algebra $n$-morphisms into a complex.

  We will discuss how the differential version of 
  the description of $p$-bundles with $p$-connection
  in terms of $p$-functors from a $p$-groupoid of
  $p$-paths to the 
  structure $p$-group(oid) corresponds to an assignment of
  dg-algebra $(n+1)$-morphisms to 
  {\v C}ech-$n$-simplices.
  In particular, the path groupoid $\P_p\of{U_i}$ 
  becomes a dg-algebra dual to
  an algebroid $\p_p\of{U_i}$
  and the structure $p$-group(oid) $G_p$ becomes a dg-algebra 
  dual to
  the structure $p$-algebra or structure $p$-algebroid $\g_p$.
  Moreover, any such assignment corresponds to an element
  in the kernel of a generalized noncommutative
  Deligne coboundary operator, which is obtained from the
  ordinary Deligne operator, by, roughly, 
  substituting the operator
  $Q$ for the de Rham differential.

  We demonstrate this explicitly for ordinary principal
  bundles, as well as for principal 2-bundles with 
  strict structure 2-group
  (which correspond to nonabelian gerbes), checking that
  the generalized Deligne closed-ness condition reproduces the
  ``infinitesimal'' version of the known cocycle conditions
  and gauge transformation laws
  for these structures.

  However, the approach presented here applies without 
  any changes to much more general situations than
  these. By choosing appropriate target dg-algebras one obtains
  the infinitesimal version of $p$-bundles with $p$-connection whose 
  ``structure group'' may be semistrict instead of strict, or even
  be a groupoid instead of a group. 
  In fact, there are semistrict Lie $p$-algebras
  which cannot be integrated to any Lie $p$-group, but which 
  can nevertheless be used in the formalism of generalized
  Deligne cohomology. 

  On the other hand, this points to a general issue that we do
  not try to address here, namely the question concerning
  what one would like to address as the process of 
  ``integrating'' a class in generalized 
  Deligne cohomology to
  a proper $p$-bundle with $p$-connection. As long as this
  question is open in the general case, 
  we cannot say how the cohomology classes of
  the generalized Deligne operator correspond exactly to
  gauge equivalence classes of locally trivialized 
  $p$-bundles with $p$-connection.

\subsection{Preliminaries}
\label{dg-algebra preliminaries}

There are some standard concepts and definitions that we should recall 
in order to fix notation and nomenclature:

\subsubsection{{\v C}ech-Simplices}

Consider a given manifold $M$ called the
{\bf base manifold}
together with a {\bf good covering}
\[
  \covering \to M
  \,.
\]
So for some countable index set $I$, $\covering$ is a collection
\[
  \covering = \bigsqcup\limits_{i \in I} U_i
\]
of open subsets $U_i \subset M$, such that $M$
is covered by these,
\[
  M = \bigcup\limits_{i \in I} U_i
  \,,
\]
and such that every nonempty finite intersection
\[
  U_{i_1 i_2 \dots i_n} \defas U_{i_1} \cap U_{i_2} \cap \dots \cap U_{i_n}
  \,,
  \hspace{1cm}\forall\; n = 1,2,\dots
\]
is contractible.

Locally trivializing a (categorified) bundle with respect to $\covering$
involves specifying transition functions and transition
functions between transition functions associated to
``{\v C}ech-simplices''. These simplices are elements of the
{\v C}ech complex:

\begin{definition}
  \label{the Cech complex}
  The $p$th {\bf {\v C}ech chain-complex} 
  \[
    C\of{\covering}
    = \bigoplus\limits_{n=0}^p
    C_n\of{\covering}
  \]
  is the free abelian group 
  generated by all tuples
  $(i_0 i_1 \dots i_n) \in I^{n+1}$,
  for $n \leq p$,
  together with the linear nilpotent {\bf {\v C}ech boundary operator}
  \begin{eqnarray*}
   \begin{array}{rcccl}
    \delta|_{C_n} \defas \delta_n &\maps& C_n\of{\covering} 
      &\to& 
     C_{n-1}\of{\covering}
    \\
    &&(i_0 i_1 \cdots i_n) 
      &\mapsto&
    -\sum\limits_{m=0}^n
    (-1)^{m}
    (i_0 i_1 \cdots \widehat{i_m} \cdots i_n)
    \,,
   \end{array}
  \end{eqnarray*}
  where, as usual, 
  $\widehat{i_m}$ indicates that the element $i_m$ is to be omitted.

  Hence we have a chain complex
  \[
    0 \stackto{0}
    C_p \stackto{\delta} C_{p-1} \stackto{\delta }
    \dots C_0 \stackto{\delta} 0  
    \,.
  \]
\end{definition}

The tuples $(i_0\dots i_n)$ are called 
{\bf {\v C}ech-$n$-simplices} and 
arbitrary linear combinations of them (over $\Z$) are called
{\bf {\v C}ech-chains}.

\begin{definition}
  \label{def source and target boundary}
  We shall write
  $C^{\pm}\of{\covering} \subset C\of{\covering}$ 
  for the subset of elements of positive or negative coeffient,
  respectively, and 
  \[
    P_\pm \maps C\of{\covering} \to C^\pm\of{\covering}
  \]
  for the obvious projection operation.
  For any given chain  
  $c \in C\of{\covering}$
  the chain $\delta c$ is of course called the {\bf boundary} of $c$
  and the elements
  \begin{eqnarray*}
    s\of{c}  &\defas& P_+\of{\delta c} 
    \\
    t\of{c}  &\defas& -P_-\of{\delta c} 
  \end{eqnarray*}
  are called the
  {\bf source boundary} and the {\bf target boundary}
  of $c$, respectively.
\end{definition}
We have
\[
  \delta c = s\of{c} - t\of{c}
  \,.
\]

For example
$(ijk)$ describes the triangle
 \begin{center}
 \begin{picture}(140,120)
  \includegraphics{triangle.eps}
  \put(-120,0){$i$}
  \put(-58,105){$j$}
  \put(2,0){$k$}
  \end{picture}
\end{center}
and $\delta(ijk) = (ik) - (ij) - (jk)$ are the  arrows
making up its boundary, with 
$s\of{ijk} = (ik)$ being the source boundary and
$t\of{ijk} = (ij) + (jk)$ being the
target boundary.

\subsubsection{Differential Graded Algebras}

\begin{definition}
  A {\bf differential graded algebra} 
  or {\bf dg-algebra} 
  $(\extd^V,\bigwedge^\bullet V^*)$
  is a graded vector space
  \[
    V^* = \bigoplus\limits_{n} V^*_n
  \]
  over some field $k$,
  with a graded commutative algebra product
  \[
    \wedge \maps V^* \times V^* \to V^*
    \,,
  \]
  together with a $k$-linear operator
  \[
    \extd^V \maps \bigwedge^n V^* \to \bigwedge^{(n+1)}V^*
  \]
  of degree $+1$ that is nilpotent
  \[
    (\extd^V)^2 = 0
  \]
  and
  that satisfies the graded Leibnitz rule
  \[
    \extd^V (\alpha\wedge \beta)
    =
    (\extd^V \alpha)\wedge\beta + (-1)^{|\alpha|}\alpha \wedge 
    (\extd^V \beta)
    \,,
  \]
 for all
 $\alpha \in \bigwedge^{|\alpha|} V^*$ and
 $\beta \in \bigwedge^{|\beta|} V^*$.
\end{definition}
Note that
\begin{eqnarray*}
  \bigwedge^1 V^*
    &=&
  V^*_1
  \\
  \bigwedge^2 V^*
    &=&
  (V^*_1 \bigwedge V^*_1) \oplus V^*_2
  \\
  \bigwedge^3 V^*
     &=&
  (V^*_1 \bigwedge V^*_1 \bigwedge V^*_1) 
   \oplus (V^*_1 \bigwedge V^*_2)
   \oplus V^*_3
  \\
  \vdots
\end{eqnarray*}

The maximal grade of the graded vector space $V^*$ 
of a dg-algebra corresponds
to the maximal dimension of nontrivial morphisms in the dual
$L_\infty$-algebra. Therefore
\begin{definition}
  \label{def: level of dg-algebra}
  We shall call a dg-algebra $(\extd^V,\bigwedge^\bullet V^*)$
  {\bf of level $p$} if
  \[
     V^*_n = 0,\,
     \forall\, n <0,\; n > p  
    \,.
  \]
\end{definition}

The nice thing about the dg-algebra description of $p$-algebroids
and $L_\infty$-algebras is that the notion of morphism between
dg-algebras is very convenient 
(\cf prop. 2 in \cite{BojowaldKotovStrobl:2004}). It
is nothing but a chain map:

\begin{definition}
  \label{def morphisms between dg-algebras}
  A {\bf morphism between dg-algebras}
  \[
    (\extd^A, \bigwedge^\bullet A^*)
    \stackto{f}
    (\extd^B, \bigwedge^\bullet B^*)	 
  \]
  is a \emph{chain map}, i.e. a linear, grade preserving map
  \[
    f \maps \bigwedge^\bullet B^* \to \bigwedge^\bullet A^*
    \,,
  \] 
  such that all these diagrams commute:
  \begin{center}
\begin{picture}(330,120)
\includegraphics{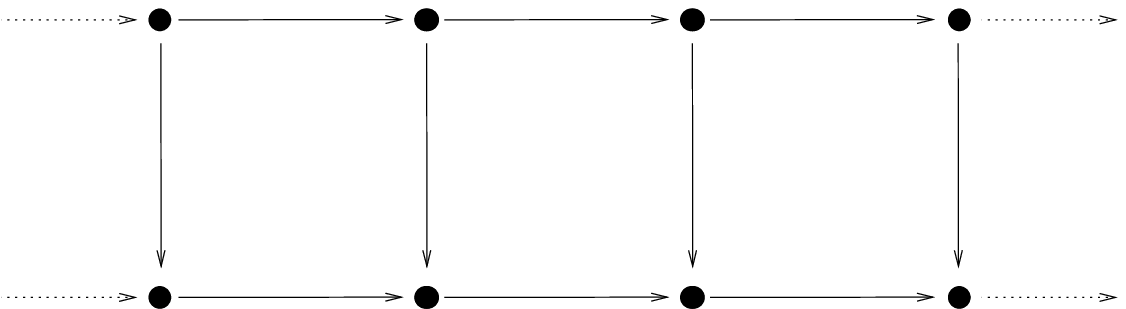}
\put(-290,92){$\bigwedge^1 B^*$}
\put(-245,87){$\extd^B$}
\put(-215,92){$\bigwedge^{2}B^*$}
\put(-168,87){$\extd^B$}
\put(-134,92){$\bigwedge^3 B^*$}
\put(-90,87){$\extd^B$}
\put(-57,92){$\bigwedge^4 B^*$}
\put(-290,-12){$\bigwedge^1 A^*$}
\put(-245,-9){$\extd^A$}
\put(-215,-12){$\bigwedge^2 A^*$}
\put(-168,-9){$\extd^A$}
\put(-134,-12){$\bigwedge^3 A^*$}
\put(-90,-9){$\extd^A$}
\put(-57,-12){$\bigwedge^4 A^*$}
\put(-275,42){$f_1$}
\put(-198,42){$f_2$}
\put(-121,42){$f_3$}
\put(-46,42){$f_4$}
\end{picture}
\end{center}
\vskip 2em
\end{definition}

This means that
\[
  f_n \circ \extd^A - \extd^B \circ f_{n+1} = 0
  \,,
  \hspace{1cm}
  \forall\; n \in \Z
  \,.
\]

For handling such morphisms
it is very convenient (and in fact crucial for
the definition of generalized Deligne cohomology
in \S\fullref{generalized Deligne hypercohomology}) 
to consider the direct sum complex:

\begin{definition}
\label{def direct sum complex}

Given complexes $(\extd^A,\bigwedge^\bullet A^*)$
and $(\extd^B,\bigwedge^\bullet B^*)$, their 
{\bf direct sum complex} is
 
\[
 \left(
  Q 
  \,,
   \bigwedge^\bullet A^* \oplus \bigwedge^\bullet B^*
 \right)
\]
with differential
\[
  Q 
   \defas
   \extd^A \oplus \extd^B
   = 
  \left[
    \begin{array}{cc}
    \extd^A & 0 \\
    0       & \extd^B
    \end{array}
  \right]
  \,.
\]
\end{definition}

We may naturally identifiy every map 
$\bigwedge^\bullet B^* \stackto{f} \bigwedge^\bullet A^*$ with a map
$
\left[
\begin{array}{cc}
  0 & f \\
  0 & 0
\end{array}
\right]
$
of the direct sum complex to itself. 

Using this, the above condition for the chain map $f$
simply says that {\bf a chain map is $Q$-closed}:
\[
  \commutator{Q}{f} = 0
  \,.
\]

A chain map is quite general a concept. We shall be interested
frequently in certain special cases of chain maps. For instance 
in those with the following property:

\begin{definition}
  \label{def: homogenizing dg-algebra morphism}
  We shall call a morphism $f$ of dg-algebras 
  \[
    (\extd^B, \bigwedge^\bullet B^*) \stackto{f}
    (\extd^A, \bigwedge^\bullet A^*)
  \]
  {\bf homogenizing}
  iff
  \[
    f\of{\bigwedge^n B^*} \subset A^*_n \subset \bigwedge^n A^*
    \,.
  \]
\end{definition}

Since 1-morphisms of dg-algebras are nothing but chain maps,
2-morphisms of dg-algebras are nothing but chain homotopies:

\begin{definition}
  \label{dg-algebra 2-morphism}
  A {\bf 2-morphisms between dg-algebra morphisms}
  \[
    \epsilon \maps f \to g
  \]
  with
  \[
    f,g \maps (\extd^B,\bigwedge^\bullet B^*) \to 
     (\extd^A,\bigwedge^\bullet A^*)
  \]
  is a {\bf chain homotopy}, i.e. a linear map
  \[
    \epsilon \maps \bigwedge^\bullet B^* \to \bigwedge^{(\bullet -1)}A^* 
  \]
  such that 
  \[
    f_n - g_n = \extd^B \circ \epsilon_{n+1} + \epsilon_n \circ \extd^A
    \,.
  \]
\end{definition}
The following diagram illustrates this situation (but note that the
triangles in this diagram are \emph{not} supposed to commute):
\begin{center}
\begin{picture}(350,120)
\includegraphics{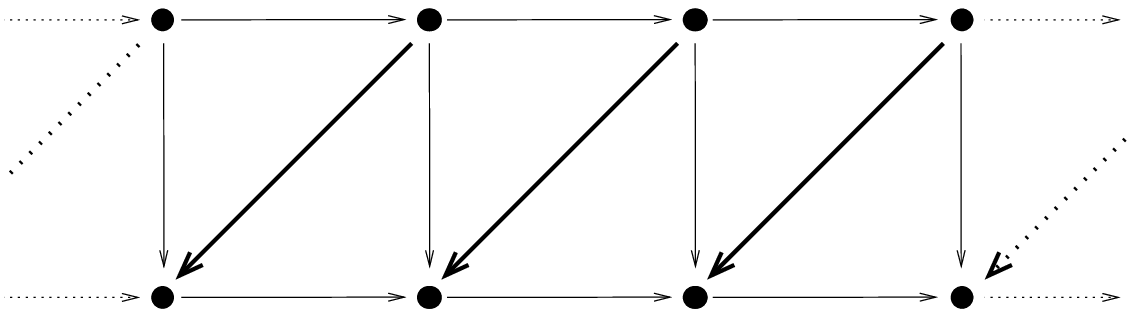}
\put(-290,92){$\bigwedge^1 B^*$}
\put(-245,87){$\extd^B$}
\put(-215,92){$\bigwedge^{2}B^*$}
\put(-168,87){$\extd^B$}
\put(-134,92){$\bigwedge^3 B^*$}
\put(-90,87){$\extd^B$}
\put(-57,92){$\bigwedge^4 B^*$}
\put(-290,-12){$\bigwedge^1 A^*$}
\put(-245,-9){$\extd^A$}
\put(-215,-12){$\bigwedge^2 A^*$}
\put(-168,-9){$\extd^A$}
\put(-134,-12){$\bigwedge^3 A^*$}
\put(-90,-9){$\extd^A$}
\put(-57,-12){$\bigwedge^4 A^*$}
\put(-275,62){$f_1 -g_1$}
\put(-198,62){$f_2-g_2$}
\put(-121,62){$f_3-g_3$}
\put(-46,62){$f_4-g_4$}
\put(-245,30){$\epsilon^1$}
\put(-168,30){$\epsilon^2$}
\put(-91,30){$\epsilon^3$}
\end{picture}
\end{center}

\vskip 2em

Using the language of the direct sum complex
\refdef{def direct sum complex},
this may be expressed by saying that
{\bf a chain homotopy is a shift by a $Q$-exact term}:
\[
  f = g + \commutator{Q}{\epsilon}
  \,.
\]
(Note that $[,]$ denotes the graded commutator and that
$\epsilon$ here is of odd degree.)

The formulation in terms of $Q$ immediately 
suggests how to define dg-algebra 
$n$-morphisms for arbitrary $n$:

\begin{definition}
  \label{def: n-morphisms between dg-algebras}
  Given dg-algebras $(\extd^A,\bigwedge^\bullet A^*)$ and
  $(\extd^B,\bigwedge^\bullet B^*)$, we say that for
  $n \geq 2$
  an {\bf $n$-morphism between dg-algebra $(n-1)$ morphisms}
  \[
    \epsilon \maps \phi \to \gamma
  \]
  with
  \[
    \phi,\gamma \maps \bigwedge^\bullet B^*
      \to 
    \bigwedge^{\bullet -n+2} A^*
  \]
  is a linear map
  \[
    \epsilon \maps \bigwedge^\bullet B^*
      \to 
    \bigwedge^{\bullet -n+1} A^*
  \]
  such that $\phi$ and $\gamma$ differ by the 
  $Q$-exact term $[Q,\epsilon]$:
  \[
    \phi - \gamma = [Q,\epsilon]
    \,.
  \]
\end{definition}

This simple definition 
of maps between dg-algebras 
in terms of {\bf $Q$-cohomology}
turns out to capture all there is
to say about $n$-morphisms between dg-algebras and hence
about $n$-morphisms between $p$-algebroids.

So we now turn to the definition of the dg-algebra
morphism to be called a local $p$-connection and the
$n$-morphisms related to that.

\newpage
\subsection{The $p$-Connection Morphism}

We have discussed
in \S\ref{the integral picture} (for $p=1,2$)
that a $p$-bundle with $p$-connection 
locally gives rise to functors
\[
  \P_p\of{U_i} \stackto{\hol_i} G_p
\]
from the $p$-path $p$-groupoid $\P_p\of{U_i}$ of a given patch $U_i$ 
to the structure $p$-group(oid) $G_p$. 
We would like to find the analogous differential version of these
functors.

Differentiating the source and target $p$-group(oid)s gives
rise to source and target $p$-algebras ($p$-algebroids). The 
$\hol_i$ functors hence should become morphisms of 
$p$-algebroids, or, dually, morphisms of the associated
dg-algebras \refdef{def morphisms between dg-algebras}.

We shall, however, \emph{not} try here to give a precise 
definition of this differentiation procedure of $p$-functors 
between $p$-groupoids. Instead, we will contend ourselves
with proposing a certain obvious
class of dg-algebra morphisms and demonstrating that the
nonabelian Deligne cohomology obtained from them
does have the right properties to be the infinitesimal
description of the integral picture described in 
\S\ref{the integral picture}.

To that end, we discuss in the following first the 
target dg-algebra $\g_p$ that is to replace the target $p$-group(oid)
$G_p$, then the source dg-algebra $\p_p\of{U_i}$ that is to replace the
source $p$-path $p$-groupoid $\P_p$ and finally the 
morphisms of dg-algebras 
\[
  \p_p\of{U_i} \stackto{\con_i} \g_p 
\]
that is to replace the $p$-holonomy $p$-functor between these.

\subsubsection{The Target dg-Algebra}

For demonstrating the consistency of the differential picture with 
the integral picture, we shall be interested in the special case where
the structure $p$-group
$G_p$ is a strict $p$-group, for $p = 1,2$.

For $p=1$ the differential version
of a strict $p$-group
is simply the ordinary Lie algebra
$\g_1 = \mathrm{Lie}\of{G_1}$. The differential version of a
strict 2-group is known \cite{BaezCrans:2003}
as a strict Lie $2$-algebra. These are specified by differential
crossed modules, whose definition is
given in def. \ref{differential crossed module} on
p. \pageref{differential crossed module}.

The dg-algebra dual to this 2-algebra is the following:

\begin{example}
 \label{example: strict Lie alg as dg-alg}
\end{example}

Given a differential crossed module 
$(\g,\h,d\alpha,dt)$
\refdef{differential crossed module},
define a graded vector space 
\[
  V^* = V_1^* \oplus V_2^* 
\] 
by
\begin{eqnarray*}
  V^*_1 &\defas& \g^*
  \\
  V^*_2 &\defas& \h^*
  \,,
\end{eqnarray*}
so that
\begin{eqnarray*}
  \bigwedge^1 V^* &=& \g^*
  \\
  \bigwedge^2 V^* &=& \h^* \oplus \bigwedge^2 \g^*
  \\
  \vdots
\end{eqnarray*}
Define on $\bigwedge^\bullet V^*$ a differential operator 
\[
  \extd^\g \maps \bigwedge^n V^* \to \bigwedge^{(n+1)}V^*
\]
by first picking a basis
\begin{eqnarray}
  \g^* &=& \langle \set{\A^a}_{a = 1\dots \dim\of{\g}}\rangle
  \\
  \h^* &=& \langle \set{{\mathbf b}^A}_{A = 1\dots \dim\of{\h}}\rangle
\end{eqnarray}
and then defining the action of $\extd^\g$ in that basis as
\begin{eqnarray*}
  \extd^\g \A^a 
    &\defas& 
   -\frac{1}{2}C^a{}_{bc}\A^b \A^c
                     - (dt)^a_A {\mathbf b}^A
  \\
  \extd^\g {\mathbf b}^A 
    &\defas& 
  -(d\alpha)^A_{aB}\A^a {\mathbf b}^B
  \,.
\end{eqnarray*}
The various tensor components here are defined 
in the obvious way as follows:
\begin{eqnarray*}
  &&
   [t_a, t_b] = C_{ab}{}^c t_c
  \\
  &&
   d\alpha\of{t_a}\of{s_B} = (d\alpha)^A_{aB} \,s_A
  \\
  &&
   dt\of{s_A} = (dt)_A^a t_a
  \,,
\end{eqnarray*}
where $t_a \in \g$ and $s_A \in \h$  are elements of the dual basis
defined by
\begin{eqnarray*}
  \A^a\of{t_b} &=& \delta^a_b
  \\
  {\mathbf b}^A\of{s_B} &=& \delta^A_B
  \,.
\end{eqnarray*}

It is straightforward to check that $(\extd^\g)^2 = 0$ if and
only if $C$, $d\alpha$ and $dt$ define a differential crossed module.
In this sense $(\extd^\g,\bigwedge^\bullet V^*)$ is the dual
incarnation of the 
differential crossed module $(\g,\h,d\alpha,dt)$.

\subsubsection{The Source dg-Algebra}

The source dg-algebra (representing the path $p$-groupoid
which involves points, paths, surfaces, volumes, etc.)
should be defined on a graded vector space $V^*$ such that
$V^*_0$ knows about points in $U_i$, $V^*_1$ about paths
in $U_i$, $V^*_2$ about surfaces, etc.

This motivates the following definition:

\begin{definition}
  \label{the path p-algebroid}
  For a given patch $U_i$ denote by 
  \[
    \p_p\of{U_i} = (\extd^\p,\bigwedge W^*)
  \]
  the dg-algebra of level $p$ \refdef{def: level of dg-algebra} 
  given by 
\[
  W_n^* 
  \defas 
  \left\lbrace
  \begin{array}{ll}
    \Gamma\of{\bigwedge^n T^*U_i} & \mbox{for $0 \leq n \leq p$}
    \\
    0 & \mbox{otherwise}
  \end{array}
   \right.
\]
(where $\Gamma\of{\bigwedge^n T^*U_i}$ denotes the space of smooth
sections of the $n$-form bundle over $U_i$)
such that for $\omega \in W^*_n$ we have
\[
  \extd^\p \omega \defas \extd \omega  \in W^*_{n+1}
  \,,
\]
where $\extd$ is the ordinary deRham operator on differential forms
 over $U_i$.
\end{definition}

Note that by this definition for instance a $p$-form regarded
as an element of $W^*_p$ is \emph{distinguished} from the same 
$p$-form regarded as an element of $\bigwedge^p W^*_1 $.
In particular
\[
  \extd^\p W^*_p = 0
\]
since $W^*_{(p+1)}$ = 0, while
\[
  \extd^\p 
  \left(
     \bigwedge^p W^*_1
  \right) 
  \subset
  \left(
    \bigwedge^{(p-1)}W^*_1
  \right) \bigwedge W^*_2
\]
need not vanish.

Let us stipulate this family of algebroids $\p_p$ as the desired
source algebroids. This will be justified by the results of the
following sections where it is shown that using this family of
algebroids the results known from the integral picture are
indeed reproduced.

\subsubsection{The Connection Morphism}
\label{the connection morphism}

With source and target algebroids in hand, we can now define the
morphism between them that shall be addressed as a $p$-connection,
serving as a differential analogue of a holonomy $p$-functor.

\begin{definition}
  \label{def p-connection}
  Given an open set $U \subset M$ and a dg-algebra 
  $\g_p = (\extd^\g, \bigwedge^\bullet V^*)$ 
  of level $p$ \refdef{def: level of dg-algebra} 
  we say that a
  dg-algebra morphism \refdef{def morphisms between dg-algebras}
  \[
    \con_U \maps \p_p\of{U} \to \g_p
  \]
  is a {\bf local $p$-connection} on $U$
  if
  \begin{itemize}
    \item
    it is homogenizing \refdef{def: homogenizing dg-algebra morphism}
    \item
    it comes from an algebra homomorphism such that
    \[
        (\con_U)_{\sum_{i=1}^n |v_i|}\of{
             v_1 v_2 \cdots v_n
          }
        = (\con_U)_{|v_1|}\of{v_1}
          \wedge 
         (\con_U)_{|v_2|}\of{v_2}
          \wedge 
          \cdots 
          \wedge 
          (\con_U)_{|v_n|}\of{v_n}
       \in 
       V^*_{\sum_{i=1}^n |v_i|}
      \,,
    \]   
  \end{itemize}
  where 
  (recall this notation from def. \ref{def morphisms between dg-algebras}) 
  $(\con_U)_{n}$ is the restriction of $\con_U$
  to $\bigwedge^n V^*$.
\end{definition}

The meaning of the first of these two conditions, and the reason
for including it, will become
clear when we discuss the $p$-\emph{curvature}
of a $p$-connection in \S\fullref{n-curvature}.

The second condition says that for specifying the action of
$\con_U$ on $\bigwedge^\bullet V^*$ it suffices to know its action on 
$V^*_0$, $V^*_1$, $V^*_2$, etc. 

(So, for instance, the connection morphism $\con_U$ of an 
infinitesimal 1-bundle (to be studied in 
\S\fullref{Infinitesimal 1-Bundles with 1-Connection}) is completely specified
by the image $\con_{U_i}\of{\A^a} = (\con_{U_i})_1\of{\A^a} = A_i^a
\in \Gamma\of{T^* U_i}$.)

\vskip 1em

This defines what we want to call a local $p$-connection,
being the differential version of a local holonomy $p$-functor.
In the integral picture we had 1-morphisms between these
holonomy $p$-functors and 2-morphisms between these 1-morphisms,
and so on. Similarly, here we have to consider morphisms between local
$p$-connections. Since a $p$-connection is a chain map,
such a morphism is a chain homotopy 
\refdef{dg-algebra 2-morphism}. But we are not interested in
arbitrary chain homotopoies, but just in those that
relate local $p$-connections that differ by an
\emph{infinitesimal} gauge transformation. 
We will now explain what this is supposed to mean, following
and building on the discussion in section 4 of
\cite{BojowaldKotovStrobl:2004}.

\subsubsection{Infinitesimal $n$-(gauge)-Transformations}
\label{infinitesimal n-transformations}

Under infinitesimal transformations between $p$-connections
we want to understand maps between elements of a given fiber of the tangent
bundle $T\mathrm{Con}$ of the space $\mathrm{Con}$ of all
$p$-connections over some open set $U$.

\begin{definition}
\label{def connections with derivatives}

If $\epsilon \mapsto \con_U^{(\epsilon)}$ is any smooth
1-parameter family of local $p$-connections, denote by
\[
  [\con_U^{(\epsilon)}]
\]
the equivalence class of such families under the equivalence relation
\[
  \con_U^{(\epsilon)} \sim \widetilde \con_U^{(\epsilon)}
  \;\;
  \Leftrightarrow
  \;\;
  \left\lbrace
  \begin{array}{ll}
    \con_U^{(0)} = \widetilde \con_U^{(0)}
  \\
  \frac{d}{d\epsilon}\con_U^{(\epsilon)}|_{\epsilon = 0}
  =
  \frac{d}{d\epsilon}\widetilde \con_U^{(\epsilon)}|_{\epsilon = 0}
  =
  \con'_U
  \end{array}
  \right.
\]
for some map
\[
  \con'_U \maps \bigwedge^\bullet V^* \to \bigwedge^{\bullet} T^* U
  \,.
\]
We write
\[
  [\con_U^{(\epsilon)}]
  \defas
  [\con_U + \epsilon \,\con'_U]\,.
\]
\end{definition}

The notion of an infinitesimal (gauge) 1-transformation is now
nothing but the concept of a chain homotopy
\refdef{dg-algebra 2-morphism} applied to $\con'_U$:

\begin{definition}
\label{def infinitesimal 1-gauge transformation}

An {\bf infinitesimal (gauge) 1-transformation} 
\[
  \Big[
    \con_U + \epsilon \, \con'_U
  \Big]  
  \stackto{l}
  \Big[
    \con_U + \epsilon\, \widetilde{\con'_U}
  \Big]  
\]
between equivalence classes of families of local $p$-connections 
\refdef{def connections with derivatives}
is a linear map
\[
  l \maps \bigwedge^\bullet V^* \to \bigwedge^{\bullet -1} T^* U
\]
such that
\[
  \widetilde{\con'_U} = \con'_U + [Q,l]
  \,.
\]
\end{definition}

So two infinitesimal 1-transformations are {\bf composable} if they correspond
to the same $\con_U$
and their {\bf composition} 
is given simply by the sum of their generators:
\[
  \Big[
    \con_U + \epsilon \, \con'_U
  \Big]
   \stackto{l_1 } 
  \Big[
    \con_U + \epsilon (\con'_U + [Q,l_1])
  \Big]
   \stackto{l_2}
  \Big[
    \con_U + \epsilon (\con'_U + [Q,l_1+l_2])
  \Big] 
  \,.
\]
In particular, the infinitesimal {\bf inverse (1-)transformation}
to $l$ is its negative $-l$.

\vskip 1em

With infinitesimal 1-transformations in hand it is easy to 
recursively define infinitesimal $n$-transformations as 
follows:

\begin{definition}
  \label{def infinitesimal $n$-transformation}

  With respect to an open set $U\subset M$ and a given 
  $p$-connection $\con_U$ \refdef{def p-connection} we say that
  for $n\geq 2$ an {\bf infinitesimal $n$-transformation}
  \[
    l_{n-1} \stackto{l_n} \tilde l_{n-1}
  \]
  between infinitesimal $(n-1)$-transformations $l_{n-1}$ and
  $\tilde l_{n-1}$ \refdef{def infinitesimal 1-gauge transformation}
  is a linear map
  \[
    l_n \maps \bigwedge^\bullet V^* 
      \to 
    \bigwedge^{\bullet -n}\Gamma\of{T^* U}
  \]
  such that
  \[
    \tilde l_{n­-1} -  l_{n-1} = [Q, l_n]
    \,.
  \]
\end{definition}

So infinitesimal $n$-transformations of local $p$-connections
are essentially nothing but $(n+1)$-morphisms of dg-algebras
\refdef{def: n-morphisms between dg-algebras}, except that
for $n=1$ we use the infinitesimal notion of morphism as given
in def. \ref{def infinitesimal 1-gauge transformation}.

In the following subsection \S\fullref{generalized Deligne hypercohomology}, 
nonabelian Deligne hypercohomology will be obtained by assigning such
infinitesimal $n$-morphisms to {\v C}ech-$n$-simplices.
In that context it will be very convenient, and is
in any case very natural, to call the
equivalence class of a $p$-connection itself
a $0$-transformation:

\begin{definition}
  \label{def 0-transformation}
  An (infinitesimal) {\bf 0-transformation}
  is nothing but an element $[\con_U + \epsilon \, \con'_U]$
  \refdef{def infinitesimal 1-gauge transformation}.
\end{definition}

One easily sees that the maps $l_n$ defining infinitesimal
$n$-transformations cannot be arbitrary linear maps, but have to
come from derivations:

\paragraph{$\con$-Derivations.}

Since we required a $p$-connection $\con_U$ to come from an algebra homomorphism
\refdef{def p-connection}
it is clear that $\con'_U$ 
in def. \refer{def connections with derivatives}
has to be what in def. 6 of
\cite{BojowaldKotovStrobl:2004} is called a
\emph{$\con_U$-Leibniz operator of degree 0} and
what we here will call a $\con_U$-derivation:

\begin{definition}
\label{Phi derivations}

Given any chain map
\[
  \Phi 
   \maps 
    \left(\extd^B,\bigwedge^\bullet B^*\right) 
    \to 
    \left(\extd^A, \bigwedge^{\bullet} A^* \right)
\]
and a linear map 
\[
  \phi \maps \bigwedge^\bullet B^* \to \bigwedge^{\bullet - n} A^*\,,
\]
$\phi$
is called a {\bf $\Phi$-derivation of degree $n$} if it
satisfies the equation
\[
  \phi\of{b_1 b_2} 
    = 
  \phi\of{b_1} \Phi\of{b_2}
  +
  (-1)^{n|b_1|}
  \Phi\of{b_1}\phi\of{b_2}
 \,.
\]
\end{definition}

In the context of the direct sum complex
\refdef{def direct sum complex}, a $\Phi$-derivation is
nothing but an ordinary derivation $\bar \phi$ composed with 
$\Phi \simeq \left[ 
  \begin{array}{cc}
    1 & \Phi \\
    0 & 0
  \end{array}
\right]$:
\[
  \phi = \bar \phi \circ \Phi
  \,.
\]
Since $\Phi$ is $Q$-closed it follows that
\[
  [Q,\phi] = 0 \Leftrightarrow [Q,\bar\phi] = 0
  \,.
\]
This means that we can work with $\Phi$-derivations 
essentially as with ordinary derivations.\footnote{ 
This relies
crucially on the property of $\Phi$ to be a chain map, 
which is the ``on-shell'' condition discussed in detail in
section 4 of \cite{BojowaldKotovStrobl:2004}.}

It follows immediately that

\begin{proposition}
\label{prop ln are con-derivations}

The generator $l_n$ of an infinitesimal $n$-transformation
\refdef{def infinitesimal $n$-transformation} 
with respect to an open set $U$ and a $p$-connection $\con_U$ 
has to be a
$\con_U$-derivation for some $i\in I$ 
\refdef{Phi derivations} of degree $n$.

\end{proposition}

It follows that the $l_n$ are completely specified by
defining their action on $V^*$. We will see examples for this
worked out in 
\S\fullref{Infinitesimal 1-Bundles with 1-Connection} and
\S\fullref{infinitesimal strict 2-bundles}.

\vskip 2em

Before discussing nonabelian Deligne hypercohomology, at last,
we should finish the discussion of local $p$-connections by
mentioning their associated $n$-curvature:

\subsubsection{$n$-Curvature}
\label{n-curvature}

It turns out that the 
{\bf curvature} of a $p$-connection 
$\con_U$ is a measure for its failure to 
constitute a certain chain map
\cite{BojowaldKotovStrobl:2004,GruetzmannStrobl:2005}:

We have defined a local $p$-connection $\con_U$ to be a
morphism of dg-algebras 
\[
  \p_p\of{U} \stackto{\con_U} \g_p
\]
with source the 
$p$-algebroid given by $\p_p\of{U}$
and target the dg-algebra $(\extd^\g, \bigwedge^\bullet V^*)$.
Every such morphism
evidently extends to a morphism
\[
  \mathfrak{dr}\of{U} \stackto{\widehat{\con_U}} \g_p
  \,,
\]
where
\[
  \mathfrak{dr}\of{U}
  \defas
  \left(
    \extd,\; \bigwedge^\bullet \Gamma\of{T^* U}
  \right)
\]
is simply the de Rham complex on $U$.

\begin{definition}
\label{def n-curvature}

The {\bf $n$-curvature} $F_U^{(n)}$ of the local $p$-connection
$\con_U$ is 
an $(n+1)$-form on $U$ taking values in the vector space
$V_n$
defined to be the $Q$-closure of $\widehat{\con_U}$
at degree $n$:
\[
  F^{(n)}_U \defas [Q,\widehat{\con_U}]|_{V^*_n} 
  \in 
  \Gamma\of{\bigwedge^{n+1} T^*U,\;V_n}
  \,.
\]
\end{definition}
Here we used the fact that $[Q,\widehat{\con_U}]$ restricts on 
$V^*_n$ to a map
\[
  V^*_n \to \Gamma\of{\bigwedge^{n+1}T^* U}
  \,,
\]
so that we can regard $F_U^{(n)}$ as an $(n+1)$-form taking values in the 
dual $V_n$ of $V^*_n$.

So for instance if $\set{\A^a}$ is a basis for $V^*_1$
and $\set{t_a}$ is the dual basis, then
\[
  F^{(1)}_U = \sum\limits_a t_a\,[Q,\widehat{\con_U}]\of{\A^a}
  \,.
\]
Examples will be discussed in more detail in
\S\fullref{The local 1-Connection Morphism} and
\S\fullref{The local 2-Connection Morphism}.

The way we have defined $\con_U$ it follows that all these
curvatures, except that for $n=p$, have to \emph{vanish}:
\[
  [Q,\con_U] = 0
  \;\;
  \Leftrightarrow
  \;\;
  F^{(n)}_U = 0\,, \hspace{.3cm}\mbox{for $0 \leq n < p$} .
\]

Had we used $\mathfrak{dr}\of{U}$ instead of $\p_p\of{U}$
as the source $p$-algebroid 
and not required $\con_U$ to be homogenizing,
then also $F^{(n=p)}_i$ would have had to vanish. 
Recall that the definition of $\p_p\of{U}$ was motivated
by the observaiton that 
we needed a dg-algebra of level $p$ \refdef{def: level of dg-algebra}
as a differential analogue of the source $p$-path $p$-groupoid.

A physical motivation for why
all curvatures except that at top level
should vanish can be found in \cite{Strobl:2004}. In 
\S\fullref{The local 2-Connection Morphism} we will see that 
this condition reproduces the condition found in
\cite{GirelliPfeiffer:2004,BaezSchreiber:2004}
of the vanishing of the ``fake curvature''
(\cf \refer{vanishing of fake curvature in central prop conc 2-connections},
p. \pageref{vanishing of fake curvature in central prop conc 2-connections}). In the integral
picture this ensures the functoriality of $\hol_U$, while
in the differential picture it ensures that $\con_U$ is 
a chain map.

The definition of curvature in terms of $Q$-closure has
two very convenient cosnequences for the formalism:

First, since gauge transformations act on $\con_i$ as additive shifts by
$Q$-exact terms, it is immediate that

\begin{proposition}
  
  The $p$-curvature $F^{(p)}$ is a globally defined $(p+1)$-form,
  i.e. there is a global
  \[
    F^{(p)} \in \Omega^{p+1}\of{M,V_p}
  \]
  such that every $F^{(p)}_U$ is the restriction of this form to $U$:
  \[
    F^{(p)}_U = (F^{(p)})|_{U}
    \,.
  \]
\end{proposition}

Second, to every $p$-curvature we immediately get a $p$-Bianchi-identity
\cite{GruetzmannStrobl:2005}:

\begin{definition}
\label{p-Bianchi identity}

  The identity
  \[
    [Q,F^{(p)}] = 0
  \]
  is called the {\bf $p$-Bianchi identity}.

\end{definition}

\newpage

\subsection{Generalized Deligne Hypercohomology}
\label{generalized Deligne hypercohomology}

Motivated by the integral picture of a $p$-bundle with
$p$-connection
(\S\fullref{the integral picture}) 
and following the considerations at the beginning of this
section, 
we are interested in associating 
infinitesimal $n$-transformations with respect to a $p$-connection 
$\con_U$
\refdef{def infinitesimal $n$-transformation} to every 
{\v C}ech-$n$-simplex \refdef{the Cech complex} such that
their source and target $(n-1)$-morphisms are those associated
to the source boundary and target boundary
\refdef{def source and target boundary}
of the $n$-simplex, respectively.

For instance, if $n=2$ we want to construct diagrams like this:
\begin{center}
 \begin{picture}(140,120)
  \includegraphics{triangle.eps}
  \put(-134,0){$(l_i)_0$}
  \put(-70,105){$(l_j)_0$}
  \put(2,0){$(l_k)_0$}
  \put(-109,50){$(l_{ij})_1$}
  \put(-25,50){$(l_{jk})_1$}
  \put(-70,-5){$(l_{ik})_1$}
  \put(-70,20){$(l_{ijk})_2$}
  \end{picture}
\end{center}
\vskip 1em 
Here, by def. \ref{def 0-transformation}, $l_0$ are
equivalence classes of local 
$p$-connections. 
The $l_1$ are infinitesimal 1-transformations 
\refdef{def infinitesimal 1-gauge transformation} between them
and the 
\[
  (l_{ijk})_2 \maps (l_{ik})_1 \to (l_{ij})_1 \circ (l_{jk})_1
\] 
is an infinitesimal 2-transformation
\refdef{def infinitesimal $n$-transformation}
between these 1-transformations.

\vskip 1em

This procedure of associating $n$-transformations to 
$n$-simplices is easily formalized and directly leads to 
the nonabelian Deligne coboundary operator:

\subsubsection{The Double Complex of Sheaves of Infinitesimal 
$n$-Transformations}

Fix a $p \in \N$
and a $p$-connection \refdef{def p-connection}
\[
  \con \defas \con_M \maps \p_p\of{M} \to \g_p
\]
on all of $M$. For any open subset $U \subset M$ write 
$\con|_{U}$ for the obvious restriction
of $\con$ to a $p$-connection $\con|_{U} \maps \p_p\of{U} \to \g_p$.
These $\con_U$ fix fibers in the tangent bundles 
\refdef{def connections with derivatives}
of the spaces of of all $p$-connections
with respect to the subsets $U \subset M$.

For each element $U_i \subset \covering$ of a good covering $\covering$ 
let
\[
  L^n\of{U_i}
\]
be the abelian group (under ordinary addition) of infinitesimal 
$n$-transformations \refdef{def infinitesimal $n$-transformation}
with respect to $\con|_{U_i}$. 
So, by prop. \refer{prop ln are con-derivations}, 
each $L^n\of{U_i}$
is a vector space of linear maps
\[
  l_n \maps \bigwedge^\bullet V^* \to \bigwedge^{\bullet -n}
  \Gamma\of{T^* U_i}
\]
that are $\con|_{U_i}$-derivations \refdef{Phi derivations} of degree $n$.

We then have sheaves $\underbar{L}^n_\con$ of 
infinitesimal $n$-transformations, defined by
\[
  \underbar{L}^n_\con\of{U_i} \defas L^n\of{U_i}
  \,.
\]
With the operator $[Q,\cdot]$ 
\refdef{def direct sum complex}
these form a bounded complex of sheaves
\[
  \underbar{L}^\bullet_\con
  \defas
  0
  \to
  \underbar{L}^p_\con
  \stackto{Q}
  \underbar{L}^{p-1}_\con
  \stackto{Q}
  \cdots
  \stackto{Q}
  \underbar{L}^{0}_\con
  \to
  0  
  \,,
\]
where we decree $L^n$ to have degree $-n$.

Recall the following standard definition of the {\v C}ech
cochain complex with respect to a given sheaf (\cf for instance 
section 1.3 of \cite{Brylinski}):

\begin{definition}
The {\bf {\v C}ech cochain complex of the sheaf}
$\underbar{L}_\con^n$ 
is the complex
\[
  0
  \to
  C^0\of{\covering,\underbar{L}^n_\con}
  \stackto{\delta}
  C^1\of{\covering,\underbar{L}^n_\con}
  \stackto{\delta}
  \cdots
  \stackto{\delta}
  C^p\of{\covering,\underbar{L}^n_\con}
  \to
  0
\]
with respect to $\covering$, defined as follows: 

The sets here are the cartesian products
\[
  C^m\of{\covering, \underbar{L}_\con^n}
  \defas
  \prod\limits_{(i_0,\dots,i_m) \in I^{m+1}}
  \underbar{L}^n_\con\of{U_{i_0\dots i_m}}
  \,,
\]
so that an element $\omega^{m,n} \in C^m\of{\covering, \underbar{L}^n_\con}$
is a map
\[
  \begin{array}{ccccc}
    & & c & \mapsto & \omega\of{c} \in L^n\of{U_{c}}
  \end{array}
\]
that assigns an infinitesimal $n$-morphism 
with respect to $\con|_{U_{i_0 \dots i_m}}$ to every
{\v C}ech $m$-simplex $c = (i_0 \dots i_m)$.

The operator $\delta$ is the dual of the {\v C}ech-boundary
operator \refdef{the Cech complex} denoted by the same symbol, 
composed with the restriction operation in the sheaf $\underbar{L}^n_\con$.
So its action on $\omega \in C^{m}\of{\covering,\underbar{L}^n_\con}$
is
\[
  (\delta\omega)\of{c}
  \defas
  \omega\of{\delta c}|_{U_{c}}
\]
for $c\in C_{m+1}\of{\covering}$ and {\v C}ech-$(m+1)$-simplex.
\end{definition}

Hence we get a {\v C}ech cochain complex assciated to $\underbar{L}^n_\con$
for every $n$. Since the $\underbar{L}^n_\con$ form a complex
themselves with respect to $Q$, the result is a double complex. 
In the standard way
(\cf \cite{Brylinski}, p. 28) we hence arrive at the 
corresponding {\v C}ech hypercohomology:

\begin{definition}
  \label{def generalized Deligne cohomology group}
  The $r$-th
  {\bf generalized (nonabelian) Deligne cohomology}
  group with respect to $\covering$ and $\con$
  is the {\v C}ech hypercohomology
  \[
    \check{H}^r\of{\covering,\underbar{L}^\bullet_\con}
    \,,
  \]
  i.e. the total cohomology of the double complex 
  $C^\bullet\of{\covering,\underbar{L}_\con^\bullet}$
  \[
    \begin{array}{cccccccc}
      && \vdots & & \vdots &&
      \\
      && \Big\uparrow Q & & \Big\uparrow Q &&
      \\
      \cdots &\stackto{\delta} & C^m\of{\covering,\underbar{L}^{n-1}_\con} & \stackto{\delta} & C^{m+1}\of{\covering,\underbar{L}^{n-1}_\con} & \stackto{\delta} & \cdots
      \\
      && \Big\uparrow Q & & \Big\uparrow Q &&
      \\
      \cdots &\stackto{\delta} & C^m\of{\covering,\underbar{L}^n_\con} & \stackto{\delta} & C^{m+1}\of{\covering,\underbar{L}^n_\con} & \stackto{\delta} & \cdots
      \\
      && \Big\uparrow Q & & \Big\uparrow Q &&
      \\
      && \vdots & & \vdots &&
    \end{array}
  \]
  with respect to the {\bf generalized (nonabelian) Deligne coboundary operator}
  \[
    D \defas \delta + (-1)^m Q
    \,,
  \]
  which is of {\bf total degee} 1 with respect to the 
  {\bf total grading}
  \[
    |C^m\of{\covering,\underbar{L}^{n}_\con}|
    =
    m - n
    \,.
  \]
\end{definition}
(Recall that the degree of $L^n$ was defined to be $-n$.)

As usual (e.g. \cite{Brylinski}, def. 1.3.9) 
this notion of generalized
Deligne cohomology 
can be made independent of the choice of cover by taking the 
direct limit over the
set of coverings, $\covering$, which is ordered under the refinement relation:
\[
  \check{H}^r\of{M,\underbar{L}^\bullet_\con}
  \defas
  \lim\limits_{\longrightarrow_\covering}
    \check{H}^r\of{\covering,\underbar{L}^\bullet_\con}
  \,.
\]

\vskip 1em

This is the central definition to be presented here. 
 
\vskip 1em

Before doing anything
with it, let us reassure ourselves that this does incorporate
ordinary Deligne cohomology as a special case.
(Ordinary Deligne
cohomology is introduced in chapter I of
\cite{Brylinski}.  
Helpful discussions can be found in section 5.2 of 
\cite{Stevenson:2000} as well as in section 2.2
of \cite{Chatterjee:1998}.)

\paragraph{Ordinary Deligne Cohomology.}

The case that the structure $p$-algebra is strict and abelian corresponds
to 
\[
  \extd^\g = 0
\] and 
\[
  V^* = V^*_{n=p}
\] 
being 1-dimensional. 

In this case the generalized Deligne cohomology 
\refdef{def generalized Deligne cohomology group} reduces to ordinary Deligne
cohomology as follows:

Denote by $\underbar{\mbox{$\Omega$}}_M^n$ the sheaf of $n$-forms on $M$. 

Since an element in $L^n_\con\of{U}$ maps 
\[
  V^*_{n=p} \to \bigwedge^{p-n} \Gamma\of{T^* U}
\]
we can identify it with a $(p-n)$-form on $U$ and hence we have a surjection
\[
  \underbar{L}^n_\con \stackto{s} \underbar{\mbox{$\Omega$}}^{p-n}_M
  \,.
\]
Noting that under this map we have
\[
  s\of{[Q,l_n]} = \extd\, s\of{l_n}
\]
it follows that the generalized Deligne double complex reduces to the
double complex $C^\bullet\of{\covering,\underbar{\mbox{$\Omega$}}_M^\bullet}$
  \[
    \begin{array}{cccccccc}
      && \vdots & & \vdots &&
      \\
      && \Big\uparrow \extd & & \Big\uparrow \extd &&
      \\
      \cdots &\stackto{\delta} & C^m\of{\covering,\underbar{\mbox{$\Omega$}}^{n+1}_M} & \stackto{\delta} & C^{m+1}\of{\covering,\underbar{\mbox{$\Omega$}}^{n+1}_M} & \stackto{\delta} & \cdots
      \\
      && \Big\uparrow \extd & & \Big\uparrow \extd &&
      \\
      \cdots &\stackto{\delta} & C^m\of{\covering,\underbar{\mbox{$\Omega$}}^n_M} & \stackto{\delta} & C^{m+1}\of{\covering,\underbar{\mbox{$\Omega$}}^n_M} & \stackto{\delta} & \cdots
      \\
      && \Big\uparrow \extd & & \Big\uparrow \extd &&
      \\
      && \vdots & & \vdots &&
    \end{array}
  \]
  with respect to the {\bf ordinary Deligne coboundary operator}
  \[
    D \defas \delta + (-1)^m \extd
    \,.
  \]
The hypercohomology of this {\v C}ech-de Rham complex is precisely
the ordinary Deligne cohomology.

\paragraph{Interpretation of Generalized Deligne Hypercohomology.}
In order to understand how the action of $D$ is related to 
the task of associating $n$-transformations to {\v C}ech-$n$-simplices,
consider the $D$-coboundary condition at degree $0$, 
i.e. consider the condition
for an element
\[
  \omega \in C^0\of{\covering,\underbar{L}^\bullet_\con}
\]
to be $D$-closed:
\begin{eqnarray*}
  && D \omega\of{c} = 0
  \\
  &\Leftrightarrow&
  \omega\of{\delta c} = (-1)^{n+1} [Q,\omega\of{c}]
\end{eqnarray*}
for $c \in C_n\of{\covering}$ any {\v C}ech-$n$-simplex and the restriction to $U_c$ on the
left being implicit.

Following def. \ref{def source and target boundary} we can split the
boundary $\delta c$ of the simplex $c$ into its source part
$s\of{c}$ and its target part $t\of{c}$ as
\[
  \delta c = s\of{c} - t\of{c}
  \,.
\]
This allows to equivalently rewrite the above condition as
\[
  \omega\of{t\of{c}} - \omega\of{s\of{c}}
  =
  (-1)^n [Q,\omega\of{c}]
  \,.
\]
Comparison of this equation with the definition of
an infinitesimal $n$-transformation of $p$-connections
in def. \ref{def infinitesimal $n$-transformation} shows that 
this says nothing but that the $n$-transformation
$\omega\of{c}$ interpolates between the two $(n-1)$-transformations
$\omega\of{s\of{c}}$ and $\omega\of{t\of{c}}$:
\[
  \omega\of{s\of{c}}
  \stackto{ \pm \omega\of{c}}
  \omega\of{t\of{c}}
  \,.
\] 

This is precisely what we are interested in. 

Therefore it is to be expected that the differential 
version of the holonomy $p$-functors
$\set{\hol_i | i \in I}$, together with the $n$-morphisms
relating them on multiple overlaps
in the integral
picture, is given by an element $\omega \in C^0\of{\covering,\underbar{L}^\bullet_\con}$
that is $D$-closed.

Different choices of trivializations of the $p$-bundle
correspond to gauge transformations of this data.
Hence consider any $\lambda \in C^{-1}\of{\covering,\underbar{L}^\bullet_\con}$ to be
be \emph{any} labeling of
{\v C}ech-$n$-simplices by infinitesimal $n+1$-transformations. 
Since $\omega + D\lambda \in C^0\of{\covering,\underbar{L}^\bullet_\con}$ is $D$-closed if
$\omega$ is, it should be true that the shift
\[
  \omega \to \omega + D\lambda
\]
is the infinitesimal version of a gauge transformation of the
holonomy $p$-functor, i.e. of a natural transformation.

\subsubsection{Infinitesimal $p$-Bundles with $p$-Connection}

All this finally motivates the following two definitions:

\begin{definition}
  \label{def infinitesimal p-bundle}
  An {\bf infinitesimal $p$-bundle with $p$-connection}
  on the base manifold $M$ relative to a given global $p$-connection
  $
    \con \maps \p_p\of{M} \to \g_p
  $
  is an element of the generalized Deligne cohomology group
  \refdef{def generalized Deligne cohomology group} 
  $
   \check{H}^0\of{\covering,\underbar{L}^\bullet_{\con}}
  $
  .
\end{definition}

\begin{definition}
  A {\bf local trivialization of an infinitesimal $p$-bundle with $p$-connection}
  on the base manifold $M$ relative to a given global $p$-connection
  $
    \con \maps \p_p\of{M} \to \g_p
  $
  is a representative 
  $\omega \in C^0\of{\covering,\underbar{L}^\bullet_\con}$
  of an element of the generalized Deligne cohomology group
  \refdef{def generalized Deligne cohomology group} 
  $
   \check{H}^0\of{\covering,\underbar{L}^\bullet_{\con}}
  $.
\end{definition}

In the next section it is shown (for $p=1,2$) that, indeed, 
the infinitesimal
version of the cocycle relations of a strict principal $p$-bundle with
$p$-connection (or equivalently of a $(p-1)$-gerbe with connection
and curving) is encoded in the condition
\[
  D\omega = 0 \,,\hspace{1cm} \omega\in C^0\of{\covering,\underbar{L}^\bullet_\con}
  \,,
\]
and that the infinitesimal version of the effect of gauge transformations
(natural transformations of the global holonomy $p$-functor)
is encoded in shifts of the form
\[
  \omega \to \omega + D\lambda \,, \hspace{1cm}\lambda \in 
    C^{-1}\of{\covering,\underbar{L}^\bullet_\con}
  \,.
\]

\newpage

\subsection{Infinitesimal 1-Bundles with 1-Connection} 	
\label{Infinitesimal 1-Bundles with 1-Connection}

First we check how ordinary 
principal bundles 
with connection 
look like in the language of dg-algebra morphisms and
nonabelian Deligne cohomology. In the next subsection
this is then generalized to 2-bundles/gerbes.

\subsubsection{The local 1-Connection Morphism}
\label{The local 1-Connection Morphism}

Let the target dg-algebra be the dual of an 
ordinary Lie (1-)algebra.
This is obtained by considering example
\ref{example: strict Lie alg as dg-alg} 
(p. \pageref{example: strict Lie alg as dg-alg})
and setting $\h = 0$.

Define on each $U_i\,, i\in I$ a 
local 1-connection \refdef{def p-connection} 
$  \con_i \maps  \p_1\of{U_i} \to \g_1 $ by:
\begin{eqnarray*}
  \begin{array}{rcccc}
  (\con_i)_1 &\maps& \bigwedge^1 V^* &\to& \Gamma\of{T^*U_i}
  \\
  && \A^a &\mapsto& A^a_i
  \end{array}
\end{eqnarray*}
Recall that we required the connection to be a homogenizing
\refdef{def: homogenizing dg-algebra morphism}
morphism of dg-algebras and that $V_n^* = 0$ for $n>1$ in 
the algebroid $\p_1\of{U_i}$ \refdef{the path p-algebroid}.
This means that $\con_i$ acts trivially on $\bigwedge^n V^*$ for $n\neq 1$.

This map is a chain map 
\refdef{def morphisms between dg-algebras}
only if $\commutator{Q}{\con_i} = 0$
\refdef{def direct sum complex}, which,
in the present case, is trivially fulfilled:
\[
  \commutator{Q}{\con_i}\of{\A^a} \in V^*_2 = 0
  \,,
\]
by the nature of $\p_1\of{U_i}$.
Recall the discussion in \S\fullref{n-curvature} for the
relevance of this simple fact.
According to the discussion there, the 1-curvature of $\con_i$ 
is
\begin{eqnarray*}
  F_i^{(1)} &=& [Q,\widehat{con_i}]|_{V^*_1}
  \\
  &=&
  F_{A_i}
  \\
  &=&
  \extd A_i + \frac{1}{2}[A_i,A_i]
  \,,
\end{eqnarray*}
as expected.

But according to \S\fullref{infinitesimal n-transformations} we are
to think of $\con_i$ only ``infinitesimally'' in order to construct an
infinitesimal 1-bundle with 1-connection from it. This means that we
are to fix a 1-connection
\[
  \con \defas \con_M \maps \p_1\of{M} \to \g_1
\] 
defined on all of $M$ and work only in the fiber of the tangent bundle of the
space of all 1-connections over $\con$. So replace
$\con_i$ by 
\[
  [\con + \epsilon\, \con'_i]
\]
with $\con'_i$ a $\con|_{U_i}$-derivation of degree 0 \refdef{Phi derivations}.

Writing
\[
  (\con + \epsilon \con'_i)\of{\A^a}
  =
  A_i^a
  =
  A^a + \epsilon A_i^{\prime a}
\]
with
$A^a \in \Gamma\of{T^* M }$ and $A_i^{\prime a} \in \Gamma\of{T^* U_i}$
we get the curvature
\[
  F_i^{(1)}
  =
  \extd A + \frac{1}{2}[A,A] + \extd A'_i + \epsilon [A'_i,A']
  \,.
\]

\subsubsection{Infinitesimal (Gauge) 1-Transformations}
\label{differential pic: gauge trafos in 1-bundles}

Now consider an infinitesimal  1-transformation 
\refdef{def infinitesimal 1-gauge transformation}
\[
  \Big[
    \con + \epsilon\, \con'_i 
  \Big]
  \stackto{l}
  \Big[
    \con + \epsilon \left(
      \con'_i 
      + [Q,l]
     \right)
  \Big]
  \,.
\]
According to 
proposition \ref{prop ln are con-derivations}
(p. \pageref{prop ln are con-derivations})
its generator 
\begin{eqnarray*}
  l \maps \bigwedge^\bullet V^* \to 
  \Gamma\of{\bigwedge^{\bullet-1}T^*U_i}
\end{eqnarray*}
has to be a $\con|_{U_i}$-derivation of degree one
and is hence
completely defined by setting
\begin{eqnarray*}
  l\of{\A^a} &\defas& -(\ln h)^a \in C^\infty\of{U_i}
  \,.
\end{eqnarray*}
Its $Q$-closure is therefore
\begin{eqnarray*}
  \commutator{Q}{l}\of{\A^a},
  &=&
  \extd l\of{\A^a} + l\of{\extd^\g \A^a}
  \\
  &=&
  -\extd (\ln h)^a + l\of{-\frac{1}{2}C^a{}_{bc}\A^b\A^c }
  \\
  &=&
  -\extd (\ln h)^a - \commutator{A}{\ln(h)}^a
  \,.
\end{eqnarray*}
This is the infinitesimal version of the 
usual gauge transformation
\[
  \tilde A_i = h A_i h^{-1} + h \extd h^{-1}
  \,.
\]
Note how the operator $Q$ takes care that the nonabelian term
$\commutator{A}{\ln(h)}$ appears.

\subsubsection{Cocycle Relations}
\label{differential pic: cocycle relations in 1-bundles}

On double intersections, $U_{ij}$, 
let 
$\con'_i|_{U_{ij}}$
and
$\con'_j|_{U_{ij}}$
be related by infinitesimal 1-transformations 
\refdef{def infinitesimal 1-gauge transformation}
\[
  [\con + \epsilon\, \con'_i]|_{U_{ij}} 
   \stackto{\g_{ij}} 
  [\con + \epsilon\, \con'_j]|_{U_{ij}}
  \,,
\] 
i.e.
\[
  \con'_j|_{U_{ij}} = \con'_i|_{U_{ij}} + [Q,\g_{ij}]
  \,,
\]
given by
\[
  \g_{ij}\of{\A^a} = - (\ln g_{ij})^a \in C^\infty\of{U_{ij}}
  \,.
\]
Require that on triple overlaps $U_{ijk}$ the diagram
 \begin{center}
 \begin{picture}(140,120)
  \includegraphics{baretriangle.eps}
  \put(-139,0){$[\con_i]$}
  \put(-70,105){$[\con_j]$}
  \put(2,0){$[\con_k]$}
  \put(-100,50){$\g_{ij}$}
  \put(-25,50){$\g_{jk}$}
  \put(-70,-5){$\g_{ik}$}
  \end{picture}
\end{center}
commutes.
This says that
\[
  \g_{ij} + \g_{jk} = \g_{ik}
  \,.
\]
In components this means that
\[
  \ln\of{g_{ij}} + \ln\of{g_{ij}} = \ln\of{g_{ik}}
  \,,
\]
which is the infinitesimal version of the familiar cocycle condition
\[
  g_{ij}g_{jk} = g_{ik}
  \,.
\]

\subsubsection{Hypercohomology Description}

We now rederive the above considerations using
nonabelian Deligne hypercohomology as defined in
\S\fullref{generalized Deligne hypercohomology}:

The nonabelian 1-bundle is described by the element 
$\omega\in \check{H}^0\of{\covering,\underbar{L}^\bullet_\con}$
given by
\begin{eqnarray*}
  \omega\of{i} &=& \con + \con'_i
  \\
  \omega\of{ij} &=& \g_{ij}
  \,.
\end{eqnarray*}
The condition that this be $D$-closed yields 
\begin{itemize}
\item
the condition that the connection $\con_i$ be a chain map:
\begin{eqnarray*}
  0 &=& (D\omega)\of{i}
  \\
  &=&
  \underbrace{(\delta\omega)\of{i}}_{=0} + (Q\omega)\of{i}
  \\
  &=&
  [Q,\omega\of{i}]
  \\
  &=&
  [Q,\con + \epsilon\,\con'_i]
\end{eqnarray*}

\item
the condition that $\g_{ij}$ is the 1-transformation
relating $\con_i$ with $\con_j$:
\begin{eqnarray*}
  0 &=&
   (D\omega)\of{ij}
  \\
  &=&
  (\delta\omega)\of{ij} + (Q\omega)\of{ij}
  \\
  &=&
  \omega\of{j} - \omega\of{i} - [Q,\omega\of{ij}]
  \\
  &=&
  \con'_j - \con'_i - [Q,\g_{ij}]
\end{eqnarray*}
(Note again that this is formally precisely as in the abelian case, but that
$Q$ correctly incorporates the nonabelian aspects, as 
discussed in \S\fullref{differential pic: gauge trafos in 1-bundles}).

\item
the cocycle condition for $\g_{ij}$:
\begin{eqnarray*}
  0 &=&
   (D\omega)\of{ijk}
  \\
  &=&
  (\delta\omega)\of{ijk} + (Q\omega)\of{ijk}
  \\
  &=&
  -\omega\of{jk} + \omega\of{ik} - \omega\of{ij} + [Q,\omega\of{ijk}]
  \\
  &=&
  \g_{ik} - \g_{ij} - \g_{jk}  - 0
  \,.
\end{eqnarray*}
\end{itemize}

Now let $\lambda\in \Omega_{-1}$ and consider the
Deligne gauge transformation
\[
  \omega \to \omega + D\lambda
\]
where
\begin{eqnarray*}
  \lambda\of{i}
  &=&
  \left(
    \begin{array}{c}
      \A^a \mapsto -(\ln h_i)^a
    \end{array}
  \right)
  \,.
\end{eqnarray*}
We have
\begin{eqnarray*}
  D\lambda\of{i}
  &=&
  [Q,\lambda\of{i}]
  \\
  D\lambda\of{ij}
  &=&
  \lambda\of{i} - \lambda\of{j} - 0
\end{eqnarray*}
and hence
\begin{eqnarray*}
  \con'_i &\to& \con'_i + [Q,\lambda\of{i}]
  \\
  \g_{ij} &\to& \g_{ij} + \lambda\of{i} - \lambda\of{j}
\end{eqnarray*}
which implies
\begin{eqnarray*}
  A_i &\to& A_i  -\extd_{A_i}\ln h_i - [A_i,\ln h_i]
  \\
  \ln g_{ij}
  &\to&
  \ln g_{ij}
  +
  \ln h_i
  -
  \ln h_j
  \,,
\end{eqnarray*}
corresponding to the finite gauge transformations
\begin{eqnarray*}
  A_i &\to& h_i A_i h_i^{-1} + h_i \extd h_i^{-1}
  \\
  g_{ij} &\to& h_i g_{ij}h_j^{-1}
  \,.
\end{eqnarray*}

This shows that nonabelian Deligne hypercohomology
completely captures the infinitesimal version of 
the cocycle conditions of locally trivialized nonabelian 
principal (1-)bundles with (1-)connections, including the rules for
gauge transformations relating different local trivializations.

\newpage
\subsection{Strict Infinitesimal 2-Bundles with 2-Connection} 	
\label{infinitesimal strict 2-bundles}

With just a little more work the above discussions of
infinitesimal 1-bundles generalizes to that of 2-bundles.
The reader should compare the infinitesimal cocycle
relations which we obtain with their finite version
listed in Prop. \ref{central proposition concerning 2-connections}
(p. \pageref{central proposition concerning 2-connections}).

\subsubsection{The local 2-Connection Morphism}
\label{The local 2-Connection Morphism}

Let the target dg-algebra $\g_2$ be the dual of a
strict Lie 2-algebra as in example 
\ref{example: strict Lie alg as dg-alg}. 
Define on each $U_i$ a local 2-connection 
$\con_i \maps \p_2\of{U_i} \to \g_2$
by the following maps:
\begin{eqnarray*}
  \con_i\of{\A^a} &=& A_i^a \in \Gamma\of{T^* U_i}
  \\
  \con_i\of{{\mathbf b}^A} &=& B^A_i \in \Gamma\of{\bigwedge^2 T^* U_i}
  \,.
\end{eqnarray*}
This defines a chain map only if $\commutator{Q}{\con_i} = 0$, which
is the case if
\begin{eqnarray*}
  0 
   &=&
   \extd \con\of{\A^a}
   -
   \con\of{\extd^\g \A^a}
  \\
  &=&
  \extd A^a_i
  +
  \frac{1}{2}C^a{}_{bc}A^b_i\wedge A^c_i + dt^a_A B^A_i
  \\
  &=&
  ( F_{A_i} + dt\of{B_i})^a
  \,.
\end{eqnarray*}
This is the vanishing of the 1-curvature \refdef{def n-curvature}
of the infinitesimal 2-bundle (\cf \S\fullref{n-curvature}).
The expression 
\begin{eqnarray*}
  \label{fake curvature of 2-bundle}
  F_i^{(1)} = F_{A_i} + dt\of{B_i}
\end{eqnarray*}
has been called the \emph{fake curvature} in
\cite{BreenMessing:2001}. It has been found in 
\cite{GirelliPfeiffer:2004,BaezSchreiber:2004} that 
the condition $F_i^{(1)} = 0$ is a necessary 
requirement for $\hol_i$ to be a 2-functor. Here it is,
analogously, a necessary condition for $\con_i$ to be a morphism of
dg-algebras.

Note that there is no condition coming from
\[
  0 = [Q,\con_i]\of{\b^A} \in W^*_3
  \,,
\]
since $W^*_3$ of $\p_2\of{U_i}$ is trivial, by 
def. \ref{the path p-algebroid}. The
2-curvature is
\begin{eqnarray*}
  F_i^{(2)} &=& [Q,\widehat{\con_i}]|_{V^*_2}
  \\
  &=&
  \extd_{A_i} B_i
  \\
  &=&
  \extd B_i + d\alpha\of{A_i}\of{B_i}
  \,,
\end{eqnarray*}
which is in general non-vanishing.

As done for 1-bundles in the previous section, we should really consider
vectors in the tangent bundle to all 2-connections. Fix a 2-connection
\[
  \con \defas \con_m \maps \p_2\of{M} \to \g_2
\]
and replace the above
$\con_i$ by 
\[
  [\con + \epsilon\, \con'_i]
\]
with $\con'_i$ a $\con|_{U_i}$-derivation of degree 0 
\refdef{Phi derivations}.
We write
\begin{eqnarray*}
  && (\con + \epsilon\, \con'_i)\of{\A^a} 
  = A^a|_{U_i} + \epsilon\, A_i^{\prime a}
  \\
  && (\con + \epsilon\, \con'_i)\of{{\mathbf b}^A} 
   = 
  B^A|_{U_i} + \epsilon\, B_i^{\prime A}
  \,.
\end{eqnarray*}

\subsubsection{Infinitesimal (Gauge) 1-Transformations}
\label{diff pic, 2-bundles, gauge transformations}

An infinitesimal 1-gauge transformation 
\refdef{def infinitesimal 1-gauge transformation} 
\[
  \Big[
    \con + \epsilon\, \con'_i 
  \Big]
  \stackto{l}
  \Big[
    \con + 
    \epsilon\, \con'_i 
    + \epsilon [Q,l]
  \Big]
\]
is specified by
\begin{eqnarray*}
  l\of{\A^a} &=& -(\ln h)^a \in C^\infty\of{M}
  \\
  l\of{{\mathbf b}^A} &=& a^A \in \Gamma\of{T^*M}
  \,.
\end{eqnarray*}
The $Q$-closure of this map is given by
\begin{eqnarray*}
  \commutator{Q}{l}\of{\A^a}
  &=&
  \extd l\of{\A^a} + l\of{\extd^\g \A^a}
  \\
  &=&
  -\extd (\ln h)^a + l\of{-\frac{1}{2}C^a{}_{bc}\A^b\A^c 
     - dt^a_A {\mathbf b}^A}
  \\
  &=&
  -\extd (\ln h)^a - \commutator{A}{(\ln h)}^a - dt\of{a}^a
\end{eqnarray*}
and
\begin{eqnarray*}
  \commutator{Q}{l}\of{{\mathbf b}^A}
  &=&
  \extd l\of{{\mathbf b}^A} + l\of{\extd^\g {\mathbf b}^A}
  \\
  &=&
  \extd a^A + l\of{-d\alpha^A_{aB}\A^a{\mathbf b}^B}
  \\
  &=&
  d\alpha\of{\ln h}\of{B}^A
  +
  \extd a^A + d\alpha\of{A}\of{a}^A
  \,.
\end{eqnarray*}

This is the infinitesimal version of 
\begin{eqnarray*}
  A
  &\to&
  h A h^{-1} + h \extd h^{-1} - dt\of{a}
  \\
  B
  &\to&
  \alpha\of{h}\of{B} + \extd_A a + a\wedge a
  \,.
\end{eqnarray*}

\subsubsection{Cocycle Relations}
\label{cocycle relations for inf. strict 2-bundles}

On each $U_{ij}$
let there be an infinitesimal 1-transformation
\begin{eqnarray*}
  \Big[
    \con + \epsilon\, \con'_i
  \Big] 
    \stackto{\g_{ij}} 
  \Big[
    \con + \epsilon \, \con'_j
  \Big]
\end{eqnarray*}
given by
\begin{eqnarray*}
  \g_{ij}\of{\A^a} &=& -\ln(g_{ij})^a
  \\
  \g_{ij}\of{{\mathbf b}^A} &=& a_{ij}^A
  \,.
\end{eqnarray*}

The differential version of the diagram 
on p. \pageref{transition modification},
which expresses how the 2-holonomy is gauge transformed
when going from $U_i$ to $U_j$ to $U_k$,
is the diagram
\begin{center}
 \begin{picture}(140,120)
  \includegraphics{triangle.eps}
  \put(-139,0){$[\con_i]$}
  \put(-70,105){$[\con_j]$}
  \put(2,0){$[\con_k]$}
  \put(-100,50){$\g_{ij}$}
  \put(-25,50){$\g_{jk}$}
  \put(-70,-5){$\g_{ik}$}
  \put(-70,24){$\f_{ijk}$}
  \end{picture}
\end{center}
\vskip 1em
This says that there is an infinitesimal
2-transformation \refdef{def infinitesimal $n$-transformation}
given by
\[
  \f_{ijk} 
    \maps 
   \bigwedge^\bullet V^* \to 
    \bigwedge^{\bullet -2} T^* U_{ijk}
\]
\begin{eqnarray*}
  \f_{ijk}\of{{\mathbf b}^A} &=& -(\ln f_{ijk})^A \in C^\infty\of{U_{ijk}}
\end{eqnarray*}
going between the 1-gauge transformations $\g_{ij}\circ \g_{jk}$ and
$\g_{ik}$. This means that the equation
\[
  \g_{ij} + \g_{jk} = \g_{ik} + [Q,\f_{ijk}]
\]
holds.

In terms of components this equation implies that
\[
  \ln g_{ij} + \ln g_{jk} = \ln g_{ik} + dt\of{\ln f_{ijk}}
\]
and
\[
  a_{ij} + a_{jk} = a_{ik} - \extd_{A_i} \ln f_{ijk}
  \,.
\]
These are indeed the infinitesimal versions of the 
known cocycylce conditions for nonabelian 2-bundles/gerbes.

The $\f_{ijk}$ have to satisfy a law 
saying that the 3-morphism inside this tetrahedron
is trivial:
\begin{center}
\begin{picture}(180,170)
  \includegraphics{3dtetrahedron.eps}
  \put(-117,56){${}_{\g_{ij}}$}
  \put(-63,56){${}_{\g_{jk}}$}
  \put(-90,-3){$\g_{ik}$}
  \put(-100,22){$\f_{ijk}$}
  \put(-137,80){$\g_{il}$}
  \put(-41,80){$\g_{kl}$}
  \put(-82,95){${}_{\g_{jl}}$}
  \put(-193,-3){$[\con_i]$}
  \put(3,-3){$[\con_k]$}
  \put(-82,67){${}_{[\con_j]}$}
  \put(-82,150){${}_{[\con_l]}$}
\end{picture}
\end{center}
\vskip 2em
There are $\f$s labeling all the four faces of this
tetrahedron, but for readability only one of them
has been indicated. When the tetrahedron is flattened
out we can write
\begin{center}
\begin{picture}(300,150)
\includegraphics{reversedtetrahedron.eps}
\put(-7,2){
\begin{picture}(300,150)
\put(-245,115){$\g_{jk}$}
\put(-65,115){$\g_{jk}$}
\put(-245,-5){$\g_{il}$}
\put(-65,-5){$\g_{il}$}
\put(-307,55){$\g_{ij}$}
\put(-127,55){$\g_{ij}$}
\put(-180,55){$\g_{kl}$}
\put(-1,55){$\g_{kl}$}
\put(-152,55){$=$}
\put(-280,30){$\g_{ik}$}
\put(-40,21){$\g_{jl}$}
\put(-58,80){$\f_{jkl}$}
\put(-82,30){$\f_{ijl}$}
\put(-234,30){$\f_{ikl}$}
\put(-254,77){$\f_{ijk}$}
\put(-320,-9){${}_{[\con_i]}$}
\put(-320,118){${}_{[\con_j]}$}
\put(-178,-9){${}_{[\con_l]}$}
\put(-178,118){${}_{[\con_k]}$}
\put(183,0){\begin{picture}(0,0)
\put(-320,-9){${}_{[\con_i]}$}
\put(-320,118){${}_{[\con_j]}$}
\put(-178,-9){${}_{[\con_l]}$}
\put(-178,118){${}_{[\con_k]}$}
\end{picture}}
\end{picture}
}
\end{picture}
\vskip 1em
\end{center}
which implies (for infinitesimal transformations)
\[
  \f_{jkl} - \f_{ikl} + \f_{ijl} - \f_{ijk} = 0
  \,.
\]
In components this says that
\[
  \ln f_{jkl} - \ln f_{ikl} + \ln f_{ijl} - \ln f_{ijk} = 0
 \,.
\]
This is once again the correct infinitesimal version of the
respective cocycle condition in a nonabelian 2-bundle/gerbe.

Of course for this last relation all effects of the nonabelian nature
of $\g$ and $\h$ are of higher order  
and hence do not appear
here. So the above equation is of the same form as in the abelian 
case.

\subsubsection{Hypercohomology Description}
\label{hypercohomology description of strict inf. 2-bundles}

We now rederive all the above considerations 
concerning 2-bundles with 2-connection using
nonabelian Deligne hypercohomology
(\cf \S\fullref{generalized Deligne hypercohomology}):

The nonabelian 2-bundle is described by the element 
$\omega\in \check{H}^0\of{\covering,\underbar{L}^\bullet_\con}$
given by
\begin{eqnarray*}
  \omega\of{i} &=& \con + \epsilon\, \con'_i
  \\
  \omega\of{ij} &=& \g_{ij}
  \\
  \omega\of{ijk} &=& \f_{ijk}
  \,.
\end{eqnarray*}
the condition that this be $D$-closed yields 
\begin{itemize}
\item
the condition that the connection $\con_i$ be a chain map:
\begin{eqnarray*}
  0 &=& (D\omega)\of{i}
  \\
  &=&
  \underbrace{(\delta\omega)\of{i}}_{=0} + (Q\omega)\of{i}
  \\
  &=&
  [Q,\omega\of{i}]
  \\
  &=&
  [Q,\con + \epsilon \, \con'_i]
\end{eqnarray*}

\item
the condition that $\g_{ij}$ is the 1-transformation
relating $\con_i$ with $\con_j$:
\begin{eqnarray*}
  0 &=&
   (D\omega)\of{ij}
  \\
  &=&
  (\delta\omega)\of{ij} + (Q\omega)\of{ij}
  \\
  &=&
  \omega\of{j} - \omega\of{i} - [Q,\omega\of{ij}]
  \\
  &=&
  \con'_j - \con'_i - [Q,\g_{ij}]
\end{eqnarray*}
(Note again that this is formally precisely as in the abelian case, but that
$[Q,\g_{ij}]$ correctly incorporates the nonabelian aspects and
would also incorporate weak aspects, if included, as 
discussed in \S\fullref{diff pic, 2-bundles, gauge transformations}).

\item
the cocycle condition for $\g_{ij}$:
\begin{eqnarray*}
  0 &=&
   (D\omega)\of{ijk}
  \\
  &=&
  (\delta\omega)\of{ijk} + (Q\omega)\of{ijk}
  \\
  &=&
  -\omega\of{jk} + \omega\of{ik} - \omega\of{ij} + [Q,\omega\of{ijk}]
  \\
  &=&
  \g_{ik} + [Q,\f_{ijk}]
  -
  \g_{ij} - \g_{jk} 
  \,.
\end{eqnarray*}

\item
the coherence law for $\f_{ijk}$:
\begin{eqnarray*}
  0 &=&
  (D\omega)\of{ijkl}
  \\
  &=&
  (\delta\omega)\of{ijkl} - (Q\omega)\of{ijkl}
  \\
  &=&
  -\omega\of{jkl} + \omega\of{ikl} - \omega\of{ijl} + \omega\of{ijk}
  \\
  &=&
  -\f\of{jkl} + \f\of{ikl} - \f\of{ijl} + \f\of{ijk}
\end{eqnarray*}

\end{itemize}

Now let $\lambda\in \Omega_{-1}$ and denote the
components of $\lambda$ as follows:
\begin{eqnarray*}
  \lambda\of{i}
  &=&
  \left(
    \begin{array}{rcl}
      \A^a &\mapsto& -(\ln h_i)^a
      \\
      {\mathbf b}^A &\mapsto& \alpha_i^A
    \end{array}
  \right)
  \\
  \lambda\of{ij}
  &=&
  \left(
    \begin{array}{rcl}
      \A^a &\mapsto& 0
      \\
      {\mathbf b}^A &\mapsto& -(\ln p_{ij})^A
    \end{array}
  \right)  
  \,.
\end{eqnarray*}
Consider the
Deligne gauge transformation
\[
  \omega \to \omega + D\lambda
  \,.
\]
We have
\begin{eqnarray*}
  D\lambda\of{i}
  &=&
  [Q,\lambda\of{i}]
  \\
  D\lambda\of{ij}
  &=&
  \lambda\of{j} - \lambda\of{i} - [Q,\lambda\of{ij}]
  \\
  D\lambda\of{ijk}
  &=&
  \lambda\of{ik} - \lambda\of{ij} - \lambda\of{jk} + [Q,\lambda\of{ijk}]  
  \,.
\end{eqnarray*}

Hence
\begin{eqnarray*}
  \con'_i &\to& \con'_i + [Q,\lambda\of{i}]
  \\
  \g_{ij} &\to& \g_{ij} + \lambda\of{i} - \lambda\of{j}
  + [Q,\lambda\of{ij}]
  \\
  \f_{ijk}
  &\to&
  \f_{ijk}
  +
  \lambda_{ik}
  -
  \lambda_{ij}
  -
  \lambda_{jk}
  +
  [Q,\lambda_{ijk}]
  \,,
\end{eqnarray*}
which in components yields
\begin{eqnarray*}
  A_i &\to& A_i -\extd_{A_i}\ln h_i + dt\of{\alpha_i}
  \\
  B_i &\to& B_i  + d\alpha\of{\ln h_i}\of{B_i} + \extd_{A_i} \alpha_i
  \\
  \ln g_{ij}
  &\to&
  \ln g_{ij}
  +
  \ln h_i 
  -
  \ln h_j
  +
  dt\of{p_{ij}}
  \\
  a_{ij}
  &\to&
  a_{ij} + \alpha_i - \alpha_j - \extd_{A} \ln p_{ij} 
  \\
  \ln f_{ijk}
  &\to&
  \ln f_{ijk}
  +
  \ln p_{ik}
  -
  \ln p_{ij}
  - 
  \ln p_{jk}
  \,.
\end{eqnarray*}

These can be checked to be the infinitesimal version of the
respective finite gauge transformation in a 2-bundle/gerbe.

In conclusion this demonstrates that nonabelian Deligne
cohomology correctly captures the infinitesimal version of the
cocycle relations and gauge transformations of a 
nonabelian gerbe or 2-bundle with 2-connection
(\cf Prop. \ref{central proposition concerning 2-connections},
p. \pageref{central proposition concerning 2-connections}). 
Note however that
the condition of vanishing fake curvature 
(\cf p. \pageref{fake curvature of 2-bundle}) has not been obtained
in the context of nonabelian gerbes. As we have discussed, this
condition is related to the existence of a notion of 2-holonomy, which
has so far not been discussed for nonabelian gerbes.

\newpage
\subsection{Semistrict Infinitesimal $\g_k$-2-Bundles with 2-Connection}
\label{Semistrict Infinitesimal gk-2-Bundles with 2-Connection}

After having convinced ourselfs 
in 
\S\fullref{Infinitesimal 1-Bundles with 1-Connection}
and
\S\fullref{infinitesimal strict 2-bundles}
that known results 
about strict 1- and 2-bundles with connection are reproduced by
nonabelian Deligne hypercohomology, let us now apply the formalism
to something new.

The differential formalism can handle 
all semistrict target Lie
$p$-algebras
(as well as semistrict Lie $p$-algebroids), 
no matter if these are integrable to Lie $p$-groups
or not. This allows for instance to consider infinitesimal
2-bundles with target algebra coming from the
family $\g_k$ of non-strict Lie 2-algebras that were introduced
in \cite{BaezCrans:2003}.
The 2-algebras $\g_k$ are non-strict only very slightly
and still give rise to very rich structures, as
shown in \cite{BaezCransSchreiberStevenson:2005}. 

\subsubsection{The local 2-Connection Morphism}

Let $\g$ be any semisimple Lie algebra and let $k\in \R$ be
any real number. Then the dg-algebra 
$(\extd^\g, \bigwedge^\bullet V^*)$
dual to 
the semistrict Lie 2-algebra $\g_k$ is given by
$V^* = V^*_1 \oplus V^*_2$ with
\begin{eqnarray*}
  V^*_1 &=& \g^* 
    =
  \langle \set{\A^a}_{a = 1,\dots \mathrm{dim}\of{\g}}\rangle
  \\
  V^*_2 
    &=& 
  (\mathrm{Lie}\of{\R})^* = \langle \set{\B} \rangle
  \,,
\end{eqnarray*}
where we have chosen a basis $\set{\A^a}_a$ 
for $V^*_1$ and a basis $\set{\B}$ for $V^*_2$.
In this basis the action of $\extd^\g$ is given by
\begin{eqnarray*}
  \extd^\g \A^a
    &\defas& 
   -\frac{1}{2}C^a{}_{bc}\A^b \A^c
  \\
  \extd^\g \B 
    &\defas& 
  - \mu_{abc} \A^a \A^b \A^c
  \,.
\end{eqnarray*}
As before in example \ref{example: strict Lie alg as dg-alg}
the tensor $C$ comes from the structure constants of $\g$.
The tensor $\mu$ on the other hand is defined by 
\[
  \mu_{abc} \defas 
  k
  \langle
    t_a,[t_b,t_c]
  \rangle
  \,,
\]
where $\set{t_a}$ is the basis dual to $\set{\A^a}$.
When we use the Killing form $\langle \cdot,\cdot\rangle$
to lower indices we can equivalently write
\begin{eqnarray*}
  \extd^\g \A^a 
    &\defas& 
   -\frac{1}{2}C^a{}_{bc}\A^b \A^c
  \\
  \extd^\g \B 
    &\defas& 
  - \frac{k}{2} \, C_{abc} \A^a \A^b \A^c
  \,.
\end{eqnarray*}
Since $C^a{}_{b[c}C^b{}_{de]} = 0$ 
(the Jacobi identity in $\g$)
this makes the nilpotency of $\extd^\g$ manifest.

We can now closely follow the discussion of strict
infinitesimal 
2-bundles in \S\fullref{infinitesimal strict 2-bundles}.

On $U_i$ the 2-connection is determined by
\begin{eqnarray*}
  \con_i\of{\A^a} &=& A_i^a \in \Gamma\of{T^* U_i}
  \\
  \con_i\of{\B} &=& B_i \in \Gamma\of{\bigwedge^2 T^* U_i}
  \,.
\end{eqnarray*}
The condition for $\con_i$ to be $Q$-closed says that
$A_i$ must be \emph{flat}:
\begin{eqnarray*}
  0 
   &=&
  \extd \con_i\of{\A^a} - \con_i\of{\extd^\g \A^a}
  \\
  &=&
  \extd A_i^a + \frac{1}{2}C^a{}_{bc} A_i^b \wedge A_i^c
  \\
  &=&
  F_{A_i}
  \,.
\end{eqnarray*}
This is the vanishing of the fake curvature known from
\S\fullref{The local 2-Connection Morphism}, only that here $dt = 0$,
so that the fake curvature is the same as the ordinary curvature of
$A_i$.

The 2-curvature of $\con_i$ (\cf \S\fullref{n-curvature})
is
\begin{eqnarray*}
  F^{(2)}|_{U_i}
  &=&
  [Q,\widehat{\con_i}]|_{V^*_2}
  \\
  &=&
  \extd \widehat{\con_i}\of{\B}
  -
  \widehat{\con_i}\of{\extd^\g \B}
  \\
  &=&
  \extd B_i
  +
  k\, \langle A_i, [A_i,A_i]\rangle
  \,.
\end{eqnarray*}

Note that $\langle A_i, [A_i,A_i]\rangle$ is proportional
to the Chern-Simons form of the flat connection $A_i$.
Furthermore note that the 2-Bianchi-identity 
\refdef{p-Bianchi identity} is
\begin{eqnarray*}
  0 &=& \extd F_i^{(2)}
  \propto
  \langle \extd A_i, [A_i,A_i]\rangle
  \,,
\end{eqnarray*}
This expression does indeed vanish since $\extd A_i = - \frac{1}{2}[A_i,A_i]$
for flat $A_i$, so that
\[
  \extd F_i^{(2)} \propto
  \langle [A_i,A_i], [A_i,A_i]\rangle  
  = 0
\]
due to the Bianchi identity.\footnote{
This should be closely related to the fact 
(indicated at the end of \cite{BaezCransSchreiberStevenson:2005})
that 2-bundles with structure algebra the strict 
2-algebra $\P_k \g$ (which is \emph{equivalent} to $\g_k$) 
can be obtained from
an ordinary  $G$-bundle only if the first Pontryagin class
$p_1/2 = \frac{1}{8\pi^2}\langle F_A, F_A\rangle$
vanishes.
}

\subsubsection{Cocycle and Gauge Transformation Relations}

Now let us study the cocycle conditions and gauge transformation
laws of infinitesimal $\g_k$-2-bundles. Recall that, as in 
the discussion of strict 2-bundles 
\S\fullref{infinitesimal strict 2-bundles},
we have
\begin{eqnarray*}
  && A_i = A|_{U_i} + \epsilon A'_i
  \\
  && B_i = B|_{U_i} + \epsilon B'_i
\end{eqnarray*}
with $A$ and $B$ globally defined. 

With a 1-transformation as in 
\S\fullref{differential pic: gauge trafos in 1-bundles}
\[
  \con_i \stackto{\g_{ij}} \con_j
\]
we have on $U_{ij}$ the relations
\begin{eqnarray*}
  A_i &=& A_j - \extd \ln g_{ij} - [A_j,(\ln g_{ij})]
  \\
  B_i &=& B_j + \extd a_{ij} + 3k \, \langle \ln g_{ij}, [A_i,A_i] \rangle
  \,.
\end{eqnarray*}

(Recall that $A$ is the globally defined 1-form around which 
the connection of the infinitesimal 2-bundle is the
``perturbation'', $A_i = A|_{U_i} + \epsilon A'_i$. See
\S\fullref{infinitesimal n-transformations}.)

On triple intersections $U_{ijk}$ a 2-transformation
\[
  \g_{ik} \stackto{\f_{ijk}} \g_{ij} + \g_{jk}
\]
as in \S\fullref{cocycle relations for inf. strict 2-bundles}
implies the laws
\begin{eqnarray*}
  &&\ln g_{ij} + \ln g_{jk} - \ln g_{ik}  = 0
  \\
  && a_{ij} + a_{jk} = a_{ik} - \extd \ln f_{ijk}
  \,.
\end{eqnarray*}
These last two are \emph{independent} of the deformation parameter $k$.
(Because $\f_{ijk}$, being a 2-transformation, 
only acts nontrivially on $\B$, while $\extd^\g$
doesn't produce any copies of $\B$.)
So the transformation laws for $A_i$ and for $\ln g_{ij}$ are
exactly those of an ordinary $G$-bundle without any 
twists introduced by $k$. 

Similarly following the previous discussion in 
\S\fullref{hypercohomology description of strict inf. 2-bundles}
one finds that gauge transformations are described by the following
equations:
\begin{eqnarray*}
  A_i &\to& A_i -\extd_{A_i} \ln h_i
  \\
  \ln g_{ij} &\to& \ln g_{ij} + \ln h_i - \ln h_j
\end{eqnarray*}
and
\begin{eqnarray*}
  B_i &\to& B_i  + \extd \alpha_i + 3k\, \langle \ln h_i, [A,A] \rangle 
  \\
  a_{ij}
  &\to&
  a_{ij} + \alpha_i - \alpha_j - \extd \ln p_{ij}
  \\
  \ln f_{ijk} 
    &\to&
  \ln f_{ijk}
  +
  \ln p_{ik}
  -
  \ln p_{ij}
  -
  \ln p_{jk}
  \,.
\end{eqnarray*}

\subsubsection{Generalized Deligne Cohomology Classes}

One can now in principle work out the Deligne cohomology classes
for $\g_k$-2-bundles. These are classes of those sets 
$\set{A_i,B_i,\ln g_{ij},a_{ij} \ln f_{ijk}}$ that satisfy the above
cocycle conditions, divided out by  the above gauge transformations. 

We will not give a full discussion of this issue
here, but make the following comments:

The conditions on $\set{A_i,\ln g_{ij}}$ are those of a
principal $\exp\of{G}$-bundle with a flat connection. Any such
bundle has to be trivial and hence it is always possible to
go to a gauge (choose a local trivialization) in which $\ln g_{ij} =0$
identically. 
In this gauge then the cocycle condition for $B_i$ becomes
\[
  B_i = B_j + \extd a_{ij} 
\]
as in a strict abelian 2-bundle/gerbe.

Once this gauge has been fixed the gauge parameter $\ln h_i$ is
restricted to satisfy $\ln h_i = \ln h_j$ on double overlaps 
$U_{ij}$. Hence it must be a globally defined function
\[
  \ln h \maps M \to \g
  \,.
\]
This again implies that the gauge transformation law for $B_i$
becomes that of an abelian gerbe up to a global shift:
\[
  B_i 
   \to
  B_i  + \extd \alpha_i + 3k\, \langle \ln h, [A,A] \rangle 
\]

The other components, $a_{ij}$ and $f_{ijk}$, satisfy the 
unmodified cocycle
relations and gauge transformation laws of an abelian gerbe
without any appearance of $k$.

So it seems one should look at classes of pairs
\[
  (A,\mathcal{G} = (f_{ijk},a_{ij},B_i))
\]
with 
$
A \in \Omega^1\of{M,\g}$ a 1-form on $\g$ satisfying 
$\extd A + \frac{1}{2}[A,A] = 0$ and $\mathcal{G}$ an abelian
gerbe on $M$. 
It seems like we need to identify two such pairs 
$  (A,\mathcal{G} = (f_{ijk},a_{ij},B_i))
$
and
  $(\tilde A,\tilde {\mathcal{G}} = (\tilde f_{ijk},\tilde a_{ij},
\tilde B_i))$
iff ${\mathcal{G}}$ and $\tilde{\mathcal{G}}$ have representatives
such that
\begin{eqnarray*}
  f_{ijk} &=& \tilde f_{ijk}
  \\
  a_{ij} &=& \tilde a_{ij}
\end{eqnarray*}
and
\[
  \extd B + k \langle A,[A,A]\rangle
  =
  \extd \tilde B + k \langle \tilde A,[\tilde A,\tilde A]\rangle
  \,.
\]
While this needs to be better understood, this discussions shows that
in principle the ``infinitesimal formalism'' including 
nonabelian Deligne hypercohomology  allows to
address issues that cannot be addressed at all in the 
``integral formalism''.

\newpage
\subsection{Strict Infinitesimal 3-Bundles with 3-Connection}
\label{Strict Infinitesimal 3-Bundles with 3-Connection}

Finally, we can check that the differential formalism 
correctly reproduces what
is known about 3-bundles with 3-connection.

Let the target dg-algebra be coming from a
strict Lie 3-algebra as in example 
\ref{example: strict Lie alg as dg-alg} and 
define a local 3-connection by the following maps:
\begin{eqnarray*}
  \con\of{\A^a} &=& A^a
  \\
  \con\of{{\mathbf b}^A} &=& B^A
  \\
  \con\of{\C^\alpha}  &=& C^\alpha
\end{eqnarray*}

Recall that we required the connection to be a homogenizing
morphism of dg-algebras and that $V_n^* = 0$ for $n>3$ in $\p_3\of{M}$.

This map is a chain map only if $\commutator{Q}{\con_3} = 0$, which
implies 
\begin{eqnarray*}
  0 
   &=&
   \extd \con\of{\A^a}
   -
   \con\of{\extd^\g \A^a}
  \\
  &=&
  \extd A^a
  +
  \frac{1}{2}C^a{}_{bc}A^b\wedge A^c + dt^a_A B^A
  \\
  &=&
  ( F_A + dt\of{B})^a
  \,.
\end{eqnarray*}
and
\begin{eqnarray*}
  0 
   &=&
   \extd \con\of{{\mathbf b}^A}
   -
   \con\of{\extd^\g {\mathbf b}^A}
  \\
  &=&
  \extd B^A
  + 
  (d\alpha_1)^A_{Ba}B^A A^a
  +
  (dt_2)^A_\alpha C^\alpha
  \\
  &=&
  (\extd_{A} B + dt_2\of{C})^A
  \,.
\end{eqnarray*}
(Note again that the relation at grade three becomes trivial due to the
nature of $\p_3$.)

These are precisely the consistency conditions
known in the integral formalism from
prop \ref{prop: conscond for local 3-holonomy} 
(p. \pageref{prop: conscond for local 3-holonomy}).

\newpage
\listoffigures

\newpage
\addcontentsline{toc}{section}{\bf References}
\bibliography{std}

\end{document}